\documentclass[longauth]{aa}
\usepackage{longtable}
\usepackage{aalongtable,lscape} 
\usepackage{pdflscape}
\usepackage{graphicx}
\usepackage{lscape}
\usepackage{txfonts}
\usepackage{natbib}
\usepackage[bookmarks=false]{hyperref}
\hypersetup{colorlinks=true, linkcolor=blue, citecolor=blue, filecolor=magenta, urlcolor=red}
\usepackage[normalem]{ulem}
\usepackage{color}

\pdfoutput=1

\begin{document} 

   \title{The \textit{Gaia}-ESO Survey: Calibrating the lithium--age relation with open clusters and associations    \thanks{Based on observations collected with ESO telescopes at the La Silla Paranal Observatory in Chile, for the \textit{Gaia}-ESO Large Public Spectroscopic Survey (188.B-3002, 193.B-0936).}   \thanks{All tables in Appendix~\ref{ap1:AppendixD} are only available in electronic form at the CDS via anonymous ftp to \url{http://cdsarc.u-strasbg.fr/} (130.79.128.5) or via \url{http://cdsweb.u-strasbg.fr/cgi-bin/qcat?J/A+A/}.} } 

   \subtitle{II. Expanded cluster sample and final membership selection.}

 \titlerunning{Calibrating the lithium--age relation II.}
 
   \author{M.L. Guti\'errez Albarr\'an
          \inst{1},
          D. Montes
           \inst{1},
           H.M. Tabernero
           \inst{1, 2},
           J.I. Gonz\'alez Hern\'andez
           \inst{3, 4},
           E. Marfil
           \inst{1, 5},
           A. Frasca 
           \inst{6},
           A. C. Lanzafame 
           \inst{7, 6},
           A. Klutsch
           \inst{8, 6},
           E. Franciosini
           \inst{9},
           S. Randich
            \inst{9},
           R. Smiljanic
             \inst{10},
           A.~J. Korn 
           \inst{11},
           G. Gilmore
           \inst{12},
           E. J. Alfaro
            \inst{13},
           T. Bensby 
             \inst{16},
           K. Biazzo
            \inst{17},
            A. Casey
            \inst{20},
           G. Carraro
           \inst{23},
           F. Damiani
           \inst{15},
           S. Feltzing
           \inst{16},
            P. François
           \inst{22},
           F. Jim\'enez Esteban
            \inst{14},
           L. Magrini
           \inst{9},
           L. Morbidelli
           \inst{9},
           L. Prisinzano
           \inst{15},
           T. Prusti
           \inst{21},
           C. C. Worley
           \inst{19},
           S. Zaggia
           \inst{18}
             \and
           GES builders
          }
   \institute {Departamento de F\'{i}sica de la Tierra y Astrof\'{i}sica and IPARCOS-UCM (Instituto de F\'{i}sica de Part\'{i}culas y del Cosmos de la UCM), Facultad de Ciencias F\'{i}sicas, Universidad Complutense de Madrid, E-28040, Madrid, Spain \email{mlgutierrez@ucm.es} \and
             Centro de Astrobiolog\'ia (CSIC/INTA), Instituto Nacional de T\'ecnica Aeroespacial, Ctra de Torrej\'on a Ajalvir, km 4, 28850 Torrej\'on de Ardoz, Madrid, Spain \and
              Instituto de Astrof\'isica de Canarias (IAC), E-38205 La Laguna, Tenerife, Spain  \and
              Universidad de la Laguna, Dept. Astrof\'isica, E-38206 La Laguna, Tenerife, Spain \and
               Hamburger Sternwarte, Gojenbergsweg 112, 21029 Hamburg, Germany \and
              INAF - Osservatorio Astrofisico di Catania, via S. Sofia, 78, 95123 Catania, Italy \and
              Universit\`a di Catania, Dipartimento di Fisica e Astronomia, Sezione Astrofisica, Via S. Sofia 78, I-95123 Catania, Italy \and
               Institut f\"ur Astronomie und Astrophysik, Eberhard Karls Universit\"at, Sand 1, D-72076 T\"ubingen, Germany  \and
              INAF - Osservatorio Astrofisico di Arcetri, Largo E. Fermi 5, 50125, Firenze, Italy \and 
              Nicolaus Copernicus Astronomical Center, Polish Academy of Sciences, ul. Bartycka 18, 00-716, Warsaw, Poland \and 
              Observational Astrophysics, Division of Astronomy and Space Physics, Department of Physics and Astronomy, Uppsala University, Box 516, SE-751 20 Uppsala, Sweden \and
               Institute of Astronomy, University of Cambridge, Madingley Road, Cambridge CB3 0HA, United Kingdom \and
               Instituto de Astrof\'isica de Andaluc\'ia (CSIC), Glorieta de la Astronom\'ia s/n, Granada 18008, Spain  \and
               Centro de Astrobiolog\'{\i}a, CSIC-INTA, Camino bajo del castillo s/n, E-28692, Villanueva de la Can\~ada, Madrid, Spain \and
               INAF - Osservatorio Astronomico di Palermo, Piazza del Parlamento, 1, 90134 Palermo, Italy  \and
               Lund Observatory, Division of Astrophysics, Department of Physics, Lund University, Box 43, SE-22100 Lund, Sweden  \and
               INAF - Rome Astronomical Observatory (OAR), Via di Frascati, 33, I-00044, Monte Porzio Catone, Italy \and
               INAF - Osservatorio Astronomico di Padova, Vicolo dell'Osservatorio 5, 35122 Padova, Italy \and
               School of Physical and Chemical Sciences --- Te Kura Mat\={u}, University of Canterbury, Private Bag 4800, Christchurch 8140, New Zealand \and
               114, 10 College Walk, School of Physics and Astronomy, Monash University VIC 3800, Australia  \and
               European Space Agency (ESA), European Space Research and Technology Centre (ESTEC), Keplerlaan 1, 2201 AZ Noordwijk, The Netherlands  \and
                GEPI, Observatoire de Paris, CNRS, Universit\'e Paris Diderot, 5 Place Jules Janssen, 92190 Meudon, France \and
              Dipartimento di Fisica e Astronomia, Universit\`a di Padova, Vicolo dell'Osservatorio 3, 35122 Padova, Italy 
              }
             
\authorrunning{Guti\'errez Albarr\'an et al.}

   \date{Received October 31, 2023; accepted January 27, 2024.}
 
  \abstract
   {The Li abundance observed in pre-main sequence and main sequence late-type stars is strongly age-dependent, but also shows a complex pattern depending on several parameters, such as rotation, chromospheric activity, and metallicity. The best way to calibrate these effects, and with the aim of studying Li as an age indicator for FGK stars, is to calibrate coeval groups of stars, such as open clusters (OCs) and associations.}
       {We present a considerable target sample of $42$ OCs and associations ---with an age range from $1$~Myr to $5$~Gyr--- observed within the \textit{Gaia}-ESO survey (GES), and using the latest data provided by GES iDR6 and the most recent release of \textit{Gaia} that was then available, EDR3. As part of this study, we update and improve the membership analysis for all $20$ OCs presented in our previous article.}
   {We perform detailed membership analyses for all target clusters to identify likely candidates, using all available parameters provided by GES, complemented with detailed bibliographical searches, and based on numerous criteria: from radial velocity distributions, to the astrometry (proper motions and parallaxes) and photometry provided by \textit{Gaia}, to gravity indicators ($\log g$ and the $\gamma$ index), [Fe/H] metallicity, and Li content in diagrams of $EW$(Li) (Li equivalent widths) versus $T_{\rm eff}$.}
   {We obtain updated lists of cluster members for the whole target sample, as well as a selection of Li-rich giant contaminants obtained as an additional result of the membership process. Each selection of cluster candidates was thoroughly contrasted with numerous existing membership studies using data from \textit{Gaia} to ensure the most robust results.}
   {These final cluster selections will be used in the third and last paper of this series, which reports the results of a comparative study characterising the observable Li dispersion in each cluster and analysing its dependence on several parameters, allowing us to calibrate a Li--age relation and obtain a series of empirical Li envelopes for key ages in our sample.}

   \keywords{Galaxy: open clusters and associations: general -- Stars: late-type -- Stars: abundances -- Techniques: spectroscopic }

\maketitle

\section{Introduction} \label{intro}

Lithium in the form of its main isotope $^7$Li (henceforth Li) is mainly observable through the spectral resonance lines of neutral Li, forming a doublet at $6707.76$~$\r{A}$ and  $6707.89$~$\r{A}$ \cite[e.g.,][]{pallavicini1992, soderblom3, lyubimkov, randich&magrini}. Li is a very fragile element that is easily destroyed in stellar interiors, burning at temperatures above $\sim$~$2.5$--$3$~x~$10^6$~K, which corresponds to the temperature at the base of the convective zone of a solar-mass star on the zero-age main sequence \citep[ZAMS; e.g.,][]{pallavicini1990, siess, magrini2021}. As a result, Li is slowly being depleted and its surface abundance decreases over time in solar-type and lower-mass FGK stars. The depletion of Li in these stars is strongly mass (colour) and age dependent: Li depletion increases with decreasing mass at a given age, and the timescales for significant Li depletion typically range from $10$--$20$~Myr in M-type stars to $\sim$~$100$~Myr in K-type stars, and $\sim$~$1$~Gyr in G-type stars \citep[e.g.,][]{jeffries, soderblom2013, lyubimkov, randich&magrini}. For their part, Li-rich giants show an additional contribution of surface Li that can only be accounted for with non-standard mixing mechanisms \citep[e.g.,][]{casey, lyubimkov, magrini2021}. As it only survives in the outer layers of a star, Li is a very sensitive indicator of youth in late-type stars, a useful tracer of stellar evolution for pre-main sequence (PMS), main sequence (MS), and post-MS stars, and is also particularly relevant for the determination of the age of stellar open clusters (OCs) and associations \citep[e.g.,][]{sestitorandich, bouvier2018, dumont2, dumont, randich&magrini}. For a more comprehensive overview of Li as an age indicator and the mechanics of Li depletion, we refer the reader to the introductory chapter of the published thesis of \citet{tesis_yo} and all references therein.

The present work is the second of a series of three papers where we study Li as an age indicator for PMS and MS FGKM late-type stars, with the ultimate aim being to calibrate an empirical Li--age relation. In this paper, we describe an expanded sample of $42$ OCs and associations covering an age range from $1$~Myr to $5$~Gyr, using data from the \textit{Gaia}-ESO Survey (GES) and from the \textit{Gaia} mission (both briefly discussed below). For all target clusters, we here constrain the cluster membership to obtain the final lists of candidate members for all target clusters, updating and expanding the work already published in \cite{yo} (hereafter, Paper I). In \cite{yo_3} (hereafter,  Paper III), presented as a companion paper to the present work, we finally reach our objective of constraining an empirical Li--age relation by first presenting a comparative study to quantify the observable Li dispersion for all target clusters and to analyse its dependence on several GES parameters, finally obtaining a set of empirical Li envelopes for several key ages in our sample. The analyses and results of this work are additionally presented in full in the published thesis of \cite{tesis_yo}.

 The \textit{Gaia}-ESO Survey \citep[GES --][]{gilmore, randich&gilmore, gilmoregaia2021, randichgaia2021, hourihane}\footnote{\url{https://www.gaia-eso.eu/}} is a large public spectroscopic survey that systematically covers all major components of the Milky Way and provides a homogeneous and ambitious overview of the distribution of the kinematics, dynamical structure, and chemical compositions in the Galaxy. The GES uses the multi-object spectrograph FLAMES (\textit{Fibre Large Array Multi Element Spectrograph}) at the Very Large Telescope (VLT at ESO, Chile) to obtain high-resolution spectra (R=$50,000$) with UVES (Ultraviolet and Visual Echelle Spectrograph), as well as medium-resolution spectra (R=$5,000$--$20,000$) with GIRAFFE \citep[e.g.,][]{smiljanic1, frasca, lanzafame, gilmoregaia2021}. The observations, which began in December 2011 and were completed in January 2018, provide high-quality, uniformly calibrated spectroscopy for about $10^{5}$ stars and a sample of $65$ (plus $18$--$20$ archive) OCs and star forming regions of all ages, metallicities, and stellar masses \citep{bragaglia2021, gilmoregaia2021, randichgaia2021}. In addition to measured spectra, this exploration uses well-defined samples from photometric surveys, such as VISTA \citep{sutherland}, 2MASS \citep{skrutskie} and a variety of photometric surveys of OCs that cover all the major components of the Galaxy \citep{gilmore, bragaglia2021}.

\textit{Gaia} \citep{gaia}\footnote{\url{https://sci.esa.int/web/gaia}} is an ongoing, ambitious global space-astrometry ESA mission that launched in 2013 and started scientific operations in mid-2014. Initially a five-year mission, in 2018 the \textit{Gaia} mission was extended to 2020, and currently \textit{Gaia} has been firmly extended until the end of 2022, with indicative approval until the end of 2025 \citep{gaiadr3}. Its primary science goal is to chart a three-dimensional map of the Galaxy, providing photometry and astrometry of unprecedented precision for most stars brighter than G=$20$ mag, and to obtain low- and medium-resolution spectroscopy data for most stars brighter than G=$17$ mag \citep{gilmore, gaia}. \textit{Gaia} provides a large stellar census that allows an extensive overview of the origin, composition, formation, and evolution of the Galaxy, producing a stereoscopic dataset that includes positions, proper motions, parallaxes, radial velocities, brightness, and astrophysical parameters of about one billion stars and other astronomical objects in the Milky Way and throughout the Local Group.

\textit{Gaia}, with its revolutionary high-precision astrometry and photometry, can be optimally complemented with GES and the superior spectroscopic capabilities of its large ground-based telescope to obtain high-quality data, providing a rich dataset yielding 3D spatial distributions, 3D kinematics, individual chemical abundances, and improved astrophysical parameters for all target objects \citep[e.g.,][]{gilmore, beccari, cantatgaudin_gaia, randich_gaia, soubiran, canovas, bossini}. We finish this introductory section by further acknowledging that, in our analysis, we also made use of several sets of rotational periods ($P_{rot}$) provided by missions such as CoRoT, \textit{Kepler}, K2, and TESS. 

  This paper is organised as follows. In Sect.~\ref{data} we describe the expanded cluster sample upon which we base our analysis. In Sect.~\ref{analysis} we discuss the expanded and improved membership cluster analysis, describing our selection criteria, which take into account radial velocities ($RV$s), \textit{Gaia} astrometry and photometry, gravity indicators, metallicity, and Li content, as well as various comparisons with previous studies from the literature using \textit{Gaia} data (Subsect.~\ref{gaia}). In Subsect.~\ref{outliers} we additionally describe the selection process for Li-rich contaminants obtained as an additional result of the membership analyses. We present the final lists of candidate members for the results of this project in Sect.~\ref{summary}.

\section{Data} \label{data}

\subsection{\textbf{GES data}} \label{spectra}

The GES consortium is structured in 19 working groups (WGs), WG0 to WG18, organised in a workflow and dedicated to different survey tasks, from data flow implementation and target selection, to the homogenisation of recommended parameters and the determination of abundances, preparation and documentation, and, finally, the delivery of the external data products to both ESO and the public archive \citep[e.g.,][]{gilmore, lanzafame, sacco, gilmoregaia2021}. Among these WGs, WG10 and WG11 are focused on the spectroscopic analysis of the GIRAFFE and UVES FGKM late-type stars, respectively \citep[e.g.,][]{gilmore, sacco}, while WG12 is dedicated to the analysis of PMS stars in the fields of young clusters using both UVES and GIRAFFE data. Each WG is divided into several nodes, providing internally consistent results with various approaches, models and methodology\footnote{Our UCM node is part of both WG11 and WG12, and we determined atmospheric parameters by employing the code \textsc{\texttt stepar} \citep{tabernero}, based on equivalent widths ($EW$s) automatizing the use of \textsc{\texttt moog} code \citep{sneden} and measured with the automatic tool \textsc{\texttt tame} \citep[Tool for Automatic Measurement of Equivalent Widths --][]{kanglee, tabernero}. Also see Paper I.}\citep{smiljanic1, lanzafame, gilmoregaia2021}. For more specific and extensive details on the methodology employed by each node, see \cite{smiljanic1} (WG11), \cite{lanzafame} (WG12), and \cite{gilmoregaia2021} (all WGs). In addition, for an in-depth overview of the GES consortium and the data reduction process for GES spectra, we further refer the reader to Paper I, and also to chapter~2 of the thesis and all references therein. 

Six analysis cycles and internal data releases have been carried out by GES (from iDR1 to iDR6), including homogenised recommended astrophysical parameters and elemental abundances derived from all the observations collected until the completion of the survey in January 2018 \citep{randichgaia2021}. Subsequent public releases to the ESO archive include DR2 (published in July 2015), DR3 (December 2016 and May 2017), DR4 (December 2020), and, finally, DR6 was released to the public in July 2023\footnote{DR6 data products available from the ESO Science Archive Portal at \url{https://archive.eso.org/scienceportal/home?data_collection=GAIAESO&publ_date=2023-07-02}}. iDR4, which was made available to the consortium in February 2016, was the data release we used in paper I for our preliminary cluster membership analysis. For all the following analysis presented in this paper we used the data provided by the sixth and last internal data release of GES (iDR6)\footnote{iDR6 data products released to the GES consortium at \url{http://ges.ast.cam.ac.uk/GESwiki/GeSDR6/DataProducts}}. iDR6 is a full release, including all observations from the beginning of the Survey until its completion in January 2018 \citep{gilmoregaia2021, franciosini2022, randichgaia2021}. Similarly to iDR5, the same line list and grid of synthetic spectra used for iDR4 were employed in this final release. This final GES catalogue provides results for $114,325$ targets and includes data on velocities, stellar parameters, and abundances for up to $32$ elements \citep{randichgaia2021}.  

 The \textit{Gaia} mission has also provided three data releases so far, from \textit{Gaia} DR1 (released in September 2016) and \textit{Gaia} DR2 (April 2018) to \textit{Gaia} DR3, which was released in June 2022, following the \textit{Gaia} Early Data Release 3 (EDR3)\footnote{See \url{https://www.cosmos.esa.int/web/gaia/earlydr3} and \url{https://www.cosmos.esa.int/web/gaia/dr3} to access \textit{Gaia} EDR3 and DR3, respectively} \citep{gaiadr1, gaiadr2, gaiadr3, gaiadr3_final}. Pending further approvals of mission extensions, additional releases will also take place, with \textit{Gaia} DR5 being currently anticipated to contain all collected data. In combination with GES iDR6, for the present project we made use of the proper motions, parallaxes and photometry provided by the latest data release then available, namely, the first installment of the third intermediate data release, \textit{Gaia} EDR3, which is based on the data of the first $34$ months of the mission. \textit{Gaia} EDR3 provides, with increased precision and accuracy, an updated source list, astrometry, and broad photometry in the \textit{G}, $G_{BP}$ and $G_{RP}$ bands \citep{gaiadr3}. 

\subsection{\textbf{Cluster sample}} \label{sample}

\begin{table*} [htp]
 \centering
 \caption{Age estimates, reddening, distance to the Sun, and GES and \text{Gaia} membership studies from the literature for the 16 star forming regions and young clusters in our sample.}
\label{table:01}  
\resizebox{\linewidth}{!}{%
\begin{tabular}{c c c c c c c}
 \noalign{\smallskip}
 \hline
 \hline
 \noalign{\smallskip}
  Cluster & Age & E(B-V)$^a$ & Distance & References & References & GES and \text{Gaia} membership  \\ 
     name  & (Myr) & (dex) & (kpc) & Ages & Distance & studies    \\
\noalign{\smallskip}
\hline
\noalign{\smallskip}
  NGC~6530 & $1$--$2$ & $0.44\pm0.10$ & $1.33$ & 3, 4, 18, 19, 20, 33 & 18, 19, 20, 21 & 2, 3, 16, 18, 19, 22 \\
  $\rho$~Oph &  $1$--$3$  & $0.76\pm0.13$ & $0.13\pm0.01$ &  1, 2, 3, 4, 5, 33, 61 & 1, 2, 6 & 1, 2, 3, 6  \\ 
  Trumpler~14 &  $1$--$3$  & $0.61\pm0.10$ & $2.90$ & 3, 4, 13, 33 & 13, 14 & 2, 3, 15, 17  \\
  Cha~I & $2$  & $0.18\pm0.08$ & $0.16\pm0.02$ & 3, 4, 7, 8, 9, 33 & 7, 8, 10, 11 & 2, 3, 7, 8, 10, 11$^b$, 12 \\
  NGC~2244 &  $4$  & $0.49\pm0.09$ & $1.59$ &  3, 23, 24, 33  & 21, 23, 24, 25 &  3, 15, 24, 26 \\
  NGC~2264 &  $4$  & $0.05\pm0.05$ & $0.76$ & 2, 3, 4, 27, 28, 29, 30, 33 & 2, 15, 25, 28, 29 & 2, 3, 15, 31, 32 \\
 $\lambda$~Ori &  $6$  & $0.09\pm0.04$ & $0.41$ & 3, 43 & 34 &  3, 15 \\
 Col~197 &  $13$  & $0.64\pm0.07$ & $0.80$--$0.90$ &  3, 5, 13, 35, 36, 37  & 13, 25, 35, 36, 37 & 3, 15 \\
 $\gamma$~Vel & $10$--$20$  & $0.04\pm0.03$ & $0.35$--$0.40$ & 2, 3, 8, 10, 38, 39, 40, 41, 42  & 8, 10, 38, 40, 41, 42 & 2, 3, 8, 10, 38, 39, 40, 41, 42, 44, 45, 46  \\
  NGC~2232 &  $18$--$32$  & $0.04\pm0.03$ & $0.32$ & 
3, 5, 43, 47, 48, 64   & 13, 25, 48 & 3, 15, 43, 48 \\
  NGC~2547 &  $20$--$45$  &  $0.06\pm0.03$ & $0.36\pm0.02$ & 2, 3, 5, 8, 32, 41, 42, 48, 49, 50, 51, 63, 64  & 2, 8, 41, 42, 48 & 
2, 3, 8, 15, 32, 42, 48, 49, 52 \\
 IC~2391 & $36\pm2$ $^c$  &  $0.03\pm0.01$ & $0.16\pm0.01$ & 3, 48, 51, 53, 64 & 2, 48, 54, 55, 56  &  2, 3, 15, 49, 52, 62  \\
 IC~2602 & $35\pm1$ $^c$  &  $0.04\pm0.02$ & $0.15\pm0.01$ & 3, 5, 48, 51, 64 & 2, 48, 56 & 2, 3, 15, 49, 52, 62 \\
 IC~4665 &  $38\pm3$ $^c$  & $0.15\pm0.02$  & $0.36\pm0.01$ &  3, 5, 48, 51, 64 & 2, 48, 57, 58 & 2, 3, 15, 48, 49, 52, 62 \\
 NGC~2451~B &  $39\pm1$ $^c$ & $0.10\pm0.03$ & $0.36$ & 3, 42, 48, 49, 51, 59, 60  & 42, 48, 59  & 2, 3, 15, 42, 48, 49 \\
 NGC~2451~A &  $44\pm2$ $^c$ & $0.02\pm0.02$ & $0.19$ & 3, 42, 48, 49, 51, 59, 60  & 42, 48, 59  & 2, 3, 15, 42, 48, 49 \\
 \noalign{\smallskip}
 \hline
\end{tabular} }
\tablefoot{
\tablefoottext{a}{References for the E(B-V) reddening adopted from \cite{jackson2021} for all clusters.}
\tablefoottext{b}{GES studies that reference the clusters and/or study them without taking membership analysis primarily into account.}
\tablefoottext{c}{Updated cluster ages using \textit{Gaia} data, as listed by \citet{bossini}.} \\
\textbf{References}. For the cluster ages, reddening and distances shown here we chose the most recent or most robust estimates for each cluster, but several other studies are further cited here, and age estimates are additionally briefly discussed in the individual notes of Appendix~\ref{ap1:AppendixB}: (1) \cite{rigliaco}; (2) \cite{spina_giants}; (3) \cite{jackson2021}; (4)  \cite{randich2020}; (5) \cite{romano}; (6) \cite{canovas}; (7) \cite{spina_cha}; (8) \cite{sacco};  (9) \cite{lopez};  (10) \cite{frasca}; (11) \cite{roccatagliata}; (12) \cite{galli}; (13) \cite{sampedro}; (14) \cite{melnik}; (15) \cite{cantatgaudin_gaia}; (16) \cite{castro_ginard}; (17) \cite{damiani2017}; (18) \cite{wright2019}; (19) \cite{damiani2019}; (20) \cite{prisinzano2005}; (21) \cite{kuhn2019}; (22) \cite{prisinzano2019}; (23) \cite{muzic}; (24) \cite{michalska}; (25) \cite{kharchenko}; (26) \cite{carrera2019};  (27) \cite{arancibia_silva}; (28) \cite{bonito2020}; (29) \cite{gillen}; (30) \cite{venuti}; (31) \cite{apellaniz}; (32) \cite{jackson}; (33) \cite{randichgaia2021}; (34) \cite{dib}; (35) \cite{dias2019}; (36) \cite{bonattobicca}; (37) \cite{vandeputte}; (38) \cite{jeffries}; (39) \cite{spina_gamma}; (40) \cite{franciosini}; (41) \cite{beccari}; (42) \cite{franciosini2021}; (43) \cite{binks2021}; (44) \cite{damiani}; (45) \cite{prisinzano}; (46) \cite{cantat2019}; (47) \cite{liu}; (48) \cite{pang}; (49) \cite{randich_gaia}; (50) \cite{sestitorandich}; (51) \cite{bossini}; (52) \cite{bravi}; (53) \cite{dumont}; (54) \cite{platais}; (55) \cite{de_silva}; (56) \cite{smiljanic2011}; (57) \cite{martinmontes}; (58) \cite{jeffries2009}; (59) \cite{silaj}; (60) \cite{netopil}; (61) \cite{kiman}; (62) \cite{gomezgarrido_tfm}; (63) \cite{oliveira}, (64) \cite{Jeffries2023_IC4665}. } 
\end{table*}

  \begin{table*} [htp]
 \centering
 \caption{Age estimates, reddening, distance to the Sun, and GES and \text{Gaia} membership studies from the literature for the 26 intermediate-age and old clusters in our sample.}
\label{table:02}  
  \resizebox{\linewidth}{!}{%
\begin{tabular}{c c c c c c c}
 \noalign{\smallskip}
 \hline
 \hline
 \noalign{\smallskip}
  Cluster & Age & E(B-V)$^a$ & Distance & References & References & GES and \text{Gaia} membership  \\ 
     name  & (Myr) & (dex) & (kpc) & Ages & Distance & studies    \\
\noalign{\smallskip}
\hline
\noalign{\smallskip}
 NGC~6405 &  $94$ & $0.14\pm0.04$ & $0.46$ & 1, 2, 3 & 2, 3, 4, 5 & 1, 3, 6 \\
 Blanco~1 &  $94\pm5$ $^c$  & $-0.01\pm0.03$ & $0.23$--$0.24$ & 1, 10, 15, 16, 17, 18, 53 & 17, 18 & 1, 18 \\
 NGC~6067 &  $120$  & $0.34\pm0.04$ & $1.4$--$1.7$ & 1, 7, 8, 9, 10, 11 & 4, 5 & 1, 6, 11 \\
 NGC~6649 &  $120$  & $1.43\pm0.05$ & $1.8\pm0.1$ & 12, 13, 14 & 12, 14 & 1, 6 \\
 NGC~2516 &  $125$--$138$  & $0.11\pm0.03$ & $0.41$ & 1, 9, 19, 21, 22 & 22, 23, 24 & 1, 6, 20, 22, 25, 26, 27, 28  \\
 NGC~6709 &  $173\pm34$ $^c$  & $0.27\pm0.02$ & $1.1$ & 1, 10, 15, 29  & 4, 29 & 1, 6 \\
 NGC~6259 &  $210$  & $0.63\pm0.09$ & $1.9$ & 1, 9, 30, 31 & 4, 29 &    1, 6 \\
 NGC~6705 & $280$ & $0.40\pm0.06$ & $1.88$ & 1, 9, 10, 26, 27 & 4, 24 & 1, 6, 26, 27, 32, 33, 34$^b$ \\
 Berkeley~30 &  $300$  & $0.51\pm0.04$ & $4.7$--$4.9$ & 1, 4, 10, 30  & 4, 30 & 1, 6 \\
 NGC~6281 &  $314$  & $0.18\pm0.02$ & $0.47$--$0.51$ & 1, 24, 30, 36 & 4, 30, 36 & 1, 6 \\
 NGC~3532 &  $399\pm5$ $^c$  & $0.05\pm0.02$ & $0.48$--$0.49$ & 1, 10, 15, 28, 35 & 4, 28 & 1, 6, 28, 36, 37 \\
 NGC~4815 & $560$ & $0.70\pm0.07$ & $2.40$--$2.90$ & 1, 26, 27, 31, 37 & 24, 37 & 1, 6, 26, 27, 32, 33, 34, 37  \\ 
 NGC~6633 & $575$ & $0.15\pm0.04$ & $0.39$ & 1, 26, 27, 31, 38 & 4, 24 & 1, 6, 20, 26, 27 \\ 
 \noalign{\smallskip}
 \hline
 \noalign{\smallskip}
NGC~2477 & $700$ & $0.31$ & $1.4$ & 39 & 3, 4, 6, 12, 39 & 6, 40    \\
Trumpler~23 & $800$ & $0.68\pm0.04$ & $2.20$ & 1, 26, 27, 31, 43 & 24, 43 & 1, 26, 27, 42\\
Berkeley~81 & $860\pm100$ & $0.85\pm0.04$ & $3.00$ & 1, 10, 26, 27, 31, 34, 45 & 24, 45  & 1, 6, 26, 27 \\ 
NGC~2355 &  $900$ & $0.13\pm0.03$ & $1.80\pm0.07$ & 1, 4, 10, 30 & 42 &  1, 6,   \\
NGC~6802 & $900$ & $0.79\pm0.06$ & $1.80$ & 1, 10, 26, 27, 31, 44 & 24 & 1, 6, 26, 27, 44 \\ 
NGC~6005 & $973\pm4$ $^c$ & $0.49\pm0.06$  & $2.70$ & 1, 15, 26, 27, 31 & 4, 24 & 1, 6, 26, 27 \\ 
Pismis~18 & $1200\pm400$ & $+0.22\pm0.04$ & $2.20$ & 27, 28, 31  & 24 & 1, 27, 28 \\ 
Melotte~71 & $1294\pm89$ $^c$ & $0.11$ & $2.2$--$3.2$ & 8, 15 & 4, 42 & 6 \\
Pismis~15 & $1300$ & $0.56\pm0.05$ & $2.6$--$2.9$ & 1, 4, 30 & 4, 30 & 1, 6 \\
Trumpler~20 & $1400$ & $0.37\pm0.03$ & $3.00$ & 1, 10, 26, 27, 41 & 24, 41 & 1, 6, 26, 27, 33, 41, 46 \\ 
Berkeley~44 & $1600$ & $0.90\pm0.07$ & $1.80$--$3.10$ & 1, 10, 26, 27, 31, 47 & 24, 47 & 1, 6, 26, 27 \\ 
NGC~2243 & $4000\pm120$ & $0.04\pm0.04$ & $4.50$ & 1, 10, 27, 31, 48, 49, 50 & 24, 32, 49 & 1, 6, 27 \\ 
M67 & $4000$--$4500$ & $0.059$ & $0.90$ & 21, 38, 51, 52 &  4, 12 & 40 \\ 
 \noalign{\smallskip}
 \hline
\end{tabular} }
 \tablefoot{
 \tablefoottext{a}{References for the E(B-V) reddening adopted from \cite{jackson2021} for all clusters except for NGC~2477 \cite{rain}.}
\tablefoottext{b}{GES studies that reference the clusters and/or study them without taking membership analysis primarily into account.} 
\tablefoottext{c}{Updated cluster ages using \textit{Gaia} data, as listed by \citet{bossini}.} \\
 \textbf{References}. For the cluster ages, reddening and distances shown here we chose the most recent or most robust estimates for each cluster. Several other studies are further cited here, and age estimates are additionally briefly discussed in the individual notes of Appendix~\ref{ap1:AppendixB}. (1) \cite{jackson2021}; (2)  \cite{kilikoglu}; (3) \cite{gao}; (4) \cite{kharchenko}; (5) \cite{melnik}; (6) \cite{cantatgaudin_gaia}; (7) \cite{frinchaboy}; (8) \cite{netopil2016}; (9) \cite{randich2020}; (10) \cite{romano}; (11) \cite{rangwal}; (12) \cite{dib}; (13) \cite{liu}; (14) \cite{alonsosantiago}; (15) \cite{bossini}; (16) \cite{pang}; (17) \cite{gillen}; (18) \cite{zhang}; (19) \cite{binks2021}; (20) \cite{randich_gaia}; (21) \cite{dumont}; (22) \cite{franciosini2021}; (23) \cite{jeffries2001}; (24) \cite{dias2002}; (25) \cite{jackson}; (26) \cite{jacobson}; (27) \cite{magrini2017}; (28) \cite{fritzewski}; (29) \cite{vandeputte}; (30) \cite{sampedro}; (31) \cite{magrini2018}; (32) \cite{magrini2014}; (33) \cite{tautvaisiene}; (34) \cite{magrini2015}; (35) \cite{dobbie}; (36) \cite{fritzewski2021}; (35) \cite{hetem}; (36) \cite{joshi}; (37) \cite{friel}; (38) \cite{sestitorandich}; (39) \cite{rain}; (40) \cite{jadhav}; (41) \cite{donati}; (42) \cite{buckner}; (43) \cite{overbeek}; (44) \cite{tang}; (45) \cite{donati2}; (46) \cite{smiljanic2016}; (47) \cite{hayesfriel}; (48) \cite{heiter}; (49) \cite{jacobson2011}; (50) \cite{frieljanes}; (51) \cite{pallavicini}; (52) \cite{richer}, (53) \cite{Jeffries2023_IC4665}. }
\end{table*}

Our present sample from GES iDR6 includes $114,325$ UVES and GIRAFFE spectra of $42$ OCs ranging in ages from $1$~Myr to $4.5$~Gyr. Here below we list the changes and improvements of the updated and improved iDR6 sample used in this work, in contrast to our former iDR4  sample of only $20$ OCs and associations, as used in Paper I:  

\begin{itemize}
 \item The number of stars in the current iDR6 sample is considerably larger than the $12,493$ UVES and GIRAFFE spectra we had at our disposal for the  preliminary cluster calibration results in Paper I. This larger number of stars in the fields of many of our sample clusters improved our membership analysis and increased the number of final candidates in more than one case. 
 
\item Several issues were also solved thanks to this larger sample, such as the fact that with our former iDR4 file only a few UVES Li values were listed for many of the intermediate-age and old clusters, especially in the $3000$--$4000$~K temperature range (M-type stars, more difficult to measure in those age ranges)\footnote{This was due to the homogenisation process in iDR4, as only WG11 and WG12 provided homogenised $EW$s, while WG10 provided only abundances. As a result, $EW$s for GIRAFFE stars in intermediate-age and old clusters were not available in this release, and in some cases, $EW$s for young clusters were also missing if the node values proved inconsistent, which happened especially for M-type stars in WG12.}. Thus, we had to make additional use of the individual Li measurements derived by the OACT (\textit{Osservatorio Astrofisico di Catania}) node, adding to our sample a number of GIRAFFE stars that had no recommended $EW$(Li) values in the iDR4 file\footnote{The iDR4 issues have since been overcome thanks to the new homogeneous analysis done in iDR6  \citep{gilmoregaia2021, randichgaia2021}.}. We note that, even considering the list of spectra that were deprecated in the working version of iDR6 released to the consortium in December 2020, we observed no such lack with our present sample, and our former member lists are as a result either typically enlarged, or at least stay with approximately the same number of candidate stars.
  
\item In addition to the larger number of spectra overall, we now count with $86$ OCs in the current iDR6, $48$ more than in iDR4. Our former iDR4 sample offered a total of $38$ clusters, $26$ of them being OCs, and the remaining $12$ of them being globular, which we discarded. We do not consider any iDR6 globular cluster in this updated analysis either, as Li cannot be used as a youth indicator in those cases.
  
\item In our current iDR6 sample, the data for the OCs suffering from the contamination of nebular lines, which can affect the $RV$ distributions and therefore the final membership analysis \citep[e.g.,][]{bonito, bonito2020}, were corrected by recalculating the $RV$s and applying an alternative different sky background subtraction \citep{gilmoregaia2021}\footnote{In general terms (this is by no means a straightforward method), an initial estimate would firstly be formed from the background emission affecting the spectrum of a target object, taking into account both astronomical and atmospheric sources. This estimated background spectrum would then be subtracted from the input science spectrum on a pixel by pixel basis, which gives rise an output spectrum consisting of emission from the science object alone.}. Thus, we were able to add to our sample a number of young and intermediate-age clusters we had to formerly discard because they exhibited high differential nebulosity (such as NGC~2264, NGC~2451~A and B, NGC~3532, NGC~6530, and Trumpler~14\footnote{The reason why this improved sky background subtraction could not be done for earlier releases to the level iDR6 has finally achieved is that, the Survey being fiber-fed, subtraction of the nebular sky background is not a straightforward procedure \citep{bonito2020}}). 
  
  \item Finally, in this study we focus on FGK stars, and so we discarded all stars with $T_{\rm eff}>7500$~K from the cluster sample in all cases. Given the nature of our membership criteria (see Subsect.~\ref{analysis}), for each cluster we also deprecated from our analysis all stars with no measurements of $RV$s (\textit{VRAD} in the iDR6 file), either $T_{\rm eff}$ (\textit{TEFF}) or the newly measured infrared photometric temperatures (\textit{TEFF\textunderscore IRFM}), and/or $EW$(Li) (be it the corrected $EW$s, \textit{EWC\textunderscore LI}, or the improved $EW$s with an additional veiling correction, taking into account the estimate of the continuum emission due to accretion, \textit{EW\textunderscore LI\textunderscore UNVEIL} \citep[e.g.,][]{lanzafame}). 
  
  \end{itemize}
  
  Out of the $86$ OCs measured in the iDR6 data, we discarded all clusters with less than $100$ stars, in order to have a sufficient minimum number of stars in each cluster to ensure a statistically significant membership analysis. On the other hand, we did not take into account any old clusters in the iDR6 file with ages older than $2$~Gyr, other than the ones we already analysed in Paper I, due to the fact that the Li content in GKM-type stars is strongly depleted in old clusters\footnote{Furthermore, these clusters being so far, the already depleted GKM stars are also too faint to be observed by GES}.
  We also only analysed four additional clusters in the $0.8$--$2$~Gyr age range, as we already counted with nine old clusters in our first sample selection with ages $0.8$--$4.5$~Gyr, and preferred rather to add more new young and intermediate-age clusters to our updated study, adding to the age ranges where Li is increasingly relevant for the analysis at hand. This criteria leaves us with a total of $42$ clusters constituting our current sample, including seven star forming regions ($1$--$6$~Myr) and nine young clusters ($10$--$50$~Myr), along with 13 intermediate-age clusters ($90$--$575$~Myr), and 13 old clusters ($0.7$--$4.5$~Gyr).

A number of membership studies have already been conducted, with potential members identified for all $42$ clusters selected in the present paper. These studies, particularly those who made use of GES data and specifically iDR6 data, were of great help to evaluate the goodness of our membership analysis by comparing our final candidates with previous membership lists. We show all these studies in Tables~\ref{table:01} and \ref{table:02}. We divided the sample clusters into groups according to age, and show them in two tables for convenience due to their length: Table~\ref{table:01} lists the young ($1$--$50$~Myr) clusters, while Table~\ref{table:02} lists the intermediate-age ($50$--$700$~Myr) and old clusters ($>700$~Myr). In the individual notes of Appendix~\ref{ap1:AppendixB}, where we detail all cluster membership results, we reference these studies in greater detail for each of the clusters. These tables also list the age estimates, distances and reddening values from the literature for all clusters, and we additionally refer to Table~\ref{table:1} for $RV$s from the literature, to Table~\ref{table:2} for proper motions ($pmra$ and $pmdec$) and parallaxes ($\pi$) from \textit{Gaia}, and to Table~\ref{table:3} for [Fe/H] metallicities from the literature.

%
\section{Updated membership analysis}  \label{analysis}
 
  \begin{table*} [htp]
 \begin{center}
 \caption{Fit parameters and $RV$ members for the target star forming regions and OCs.}
 \label{table:1}   
\resizebox{\textwidth}{!}{ \begin{tabular}{c r r c c c r c} 
\hline
\hline
\noalign{\smallskip}

 Cluster$^a$ & $RV$ $^b$ & \multicolumn{2}{c}{2$\sigma$ clipping} & 2$\sigma$ membership & No.  $RV$  & \multicolumn{2}{c}{Final fit of member list}   \\ [1 ex] 
     &  (km~s~$^{-1}$)   & $\langle RV \rangle$ (km~s~$^{-1}$) &  $\sigma$ (km~s~$^{-1}$) & intervals & members & $\langle RV \rangle$ (km~s~$^{-1}$) & $\sigma$  (km~s~$^{-1}$)   \\
\noalign{\smallskip}
\hline
\noalign{\smallskip}
  NGC~6530 & $-1.9\pm7.2$ $^1$ &  $0.3$ & $2.1$ & [$-3.8$, $4.6$]  & 477  &  $0.0$  & $2.5$  \\ 
 $\rho$~Oph & $-7.0\pm0.2$ $^2$ & $-$6.4 & $2.1$ & [$-$10.6, $-$2.2]  & 48 & $-$6.6 & 1.0  \\ 
 Trumpler~14 & $-10.5\pm6.9$ $^3$ & $-7.9$ & $2.9$ & [$-13.7$, $-2.1$]  & 231 & $-8.4$ & $3.0$  \\ 
 Cha~I & $14.6\pm1.2$ $^4$ & 15.7 & 1.1 & [13.5, 17.9] & 102 & 15.6 & 1.1 \\
 NGC~2244 & $33.6\pm0.7$  $^5$ &  $30.7$ & 2.7 & [$25.3$, $36.1$]  & 143 & 30.5 & 2.7  \\ 
 NGC~2264 & $20.2\pm6.4$ $^3$ &  20.2 & 2.5 & [$15.2$, $25.2$]  & 618 & 19.9 & 2.2  \\ 
 $\lambda$~Ori & $27.5\pm0.4$ $^5$ &  27.1 & 1.0 & [$25.1$, $29.1$]  & 208 & 27.3 & 1.0 \\ 
 Col~197 & $35.8\pm2.3$ $^3$ &  20.8 & 1.1 & [$18.6$, $23.0$]  & 124 & 20.7 & 1.1 \\
 NGC~2232 & $25.4\pm0.9$ $^3$ &  29.7 & 14.1 & [$-0.2$, $56.2$]  & 750 & 25.3 & 0.6 \\
 IC~2391 & $15.3\pm0.2$ $^3$ & 15.4 & 2.3 & [10.5, 19.7] & 55 & 14.6 & 0.7  \\
 IC~2602 & $17.4\pm0.2$ $^3$ & 13.2 & 10.2 & [$-$8.0, 34.0] & 309 & 17.5 & 0.6 \\
 IC~4665 & $-14.4\pm0.8$ $^3$ & $-$13.4 & 14.8 & [$-$43.0, 16.2] & 233 & $-$14.0 & 0.7  \\
 NGC~6405 & $-9.2\pm0.8$ $^6$ & $-9.0$ & 6.5 & [$-22.0$, $4.0$]  & 251 & $-8.8$ & 0.8 \\
 Blanco~1 & $5.8\pm0.1$ $^6$ & 5.8 & 0.7 & [$4.4$, $7.2$] & 141 & 5.8 &  0.6 \\
 NGC~6067 & $-39.9\pm0.2$ $^1$ & $-38.1$ & 3.3 & [$-44.7$, $-31.5$]  & 209 & $-37.8$ & 2.0  \\
 NGC~6649 & $-8.9\pm0.5$ $^3$ & $-10.3$ & 3.8 & [$-18.7$, $-1.9$]  & 32 & \dots $^c$ & \dots \\
 NGC~2516 & $23.8\pm0.2$ $^3$ & 23.6 & 0.7 & [22.3, 25.1] & 460 & 23.8 & 0.8 \\ 
 NGC~6709 & $-7.8\pm5$ $^1$ & $-7.7$ & 10.3 & [$-28.3$, $12.9$]  &  322 & $-9.4$ & 0.7 \\
 NGC~6259 & $-32.8\pm0.4$ $^3$ & $-32.2$ & 3.3 & [$-38.8$, $-25.6$]  & 125 & $-32.5$ & 2.5 \\
 NGC~6705 & $36.0\pm0.2$ $^3$ & 35.6 & 1.9 & [31.8, 39.4] & 391 & 35.4 & 1.8 \\
 Berkeley~30 & \dots $^d$ & 47.7 & 3.6 & [$40.5$, $54.9$]  & 78 & 48.1 &  3.0 \\
 NGC~6281 & $-5.0\pm0.1$ $^3$ & $-4.8$ & 1.9 & [$-8.6$, $-1.0$]  & 82  & $-4.5$ & 0.7  \\
 NGC~3532 & $5.4\pm0.2$ $^3$ & 5.3 & 0.9 & [$3.5$, $7.1$]  & 518 & 5.4 & 0.9 \\
 NGC~4815 & $-29.8\pm0.3$ $^3$ & $-29.0$ & 3.4 & [$-36.4$, $-22.8$] & 68 & $-27.6$ & 4.0 \\ 
 NGC~6633 & $-28.6\pm0.1$ $^3$ & $-22.4$ & 13.2 & [$-49.0$, 0.0]$^e$ & 617 & $-28.3$ & 0.9 \\
 NGC~2477 & $7.9\pm0.1$ $^3$ & 8.1 & 1.0 & [$6.1$, $10.1$]  & 86 & 7.2 & 2.1 \\
 Trumpler~23 & $-61.4\pm0.5$ $^3$ & $-61.3$ & 1.8 & [$-57.7$, $-64.9$] & 51 & $-62.0$ & 1.5 \\ 
 Berkeley~81 & $50.0\pm0.7$ $^3$ & 47.7 & 2.2 & [43.3, 52.1] & 69 & 48.2 & 0.3 \\ 
 NGC~2355 & $36.9\pm0.7$ $^3$ & 36.2 & 0.9 & [$34.4$, $37.0$]  & 119 & 36.3 & 0.9  \\
 NGC~6802 & $11.8\pm0.4$ $^3$ & 13.0 & 1.8 & [9.6, 16.6] & 77 & 12.0 & 0.7 \\
 NGC~6005 & $-25.6\pm0.5$ $^3$ & $-25.5$ & 3.2 & [$-31.9$, $-19.1$] & 174 & $-24.6$ & 1.1 \\ 
 Pismis 18 & $-28.5\pm0.6$ $^3$ & $-29.7$ & 2.8 & [$-35.3$, $-24.1$] & 41 & $-28.2$ & 0.8 \\ 
 Melotte~71 & $51.3\pm0.4$ $^3$ & 51.6 & 0.7 & [$50.2$, $53.0$]  &  71 & 51.1 & 0.4  \\
 Pismis~15 & $36.3\pm0.7$ $^3$ & 35.0 & 0.1 & [$32.6$, $37.4$]  & 93 & 35.2 & 0.8  \\
 Trumpler~20 & $-39.8\pm0.2$ $^3$ & $-39.8$ & 1.4 & [$-42.6$, $-37$] & 451 & $-40.0$ & 1.2 \\
 Berkeley~44 & $-7.6\pm0.4$ $^3$ & $-8.8$ &  0.4 & [$-10.4$, $-7.2$] & 39 & $-8.8$  & 0.7 \\
 NGC~2243 & $59.6\pm0.5$ $^3$ & 59.5 & 0.6 & [58.3, 60.7] & 469 & 59.6 & 0.6 \\
 M67 & $34.1\pm0.1$ $^3$ & 34.0 & 0.8 &  [32.4, 35.6] & 110 & 33.9 & 0.8 \\ [1ex] 
 \noalign{\smallskip}
 \hline
 \end{tabular} }
  \tablefoot{
  \tablefoottext{a}{As stated earlier in the text, regarding the clusters $\gamma$~Vel, NGC~2547 and NGC~2451 A and B, we directly used the selections obtained by several studies listed in Table~\ref{table:01}.}
\tablefoottext{b}{References for the mean cluster $RV$s: 
 (1) \cite{conrad}; (2) \cite{rigliaco}; (3) \cite{soubiran}; (4) \cite{lopez}; (5) \cite{carrera2019}; (6) \cite{gaiadr2}.}
\tablefoottext{c}{Final selection only consists of two members, final fit is not possible.}
\tablefoottext{d}{We could not find any prior $RV$ measurements from the literature for this cluster.}
\tablefoottext{e}{The 2$\sigma$ membership interval for this cluster has been obtained after excluding a large contaminant population in the middle of the distribution, as detailed in Appendix~\ref{ap1:AppendixB}.}
}
 \end{center}
\end{table*}

\begin{table*} [htp]
  \begin{center}
 \caption{Fit parameters and parallax membership for the target star forming regions and OCs.}
 \label{table:2}     
 \resizebox{\textwidth}{!}{ \begin{tabular}{c c c c c c c c c} 
\hline
\hline
\noalign{\smallskip}

 Cluster$^a$ & $pmra$ $^b$ & $pmdec$ $^b$ & Parallax ($\pi$) $^b$ & \multicolumn{2}{c}{2$\sigma$ clipping} & 2$\sigma$ membership & \multicolumn{2}{c}{Final fit of member list}   \\ [1 ex] 
     &  (mas~yr$^{-1}$)   & (mas~yr$^{-1}$)  &  (mas) & $\langle \pi \rangle$ (mas) &  $\sigma$ (mas) & intervals & $\langle \pi \rangle$ (mas) & $\sigma$  (mas)   \\
\noalign{\smallskip}
\hline
\noalign{\smallskip}
  NGC~6530 & $1.32\pm0.08$ $^1$ & $-2.07\pm0.08$ $^1$ & $0.750$ $^2$ & $0.770$ & $0.060$ & [$0.65$, $0.89$]  &  $0.769$  & 0.062  \\ 
 $\rho$~Oph & $-1.70\pm0.07$ $^1$ & $0.20\pm0.07$ $^1$ & $7.692$ $^1$ & $7.237$ & $0.120$ & [$7.001$, $7.481$]  & 7.250  & 0.116  \\ 
 Trumpler~14 & $-6.54\pm0.07$ $^1$ & $2.06\pm0.07$ $^1$ & $0.340$ $^1$ & $0.390$ & $0.050$ & [$0.290$, $0.490$]  & 0.395  & 0.066  \\ 
 Cha~I & $-19.0\pm5.00$ $^3$ & $2.00\pm4.00$ $^3$ & $6.250$ $^1$ & $5.240$ & $0.080$ & [$5.080$, $5.400$]  & 5.247 & 0.087 \\ 
 NGC~2244 & $-1.70\pm0.07$ $^1$ & $0.20\pm0.07$ $^1$ & $0.590$--$0.710$ $^1$ & $0.670$ & $0.060$ & [$0.550$, $0.790$]  &  0.665  & 0.071  \\ 
 NGC~2264 & $-1.76\pm0.08$ $^1$ & $-3.72\pm0.07$ $^1$ & $1.250$ $^1$ & $1.380$ & $0.040$ & [$1.200$, $1.600$]  & 1.388 & 0.063  \\  
 $\lambda$~Ori & $1.19\pm0.51$ $^4$ & $-2.12\pm0.39$ $^4$ & $2.462\pm0.124$ $^4$ & $2.500$ & $0.048$ & [$2.404$, $2.596$]  & 2.492 & 0.060 \\ 
 Col~197 & $-5.81\pm0.31$ $^4$ & $3.93\pm0.39$ $^4$ & $1.135$ $^1$ & $1.020$ & $0.050$ & [$0.930$, $1.130$]  &  1.023 & 0.059 \\ 
 NGC~2232 & $-4.70$ $^4$ & $-1.80$ $^4$ & $3.067\pm0.099$ $^4$ & $3.100$ & $0.060$ & [$2.980$, $3.220$]  & 3.111  & 0.058  \\ 
 IC~2391 & $-24.64\pm0.88$ $^4$ & $23.32\pm0.73$ $^4$ & $6.585$ $^5$ & $6.620$ & $0.070$ & [$6.490$, $6.790$]  & 6.613 & 0.086  \\ 
 IC~2602 & $-17.58\pm0.83$ $^4$ & $10.70\pm0.90$ $^4$ & $6.758$ $^5$ & $6.610$ & $0.080$ & [$6.460$, $6.780$]  & 6.629  & 0.084  \\ 
 IC~4665 & $-0.91\pm0.28$ $^4$ & $-8.52\pm0.28$ $^4$ & $2.872$ $^5$ & $2.900$ & $0.060$ & [$2.780$, $3.020$]  & 2.887 & 0.077  \\ 
 NGC~6405 & $1.31\pm0.34$ $^4$ & $-5.85\pm0.34$ $^4$ & $2.172$ $^4$ & $2.190$ & $0.020$ & [$2.150$, $2.230$]  & 2.185  & 0.022  \\  
 Blanco~1 & $18.74\pm0.43$ $^4$ & $2.60\pm0.44$ $^4$ & $4.210\pm0.120$ $^4$ & $4.200$ & $0.070$ & [$4.060$, $4.340$]  & 4.225 & 0.062  \\  
 NGC~6067 & $-1.91\pm0.12$ $^4$ & $-2.59\pm0.12$ $^4$ & $0.443\pm0.065$ $^4$ & $0.470$ & $0.028$ & [$0.416$, $0.528$]  & 0.472 & 0.024  \\  
 NGC~6649 & $-0.01\pm0.18$ $^4$ & $-0.06\pm0.18$ $^4$ & $0.467\pm0.087$ $^4$ & $0.491$ & $0.060$ & [$0.371$, $0.611$]  &  \dots $^c$  & \dots  \\ 
 NGC~2516 & $-4.75\pm0.44$ $^4$ & $11.22\pm0.35$ $^4$ & $2.417\pm0.045$ $^4$ & $2.429$ & $0.031$ & [$2.367$, $2.491$]  & 2.428 & 0.033 \\  
 NGC~6709 & $-1.76\pm0.08$ $^1$ & $-3.72\pm0.07$ $^1$ & $1.250$ $^1$ & $1.380$ & $0.040$ & [$1.200$, $1.600$] & 0.916 & 0.033 \\  
 NGC~6259 & $-1.02\pm0.13$ $^4$ & $-2.89\pm0.12$ $^4$ & $0.408\pm0.057$ $^4$ & $0.432$ & $0.061$ & [$0.310$, $0.550$]  & 0.463 & 0.034 \\  
 NGC~6705 & $-1.57\pm0.16$ $^4$ & $-4.14\pm0.17$ $^4$ & $0.427\pm0.083$ $^4$ & $0.411$ & $0.039$ & [$0.333$, $0.489$]  & 0.404 & 0.054 \\  
 Berkeley~30 & $-0.25\pm0.17$ $^4$ & $-0.33\pm0.13$ $^4$ & $0.164$ $^4$ & $0.209$ & $0.138$ & [$-0.074$, $0.478$]  & 0.201 & 0.144 \\  
 NGC~6281 & $-1.86\pm0.30$ $^4$ & $-4.02\pm0.24$ $^4$ & $1.873\pm0.080$ $^4$ & $1.880$ & $0.030$ & [$1.820$, $1.940$]  & 1.889 & 0.056 \\  
 NGC~3532 & $-10.39\pm0.40$ $^4$ & $5.18\pm0.40$ $^4$ & $2.066\pm0.062$ $^4$ & $2.086$ & $0.025$ & [$2.035$, $2.136$]  &  2.085 & 0.024 \\  
 NGC~4815 & $-5.76\pm0.11$ $^4$ & $-0.96\pm0.10$ $^4$ & $0.261\pm0.062$ $^4$ & $0.294$ & $0.076$ & [$0.142$, $0.446$]  &  0.294  &  0.084 \\   
 NGC~6633 & $1.20\pm0.33$ $^4$ & $-1.81\pm0.30$ $^4$ & $2.525\pm0.073$ $^4$ & $2.533$ & $0.045$ & [$-0.448$, $3.432$]$^d$  & 2.545 & 0.051  \\  
 NGC~2477 & $-2.45\pm0.01$ $^4$ & $0.88\pm0.01$ $^4$ & $0.665\pm0.037$ $^4$ & $0.688$ & $0.030$ & [$0.628$, $0.784$]  &  0.694  & 0.029 \\  
 Trumpler~23 & $-4.18\pm0.12$ $^4$ & $-4.75\pm0.11$ $^4$ & $0.352\pm0.059$ $^4$ & $0.353$ & $0.040$ & [$0.273$, $0.433$]  & 0.347 & 0.043 \\ 
 Berkeley~81 & $-1.20\pm0.16$ $^4$ & $-1.83\pm0.16$ $^4$ & $0.254\pm0.090$ $^4$ & $0.261$ & $0.054$ & [$0.153$, $0.369$]  & 0.263 & 0.048 \\  
 NGC~2355 & $-3.80\pm0.14$ $^4$ & $-1.09\pm0.13$ $^4$ & $0.497\pm0.056$ $^4$ & $0.530$ & $0.022$ & [$0.486$, $0.574$]  & 0.529 & 0.015 \\  
 NGC~6802 & $-2.81\pm0.11$ $^4$ & $-6.44\pm0.13$ $^4$ & $0.309\pm0.067$ $^4$ & $0.335$ & $0.028$ & [$0.279$, $0.391$]  &  0.349 & 0.053 \\ 
 NGC~6005 & $-4.01\pm0.12$ $^4$ & $-3.81\pm0.12$ $^4$ & $0.362\pm0.060$ $^4$ & $0.390$ & $0.051$ & [$0.288$, $0.492$]  & 0.327 & 0.135 \\ 
 Pismis 18 & $5.66\pm0.11$ $^4$ & $-2.29\pm0.11$ $^4$ & $0.332\pm0.052$ $^4$ & $0.349$ & $0.022$ & [$0.305$, $0.393$]  & 0.342 & 0.100 \\ 
 Melotte~71 & $-2.45\pm0.12$ $^4$ & $4.21\pm0.11$ $^4$ & $0.434\pm0.056$ $^4$ & $0.443$ & $0.026$ & [$0.391$, $0.495$]  & 0.348 & 0.100 \\ 
 Pismis~15 & $-5.30\pm0.10$ $^4$ & $3.33\pm0.12$ $^4$ & $0.354\pm0.049$ $^4$ & $0.387$ & $0.061$ & [$0.265$, $0.509$]  & 0.401 &  0.058 \\ 
 Trumpler~20 & $-7.09\pm0.09$ $^4$ & $0.18\pm0.09$ $^4$ & $0.251\pm0.045$ $^4$ & $0.272$ & $0.030$ & [$0.212$, $0.332$]  & 0.285 & 0.048 \\ 
 Berkeley~44 & $0.01\pm0.13$ $^4$ & $-2.83\pm0.14$ $^4$ & $0.303\pm0.060$ $^4$ & $0.322$ & $0.042$ & [$0.238$, $0.406$]  & 0.341 & 0.062 \\ 
 NGC~2243 & $-1.28\pm0.13$ $^4$ & $5.49\pm0.13$ $^4$ & $0.211\pm0.060$ $^4$ & $0.225$ & $0.052$ & [$0.121$, $0.329$]  &  0.220 & 0.048 \\ 
 M67 & $-10.97\pm0.24$ $^4$ & $-2.95\pm0.24$ $^4$ & $1.135\pm0.051$ $^4$ & $1.161$ & $0.026$ & [$1.110$, $1.214$] & 1.162  & 0.026 \\ [1ex] 
 \noalign{\smallskip}
 \hline
 \end{tabular} } 
  \tablefoot{
\tablefoottext{a}{As stated earlier in the text, regarding the clusters $\gamma$~Vel, NGC~2547 and NGC~2451 A and B, we directly used the selections obtained by several studies listed in Table~\ref{table:01}.}
\tablefoottext{b}{References for the mean cluster proper motions ($pmra$ and $pmdec$) and parallaxes: (1) \cite{kuhn2019}; (2) \cite{wright2019}; (3) \cite{lopez}; (4) \cite{cantatgaudin_gaia}; (5) \cite{kounkel}.}
\tablefoottext{c}{Final selection only consists of two members, final fit is not possible.}
\tablefoottext{d}{The 2$\sigma$ membership interval for this cluster has been obtained after excluding a large contaminant population in the middle of the distribution, as detailed in Appendix~\ref{ap1:AppendixB}.}
}
\end{center}
\end{table*}

\begin{table*} [htp]
\begin{center}
 \caption{Fit parameters and metallicity membership for the target star forming regions and OCs.}
 \label{table:3}     
 \resizebox{\textwidth}{!}{ \begin{tabular}{c r c c c c c c} 
\hline
\hline
\noalign{\smallskip}

 Cluster$^a$ & [Fe/H] & References & \multicolumn{2}{c}{3$\sigma$ clipping} &  3$\sigma$ membership  & \multicolumn{2}{c}{Final fit of member list}      \\ [1 ex] 
 &  (dex)   & [Fe/H] &  $\langle [Fe/H] \rangle$ (dex) &  $\sigma$ (dex) & intervals$^d$ & $\langle [Fe/H] \rangle$ (dex) & $\sigma$ (dex) \\
\noalign{\smallskip}
\hline
\noalign{\smallskip}
 NGC~6530 & $-0.04\pm0.01$ & 1 & $0.00$ & 0.08 & [$-0.24$, $0.24$]  &  $-0.01$  & $0.09$  \\
 $\rho$~Oph & $-0.08\pm0.02$ & 2 & 0.02 & 0.09 & [$-0.25$, $0.29$] & -0.03 & 0.20  \\
 Trumpler~14 & $-0.03\pm0.02$ & 1 & $0.00$ & $0.06$ & [$-0.18$, $0.18$] & $-0.02$ & $0.07$  \\ 
 Cha~I & $-0.07\pm0.04$ & 2, 3 & $-0.02$ & 0.04 & [$-0.15$, $0.10$] & $-0.04$ & 0.05 \\
  NGC~2244 & $-0.23$ & 4 &  $-0.04$ & 0.05 & [$-0.20$, $0.11$]  & $-0.05$ $^b$ & 0.05  \\ 
  NGC~2264 & $-0.0 6\pm0.01$ & 1, 2, 5 & $-0.03$ & 0.06 & [$-0.21$, $0.15$] & $-0.03$ & 0.06  \\ 
  $\lambda$~Ori & $-0.01$ & 4 & $-0.04$ & 0.07 & [$-0.25$, $0.17$]  & $-0.03$ & 0.06 \\ 
  Col~197 & \dots $^c$ & \dots & 0.02 & 0.06 & [$-0.16$, $0.20$] & 0.01 & 0.06 \\
  NGC~2232 & $0.04\pm0.01$ & 4 & 0.02 & 0.04 & [$-0.15$, $0.15$]  & 0.00 & 0.05 \\
  IC~2391 & $-0.03\pm0.02$ & 1, 2, 6, 7, 8 & $-0.03$ & 0.03 & [$-0.10$, $0.10$]  & $-0.03$ & 0.09  \\
  IC~2602 & $-0.02\pm0.02$ & 2, 7, 8 & 0.00 & 0.07 & [$-0.21$, $0.21$] & 0.00 & 0.07 \\
  IC~4665 & $0.00\pm0.02$ & 2 & $-0.01$ & 0.09 & [$-0.19$, 0.17]  & 0.00 & 0.01  \\
  NGC~6405 & $0.07\pm0.03$ & 9, 10 & 0.03 & 0.12 & [$-0.33$, $0.39$]  & 0.00 & 0.06 \\
  Blanco~1 & 0.00 & 9, 12 & 0.01 & 0.02 & [$-0.05$, $0.07$] & 0.01 &  0.02 \\
  NGC~6067 & $0.20\pm0.08$ & 9, 11 & 0.11 & 0.09 & [$-0.16$, $0.38$]  & $0.07$ & 0.10  \\
  NGC~6649 & \dots $^c$ & \dots & 0.03 & 0.19 & [$-0.54$, $0.60$] & \dots $^e$ & \dots \\
  NGC~2516 & $-0.02\pm0.01$ & 5, 13 & 0.02 & 0.05 & [$-0.13$, $0.17$]  & 0.00 & 0.06 \\
  NGC~6709 & \dots $^c$ & \dots & 0.00 & 0.17 & [$-0.51$, $0.51$] & $-0.08$ & 0.07  \\
  NGC~6259 & $0.21\pm0.04$ & 11, 14 & 0.18 & 0.09 & [$-0.09$, $0.45$] & 0.17 & 0.10 \\
  NGC~6705 & $0.16\pm0.04$ & 11, 13 & 0.06 & 0.08 & [$-0.18$, $0.30$]  & 0.06 & 0.08 \\
  Berkeley~30 & $0.10$ & 15 &  $-0.19$ & 0.18 & [$-0.73$, $0.35$] & $-0.20$ & 0.19 \\
  NGC~6281 & $0.06$ & 9, 17 & 0.07 & 0.20 & [$-0.53$, $0.67$] & 0.08 & 0.15  \\
  NGC~3532 & $-0.07\pm0.10$ & 16 & 0.02 & 0.06 & [$-0.16$, $0.20$] & 0.02 & 0.07 \\
  NGC~4815 & $0.11\pm0.01$ & 11, 13, 18 & 0.07 & 0.20 & [$-0.53$, $0.67$] & 0.01 & 0.08 \\ 
  NGC~6633 & $-0.01\pm0.11$ & 11, 13, 19 & $-0.01$ & 0.19 & [$-0.58$, $0.56$]  & 0.03 & 0.15 \\
  NGC~2477 & $0.07\pm0.03$ & 9, 17, 20 & 0.10 & 0.04 & [$-0.02$, $0.22$] & 0.14 & 0.04 \\
  Trumpler~23 & $0.21\pm0.04$ & 11, 13, 21 & 0.16 & 0.10 & [$-0.14$, $0.46$]  & 0.19 & 0.07 \\ 
  Berkeley~81 & $0.22\pm0.07$ & 11, 13 & 0.18 & 0.11 & [$-0.15$, $0.51$] & 0.18 & 0.10 \\ 
  NGC~2355 & $-0.11$ & 4 & $-0.12$ & 0.05 &[$-0.27$, $0.03$] & $-0.13$ & 0.05   \\
  NGC~6802 & $0.10\pm0.02$ & 11, 13, 22 & 0.03 & 0.15 & [$-0.42$, $0.48$] & 0.05 & 0.13 \\ 
  NGC~6005 & $0.19\pm0.02$ & 11, 13 & 0.14 & 0.10 & [$-0.16$, $0.44$] & 0.19 & 0.08 \\ 
  Pismis~18 & $0.22\pm0.04$ & 11, 13 & 0.06 & 0.11 & [$-0.27$, $0.39$] & 0.13 $^b$ & 0.02 \\ 
  Melotte~71 & $-0.27$ & 9, 12 & $-0.33$ & 0.24 & [$-1.08$, $0.42$]  & $-0.09$ & 0.01 \\
  Pismis~15 & $0.01\pm0.01$ & 23 & $-0.10$ & 0.22 & [$-0.76$, $0.56$] & $-0.07$ & 0.11  \\
  Trumpler~20 & $0.10\pm0.05$ & 11, 13 & 0.06 & 0.10 & [$-0.24$, $0.36$] & 0.08 & 0.10 \\
  Berkeley~44 & $0.27\pm0.06$ & 11, 13 & 0.12 & 0.09 & [$-0.15$, $0.39$] & 0.12 & 0.09 \\ 
  NGC~2243 & $-0.38\pm0.04$ & 11, 13, 17 & $-0.57$ & 0.13 & [$-0.96$, $-0.18$]  & $-0.61$ & 0.13 \\
  M67 & $-0.01\pm0.04$ & 11, 17, 19, 24 & 0.00 & 0.05 &  [$-0.15$, $0.15$]  & 0.00 & 0.05 \\ 
  [1ex] 
 \noalign{\smallskip}
 \hline
 \end{tabular} } 
 \tablefoot{
\tablefoottext{a}{For $\gamma$~Vel, NGC~2547 and NGC~2451 A and B, we used the selections listed in Table~\ref{table:01}.}
\tablefoottext{b}{For these clusters, the final  mean [Fe/H] values deviate more appreciably from the literature values.}
\tablefoottext{c}{We could not find any prior measurements from the literature for these clusters.}
\tablefoottext{d}{In the individual notes of Appendix~\ref{ap1:AppendixB} we also list the number of stars that we accepted as robust members despite deviating from these 3$\sigma$ intervals.} 
\tablefoottext{e}{Final selection only consists of two members, final fit is not possible.} \\
 \textbf{[Fe/H] references}: Shown here are the most recent or robust estimates for each cluster, several studies are further cited in Appendix~\ref{ap1:AppendixB}: (1) \cite{randich_gaia}; (2) \cite{spina_giants}; (3) \cite{sacco}; (4) \cite{carrera2019}; (5) \cite{binks2021}; (6) \cite{de_silva}; (7) \cite{dumont}; (8) \cite{smiljanic2011}; (9) \cite{netopil}; (10) \cite{kilikoglu}; (11) \cite{magrini2018}; (12) \cite{bossini}; (13) \cite{jacobson}; (14) \cite{casali}; (15) \cite{paunzen}; (16) \cite{fritzewski}; (17) \cite{heiter}; (18) \cite{friel}; (19) \cite{sestitorandich}; (20) \cite{rain}; (21) \cite{overbeek}; (22) \cite{tang}; (23) \cite{carraro}; (24) \cite{liu}. 
}
\end{center}
\end{table*}

To obtain final lists of candidate members for the $42$ clusters in our sample, we conducted a homogeneous and coherent analysis of their membership according to several criteria, which we now detail below: 

\begin{itemize}
  \item \textbf{$RV$ analysis}
  (full description in Paper I and the published thesis): We studied the radial velocity distribution of each cluster with the aid of the {\texttt RStudio} environment for R by applying a two-sigma clipping procedure on the median, and then adopting a 2$\sigma$ limit about the cluster mean yielded by the Gaussian fit to identify the most likely kinematic members. Depending on the specific cluster, this 2$\sigma$ limit was sometimes enlarged to 3$\sigma$ to further consider marginal members, as detailed in Appendix~\ref{ap1:AppendixB}. In Table~\ref{table:1} we show the fit parameters (mean values, associated dispersions and 2$\sigma$ intervals) and the number of kinematic members resulting from the updated study of the distributions of $RV$ for the cluster sample. We additionally report updated reference values estimated from the literature, and also present in the last two columns of the table the mean values and associated dispersions of all final members of each cluster, after applying all membership criteria and concluding our membership analysis.
  \item  \textbf{\textit{Gaia} astrometry} (Subsect.~\ref{astrometry}): As the next step we reinforced the kinematic selection and obtain lists of probable astrometric members with the aid of the proper motions and parallaxes provided by \textit{Gaia} EDR3, analysing the locus of probable members in $pmra$-versus-$pmdec$ diagrams, and studying the parallax distributions with a Gaussian fit, similarly to the case of $RV$s. See Table~\ref{table:2} for the results of the study of parallax distributions for the cluster sample, including all fit parameters as well as the mean values and associated dispersions of all final candidates for each cluster, after applying all criteria. \\
  
 \item  \textbf{\textit{Gaia} photometry} (Subsect.~\ref{cmd}): We also made use of \textit{Gaia} EDR3 photometry to analyse the goodness of the astrometric candidates and discard field contaminants in $G$-versus-$G_{BP-RP}$ colour-magnitude diagrams (CMDs). \\
    
 \item \textbf{Gravity indicators} 
 (full description in Paper I and the thesis): We used the Kiel ($\log g$-versus-$T_{\rm eff}$) diagram to identify and discard evolved outliers, such as Li-rich giant stars (see Subsect.~\ref{outliers}) and other field contaminants. For all clusters we made use of the PARSEC isochrones \citep{bressan}, with Z=0.019, and ages ranging from 50~Myr to 5~Gyr. In the case of young clusters, we mainly used the $\gamma$ index defined by \cite{damiani} to effectively discard giant contaminants as a first step of the membership process. Regarding the order of criteria, for the young clusters we discarded all giant contaminants using the $\gamma$ index before performing the $RV$ and astrometric analyses, due to their appreciable field contamination. \\
 \item \textbf{Metallicity}
 (full description in Paper I and the thesis): Using the spectroscopic index [Fe/H] as a proxy of metallicity, an analysis of the metallicity distributions for each cluster (including all stars in the field before applying any other membership criteria) provided further confirmation of the membership of the candidate selections, and it was also taken into account to identify occasional rogue outliers that had been missed with all prior criteria. In Table~\ref{table:3} we present the updated results of the study of metallicity for the cluster sample, including all fit parameters, as well as the mean values and associated dispersions of all final candidates after concluding the membership analysis. We note that, in contrast to the study of $RV$s and parallaxes, we decided to adopt 3$\sigma$ membership intervals, instead of a 2$\sigma$ limit, as the starting point to identify the most likely metallicity members. We also note that we also accepted as final cluster candidates a number of marginal metallicity members deviating moderately from the starting 3$\sigma$ limits, as they fully fulfilled the rest of criteria in our analysis. We refer the reader to the individual notes of Appendix~\ref{ap1:AppendixB} for more details on these marginal stars for each cluster, as well as listing, in the case of some of the clusters, a smaller number of stars that were similarly accepted as robust cluster members, despite having [Fe/H] values that deviated more appreciably from the 3$\sigma$ limit. It is thus apt to note that, in contrast to the most robust criteria in our analysis (particularly \textit{Gaia} astrometry and CMDs), we consider metallicity to be among our less restrictive criteria and selected the final cluster members accordingly. This takes into account the inherent higher uncertainties related to [Fe/H] values derived from GIRAFFE spectra, even after the improvements achieved for the iDR6 data \citep[e.g.,][]{spina_gamma, gilmoregaia2021}, as well as the fact that the variations in stellar abundances also depend on several astrophysical factors and processes, such as atomic diffusion or the influence of planets \citep[e.g.,][]{oh2018, liu, yong}. Finally, stellar Li itself also depends on metallicity (a dependence that will be further analysed in Paper III) \citep[e.g.,][]{jeffries2014, randich&magrini, Limet_2023}, which makes it important not to discard those cluster members that are iron-rich or iron-poor. \\

   \item \textbf{Li content} 
   (full description in Paper I and the thesis): Probable candidates fulfilling the rest of criteria were considered Li members according to their locus in the $EW$(Li)-versus-$T_{\rm eff}$ diagrams, using as guides the upper Li envelope of IC~2602 (35~Myr) \citep{montes, lopezsantiago2003}, the upper \citep{neuhaeuser} and lower \citep{soderblom2} envelopes of the Pleiades ($78$--$125$~Myr), and the upper envelope of the Hyades ($750$~Myr) \citep{soderblom2}. As this work revolves around the calibration of Li and its observable dispersion, this is one of the final criteria so as to already count with a robust list of probable cluster members and add the least bias to our study. We refer to the individual notes of Appendix~\ref{ap1:AppendixB} to illustrate some examples of the bias that we could add by filtering candidates on the basis of Li before other criteria. In addition, for the star forming regions in our sample we additionally considered as cluster candidates all kinematic and astrometric members that are strong accretors with H$\alpha$10\%~$>270$--$300$~km~s$^{-1}$, disregardless of their Li content \citep{frasca, sacco2, bonito2020}\footnote{A tracer of accretion and youth indicator in young PMS stars, H$\alpha$10\% refers to the width of the H$\alpha$ emission line at 10$\%$ peak intensity.}. This is due to the enhancement and/or underestimation that can be caused by strong accretion (for more details, see Paper III).
 \item \textbf{Other \textit{Gaia} membership studies} (Subsect.~\ref{gaia}): Finally, we made use of additional studies conducted from \textit{Gaia} DR1, DR2 and EDR3 data \citep[e.g.,][]{cantatgaudin_gaia, randich_gaia, soubiran, bossini, jackson2021} to better confirm the robustness of our final candidate selections. These studies were also of great help to decide on the membership of marginal members according to one of more criteria.
\end{itemize}

In this work, we improve and update all criteria presented and described in Paper I (namely, the analysis of $RV$ distributions, surface gravity indicators, Li content, metallicity and comparison with several \textit{Gaia} studies from the literature), as well as using the astrometry and photometry from \textit{Gaia} EDR3 as new criteria, upon which we further elaborate in Subsects.~\ref{astrometry} and~\ref{cmd} in this section below. We additionally note that in our final analysis we modify the order of the criteria applied in the membership analysis in some cases, as briefly mentioned above, and we also improve on several of our methods in comparison to our analysis in Paper I. A complete in-depth description of all updated and improved criteria in our final membership analysis can be found in Chapter~2 of the thesis, and we also refer the reader to the individual notes of Appendix~\ref{ap1:AppendixB}, where we offer a detailed discussion of the membership analysis for all clusters in the sample. Finally, we also refer to Appendix~\ref{ap1:AppendixC}, where we show the individual figures for each cluster corresponding to most of the membership criteria.

We also note that for all clusters, we identified and discarded a series of binary stars, both SB1 (single line spectroscopic binaries) and SB2/3/4 (double and multiple line spectroscopic binaries), which can add significant contamination to our analysis. SB1s were excluded from our kinematic analysis, as they can strongly affect the observed $RV$ distributions, but we decided to include those SB1 stars that had not already been discarded after applying all kinematic and astrometric filters for the rest of our membership criteria, seeing as Li measurements are not affected by SB1 binaries, and they could still be of interest for our analysis. On the other hand, SB2/3/4s were fully discarded from our data sample for all clusters and membership cirteria. All binary stars were identified using the iDR6 data release metadata (via the column \textit{PECULI}) \citep[e.g.,][]{gilmoregaia2021}, as well as existing studies \citep{merle, merle2}. SB1 and SB2/3/4 stars are listed in the long tables of Appendix~\ref{ap1:AppendixD}.

  \subsection{\textbf{Proper motions and parallaxes}} \label{astrometry} 
 
\begin{figure} [h!]
   \centering
   \includegraphics[width=0.9\linewidth]{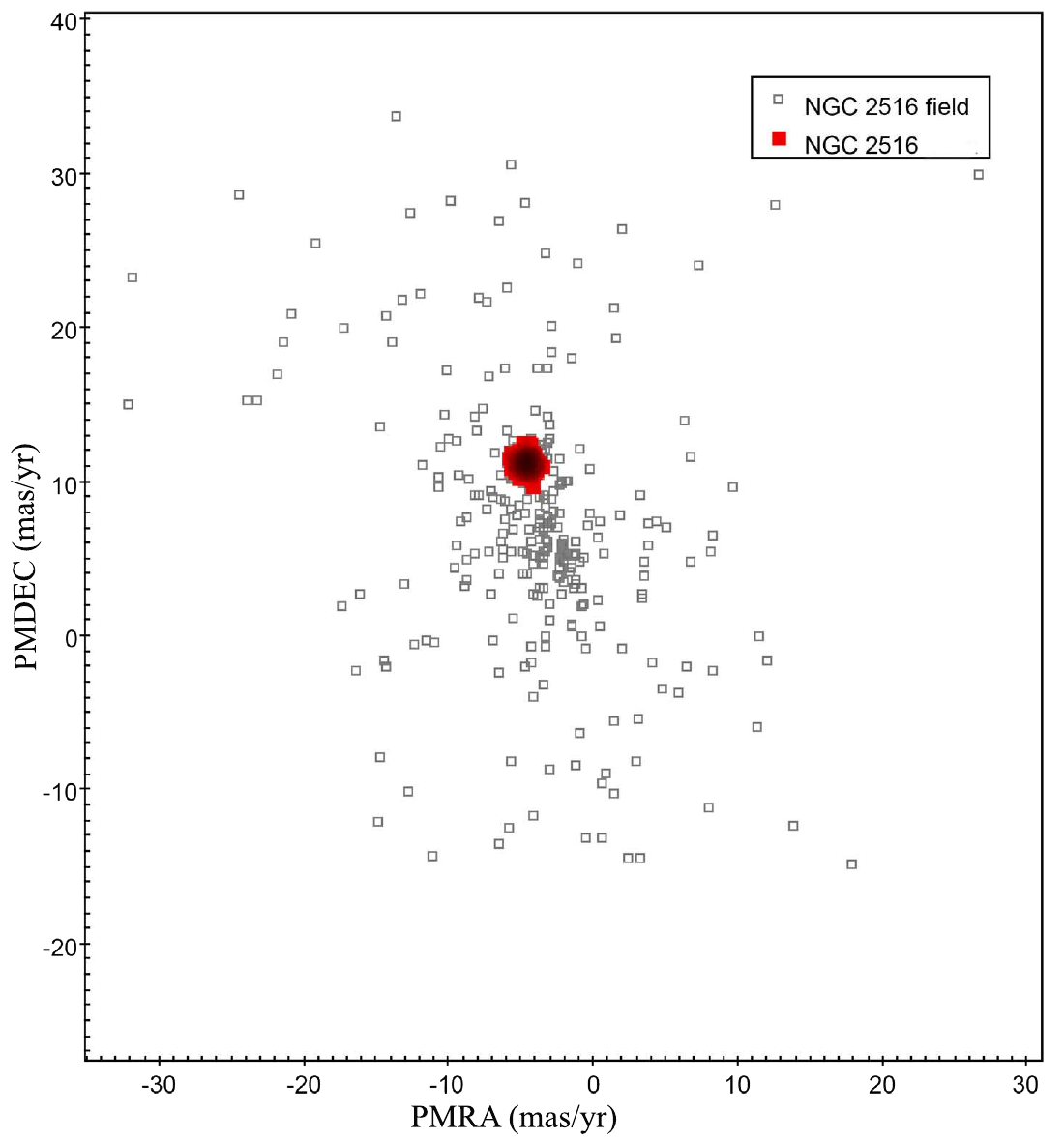}
  \caption[$pmra$-versus-$pmdec$ proper motions diagram for NGC~2516.]{$pmra$-versus-$pmdec$ proper motion diagram showing the final candidate selection (red squares) for NGC~2516, a $125$--$138$~Myr intermediate-age cluster.}
            \label{PMS}
    \end{figure}
    
In our earlier cluster calibration work using iDR4 data as described in Paper I, we made use of several GES membership studies using \textit{Gaia} DR1 and DR2 to reinforce our final candidate selections, thus making use in an indirect way of the precision and robustness of the data provided by \textit{Gaia}. One of the main improvements of the present analysis is to use the proper motions and parallaxes provided by \textit{Gaia} EDR3 to identify robust astrometric candidates among the kinematic selection. 

In order to be able to use \textit{Gaia} data for our cluster sample, we crossmatched the GES iDR6 files with \textit{Gaia} EDR3 by making use of the \textit{CDS Upload X-Match} functionality in TOPCAT\footnote{\url{http://www.star.bris.ac.uk/~mbt/topcat/}}, which allows users to join a local table with any table provided by VizieR\footnote{\url{https://vizier.cds.unistra.fr/viz-bin/VizieR}}, or SIMBAD\footnote{\url{http://simbad.u-strasbg.fr/simbad/}} \citep{simbad, topcat}. We chose a search radius of $5$~arcsec for all clusters in our sample, and did not encounter any problems with duplicates. We then constrained all the \textit{Gaia} measurements according to a series of quality indicators in order to ensure that the proper motions, parallaxes and photometry we used throughout the membership analysis are of sufficient quality regarding their precision, reliability and consistency \citep{lindegren2, lindegren2018}. We here used the renormalized unit weight error \citep[RUWE --][]{lindegren2018, gaiadr3}, only considering for the astrometric study of each cluster the stars with RUWE$<1.4$ \citep{lindegren2, tfm_andres}\footnote{We refer to the following studies for more details on  \textit{Gaia} DR2 and EDR3 data processing and quality indicators \citep{cantatgaudin_gaia, gaiadr2_2, gaiadr2, lindegren2, lindegren2018, randich_gaia, gaiadr3, gaiadr3_2, gaiadr3_3}} (also see Subsect.~\ref{cmd} below for the quality indicators applied to \textit{Gaia} photometry).  

We studied the astrometric goodness of the kinematic members whose associated \textit{Gaia} data fulfilled all these quality indicators by first plotting them in $pmra$-versus-$pmdec$ diagrams and analysing the locus of probable members. As reference in order to identify this location for each cluster, we used the estimated proper motions from the literature listed in Table~\ref{table:2}. As we can see in the example of Fig.~\ref{PMS} for intermediate-age cluster NGC~2516 ($125$--$138$~Myr), all astrometric members can be found clustered together in the expected location in the $pmra$-versus-$pmdec$ diagram as estimated by the literature. This is one of our most restrictive criteria, and so we discarded from our analysis all the kinematic members that strayed significantly from the expected locus. As a limit, we applied the criterion of studies such as \cite{cantatgaudin_gaia} to select sources with proper motions within a maximum of $0.5$~mas~yr$^{-1}$ of the centroid. 

This criterion, combining precise measurements and accurate literature estimates, allowed us to optimally reinforce our list of kinematic members. We were also able to ascertain the astrometric membership of any marginal kinematic stars by studying their position in the proper motions diagrams, fully accepting them as potential candidates if they proved to be robust astrometric members. On the other hand, there were also a number of stars in each cluster to which we were not able to apply astrometric criteria, be it because there were no \textit{Gaia} data available for those GES stars, or because the crossmatched \textit{Gaia} data was filtered out with quality indicators. In those cases, we analysed the stars with the same criteria we used in Paper I (kinematics, gravity indicators, metallicity, and Li). We refer the reader to the individual notes of Appendix~\ref{ap1:AppendixB} for more details on the individual decisions for each cluster in these cases. 

    \begin{figure}[h!]
  \centering
\includegraphics[width=0.95\linewidth]{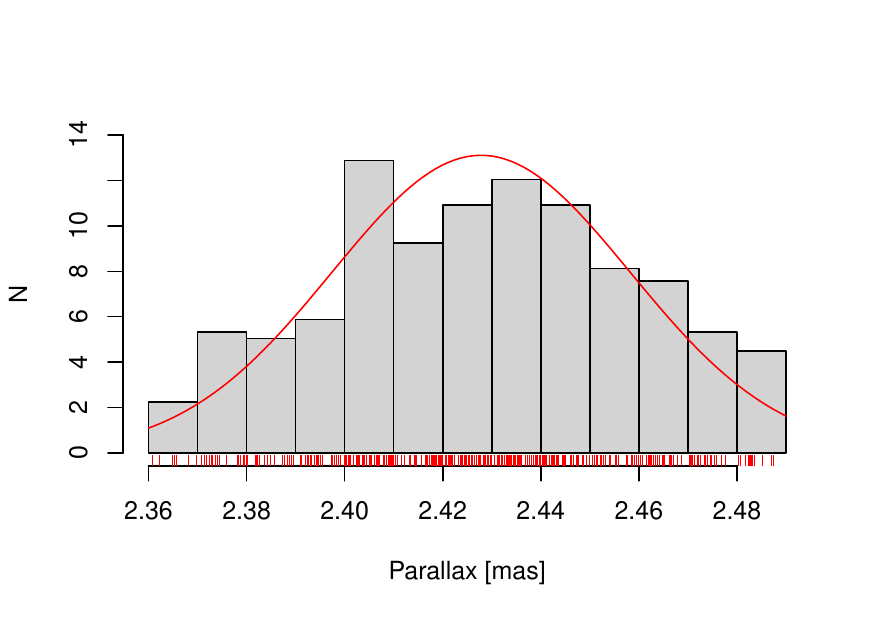} 

\includegraphics[width=0.95\linewidth]{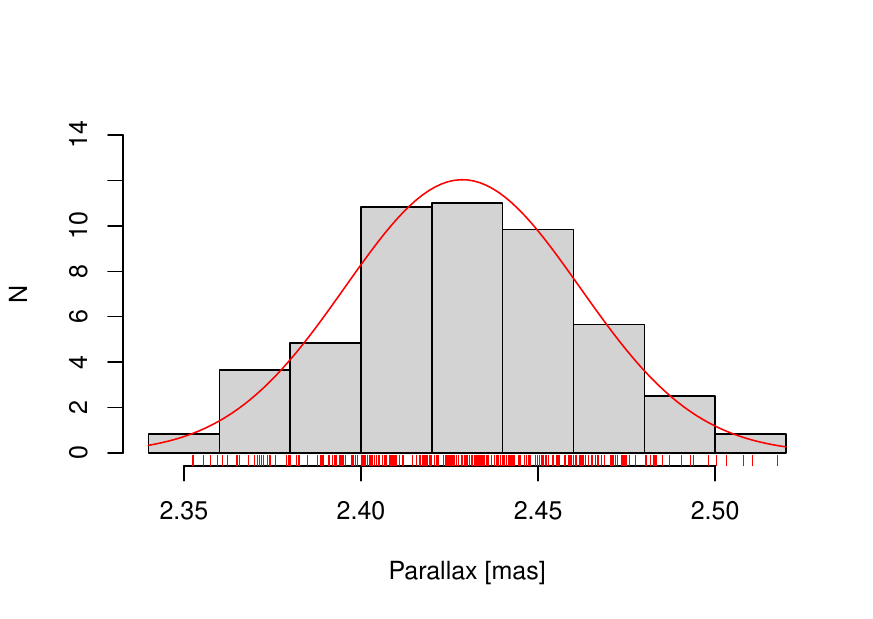}
\caption[Distributions of parallaxes for NGC~2516.]{Distributions of parallaxes and Gaussian fits for the intermediate-age cluster NGC~2516 ($125$--$138$~Myr). We display the histogram both for sources resulting from the 2$\sigma$ clipping procedure on all the GES sources in this field (top panel), and for likely cluster members after applying all of our membership criteria (bottom panel).}
\label{parallax}
\end{figure}

The next step in the astrometric analysis was to study the distributions of parallaxes ($\pi$) for each cluster. Similarly to the analysis of $RV$ and metallicity distributions (as fully discussed in Paper I and the thesis), we fitted the initial parallax distributions for all clusters in the sample, making use of Gaussian curves, applying a 2$\sigma$ clipping procedure on the median, and adopting a 2$\sigma$ limit about the cluster mean yielded by the Gaussian fit to obtain the most likely parallax members. As was also the case with our $RV$ and metallicity selections, we also considered a series of marginal parallax members that fully fulfilled the kinematic and proper motions criteria, and for these borderline members we similarly chose to enlarge the obtained 2$\sigma$ intervals up to a certain threshold to assess them, typically resulting in a slightly larger 3$\sigma$ interval (the exact value depends on each cluster, as is detailed in Appendix~\ref{ap1:AppendixB}).  

In Table~\ref{table:2}, we list all the mean parallax values, their associated dispersions and the resulting 2$\sigma$ intervals for all $38$ clusters analysed, alongside the estimated parallax values from the literature. The updated results for study of the distributions of $RV$ and metallicity are also summarised in Tables~\ref{table:1} and~\ref{table:3}, respectively (see Paper I and Chapter 2 of the thesis). As already discussed in Paper I for the kinematic analysis, the candidate selections for clusters $\gamma$~Vel, NGC~2547 and NGC~2451~A and B were taken from several prior membership studies from the literature. However, in this case we also improved the final selections for these clusters by discarding several spurious stars that deviated appreciably in regards to our astrometry criteria. As in the case of $RV$ and metallicity distributions (similarly further discussed in both Paper I and the published thesis), we also fitted the parallax distributions of our final lists of candidates for each cluster, and compared the obtained final central mean parallaxes and their associated dispersions with those available in the literature for each cluster, finding all of our final estimates to be in agreement. Figure~\ref{parallax} additionally shows an example of the parallax distribution analysis for intermediate-age cluster NGC~2516, comparing the initial fit following the 2$\sigma$ clipping procedure, from which we obtain a preliminary 2$\sigma$ membership interval, with the final distribution of the parallaxes for the final candidates for the cluster. 

We consider astrometry to be one of the most restrictive criteria in our analysis due to its reliability and precision, and thus we fully discarded from our analysis all stars that proved to be non-members according to proper motions, and all stars that fulfilled the proper motions criterion but proved to be non-members according to the study of parallaxes were similarly fully discarded. In contrast to this, other criteria such as gravity indicators and metallicity being less robust, we chose not to make them as restrictive in order to obtain the most probable lists of candidate members (as discussed in more detail in both Paper I and Chapter 2 of the thesis).
 
  \subsection{\textbf{Colour--magnitude diagrams}} \label{cmd} 

\begin{figure} [h!]
   \centering
   \includegraphics[width=0.95\linewidth]{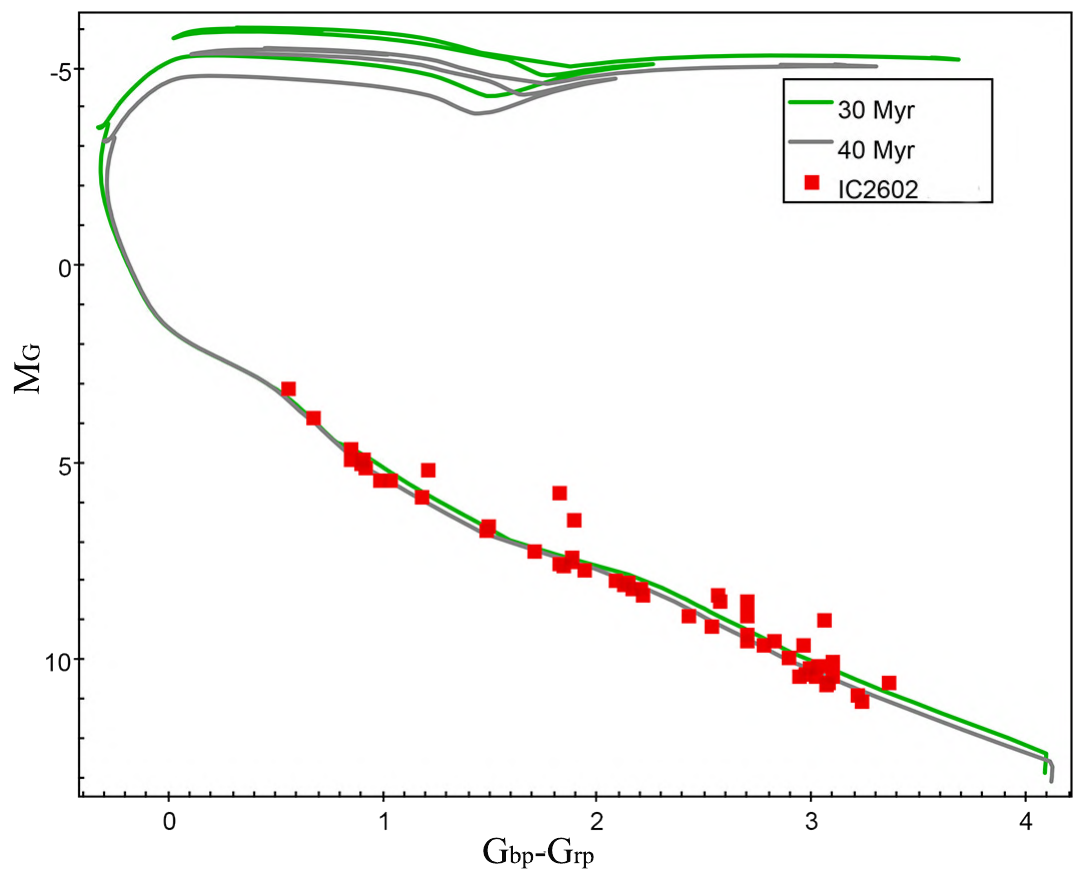}
  \caption[CMD showing the final candidate selection of IC~2602.]{CMD showing the final candidate selection (red squares) of IC~2602, a 35~Myr-old cluster. We overplot the PARSEC isochrones with Z=0.019 for 30~Myr (green curve) and 40~Myr (grey curve).}
            \label{CMD}
    \end{figure}

We made use of the photometry in the $G$, $G_{BP}$, and $G_{RP}$ bands provided by \textit{Gaia} EDR3 to reinforce the selections of kinematic and astrometric candidates and discard field contaminants by means of $M_{G}$-versus-$G_{BP}$-$G_{RP}$ colour--magnitude diagrams \citep[CMDs; e.g.,][]{gaiadr3_2}. The absolute magnitudes $M_{G}$ were calculated using the apparent magnitudes $G$ measured by \textit{Gaia}, as well as the parallaxes ($\pi$) measured in milliarcseconds (mas). In addition to the criterion of RUWE$<1.4$, already applied to astrometric criteria in Subsect.~\ref{astrometry}, to ensure the most precise dataset possible, we used a series of additional quality indicators to the \textit{Gaia} photometry, applying the following filters on the relative magnitude errors on the \textit{G}, $G_{BP}$ and $G_{RP}$ photometric bands: $\sigma_{G}<0.022$~mag, $\sigma_{RP}<0.054$~mag, and $\sigma_{BP}<0.054$~mag \citep{gaiadr2_2, lindegren2018, tfm_christian, tfm_andres, gaiadr3}. 

The use of \textit{Gaia} photometry in CMDs offers a more precise and robust way to assess the membership of our astrometric candidates and discard field contaminants than criteria based on spectroscopic gravity indicators, such as the analysis of the potential candidate list in Kiel diagrams (the $T_{\rm eff}$-versus-$\log g$ plane), or using the $\gamma$ index for the young  clusters in the sample (both criteria are described in detail in Paper I and the thesis). The reason why Kiel diagrams prove to be less reliable is that $\log g$ values are generally less precise, and also scarcer in our sample, and so, we consider the analysis of CMDs to be a far more restrictive and reliable criterion than gravity indicators\footnote{Similarly to surface gravity indicators, [Fe/H] metallicity is not as robust a criterion as kinematics, astrometry and CMDs, as the [Fe/H] values derived from GIRAFFE spectra are widely dispersed and subject to larger uncertainties} (as further discussed in Paper I and Chapter 2 of the thesis). We encountered several cases, to give an example, where a target star fully fulfilled the CMD criterion but appeared to deviate more appreciably in the Kiel diagram, and in such cases we relied more heavily on our CMD analysis due to its superior precision and reliability. Similarly to the astrometric criteria described above, we fully discarded any astrometric candidates from the analysis when they deviated significantly from the expected trend in the CMD in a way that could not be explained by the existing inherent dispersion among the cluster members. We note, however, that most of our astrometric selections already being quite robust, these spurious astrometric candidates were not particularly common in our analysis. 

For all clusters we made use of the PARSEC CMD isochrones \citep{bressan, tfm_andres}\footnote{\url{http://stev.oapd.inaf.it/cgi-bin/cmd}}, choosing the \textit{Gaia} EDR3 photometric system, for ages ranging from 1~Myr to 5~Gyr, and with a metal fraction of Z=0.019 (except for the very low metallicity old cluster NGC~2243, where we used isochrones with Z=0.006). We also took the interstellar reddening and extinction into account when obtaining the isochrones for each cluster age by applying the corresponding $A_{V}$ (mag) extinction values. For each cluster, we calculated the $A_{V}$ extinction in the $V$ band by means of the expression $A_{V}= 3.2E(B-V)$, using the $E(B-V)$ reddening given by \citep{jackson2021} and listed in Tables~\ref{table:01} and ~\ref{table:02}. As an example, in Fig.~\ref{CMD} we present the CMD for the cluster IC~2602, a 35~Myr old young cluster, while the CMDs for all clusters in the sample are shown in Appendix~\ref{ap1:AppendixC}.  

\subsection{\textbf{Comparison with \textit{Gaia} studies}} \label{gaia}

As part of the preliminary work on the membership analysis of the $42$ OCs in the sample, we did an extensive research on each cluster, and listed all available membership studies from the literature that included lists of candidates. In Appendix~\ref{ap1:AppendixB} we delve in detail on the comparisons between all these membership studies and our own final cluster selections. Among these, several membership studies were particularly relevant for our work, conducted from \textit{Gaia} DR1 \citep{randich_gaia}, DR2 \citep{cantatgaudin_gaia, soubiran, bossini} and EDR3 data \citep{jackson2021}. 

We adopted the ages revised by \cite{bossini} for nine of our sample clusters, and the $RV$s from \cite{soubiran} for 28 of them, as reported in Tables~\ref{table:01} and \ref{table:02} for the ages, and Table~\ref{table:1} for the $RV$s, respectively. We also observed that, judging by the measured \textit{Gaia} ages by \cite{bossini}, as well as the empirical Li envelopes constructed by using our cluster candidates (see Paper III) it is possible that some of the former age estimates for the preselected intermediate-age and old clusters could be overestimates - NGC~6005, for example, had a former age estimate of 1.2~Gyr, while \cite{bossini} gives a lower age of $973\pm5$~Myr, which is more in accordance with the Li envelope of our candidate selection. However, we note that we also decided not to use the age estimates by \cite{bossini} for two clusters (NGC~2516 and NGC~6633). The reason for this is that we believe these ages to be overestimates as well, judging by both more recent age estimations, and once again our own candidate selections and obtained empirical Li envelopes (see Appendix~\ref{ap1:AppendixB} for more details on the sample cluster ages). 

We used the other three studies cited above \citep{cantatgaudin_gaia, randich_gaia, jackson2021} as an additional tool to assess our own selections after concluding all membership analyses and applying all the criteria discussed in this section. All these works were of great aid to confirm and reinforce the robustness of our final candidate selections, and were also markedly useful to aid in the confirmation of marginal members in those cases when the membership criteria were not sufficient to fully confirm their membership to the clusters. For the preliminary work published in Paper I,  we mainly made use of the first two studies, conducted from \textit{Gaia} DR1 and DR2 \citep{cantatgaudin_gaia, randich_gaia}, and for the final version of this work extensive use was made of the data in \cite{jackson2021} in order to asses our final member selections. We also note the interest of later membership studies that were published after the completion of the present work \cite{franciosini2022, prisinzano2022}.

\cite{jackson2021} combined GES iDR6 spectroscopic data with the astrometry provided by \textit{Gaia} EDR3 to assign membership probabilities for a target sample of $63$ OCs (including $39$ out of $42$ of the clusters in our sample (all except for NGC~2477, Melotte~71 and M67), as well as $7$ globular clusters. Out of $43211$ targets, \cite{jackson2021} listed $13985$ as highly probable cluster members, with $P>0.9$, and an average membership probability of $0.993$. Similarly to \cite{cantatgaudin_gaia}, the membership selection is purely kinematic and independent of photometry and chemistry and the final selection catalogues from both studies can be successfully combined with other photometric and spectroscopic criteria from GES.

 For each of the clusters considered in these studies, see the individual notes in Appendix~\ref{ap1:AppendixB} for in-depth details regarding the comparison between the candidates listed in these \textit{Gaia} studies and our final member selections. In the online tables described in Appendix~\ref{ap1:AppendixD} we also include for reference these \textit{Gaia} membership selections alongside the columns listing the results of our membership analysis criteria. 


\subsection{\textbf{Identification of Li-rich giant contaminants}}  \label{outliers}

\begin{figure}[h!]
 \centering
\includegraphics[width=0.85\linewidth]{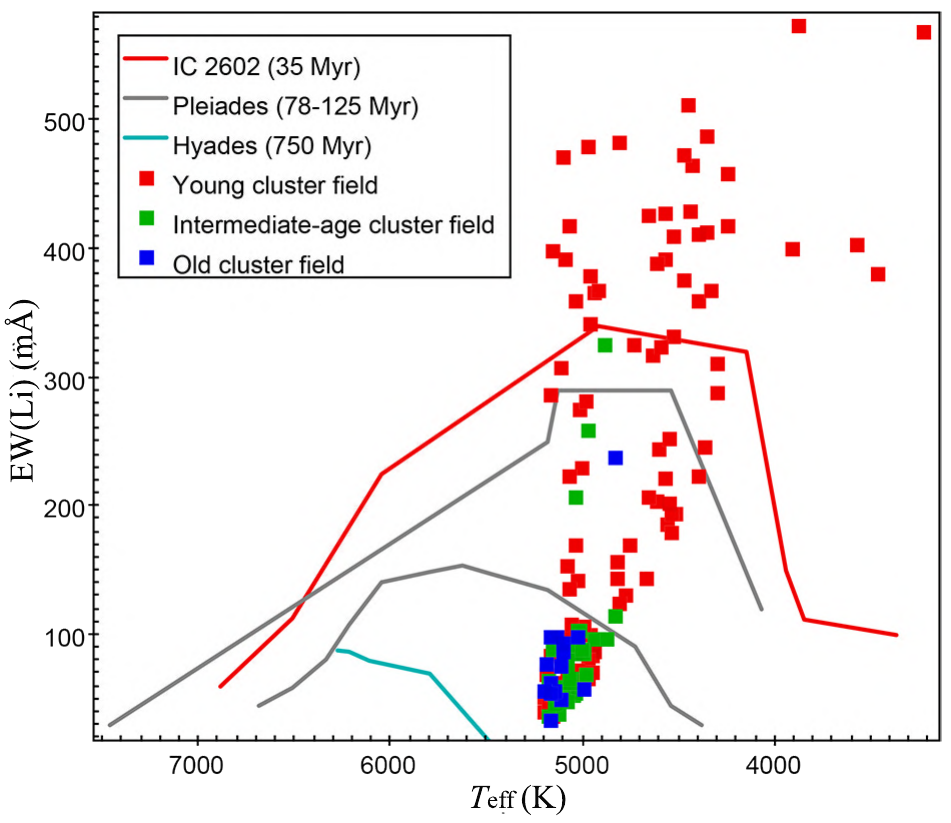} 
\caption[$EW$(Li)-versus-$T_{\rm eff}$ diagram for the Li-rich giant outliers.]{$EW$(Li)-versus-$T_{\rm eff}$ diagram for the Li-rich giant outliers in the field of the young (red squares), intermediate-age (green squares), and old clusters (blue squares). Also shown are the upper envelope of $EW$(Li) for IC~2602 ($35$~Myr, red), the upper and lower envelopes of the Pleiades ($78$--$125$~Myr, grey), and the upper envelope of the Hyades ($750$~Myr, turquoise).}
\label{Outliers1}
\end{figure}

   \begin{figure}[h!]
  \centering
\includegraphics[width=0.95\linewidth]{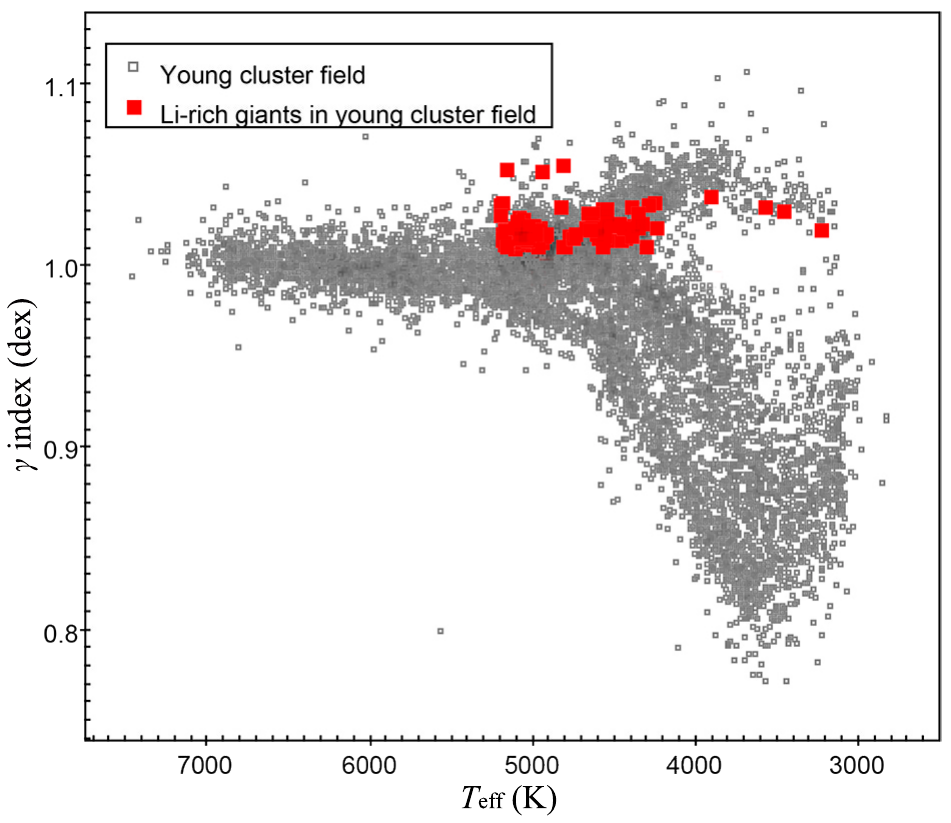}

\includegraphics[width=0.9\linewidth]{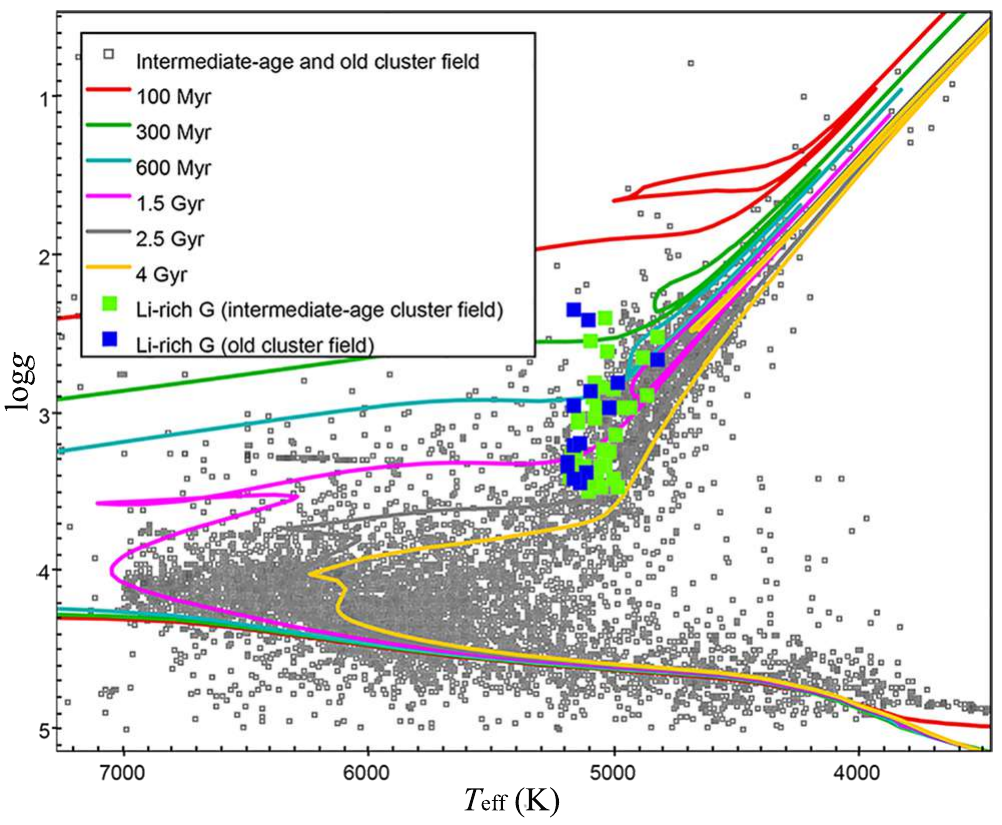}
\caption[Li-rich giant outliers in Kiel and $\gamma$ index-versus-$T_{\rm eff}$ diagrams.]{$\gamma$ index and $\log g$ as a function of $T_{\rm eff}$ for the Li-rich giant outliers (red squares) in the field of the young (top), and intermediate-age (green squares), and old (blue squares) clusters (bottom).}
\label{Outliers2}
\end{figure}

As discussed in more detail in Paper I, gravity indicators help identify giant contaminants in the field of the clusters by plotting the sample stars in the Kiel diagram and the ($\gamma$, $T_{\rm eff}$) plane. Given their interest\footnote{Comprising 1–2$\%$ of FGK giants and supergiants, most are still not well understood, requiring non-standard evolution models to account for the fresh Li detected on their surface (see the thesis and references therein).} \citep[e.g.,][]{smiljanic_giants, magrini2021}, as a parallel result of the membership analysis we also listed some of these outliers for future study, specifically potential Li-rich giants with \textit{A}(Li)$>1.5$. We consider as likely giants any source with $\log g<3.5$ \citep{spina_cha, spina_giants} and/or with $\gamma>1.01$ \citep{damiani, sacco, casey, spina_giants}. We also consider Li-rich giants to have $T_{\rm eff}<5200$~K \citep{casey, spina_giants} and, in the case of stars in the field of young clusters, a lack of H$\alpha$ emission, given that this is a youth indicator for PMS stars \citep{casey}. In Figs.~\ref{Outliers1} and~\ref{Outliers2} we show all Li-rich giant outliers obtained in the field of each of the $42$ clusters in our sample as a result of the membership analysis in diagrams of $EW$(Li), $\log g$, and $\gamma$ as a function of $T_{\rm eff}$, for both the young clusters and the intermediate-age and old clusters in the sample. All Li-rich giants are also listed in Tables~\ref{table:5} and \ref{table:6} in Sect.~\ref{candidates}, as well as in the long tables of Appendix~\ref{ap1:AppendixC}.  

We note that the classification of Li-rich giant stars in this work is only preliminary. We find a large number of potential Li-rich giants in the field of some clusters (e.g., IC~2602) and, while these stars fulfil the adopted criteria ($T_{\rm eff}<5200$~K and \textit{A}(Li)$>1.5$), given the rare nature of these objects, further confirmation would be required to accept them as \textit{bona fide} Li-rich giants. It is also worth noting that we selected all Li-rich giant outliers according to the filters on gravity criteria ($\gamma$ index for young clusters, and $\log g$ for intermediate-age and old clusters) detailed above. In the case of several clusters, we do find some inconsistencies when plotting Li-rich giant outliers in CMDs, with the preselected Li-rich giants sometimes appearing to be non-giants according to photometric data. We detail these instances in the individual notes of Appendix~\ref{ap1:AppendixB}. For the moment, we decided to limit our classification of Li-rich giant contaminants to gravity indicators, as was also done in all cited instances in the literature, but this issue does reinforce the necessity to further confirm the goodness of all selected Li-rich giant outliers in this study. 

         \begin{figure} [htp]
   \centering
\includegraphics[width=0.90\linewidth]{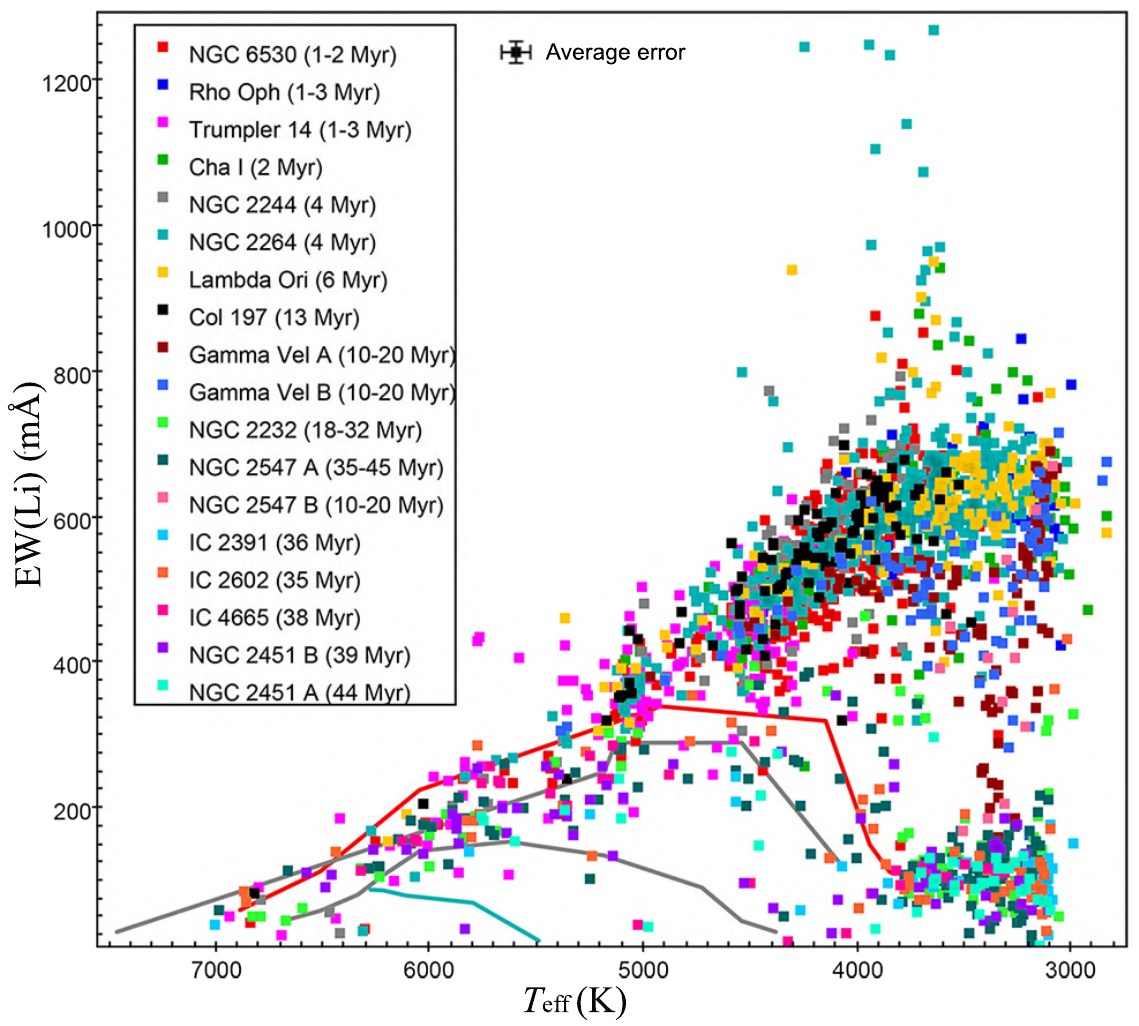}
\includegraphics[width=0.90\linewidth]{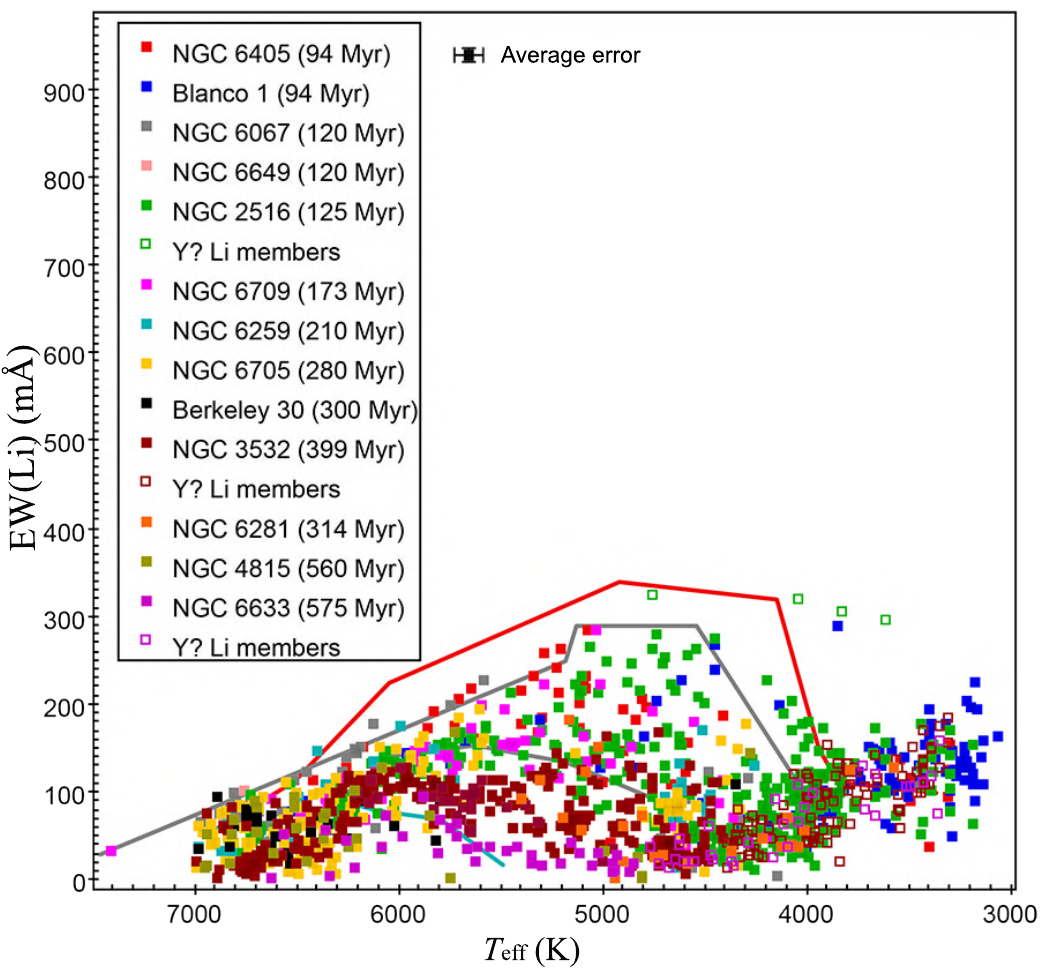}
\includegraphics[width=0.90\linewidth]{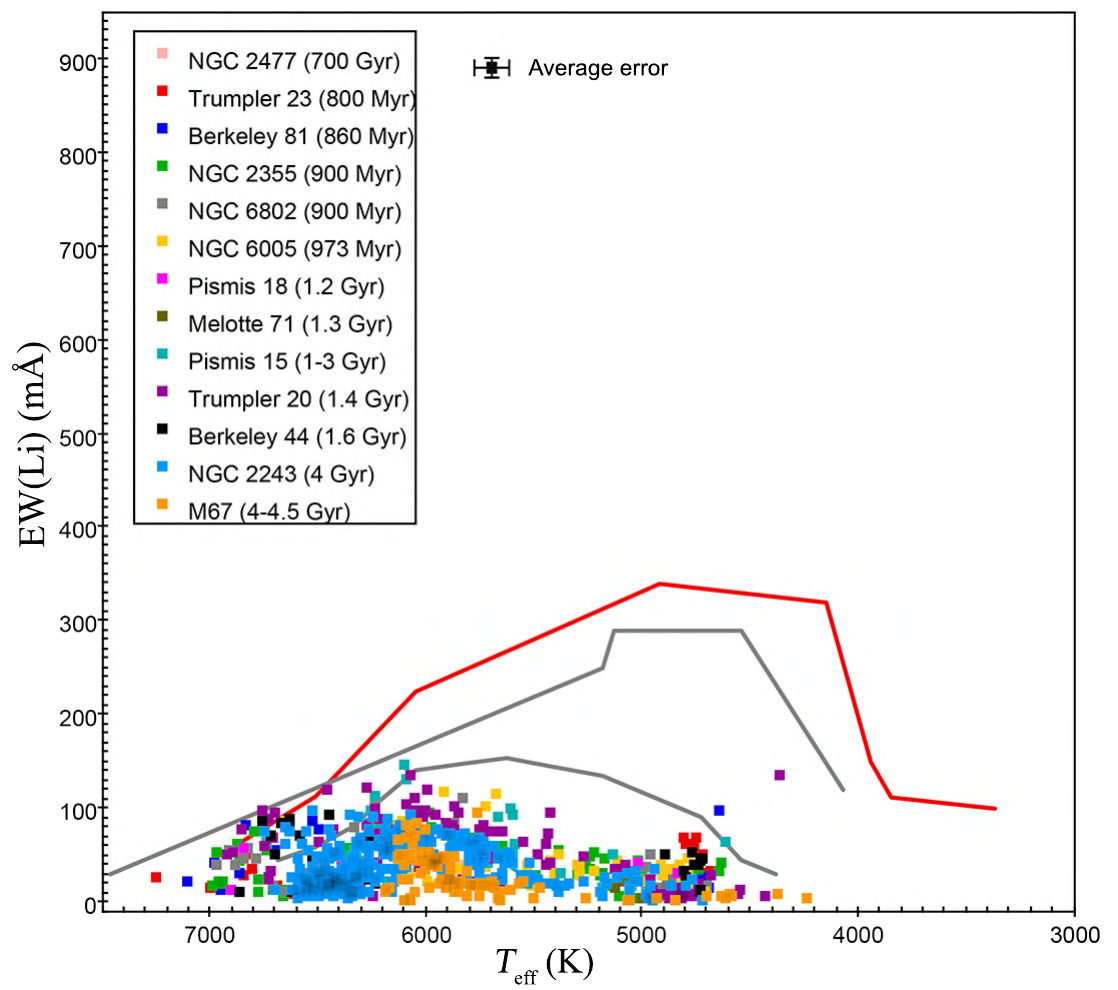} 
\caption{$EW$(Li)-versus-$T_{\rm eff}$ diagrams for the candidate members of the young ($1$--$50$~Myr; top panel), intermediate-age ($50$--$700$~Myr; middle panel), and old clusters ($>700$~Myr; bottom panel). Open squares indicate improbable $EW$(Li) values for some members. Average errors in $T_{\rm eff}$ (K) and $EW$(Li) ($m\r{A}$) are also shown. Reference envelopes for IC~2602, Pleiades and Hyades shown as described in Fig.~\ref{Outliers1}.}
           \label{Final_Li}
   \end{figure}
   
              \begin{figure} [htp]
   \centering
 \includegraphics[width=0.90\linewidth]{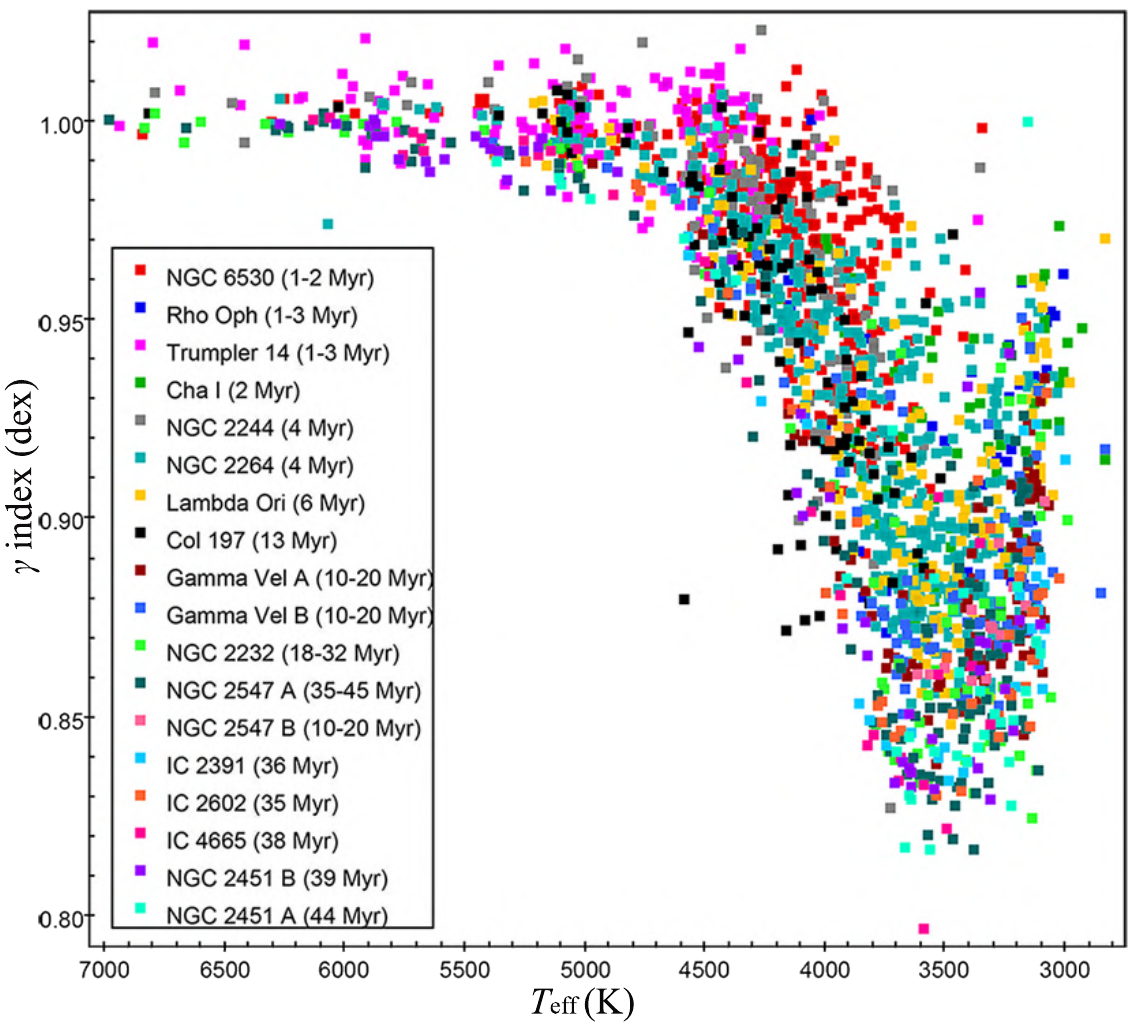}
\includegraphics[width=0.90\linewidth]{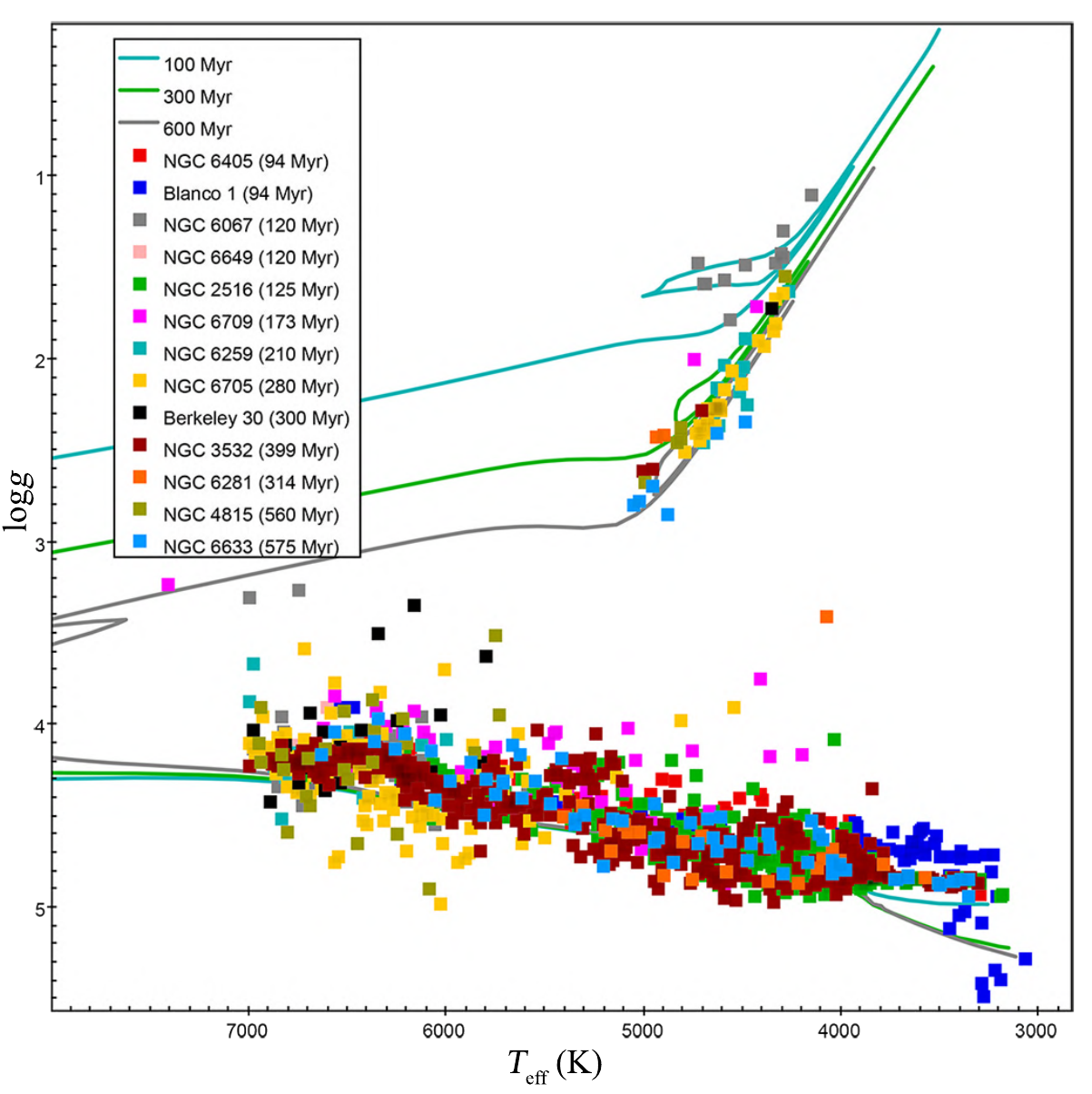}
\includegraphics[width=0.90\linewidth]{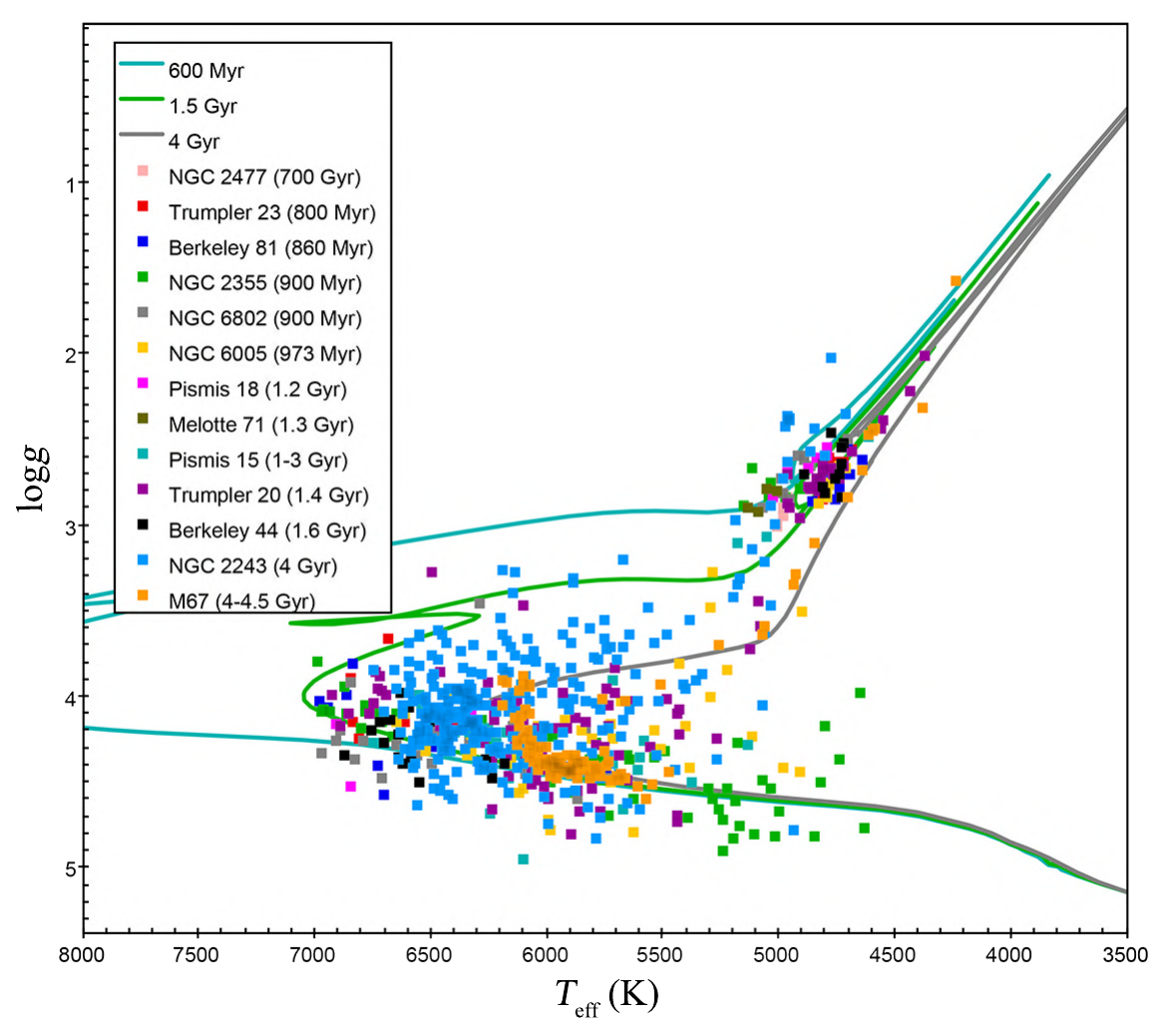} 
\caption{Gravity index $\gamma$ as a function of $T_{\rm eff}$ for the members for all young clusters of the sample (top panel), and Kiel diagrams for all intermediate-age and old clusters for the candidate members of the intermediate-age ($50$--$700$~Myr; middle panel) and old clusters ($>700$~Myr; bottom panel). We overplot the PARSEC isochrones in a similar age range, with a metallicity of Z=0.019.}
             \label{Final_gravity}
    \end{figure}

%

\section{\textbf{Cluster member selections}} \label{candidates}
         
      \begin{table*} [htp] 
       \begin{center}
       \small
      \caption[Main results for the 42 star forming regions and OCs analysed.]{Main results for the 42 OCs analysed, indicating, for each cluster, the number of stars from the sample detected in UVES and GIRAFFE; the number of stars with $EW$(Li) values; the number of stars selected as candidate members (including $RV$, astrometric and Li members, as well as the final members); and the number of Li-rich giant contaminants.}
         \label{table:5}
              \resizebox{\textwidth}{!}{  \begin{tabular}{cccccccccc} 
            \hline
             \hline
             \noalign{\smallskip}
 Cluster$^a$ & \multicolumn{2}{c}{UVES}  & \multicolumn{2}{c}{GIRAFFE}  & \multicolumn{4}{c}{Membership} & Li-rich G  \\ 
         & \small All stars & \small With Li & \small All stars & \small With Li &  \small $RV$ &  \small \textit{Gaia} & \small Li & \small Final  & outliers  \\
 \noalign{\smallskip}
\hline
\noalign{\smallskip}
  NGC~6530 & 52 & 5 & 1931 & 1325 & 470 & 200 & 359 & 343 & 11   \\
  $\rho$~Oph & 23 & 20 & 288 & 277 & 48 & 29 & 37 & 44 &  2   \\ 
  Trumpler~14 & 43 & 11 & 1859 & 1045 & 228 & 55 & 165 & 159 & 11   \\
  Cha~I & 47 & 36 & 660 & 623 & 102 & 54 & 88 & 87 & 6   \\
  NGC~2244 & 8 & 6 & 444 & 385 & 143 & 79 & 123 & 116 & 1  \\
  NGC~2264 & 113 & 70 & 1740 & 1539 & 621 & 423 & 507 & 503 & 10   \\
  $\lambda$~Ori & 116 & 103 & 720 & 675 & 207 & 142 & 163 & 161 & 3  \\
  Col~197 & 8 & 3 & 401 & 363 & 123 & 86 & 104 & 92 & 7   \\
  $\gamma$~Vel &  79 & 50 & 1183 & 1140 & \dots & \dots & \dots & 234 & 1  \\
  NGC~2232 & 47 & 28 & 1822 & 1722 & 750 & 56 & 84 & 68 & 9   \\
  NGC~2547 & 54 & 33 & 423 & 372 & \dots & \dots & \dots & 151 & 0 \\
  IC~2391 & 48 & 26 & 386 & 374 & 56 & 27 & 37 & 35 & 6  \\
  IC~2602 & 131 & 89 & 1721 & 1651 & 309 & 43 & 59 & 55 & 24   \\
  IC~4665 & 34 & 30 & 533 & 514 & 233 & 29 & 51 & 33 &  0 \\
  NGC~2451~A/B & 90 & 70 & 1566 & 1537 & \dots & \dots & \dots & 106 & 10   \\
 \noalign{\smallskip}
\hline
\noalign{\smallskip}
 NGC~6405 & 21 & 12 & 680 & 486 & 251 & 80 & 53 & 51 & 1   \\
 Blanco~1 & 37 & 31 & 426 & 373 & 142 & 119 & 101 & 98 & 0   \\
 NGC~6067 & 27 & 16 & 753 & 327 & 209 & 126 & 60 & 56 & 0   \\
 NGC~6649 & 6 & 0 & 277 & 62 & 42 & 21 & 4 & 2 & 0   \\
 NGC~2516 & 51 & 32 & 708 & 645 & 460 & 378 & 379 & 376 & 0  \\
 NGC~6709 & 10 & 10 & 720 & 590 & 322 & 71 & 53 & 49 & 10   \\
 NGC~6259 & 16 & 14 & 478 & 264 & 125 & 71 & 39 & 35 & 0  \\
 NGC~6705 & 49 & 31 & 1017 & 579 & 391 & 313 & 142 & 139 & 0 \\
 Berkeley~30 & 14 & 13 & 318 & 144 & 78 & 44 & 24 & 24 & 0   \\
 NGC~6281 & 16 & 7 & 304 & 214 & 82 & 38 & 23 & 23 & 1   \\
 NGC~3532 & 67 & 51 & 1094 & 809 & 518 & 411 & 323 & 384 & 11   \\
 NGC~4815 & 14 & 12 & 204 & 92 & 68 & 50 & 30 & 29 & 0 \\ 
 NGC~6633 & 57 & 38 & 1605 & 1463 & 617 & 35 & 62 & 67 & 9  \\ 
 \noalign{\smallskip}
\hline
\noalign{\smallskip}
 NGC~2477 & 11 & 10 & 114 & 0 & 86 & 71 & 9 & 9 & 0   \\
 Trumpler~23 & 16 & 15 & 165 & 70 & 51 & 41 & 25 & 23 & 0   \\
 Berkeley~81 & 14 & 14 & 265 & 159 & 69 & 42 & 25 & 24 & 0  \\
 NGC~2355 & 11 & 11 & 197 & 149 & 119 & 129 & 87 & 86 & 0   \\
 NGC~6802 & 13 & 13 & 184 & 69 & 77 & 51 & 36 & 32 & 2 \\ 
 NGC~6005  & 19 & 19 & 541 & 298 & 174 & 112 & 55 & 49 & 3 \\ 
 Pismis~18 & 10 & 10 & 134 & 72 & 41 & 24 & 12 & 10 & 0  \\ 
 Melotte~71 & 9 & 9 & 111 & 0 & 71 & 64 & 4 & 4 & 0  \\
 Pismis~15 & 11 & 11 & 322 & 201 & 91 & 66 & 33 & 31 & 1 \\
 Trumpler~20 & 29 & 26 & 1184 & 404 & 451 & 367 & 116 & 104 & 0 \\ 
 Berkeley~44 & 7 & 7 & 86 & 73 & 39 & 33 & 31 & 30 & 0  \\ 
 NGC~2243 & 27 & 27 & 634 & 576 & 469 & 446 & 289 & 289 & 8  \\
 M67 & 42 & 36 & 95 & 85 & 109 & 89 & 96 & 96 & 0  \\ 
 [1ex] 
 \hline
 \end{tabular} }
  \tablefoot{
\tablefoottext{a}{Regarding the clusters $\gamma$~Vel, NGC~2547 and NGC~2451 A and B, we directly used the selections obtained by several studies listed in Table~\ref{table:01}.}  }
 \end{center}
\end{table*}

\begin{table*} [htp] 
         \centering
         \fontsize{5}{4}\selectfont
      \caption[]{Results for the $42$ star forming regions and OCs in the sample, indicating, for each cluster: all stars (both UVES and GIRAFFE) from the GES sample and the number of stars detected with measured $EW$(Li) values, the number of stars selected as candidate members, and Li-rich giant field contaminants. Regarding member stars, we provide their percentages with respect to all GES stars and to those with a $EW$(Li) measurement in the field of each cluster.}
         \label{table:6}
         \resizebox{\linewidth}{!}{%
         \begin{tabular}{c c c c c c c c} 
            \hline
             \hline
             \noalign{\smallskip}
Cluster & \multicolumn{2}{c}{iDR6 stars} &  \multicolumn{3}{c}{Members}  & \multicolumn{2}{c}{Li-rich giants} \\ 
         & All & With Li & $\#$ & $\%$(All) & $\%$(with Li)  & $\#$  & $\%$(All) \\
         
 \noalign{\smallskip}
\hline
\noalign{\smallskip}
NGC~6530 & 1983 & 1330  & 343 & 17.3 & 25.8     & 11 &  0.6     \\
$\rho$~Oph      & 311 & 297 &   44      & 13.8 & 14.5 & 2 & 0.6         \\
Trumpler~14     & 1902 & 1056 & 159     & 8.4 & 15.1 & 11 & 0.6                  \\
Cha~I   & 707 & 659      &      87      & 12.3 & 13.2 & 6 & 0.8  \\
NGC~2244 &      452     & 391 & 116     & 25.7 & 29.7 & 1 &     0.2     \\
NGC~2264 &      1853 & 1609     & 503 & 27.1 & 31.3 & 10 & 0.5  \\
$\lambda$~Ori   &       836     & 778 & 161     & 19.3 & 20.7 & 3 &     0.4         \\
Col~197 &       409     & 366   & 92 &  22.5 &  25.1 & 7 & 1.7  \\
$\gamma$~Vel &  1262 & 1190 & 234 &     18.5 &  19.7 & 1 & 0.1  \\
NGC~2232 &      1869 & 1750     & 68 &  3.6     & 3.9 & 9 &     0.5   \\
NGC~2547 & 477 & 405 &  151     & 31.7 & 37.3 & 0 & 0.0         \\
IC~2391 & 434 & 400     & 35 & 8.1 & 8.8 & 6 & 1.4      \\
IC~2602 & 1852 & 1740 & 55 & 3.0 &      3.2     & 24 & 1.3      \\
IC~4665 & 567 & 544     & 33 &  5.8     & 6.1 & 0 & 0.0  \\
NGC~2451~A/B  &  1656 & 1607 &  106 &   6.4     &       6.6 & 10 & 0.6  \\
 \noalign{\smallskip}
\hline
\noalign{\smallskip}
NGC~6405 & 701 & 498 & 51 &     7.3 & 10.2 & 1 & 0.1 \\
Blanco~1 & 463 & 404 & 96 &     20.7 & 23.8     & 0 & 0.0 \\
NGC~6067 & 780 & 343 & 56 &     7.2     & 16.3 & 0 & 0.0 \\
NGC~6649 & 283 & 62 & 2 & 0.7 & 3.2 & 0 & 0.0 \\
NGC~2516 & 759 & 677 & 376 & 49.5 &     55.5 & 0 & 0.0 \\
NGC~6709 & 730 & 600 & 49 & 6.7 & 8.2 & 10 & 1.4 \\
NGC~6259 & 494 & 278 & 35 &     7.1     & 12.6 & 0 & 0.0 \\
NGC~6705 & 1066 & 610 & 139     & 13.0 & 22.8 & 0 & 0.0 \\
Berkeley~30     & 332 & 157     & 24 & 7.2 & 15.3 & 0 & 0.0 \\
NGC~6281 & 320 & 221 & 23 & 7.2 & 10.4 & 1 & 0.3 \\
NGC~3532 & 1145 & 860 & 384 & 33.5 & 44.7 & 11 & 1.0 \\
NGC~4815 & 218 & 104 & 29 &     13.3 & 27.9 & 0 & 0.0 \\
NGC~6633 & 1662 & 1501 & 67 & 4.0 &     4.5 & 9 & 0.5 \\
\noalign{\smallskip}
\hline
\noalign{\smallskip}
NGC~2477 & 125 & 10 & 9 & 7.2 & 90.0 & 0 & 0.0 \\
Trumpler~23     & 165 & 85 & 23 & 13.9 & 27.1 & 0 & 0.0 \\
Berkeley~81     & 279 & 173 & 24 & 8.6 & 13.9 & 0 & 0.0 \\
NGC~2355 & 208 & 160 & 86 & 41.4 & 53.8 & 0 & 0.0 \\
NGC~6802 & 197 & 82 & 32 & 16.2 & 39.0 & 2 & 1.0 \\
NGC~6005 & 560 & 317 & 49 &     8.8 & 15.5 & 3 & 0.5 \\
Pismis~18 &     142     & 82 & 10 & 7.0 & 12.2 & 0 & 0.0 \\
Melotte~71 & 120 & 9 & 4 & 3.4 & 44.5 & 0 & 0.0 \\
Pismis~15 & 333 & 211 & 31 & 9.3 & 14.7 & 1 & 0.3 \\
Trumpler~20     & 1213 & 430 & 104 & 8.6 & 24.2 & 0 & 0.0 \\
Berkeley~44     & 93 & 80 &     30 & 32.3 & 37.5 & 0 & 0.0 \\
NGC~2243 & 661 & 603 & 289 & 43.7 & 48.0 & 8 & 1.2 \\
M67     & 131 & 121 & 96 & 73.3 & 79.3 & 0 & 0.0 \\
 \hline
 \end{tabular} }
\end{table*}

We finish this section with the final results from the membership analysis of all $42$ sample clusters, as summarised in Table~\ref{table:5}. For each cluster, we report \textit{i}) the number of stars from the iDR6 sample observed with both UVES and GIRAFFE; \textit{ii}) those with measured values of $EW$(Li); \textit{iii}) the number of stars selected as candidate members; and \textit{iv}) the number of Li-rich giant outliers obtained as a parallel result during the membership analysis. Readers are directed to the individual notes of Appendix~\ref{ap1:AppendixB}, where we offer a detailed and in-depth discussion of the membership analysis for each cluster, as well as commenting on features of interest regarding individual stars in the selection, and comparing our candidate lists with former membership studies. The full tables resulting from our membership analysis, divided into young, intermediate-age and old age ranges, are provided in online form, and described in Appendix~\ref{ap1:AppendixD}. In these tables we list all membership criteria in our analysis and the final selections of candidate members for each of the $42$ OCs analysed. 
  
 We also show our final selections in the following figures: Figure~\ref{Final_Li} shows the $EW$(Li)-versus-$T_{\rm eff}$ diagrams subdivided into young, intermediate-age, and old clusters. These figures also include for reference the upper envelope of $EW$(Li) for IC~2602 ($35$~Myr, in red), the upper and lower envelopes of the Pleiades ($78$--$125$~Myr, in grey), and the upper envelope of the Hyades ($750$~Myr, turquoise). In Figure~\ref{Final_Li} we also show the representative average errors in $T_{\rm eff}$ and $EW$(Li) for all members of the young, intermediate-age and old clusters. These average errors amount to 68~K and 12~m$\r{A}$ for the young clusters; 69~K and 7~m$\r{A}$ for the intermediate-age clusters; and 65~K and 8~m$\r{A}$ for the old clusters. On the other hand, Fig.~\ref{Final_gravity} shows the $\gamma$-versus-$T_{\rm eff}$ diagram for the young clusters in our sample, as well as the Kiel diagram for all intermediate-age and old clusters in the sample. Additionally, Appendix~\ref{ap1:AppendixC} shows all the individual figures for the cluster sample, including both candidate members as well as Li-rich contaminants of interest. 
  
 Finally, in Table~\ref{table:6} we show some further results of our membership analysis for the $42$ sample clusters. As in Table~\ref{table:5}, we show the number of stars in the field of each cluster from the initial iDR6 sample and the number of candidate stars for all clusters, as well as the Li-rich giant contaminants. With these results we derived percentages of the candidate members and outlier contaminants, which can be used to rank the target clusters and different age ranges in terms of the percentage of candidate members and contaminants identified in each case. Regarding the cluster members, these percentages are considered firstly with respect to all stars in the field of each cluster, and also with respect to all stars that present Li in the initial sample. However, we only present percentages for the Li-rich giant outliers with respect to all stars in the field, given that these were selected taking \textit{A}(Li) and not $EW$(Li) into account.

          
\section{Summary} \label{summary}

This work is a large-scale project, which we started in Paper I \citep{yo} and concluded in Paper III \citep{yo_3}, with the main objective of studying Li as an age indicator and calibrating a Li--age relation for PMS, ZAMS, and early MS FGK late-type stars. With a considerably expanded sample of $42$ OCs spanning a wide age range from $1$--$3$~Myr to $4$--$4.5$~Gyr, in the present work we made use of the most recent available GES-derived data provided by iDR6, as well as of the high-precision data provided by \textit{Gaia} EDR3, in order to conduct a thorough membership analysis and obtain updated and expanded selections of candidates for all target clusters. We summarise our analysis and the main results of this work as follows:

\begin{itemize}
      \item With the present work, we considerably
enlarged and improved our target sample, with $42$ young, intermediate-age and old clusters provided by GES iDR6 at our disposal, as compared to our former sample of $20$ clusters using data from GES iDR4 from Paper I. The new homogeneous analysis done in iDR6 also offers marked improvements to both our cluster sample and our consequent analysis, including a significantly larger number of stars measured in the field of many of the clusters, the recalculation of parameters for higher accuracy, and accounting for the contamination of background nebular lines (as detailed in Sect.~\ref{data}). 
      
      \item We carried out extensive preliminary bibliographical research on each of the $42$ sample clusters, compiling a thorough selection of literature data on ages, distances, and reddening values (see Tables~\ref{table:01} and~\ref{table:02} in Sect.~\ref{data}), as well as $RV$s (Table~\ref{table:1} in Sect.~\ref{analysis}), proper motions, parallaxes (Table~\ref{table:2} in Sect.~\ref{analysis}), [Fe/H] metallicity (Table~\ref{table:3} in Sect.~\ref{analysis}), previous Li measurements, and numerous existing membership studies (as also presented in Tables~\ref{table:01} and~\ref{table:02}, as well as in Appendix~\ref{ap1:AppendixB}). 
      
      \item We performed exhaustive and detailed membership analyses (see Sect.~\ref{analysis}), firstly studying the $RV$ distributions to obtain likely kinematic candidates, and complementing this with a thorough study of the proper motions and parallaxes provided by \textit{Gaia}, a marked improvement in regards to our former analyses in Paper I. Gravity indicators such as $\log g$ and the $\gamma$ index were helpful in order to discard field giant contaminants and confirm the membership of the astrometric selections, also significantly improved by \textit{Gaia} photometry in CMDs. We complemented all these criteria with a study of the distribution of [Fe/H] metallicity, which helped reinforce our cluster selections and discard further rogue contaminants. Finally, we used Li as a final criterion by plotting the candidates in $EW$(Li) versus $T_{\rm eff}$ diagrams.

      \item We obtained selections of robust candidates for all $42$ sample clusters, some of which had not been previously studied in detail by the GES consortium (see Table~\ref{table:5} in Sect.~\ref{candidates}), and we also discuss which individual clusters and age ranges present the highest percentages of members in our target sample in Table~\ref{table:6}. All individual figures are displayed in Appendix~\ref{ap1:AppendixC}, and this paper also includes descriptions for the associated online long tables of results, described in Appendix~\ref{ap1:AppendixD}, in which we list the final candidates for all clusters and specify which membership criteria are fulfilled by each of the target stars for each cluster.

      \item We made use of a number of studies from the literature conducted from \textit{Gaia} DR1, DR2, and EDR3 data to do an in-depth assessment of our candidate selections after concluding our membership analysis, typically obtaining coherent results, as we found our lists of cluster members to be in general agreement with these previous GES studies (see Appendix~\ref{ap1:AppendixB}). In addition, we also fitted the distributions of $RV$, parallaxes, and [Fe/H] metallicity values of all final cluster selections, as displayed in Tables~\ref{table:1} (for $RV$s),~\ref{table:2} (for parallaxes), and~\ref{table:3} (for metallicity), obtaining mean values that are consistent with the literature for the majority of clusters, with very few exceptions.

     \item Given their importance for our understanding of stellar Li, as an additional result of our membership analysis we also selected a series of preliminary Li-rich giant outliers for $23$ out of the $42$  target clusters (see Sect.~\ref{outliers}, as well as Table~\ref{table:5} in Sect.~\ref{candidates} and Appendix~\ref{ap1:AppendixD}). Given the scarcity of these objects, we require further confirmation to accept all stars listed as Li-rich giant contaminants in this study.
      
      \item The selections of cluster candidates obtained by means of recent GES and \textit{Gaia} data can be used by the scientific community in multiple applications, from the calibration of a Li--age relation ---the aim of this particular work--- to further analysis of (the same or other) cluster members, or the characterisation of a variety of stellar parameters and other areas of interest regarding OCs and the formation and evolution of stars in the Galaxy. 

       \item In the companion paper to this work, Paper III, we conclude our calibration of an empirical Li--age relation and obtain a set of empirical Li envelopes for several key ages in our sample. This final paper will also describe our future work.
\end{itemize}
  
\begin{acknowledgements}

We acknowledge financial support from the Universidad Complutense de Madrid (UCM) and the Agencia Estatal de Investigaci\'on (AEI/10.13039/501100011033) of the Ministerio de Ciencia e Innovaci\'on and the ERDF ``A way of making Europe'' through projects 
PID2019-109522GB-C5[4] and PID2022-137241NBC4[4].
We acknowledge the support from INAF and Ministero dell' Istruzione, dell' Universit\'a' e della Ricerca (MIUR) in the form of the grant "Premiale VLT 2012". 
T.B. was supported by grant No. 2018-04857 from the Swedish Research Council.
J.I.G.H. acknowledges financial support from the Spanish Ministry of Science, Innovation and Universities (MICIU) under the 2003 Ram\'on y Cajal program RYC-2013-14875, and also from the Spanish Ministry project MICIU AYA2017-86389-P.  
E.J.A. acknowledges financial support from the State Agency for Research of the Spanish MCIU through the “Center of Excellence Severo Ochoa" award to the Instituto de Astrof\'sica de Andaluc\'a (CEX2021- 001131-S).
E. M. acknowledges financial support through a "Margarita Salas" postdoctoral fellowship from Universidad Complutense de Madrid (CT18/22), funded by the Spanish Ministerio de Universidades with NextGeneration EU funds.
F.J.E. acknowledges support from ESA through the Faculty of the European Space Astronomy Centre (ESAC) - Funding reference 4000139151/22/ES/CM.
Based on data products from observations made with ESO Telescopes at the La Silla Paranal Observatory under programme focusID 188.B-3002. 
These data products have been processed by the Cambridge Astronomy Survey Unit (CASU) at the Institute of Astronomy, University of Cambridge, and by the FLAMES/UVES reduction team at INAF--Osservatorio Astrofisico di Arcetri. 
These data have been obtained from the GES Data Archive, prepared and hosted by the Wide Field Astronomy Unit, Institute for Astronomy, University of Edinburgh, which is funded by the UK Science and Technology Facilities Council.
The results presented here benefit from discussions held during GES workshops and conferences supported by the ESF (European Science Foundation) through the GREAT Research Network Programme. 
This work has made use of data from the European Space Agency (ESA) mission {\it Gaia} (\url{https://www.cosmos.esa.int/gaia}), processed by the {\it Gaia} Data Processing and Analysis Consortium (DPAC, \url{https://www.cosmos.esa.int/web/gaia/dpac/consortium}). Funding for the DPAC has been provided by national institutions, in particular the institutions participating in the {\it Gaia} Multilateral Agreement. 
This publication makes use of the VizieR database \citep{vizier} and the SIMBAD database \citep{simbad}, both operated at CDS, Centre de Donn\'ees astronomiques de Strasbourg, France. 
This research also made use of the WEBDA database, operated at the Department of Theoretical Physics and Astrophysics of the Masaryk University, and the interactive graphical viewer and editor for tabular data TOPCAT \citep{topcat}. 
For the analysis of the distributions of $RV$ and metallicity we used RStudio Team (2015). Integrated Development for R. RStudio, Inc., Boston, MA (\url{http://www.rstudio.com/}). Finally, we would like to thank the anonymous referee for helpful comments and suggestions.
\end{acknowledgements}


\bibliographystyle{aa} 

\begin{appendix}



%
%

\section{Cluster selections: Individual notes}
\label{ap1:AppendixB}

In this appendix we present an in depth account of each of the $42$ clusters in the sample, including general information about the different age estimations of the cluster sample, as well as a detailed discussion on membership analysis and final selections from Sect.~\ref{analysis}, especially featuring relevant particular cases, as well as a comparison of the final candidates with a series of studies from the literature. In Paper III, we further comment on particular cases in regards to the study of rotation, activity and metallicity. We additionally refer to the individual figures of Sect.~\ref{analysis} in Appendix~\ref{ap1:AppendixC}.

\subsection{Star forming regions (age~$\leq6$~Myr) and young OCs (age~$\leq50$~Myr)}

\textbf{$\ast$ NGC~6530}  

NGC~6530 is a $1$--$2$~Myr star forming region \citep{prisinzano2005, damiani2019, wright2019, randich2020, jackson2021} associated to the HII region M8 (Lagoon Nebula).

Of the final 359 Li candidates for NGC~6530 we discarded 16 stars according to the following criterion: We decided to discard all stars which we could not analyse using the astrometric and photometric \textit{Gaia} data (be it because there were not \textit{Gaia} data available for those stars, or because said data did not pass the quality indicators) if they were additionally listed as clear non-members by \cite{jackson2021}. All of these 16 stars are listed as clear non-members with a probability of $0.00$--$0.16$, and one of them also has a metallicity value deviating appreciably from the mean of the cluster. As a result of the analysis of NGC~6530 we found 14 strong accreting stars, all of which are Li members and classified as high-probability members (P $>0.90$) by \cite{jackson2021}. 
As for kinematics, 44 stars in our final selection were $RV$ non-members according to the initial 2$\sigma$ interval, with $RV$s deviating up to $13.0$--$15.0$~km~s$^{-1}$ from the mean of the cluster. We accepted the stars which deviate from the limits established by the 2$\sigma$ interval by $\sim$10~km~s$^{-1}$ given that they fulfil the rest of membership criteria (especially regarding available \textit{Gaia} data, as well as gravity indicators and lithium) (see Paper I). All of these stars are additionally listed as high-probability members by several studies \citep[e.g, ][]{damiani2019, prisinzano2019, wright2019, jackson2021}. On the other hand, we also accepted as final candidates four stars with no measured $RV$ values in iDR6, due to the fact that they fully fulfil the astrometry criteria (as well as the other criteria), and are also listed as members by these studies \citep[e.g, ][]{damiani2019, prisinzano2019, wright2019, jackson2021}. Finally, regarding metallicity, 26 stars in our final selection are marginal members which deviate moderately from the limits of the initial 3$\sigma$ membership interval, with values up to $0.40$--$0.80$~dex from the mean of the cluster. As discussed in Sect.~\ref{analysis} (also see Paper I), we accepted these marginal stars as final cluster members, seeing as they fulfil all prior criteria (including the more robust criteria of cinematics and astrometry), and all of these stars are also listed as high-probability members by various studies \citep[e.g, ][]{damiani2019, prisinzano2019, wright2019, jackson2021}. In addition, five of the stars in our final selection additionally present [Fe/H] values which deviate more appreciably from the 3$\sigma$ limit, up to $0.93$--$1.68$~dex from the cluster mean. As further discussed in both Paper I and the updated analysis presented in the thesis, however, we also listed these stars as final candidates, given that they fully fulfil the rest of criteria, including the more restrictive criteria, and they are all additionally listed as members by \cite{jackson2021} with P=$0.99$. 

In regards to previous selections, we found the following number of common stars with our selection: 52 stars in \cite{prisinzano2007} and \cite{prisinzano2012}, 148 in \cite{spina_giants}, 255 in \cite{damiani2019}, 320 in \cite{prisinzano2019}, 286 in \cite{wright2019}, and one in \cite{castro_ginard}. Finally, \cite{jackson2021} includes 331 out of our 343 candidates. Of the remaining 12 stars, 10 of them were not included in their analysis, and two of them are listed with probabilities of $0.26$--$0.33$ (\cite{jackson2021} considers high-probability members to have P $>0.90$ and definite non-members to have P$<0.10$). We accepted these two stars given that they fulfilled all of our available membership criteria, and were furthermore included in several other studies \citep[e.g, ][]{damiani2019, prisinzano2019, wright2019}. As already discussed in Subsect.~\ref{outliers}, we also note that we seem to find some inconsistencies when plotting the Li-rich giant outliers selected in the field of this cluster in the CMD diagram, with a couple of them appearing among the cluster candidates. However, when plotting them in the $\gamma$ index-versus-$T_{\rm eff}$ diagram the expected clear distinction is found between the non-giant cluster candidates and the Li-rich giant outliers. \\

\textbf{$\ast$ Rho~Oph}  


The $\rho$~Ophiuchi ($\rho$~Oph) molecular cloud complex, consisting of two major regions of dense gas and dust (L1688 and L1689), is an $1$--$3$~Myr star forming region \citep{barsony, rigliaco, spina_giants, canovas, randich2020, kiman, jackson2021}. 

In addition to the final 37 Li members, for $\rho$~Oph we also considered for completeness seven strong accreting stars which presented no values of $EW$(Li) in the iDR6 file.  We also note that one of the stars in our final selection (16270456-2442140) presents a [Fe/H] value which deviates appreciably from the 3$\sigma$ limit ($0.45$~dex from the cluster mean). We listed it as a final candidate given that it fully fulfils all the other criteria, and is additionally listed as a member by \cite{canovas} and \cite{jackson2021} (with P=$0.99$). Of the final members in $\rho$~Oph, 29 belong to the star forming region L1688 \citep{rigliaco}, and 15 of them are also strong accretors with H$\alpha$10\%~$>270$--$300$~km~s$^{-1}$. Of these 15 accreting stars, only one (16273311-2441152) does not pass our gravity criteria. This star, with $\gamma$=1.022 and \textit{A}(Li)=3.2, could be listed as a potential Li-rich giant contaminant, but due to the fact that it is also a strong accretor (with H$\alpha$10\%~$=469.13$~km~s$^{-1}$, we counted it as a likely member of $\rho$~Oph, also in accordance with \cite{rigliaco} and \cite{jackson2021}. Regarding other membership studies from the literature, \cite{rigliaco} has 42 of our final candidates in their selection, while \cite{ducourant} lists 10 common members, \cite{spina_giants} also lists 10 common stars for $\rho$~Oph, and \cite{canovas} obtained 35 stars in common with our selection. Finally, we have 42 stars in common with \cite{jackson2021}. The remaining two stars in our selection for $\rho$~Oph were not included in the analysis of \cite{jackson2021}, but are however listed as members by \cite{rigliaco} and fulfil all our membership criteria. \\

\textbf{$\ast$ Trumpler~14}  


Trumpler~14 is a $1$--$3$~Myr star forming region \citep{sampedro, randich2020, jackson2021} associated to the Carina Nebula, one of the most massive HII regions in the Galaxy. 
Of the final 165 Li candidates in Trumpler~14 we discarded six stars we could not analyse using the astrometric and photometric \textit{Gaia} data, and they were additionally listed as non-members by \cite{jackson2021} (most of them are clear non-members with a probability of $0.00$, one of them is listed with P=$0.27$, but it is similarly discarded). One of these six stars also has a metallicity value deviating appreciably from the mean of the cluster. We also found five strong accreting stars among our final members, all of which are Li members and classified as high-probability members by \cite{jackson2021}. We also note that we discarded an F-type star with an $EW$(Li) that seemed to be higher than we would expect for a star forming region ($150$~m$\r{A}$). This star had no recorded measurements of either accretion or H$\alpha$ that could perhaps explain its $EW$(Li), and it was additionally listed as a non-member by \cite{damiani2017}. Regarding kinematics, five stars in our final selection were $RV$ non-members according to the initial 2$\sigma$ interval, with $RV$s deviating up to $13.0$--$15.0$~km~s$^{-1}$ from the mean of the cluster. We accepted the stars which deviate from the limits established by the 2$\sigma$ interval by $\sim$10~km~s$^{-1}$, given that they fulfil the rest of membership criteria (especially regarding available \textit{Gaia} data, as well as gravity indicators and lithium. All of these stars are additionally listed as high-probability members by \cite{jackson2021}. In addition, we also accepted as final candidates 24 stars with no measured $RV$ values in iDR6, due to the fact that they fully fulfil the astrometry criteria (as well as the other criteria), and are also listed as members by \cite{jackson2021} with high probabilities in the range of $0.85$--$1.00$. Finally, we also accepted as final cluster candidates 11 marginal members deviating from the 3$\sigma$ metallicity limit up to $0.26$~dex from the mean of the cluster. All of these stars fulfil all other criteria, and 10 of them are listed as high-probability members by \cite{jackson2021} (with P=$0.86$--$0.99$).

Regarding other membership selections from the literature, we found the following number of common stars with our selection: 63 stars in \cite{spina_gamma}, as well as 65 in \cite{damiani2017}, 51 stars in \cite{cantatgaudin_gaia}, and one star in \cite{castro_ginard}. Finally, \cite{jackson2021} includes 153 out of our 159 candidates (the remaining six were not included in their analysis). We also note that \cite{jackson2021} includes several stars which we initially discarded for not fulfilling our gravity criteria of $\gamma<1.01$ for non-giants (see Subsect.~\ref{outliers}). The reason for this is that \cite{jackson2021} considers a less restrictive criterion of $\gamma<1.33$. We also note that we seem to find some inconsistencies when plotting the Li-rich giant outliers selected in the field of this cluster in the CMD diagram, with eight of them appearing among the cluster candidates. However, when plotting them in the $\gamma$ index-versus-$T_{\rm eff}$ diagram the expected distinction is found between the non-giant cluster candidates and the Li-rich giant outliers. \\

\textbf{$\ast$ Chamaeleon~I}  

Chamaeleon~I (Cha~I) is a $\sim$2 Myr star forming region \citep{lopez, spina_cha, sacco, randich2020, jackson2021}, composed of two subclusters \citep{luhman_2, sacco2, roccatagliata}, with a shift in velocity of $\sim$1~km~s$^{-1}$, and ages in the range of $5$--$6$~Myr and $3$--$4$~Myr, respectively. 

Of the final 88 Li candidates for Cha~I we discarded two stars: The first one (11080297-7738425) had no \textit{Gaia} data available (thus, astrometry could not be analysed) and its [Fe/H] value deviated appreciably from the mean of the distribution, beyond the extended 3$\sigma$ membership interval. The second star (11095340-7634255), also a non-member according to metallicity, we considered as a final non-member because we could not analyse this star using the astrometric and photometric \textit{Gaia} data and it was additionally listed as a clear non-member with a probability of $0.00$ by \cite{jackson2021}. 
As a result of the analysis of Cha~I we found 35 strong accreting stars, of which only one (11104959-7717517) is not a Li member, probably due to possible veiling suppressing the absorption Li line. We classified this strong accretor as an additional likely member, and it is additionally considered as a member of Cha~I by \cite{jackson2021}. We also decided to classify as final cluster members four very young active stars which exhibit little dispersion with respect to the rest of candidates, but also present values of $EW$(Li) which are slightly larger than in the case of most member stars ($841$--$948$~m$\r{A}$). Two of them, the ones with the largest $EW$(Li)s, are additionally strong accretors with H$\alpha$10\%~$>270$--$300$~km~s$^{-1}$, and so their larger $EW$(Li)s are a potential effect of either accretion-induced enhancement, or, more probably in this case, a result of a strong veiling correction, where derived $EW$s may be too high if the veiling factor is overestimated. For more details on chromospheric activity and accretion mechanisms in regards to Li depletion, we refer the reader to Paper III. We also note that three of the stars in our final selection were $RV$ non-members according to the initial 2$\sigma$ interval, with $RV$s deviating up to $3.7$--$10.2$~km~s$^{-1}$ from the mean of the cluster. We accepted the stars which deviate from the limits established by the 2$\sigma$ interval by $\sim$10~km~s$^{-1}$, given that they fulfil the rest of membership criteria (especially regarding gravity indicators and lithium, as only one of them has available \textit{Gaia} data), and all three of them are additionally listed as candidates by several studies \citep[e.g, ][]{esplin, sacco2, galli, jackson2021}. Regarding metallicity, we also accepted seven marginal members (three of which are also strong accreting stars), deviating up to $0.45$~dex from the mean of the cluster; as well as one additional star (11091172-7729124) which deviated more appreciably from the 3$\sigma$ limit ($0.80$~dex from the cluster mean). All of these stars fulfil all other criteria, and are listed as high-probability members with P=$0.99$ by \cite{jackson2021}.

Regarding previous selections, we found the following number of common stars with our selection: nine stars in \cite{robrade_schmitt}, 23 in \cite{luhman}, and 47 in \cite{lopez}. We found 12 common stars in \cite{spina_cha}, all UVES members in their selection, except for 10555973-7724399 and 11092378-7623207, for which several parameters are not released in the iDR6 catalogue. We also found 88 of our candidate stars in the member list of \cite{esplin}, 75 in \cite{sacco2}, and 66 in \cite{galli}. Finally, \cite{jackson2021} includes 82 out of our 87 candidates (the remaining five were not included in their analysis).  Regarding field contaminants, one Li-rich giant (11000515-7623259) is listed in \cite{casey}. \\

\textbf{$\ast$ NGC~2244} 


NGC~2244 is a $2$--$4$~Myr star forming region \citep{michalska, muzic, jackson2021}, associated with the Rosette Nebula in the Perseus Arm of the Galaxy. 

Of the final 123 Li candidates for NGC~2244, we discarded 11 stars which we could not analyse using the astrometric and photometric \textit{Gaia} data, and which were additionally listed as non-members by \cite{jackson2021} (all but one are definite non-members with a probability of $0.00$, while the remaining one is listed with P=$0.29$, and is similarly discarded).We also found 10 strong accreting stars among our final members, and all but one are Li members and classified as high-probability members by \cite{jackson2021} (the remaining one has no measured $EW$(Li) in DR6 and is not included in \cite{jackson2021}). Regarding kinematics, one star in our final selection was a $RV$ non-member according to the initial 2$\sigma$ interval, with a $RV$ deviating $6.0$~km~s$^{-1}$ from the mean of the cluster. We accepted it given that it fulfilled the rest of membership criteria, and it was also listed as a high-probability member by \cite{jackson2021}. On the other hand, we also accepted as final candidates nine stars with no measured $RV$ values in iDR6, due to the fact that they fully fulfil the astrometry criteria (as well as the other criteria), and are also listed as members by \cite{jackson2021} with probabilities in the range of $0.53$--$0.99$. As for metallicity, we accepted 22 marginal members which deviate up to $0.35$~dex from the mean of the cluster, and also one additional star (06312952+0454342) deviating more appreciably from the 3$\sigma$ limit ($0.45$~dex from the cluster mean). All of these stars fulfil the rest of criteria, and are listed as high-probability members with P=$0.99$ \cite{jackson2021}. Regarding other membership selections from the literature, we found the following number of common stars with our selection: 78 stars in \cite{cantatgaudin_gaia}, one star in \cite{carrera2019}, and 39 stars in \cite{michalska}. Finally, \cite{jackson2021} includes 114 out of our 116 candidates (the remaining two were not included in their analysis). We also note that we included several stars with a marginal metallicity, deviating slightly from the 2$\sigma$ interval, as they fulfilled the rest of criteria and were furthermore reinforced by high probabilities listed by \cite{jackson2021}. \\

\textbf{$\ast$ NGC~2264} 
\label{ap1:NGC2264}


NGC~2264 is a $4$~Myr star forming region \citep{jackson2021}, the dominant component of the Mon OB1 association in the Monoceros constellation. Several studies list ages in the $1$--$5$~Myr range \citep{spina_giants, cantatgaudin_gaia, venuti, arancibia_silva, bonito2020, gillen, randich2020}. 

Of the final 507(535) Li candidates for NGC~2264, we discarded 28 stars on the basis of not being able to analyse them using astrometric and photometric \textit{Gaia} data, and all of these 28 stars were additionally listed as non-members by \cite{jackson2021} (most are clear non-members with a probability of $0.00$, while a few are listed with probabilities up to P=$0.35$, and are similarly discarded). In the analysis of  NGC~2264 we found 144 strong accreting stars with H$\alpha$10\%~$>270$--$300$~km~s$^{-1}$ among our final members 
, and a high number of the remaining candidates show considerable values of accretion as well, albeit under that limit. Of these 144 strong accretors, all are Li members and classified as high-probability members by \cite{jackson2021}, except for 22 stars which show higher levels of $EW$(Li)s than expected (marked as `Y?' in the table of Appendix~\ref{ap1:AppendixD}), and two stars which do not appear to be Li members, but are similarly included as final members due to their strong accretion. These two stars furthermore fulfil the rest of our criteria and are included as high probability members by \cite{jackson2021}. As was also discussed earlier with Cha~I, we also considered as members all of the 28 strong accretors which present values of $EW$(Li) notably larger than the rest of candidates ($800$--$1270$~m$\r{A}$). Furthermore, 20 of them are listed as high probability members by \cite{jackson2021} and several are also included in several others membership studies \cite[e.g, ][]{lim, cantatgaudin_gaia, apellaniz, venuti}. Also see Paper III for more details on the influence of accretion for this cluster specifically.
As for kinematics, three star in our final selection were $RV$ non-members according to the initial 2$\sigma$ interval, with $RV$s deviating up to $9.0$~km~s$^{-1}$ from the mean of the cluster. As in the case of other cluster analyses, we accepted them as final members, as they fulfilled the rest of membership criteria, and were also listed as high-probability members by \cite{jackson2021}. On the other hand, we also accepted as final candidates 22 stars with no measured $RV$ values in iDR6, due to the fact that they fully fulfilled all astrometric criteria (as well as the other criteria), and are also listed as members by \cite{jackson2021} with probabilities in the range of $0.67$--$0.99$. Finally, we also accepted 12 marginal metallicity members (three of which are also strong accreting stars), deviating from the 3$\sigma$ limit up to $0.20$--$0.40$~dex from the mean of the cluster. All of these stars fulfil all other criteria, and are listed as high-probability members with P=$0.96$--$0.99$ by \cite{jackson2021}.

Regarding previous membership selections from the literature, we found the following number of common stars with our selection in several studies: 197 stars in \cite{jackson}, 69 stars in \cite{lim}, 130 stars in \cite{spina_giants}, 116 stars in \cite{cantatgaudin_gaia}, 198 stars in \cite{apellaniz}, and 226 stars in \cite{venuti}. Finally, \cite{jackson2021} includes 479 out of our 503 candidates. Of the remaining candidates in our selection, 21 were not included in their analysis, and the other four (which fully fulfil all our criteria) are listed with probabilities in the range of $0.29$--$0.38$ by \cite{jackson2021}. We also note that we seem to find some inconsistencies when plotting the Li-rich giant outliers selected in the field of this cluster in the CMD diagram, with one of them appearing among the cluster candidates. However, when plotting them in the $\gamma$ index-versus-$T_{\rm eff}$ diagram the expected distinction is found between the non-giant cluster candidates and the Li-rich giant outliers. \\
 
\textbf{$\ast$ $\lambda$~Ori} 


$\lambda$~Ori (Collinder~69) is a young cluster with an age range of $5$--$12$~Myr \citep{jackson2021, binks2021}, the oldest of the associations included in the $\lambda$~Ori star forming region \citep{barrado2011}. 

Of the final 163 Li candidates we discarded two stars because we were not able to analyse them using astrometric and photometric \textit{Gaia} data, and they were listed as non-members with P=$0.00$--$0.01$ by \cite{jackson2021}. In the analysis we also found 34 strong accreting stars among our final candidates. Of these 34 strong accretors, all of them are Li members, and 17 are also listed as high-probability members by \cite{jackson2021} (the rest are not included in their study). In addition, we list four stars with $EW$(Li) values slightly larger than the rest of candidates ($870$--$950$~m$\r{A}$). Similarly to the cases of Cha~I and NGC~2264, we consider them as members given that they exhibit little dispersion with respect to the rest of candidates, and we explain their larger apparent values of Li for being young active stars, with high values of both accretion, and in some cases also chromospheric H$\alpha$. Two of them, the ones with the largest $EW$(Li) values, are additionally strong accretors with H$\alpha$10\%~$>270$--$300$~km~s$^{-1}$, and so, once again, their larger $EW$(Li)s could also be an effect of either accretion-induced enhancement or the result of a strong veiling correction. As for kinematics, 105 stars in our final selection were $RV$ non-members according to the initial 2$\sigma$ interval, with $RV$s deviating $7$--$10$~km~s$^{-1}$ from the mean of the cluster. As in the case of other cluster analyses, we accepted them as final members, as they fulfilled the rest of membership criteria, and all stars except for 14 (which are not included in their study) were also listed as high-probability members by \cite{jackson2021}. Regarding previous membership selections from the literature, we found the following number of common stars with our selection in several studies: 10 stars in \cite{sacco2008}, 70 stars in \cite{hernandez}, 30 stars in \cite{barrado2011}, 16 stars in \cite{franciosini2011}, 51 stars in \cite{bayo2012}, and 94 stars in \cite{cantatgaudin_gaia}. Finally, \cite{jackson2021} includes 116 out of our 161 candidates (the 45 remaining stars were not included in their study). \\

\textbf{$\ast$ Col~197} 


Collinder~197 (Col~197 or  Cr~197) is a low-population young cluster with an age of $12$--$14$~Myr \citep{bonattobicca, vandeputte, sampedro, dias2019, jackson2021, romano}, embedded in the HII region of Gum~15. 

Similarly to other clusters, of the final 104 Li candidates for Col~197 we discarded 11 stars as we were not able to analyse them using astrometric and photometric \textit{Gaia} data, and they were listed as non-members with P=$0.00$--$0.01$ by \cite{jackson2021}. We also decided to discard one star, in spite of fulfilling all our criteria, for being listed as a non-member by both \cite{cantatgaudin_gaia} and \cite{jackson2021}. In addition, we also accepted as final candidates six stars with no measured $RV$ values in iDR6, due to the fact that they fully fulfil the astrometry criteria (as well as the other criteria), and are also listed as members by \cite{jackson2021} with probabilities in the range of $0.70$--$1.00$. In the analysis we also found 11 strong accreting stars among our final candidates. Of these 11 strong accretors, all of them are Li members, and eight are also listed as members by \cite{jackson2021}. Finally, we also accepted 10 marginal metallicity members (one of which is also a strong accretor), deviating from the 3$\sigma$ limit up to $0.38$--$0.50$~dex from the mean of the cluster. All of these stars fulfil all other criteria, and are listed as members by \cite{cantatgaudin_gaia} and/or \cite{jackson2021} (in the case of 5 of these stars, with P=$0.99$).

Regarding previous membership selections from the literature, we found 79 common stars in \cite{cantatgaudin_gaia}, and \cite{jackson2021} includes 65 out of our 92 candidates. Of the remaining candidates in our selection, eight stars were not included in their study, and in an untypical way in all our membership comparisons with this work, we found the remaining 19 stars listed as non-members with P=$0.00$--$0.02$. All of these 19 stars seem to fulfil all of our criteria, including \textit{Gaia}, so we did not discard them from our final selection, but the disagreement is worth noting, even more so considering that this is the only cluster in our sample of $42$ clusters whose final $RV$ mean estimation additionally does not agree with the values given by the literature.\\

\textbf{$\ast$ Vela OB2 association: $\gamma$~Velorum and NGC~2547} 
\label{ap1:NGC2547}

The $\gamma$~Velorum ($\gamma$~Vel) and NGC~2547 clusters are two of the PMS clusters around the Vela OB2 association ($\gamma^2$~Vel), each of them composed of two kinematically distinct populations. The age of both $\gamma$~Vel~A and B is typically listed as $10$--$20$~Myr, with $\gamma$~Vel~A being $1$--$2$~Myr older than $\gamma$~Vel~B \citep[e.g.,][]{jeffries, spina_gamma, frasca, sacco, sacco2, spina_giants, beccari, franciosini, armstrong, jackson2021}. The age of NGC~2547 is listed as either $35$--$45$~Myr (with \cite{Jeffries2023_IC4665} recently obtaining an age of $39.2_{-1.3}^{+1.6}$~Myr), or a slightly younger range of $20$--$35$~Myr. NGC~2547~B is younger than NGC~2547~A, and coeval with $\gamma$~Vel, with an age of $10$--$20$~Myr \citep[e.g., ][]{oliveira, soderblom2013, sacco2,  spina_giants, beccari, randich_gaia, bossini, jackson2021, pang, romano}.

The cluster membership selections for both $\gamma$~Vel and NGC~2547 consist of 234 stars in $\gamma$~Vel (99 in $\gamma$~Vel~A and 84 in $\gamma$~Vel~B, as well as 51 additional candidate stars that are not associated to a specific population), and 151 stars in NGC~2547 (138 in NGC~2547~A and 13 in NGC~2547~B). As mentioned in Sect.~\ref{data}, for this study we used the membership selections obtained by a series of former studies in the literature: Regarding $\gamma$~Vel, we first used \cite{jeffries}, the study specifying the members of $\gamma$~Vel~A and ~B, as well as a series of other GES studies \citep{damiani, spina_gamma, frasca, prisinzano, cantat2019, jackson2021}. For NGC~2547 we similarly used several membership studies from the literature, some of which list the membership probabilities of each candidate to Pop.~A/B \cite{sacco, randich_gaia, cantatgaudin_gaia}, while others do not \cite{spina_giants, jackson2021}. \textit{Gaia} studies \cite{randich_gaia}, \cite{cantatgaudin_gaia} and \cite{jackson2021} offer updated membership probabilities for $\gamma$~Vel~A \cite{jackson2021} and NGC~2547 (all three of them). \cite{randich_gaia} listed a series of new members with respect to \cite{sacco}. Although their membership probabilities for each population are generally consistent with each other, a small number of stars were associated with different populations in \cite{sacco} and \cite{randich_gaia}. In this case, we adopted the membership from \cite{randich_gaia}, as this is the most recent study of the two. For NGC~2547 we also compared the member stars in \cite{sacco}, \cite{cantatgaudin_gaia} and \cite{randich_gaia} with the candidate members in \cite{bravi}. For all our adopted candidates, \cite{bravi} found high probabilities, namely ranging from 60 to $100$\%, of being $RV$ members of the cluster. Finally, \cite{jackson2021} offer updated probabilities for both clusters, and we decided to also prioritize the membership probabilities of this work with respect to earlier studies, for those cases were they might appear to disagree.

While we mainly used the membership lists offered by the aforementioned works, we also note that we did make use of the \textit{Gaia} EDR3 data to analyse the proper motions, parallaxes and the position in the CMD for all stars marked as members for $\gamma$~Vel and NGC~2547, in order to improve the final selections. We thus discarded a series of stars which deviated appreciably from the locus of the rest of members in the $pmra$-versus-$pmdec$ diagram, as well as those stars which proved to be parallax non-members, and those which also deviated appreciably from the rest in the CMDs. As to field contaminants, we also find two of the $\gamma$~Vel Li-rich giants in our list (J08095783-4701385 and J08102116-4740125), as well as one of the Li-rich giants from NGC~2547 (J08110403-4852137), in \cite{casey}. Another one, (J08110403-4658057) in $\gamma$~Vel, is listed in \cite{smiljanic_giants}. \\

\textbf{$\ast$ NGC~2232} 


NGC~2232 is a young cluster in Monoceros with an estimated age range of $18$--$38$~Myr, with the most recent studies veering towards younger ages in the $18$--$25$~Myr range \citep{liu, jackson2021, pang, romano, binks2021}. \cite{Jeffries2023_IC4665} recently obtained an age of $29.3_{-1.0}^{+1.2}$~Myr.  

Of the final 68 Li candidates for NGC~2232 we discarded 16 stars, listed as non-members with P=$0.00$--$0.01$ by \cite{jackson2021}, and not including any astrometric or photometric \textit{Gaia} data. One of them additionally has [Fe/H] value deviating appreciably from the mean of the cluster, beyond the extended 3$\sigma$ membership interval. In our final selection we note as a particular case a star with a $EW$(Li) value which is higher than the rest of candidates ($841$--$948$~m$\r{A}$), but without exhibiting a large dispersion with respect to the rest of the candidates. We consider it as a final member given that it fulfilled the other criteria as well as being listed in several studies as a candidate of the cluster \citep{cantatgaudin_gaia, jackson2021, binks2021}. As for metallicity, we accepted one marginal member deviating $0.24$~dex from the mean of the cluster, as it fulfilled all other criteria, and is listed as a high-probability member with P=$0.98$ by \cite{jackson2021}.
Regarding previous membership selections from the literature, we found 46 stars in common with \cite{cantatgaudin_gaia}, as well as 49 stars in \cite{pang}, and 58 stars in \cite{binks2021}. Finally, \cite{jackson2021} includes 56 out of our 68 candidates (the remaining stars were not included in their study, except for one which was listed with P=$0.30$). \\

\textbf{$\ast$ IC~2391} 
\label{ap1:IC2391}


IC~2391 is an open cluster in the constellation Vela, with an age in the  $20$--$60$~Myr range \citep[e.g., ][]{stauffer, randich2001, barrado3, platais, smiljanic2011, de_silva}. \cite{bossini} gave an age of $36\pm2$~Myr using \textit{Gaia} data, while LDB estimates result in an age range of $50$--$58$~Myr, and \cite{Jeffries2023_IC4665} recently proposed an age of $60.1_{-9.2}^{+26.8}$~Myr. 

Of the final 37 Li candidates for IC~2391, we discarded two stars from the final member selection. These stars could not be analysed with any \textit{Gaia} data (gravity, $RV$ and Li were the only criteria available), and they were additionally classified as non-members with a probability of $0.00$ by \cite{jackson2021}.  We also note that we added one star as part of our final selection which initially was a $RV$ non-member according to the initial 2$\sigma$ interval, with a $RV$ deviating $6.5$~km~s$^{-1}$ from the mean of the cluster. We accepted it as it deviated by less than $10$~km~s$^{-1}$ from the mean $RV$, fulfilled the rest of criteria and is listed as a member by \cite{jackson2021}. Finally, we accepted a  marginal metallicity member deviating $0.18$~dex from the mean of the cluster, as well as one additional star (08384138-5255039) which deviated more appreciably from the 3$\sigma$ limit ($0.30$~dex from the cluster mean). Both of these stars fulfil all other criteria, and are listed as members by several studies \citep[e.g, ][]{randich_gaia, bravi, jackson2021}.

Regarding previous selections, we found the following number of common candidates in the membership lists of these non-GES studies: 19 stars in \cite{barrado2} and \cite{barrado3}; three stars in \cite{randich2001}; four stars in \cite{dodd}; six stars in \cite{platais}, \cite{messina} and \cite{de_silva}; and three stars in \cite{elliott}. We also refer to \cite{gomezgarrido_tfm} and \cite{gomezgarrido} for a membership study of this cluster, alongside IC~2602 and IC~4665. Regarding GES studies, we firstly found four common stars in \cite{spina_giants} and 14 stars in \cite{bravi}. The latter study derived $RV$ membership probabilities and lists of candidate members for this cluster (alongside IC~2602, IC~4665 and NGC~2547) using iDR4 data. Comparing our final selection of 35 members with the list of 53 candidate stars of \cite{bravi}, we find 30 kinematic candidates and 14 final members in common (all of the common members stars have high $RV$ membership probabilities of at least 0.95, except for two stars in the $0.7$--$0.9$ range). We note that, for many stars \cite{bravi} used values of $EW$(Li) and/or $T_{\rm eff}$ which were derived from one of the WG12 nodes and do not appear in our sample. As a result, we excluded these stars from our membership analysis and only consider those stars in \cite{bravi} with $EW$(Li) values in our iDR6 sample. Our mean $RV$ and $\sigma$ for IC~2391 also agree with the estimates in \cite{bravi} (see Table~\ref{table:1}). Finally, a series of more recent studies using both GES and \textit{Gaia} data also list updated membership probabilities for IC~2391 \citep{cantatgaudin_gaia, randich_gaia, jackson2021, pang}. We have 28 common stars as members with \cite{randich_gaia}, as well as 24 stars with \cite{cantatgaudin_gaia}, 29 with \cite{jackson2021} (the remaining six were not included in their analysis), and 23 with \cite{pang}. As we mention in Subsect.~\ref{gaia}, in this study we relied more heavily on \cite{jackson2021}, the most extensive and one of the most recent studies using both GES and \textit{Gaia} data. Thus, we decided to discard as a final member a marginal star according to our analysis given that \cite{jackson2021} listed it as a non-member, even though \cite{randich_gaia} considered it a candidate. To finish, for completeness we also mention here a series of recent studies that used \textit{Gaia}-DR2 data to study the spacial-kinematic distribution and cluster membership of IC 2391 \citep{postnikova2, postnikova, vereshchagin}. \\

\textbf{$\ast$ IC~2602} 
\label{ap1:IC2602}


IC~2602 is an open cluster in the constellation Carina, with an age of $35\pm1$~Myr according to \cite{bossini}, while previous studies found ages ranging from $30$--$67$~Myr \citep{randich_1997, stauffer, randich2001, smiljanic2011, jackson2021, pang, romano}. Ages in the range of $46$--$53$~Myr were also derived from the LDB method \citep{dobbie, Jeffries2023_IC4665}, and \cite{Jeffries2023_IC4665} proposed an slightly older estimation of $45.8_{-3.3}^{+4.7}$~Myr for this cluster. 

Of the final 59 Li candidates for IC~2602, we discarded four stars from the final member selection. Two of them were also discarded via the criteria of $RV$ and metallicity. In addition, similarly to other clusters, all four of these stars could not be analysed with any \textit{Gaia} data (gravity, $RV$, Li, and sometimes metallicity were the only criteria available), and they were additionally classified as non-members with a probability of $0.00$ by \cite{jackson2021} (all four stars were also listed as non-members by \cite{randich_gaia}). Finally, we accepted two  marginal metallicity members, deviating moderately from the 3$\sigma$ limit ($0.30$--$0.40$~dex from the mean of the cluster), both fulfilling all other criteria. One of them (10452826-6413450) is also listed as a high-probability member with P=$0.99$ by \cite{jackson2021}.

Comparing our selection with former studies, we found three stars in common with non-GES \cite{randich2001}, and we also reinforced our membership selection with 28 additional members of IC~2602 with measured $EW$(Li)s listed by this study. We also made use of 44 additional members with $EW$(Li) measurements from  \cite{jeffries2009}. All these member stars which were not measured by GES were  of particular interest to construct the empirical lithium envelope for IC~2602 (with the envelope of \cite{montes} as a base) as well, especially in the region of the LDB (Lithium Depletion Boundary, see Paper III). We also refer to \cite{gomezgarrido_tfm} and \cite{gomezgarrido} for a membership study of this cluster, alongside IC~2391 and IC~4665. Regarding GES studies, we firstly found 11 common stars in \cite{spina_giants}, as well as 55 kinematic candidates and 27 final members in common with the list of 101 candidates in \cite{bravi}. As in the case of IC~2391, many of the stars in \cite{bravi} have no $EW$(Li) values in the iDR6 sample, and therefore we excluded them in our own membership analysis.The mean $RV$ and $\sigma$ derived in \cite{bravi} for IC~2602 are also in agreement with the ones obtained in this paper. As for GES studies that include \textit{Gaia} data, we have 36 common stars listed as members with \cite{randich_gaia}, as well as 29 stars with \cite{cantatgaudin_gaia}, and 49 with \cite{jackson2021} (the remaining six were not included in their analysis). \\

\textbf{$\ast$ IC~4665} 
\label{ap1:IC4665}


IC~4665 is an open cluster in the constellation Ophiuchus. \cite{bossini} calculated an age of $38\pm3$~Myr for this cluster, in agreement with the $35$--$43$~Myr range reported in several studies \citep{martinmontes, jeffries2001, jackson2021, pang, romano}. However, while they also cite an LDB age of $\sim$~$32$~Myr, \cite{Jeffries2023_IC4665} recently obtained a slightly older age of $52.9_{-5.2}^{+6.8}$~Myr for this cluster, proposing that the published LDB age should be interpreted as a lower limit. 

Of the final 51 Li candidates for IC~4665, we discarded 18 stars from the final member selection. Similarly to the case of other clusters, all of these stars could not be analysed with any \textit{Gaia} data, and they were all classified as non-members by \cite{jackson2021} (15 stars with a probability of $0.00$, the remaining three in the potential but not definite non-member range of $0.17$-$0.49$). \cite{randich_gaia} also lists seven of these 18 stars as non-members (the rest were not considered in their analysis). 

Regarding previous selections, we found the following number of common candidates in the membership lists of these non-GES studies: six stars in \cite{dewit} and \cite{jeffries2009}; 11 stars in \cite{manzi}; and 17 stars in \cite{lodieu}. We also reinforced our membership selection with 23 additional members of IC~4665 with measured $EW$(Li)s listed by \cite{manzi}. We also refer to \cite{gomezgarrido_tfm} and \cite{gomezgarrido} for a membership study of this cluster, alongside IC~2391 and IC~2602. As for GES studies, we found 15 common stars in \cite{spina_giants}, as well as 30 kinematic candidates and 19 final members in common with the list of 122 candidates in \cite{bravi}. Our mean $RV$ and $\sigma$ are also consistent with the previous estimates in this work. Finally, regarding GES studies that include \textit{Gaia} data, we have 11 common stars listed as members with \cite{randich_gaia}, as well as 16 stars with \cite{cantatgaudin_gaia}, 23 stars with \cite{pang}, and 29 candidate stars in common with \cite{jackson2021}. In this latter comparison, three out of the remaining four stars were not included in the analysis of cite{jackson2021}, and the other one was listed with a probability of $0.14$, but seeing as it fully fulfilled all our criteria, including \textit{Gaia} astrometry, we accepted it as a final member. In spite of using the same data sets and very similar membership criteria, it is probable that the reason for these individual deviations between our two analyses originated from the existing differences regarding, for example, criteria limits or the weight of the factors implicated in the decision to accept or discard stars (especially if they seem to be marginal stars according to the individual membership criteria) as final candidates. \\

\textbf{$\ast$ NGC~2451~A and B} 


NGC~2451~A and NGC~2451~B are two open clusters projected along the same line of sight. The age of both NGC~2547~A and B is generally listed in the $50$--$80$~Myr range \citep[e.g., ][]{balog, netopil, silaj, randich_gaia, franciosini2021, jackson2021, pang}, while individual ages are also listed, such as the estimations of \cite{bossini}, who calculated an age of $44\pm2$~Myr for NGC~2547~A and $39\pm1$ for NGC~2547~B. Thus, the nearer cluster NGC~2547~A seems to be slightly older than the more distant cluster NGC~2547~B \cite{hunsch}. 

The cluster membership selections for both NGC~2451~A and NGC~2451~B consist of 42 stars in NGC~2451~A and 64 stars in NGC~2451~B. For this study we used the membership selections obtained by a series of former studies in the literature: \cite{silaj} only gives a candidate list for NGC~2451~A, while the rest include final member lists for both NGC~2451~A and NGC~2451~B (indicated as 'A' and 'B' in the table of Appendix~\ref{ap1:AppendixD}) \citep{spina_giants, randich_gaia, cantatgaudin_gaia, jackson2021}. \textit{Gaia} studies \cite{randich_gaia}, \cite{cantatgaudin_gaia} and \cite{jackson2021} offer updated membership probabilities for NGC~2451~A and NGC~2451~B. Although their membership probabilities for each population are generally consistent with each other, a small number of stars were considered as members by  \cite{randich_gaia} and \cite{cantatgaudin_gaia} but were listed as definite non-members in \cite{jackson2021}. In these cases, we adopted the membership from \cite{jackson2021}, as this is the most recent study. As seen with other clusters, we decided to prioritize the membership probabilities of \cite{jackson2021} with respect to earlier studies, for those cases were they might appear to disagree. While we mainly used the membership lists offered by the aforementioned works, we also note that we did make use of the \textit{Gaia} EDR3 data to analyse the proper motions, parallaxes and the position in the CMD for all stars marked as members for NGC~2451~A/B, in order to improve the final selections. We thus discarded a series of stars which deviated appreciably from the locus of the rest of members in the $pmra$-versus-$pmdec$ diagram, as well as those stars which proved to be parallax non-members, and those which also deviated appreciably from the rest in the CMDs. Most of these discarded stars were also listed as non-members by \cite{jackson2021}, in contrast to them being considered as members by earlier studies, which reinforced our decision to discard them on the basis of our astrometric criteria.\\

\subsection{Intermediate-age clusters (age$=50$--$700$~Myr)}

\textbf{$\ast$ NGC~6405} 


NGC~6405 (also known as M6) is a $80$--$100$~Myr cluster \citep{kilikoglu, netopil2016, gao, jackson2021}, located between the local arm and the Sagittarius arm of the Galaxy. 

Of the final 53 Li candidates for NGC~6405, we discarded two stars: one of them, additionally listed as a clear non-member by \cite{jackson2021}, we could not analyse using the astrometric and photometric \textit{Gaia} data. The remaining one was similarly listed as a non-member by both \cite{cantatgaudin_gaia} and \cite{jackson2021}, and it was also a marginal non-member according to our parallax criterion. We also accepted four marginal metallicity members, deviating from the 3$\sigma$ limit up to $0.35$~dex from the mean of the cluster. All of these stars fulfil all other criteria, and are listed as high-probability members with P=$0.89$--$0.99$ by \cite{jackson2021}. Regarding other membership selections from the literature, we found 46 stars in common with \cite{cantatgaudin_gaia}, and 45 star in \cite{gao}. \cite{jackson2021} includes 50 out of our 51 candidates. The remaining one is listed with P=$0.28$. Seeing as it fully fulfilled all our criteria, including \textit{Gaia} astrometry, we accepted it as a final member. \\

\textbf{$\ast$ Blanco~1} 


Blanco~1 is a cluster located towards the South Galactic Pole, with an age in the range $90$--$150$~Myr \citep{gillen, zhang, jackson2021, pang, romano}. \cite{bossini} calculated an age estimation of $94\pm5$~Myr, while studies based on the LDB method give slighter higher ages: $114\pm10$~Myr \citep{juarez} and $132\pm24$~Myr \cite{cargile}. On the other hand, \cite{Jeffries2023_IC4665} recently proposed a younger age of $73.5_{-6.9}^{+12.9}$~Myr for this cluster. 

Of the final 101 Li candidates for Blanco~I, we discarded three stars because we were not able to analyse them using astrometric and photometric \textit{Gaia} data, and they were listed as non-members with P=$0.00$ by \cite{jackson2021}. As for kinematics, nine stars in our final selection were $RV$ non-members according to the initial 2$\sigma$ interval, with $RV$s deviating $4$--$8$~km~s$^{-1}$ from the mean of the cluster. As in the case of other cluster analyses, we accepted them as final members, and all stars were also listed as high-probability members by \cite{jackson2021}. On the other hand, we also accepted as final candidates two stars with no measured $RV$ values in iDR6, due to the fact that they fully fulfil the astrometry criteria (as well as the other criteria), and are also listed as members by several studies \citep{platais2011, zhang, jackson2021}. Finally, we also accepted seven marginal metallicity members, deviating from the 3$\sigma$ limit up to $0.16$~dex from the mean of the cluster. All of these stars fulfil all other criteria, and are listed as high-probability members by \cite{jackson2021}, with P=$0.99$--$1.00$.

Regarding previous membership selections from the literature, we found 58 stars in common with \cite{platais2011}, one star in \cite{juarez}, and 80 stars in \cite{zhang}. Finally, \cite{jackson2021} includes 92 out of our 98 candidates. Four of the remaining ones were not included in their study, one is listed with P=$0.37$, and the last one is listed as a non-member with P=$0.00$. The latter two stars fully fulfil all our membership criteria, so we accepted them as final members. As in the case of IC~4665, it is probable that the reason for these individual deviations between our two analyses originated from the existing differences regarding criteria limits or the weight of the factors implicated in the decision to accept or discard stars. \\

\textbf{$\ast$ NGC~6067} 


NGC~6067 is an open cluster superimposed on the Norma star cloud, with an age in a $60$--$150$~Myr range, with the most recent studies citing ages of $100$--$120$~Myr \citep{frinchaboy, netopil2016, alonsosantiago2017, magrini2018, rangwal, randich2020, jackson2021, romano}. 

Of the final 60 Li candidates for NGC~6067, we discarded four stars with no astrometric and photometric \textit{Gaia} data, further listed as non-members with P=$0.00$ (for two stars) and P=$0.19$--$0.36$ by \cite{jackson2021}. As for kinematics, two stars in our final selection were $RV$ non-members according to the initial 2$\sigma$ interval, with $RV$s deviating $13$--$14$~km~s$^{-1}$ from the cluster mean. Listed as high-probability members by \cite{jackson2021} and fulfilling the rest of criteria, we accepted them as final members. Regarding metallicity, we also accepted six marginal members (deviating up to $0.20$~dex from the mean of the cluster), as well as one additional star (16130269-5408556) which fulfilled all criteria except for metallicity, deviating more appreciably from the 3$\sigma$ limit ($0.36$~dex from the cluster mean). All of these stars are further listed as high-probability members by \cite{jackson2021}, with P=$0.97$--$0.99$. As for previous membership selections from the literature, we found three stars in common with \cite{frinchaboy}, eight stars in \cite{alonsosantiago2017}, and 42 stars in \cite{cantatgaudin_gaia}. \cite{jackson2021} includes all 56 of our final candidates. \\

\textbf{$\ast$ NGC~6649} 


NGC~6649 is a heavily-reddened (E(B-V)=$1.43\pm0.05$~Myr) open cluster located in the first Galactic quadrant with an age in a $50$--$126$~Myr range, close to the age of the Pleiades, with the most recent studies citing ages of $120$~Myr \citep{kharchenko, dib, liu, alonsosantiago, jackson2021}.

Of the final 3 Li candidates for NGC~6649, we discarded one star with no astrometric and photometric \textit{Gaia} data, also listed as a non-member with P=$0.00$ by \cite{jackson2021}. As for kinematics, one star in our final selection was a $RV$ non-member according to the initial 2$\sigma$ interval, with a $RV$ value deviating $10$~km~s$^{-1}$ from the mean of the cluster. Similarly to other clusters in this case, we accepted it as a final member, and it is also listed as a high-probability member by \cite{jackson2021}. Regarding previous membership selections from the literature, we found no stars in common with \cite{cantatgaudin_gaia}, and one star out of our two final candidates in common with \cite{jackson2021} (the other one was not included in their study). \\ 

\textbf{$\ast$ NGC~2516} 

\label{ap1:NGC2516}


NGC~2516 is an open cluster in the southern constellation of Carina. The age of $251\pm3$~Myr in \cite{bossini} is older than the previous age estimates ranging from $110$ to $150$~Myr \citep[e.g., ][]{jacobson, magrini2017, fritzewski2020, randich2020, dumont, franciosini2021, jackson2021, binks2021}. 
 
Of the final 379(383) Li candidates for NGC~2516, we discarded seven stars which we were not able to analyse using astrometric and photometric \textit{Gaia} data, and all of them were further listed as non-members with P=$0.00$--$0.10$ by \cite{jackson2021}. The four stars we classified as possible Li members (marked as `Y?' in the table of Appendix~\ref{ap1:AppendixD}) present values of $EW$(Li) slightly larger than the rest of candidates ($300$--$325$~m$\r{A}$), especially regarding what we could expect from the age of the cluster. However, they do not exhibit appreciable dispersion with respect to the other members, all four stars fulfil the rest of membership criteria, and all of them are also listed as high probability members by \cite{jackson}, and so we accepted them as final members of the cluster. As well as possible scatter due to errors in the measures, we could also explain the higher observable values of $EW$(Li) of these stars considering the fact that they also exhibit either higher values of rotation and/or chromospheric activity, and thus, a potential slower rate of Li depletion (see more on rotation and activity in regards to Li depletion in Paper III, where we analyse in detail the dependence with several parameters). As for kinematics, 23 stars in our final selection were $RV$ non-members according to the initial 2$\sigma$ interval, with $RV$s deviating $5$--$8$~km~s$^{-1}$ from the mean of the cluster. Fulfilling the rest of cirteria and listed as high-probability members with P=$0.99$--$1.00$ by \cite{jackson2021}, we accepted all of these stars as members. In addition, we accepted as final members 63 marginal metallicity members, deviating $0.12$--$0.30$~dex from the mean of the cluster, as well as eight additional stars with values which deviate more appreciably from the 3$\sigma$ limit, up to $0.40$~dex from the cluster mean. As with prior clusters, all of these stars fulfil the rest of criteria and are also listed as members by several studies \citep{randich_gaia, cantatgaudin_gaia}. Finally, all but one of these stars are further listed as high-probability members by \cite{jackson2021}, with P=$0.98$--$1.00$.

Regarding previous membership selections from the literature, we found the following number of common stars with our selection in a series of both non-GES and GES studies: Firstly, we found 326 stars in \cite{jeffries2001}, 25 stars in \cite{terndrup}, 55 stars in \cite{irwin}, 44 stars in \cite{jacksonjeffries}, six stars in \cite{wright2011}, and two stars in \cite{heiter}. As for GES studies, we found 56 common stars in \cite{jacobson}, 15 stars in \cite{jackson}, 13 stars in \cite{magrini2017}, 49 stars in \cite{bailey}, 210 stars in \cite{cantatgaudin_gaia}, 332 stars in \cite{randich_gaia}, and 226 stars in \cite{fritzewski}. Finally, \cite{jackson2021} includes 373 out of our 379 candidates (the remaining six stars were not included in their analysis).  We also note that, in order to further reinforce the membership of the stars in the field of this cluster, we also made use of the additional non-GES members with $EW$(Li) values of \cite{jeffries1998}. 

\textbf{$\ast$ NGC~6709} 


NGC~6709 is an open cluster situated towards the center of the Galaxy in the constellation Aquila, with an age in a $150$--$190$~Myr range \citep{subramaniam, vandeputte, jackson2021, romano}. \cite{bossini} calculated an age estimation of $173\pm34$~Myr. 

Of the final 53 Li candidates for NGC~6709, we discarded four stars with no astrometric and photometric \textit{Gaia} data,  all four of them also listed as non-members with P=$0.00$ by \cite{jackson2021}. Regarding previous membership selections from the literature, we found 45 stars in common with \cite{cantatgaudin_gaia}, and \cite{jackson2021} includes 46 out of our 49 of final candidates. Of the remaining three, two stars were not included in their study, and the last one is listed with P=$0.31$. Given that said star fully fulfils our membership criteria, including astrometry, we included it in our final selection. We also note that we seem to find some inconsistencies when plotting the Li-rich giant outliers selected in the field of this cluster in the CMD diagram, with two of them, for example, appearing very close among the non-giant cluster candidates. However, when plotting them in the Kiel diagram the expected distinction is found between the non-giant cluster candidates and the Li-rich giant outliers. \\

\textbf{$\ast$ NGC~6259} 


NGC~6259 is an open cluster with an age of $210$~Myr \citep{sampedro, magrini2018, casali, randich2020, jackson2021}. 

Of the final 39 Li candidates for NGC~6259, we discarded four stars with no astrometric or photometric data, additionally listed as non-members with P=$0.00$ by \cite{jackson2021}. We also accepted two marginal metallicity members (deviating up to $0.28$~dex from the mean of the cluster), which fulfilled all other criteria and were listed as members by \cite{jackson2021}. Regarding previous membership selections from the literature, we found 20 stars in common with \cite{cantatgaudin_gaia}, and 34 out of our 35 members in \cite{jackson2021}. The remaining star in our final selection is listed in the latter study with P=$0.32$. Seeing as it fully fulfils all our membership criteria, we accepted it as a final member. \\

\textbf{$\ast$ NGC~6705} 


NGC~6705 (also known as M11) is a rich open cluster in the constellation Scutum, with an estimated age of $280$--$300$~Myr \citep{jacobson, magrini2017, randich2020, jackson2021, romano}.

Of the final 142 Li candidates for NGC~6705, we discarded three stars with no photometric or astrometric data, also listed as non-members by \cite{jackson2021}. As for kinematics, eight stars in our final selection were $RV$ non-members according to the initial 2$\sigma$ interval, with $RV$s deviating up to $10$--$13$~km~s$^{-1}$ from the mean of the cluster. Also listed as high-probability members by \cite{jackson2021}, as in the case of other clusters, we accepted all of these stars as final members. In addition, we also accepted 14 marginal metallicity members (deviating up to $0.30$--$0.40$~dex from the mean of the cluster), which fulfilled the rest of criteria, as well as one additional star (18504954-0614585) which fulfilled all criteria except for $\log g$ and metallicity, deviating appreciably from the starting 3$\sigma$ limit ($0.95$~dex from the cluster mean). All of these stars are further listed as high-probability members by \cite{jackson2021}, with P=$0.99$--$1.00$.

Comparing our final selection with existing membership studies from the literature for this cluster, we found the following number of common stars with a number of studies: Firstly, we found 21 common stars in \cite{magrini2014}, as well as 27 stars in \cite{tautvaisiene} and \cite{jacobson}, 15 stars in \cite{sampedro} and \cite{magrini2017}, and 44 stars in \cite{cantatgaudin_gaia} and 11 in \cite{casamiquela}. Finally, \cite{jackson2021} includes 136 out of our final 139 candidates (the remaining three stars in our selection were not included in their study). \\

\textbf{$\ast$ Berkeley~30} 


Berkeley~30 is an open cluster with an age of $300$--$313$~Myr \citep{kharchenko, sampedro, jackson2021, romano}.

Among the final 24 Li candidates for this cluster, we included one marginal metallicity member (deviating $0.20$~dex from the mean of the cluster), as well as two stars with [Fe/H] values deviating more appreciably from the 3$\sigma$ limit ($0.83$~dex from the cluster mean). All three of them are additionally listed as members by \cite{cantatgaudin_gaia}, and by \cite{jackson2021}, with P=$0.94$--$0.99$. Regarding previous membership selections from the literature, we found five  stars in common with \cite{cantatgaudin_gaia}, and \cite{jackson2021} includes all 24 of our final candidates.\\

\textbf{$\ast$ NGC~6281} 


NGC~6281 is an open cluster with an age of $314$--$316$~Myr \citep{dias2002, joshi, sampedro, jackson2021}. 

Among the final 24 Li candidates for NGC~6281, we accepted as a final candidate one star, listed as a high-probability member by \cite{jackson2021}, in spite of being a $RV$ non-member according to the initial 2$\sigma$ interval, with a $RV$ value deviating $5$~km~s$^{-1}$ from the mean of the cluster. Regarding previous membership selections from the literature, we found the following number of stars in common with several studies: Firstly, we found two  stars in common with \cite{frinchaboy} and \cite{heiter}, as well as 19 common stars in \cite{cantatgaudin_gaia}. Finally, \cite{jackson2021} includes all 23 of our final candidates. \\

\textbf{$\ast$ NGC~3532} 
\label{ap1:NG3532}


NGC~3532 is a very rich southern open cluster in a crowded Galactic field in Carina. Intermediate in age between the Pleiades ($78$--$125$~Myr) and the Hyades ($750$~Myr), it has an age estimate of $399\pm5$~Myr according to \cite{bossini}, while other studies give an age range of $300$--$399$~Myr \citep{dobbie, fritzewski, jackson2021, romano}. 

Of the final 323 Li candidates for NGC~3532, we discarded a small number of stars according to the following criteria: Firstly, we discarded seven stars which we were not able to analyse using astrometric and photometric \textit{Gaia} data, and which were further listed as non-members by \cite{jackson2021}. We also discarded one SB2 star, as well as another star we also could not analyse with astrometric criteria and deviated appreciably in the Kiel diagram. Lastly, we discarded two final stars with no \textit{Gaia} data analysis available which was marked as a non-member by \cite{fritzewski}. As for kinematics, we note that 42 stars in our final selection were $RV$ non-members according to the initial 2$\sigma$ interval, with $RV$s deviating up to $8$--$11$~km~s$^{-1}$ from the mean of the cluster. As in the case of other cluster analyses, we accepted all of them as final members, and all 42 stars were also listed as high-probability members by \cite{jackson2021}. Regarding metallicity, we accepted as final candidates 76 marginal members (deviating up to $0.48$~dex from the mean of the cluster), as well as two additional stars which deviate more appreciably from the 3$\sigma$ limit, up to $0.83$~dex from the cluster mean. All of these stars fulfil all other criteria, and are further listed as members by several studies \cite[e.g.,][]{cantatgaudin_gaia, fritzewski, fritzewski2021}, with all but eight of these stars being also considered as high-probability members by \cite{jackson2021}, with P=$0.98$--$1.00$.

Regarding previous membership selections from the literature, we found the following number of stars in common with these studies: Firstly, we found one star in common with \cite{heiter}, as well as 310 stars in \cite{cantatgaudin_gaia}, 216 stars in \cite{fritzewski}, 27 stars in \cite{hetem}, and 132 stars in \cite{fritzewski2021}. Finally, \cite{jackson2021} includes 382 out of our 384 final candidates (the remaining two stars in our final selection were not included in their study).

We will now discuss the 72 stars that we classified as possible Li members, marked as `Y?' in the table of Appendix~\ref{ap1:AppendixD} and represented as open squares in all $EW$(Li)-versus-$T_{\rm eff}$ diagrams to differentiate them from the rest of the candidates (see Appendix~\ref{ap1:AppendixC}). These 72 stars are late-K and M-type stars with $T_{\rm eff}<$~4100~K, and all of them consistently present values of $EW$(Li) which are considerably larger than we would expect for KM stars in NGC~3532, a cluster with an age of $300$--$399$~Myr. However, similarly to other clusters such as NGC~6633 (see below), the apparently overestimated $EW$(Li) values of these stars can be explained by taking into account the inherent difficulty of obtaining accurate Li measurements for KM type stars in this age range: At low temperatures ($T_{\rm eff}$~$\leq~4250$), the $EW$s can be thought of as pseudo-$EW$s derived by integrating over an interval which depends on $v$sin$i$ values, and so they include several additional components. As a result, Li measurements for this age range do not go to zero even when Li would already be depleted, and they furthermore actually increase with decreasing $T_{\rm eff}$ as the additional components become stronger \citep{gilmoregaia2021, franciosini2022, randichgaia2021}. And, while we see such $EW$s for KM stars very clearly in intermediate-age clusters such as NGC~3532 and NGC~6633, all clusters reaching such low $T_{\rm eff}$s can show this effect, starting with younger clusters such as NGC~2547~A and NGC~2451~A/B.

Given that the $EW$(Li) values of these stars are for these reasons not truly representative of the age of this cluster, we decided not to take them into account when creating the empirical envelope for this cluster (see Paper III). However, we also consider all these stars to be robust members of NGC~3532, given that i) They do not exhibit appreciable dispersion with respect to each other or the other members, making it highly improbable for all 93 stars are spurious outliers; ii) All of them consistently fulfil the rest of membership criteria (including kinematics, astrometry, gravity and metallicity criteria); and iii) All of them are listed as high probability members by \cite{jackson} with P=$0.97$--$1.00$, and several appear as candidates in other studies as well \citep{cantatgaudin_gaia, fritzewski, fritzewski2021}. 

Another factor with which we could potentially explain some of the higher $EW$(Li) values for these stars involves the study of their levels of rotation and activity. It is possible to explain higher observable values of $EW$(Li) considering the fact that they might exhibit either higher values of rotation and/or higher values of chromospheric activity, as we already saw in the case of clusters such as NGC~2516, where undepleted members due to magnetic activity were found both in the literature and possibly in our final selection as well (see above). Due to the Li-rotation and Li-activity anti-correlations, these stars would consequently deplete lithium in a slower rate than their less active and/or slower rotating counterparts, thus exhibiting higher values of $EW$(Li) as a result. Studies such as \cite{pallavicini1990} concluded that, while less common, this effect can also be observed in K-type stars in older clusters with intermediate ages between the Pleiades ($78$--$125$~Myr) and the Hyades ($750$~Myr), such as both NGC~3532 (399~Myr) and NGC~6633 (575~Myr).

This cluster does indeed include a series of K-type stars exhibiting high rotation and chromospheric activity, and thus, some of these faster rotators and the more active stars could have consequently depleted Li more slowly, resulting in higher $EW$(Li)s than would be expected for late K-type stars at this age. However, while this may be a factor for some of the 93 KM stars here discussed, not all of them exhibit sufficiently high rotations and activity to be able to explain their observable Li as a consequence of these effects, many of them  actually being among the slowest rotating stars in the final selection for this cluster. And so, for clusters such as NGC~3532 and NGC~6633 (see below), we believe that the most probable explanation for the $EW$s of KM members points to the GES measurement process detailed above. \\

\textbf{$\ast$ NGC~4815} 


NGC~4815 is a heavily populated open cluster in the constellation of Musca, with an age of $500$--$570$~Myr \citep{friel, jacobson, magrini2017, magrini2018, jackson2021}.

Similarly to our analysis of others clusters in the sample, of the final 30 Li candidates we discarded one star for not being able to analyse it using astrometric and photometric \textit{Gaia} data, and because it was additionally listed as a non-member by \cite{jackson2021}. As for kinematics, we accepted five stars in our final selection (also listed as high-probability members by \cite{jackson2021}) as final members in spite of being $RV$ non-members, with $RV$s deviating up to $14$~km~s$^{-1}$ from the mean of the cluster. Regarding previous membership selections from the literature, we found the following number of stars in common with several studies: Firstly, we found five  stars in common with \cite{friel}, \cite{magrini2014}, \cite{tautvaisiene} and  \cite{jacobson}, as well as three common stars in \cite{magrini2017}, and seven stars in \cite{cantatgaudin_gaia}. Finally, \cite{jackson2021} includes all 29 of our final candidates.  We also note that two of the final candidates in our sample (12572442-6455173 and 12575818-6459323) show higher values of $EW$(Li) compared to the other candidates, showing more dispersion with respect to the rest of the cluster members as well. Fulfilling the rest of criteria, these stars seem to be Li-rich (non-giant) members, and are included as candidates of the cluster by a series of previous studies \citep{magrini2014, tautvaisiene, jacobson, jackson2021}. \\

\textbf{$\ast$ NGC~6633} 

    \begin{figure} [h!]
   \centering
   \includegraphics[width=0.8\linewidth]{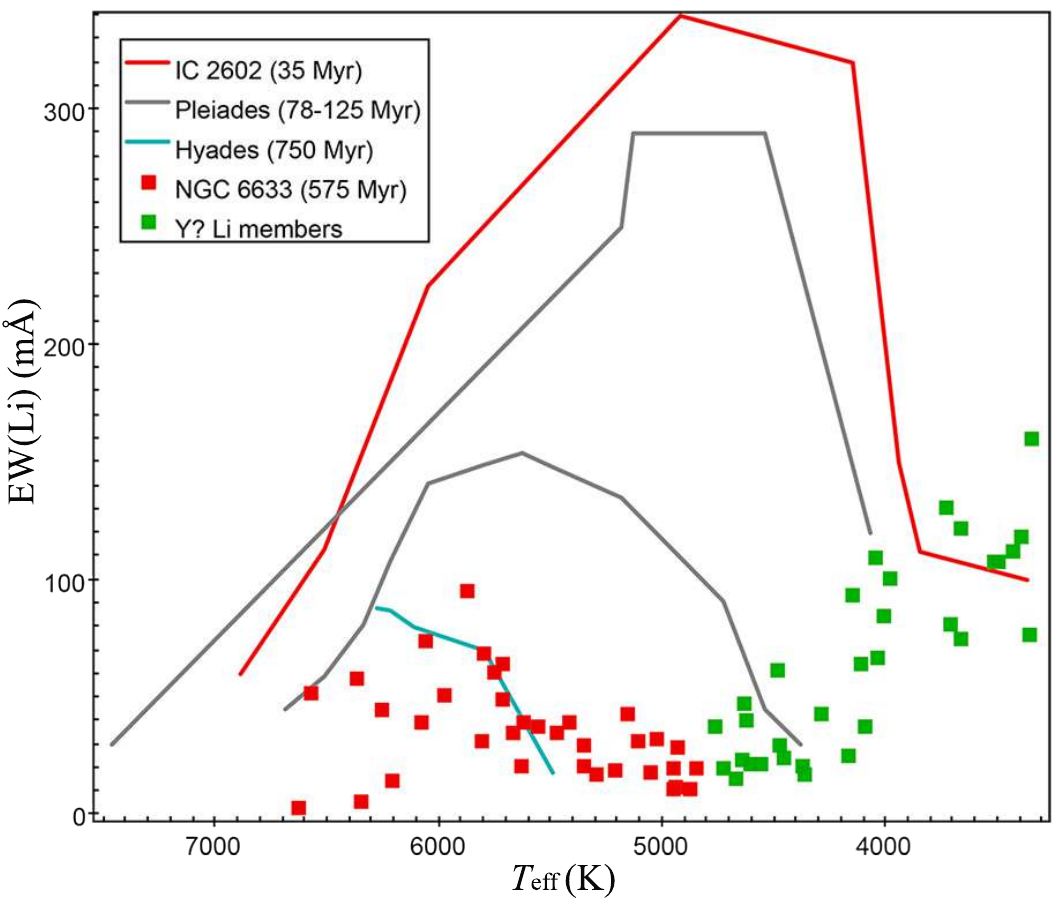}
  \caption[$EW$(Li)-versus-$T_{\rm eff}$ diagram for NGC~6633.]{$EW$(Li)-versus-$T_{\rm eff}$ diagram showing the final candidate selection (red squares) for NGC~6633, a 575~Myr-old intermediate-age cluster. In green squares we mark a series of late-K and M stars with considerably higher values of $EW$(Li) than would be expected for this cluster, but still listed as final candidates of the cluster.}
            \label{Li2}
    \end{figure}

NGC~6633 is an open cluster in the constellation Ophiuchus. The age estimations for this cluster span an age range of $575$--$773$~Myr \citep[e.g.,][]{umezu, jeffries2002, sestitorandich, jacobson, magrini2017}. While \cite{bossini} gave an estimation of $773_{-10}^{+50}$~Myr (older than the age of the Hyades, $750$~Myr), we have here adopted the younger estimation of $575$~Myr in \cite{jackson2021}.

Of the final 62 Li candidates for NGC~6633, we discarded 14 stars with no astrometric and photometric \textit{Gaia} data, further listed as non-members by \cite{jackson2021}. Regarding kinematics and astrometry, we note that both the $RV$ and parallax distributions for NGC~6633 revealed a large contaminant population in the middle of the distribution, which could not be discarded with the aid of the 2$\sigma$ clipping procedure (as we have done in the case of the remaining $41$ clusters in the sample to identify contaminants in the tails of the distribution). The presence of this contaminant population significantly affected the mean $RV$ and mean parallax rendered by the Gaussian fits, and also gave very high final dispersions even after the final convergence of the clipping procedure was reached. When comparing the final fits with the literature values (see Tables~\ref{table:1} and \ref{table:2}), we saw that taking all the contaminant $RV$s and parallaxes in the middle of the distributions into consideration caused the mean $RV$ and parallax to deviate considerably from the reference estimates for this cluster and effectively affected the final cluster candidate selection well. For this reason, we decided to manually filter out this contaminant population before re-analysing the $RV$ and parallax distributions. In this way, we obtained mean values fully consistent with the literature estimations. This step was vital in order to obtain a probable final list of candidate members for this cluster, seeing as we found that some of the stars in the contaminant population even seemed to fulfil the rest of criteria in spite of clearly not being part of NGC~6633 according to the kinematic and astrometric distributions (which is why kinematics and astrometry are the first criteria applied and the most restrictive in our analysis). 

Regarding previous membership selections from the literature, we found the following number of stars in common with these studies: Firstly, we found seven stars in common with \cite{heiter}, as well as eight stars in \cite{jacobson}, six stars in \cite{magrini2017}, 24 stars in \cite{sampedro}, 22 stars in \cite{cantatgaudin_gaia}, and 32 stars in \cite{randich_gaia}. Finally, \cite{jackson2021} includes only 19 out of our 67 final candidates, and all the remaining stars in our final selection were not included in their study. We also note that we reinforced the membership of our final selection by making use of the additional members with $EW$(Li) in \cite{jeffries1997}, \cite{heiter}, \cite{magrini2017}, and \cite{sampedro}. Finally, regarding non-members, one of the Li-rich giants in our list (J18265248+0627259) is listed in \cite{smiljanic_giants}.

For these cluster we have classified 19 stars as possible Li members, marked as `Y?' in the table of Appendix~\ref{ap1:AppendixD} and indicated as open squares in all $EW$(Li)-versus-$T_{\rm eff}$ diagrams to differentiate them from the rest of the candidates (see Subsect.~\ref{candidates} and Appendix~\ref{ap1:AppendixC}). As showcased in Fig.~\ref{Li2}, these stars are late K and M type stars with $T_{\rm eff}<$~$4800$--$5000$~K, and all of them consistently present values of $EW$(Li) which are considerably larger than we would expect for KM stars in NGC~6633 ($575$--$630$~Myr). As such,  we decided not to take any of these 19 stars into account when creating the empirical envelope for this cluster. We here refer to the detailed discussion in the individual note of NGC~3532, another intermediate-age cluster that notably showed the same issue with a larger number of stars, showcasing higher Li values than would be expected for KM member stars as a result of the GES measurement process to obtain $EW$(Li)s at very low temperatures. Similarly to NGC~3532, we consider all these 19 stars to be robust cluster members, consistently fulfilling the rest of membership criteria, with six of them listed as high probability members by \cite{jackson} with P=$0.87$--$1.00$ (the rest of them were not included in their study), and several appearing as candidates in other studies as well \citep{cantatgaudin_gaia, randich_gaia}. Even though we believe that this factor cannot fully explain the higher $EW$s of all of these 19 stars, we also note that some of them also exhibit high rotation (the iDR6 sample for this cluster does not offer any values of H$\alpha$), and these faster rotators certainly could have consequently depleted Li more slowly, resulting in somewhat higher $EW$(Li)s.

Regarding the age of NGC~6633, we note that the age estimations for this cluster vary  depending on the study, spanning an age range of $575$--$773$~Myr: from $600$--$630$~Myr \citep{sestitorandich, jacobson, magrini2017}, to the estimation of $773_{-10}^{+50}$~Myr by \cite{bossini} (older than the age of the Hyades, $750$~Myr), to $575$~Myr in the recently published paper by \cite{jackson2021} (younger than the age of the Hyades). On the other hand, \cite{umezu} and \cite{jeffries2002} considered a very similar age to the Hyades for NGC~6633, and concluded that the lower metallicity of this cluster, $-0.10$--$-0.01$ dex \citep{jeffries2002, jacobson, magrini2018}, was a factor that could explain why the $EW$(Li) envelope for NGC~6633 lied above the Hyades, and thus, the fact that, according to the study, Li was being depleted at a slower pace in the case of NGC~6633. We personally interpreted the dependence of Li depletion with cluster metallicity differently, as we adopted an appreciably younger age of $575$~Myr for NGC~6633, as reported by \cite{bossini} making use of \textit{Gaia} DR2 data. We believe that this younger age is in agreement with our final selection for this cluster, and explain the fact that its upper Li envelope lies slightly above the Hyades due to NGC~6633 being younger by at least $150$~Myr, and not as a result of being more metal-poor. We thus suggest that the metallicity of NGC~6633 has no observable effect on the position of its Li envelope in comparison to the envelope for the Hyades cluster as these clusters do not seem to be coeval. \\

\subsection{Old clusters (age $>700$~Myr)}

\textbf{$\ast$ NGC~2477} 


NGC~2477 is one of the richest open clusters in the Southern sky with an age of $0.7$--$1.0$~Gyr \citep{gao, rain}.

Among the nine Li candidates for NGC~2477, we note that we accepted as a final candidate one star with no measured $RV$ in iDR6, due to the fact that it fully fulfilled the astrometry criteria (as well as the other criteria), and was also listed as a member by \cite{cantatgaudin_gaia}. We also accepted one marginal metallicity member (07521722-3831323) deviating $0.20$~dex from the mean of the cluster, as it fulfilled all other criteria, and is listed as a member by \cite{cantatgaudin_gaia}. Regarding previous membership selections from the literature, we found three stars in common with \cite{eigenbrod}, 10 stars in \cite{cantatgaudin_gaia}, and seven stars in \cite{jadhav}. We note that this is one of the three clusters out of our sample of 42 clusters which is not included in \cite{jackson2021}. \\

\textbf{$\ast$ Trumpler~23} 


Trumpler~23 is a  well-populated open cluster with an age of about 800~Myr \citep{jacobson, magrini2017, magrini2018, overbeek, jackson2021}. 

Of the final 25 Li candidates for Trumpler~23, we discarded two stars with no astrometric and photometric \textit{Gaia} data, further listed as non-members with P=$0.00$ by \cite{jackson2021}. As for kinematics, we also listed two stars as final candidates, $RV$ non-members according to the initial 2$\sigma$ interval, with $RV$s deviating up to $6$~km~s$^{-1}$ from the mean of the cluster, but fulfilling the rest of criteria and further listed as high-probability members by \cite{jackson2021}. Regarding previous membership selections from the literature, we firstly found nine stars in common with \cite{jacobson}, as well as 10 stars in \cite{magrini2017} and \cite{overbeek}, and 14 stars in \cite{sampedro}. Finally, \cite{jackson2021} includes all 23 of our final candidates.  We note that, in spite of being listed as a non-member by \cite{overbeek}, we considered the UVES star 16004025-5329439 as a candidate of Trumpler~23, as opposed to a Li-rich giant non-member, given that this star fulfilled all our membership criteria, and was listed as a member by several studies, such as \cite{magrini2017} and \cite{jackson2021}.  \\

\textbf{$\ast$ Berkeley~81} 


Berkeley~81 (also known as Br~81 or Be~81) in an open cluster with an age of $0.75$--$1$~Gyr \citep{donati2, jacobson, magrini2017, magrini2018, jackson2021, romano}. 

Similarly to other clusters in this range, of the final 25 Li candidates for Berkeley~81, we discarded one star with no astrometric or photometric \textit{Gaia} data (further listed as a non-member by \cite{jackson2021}). We also considered one UVES Li-rich giant star (19014498-0027496) as an additional candidate instead of a giant contaminant, not only because it fulfilled all of our membership criteria, but also because other studies \citep{jacobson, magrini2017} considered it to be a member of this cluster. As for kinematics, two stars in our final selection were $RV$ non-members according to the initial 2$\sigma$ interval, with $RV$s deviating $11$~km~s$^{-1}$ from the mean of the cluster. Also listed as high-probability members by \cite{jackson2021} and fulfilling all other criteria, we accepted them as candidates. In addition, we also accepted one star (19013633-0024141), listed as a member with P=$0.99$ by \cite{jackson2021} and fulfilling all criteria except for metallicity, as it deviates more appreciably from the 3$\sigma$ limit ($0.43$~dex from the cluster mean). On the other hand, another star we discarded for deviating appreciably from the [Fe/H] of the cluster, as it was not included in any study, and we also could not analyse this star using astrometry. Regarding previous membership selections from the literature, we firstly found seven stars in common with \cite{magrini2015}, as well as 13 stars in \cite{jacobson} and \cite{magrini2017}, 10 stars in \cite{sampedro}, and three stars in \cite{cantatgaudin_gaia}. Finally, \cite{jackson2021} includes all 24 of our final candidates. \\

\textbf{$\ast$ NGC~2355} 

\label{ap1:NGC2355}


NGC~2355 is an open cluster in the outer part of the Galactic disc with an age of $0.8$--$1$~Gyr \citep{kharchenko, sampedro, jackson2021, romano}. 

Of the final 87 Li candidates for NGC~2355, we discarded one star, listed as a non-member by \cite{jackson2021} and with no astrometric or photometric data available, as well as not fulfilling the gravity criteria. As for kinematics, 23 stars in our final selection were $RV$ non-members according to the initial 2$\sigma$ interval, with $RV$s deviating up to $6$--$8$~km~s$^{-1}$ from the mean of the cluster. Fulfilling the rest of membership criteria, and also listed as high-probability members by \cite{jackson2021} with P=$0.99$--$1.00$, we accepted them as candidates as in the case of other clusters. Regarding metallicity, we accepted 10 marginal members, as well as two stars which fulfilled all criteria except for metallicity, deviating more appreciably from the 3$\sigma$ limit (up to $0.25$~dex from the cluster mean). All of these stars are further listed as high-probability members by \cite{jackson2021}, with P=$1.00$. Regarding previous membership selections from the literature, we found 64 stars in common with \cite{cantatgaudin_gaia}, and \cite{jackson2021} includes 84 of our 86 final candidates (the remaining two stars in our selection were not included in their study). \\

\textbf{$\ast$ NGC~6802} 


NGC~6802 is an inner disc open cluster in the constellation Fuchs, with an age of $0.9$--$1.0$~Gyr \citep{jacobson, magrini2017, tang, magrini2018, jackson2021, romano}.

Of the final 36 Li candidates for NGC~6802, we firstly discarded four stars listed as non-members by \cite{jackson2021} and without any astrometric or photometric \textit{Gaia} data available. As for kinematics, we accepted six stars in our final selection which were $RV$ non-members according to the initial 2$\sigma$ interval, listed as high-probability members by \cite{jackson2021} with P=$0.94$--$0.99$ and fulfilling all criteria except for the fact that their $RV$s deviated up to $4$--$17$~km~s$^{-1}$ from the mean of the cluster. We also note that the final selection of this cluster includes a couple of stars which present values of $EW$(Li) slightly larger than the rest of candidates ($112$--$113$~m$\r{A}$). These two stars fulfil the rest of membership criteria, and one of them is also listed as a high probability member by \cite{jackson} (the other one is not included in their study), and so we accepted them as final members of the cluster. Furthermore, we can also explain the higher observable values of $EW$(Li) of these stars considering the fact that they potentially experimented a lower rate of Li depletion due to them exhibiting values of rotation that are in the middle of the scale for this cluster (also see Paper III). Regarding previous membership selections from the literature, we found the following stars in common with several studies: Firstly, we found eight common stars with \cite{jacobson} and \cite{magrini2017}, as well as 10 stars in \cite{sampedro}, 18 stars in \cite{tang}, and five common stars in \cite{cantatgaudin_gaia}. \cite{jackson2021} includes 31 of our 32 final candidates (the remaining star in our selection was not included in their study). Finally, one of the Li-rich giants in our list (J19304281+2016107) is also listed in \cite{casey}. \\

\textbf{$\ast$ NGC~6005} 


NGC~6005 is a Southern open cluster in the constellation Norma with an age of $0.97$--$1.20$~Gyr \citep{jacobson, magrini2017, bossini, jackson2021}. \cite{bossini} calculated an age of $973\pm4$~Myr for this cluster. 

Of the final 55 Li candidates for NGC~6005, we discarded five stars which we were not able to analyse using astrometric and photometric \textit{Gaia} data, and were further listed as non-members by \cite{jackson2021}. We also discarded another star which fulfilled all criteria except for metallicity. In contrast to our decision to accept stars as final candidates if they fulfil the rest of criteria apart from metallicity or $\log g$, especially the more restrictive astrometric criteria, in this case we discarded it from the final selection on the basis of not being able to analyse it with \textit{Gaia} data, and additionally for not being included by \cite{jackson2021}. As for kinematics, we accepted two stars which were $RV$ non-members according to the initial 2$\sigma$ interval, with $RV$s deviating up to $7$--$11$~km~s$^{-1}$ from the mean of the cluster, but fulfilled all other criteria and were also listed as members of the cluster by \cite{jackson2021} with P=$0.90$--$0.91$. We also note that the final selection of this cluster includes three stars which, even while presenting values of $EW$(Li) slightly larger than the rest of candidates ($102$--$118$~m$\r{A}$), fulfil the rest of membership criteria, are listed as high probability members by \cite{jackson}, and could also be experiencing a slower rate of Li depletion caused by rotation values that are in the middle of the scale for this cluster. Finally, we also accepted eight marginal members (deviating up to $0.30$--$0.50$~dex from the mean of the cluster), as they fulfilled all other criteria, and are listed as high-probability members by \cite{jackson2021}, with P=$0.90$--$0.99$.

Regarding previous membership selections from the literature, we found the following stars in common with several studies: Firstly, we found 12 common stars with \cite{jacobson}, as well as seven stars in \cite{magrini2017}, 11 stars in \cite{sampedro}, and 12 common stars in \cite{cantatgaudin_gaia}. \cite{jackson2021} includes all 49 of our final candidates. As already discussed in Sect.~\ref{outliers}, we also note that we seem to find some inconsistencies when plotting the Li-rich giant outliers selected in the field of this cluster in the CMD diagram, with all three of them appearing among the non-giant cluster candidates. However, when plotting them in the Kiel diagram the expected distinction is found between the non-giant cluster candidates and the Li-rich giant outliers. \\ 

\textbf{$\ast$ Pismis~18} 


Pismis~18 is a Southern open cluster with an age of 1.2~Gyr \citep{jacobson, magrini2017, magrini2018, jackson2021}. 

Of the final 12 Li candidates for Pismis~18, we firstly accepted one star in our final selection which fulfilled all criteria listed as  a member of the cluster by \cite{jackson2021} with P=$0.98$, but was a $RV$ non-member according to the initial 2$\sigma$ interval, with a $RV$ deviating up to $6$~km~s$^{-1}$ from the mean of the cluster. We also discarded two stars with no astrometric and photometric \textit{Gaia} data available, further listed as non-members by \cite{jackson2021}.  Regarding previous membership selections from the literature, we found the following stars in common with several studies: Firstly, we found five common stars with \cite{jacobson}, as well as three stars in \cite{magrini2017}, and all of our 10 stars are listed in \cite{hatzdimitriou}. \cite{jackson2021} also includes all 10 of our final candidates. \\

\textbf{$\ast$ Melotte~71} 


Melotte~71 (also known as Mel~71) is an open cluster between the inner and outer disc. \cite{bossini} calculated an age of $1294\pm89$~Myr for this cluster, while \cite{netopil2016} listed an age of $0.7$~Gyr. 

Regarding previous membership selections from the literature for Melotte~71, we found three stars in common with \cite{sampedro}, and four common stars in \cite{cantatgaudin_gaia}. We note that this is one of the three clusters out of our sample of 42 clusters which is not included in \cite{jackson2021}. \\

\textbf{$\ast$ Pismis~15} 


Pismis~15 is an open cluster with an age of $1.3$~Gyr \citep{carraro, kharchenko, sampedro, liu, jackson2021}. 

Among the final 33 Li candidates for Pismis~15, we accepted a star which fulfilled all criteria except for metallicity, deviating more appreciably from the 3$\sigma$ limit ($0.66$~dex from the cluster mean). This star is further listed as a member by \cite{jackson2021}, with P=$0.99$. We also discarded two stars with no astrometric or photometric \textit{Gaia} data, listed as non-members by \cite{jackson2021}. We also note that the final selection of this cluster includes eight stars which present values of $EW$(Li) slightly larger than the rest of candidates ($92$--$147$~m$\r{A}$), albeit without exhibiting appreciable dispersion with respect to the other members, fulfilling the rest of membership criteria, and also being listed as high probability members by \cite{jackson} with P=$0.98$--$0.99$. Furthermore, a couple of these stars also exhibit high values of rotation according to the range observed for this cluster. The effects of rotation could thus also explain the higher values of $EW$(Li) of these stars, as we could explain them as members of Pismis~15 exhibiting a slightly slower rate of Li depletion as a result of being faster rotators. Regarding previous membership selections from the literature, we found five common stars with \cite{sampedro}, as well as 15 stars in \cite{cantatgaudin_gaia}. \cite{jackson2021} includes all 31 of our final candidates. \\

\textbf{$\ast$ Trumpler~20} 

\label{ap1:Trumpler20}


Trumpler~20 is an open cluster located towards the Galactic centre, with an age of $1.4$--$1.8$~Gyr \citep{donati, jacobson, magrini2017, jackson2021, romano}. 

Of the final 116 Li candidates for Trumpler~20, we firstly discarded 12 stars, listed as non-members by \cite{jackson2021}, with no astrometric \textit{Gaia} data available; as well as another star which fulfilled all criteria except for metallicity (but lacked \textit{Gaia} data), and was listed as a definite non-member by \cite{jackson2021}. As for kinematics, we accepted five stars, all of them listed as members of the cluster by \cite{jackson2021} with P=$0.87$--$0.98$, albeit being $RV$ non-members according to the initial 2$\sigma$ interval, with $RV$s deviating up to $10$~km~s$^{-1}$ from the mean of the cluster. As for metallicity, we also accepted 10 marginal members (deviating up to $0.20$--$0.40$~dex from the mean of the cluster). All of these stars fulfilled the rest of criteria, and all are further listed as members by \cite{jackson2021}, with P=$0.87$--$0.99$. 

Thee final selection for this cluster also includes a number of stars which present values of $EW$(Li) somewhat larger than the rest of candidates ($90$--$120$~m$\r{A}$), albeit without showing an appreciable dispersion with respect to the other members, and also fulfilling all other criteria, and listed as high probability members by \cite{jackson} with P=$0.89$--$.99$. Furthermore, we may also explain the higher observable values of $EW$(Li) of several of these stars seeing as they also exhibit values of rotation that are in the middle and high end of the scale for this cluster, thus potentially making them cluster members that have experienced a slower rate of Li depletion as a result (see Paper III). Finally, we also find one star (12400449-6036566) which also exhibits a high $EW$(Li) ($137$~m$\r{A}$) and stands apart from the rest of the candidates. We listed this star as a Li-rich candidate of the cluster. This star fulfilled the rest of our criteria, and it is also listed as a Li-rich member by \cite{smiljanic2016}. 

Regarding previous membership selections from the literature, we found the following stars in common with several studies: Firstly, we found 97 common stars with \cite{donati}, as well as 23 stars in \cite{tautvaisiene}, 23 stars in \cite{jacobson}, 23 stars in \cite{smiljanic2016}, 17 stars in \cite{magrini2017}, 31 stars in \cite{sampedro}, and 42 stars in \cite{cantatgaudin_gaia}. Finally, \cite{jackson2021} includes all but one of our 104 final candidates. The remaining star in our selection fulfils all of our criteria in spite of being listed as a non-member in \cite{jackson2021}. Once again, the reason for these individual deviations between our two analyses probably originates from the existing differences regarding criteria limits or the weight of the factors implicated in the decision to accept or discard stars as final candidates. \\

\textbf{$\ast$ Berkeley~44} 


Berkeley~44 (also known as Br~44) is an open cluster with a dense background field and an age of  $1.4$--$1.6$~Gyr \citep{hayesfriel, jacobson, magrini2017, magrini2018, jackson2021, romano}. 

Of the final 31 Li candidates for Berkeley~44, we discarded one star which we could not analyse with \textit{Gaia} astrometric data, also listed as a non-member by \cite{jackson2021}. On the other hand, we accepted five stars, listed as members of the cluster by \cite{jackson2021} with P=$0.86$--$0.99$ and fulfilling all other critera, which were $RV$ non-members according to the initial 2$\sigma$ interval, with $RV$s deviating up to $3$--$4$~km~s$^{-1}$ from the mean of the cluster. We also accepted one marginal metallicity member (deviating $0.29$~dex from the mean of the cluster), as it fulfilled all other criteria and is listed as a high-probability member by \cite{jackson2021}, with P=$1.00$. Regarding previous membership selections from the literature, we found the following stars in common with several studies: Firstly, we found three common stars with \cite{jacobson} and \cite{magrini2017}, as well as eight stars in \cite{sampedro}, and 20 stars in \cite{cantatgaudin_gaia}. Finally, \cite{jackson2021} includes all 30 of our final candidates. \\

\textbf{$\ast$  NGC~2243} 

\label{ap1:NGC2243}


NGC~2243 is a rich open cluster in the constellation of Canis Major, with an age in the range of $3.8$--$4.4$~Gyr \citep{richer, jacobson2011, heiter, magrini2017, magrini2018, jackson2021, romano}. NGC~2243 is one of the most metal-poor clusters known, with a [Fe/H] value of $-0.38\pm0.04$ dex \citep{magrini2017, magrini2018}. We note that for the individual figures of this cluster (see Appendix~\ref{ap1:AppendixC}),  due to the low metallicity of this cluster we considered PARSEC isochrones with Z=0.006 instead of the usual near-solar metallicity of Z=0.019 we used for the rest of the clusters in our sample. 

Regarding kinematics, 36 stars in our final selection for NGC~2243 were $RV$ non-members according to the initial 2$\sigma$ interval, with $RV$s deviating up to $8$~km~s$^{-1}$ from the mean of the cluster. As in the case of other cluster analyses, we accepted both of them as final members, and both stars were also listed as members of the cluster by \cite{jackson2021} with P=$0.93$--$1.00$. We also accepted 13 marginal metallicity members (deviating up to $0.50$~dex from the mean of the cluster), as well as three additional stars which deviated more appreciably from the 3$\sigma$ limit (up to $0.60$--$0.80$~dex from the cluster mean). All of these stars fulfilled all other criteria and all but one are further listed as high-probability members by \cite{jackson2021}, with P=$0.99$--$1.00$.

Regarding previous membership selections from the literature, we found the following stars in common with several studies: Firstly, we found nine common stars with \cite{jacobson2011}, as well as one star in \cite{heiter}, 13 stars in \cite{magrini2017}, 42 stars in \cite{sampedro}, and 178 stars in \cite{cantatgaudin_gaia}. Finally, \cite{jackson2021} includes all but one of our 289 of our final candidates (the remaining star in our selection was not included in their study). Given that NGC~2243 and M67 are very close age-wise, we also note that we additionally made use of the Li envelope created by our candidate selection for M67 (see Paper III, as well as former M67 attested members, to help confirm the membership of our selection for NGC~2243. We also note that we seem to find some inconsistencies when plotting the Li-rich giant outliers selected in the field of this cluster in the CMD diagram, with five of them appearing among the non-giant cluster candidates. However, when plotting them in the Kiel diagram the expected distinction is found between the non-giant cluster candidates and the Li-rich giant outliers.  \\

\textbf{$\ast$ M67} 

\label{ap1:M67}


M67 (also known as Messier 67 or NGC~2682) is an open cluster in the constellation of Cancer, with an age in the range of $3.6$--$4.5$~Gyr, an age close to that of the Sun \citep{balachandran, richer, pallavicini, sestitorandich}. 

Eight stars in our final selection for M67 were $RV$ non-members according to the initial 2$\sigma$ interval, with $RV$s deviating up to $4$--$9$~km~s$^{-1}$ from the mean of the cluster, but were accepted as final members and also listed as definite candidates by \cite{jackson2021} with P=$0.99$. We also accepted one marginal metallicity member (08511748+1145225), which deviates $0.14$~dex from the cluster mean but fulfilled all other criteria, also being listed as a high-probability member by \cite{jackson2021}, with P=$1.00$. Regarding previous membership selections from the literature, we found the following stars in common with several studies: Firstly, we found 65 common stars with \cite{pace}, as well as 50 stars in \cite{pasquini2012}, two stars in \cite{carlberg}, 65 stars in \cite{geller}, 55 stars in \cite{brucalassi}, 74 stars in \cite{sampedro}, 29 stars in \cite{jadhav}, and seven stars in \cite{ilin}. Finally, \cite{jackson2021} includes all 96 of our final candidates. We also note that in order to reinforce the membership of our GES selection we made use of several lists of M67 candidates from a series of non-GES studies \citep{hobbs, balachandran, pallavicini1997, pasquini1997, jones, randich2002}. \\



%
%

\section{Cluster membership analysis: Individual figures}
\label{ap1:AppendixC}

This appendix includes individual figures for several membership criteria discussed in Sect.~\ref{analysis}, for all $42$ clusters in the sample. For the final selection of candidate members in each cluster, we show the final $RV$ distribution and parallax distributions, the $pmra$-versus-$pmdec$ proper motions diagram, the \textit{Gaia} CMD, the $\gamma$-versus-$T_{\rm eff}$ and/or Kiel diagrams, and the $EW$(Li)-versus-$T_{\rm eff}$ diagram. When available, strong accreting members and Li-rich giant contaminants are further plotted alongside the candidate members. Also shown in all $EW$(Li)-versus-$T_{\rm eff}$ diagrams are the upper envelope of $EW$(Li) for IC~2602 ($35$~Myr, red), the upper and lower envelopes of the Pleiades ($78$--$125$~Myr, grey), and the upper envelope of the Hyades ($750$~Myr, turquoise).

\clearpage 

\subsection{NGC~6530}


   \begin{figure} [htp]
   \centering
\includegraphics[width=1\linewidth, height=5cm]{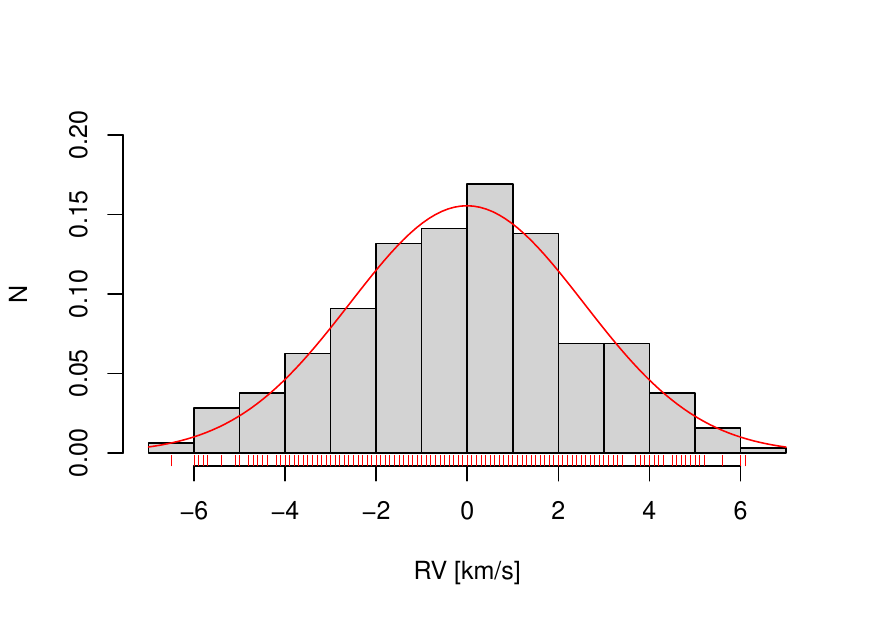}
\caption{$RV$ distribution for NGC~6530.}
             \label{fig:1}
    \end{figure}
    
           \begin{figure} [htp]
   \centering
\includegraphics[width=1\linewidth, height=5cm]{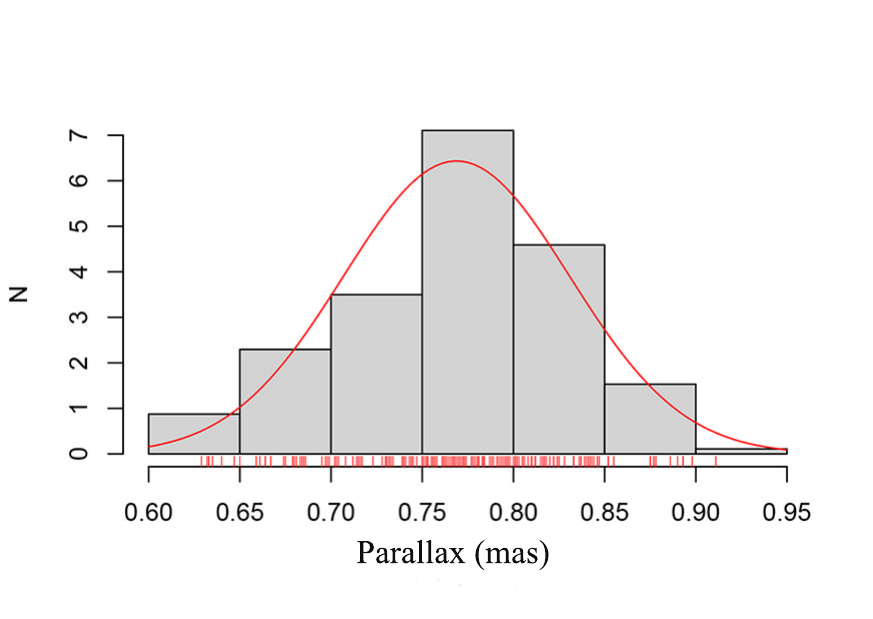}
\caption{Parallax distribution for NGC~6530.}
             \label{fig:2}
    \end{figure}
    
               \begin{figure} [htp]
   \centering
   \includegraphics[width=0.9\linewidth]{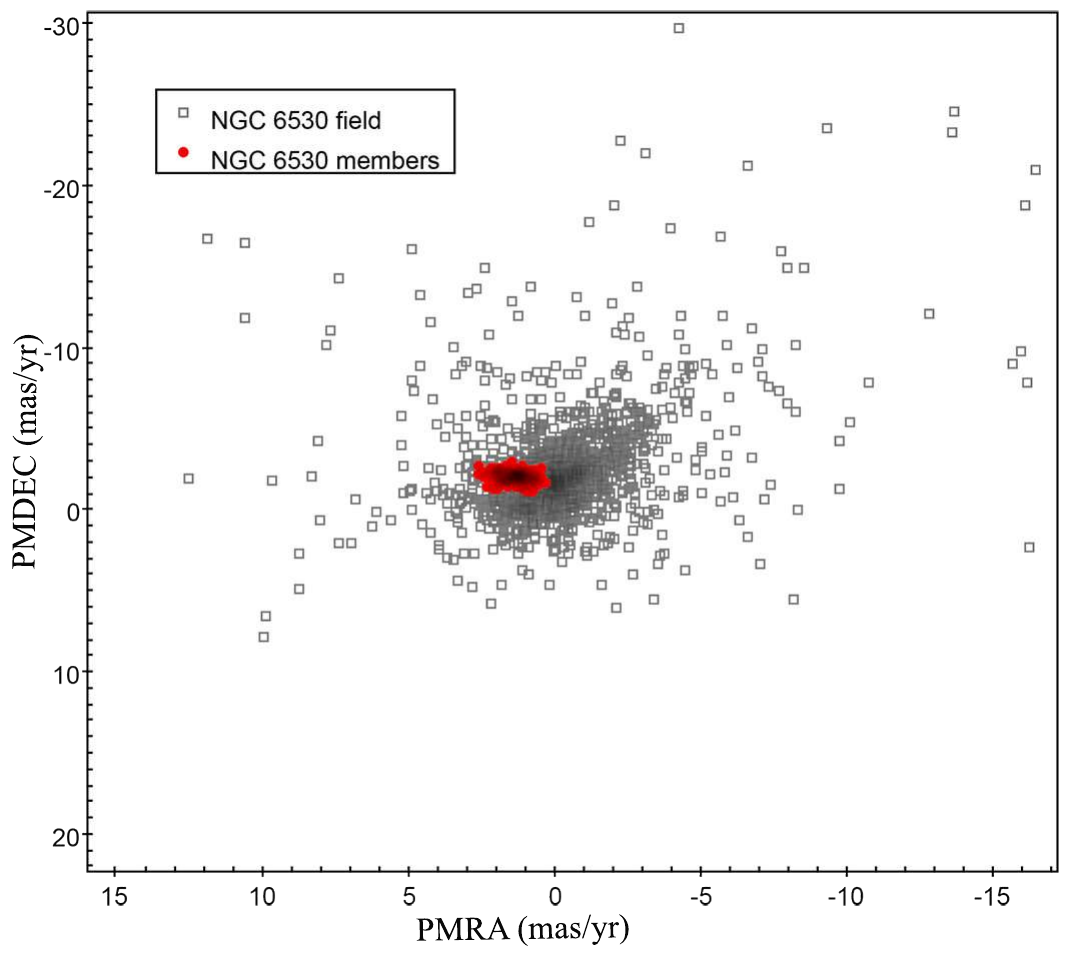}
   \caption{PMs diagram for NGC~6530.}
             \label{fig:3}
    \end{figure}
    
     \begin{figure} [htp]
   \centering
   \includegraphics[width=0.8\linewidth, height=7cm]{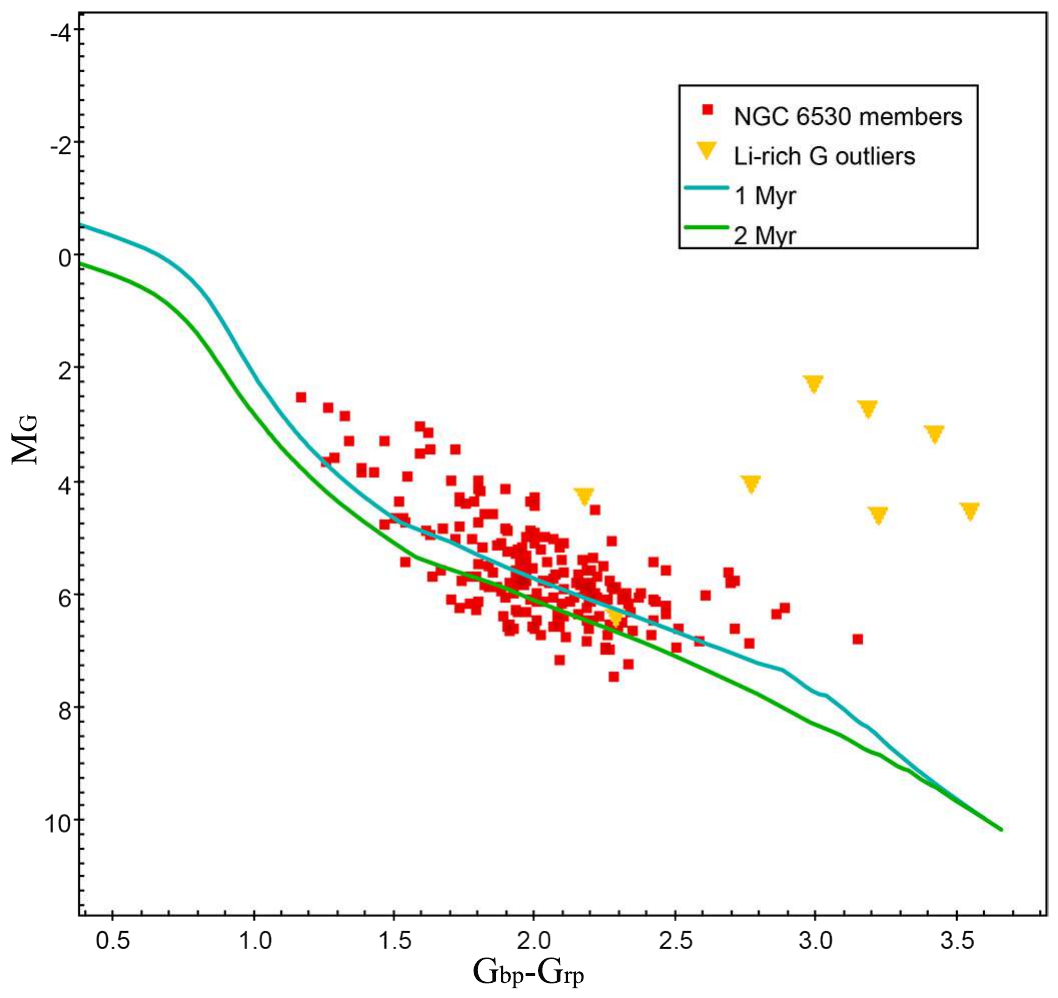}
   \caption{CMD for NGC~6530.}
             \label{fig:4}
    \end{figure}
    
         \begin{figure} [htp]
   \centering
   \includegraphics[width=0.8\linewidth, height=7cm]{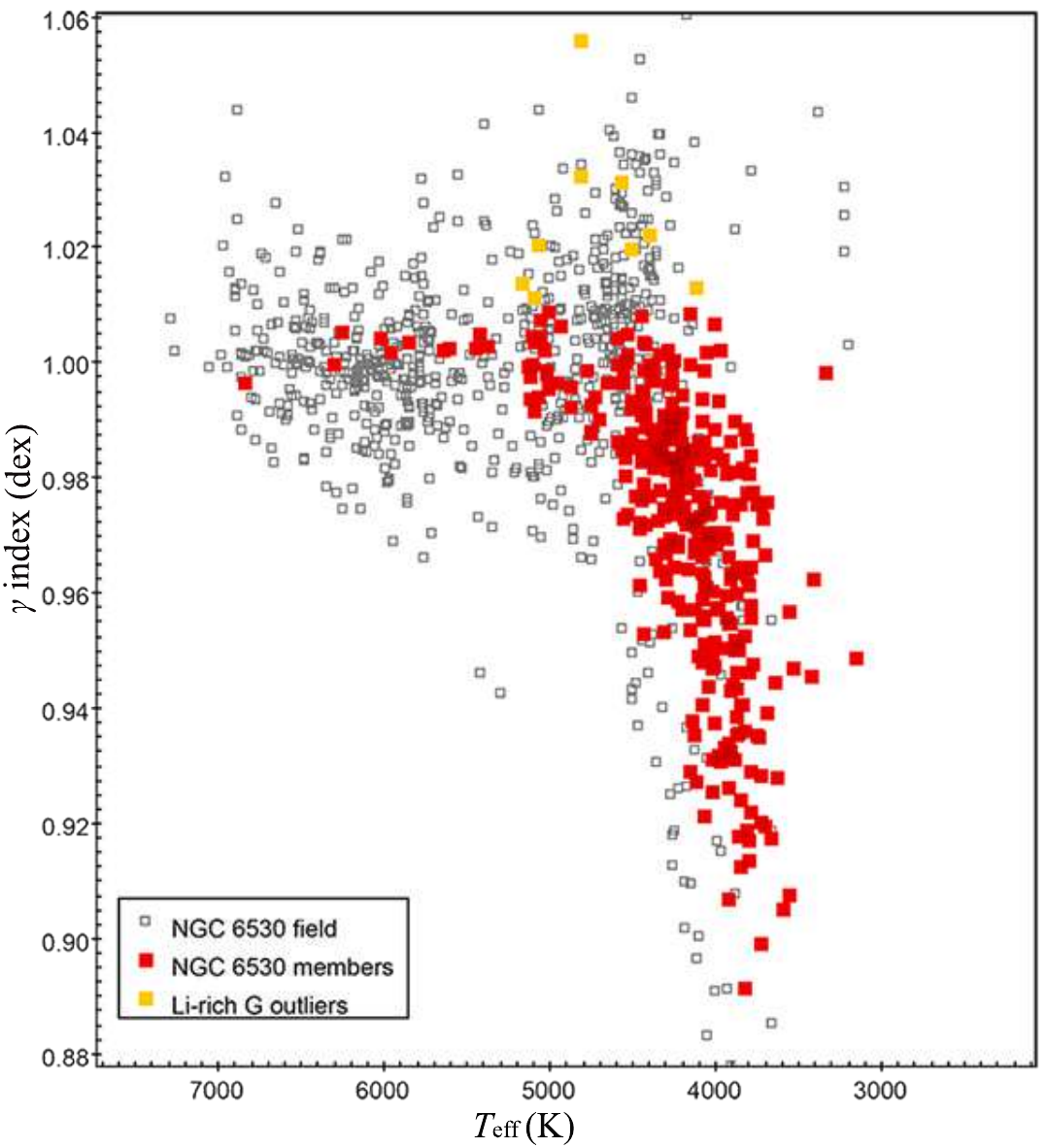}
   \caption{$\gamma$ index-versus-$T_{\rm eff}$ diagram for NGC~6530.}
             \label{fig:5}
    \end{figure}

  \begin{figure} [htp]
   \centering
 \includegraphics[width=0.8\linewidth]{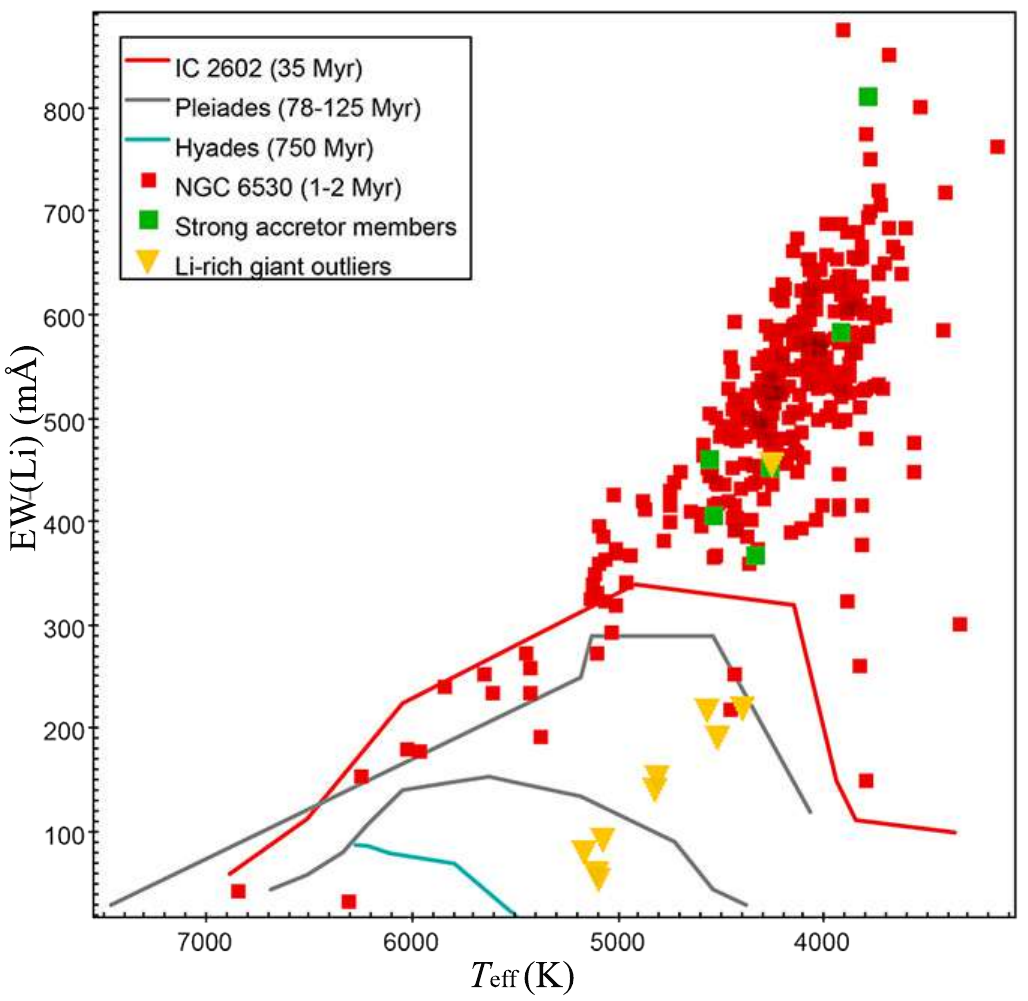} 
\caption{$EW$(Li)-versus-$T_{\rm eff}$ diagram for NGC~6530.}
             \label{fig:6}
    \end{figure}

\clearpage

\subsection{Rho~Oph}

   \begin{figure} [htp]
   \centering
\includegraphics[width=1\linewidth, height=5cm]{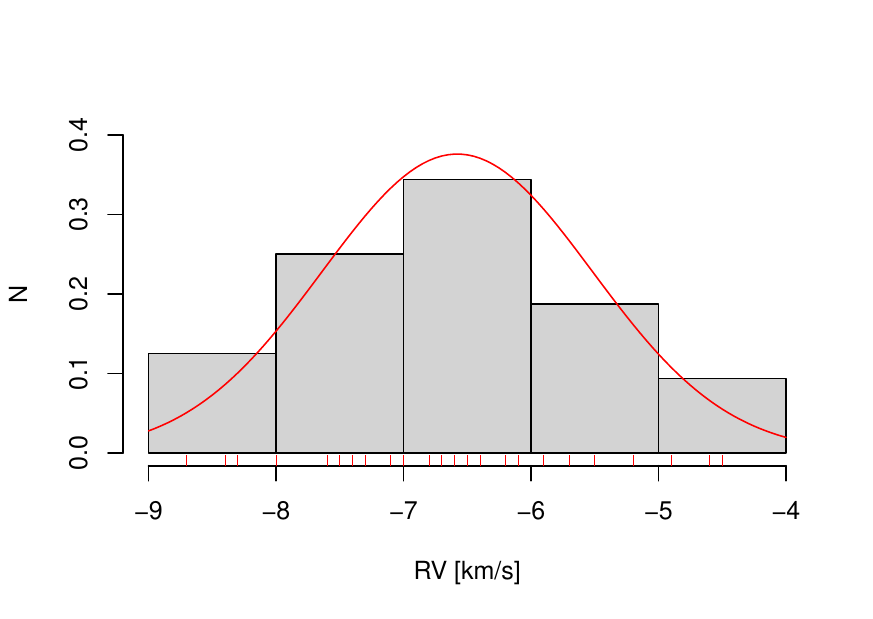}
\caption{$RV$ distribution for Rho~Oph.}
             \label{fig:7}
    \end{figure}
    
           \begin{figure} [htp]
   \centering
\includegraphics[width=1\linewidth, height=5cm]{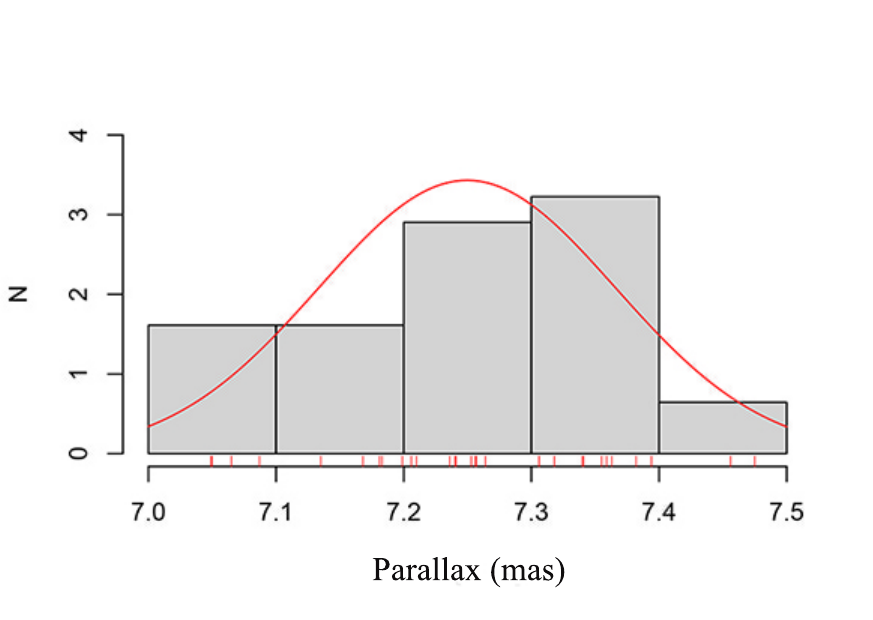}
\caption{Parallax distribution for $\rho$~Oph}
             \label{fig:8}
    \end{figure}
    
               \begin{figure} [htp]
   \centering
   \includegraphics[width=0.9\linewidth]{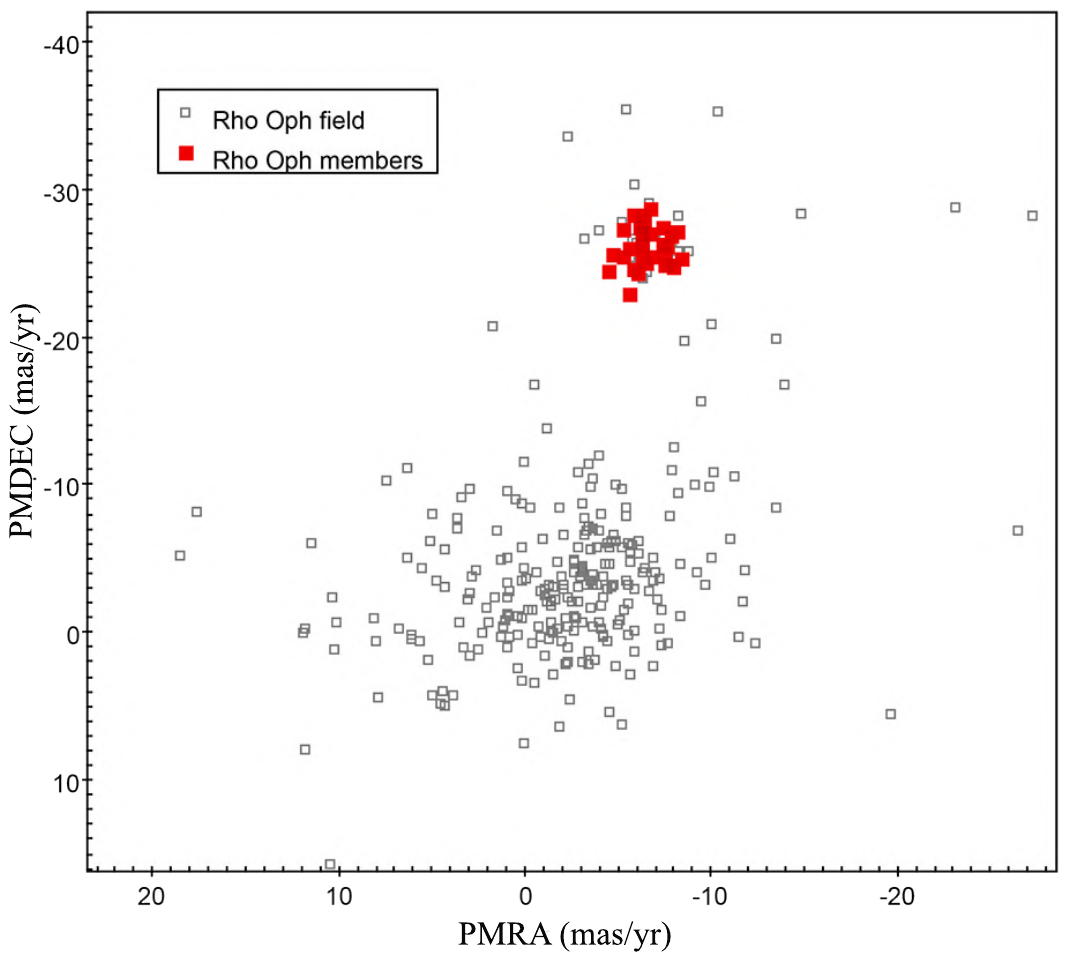}
   \caption{PMs diagram for $\rho$~Oph.}
             \label{fig:9}
    \end{figure}
    
     \begin{figure} [htp]
   \centering
   \includegraphics[width=0.8\linewidth, height=7cm]{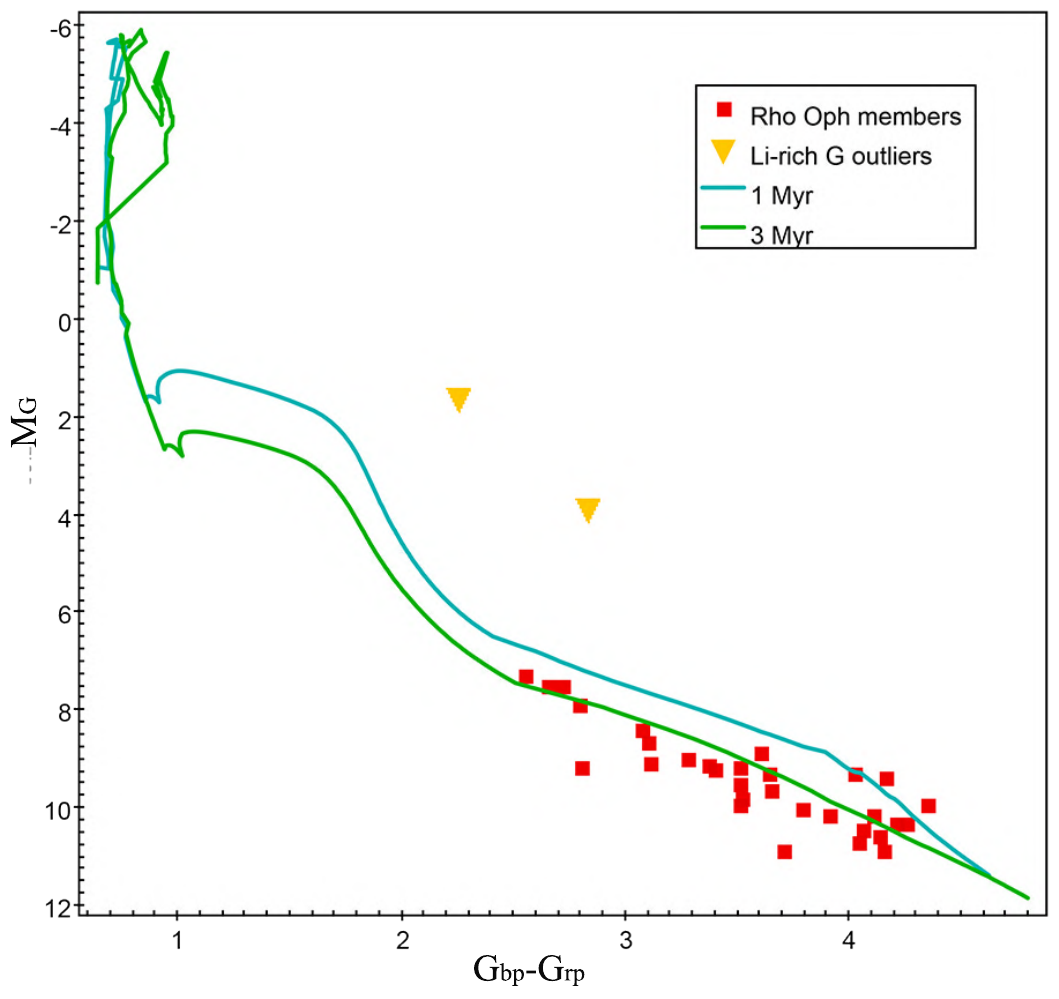}
   \caption{CMD for $\rho$~Oph.}
             \label{fig:10}
    \end{figure}
    
         \begin{figure} [htp]
   \centering
   \includegraphics[width=0.8\linewidth, height=7cm]{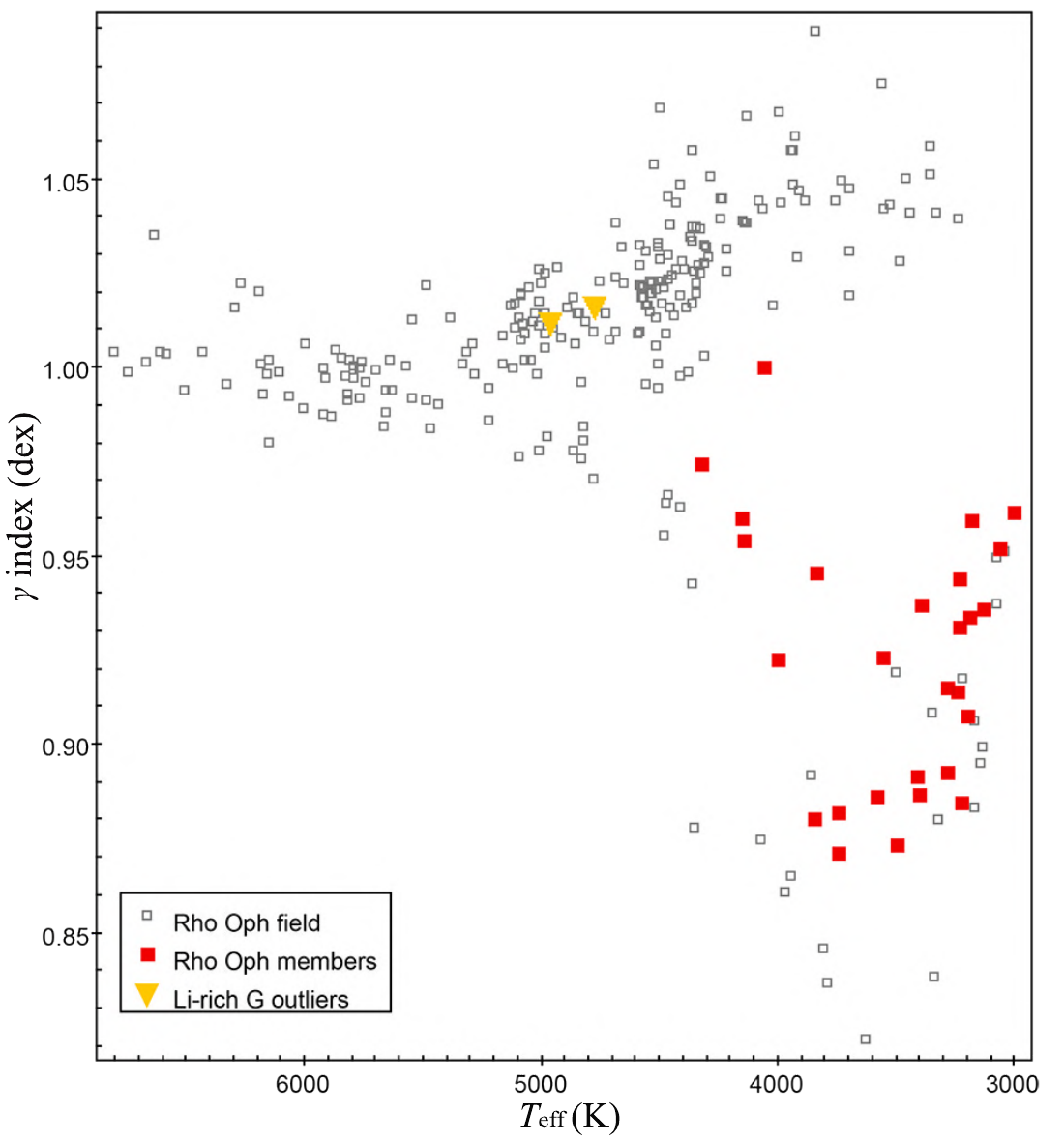}
   \caption{$\gamma$ index-versus-$T_{\rm eff}$ diagram for $\rho$~Oph.}
             \label{fig:11}
    \end{figure}

  \begin{figure} [htp]
   \centering
 \includegraphics[width=0.8\linewidth]{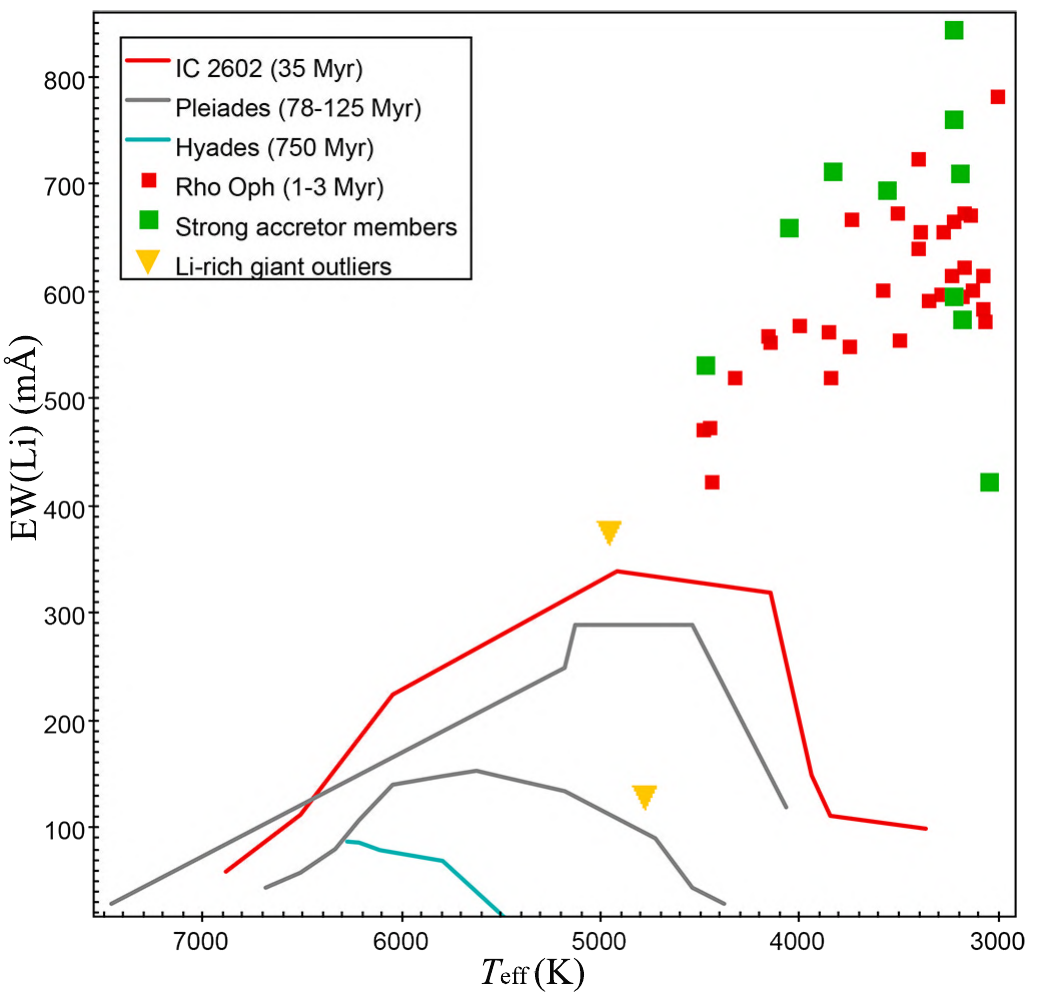} 
\caption{$EW$(Li)-versus-$T_{\rm eff}$ diagram for $\rho$~Oph.}
             \label{fig:12}
    \end{figure}

\clearpage

\subsection{Trumpler~14}

   \begin{figure} [htp]
   \centering
\includegraphics[width=1\linewidth, height=5cm]{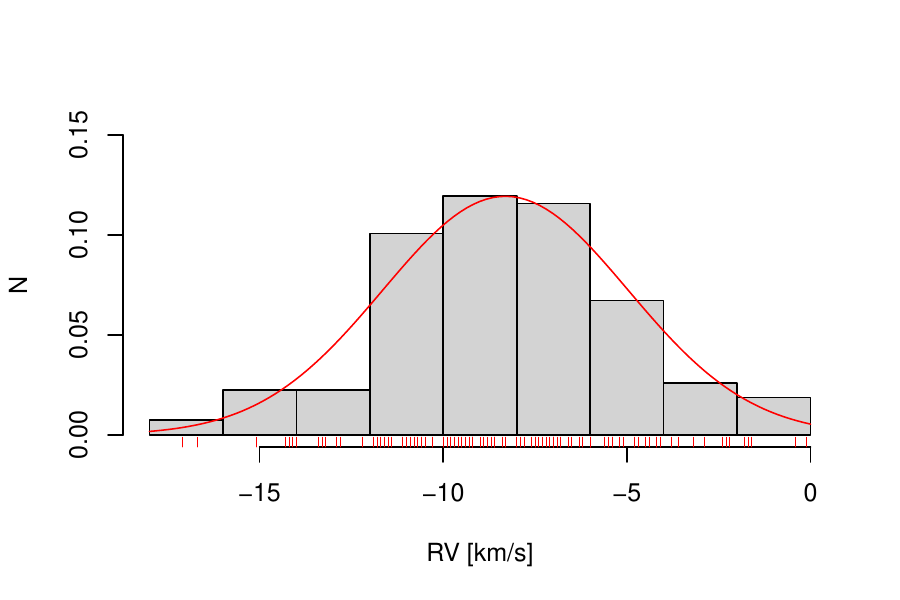}
\caption{$RV$ distribution for Trumpler~14.}
             \label{fig:13}
    \end{figure}
    
           \begin{figure} [htp]
   \centering
\includegraphics[width=1\linewidth, height=5cm]{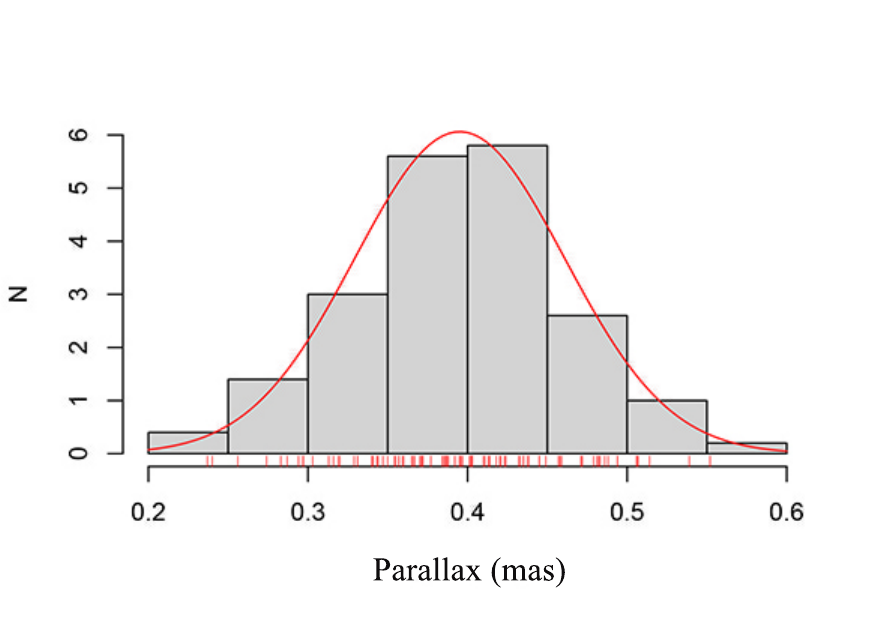}
\caption{Parallax distribution for Trumpler~14.}
             \label{fig:14}
    \end{figure}
    
               \begin{figure} [htp]
   \centering
   \includegraphics[width=0.9\linewidth]{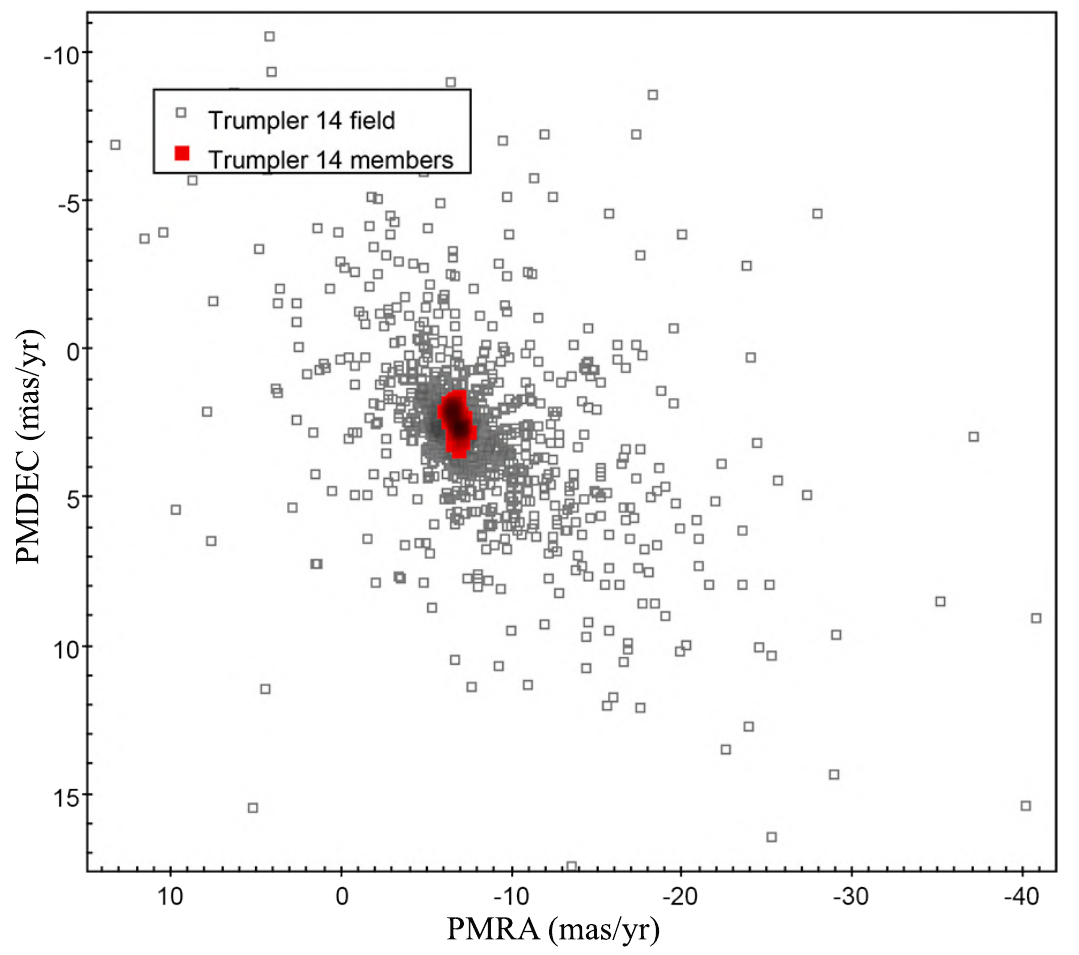}
   \caption{PMs diagram for Trumpler~14.}
             \label{fig:15}
    \end{figure}
    
     \begin{figure} [htp]
   \centering
   \includegraphics[width=0.8\linewidth, height=7cm]{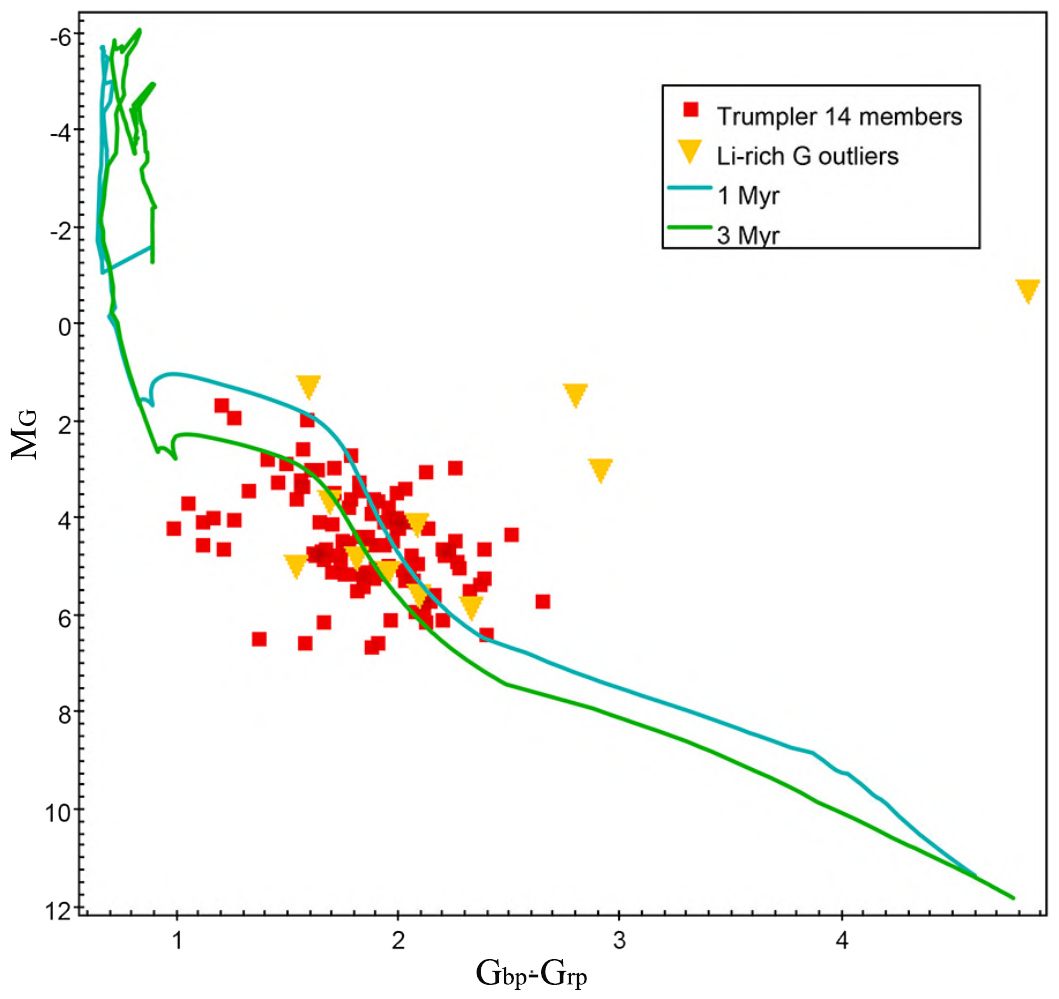}
   \caption{CMD for Trumpler~14.}
             \label{fig:16}
    \end{figure}
    
         \begin{figure} [htp]
   \centering
   \includegraphics[width=0.8\linewidth, height=7cm]{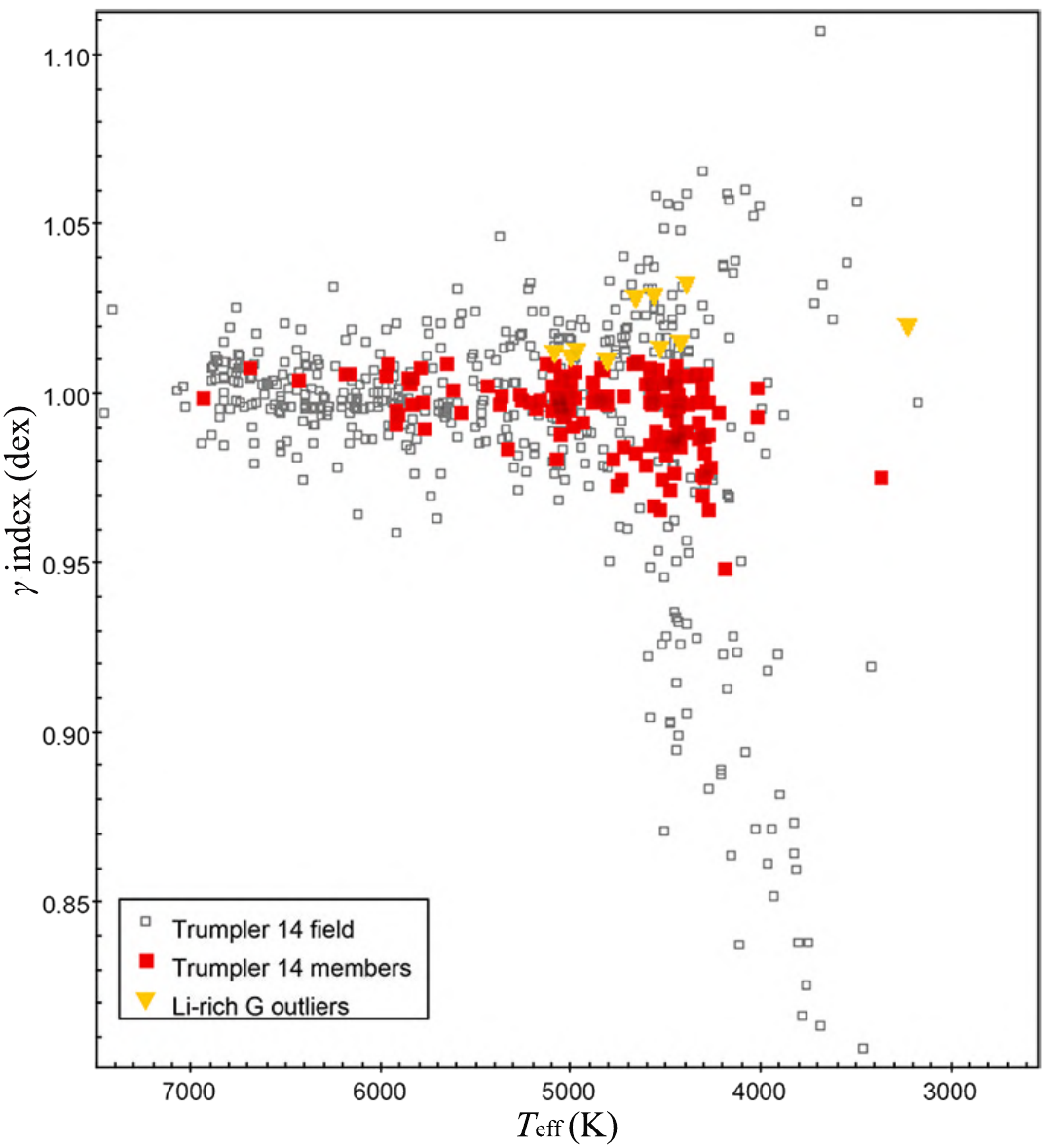}
   \caption{$\gamma$ index-versus-$T_{\rm eff}$ diagram for Trumpler~14.}
             \label{fig:17}
    \end{figure}

  \begin{figure} [htp]
   \centering
 \includegraphics[width=0.8\linewidth]{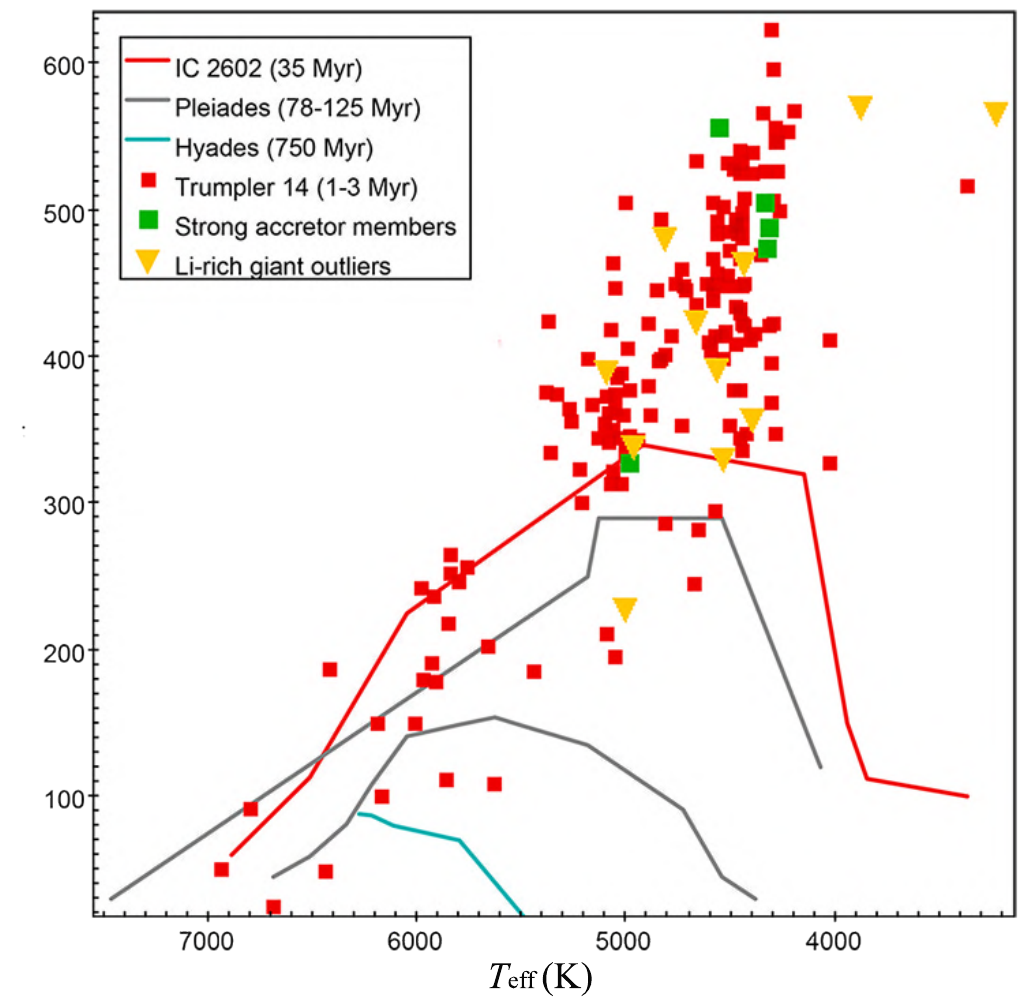} 
\caption{$EW$(Li)-versus-$T_{\rm eff}$ diagram for Trumpler~14.}
             \label{fig:18}
    \end{figure}

\clearpage

\subsection{Cha~I}

   \begin{figure} [htp]
   \centering
\includegraphics[width=1\linewidth, height=5cm]{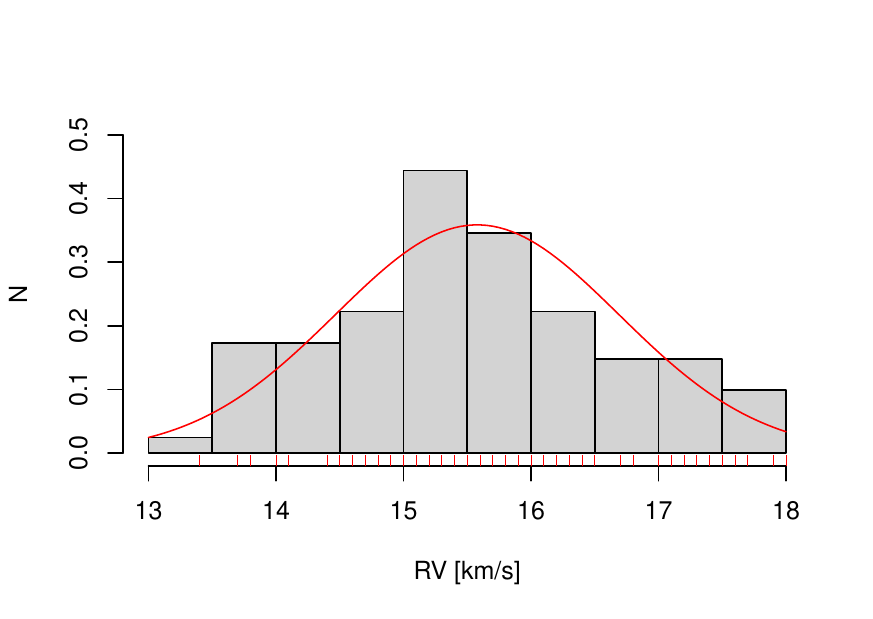}
\caption{$RV$ distribution for Cha~I.}
             \label{fig:19}
    \end{figure}
    
           \begin{figure} [htp]
   \centering
\includegraphics[width=1\linewidth, height=5cm]{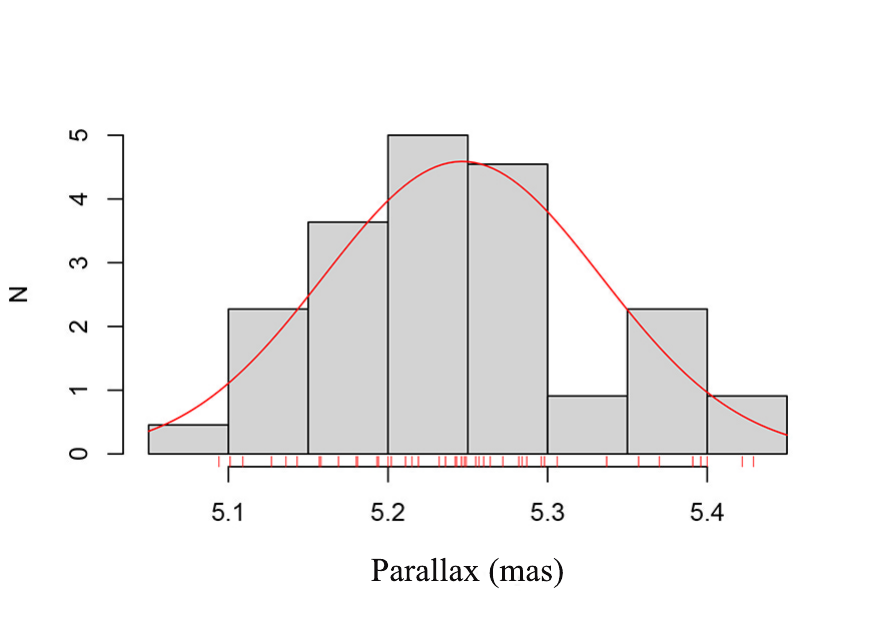}
\caption{Parallax distribution for Cha~I.}
             \label{fig:20}
    \end{figure}
    
               \begin{figure} [htp]
   \centering
   \includegraphics[width=0.9\linewidth]{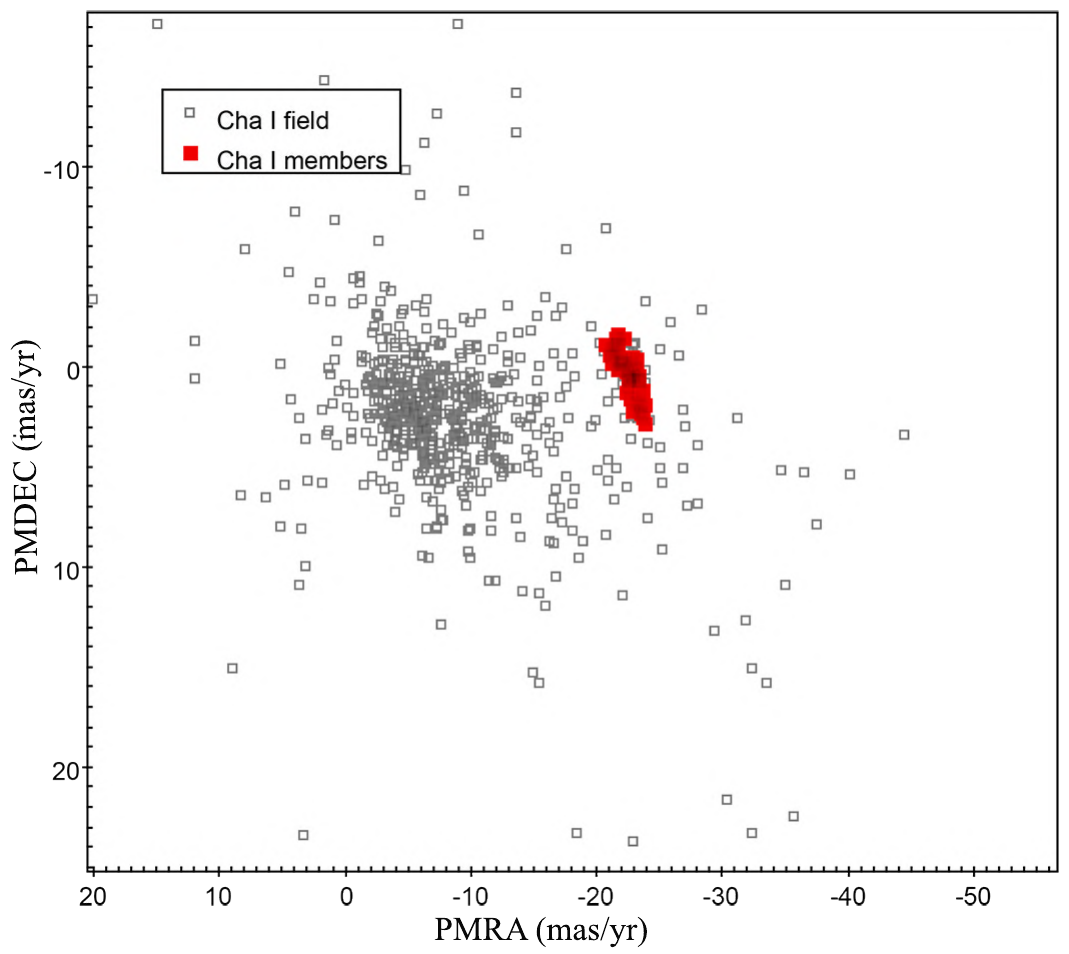}
   \caption{PMs diagram for Cha~I.}
             \label{fig:21}
    \end{figure}
    
     \begin{figure} [htp]
   \centering
   \includegraphics[width=0.8\linewidth, height=7cm]{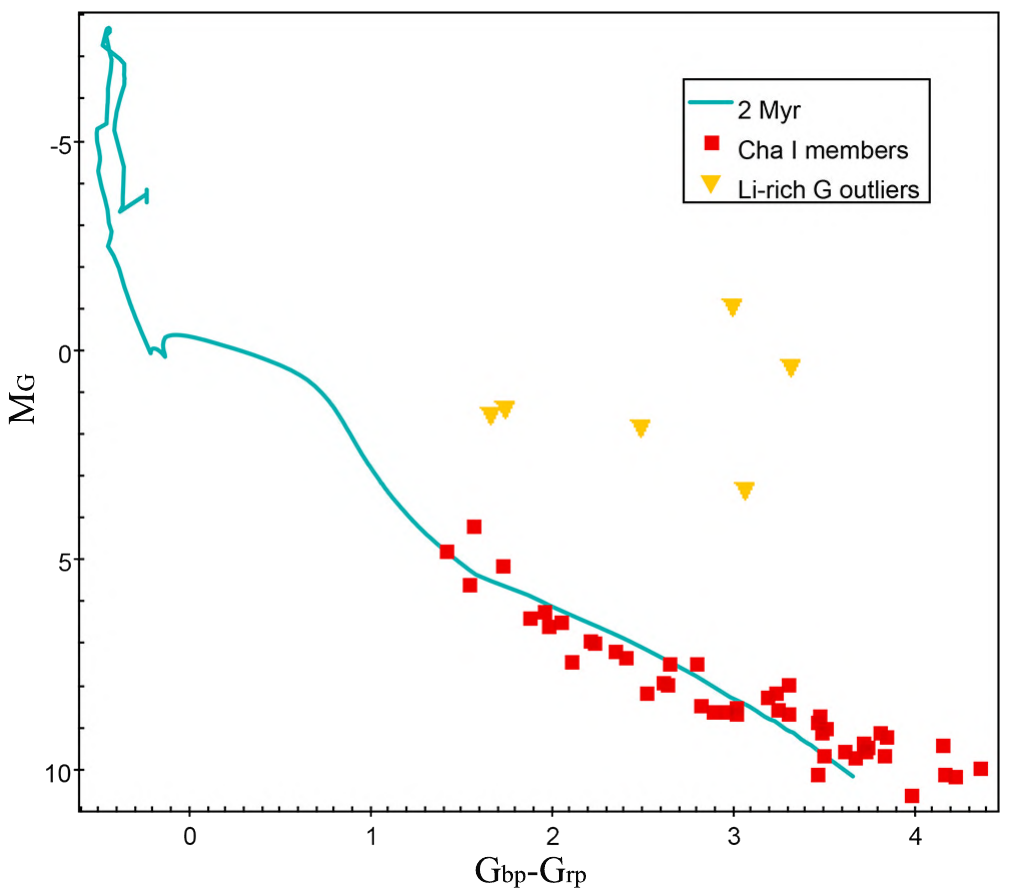}
   \caption{CMD for Cha~I.}
             \label{fig:22}
    \end{figure}
    
         \begin{figure} [htp]
   \centering
   \includegraphics[width=0.8\linewidth, height=7cm]{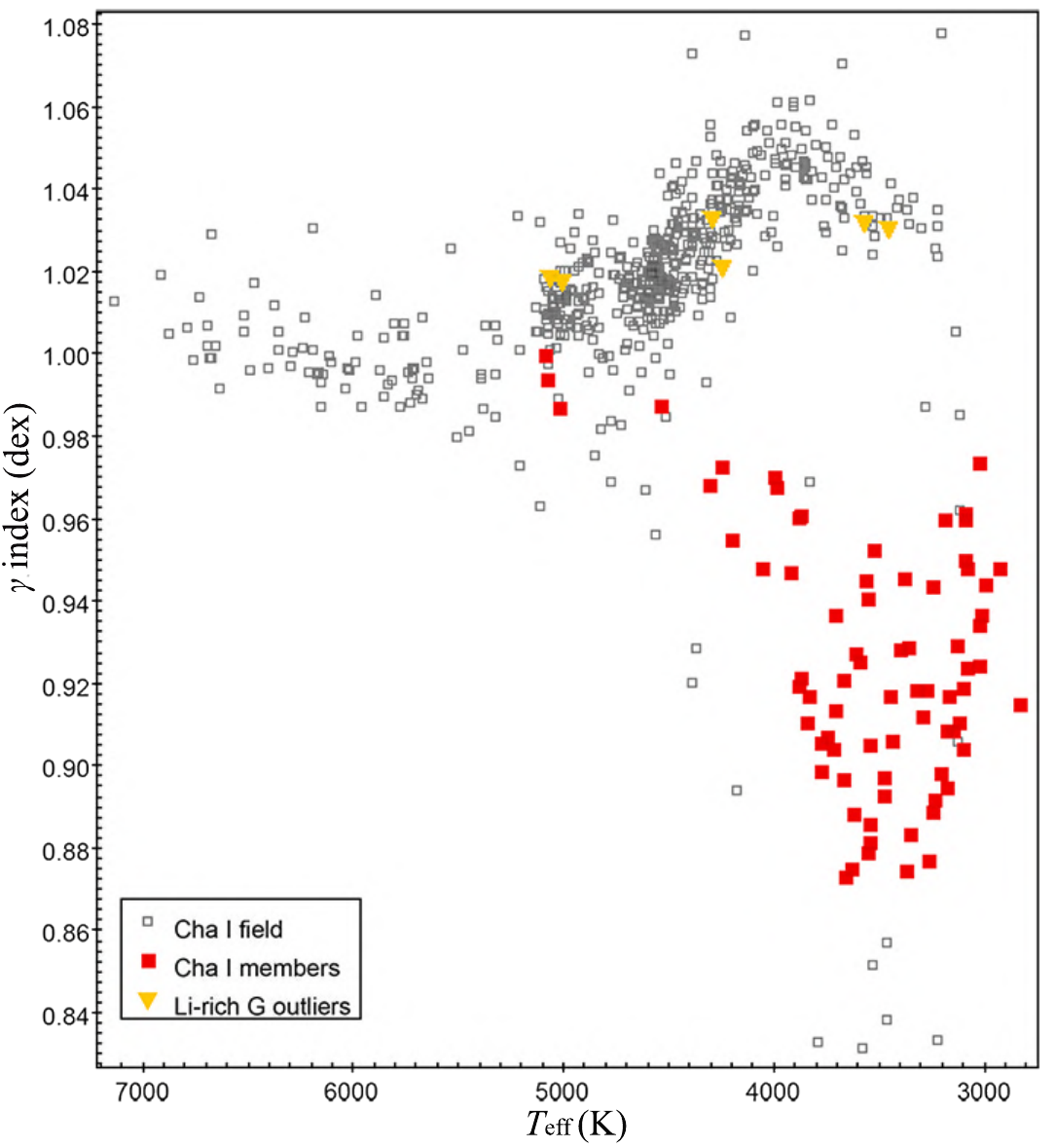}
   \caption{$\gamma$ index-versus-$T_{\rm eff}$ diagram for Cha~I.}
             \label{fig:23}
    \end{figure}

  \begin{figure} [htp]
   \centering
 \includegraphics[width=0.8\linewidth]{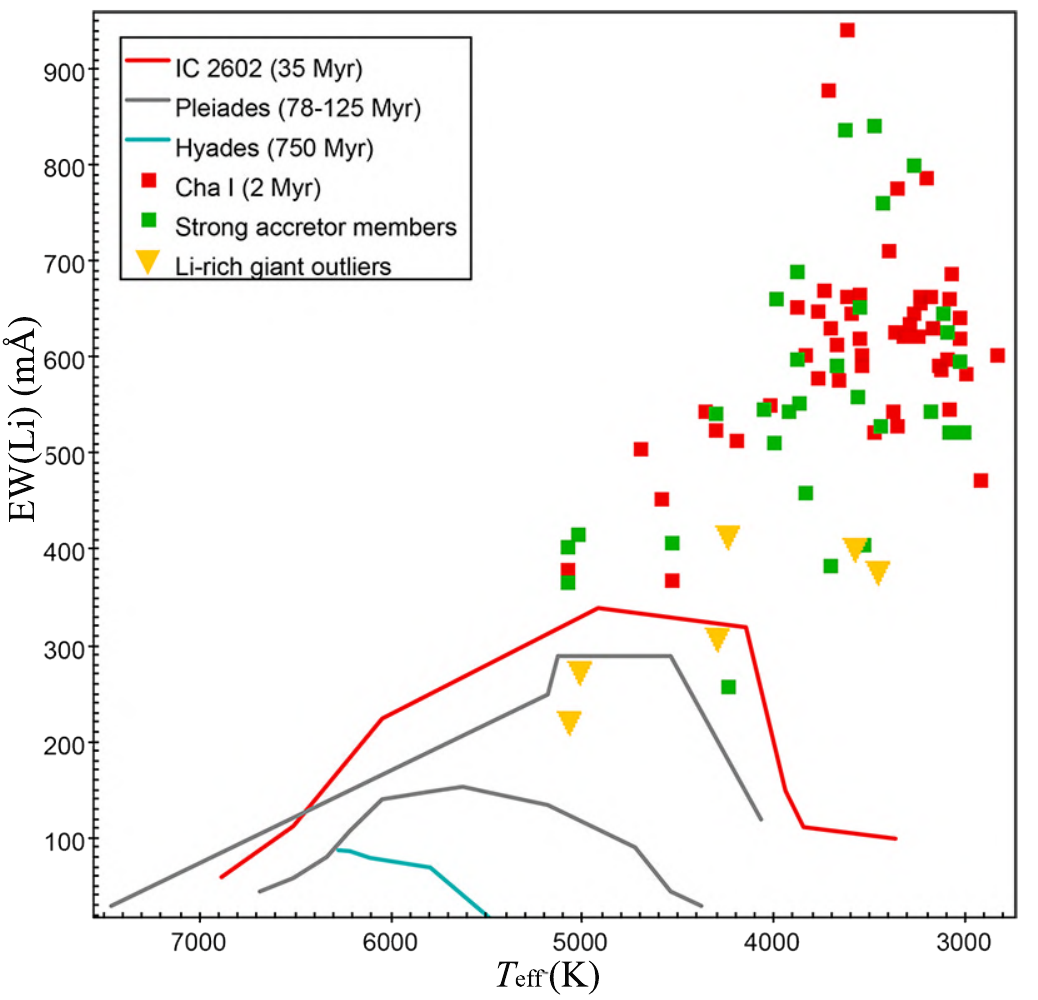} 
\caption{$EW$(Li)-versus-$T_{\rm eff}$ diagram for Cha~I.}
             \label{fig:24}
    \end{figure}

\clearpage

\subsection{NGC~2244}

   \begin{figure} [htp]
   \centering
\includegraphics[width=1\linewidth, height=5cm]{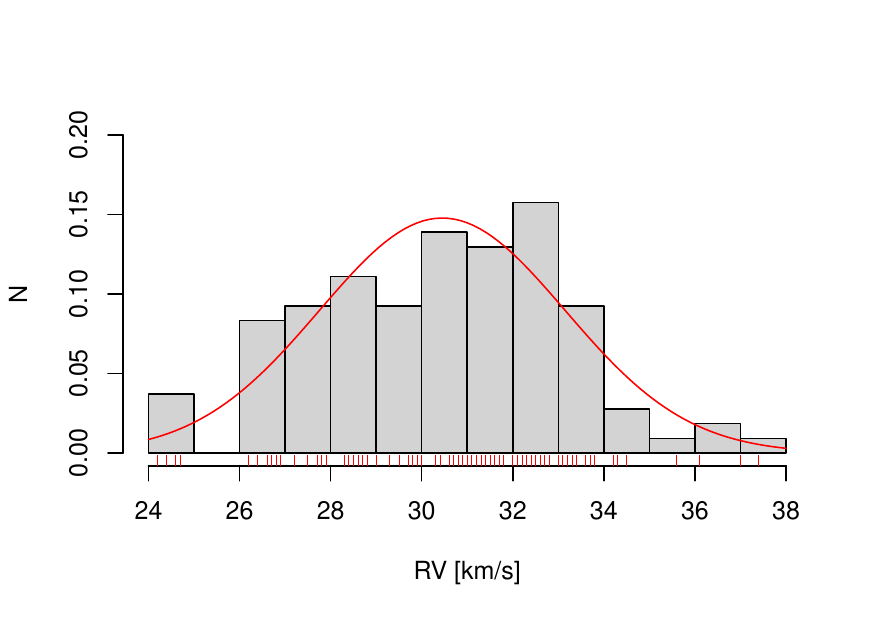}
\caption{$RV$ distribution for NGC~2244.}
             \label{fig:25}
    \end{figure}
    
           \begin{figure} [htp]
   \centering
\includegraphics[width=1\linewidth, height=5cm]{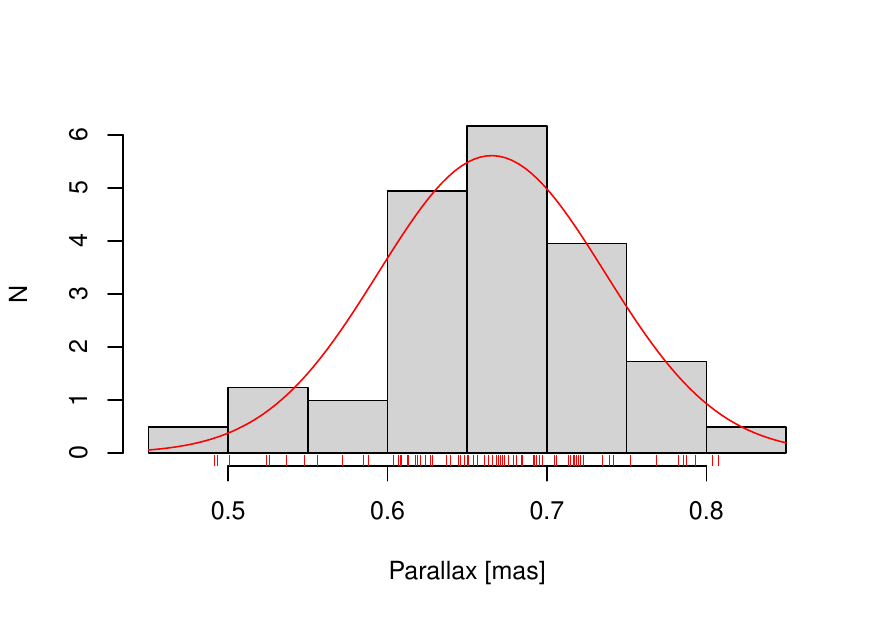}
\caption{Parallax distribution for NGC~2244.}
             \label{fig:26}
    \end{figure}
    
               \begin{figure} [htp]
   \centering
   \includegraphics[width=0.9\linewidth]{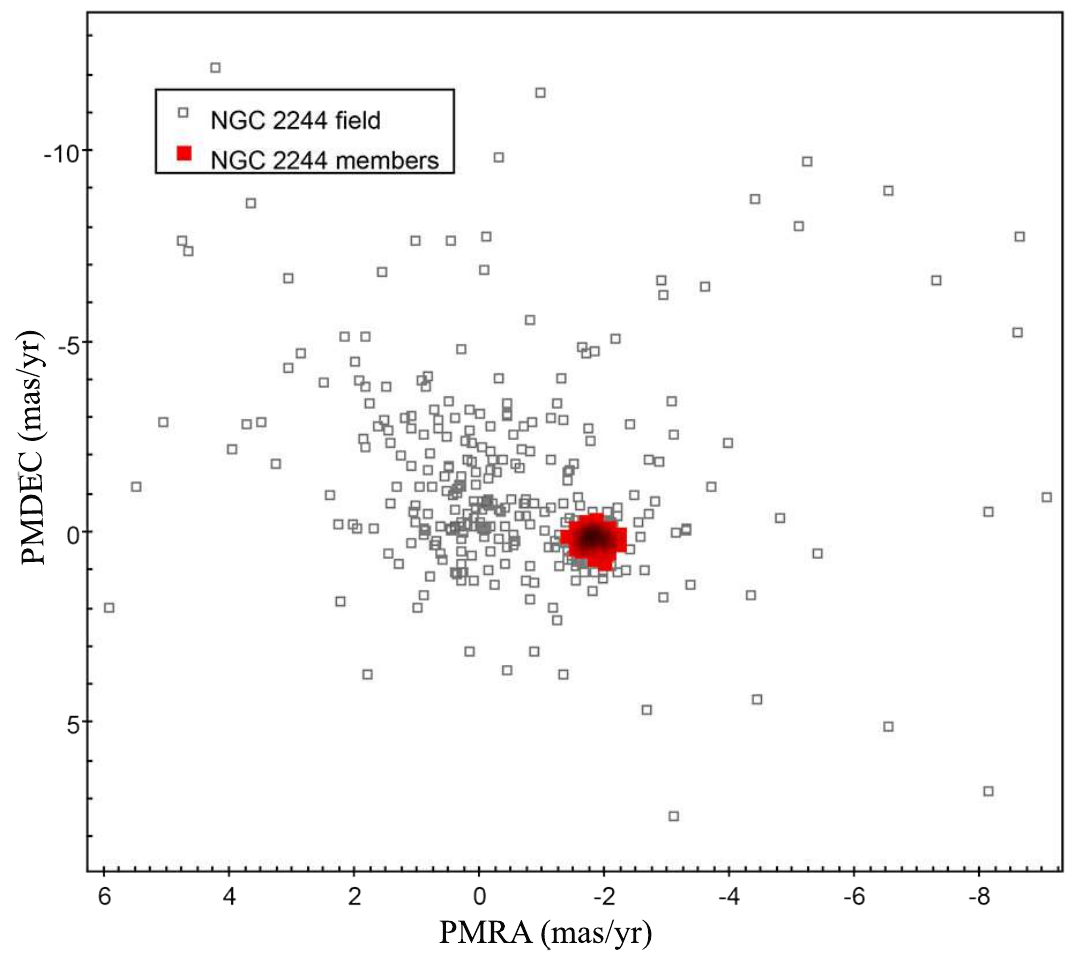}
   \caption{PMs diagram for NGC~2244.}
             \label{fig:27}
    \end{figure}
    
     \begin{figure} [htp]
   \centering
   \includegraphics[width=0.8\linewidth, height=7cm]{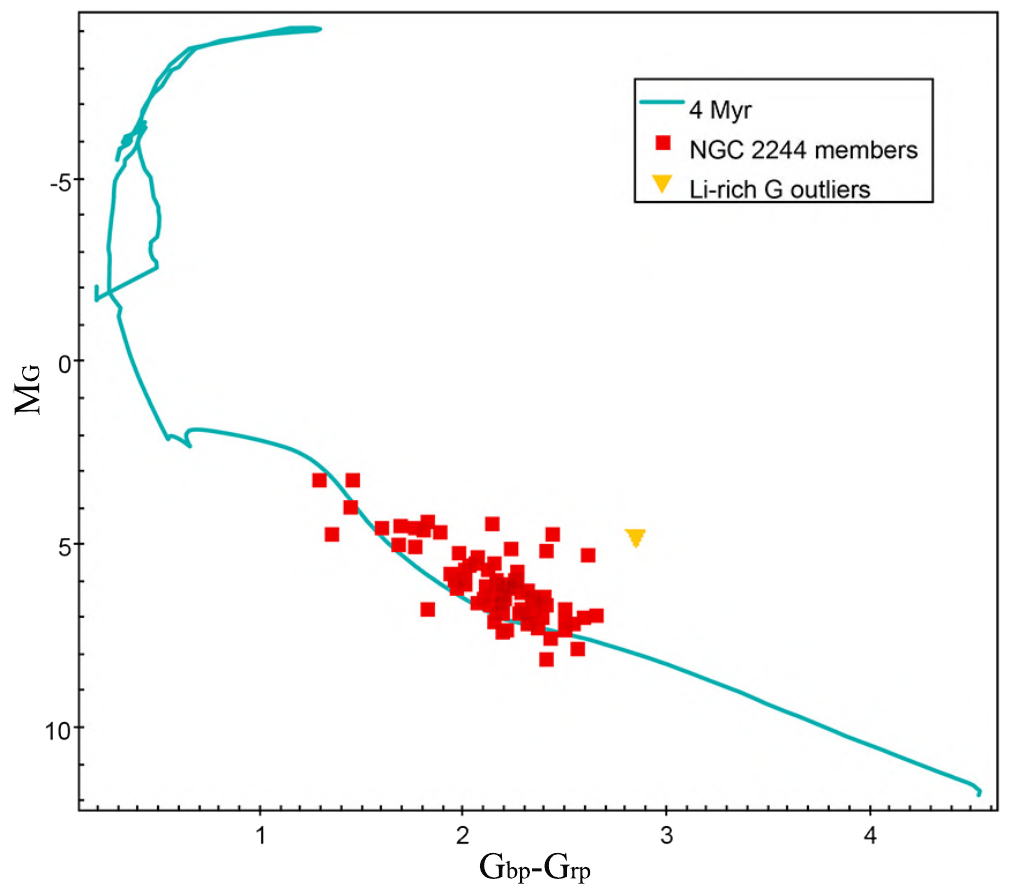}
   \caption{CMD for NGC~2244.}
             \label{fig:28}
    \end{figure}
    
         \begin{figure} [htp]
   \centering
   \includegraphics[width=0.8\linewidth, height=7cm]{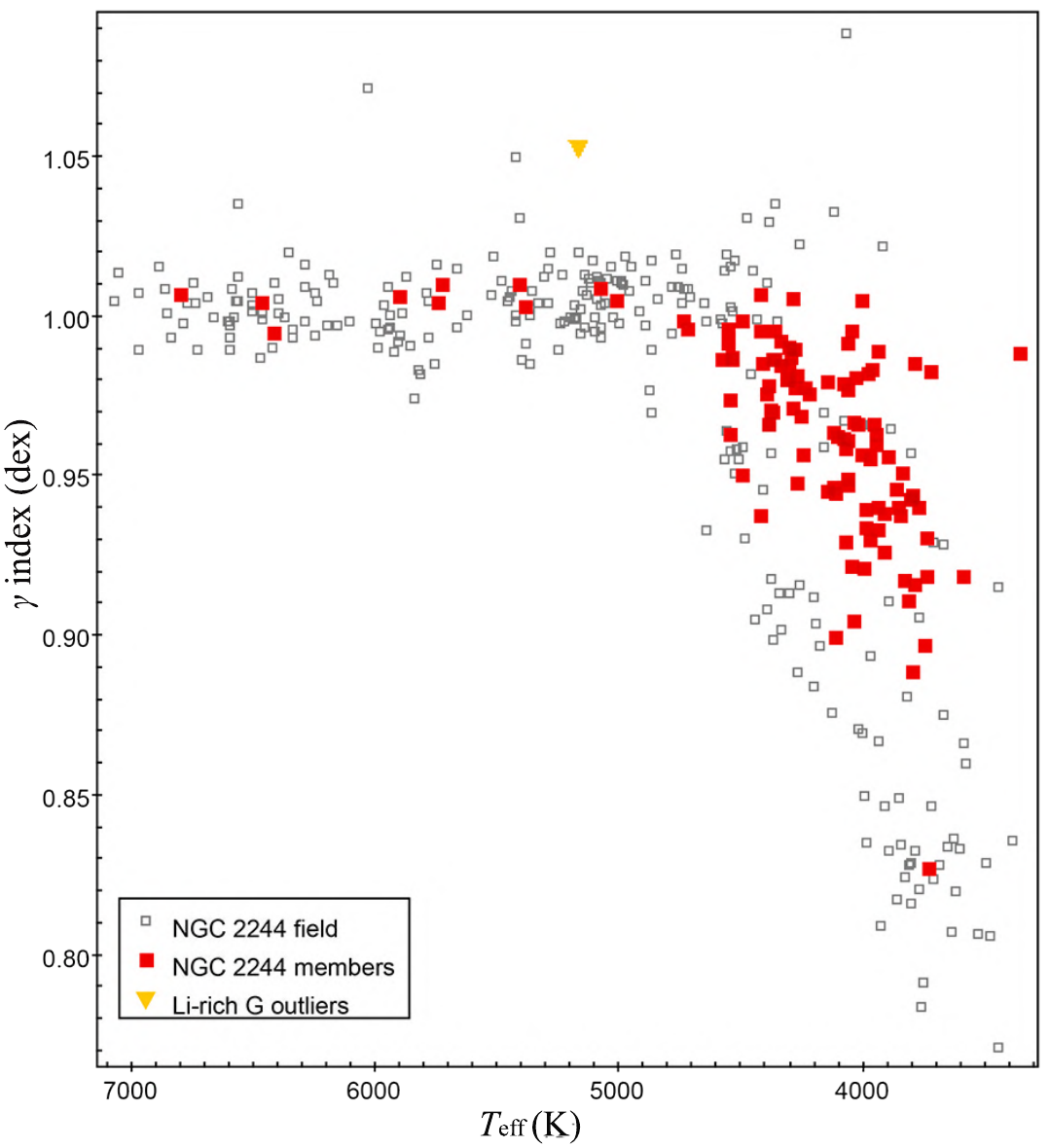}
   \caption{$\gamma$ index-versus-$T_{\rm eff}$ diagram for NGC~2244.}
             \label{fig:29}
    \end{figure}

  \begin{figure} [htp]
   \centering
 \includegraphics[width=0.8\linewidth]{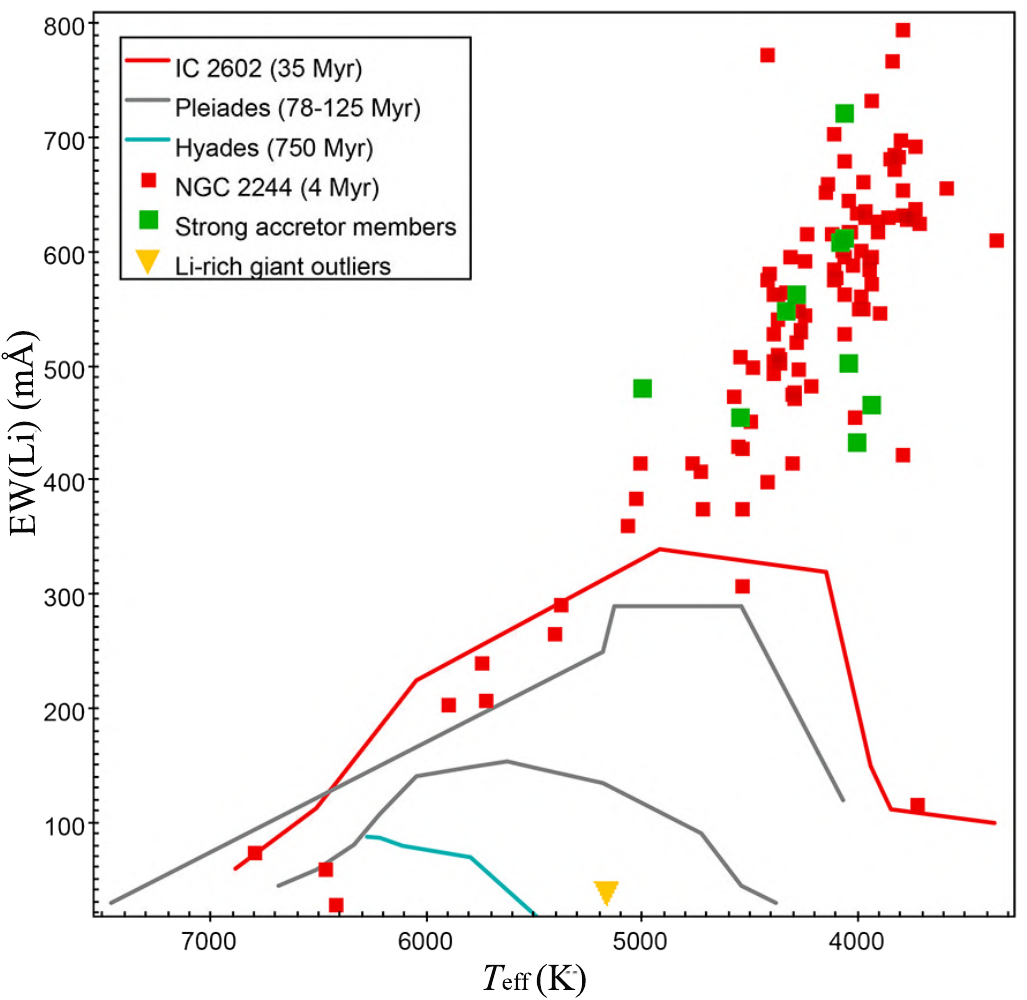} 
\caption{$EW$(Li)-versus-$T_{\rm eff}$ diagram for NGC~2244.}
             \label{fig:30}
    \end{figure}

\clearpage

\subsection{NGC~2264}

 \begin{figure} [htp]
   \centering
\includegraphics[width=1\linewidth, height=5cm]{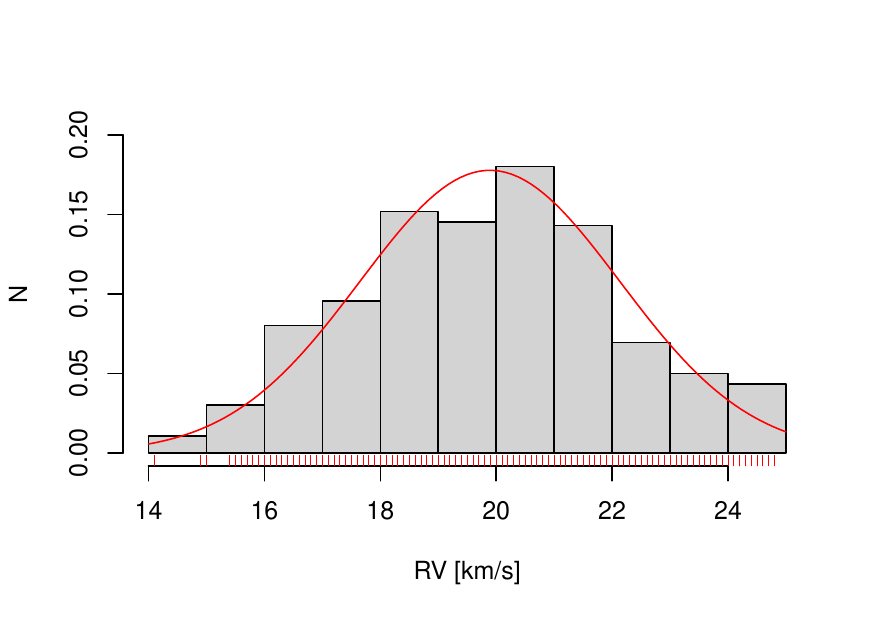}
\caption{$RV$ distribution for NGC~2264.}
             \label{fig:31}
    \end{figure}
    
           \begin{figure} [htp]
   \centering
\includegraphics[width=1\linewidth, height=5cm]{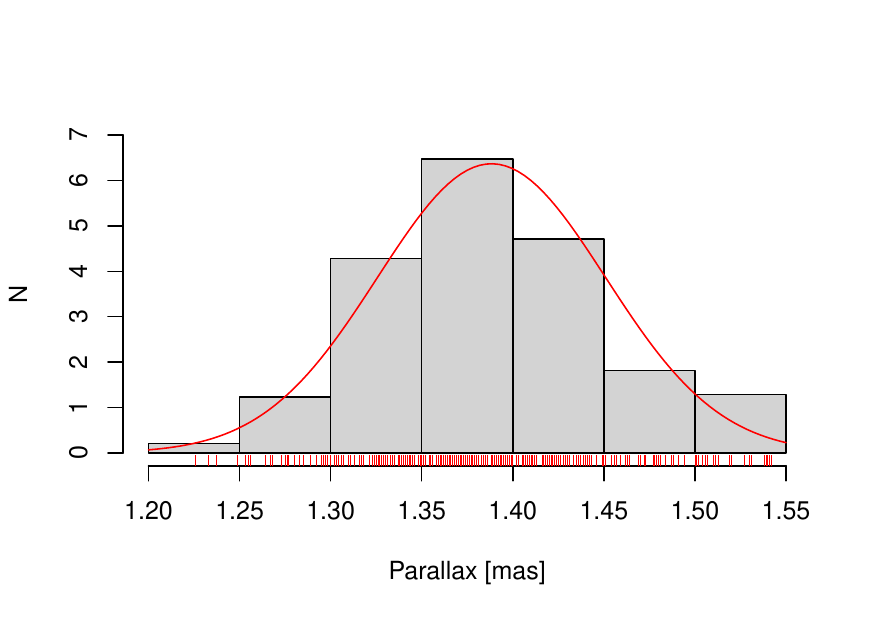}
\caption{Parallax distribution for NGC~2264.}
             \label{fig:32}
    \end{figure}
    
               \begin{figure} [htp]
   \centering
   \includegraphics[width=0.9\linewidth]{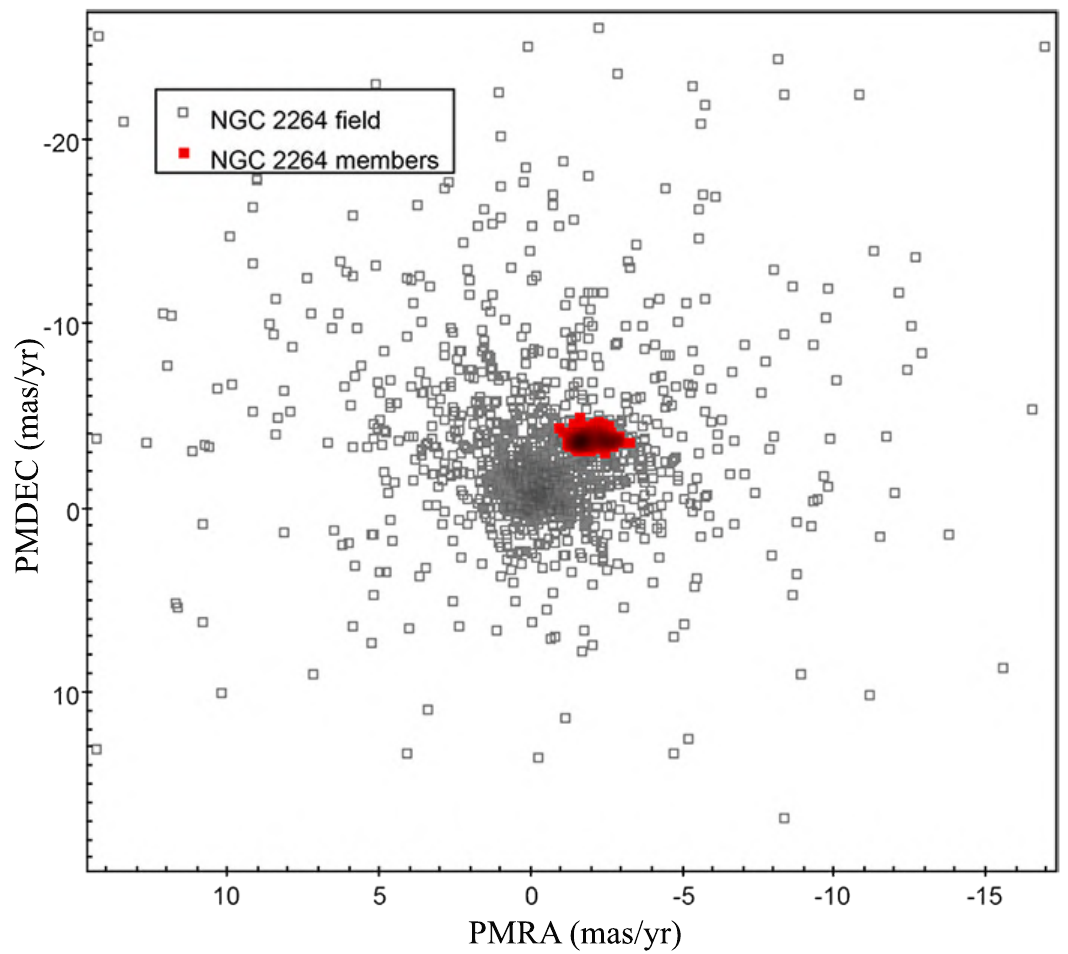}
   \caption{PMs diagram for NGC~2264.}
             \label{fig:33}
    \end{figure}
    
     \begin{figure} [htp]
   \centering
   \includegraphics[width=0.8\linewidth, height=7cm]{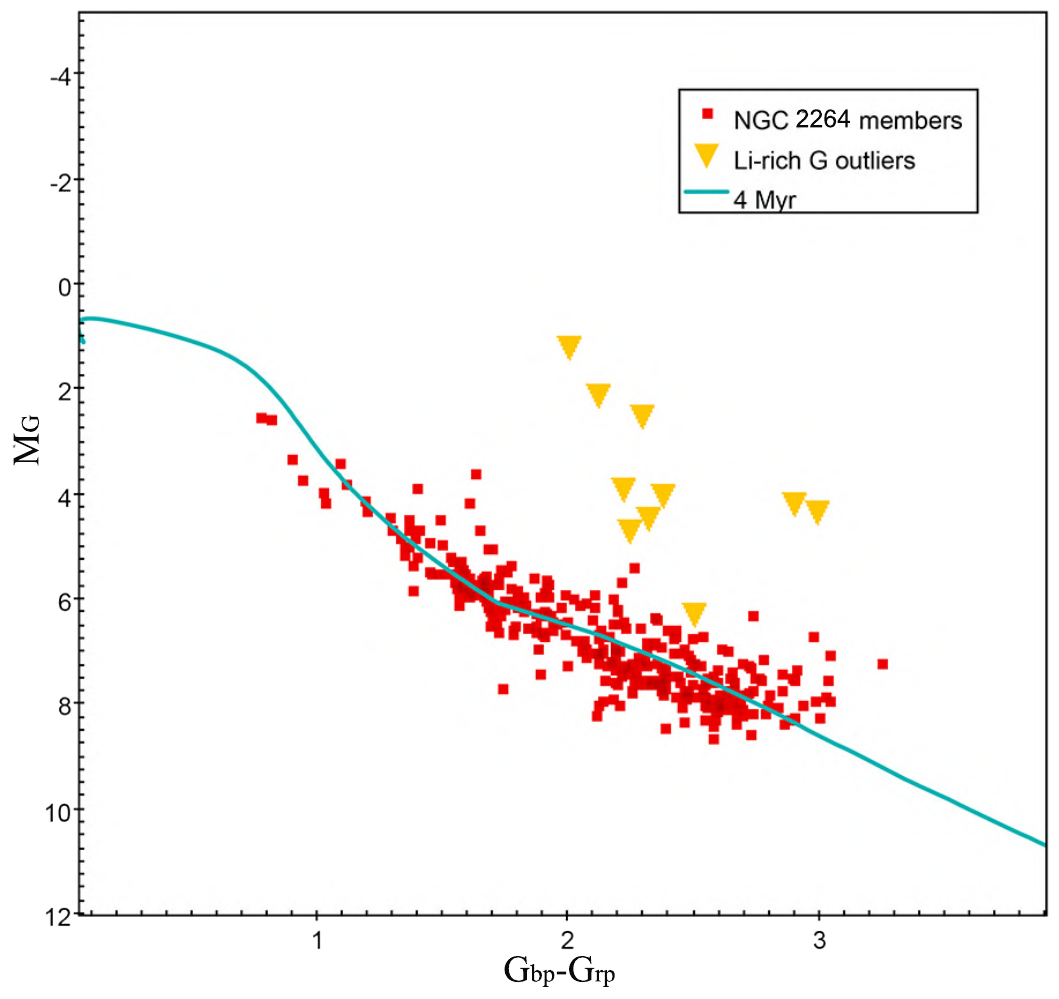}
   \caption{CMD for NGC~2264.}
             \label{fig:34}
    \end{figure}
    
         \begin{figure} [htp]
   \centering
   \includegraphics[width=0.8\linewidth, height=7cm]{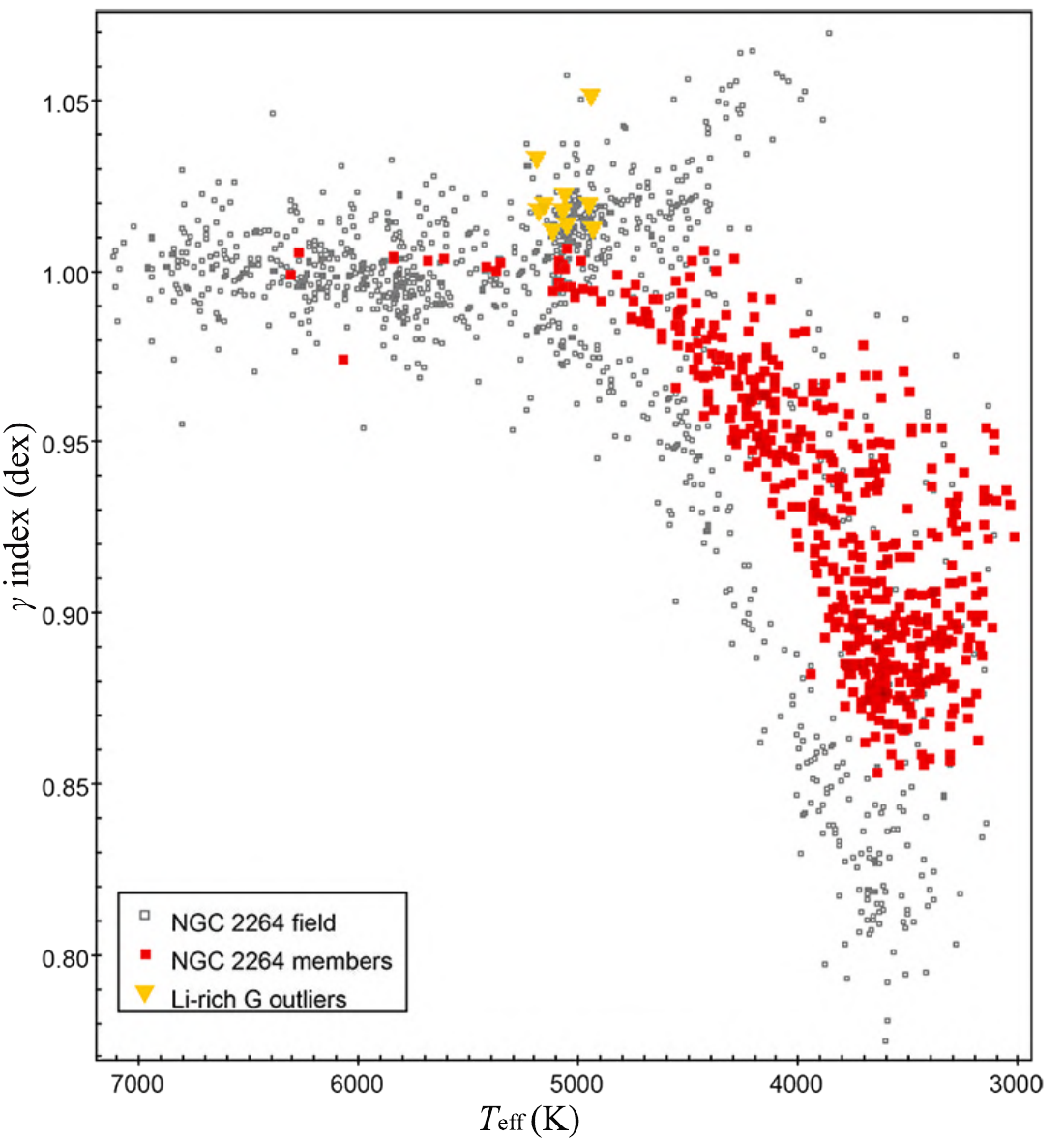}
   \caption{$\gamma$ index-versus-$T_{\rm eff}$ diagram for NGC~2264.}
             \label{fig:35}
    \end{figure}

  \begin{figure} [htp]
   \centering
 \includegraphics[width=0.8\linewidth]{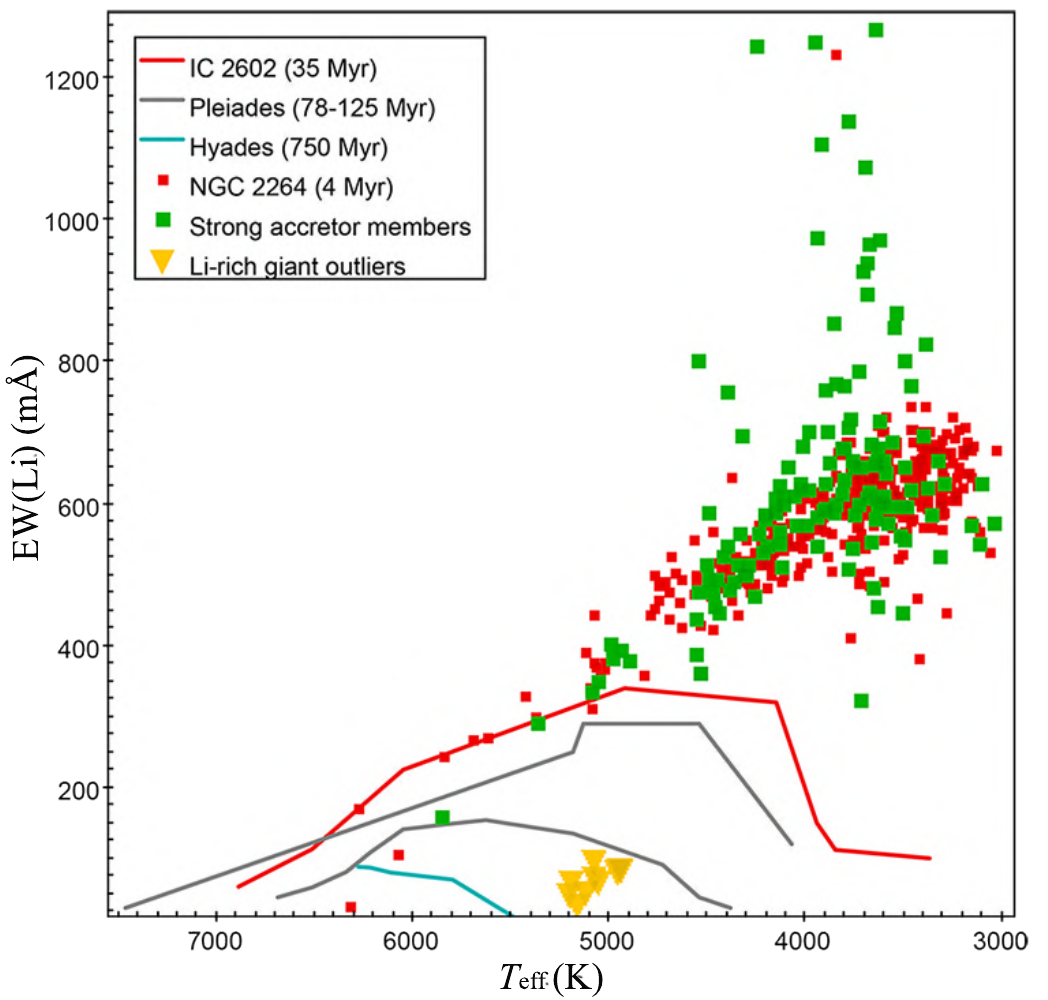} 
\caption{$EW$(Li)-versus-$T_{\rm eff}$ diagram for NGC~2264.}
             \label{fig:36}
    \end{figure}

\clearpage

\subsection{$\lambda$~Ori}

 \begin{figure} [htp]
   \centering
\includegraphics[width=1\linewidth, height=5cm]{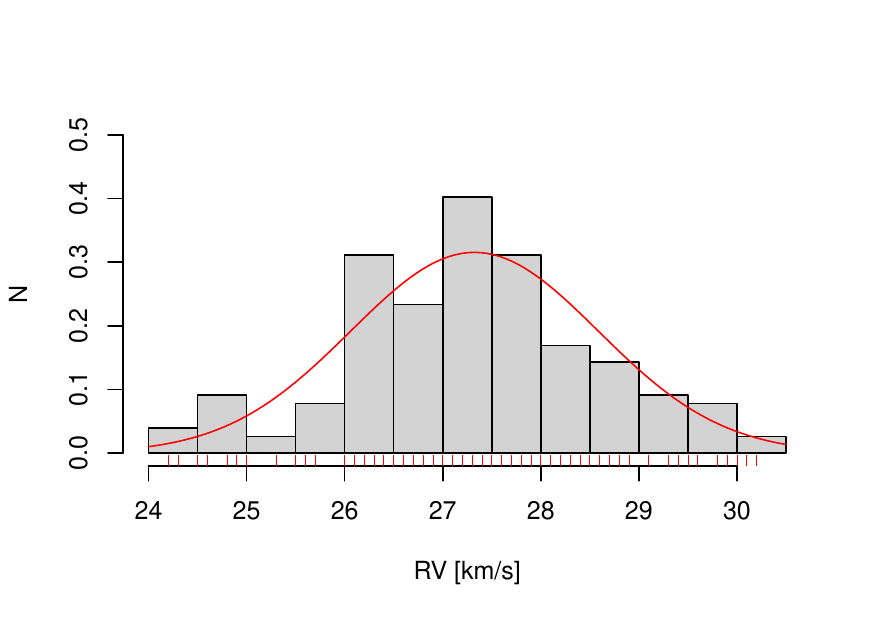}
\caption{$RV$ distribution for $\lambda$~Ori.}
             \label{fig:37}
    \end{figure}
    
           \begin{figure} [htp]
   \centering
\includegraphics[width=1\linewidth, height=5cm]{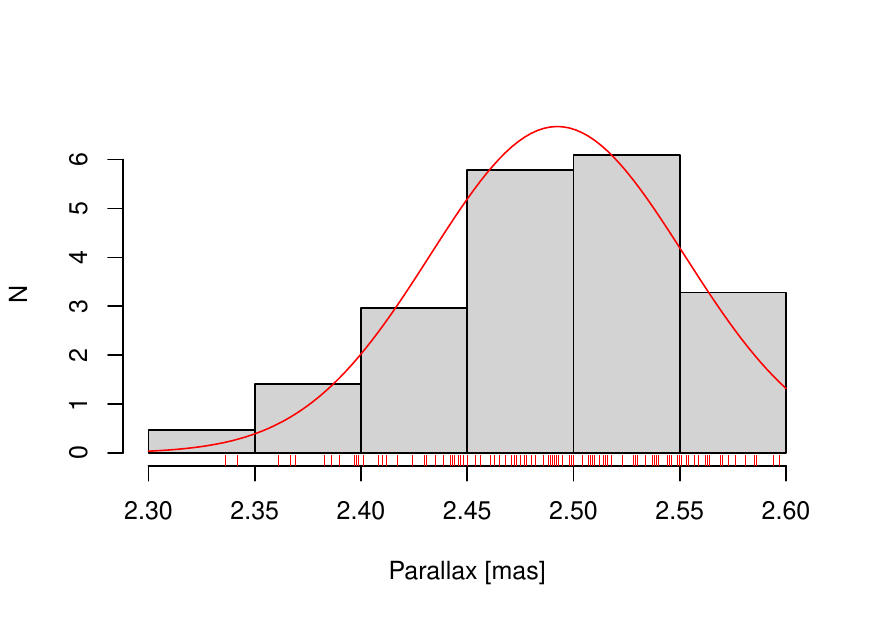}
\caption{Parallax distribution for $\lambda$~Ori.}
             \label{fig:38}
    \end{figure}
    
               \begin{figure} [htp]
   \centering
   \includegraphics[width=0.9\linewidth]{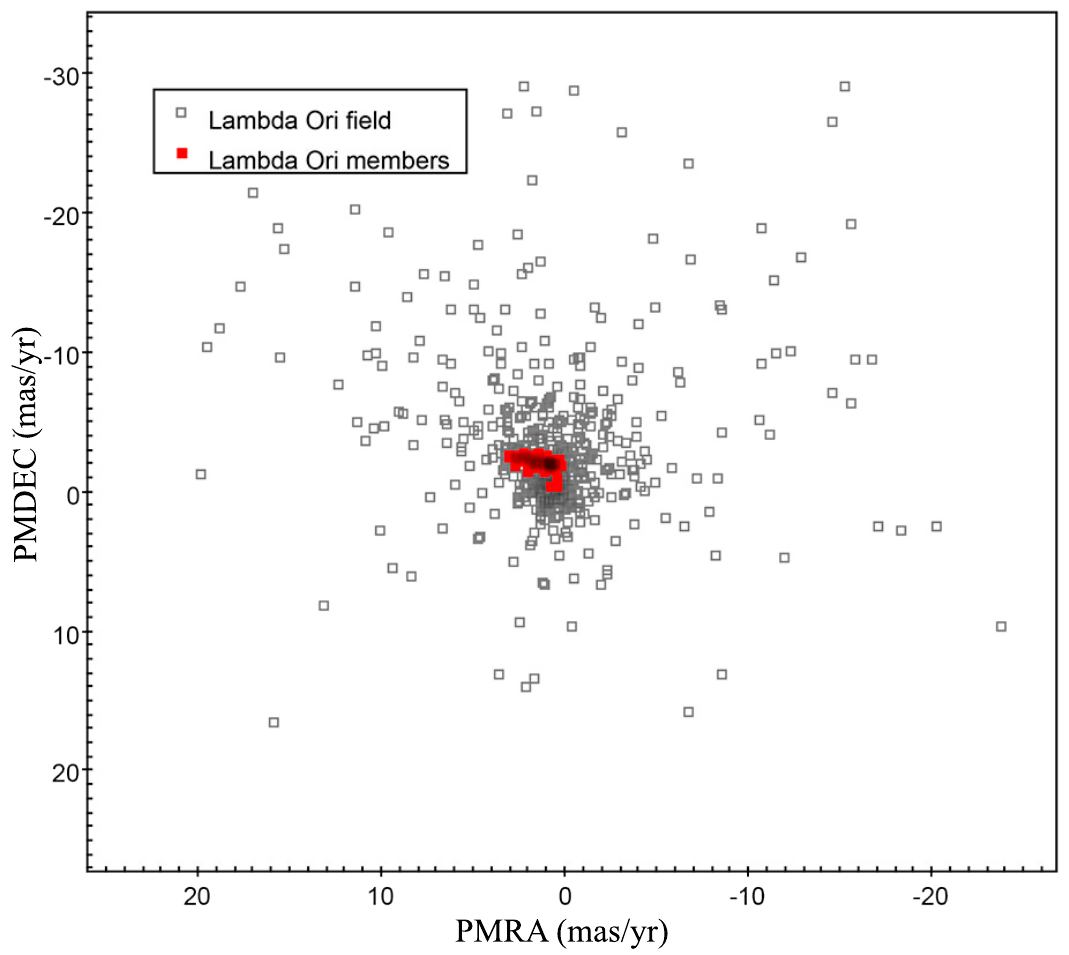}
   \caption{PMs diagram for $\lambda$~Ori.}
             \label{fig:39}
    \end{figure}
    
     \begin{figure} [htp]
   \centering
   \includegraphics[width=0.8\linewidth, height=7cm]{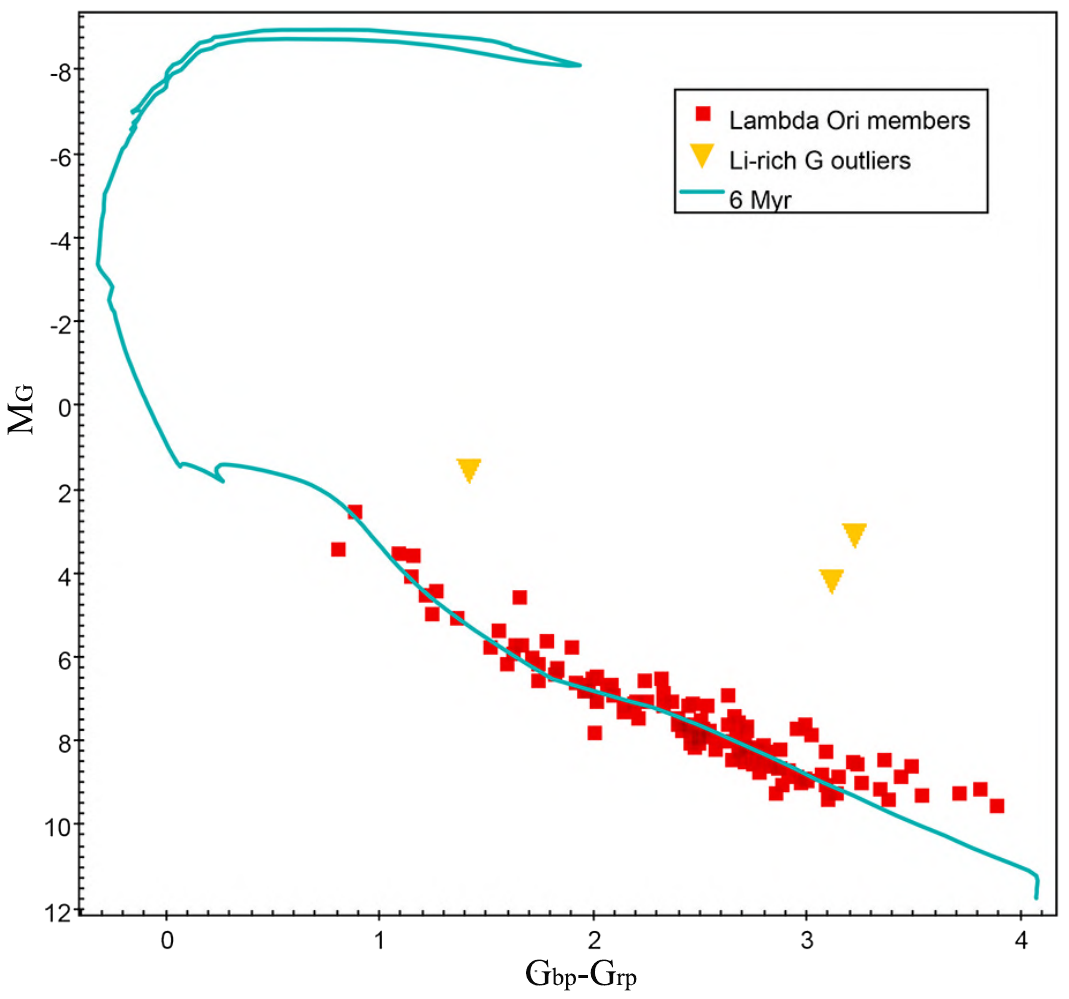}
   \caption{CMD for $\lambda$~Ori.}
             \label{fig:40}
    \end{figure}
    
         \begin{figure} [htp]
   \centering
   \includegraphics[width=0.8\linewidth, height=7cm]{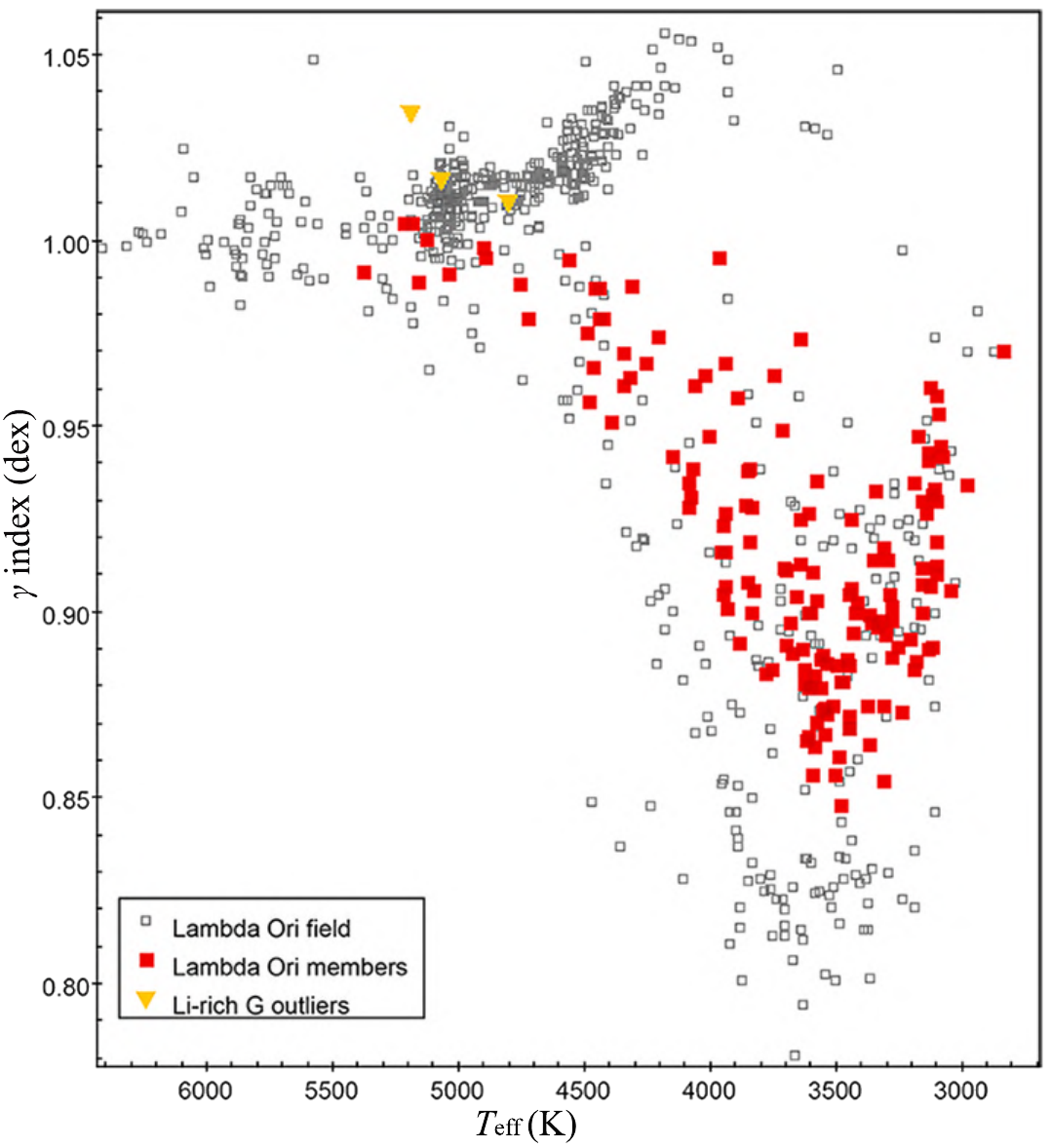}
   \caption{$\gamma$ index-versus-$T_{\rm eff}$ diagram for $\lambda$~Ori.}
             \label{fig:41}
    \end{figure}

  \begin{figure} [htp]
   \centering
 \includegraphics[width=0.8\linewidth]{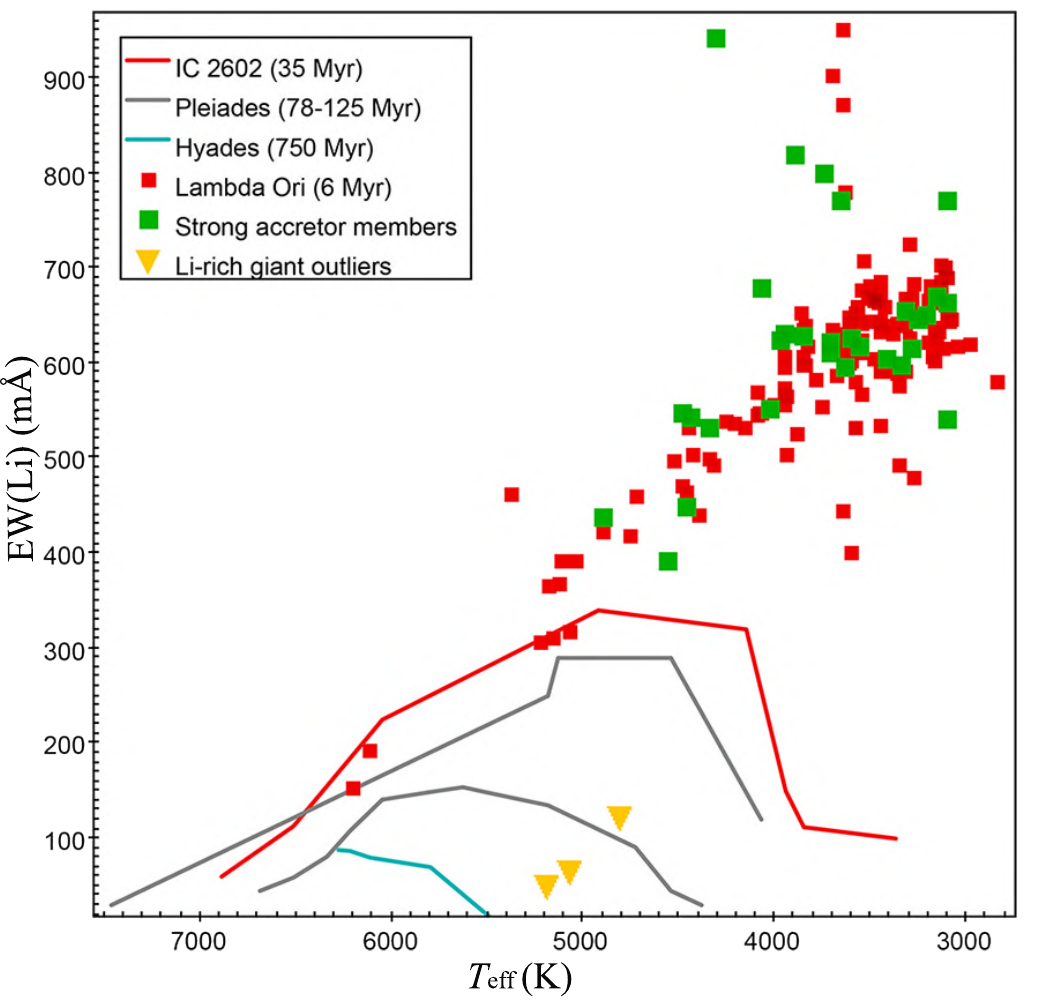} 
\caption{$EW$(Li)-versus-$T_{\rm eff}$ diagram for $\lambda$~Ori.}
             \label{fig:42}
    \end{figure}

\clearpage

\subsection{Col~197}

 \begin{figure} [htp]
   \centering
\includegraphics[width=1\linewidth, height=5cm]{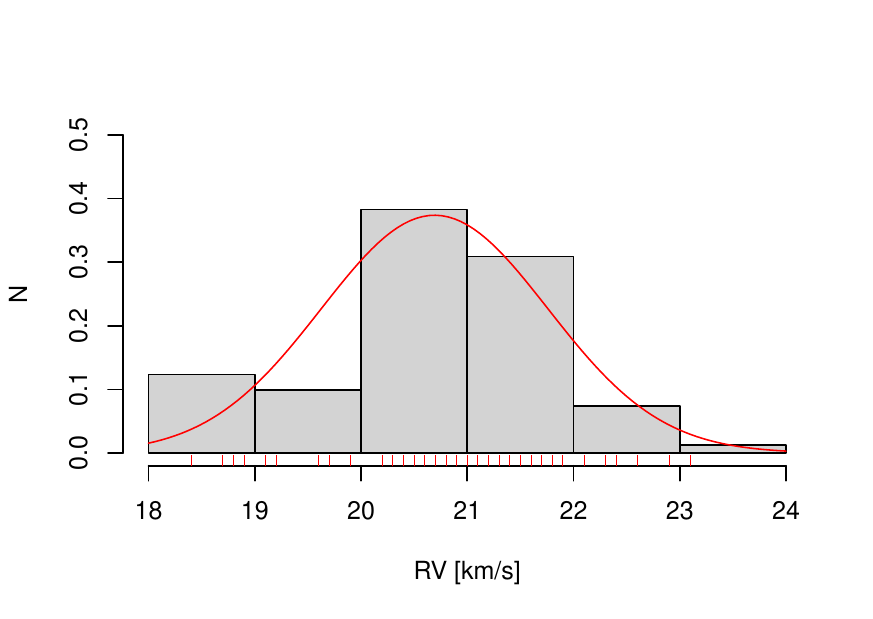}
\caption{$RV$ distribution for Col~197.}
             \label{fig:43}
    \end{figure}
    
           \begin{figure} [htp]
   \centering
\includegraphics[width=1\linewidth, height=5cm]{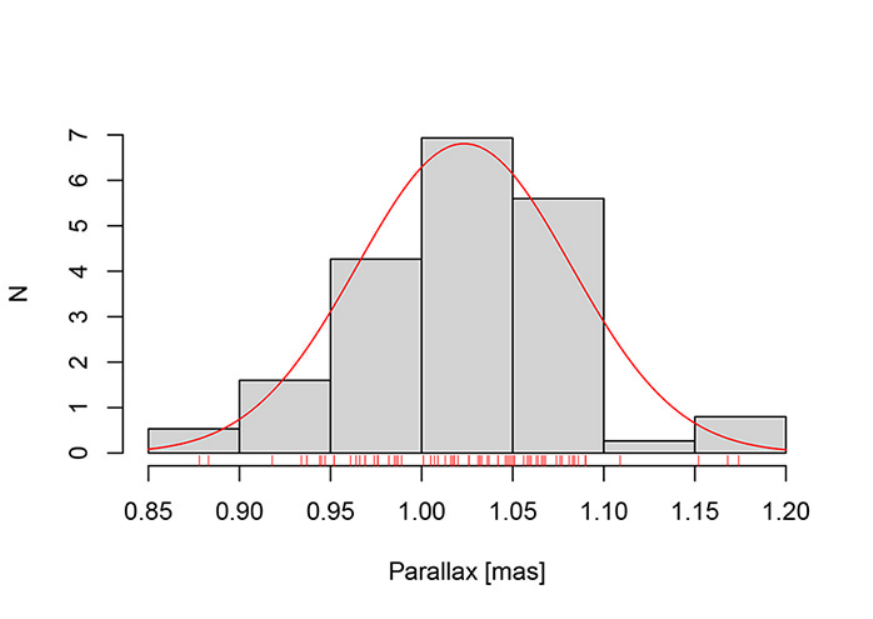}
\caption{Parallax distribution for Col~197.}
             \label{fig:44}
    \end{figure}
    
               \begin{figure} [htp]
   \centering
   \includegraphics[width=0.9\linewidth]{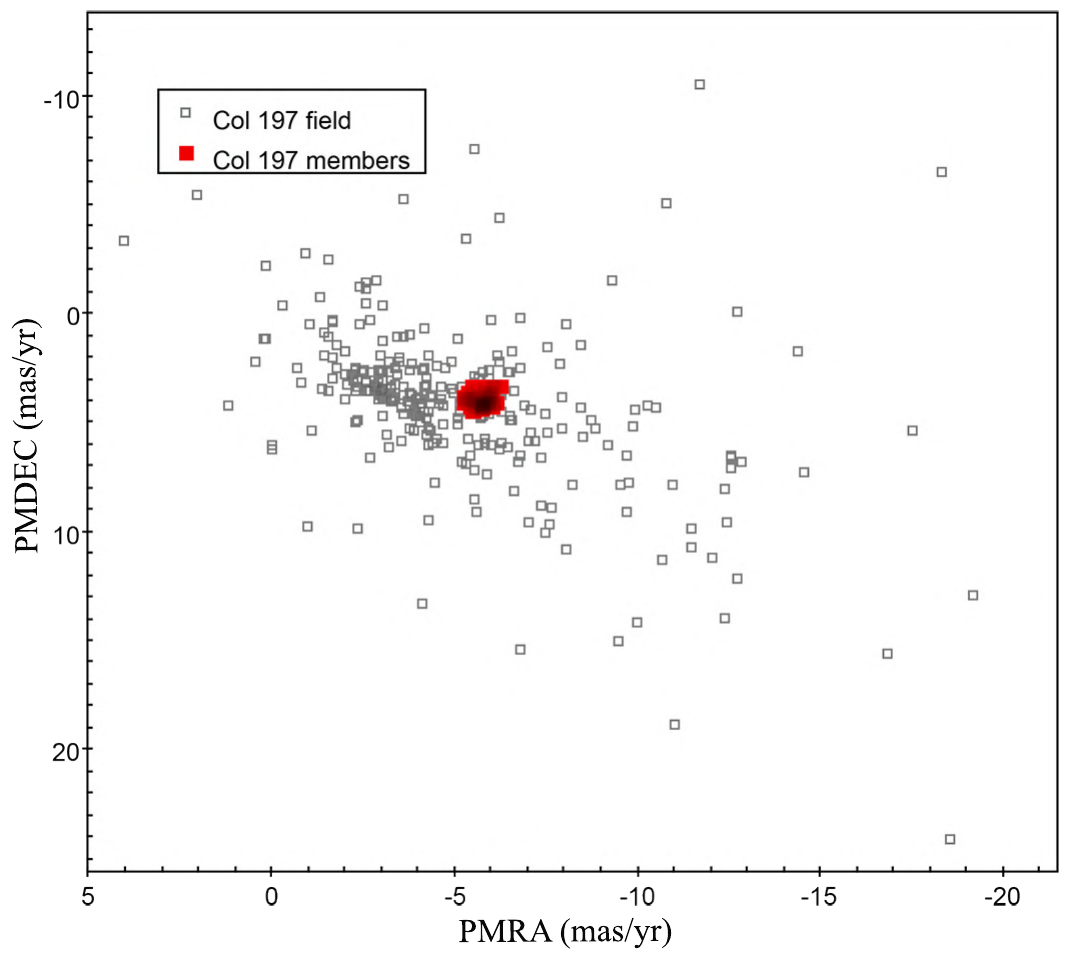}
   \caption{PMs diagram for Col~197.}
             \label{fig:45}
    \end{figure}
    
     \begin{figure} [htp]
   \centering
   \includegraphics[width=0.8\linewidth, height=7cm]{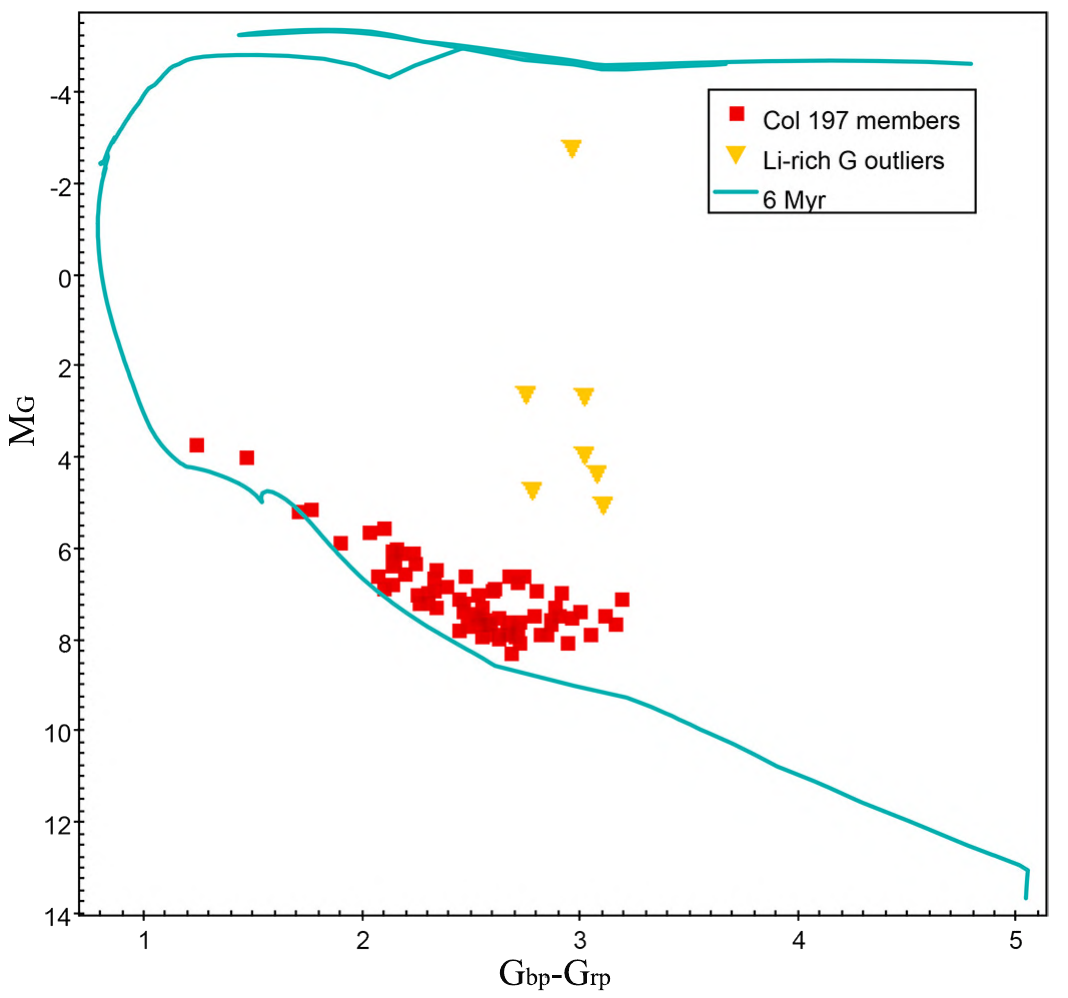}
   \caption{CMD for Col~197.}
             \label{fig:46}
    \end{figure}
    
         \begin{figure} [htp]
   \centering
   \includegraphics[width=0.8\linewidth, height=7cm]{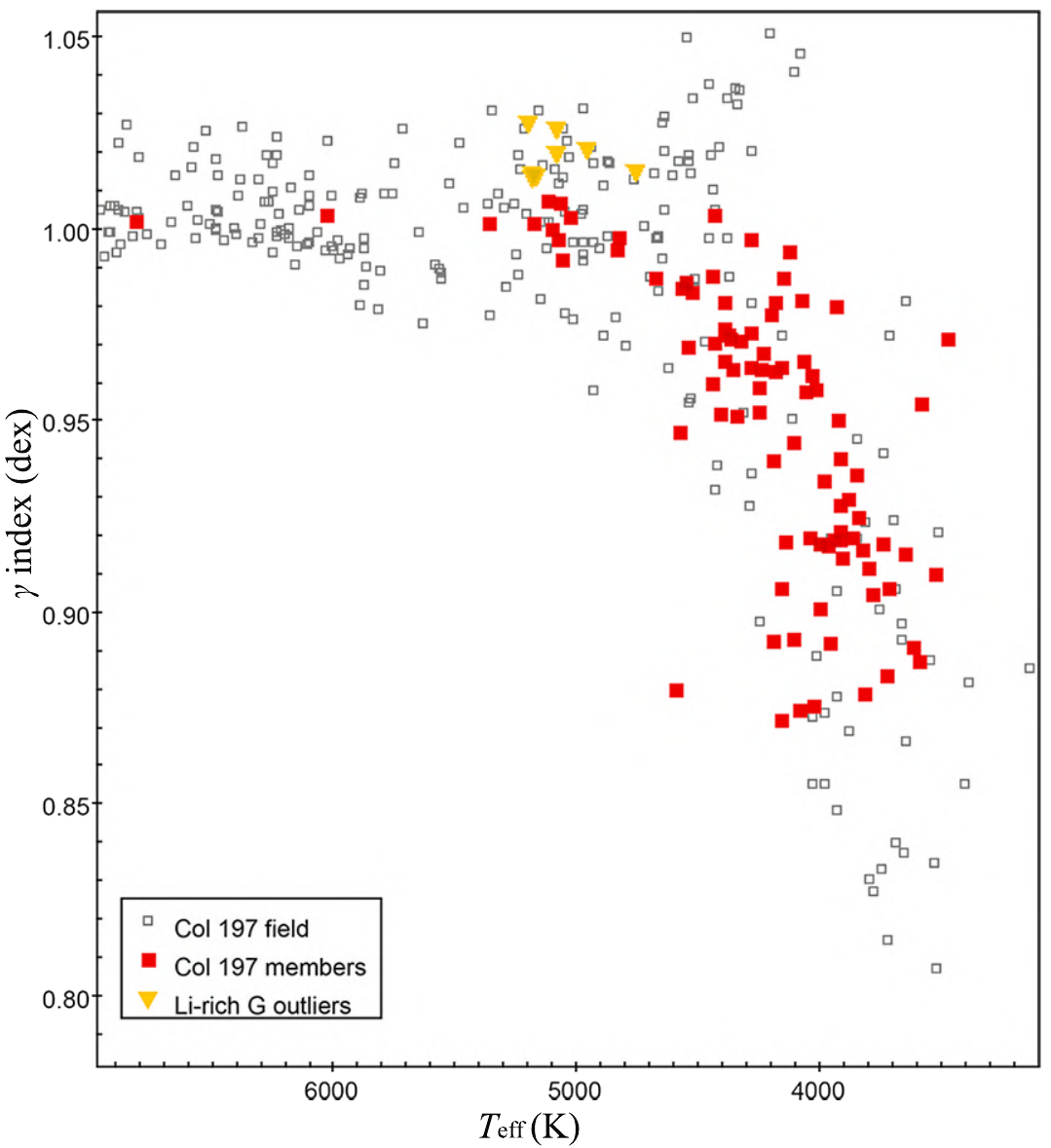}
   \caption{$\gamma$ index-versus-$T_{\rm eff}$ diagram for Col~197.}
             \label{fig:47}
    \end{figure}

  \begin{figure} [htp]
   \centering
 \includegraphics[width=0.8\linewidth]{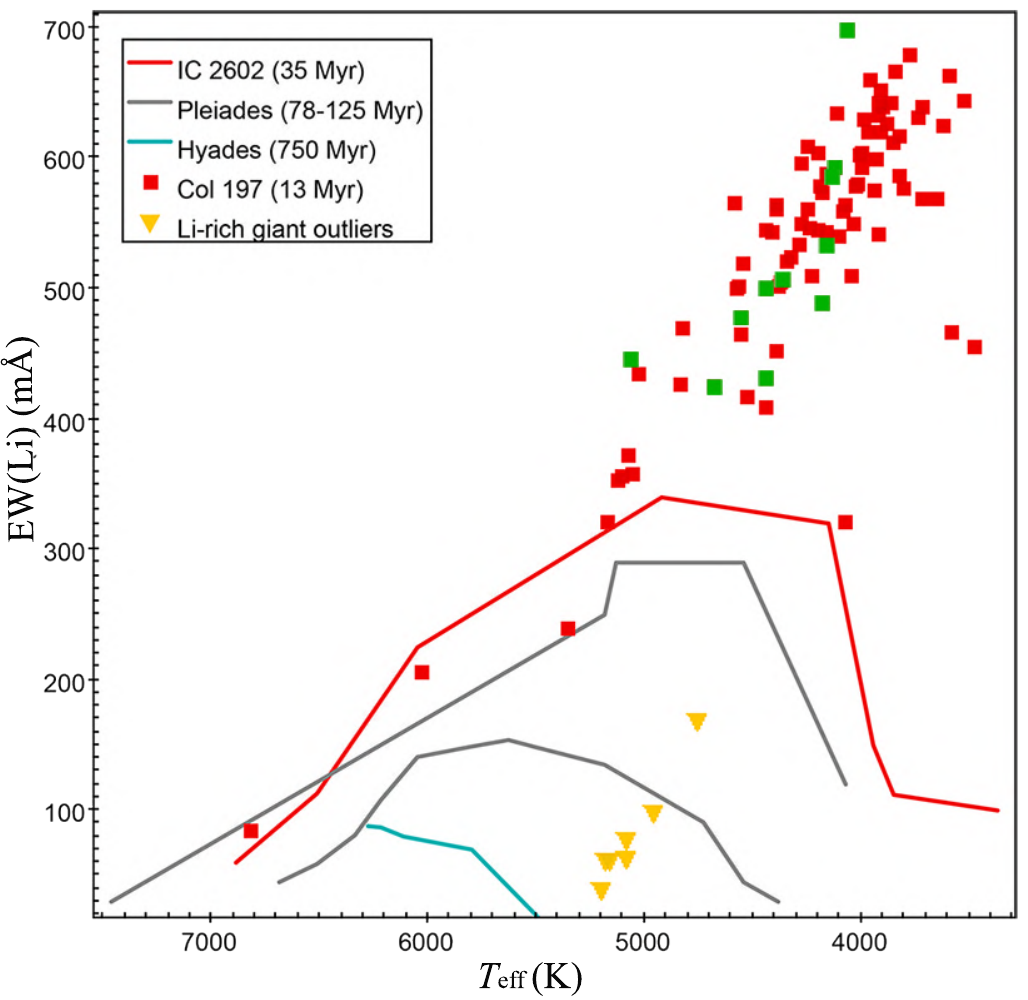} 
\caption{$EW$(Li)-versus-$T_{\rm eff}$ diagram for Col~197.}
             \label{fig:48}
    \end{figure}

\clearpage

\subsection{$\gamma$~Vel}
    
               \begin{figure} [htp]
   \centering
   \includegraphics[width=0.9\linewidth]{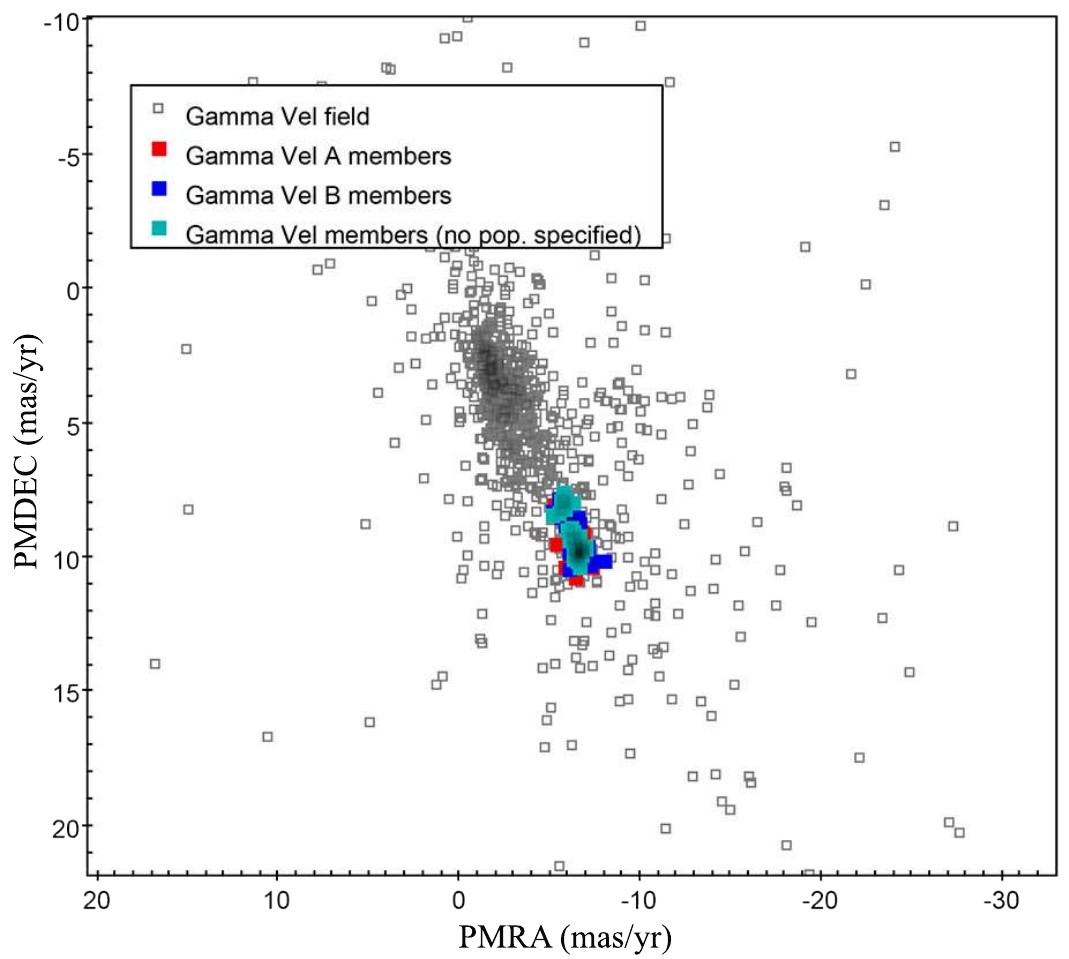}
   \caption{PMs diagram for $\gamma$~Vel.}
             \label{fig:49}
    \end{figure}
    
     \begin{figure} [htp]
   \centering
   \includegraphics[width=0.9\linewidth, height=7cm]{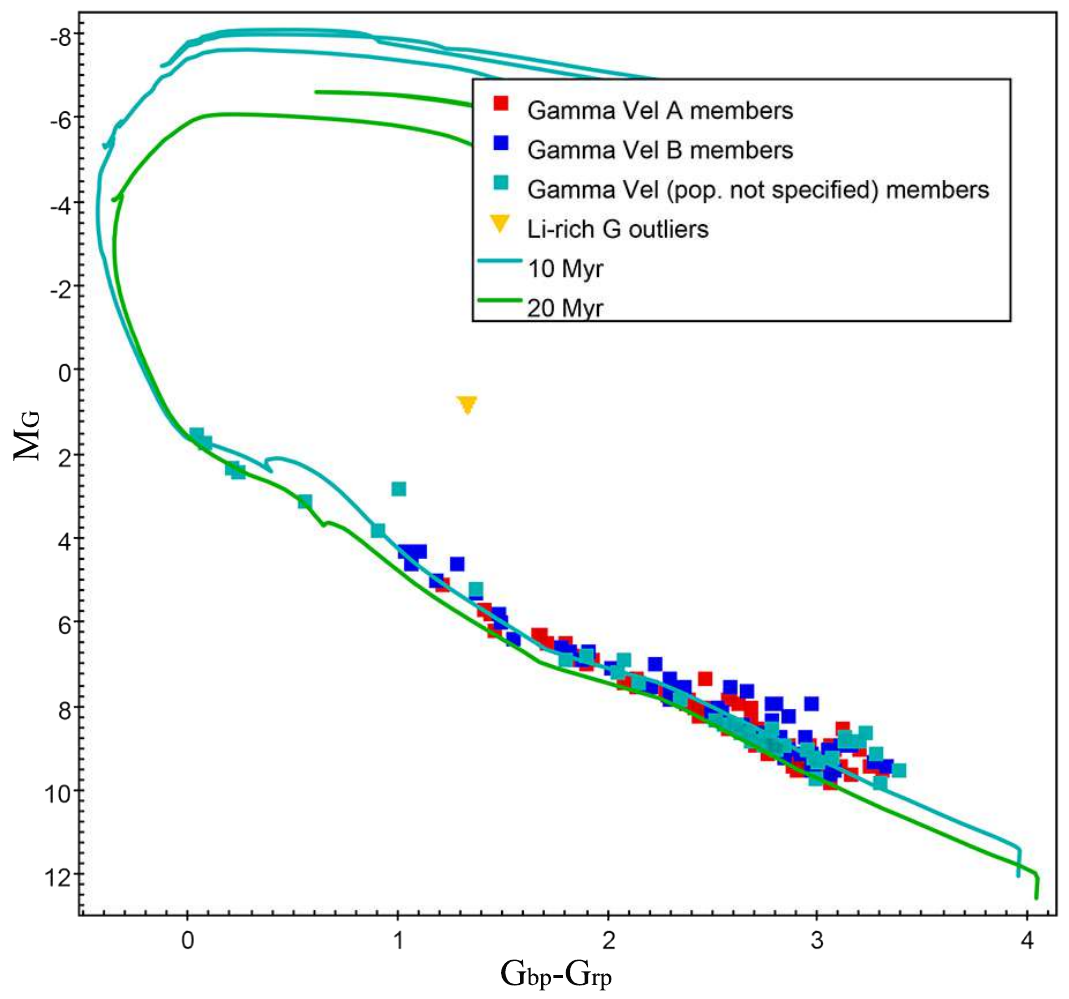}
   \caption{CMD for $\gamma$~Vel.}
             \label{fig:50}
    \end{figure}
    
         \begin{figure} [htp]
   \centering
   \includegraphics[width=0.9\linewidth, height=8cm]{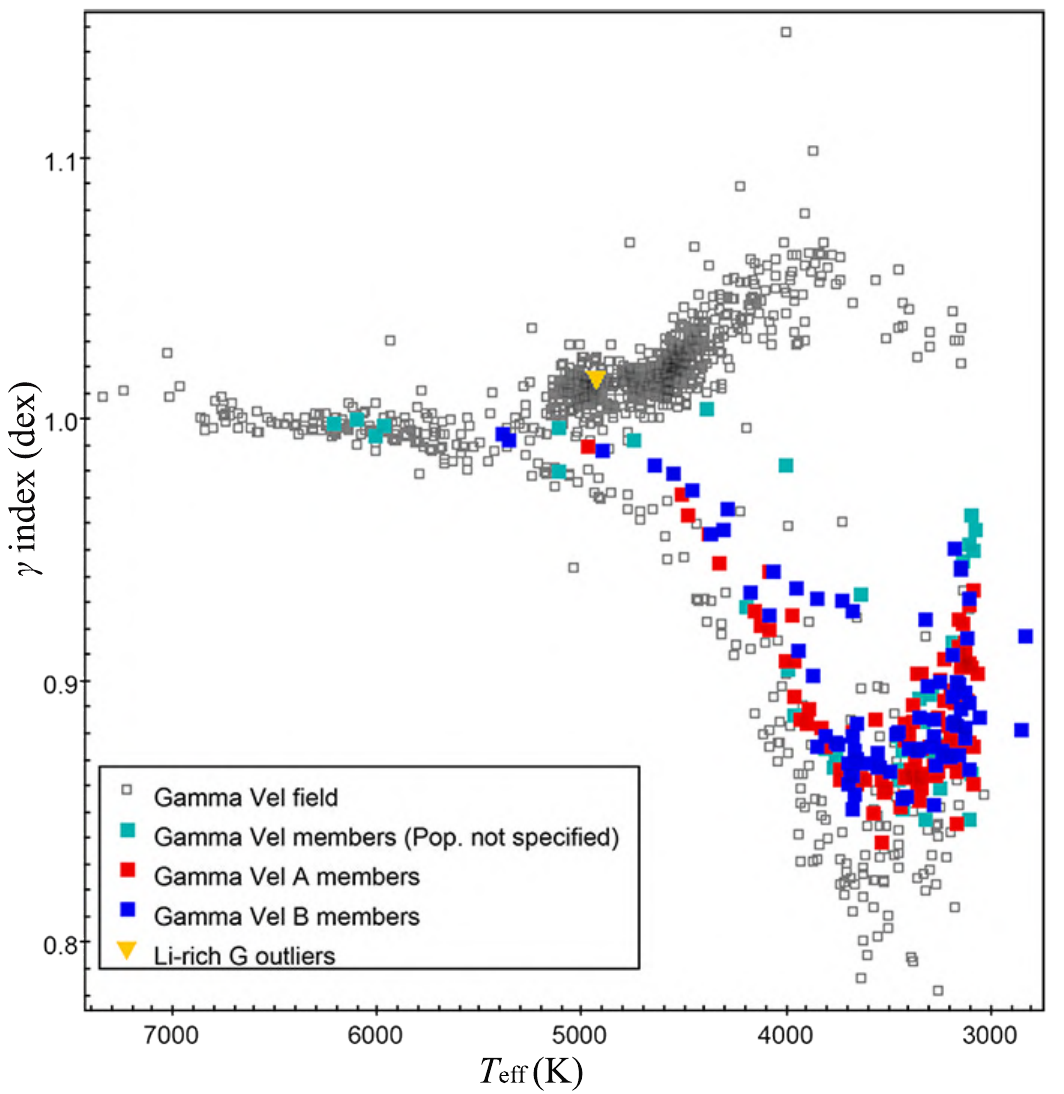}
   \caption{$\gamma$ index-versus-$T_{\rm eff}$ diagram for $\gamma$~Vel.}
             \label{fig:51}
    \end{figure}

  \begin{figure} [htp]
   \centering
 \includegraphics[width=0.9\linewidth]{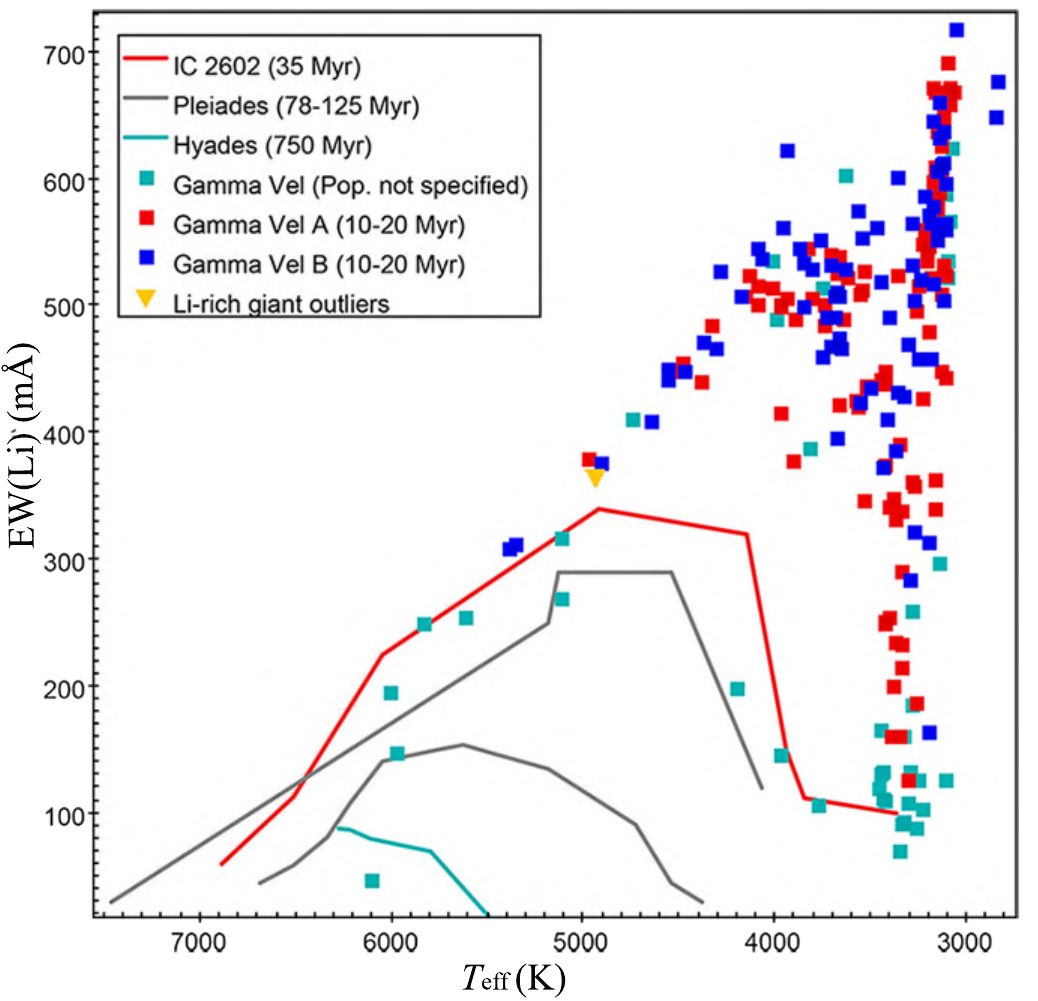} 
\caption{$EW$(Li)-versus-$T_{\rm eff}$ diagram for $\gamma$~Vel.}
             \label{fig:52}
    \end{figure}

\clearpage

\subsection{NGC~2232}

 \begin{figure} [htp]
   \centering
\includegraphics[width=1\linewidth, height=5cm]{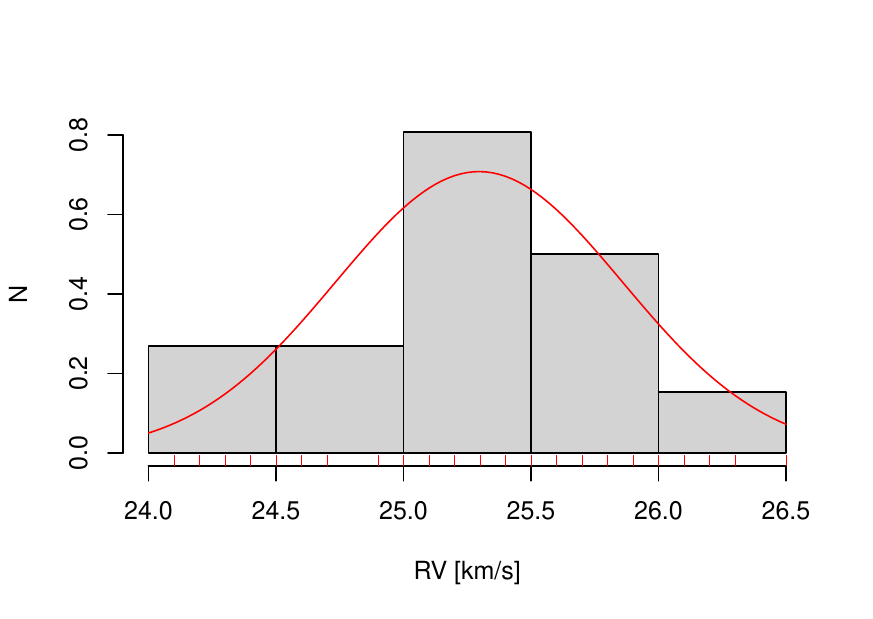}
\caption{$RV$ distribution for NGC~2232.}
             \label{fig:53}
    \end{figure}
    
           \begin{figure} [htp]
   \centering
\includegraphics[width=1\linewidth, height=5cm]{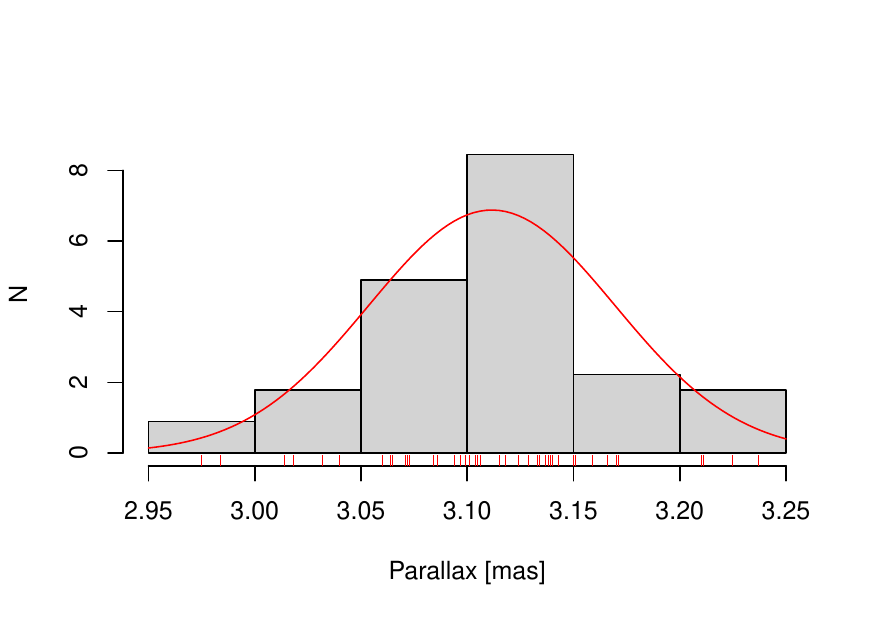}
\caption{Parallax distribution for NGC~2232.}
             \label{fig:54}
    \end{figure}

               \begin{figure} [htp]
   \centering
   \includegraphics[width=0.9\linewidth]{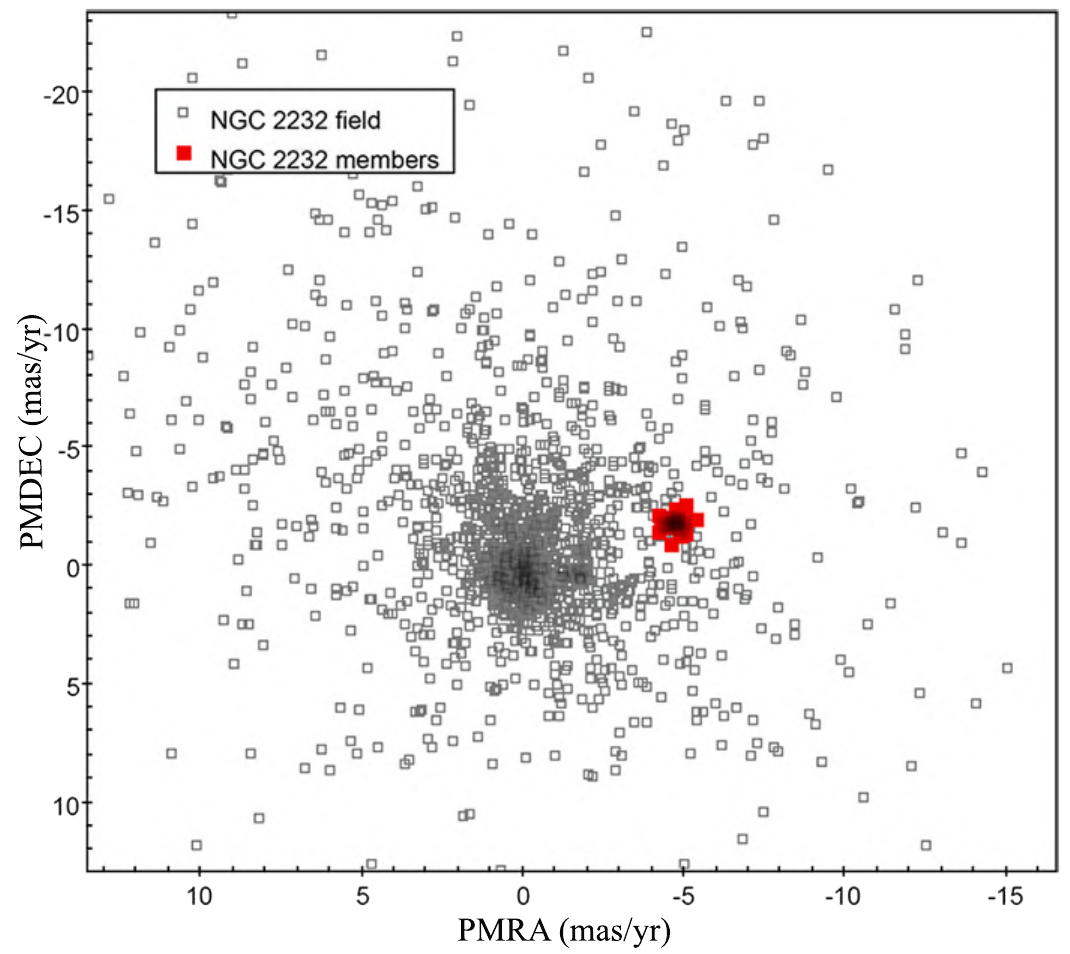}
   \caption{PMs diagram for NGC~2232.}
             \label{fig:55}
    \end{figure}
    
     \begin{figure} [htp]
   \centering
   \includegraphics[width=0.8\linewidth, height=7cm]{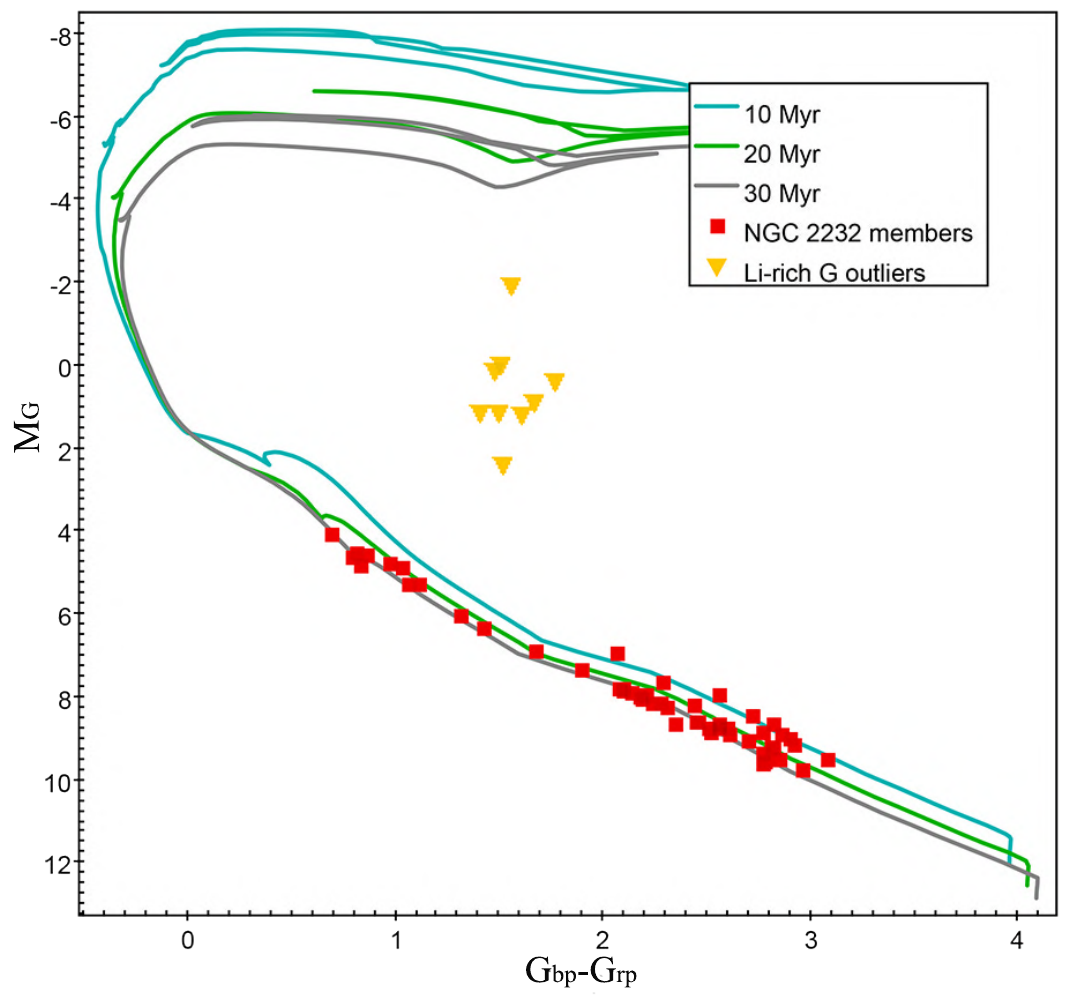}
   \caption{CMD for NGC~2232.}
             \label{fig:56}
    \end{figure}
    
         \begin{figure} [htp]
   \centering
   \includegraphics[width=0.8\linewidth, height=7cm]{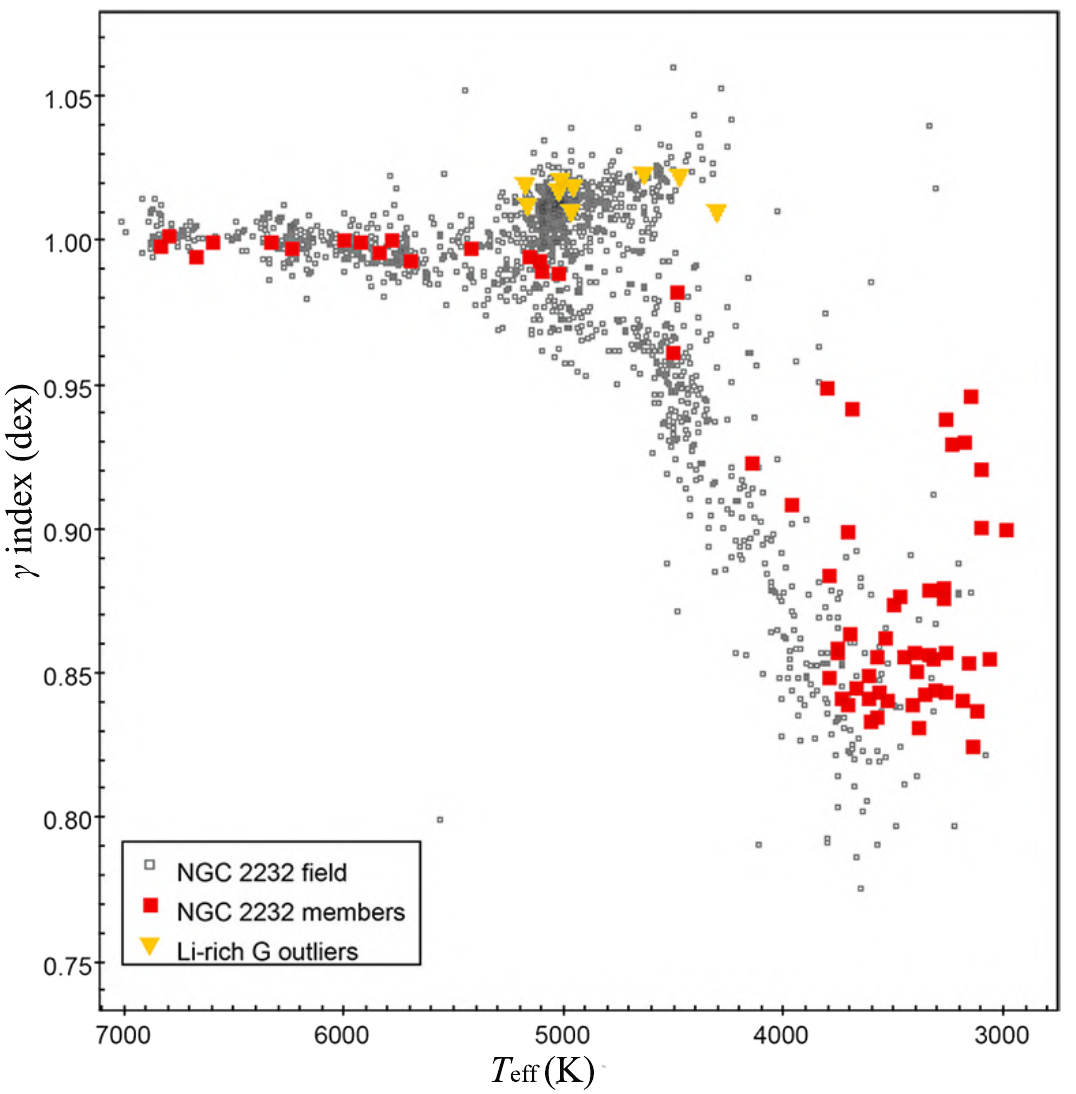}
   \caption{$\gamma$ index-versus-$T_{\rm eff}$ diagram for NGC~2232.}
             \label{fig:56}
    \end{figure}

  \begin{figure} [htp]
   \centering
 \includegraphics[width=0.8\linewidth]{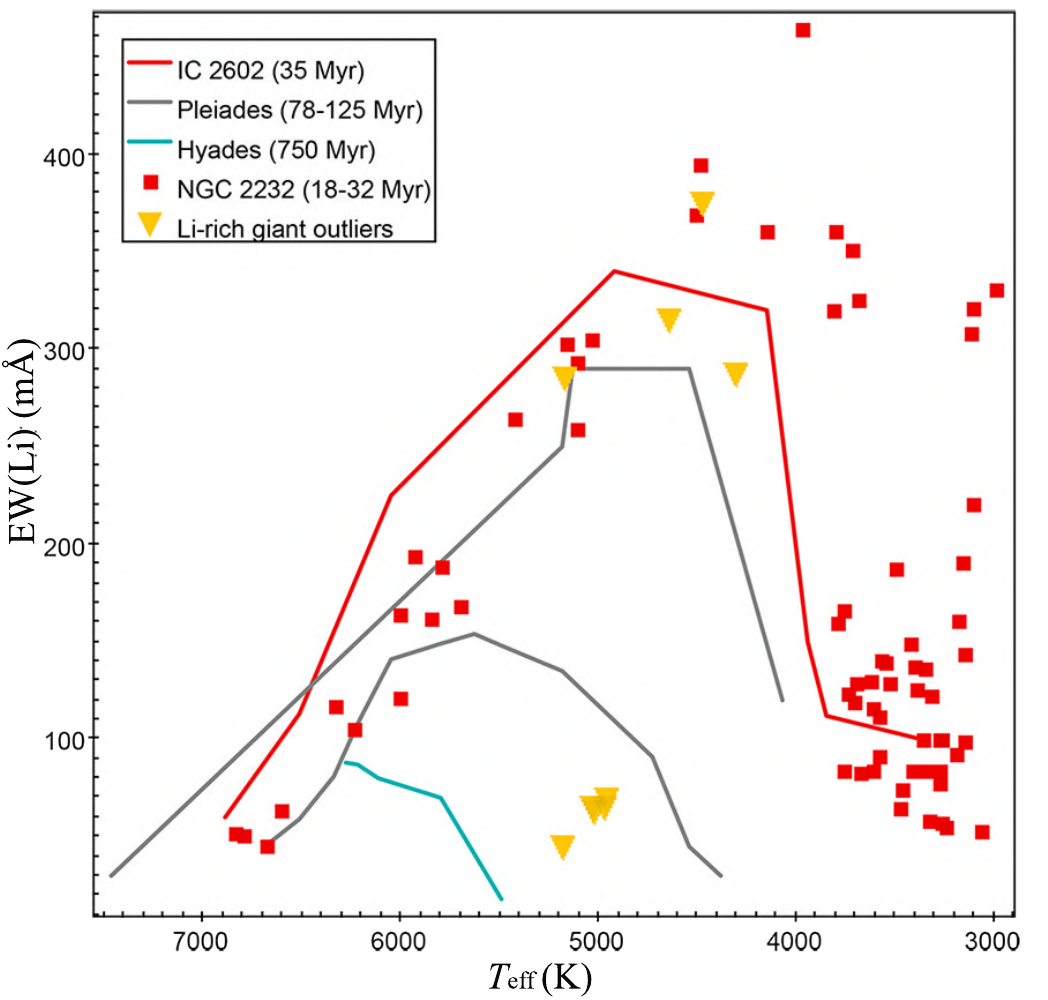} 
\caption{$EW$(Li)-versus-$T_{\rm eff}$ diagram for NGC~2232.}
             \label{fig:57}
    \end{figure}

\clearpage

\subsection{NGC~2547}

               \begin{figure} [htp]
   \centering
   \includegraphics[width=0.9\linewidth]{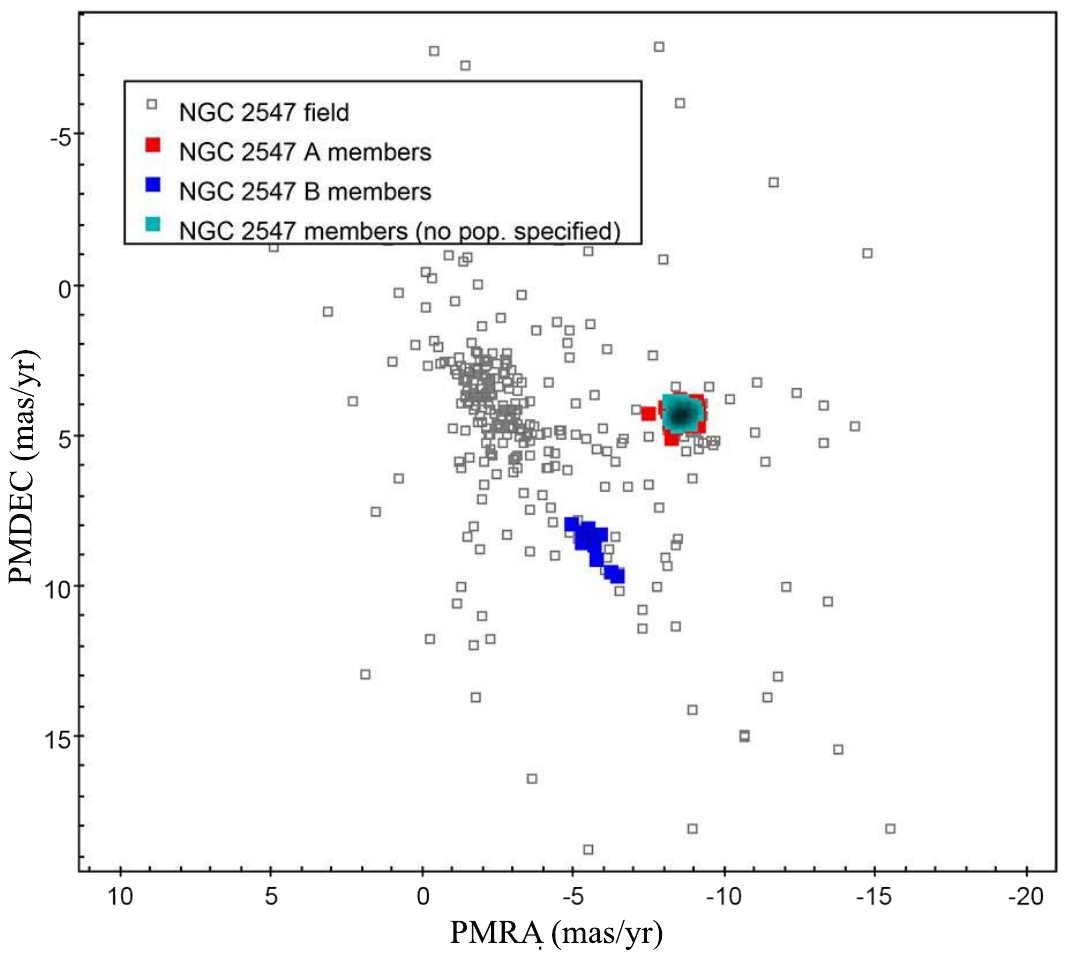}
   \caption{PMs diagram for NGC~2547.}
             \label{fig:58}
    \end{figure}
    
     \begin{figure} [htp]
   \centering
   \includegraphics[width=0.9\linewidth, height=7cm]{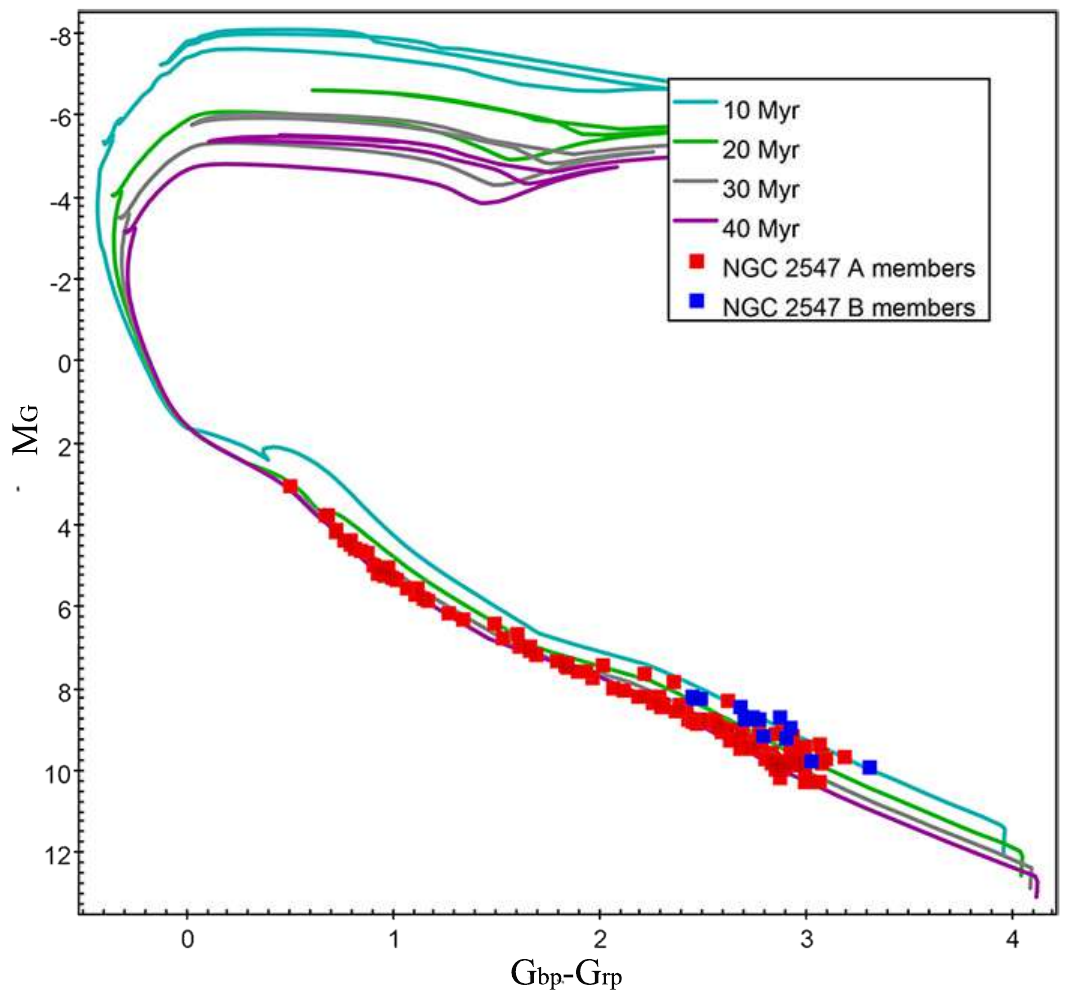}
   \caption{CMD for NGC~2547.}
             \label{fig:59}
    \end{figure}
    
         \begin{figure} [htp]
   \centering
   \includegraphics[width=0.9\linewidth, height=8cm]{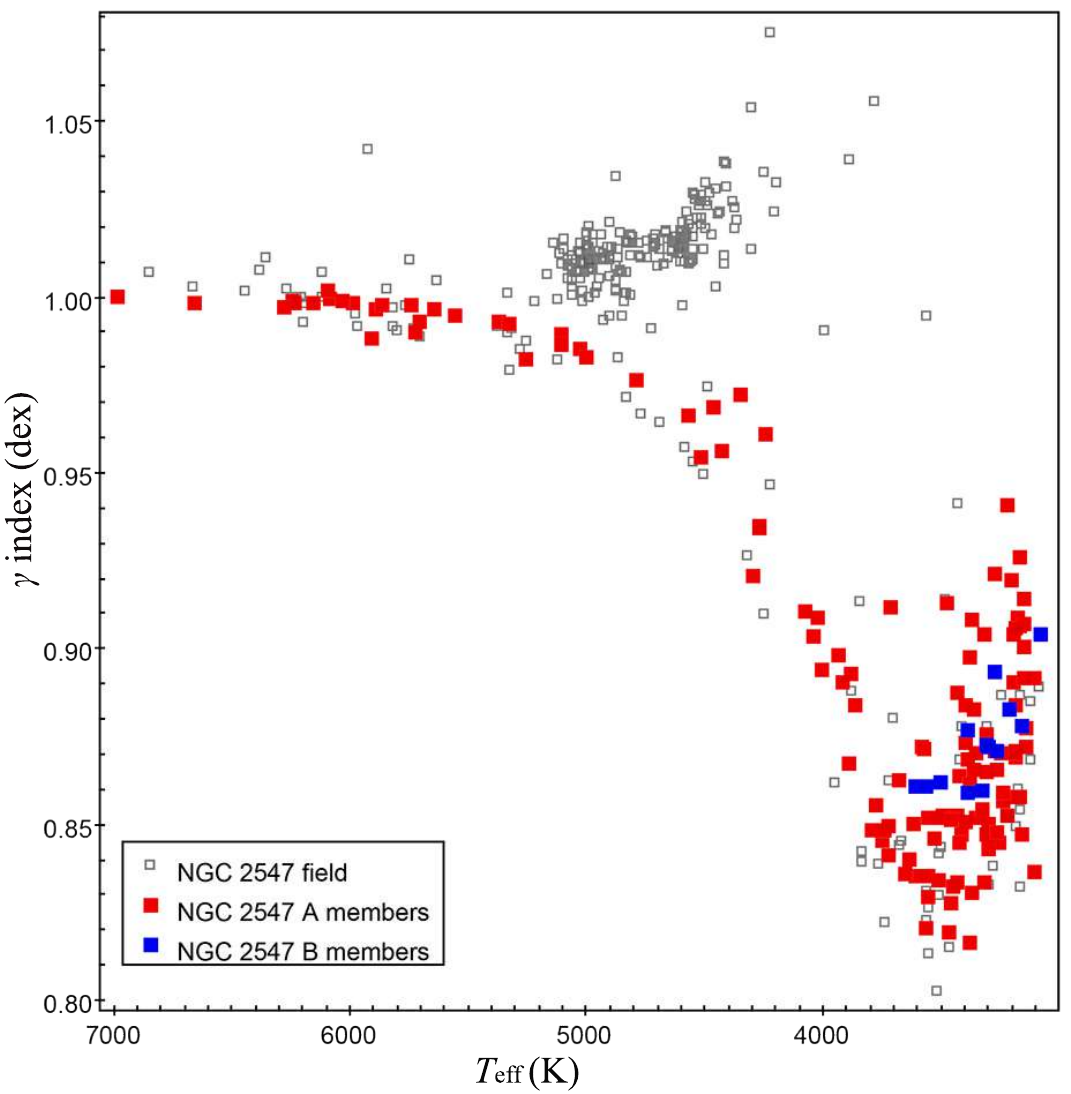}
   \caption{$\gamma$ index-versus-$T_{\rm eff}$ diagram for NGC~2547.}
             \label{fig:60}
    \end{figure}

  \begin{figure} [htp]
   \centering
 \includegraphics[width=0.9\linewidth]{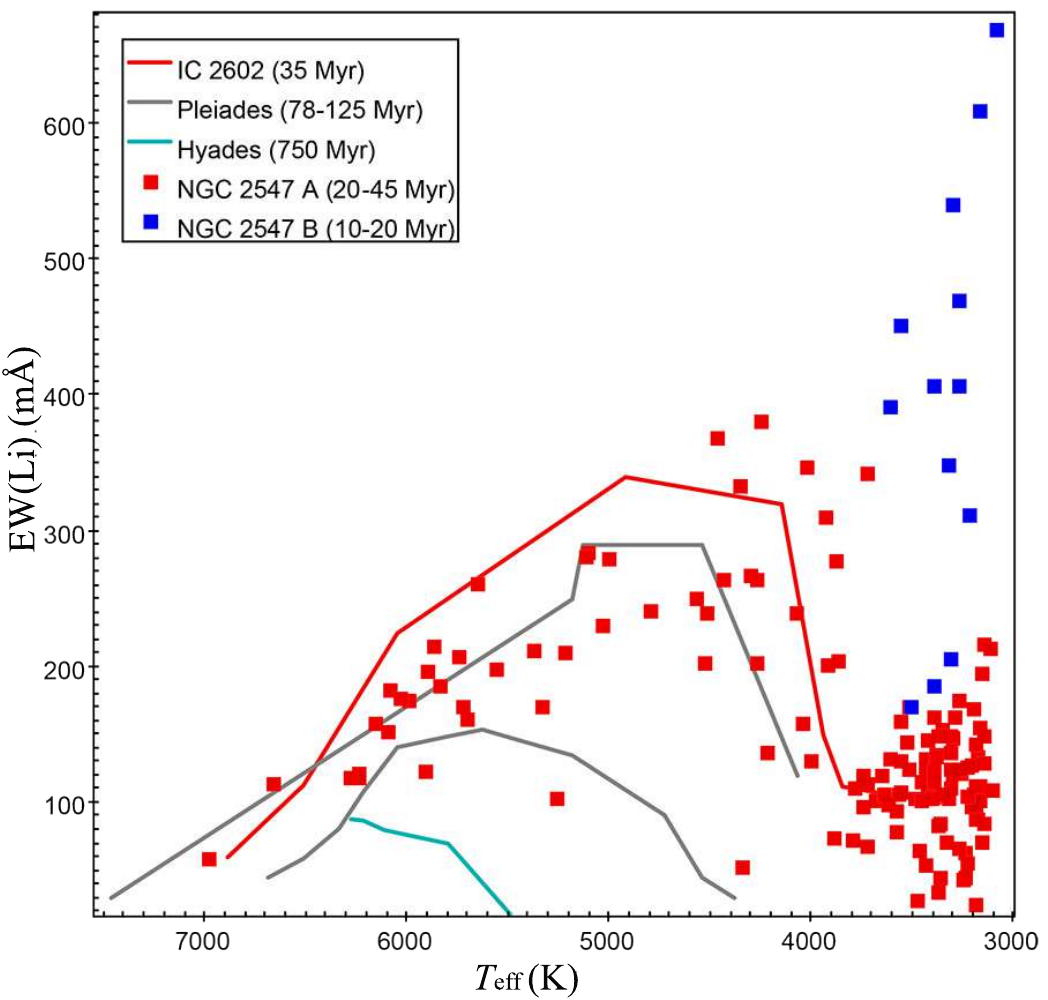} 
\caption{$EW$(Li)-versus-$T_{\rm eff}$ diagram for NGC~2547.}
             \label{fig:61}
    \end{figure}

\clearpage

\subsection{IC~2391}

 \begin{figure} [htp]
   \centering
\includegraphics[width=1\linewidth, height=5cm]{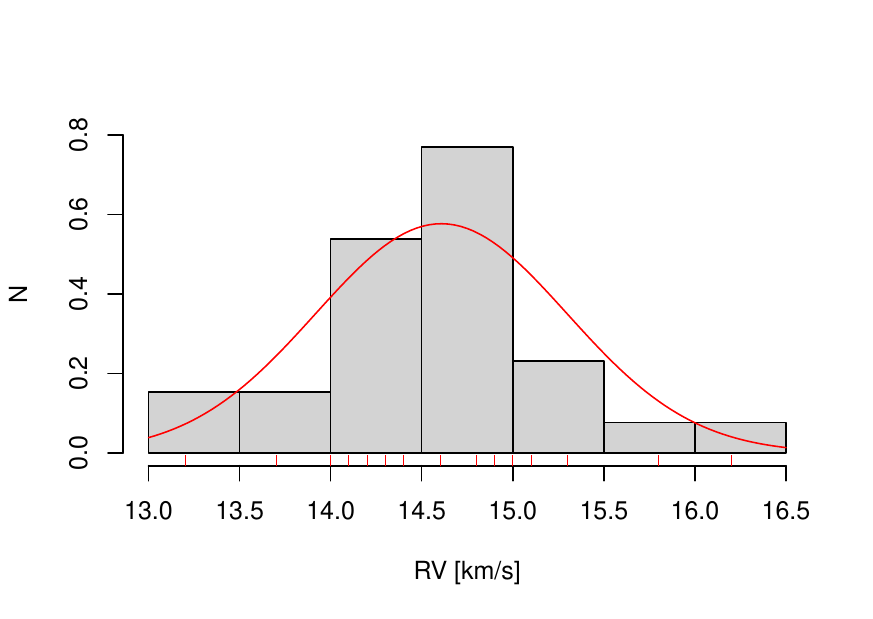}
\caption{$RV$ distribution for IC~2391.}
             \label{fig:62}
    \end{figure}
    
           \begin{figure} [htp]
   \centering
\includegraphics[width=1\linewidth, height=5cm]{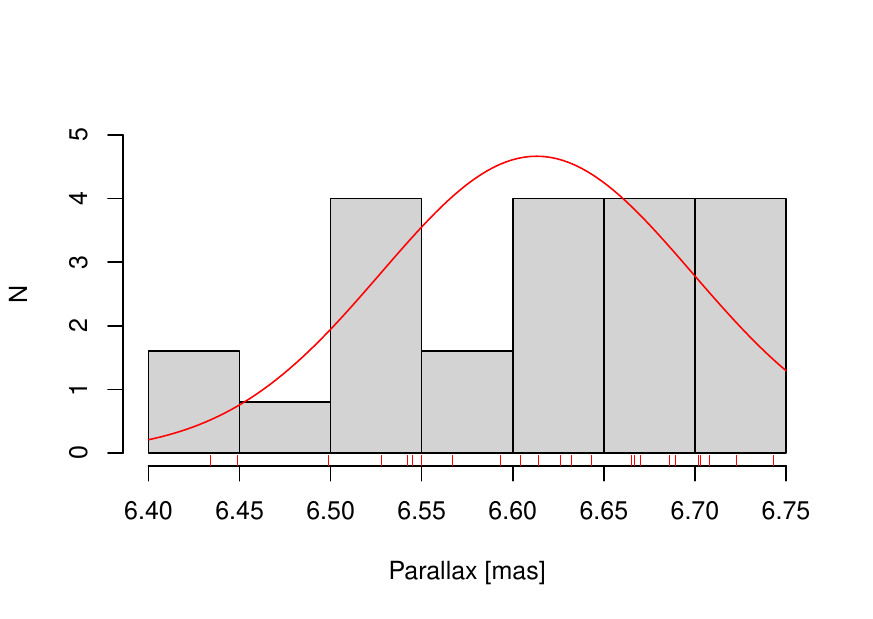}
\caption{Parallax distribution for IC~2391.}
             \label{fig:63}
    \end{figure}

               \begin{figure} [htp]
   \centering
   \includegraphics[width=0.9\linewidth]{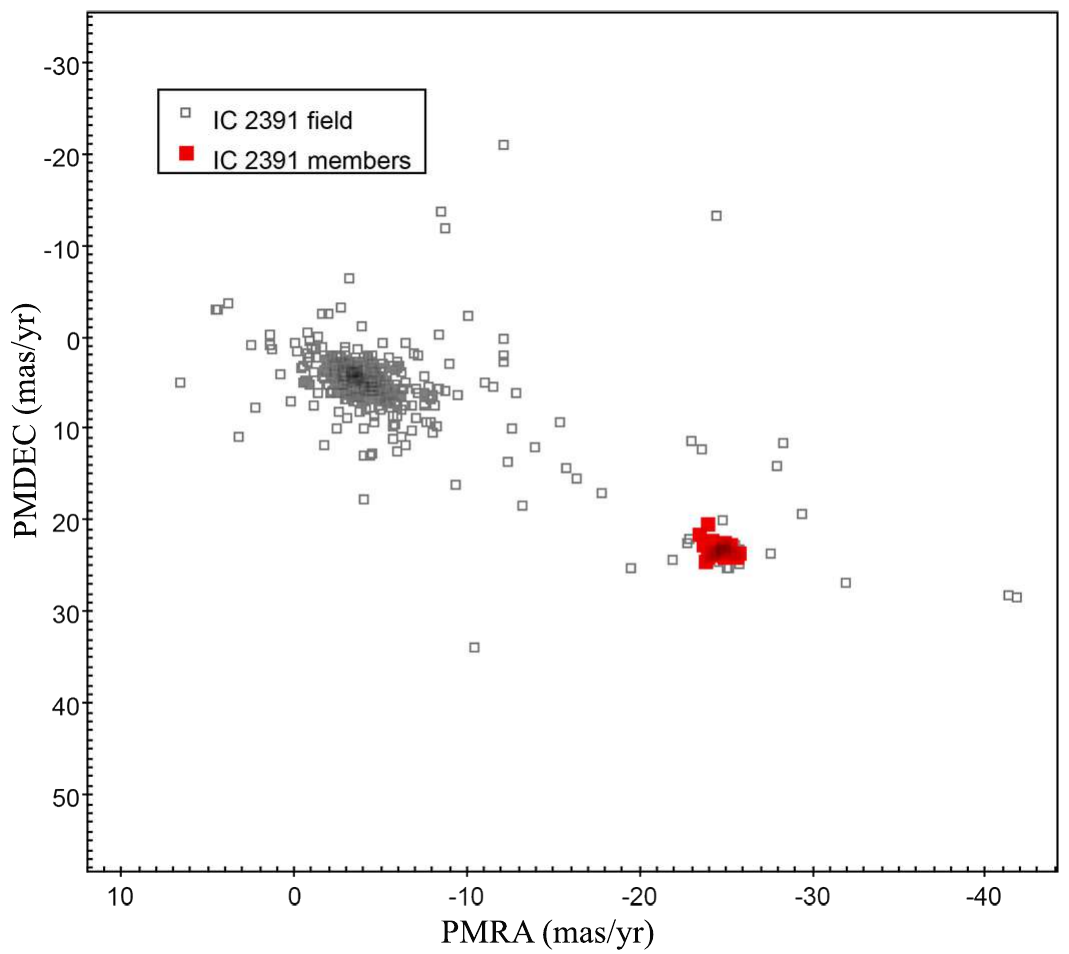}
   \caption{PMs diagram for IC~2391.}
             \label{fig:64}
    \end{figure}
    
     \begin{figure} [htp]
   \centering
   \includegraphics[width=0.8\linewidth, height=7cm]{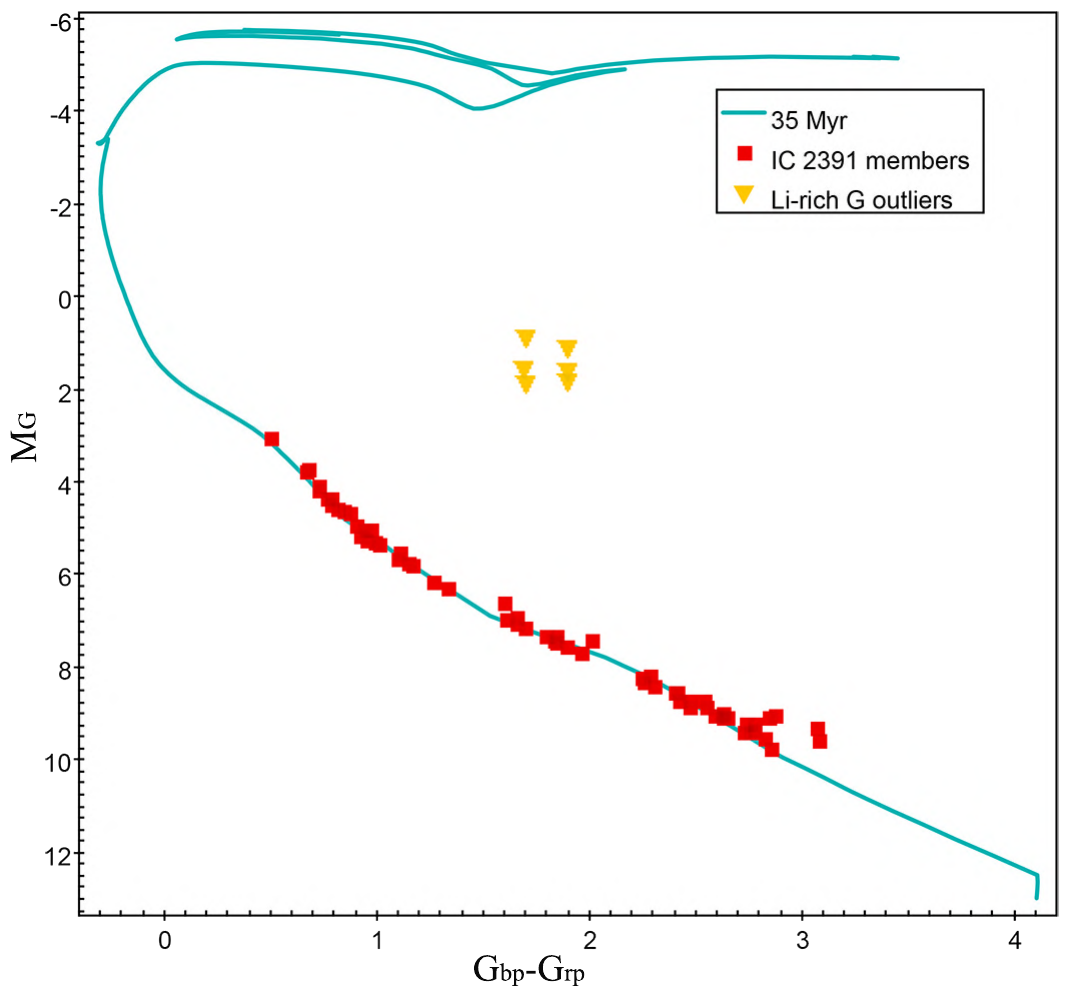}
   \caption{CMD for IC~2391.}
             \label{fig:65}
    \end{figure}
    
         \begin{figure} [htp]
   \centering
   \includegraphics[width=0.8\linewidth, height=7cm]{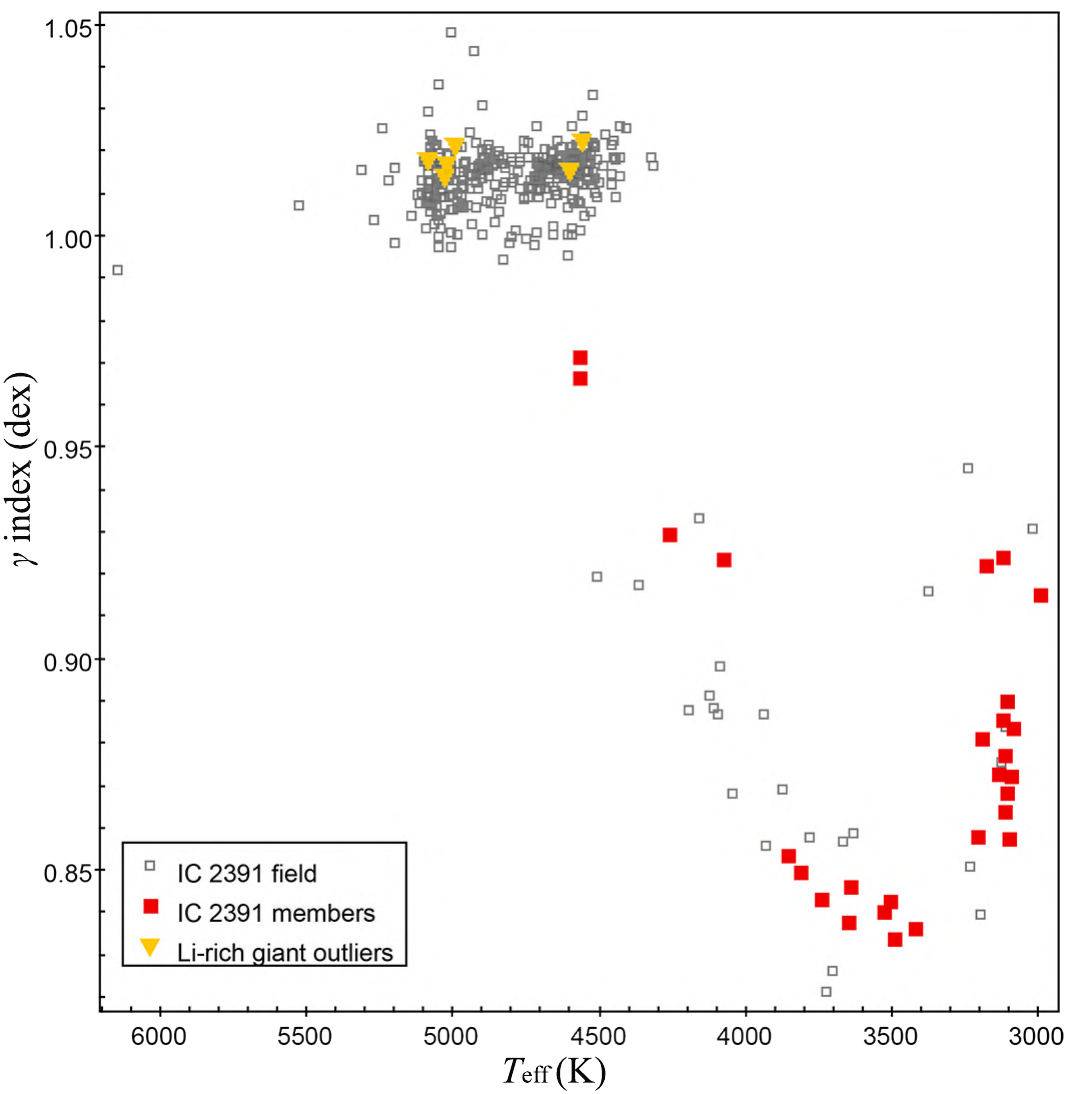}
   \caption{$\gamma$ index-versus-$T_{\rm eff}$ diagram for IC~2391.}
             \label{fig:66}
    \end{figure}

  \begin{figure} [htp]
   \centering
 \includegraphics[width=0.8\linewidth]{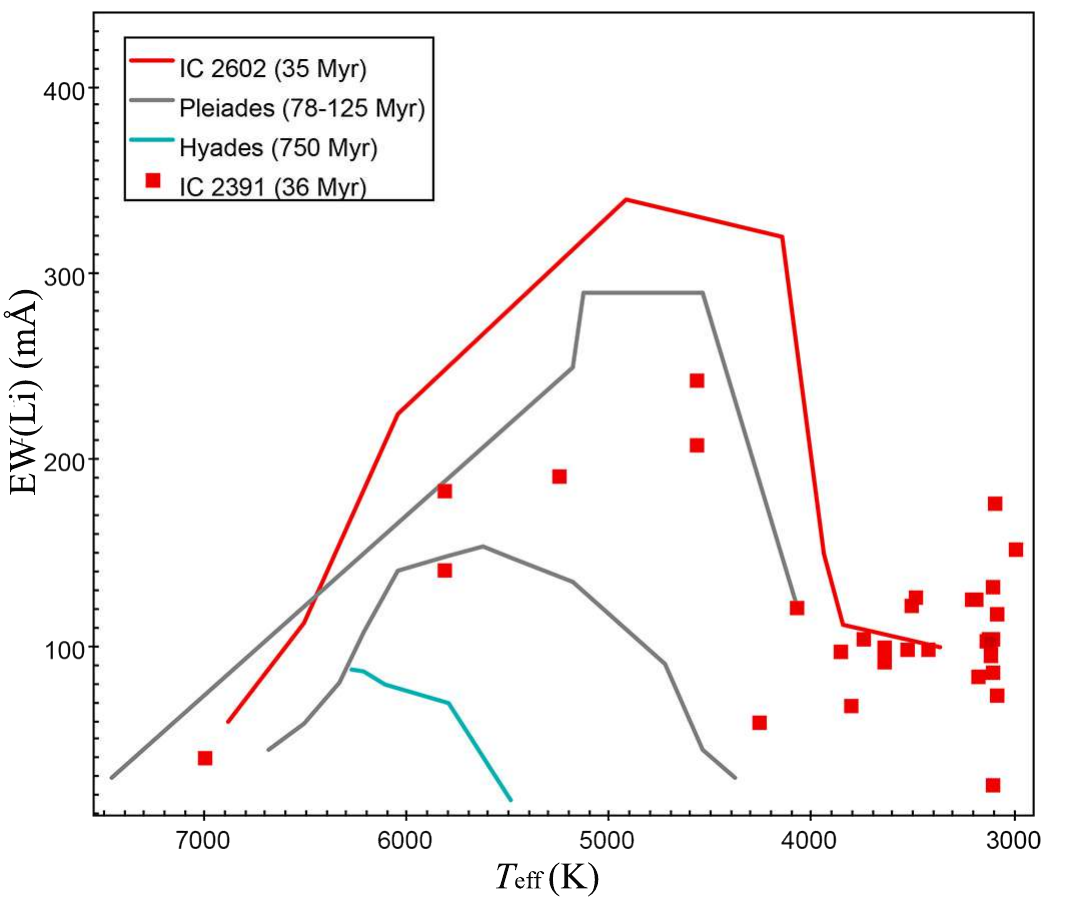} 
\caption{$EW$(Li)-versus-$T_{\rm eff}$ diagram for IC~2391.}
             \label{fig:67}
    \end{figure}

\clearpage

\subsection{IC~2602}

 \begin{figure} [htp]
   \centering
\includegraphics[width=1\linewidth, height=5cm]{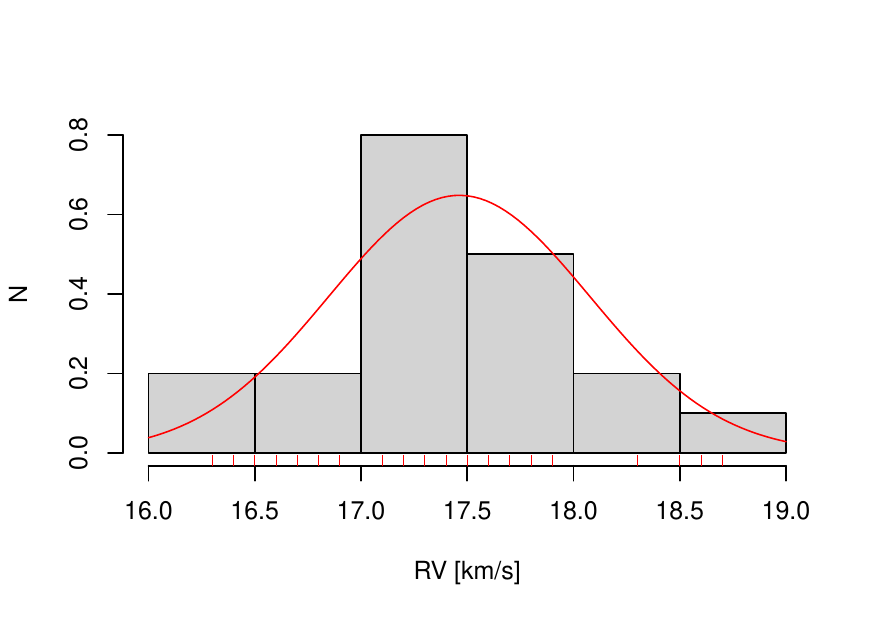}
\caption{$RV$ distribution for IC~2602.}
             \label{fig:68}
    \end{figure}
    
           \begin{figure} [htp]
   \centering
\includegraphics[width=1\linewidth, height=5cm]{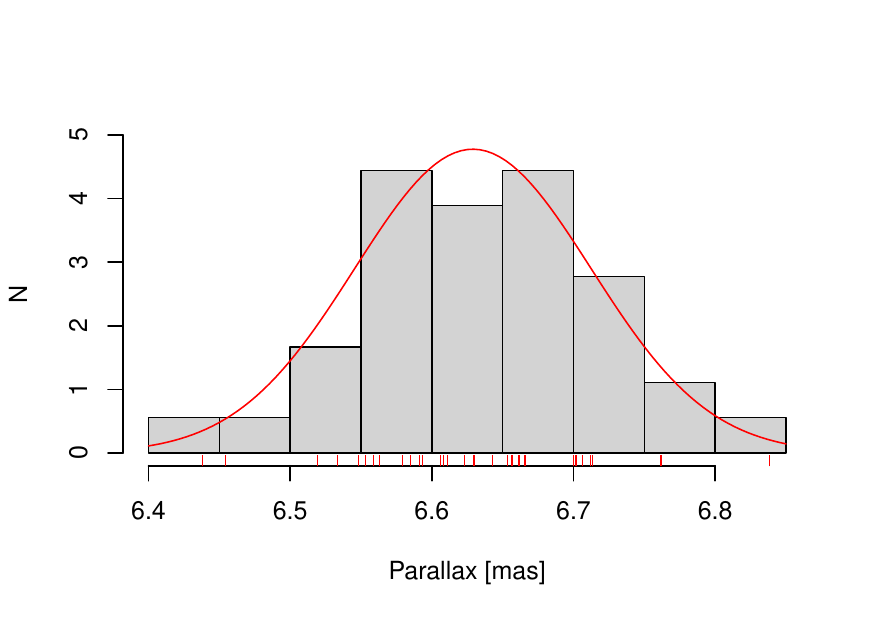}
\caption{Parallax distribution for IC~2602.}
             \label{fig:69}
    \end{figure}

               \begin{figure} [htp]
   \centering
   \includegraphics[width=0.9\linewidth]{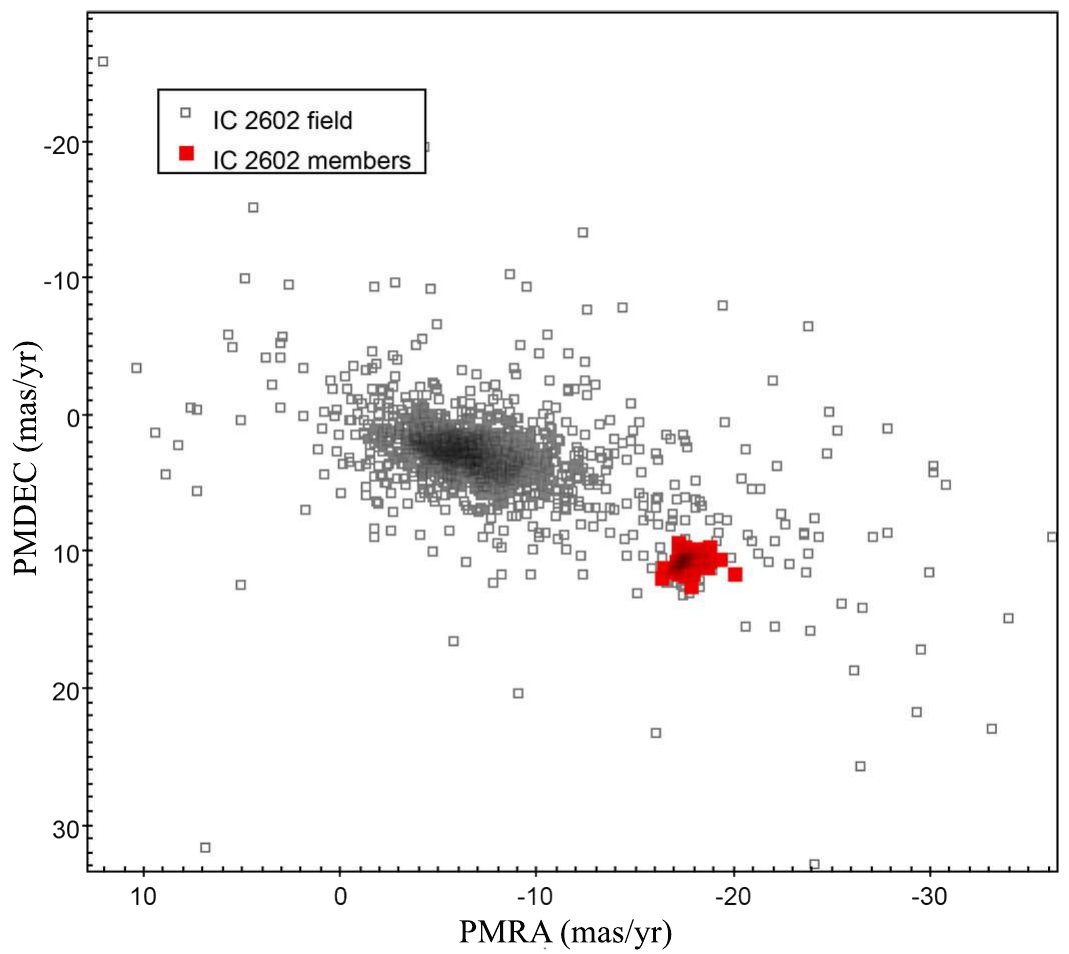}
   \caption{PMs diagram for IC~2602.}
             \label{fig:70}
    \end{figure}
    
     \begin{figure} [htp]
   \centering
   \includegraphics[width=0.8\linewidth, height=7cm]{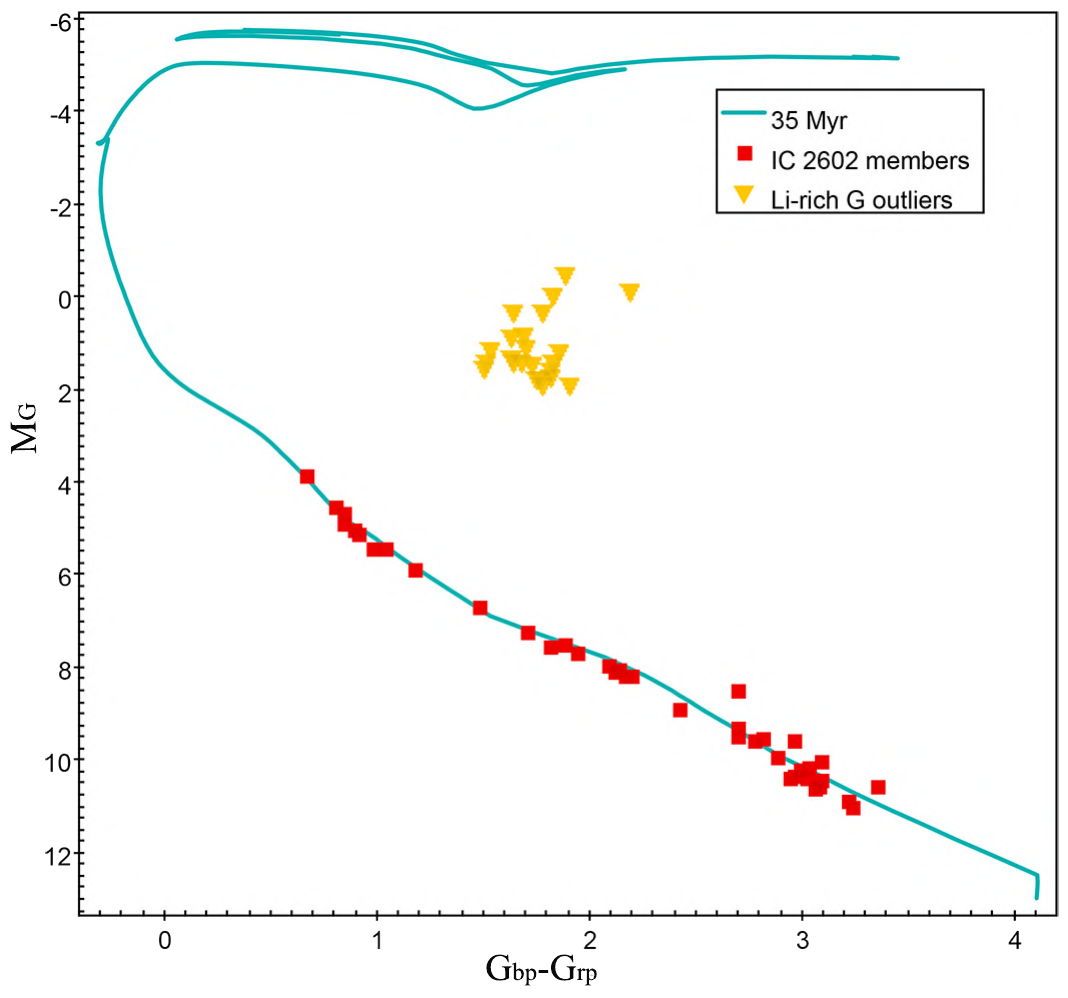}
   \caption{CMD for IC~2602.}
             \label{fig:71}
    \end{figure}
    
         \begin{figure} [htp]
   \centering
   \includegraphics[width=0.8\linewidth, height=7cm]{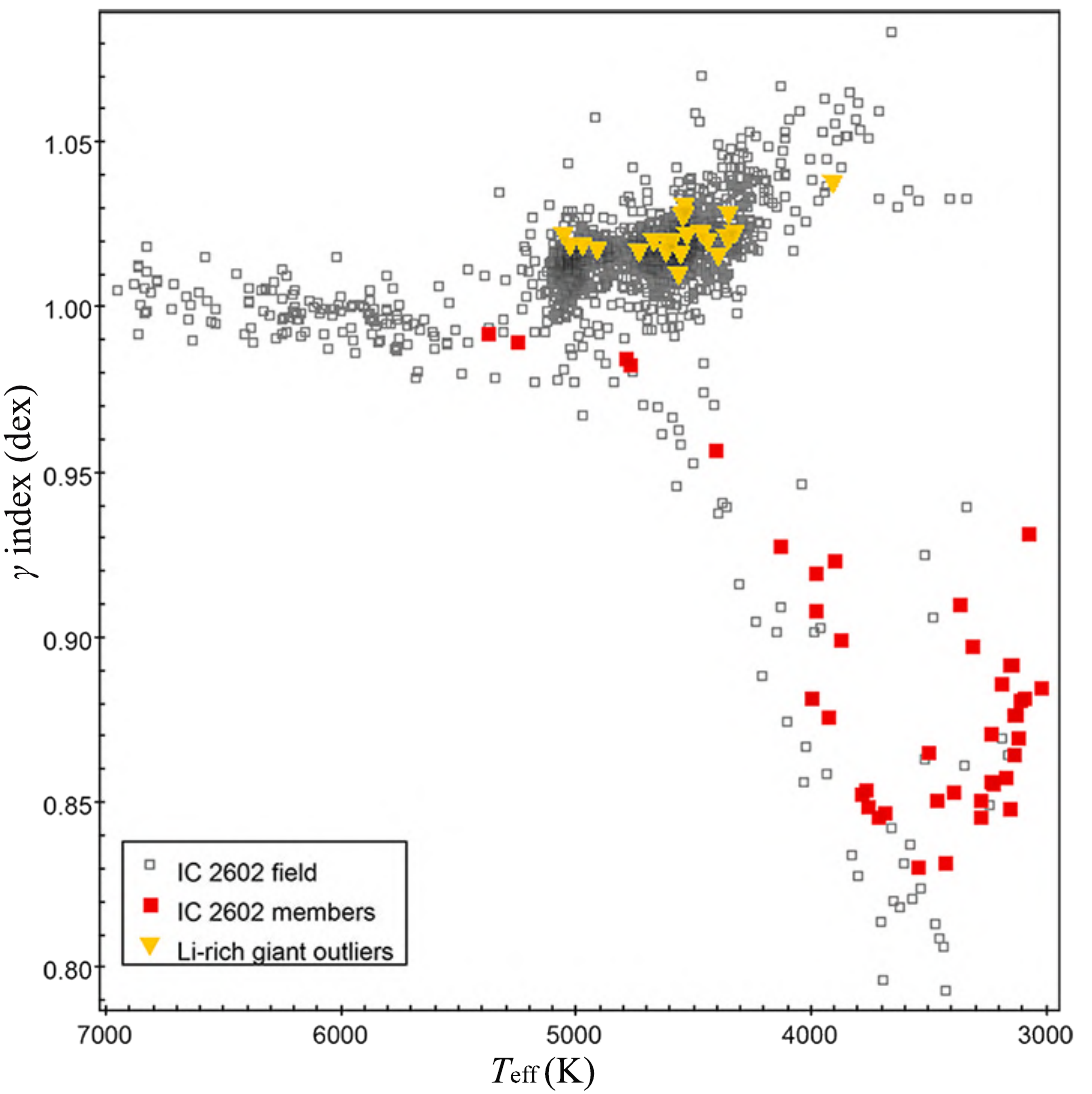}
   \caption{$\gamma$ index-versus-$T_{\rm eff}$ diagram for IC~2602.}
             \label{fig:72}
    \end{figure}

  \begin{figure} [htp]
   \centering
 \includegraphics[width=0.9\linewidth]{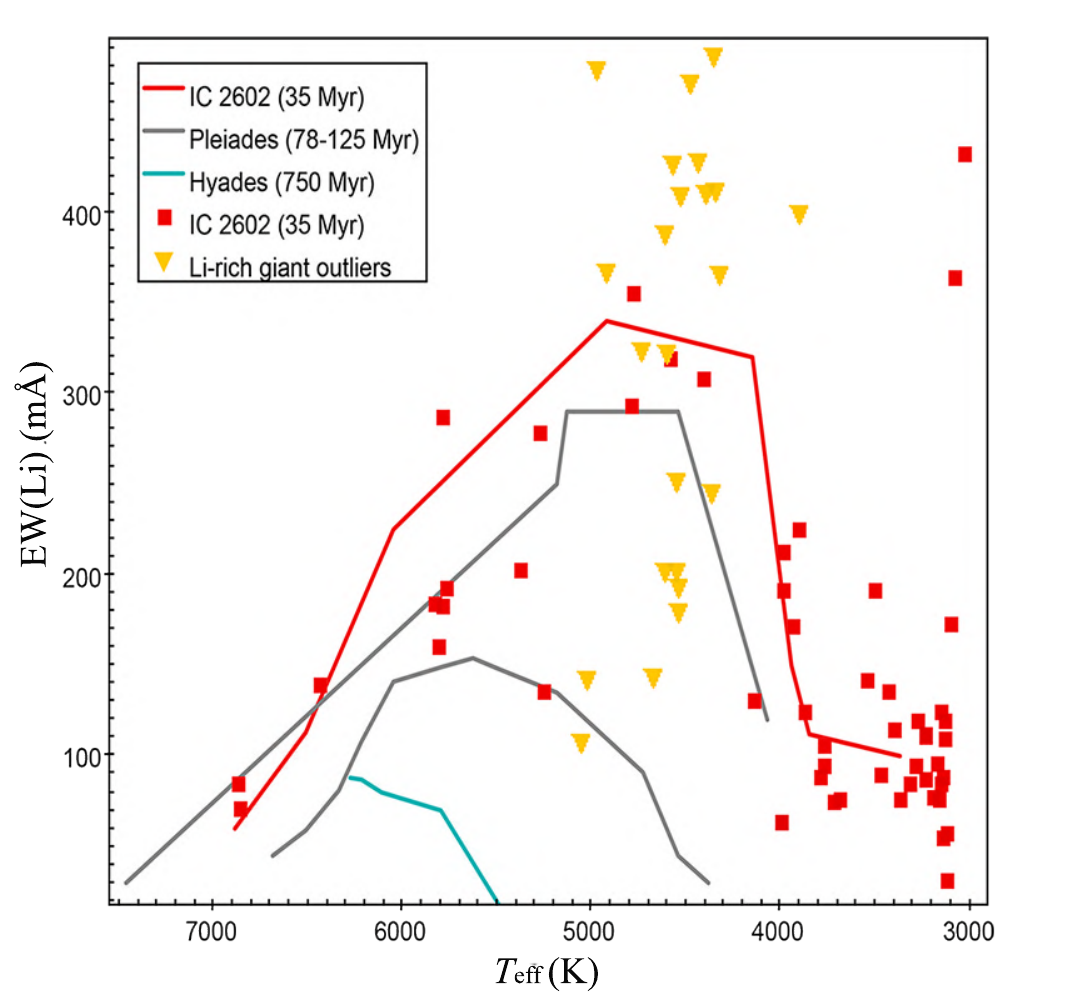} 
\caption{$EW$(Li)-versus-$T_{\rm eff}$ diagram for IC~2602.}
             \label{fig:73}
    \end{figure}

\clearpage

\subsection{IC~4665}

 \begin{figure} [htp]
   \centering
\includegraphics[width=0.9\linewidth, height=5cm]{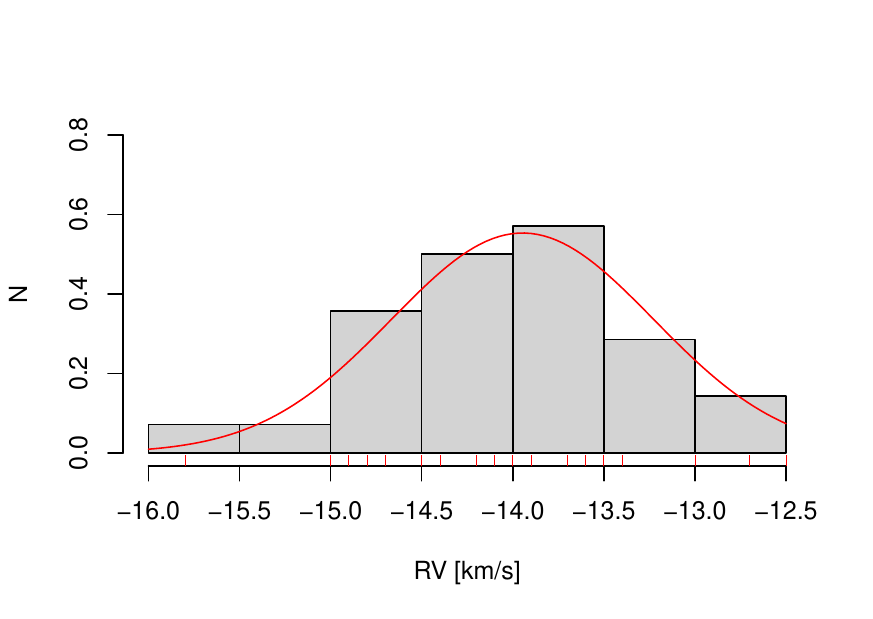}
\caption{$RV$ distribution for IC~4665.}
             \label{fig:74}
    \end{figure}
    
           \begin{figure} [htp]
   \centering
\includegraphics[width=0.9\linewidth, height=5cm]{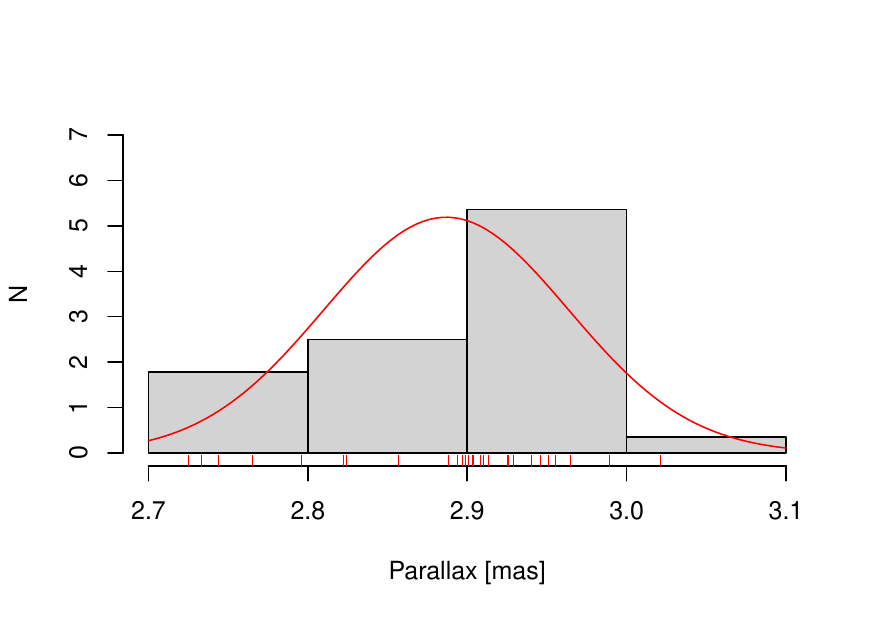}
\caption{Parallax distribution for IC~4665.}
             \label{fig:75}
    \end{figure}

               \begin{figure} [htp]
   \centering
   \includegraphics[width=0.9\linewidth]{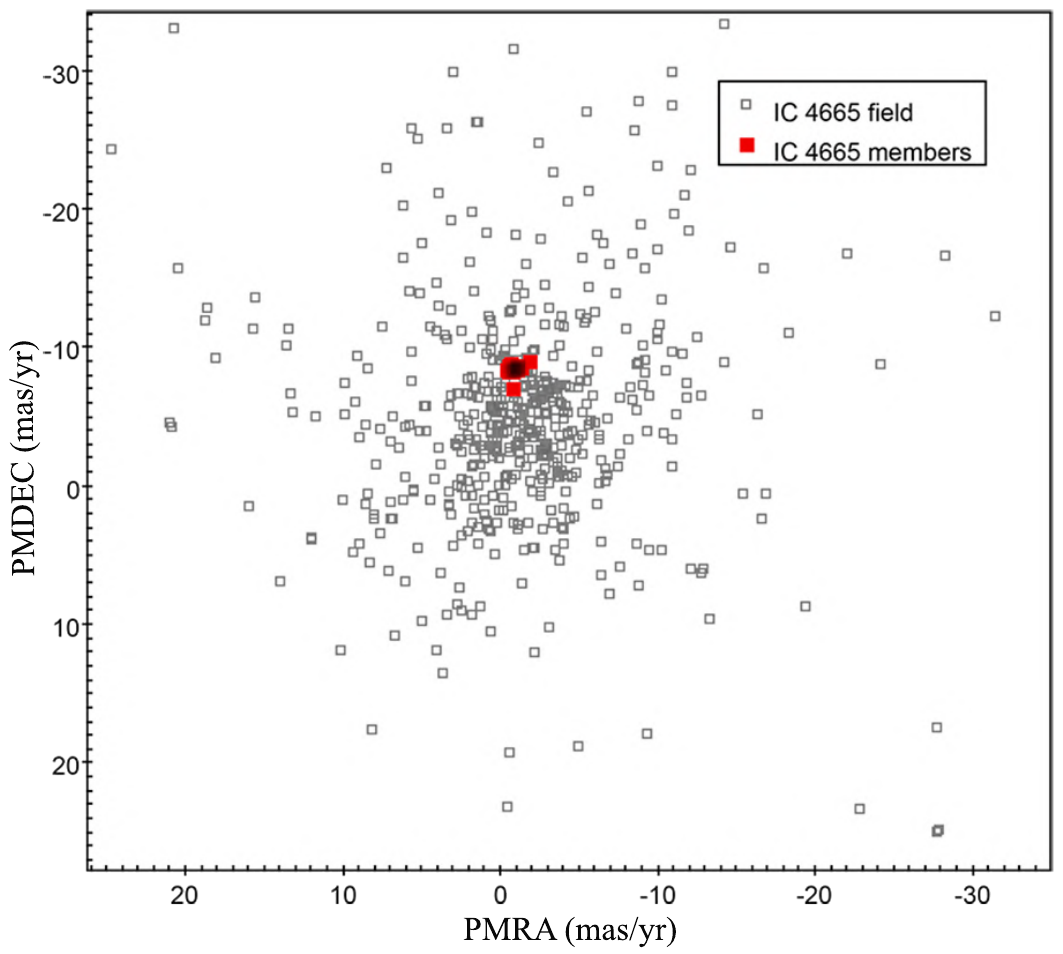}
   \caption{PMs diagram for IC~4665.}
             \label{fig:76}
    \end{figure}
    
     \begin{figure} [htp]
   \centering
   \includegraphics[width=0.8\linewidth, height=7cm]{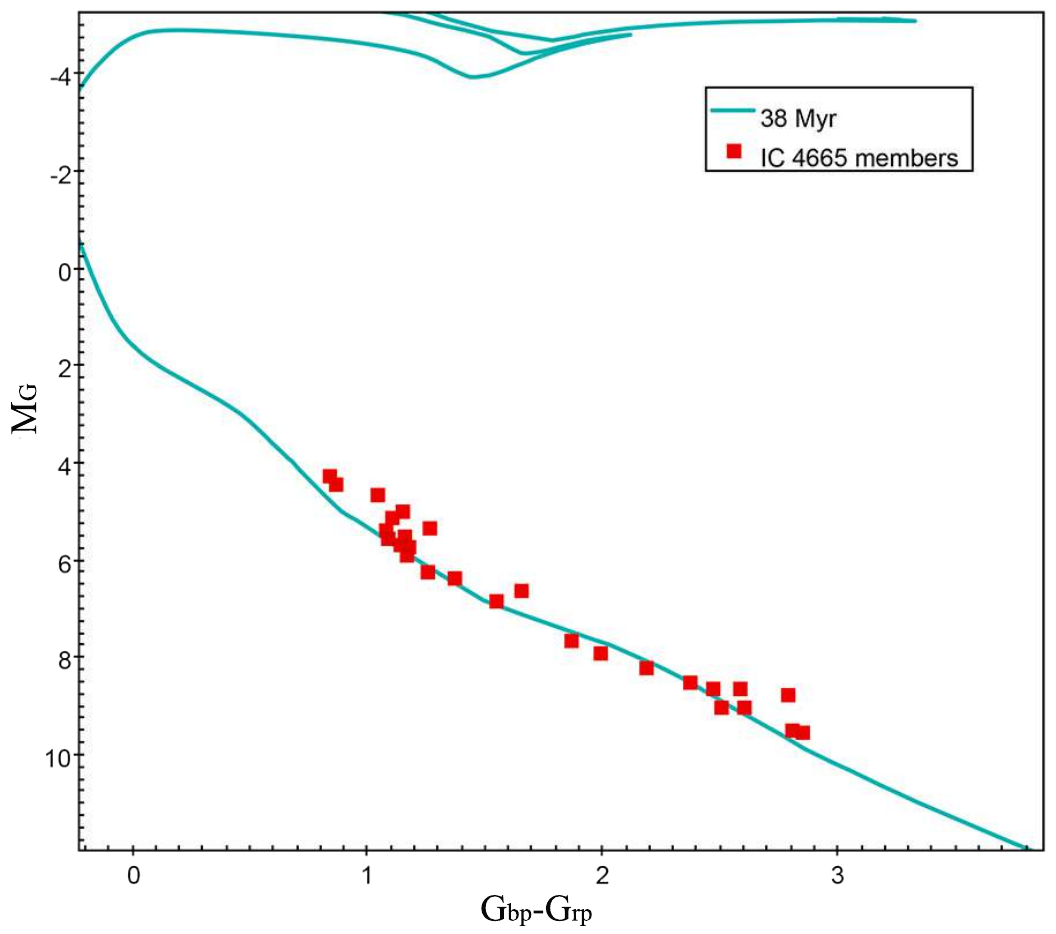}
   \caption{CMD for IC~4665.}
             \label{fig:77}
    \end{figure}
    
         \begin{figure} [htp]
   \centering
   \includegraphics[width=0.8\linewidth, height=7cm]{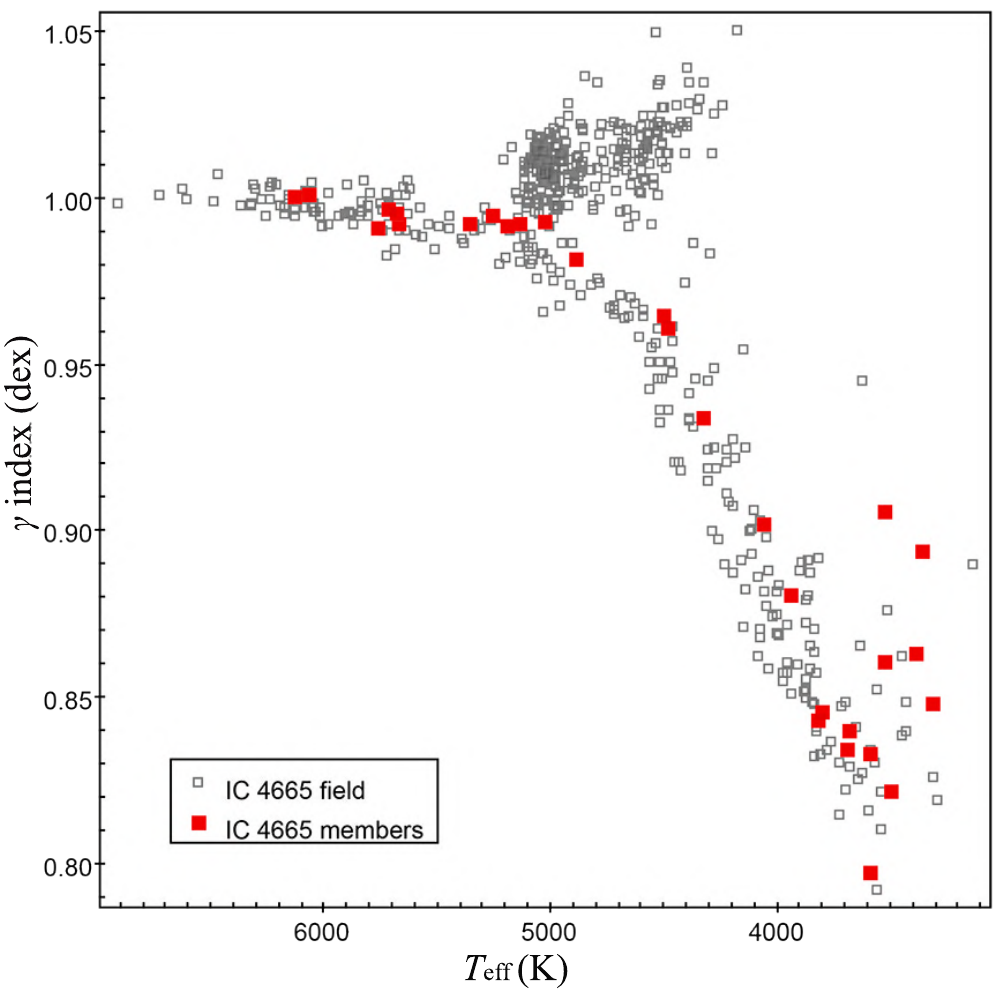}
   \caption{$\gamma$ index-versus-$T_{\rm eff}$ diagram for IC~4665.}
             \label{fig:78}
    \end{figure}

  \begin{figure} [htp]
   \centering
 \includegraphics[width=0.8\linewidth]{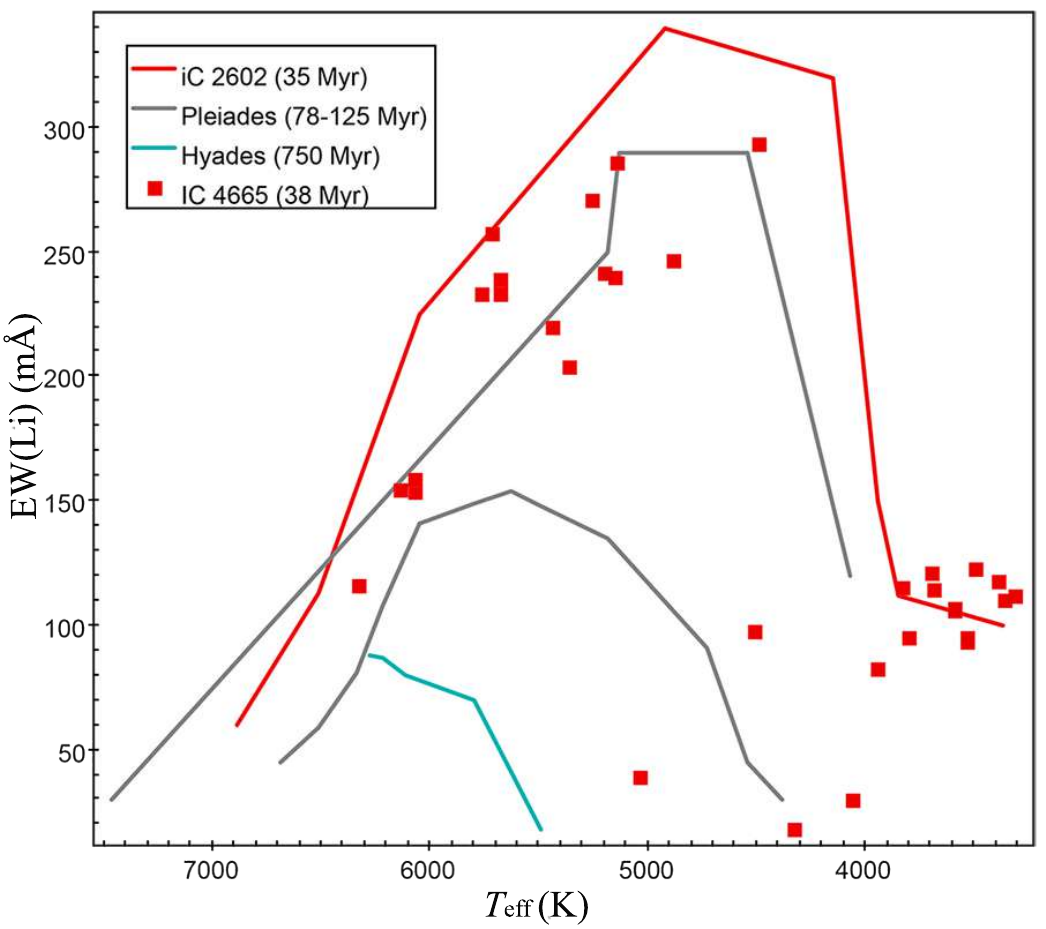} 
\caption{$EW$(Li)-versus-$T_{\rm eff}$ diagram for IC~4665.}
             \label{fig:79}
    \end{figure}

\clearpage

\subsection{NGC~2451}

     \begin{figure} [htp]
   \centering
   \includegraphics[width=0.9\linewidth]{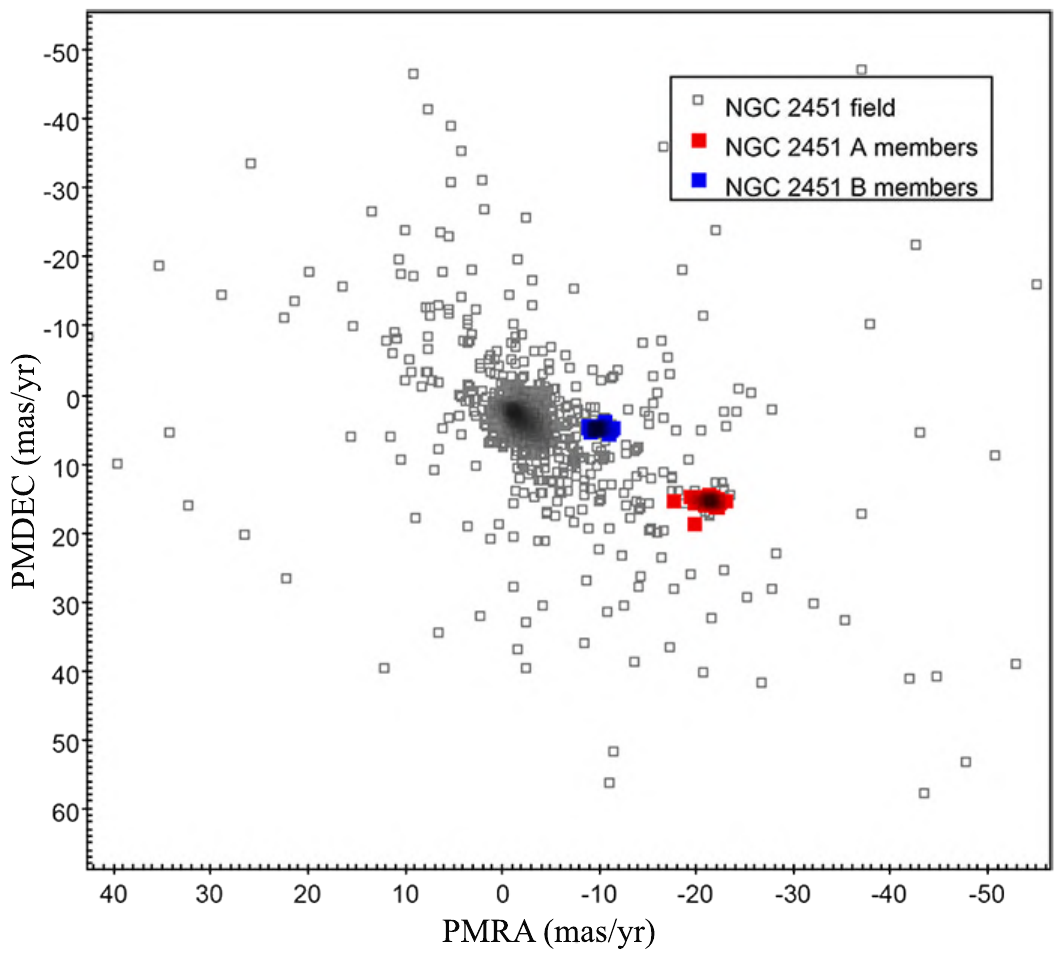}
   \caption{PMs diagram for NGC~2451.}
             \label{fig:80}
    \end{figure}
    
     \begin{figure} [htp]
   \centering
   \includegraphics[width=0.9\linewidth, height=7cm]{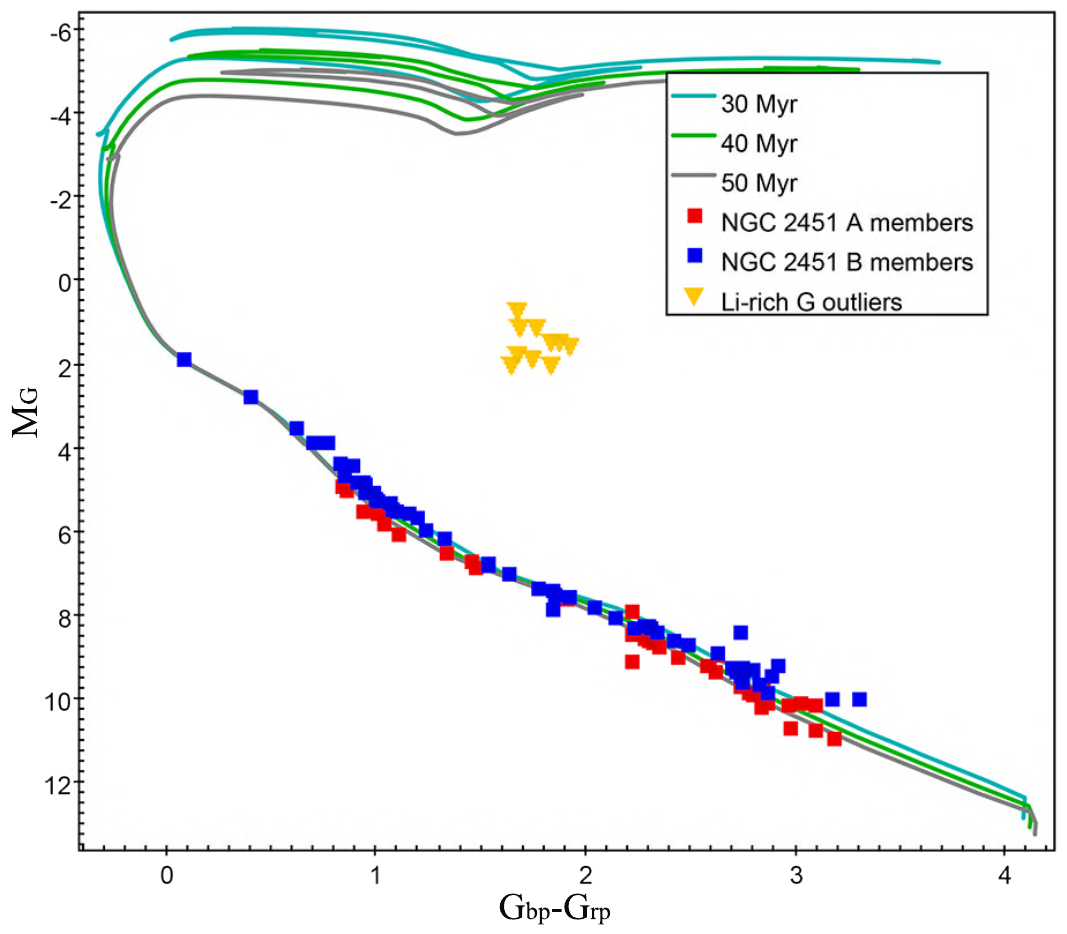}
   \caption{CMD for NGC~2451.}
             \label{fig:81}
    \end{figure}
    
         \begin{figure} [htp]
   \centering
   \includegraphics[width=0.9\linewidth, height=8cm]{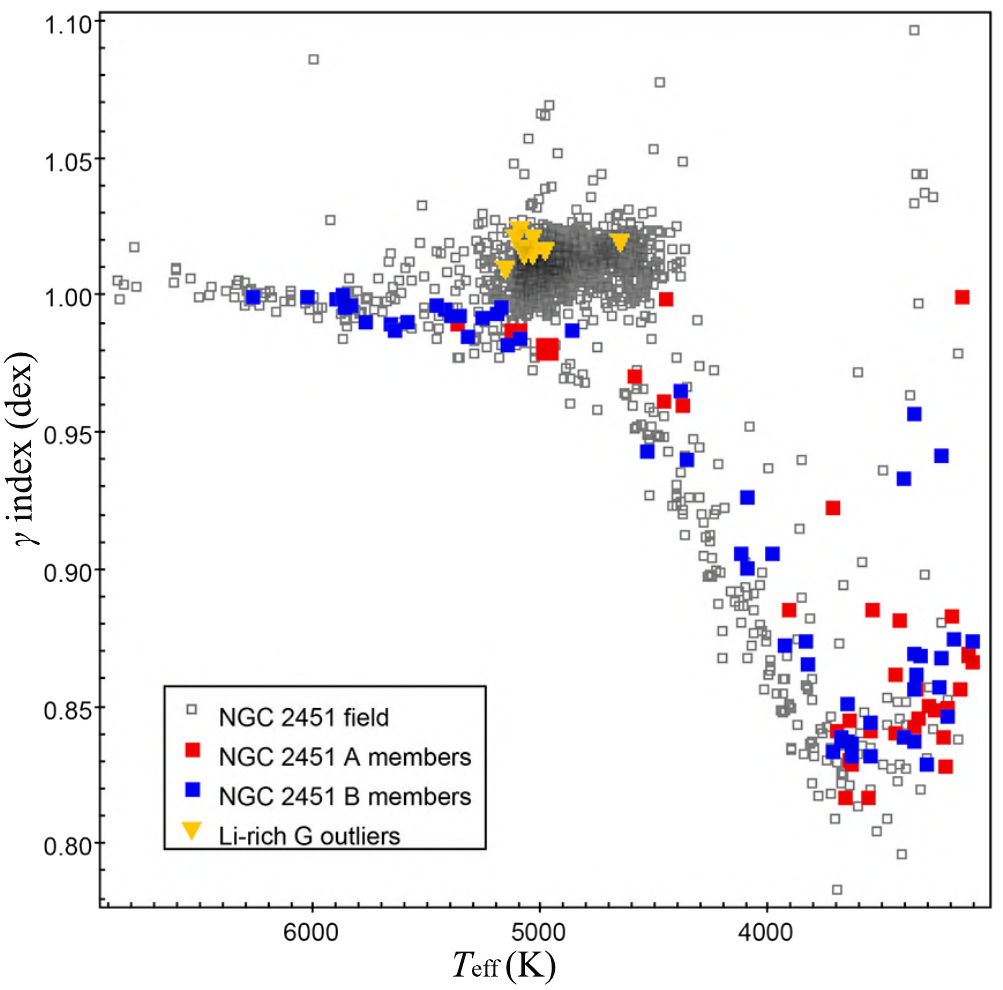}
   \caption{$\gamma$ index-versus-$T_{\rm eff}$ diagram for NGC~2451.}
             \label{fig:82}
    \end{figure}

  \begin{figure} [htp]
   \centering
 \includegraphics[width=0.9\linewidth]{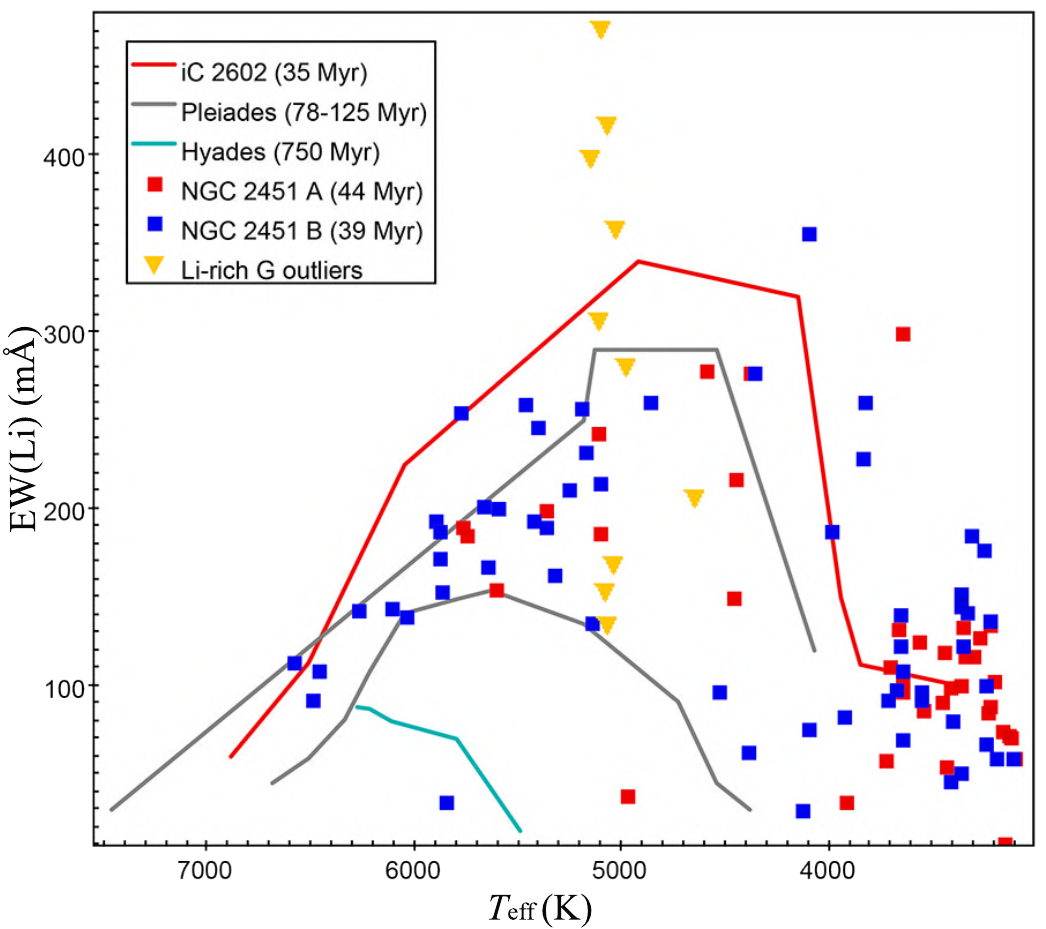} 
\caption{$EW$(Li)-versus-$T_{\rm eff}$ diagram for NGC~2451.}
             \label{fig:82}
    \end{figure}

\clearpage

\subsection{NGC~6405}

 \begin{figure} [htp]
   \centering
\includegraphics[width=0.9\linewidth, height=5cm]{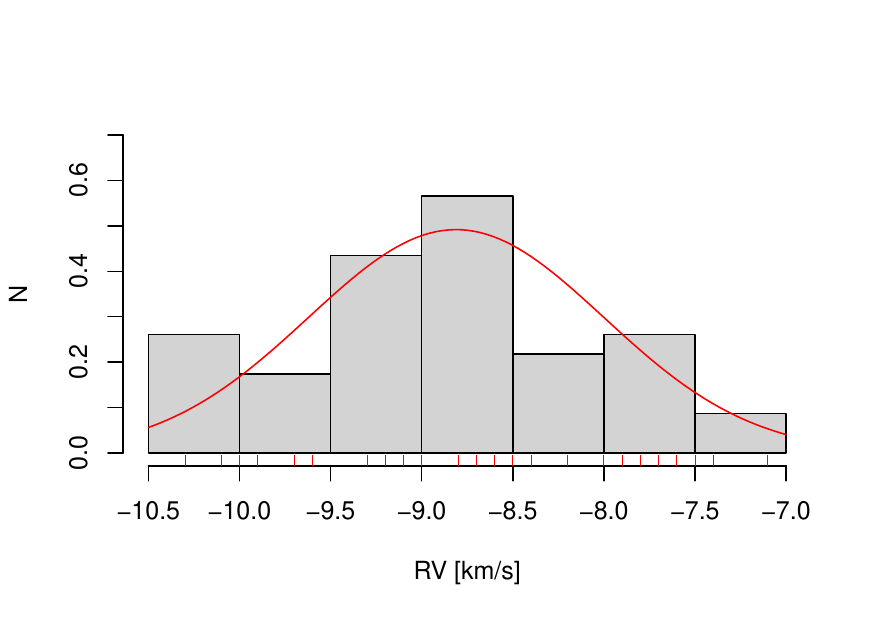}
\caption{$RV$ distribution for NGC~6405.}
             \label{fig:83}
    \end{figure}
    
           \begin{figure} [htp]
   \centering
\includegraphics[width=0.9\linewidth, height=5cm]{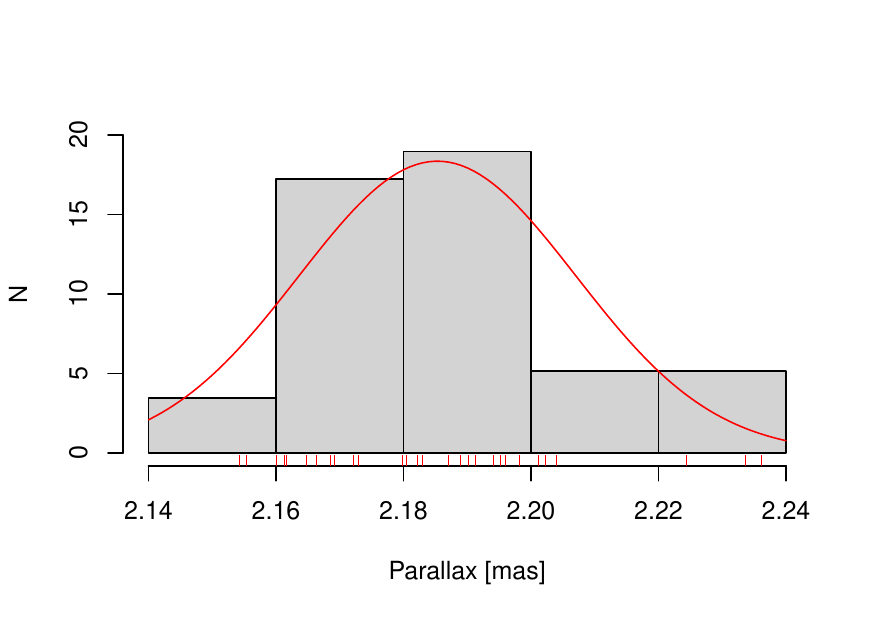}
\caption{Parallax distribution for NGC~6405.}
             \label{fig:84}
    \end{figure}

               \begin{figure} [htp]
   \centering
   \includegraphics[width=0.9\linewidth]{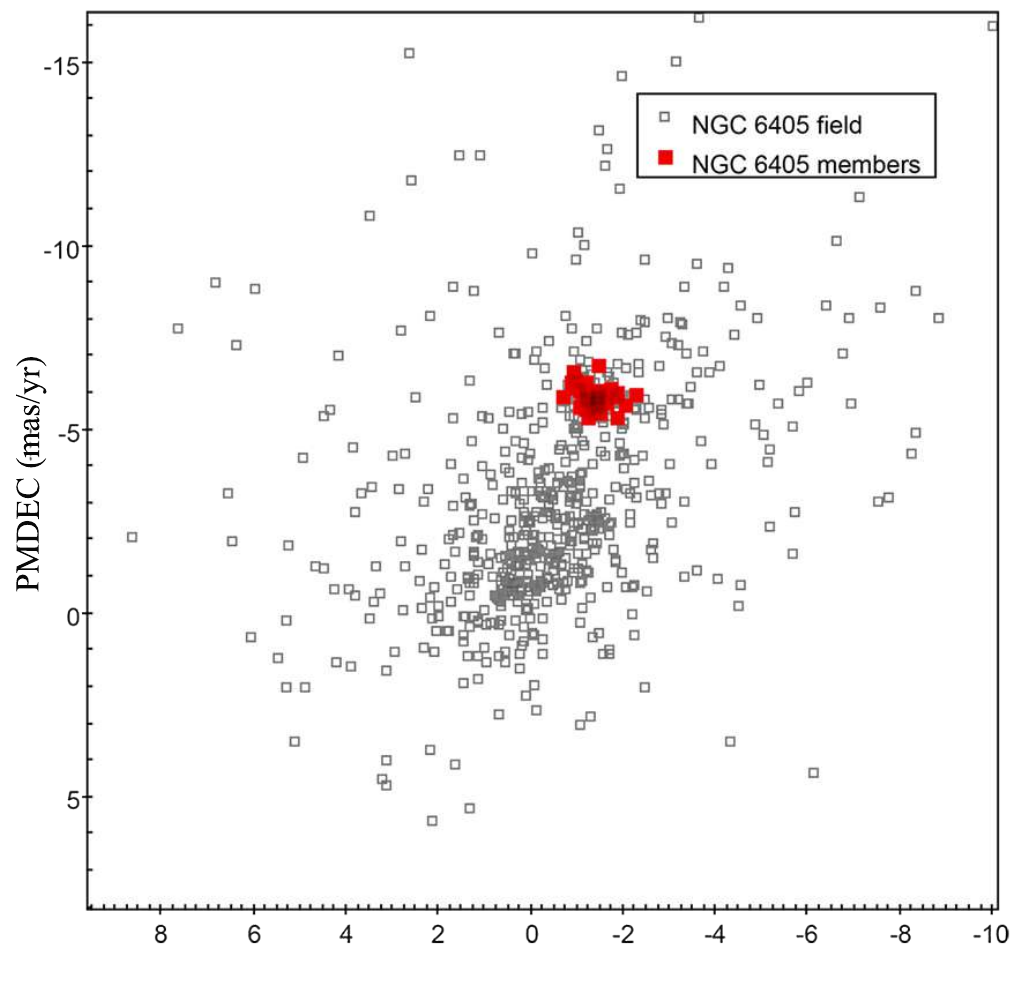}
   \caption{PMs diagram for NGC~6405.}
             \label{fig:85}
    \end{figure}
    
     \begin{figure} [htp]
   \centering
   \includegraphics[width=0.9\linewidth, height=7cm]{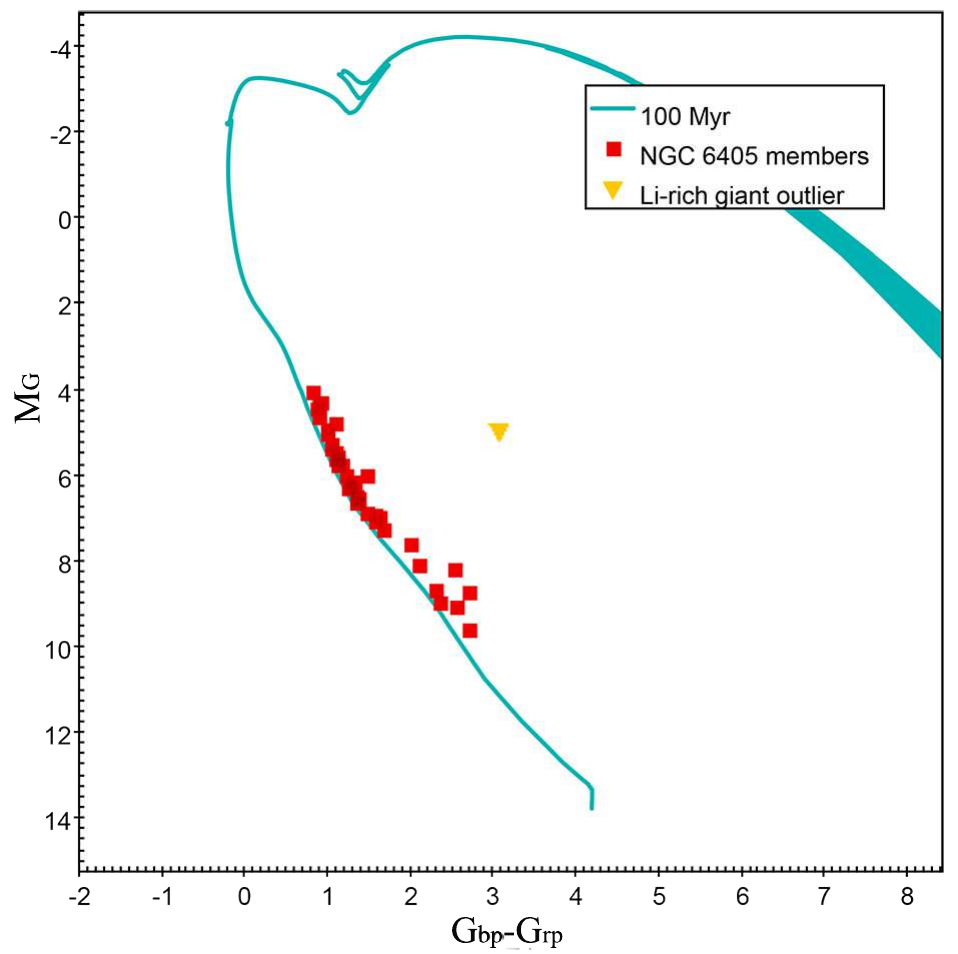}
   \caption{CMD for NGC~6405.}
             \label{fig:86}
    \end{figure}
    
      \begin{figure} [htp]
   \centering
 \includegraphics[width=0.9\linewidth]{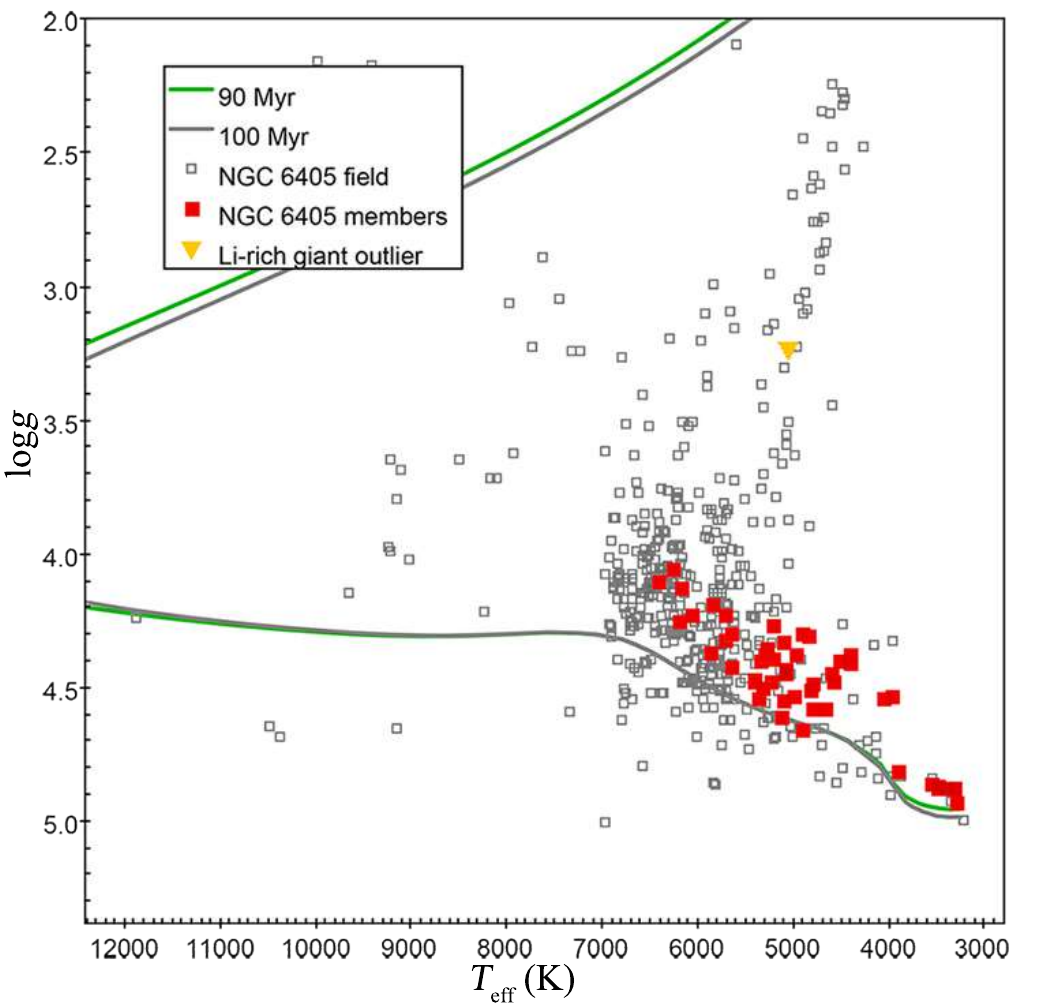} 
\caption{Kiel diagram for NGC~6405.}
             \label{fig:87}
    \end{figure}

  \begin{figure} [htp]
   \centering
 \includegraphics[width=0.9\linewidth]{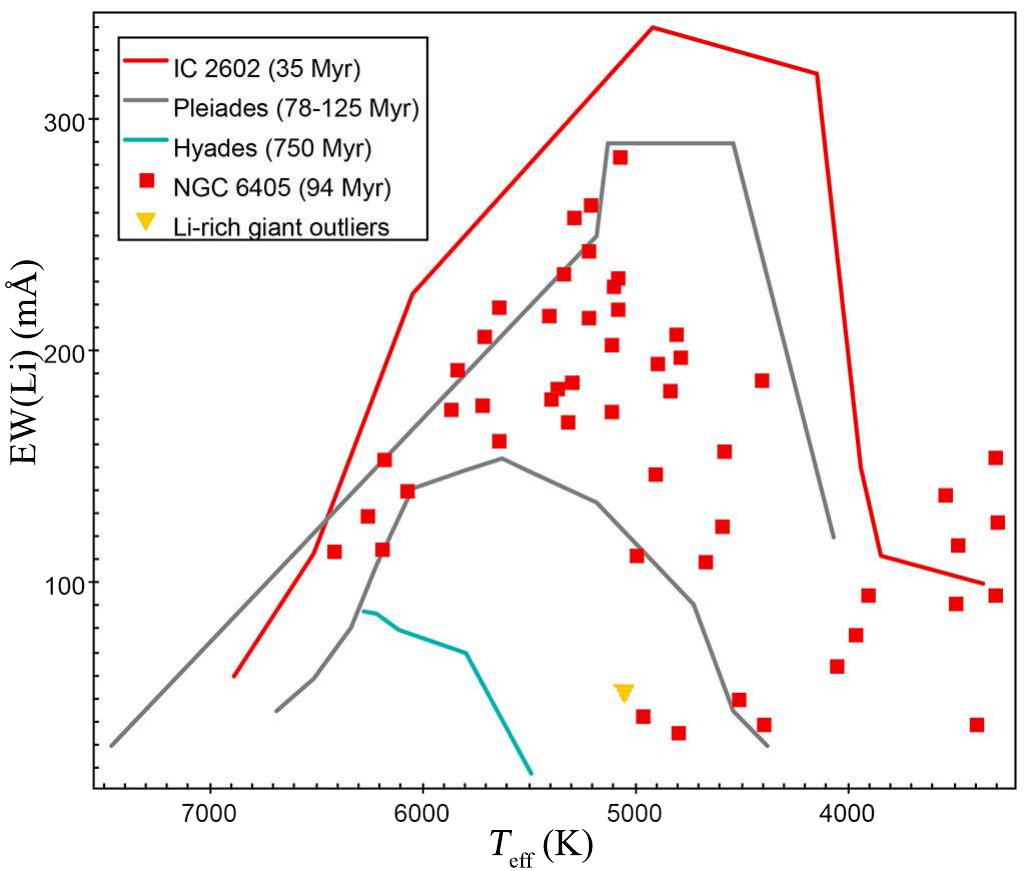} 
\caption{$EW$(Li)-versus-$T_{\rm eff}$ diagram for NGC~6405.}
             \label{fig:88}
    \end{figure}

\clearpage

\subsection{Blanco~1}

 \begin{figure} [htp]
   \centering
\includegraphics[width=0.9\linewidth, height=5cm]{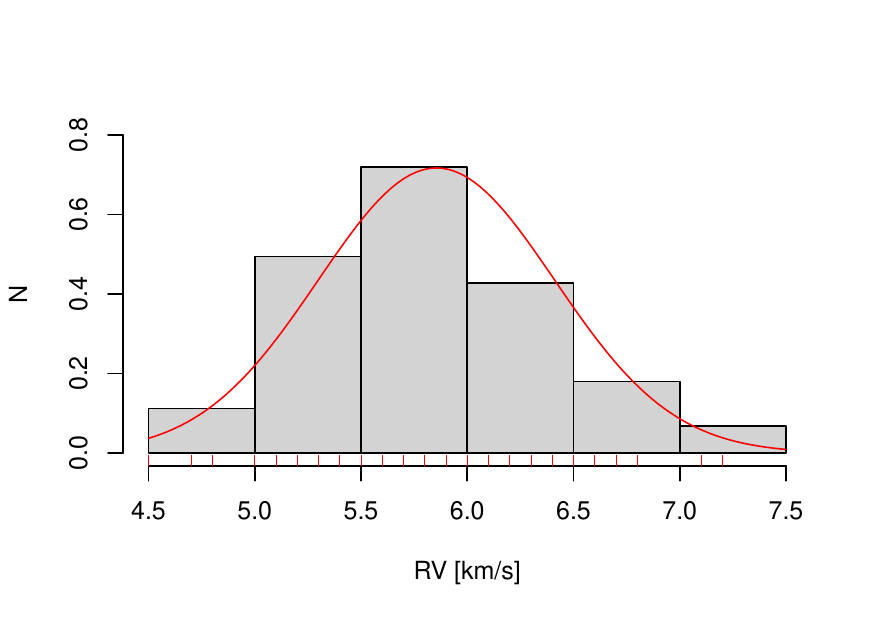}
\caption{$RV$ distribution for Blanco~1.}
             \label{fig:89}
    \end{figure}
    
           \begin{figure} [htp]
   \centering
\includegraphics[width=0.9\linewidth, height=5cm]{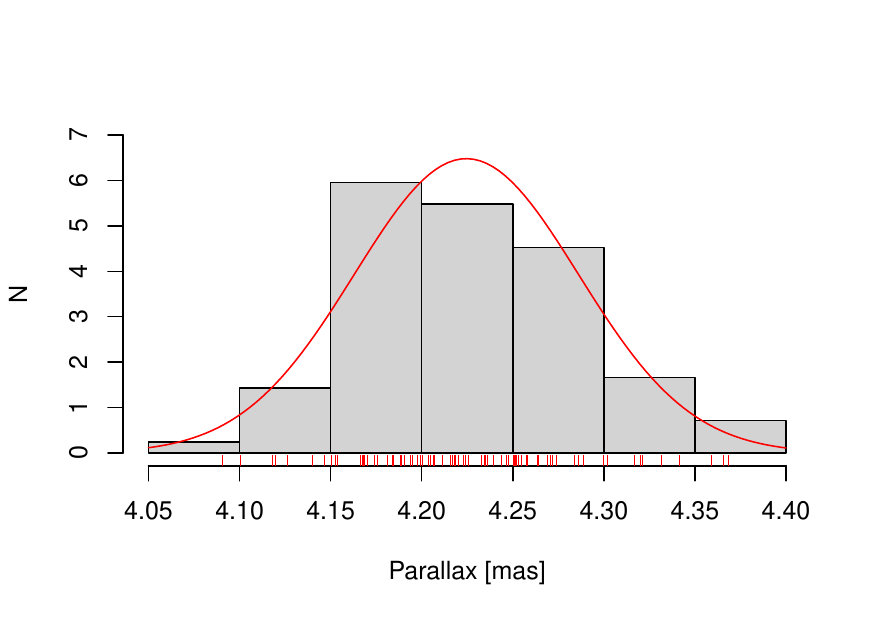}
\caption{Parallax distribution for Blanco~1.}
             \label{fig:90}
    \end{figure}

               \begin{figure} [htp]
   \centering
   \includegraphics[width=0.9\linewidth]{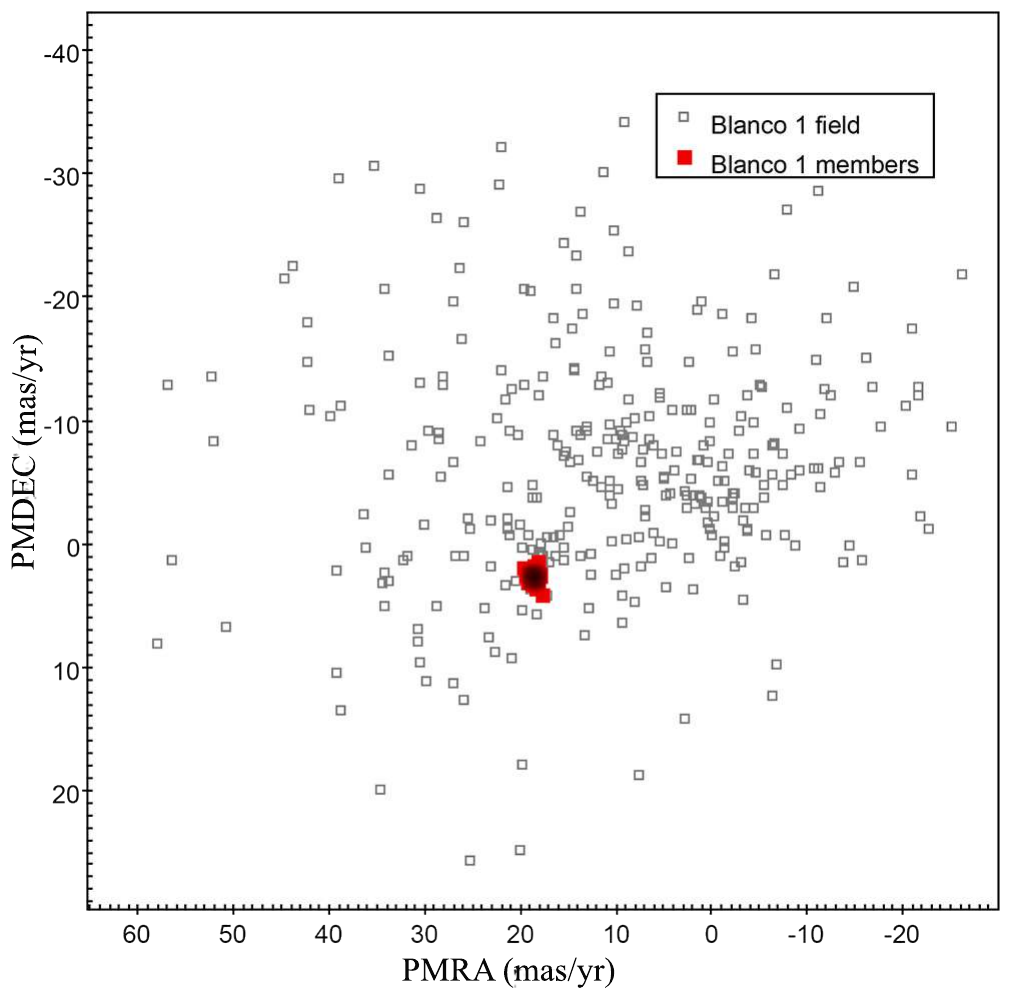}
   \caption{PMs diagram for Blanco~1.}
             \label{fig:91}
    \end{figure}
    
     \begin{figure} [htp]
   \centering
   \includegraphics[width=0.9\linewidth, height=7cm]{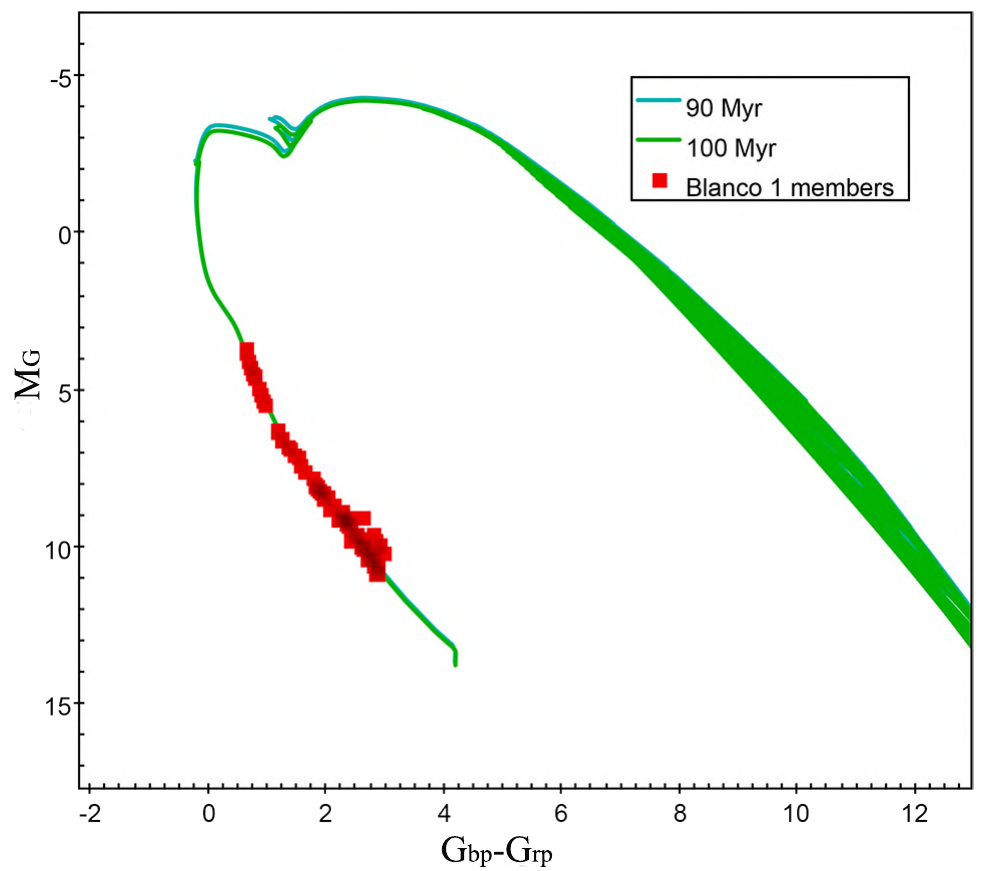}
   \caption{CMD for Blanco~1.}
             \label{fig:92}
    \end{figure}
    
      \begin{figure} [htp]
   \centering
 \includegraphics[width=0.9\linewidth]{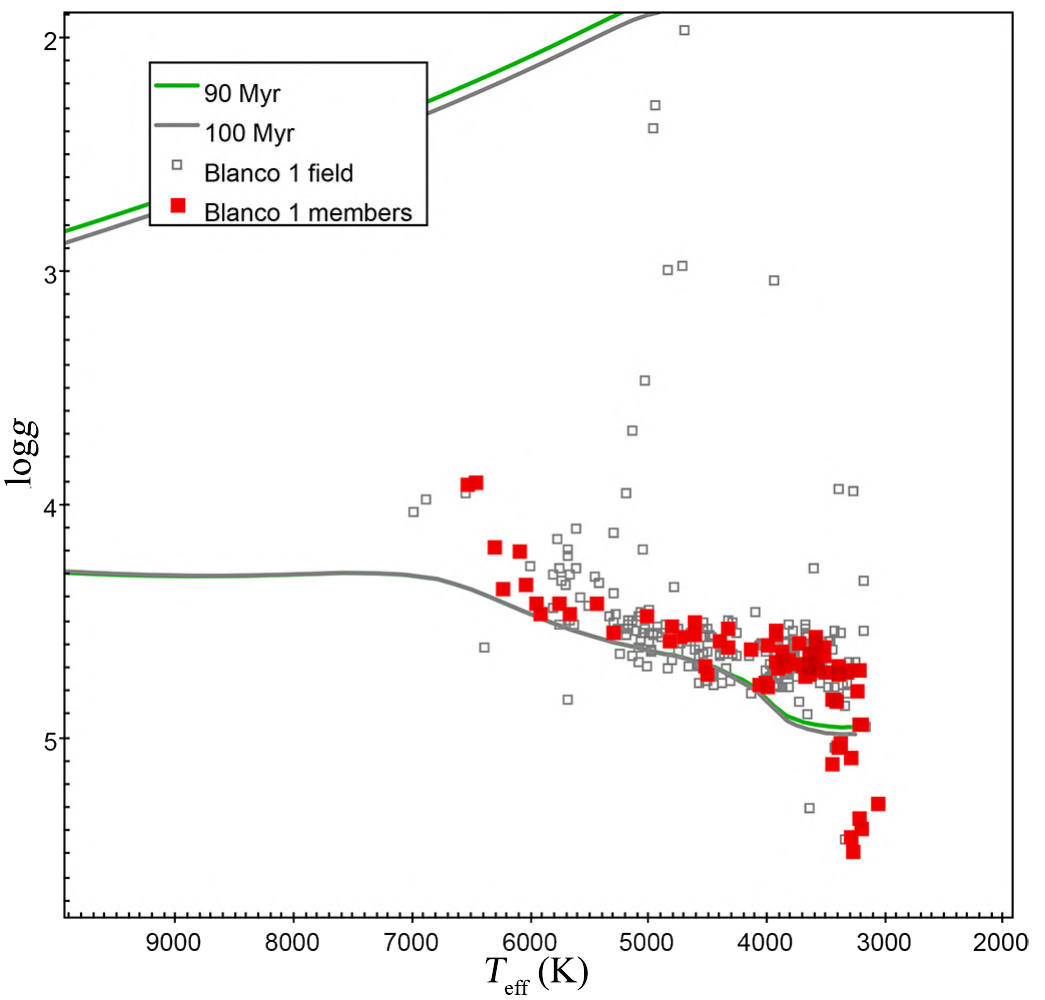} 
\caption{Kiel diagram for Blanco~1.}
             \label{fig:93}
    \end{figure}

  \begin{figure} [htp]
   \centering
 \includegraphics[width=0.9\linewidth]{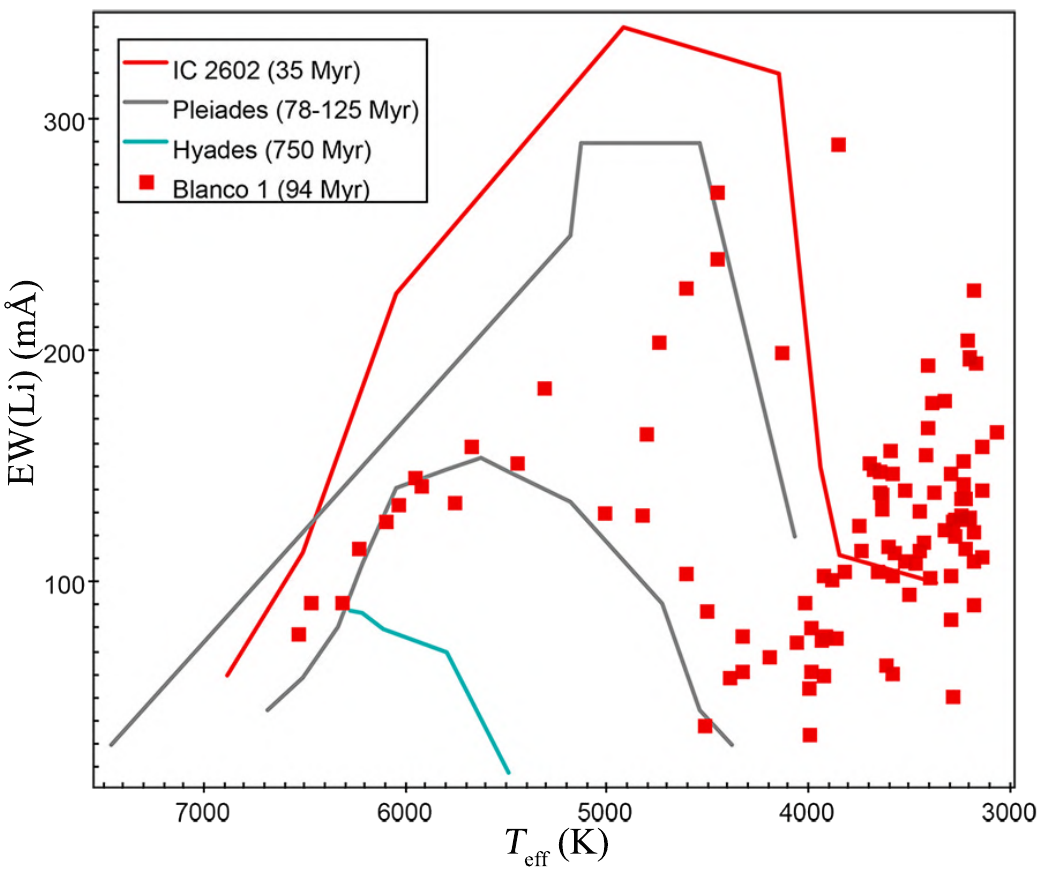} 
\caption{$EW$(Li)-versus-$T_{\rm eff}$ diagram for Blanco~1.}
             \label{fig:94}
    \end{figure}
    
    \clearpage

\subsection{NGC~6067}

 \begin{figure} [htp]
   \centering
\includegraphics[width=0.9\linewidth, height=5cm]{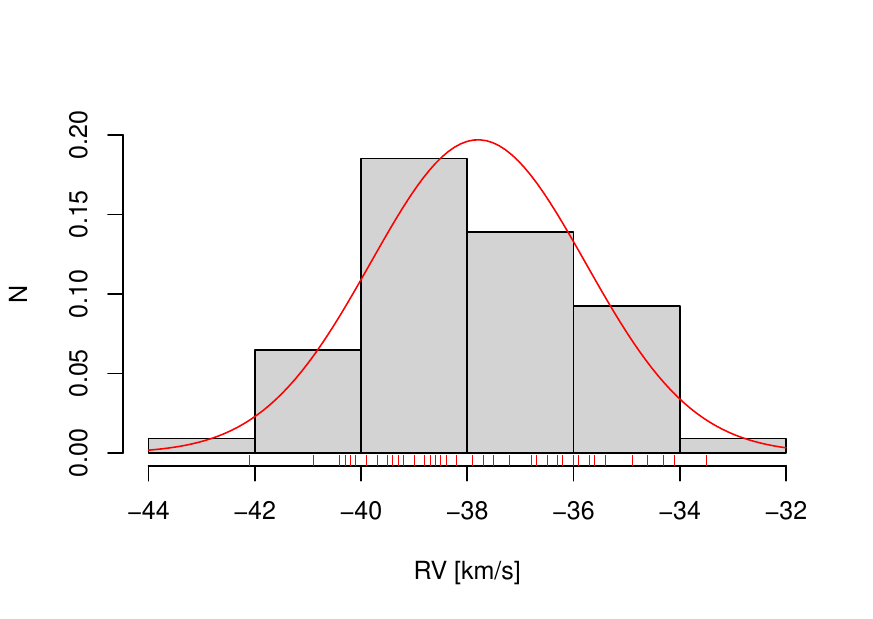}
\caption{$RV$ distribution for NGC~6067.}
             \label{fig:95}
    \end{figure}
    
           \begin{figure} [htp]
   \centering
\includegraphics[width=0.9\linewidth, height=5cm]{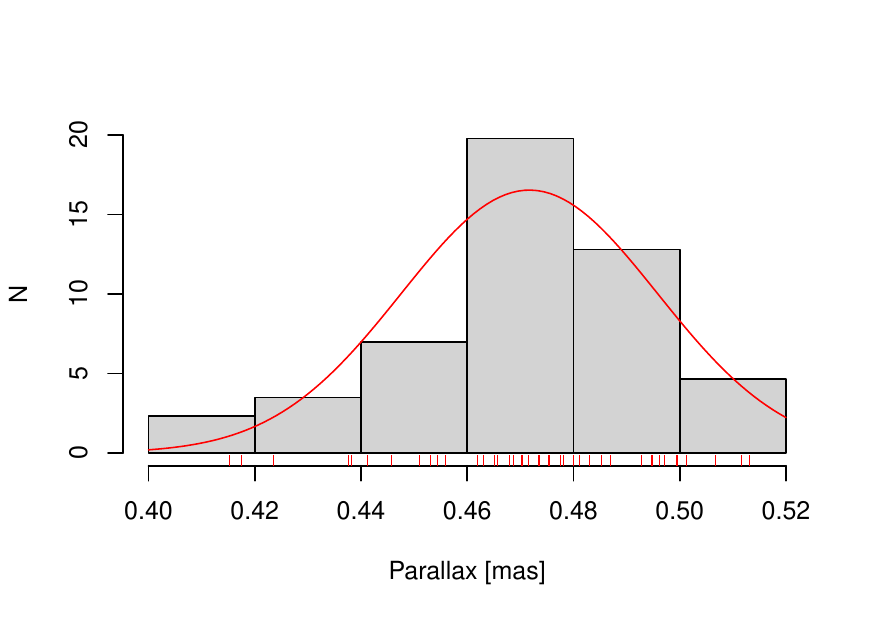}
\caption{Parallax distribution for NGC~6067.}
             \label{fig:96}
    \end{figure}

               \begin{figure} [htp]
   \centering
   \includegraphics[width=0.8\linewidth]{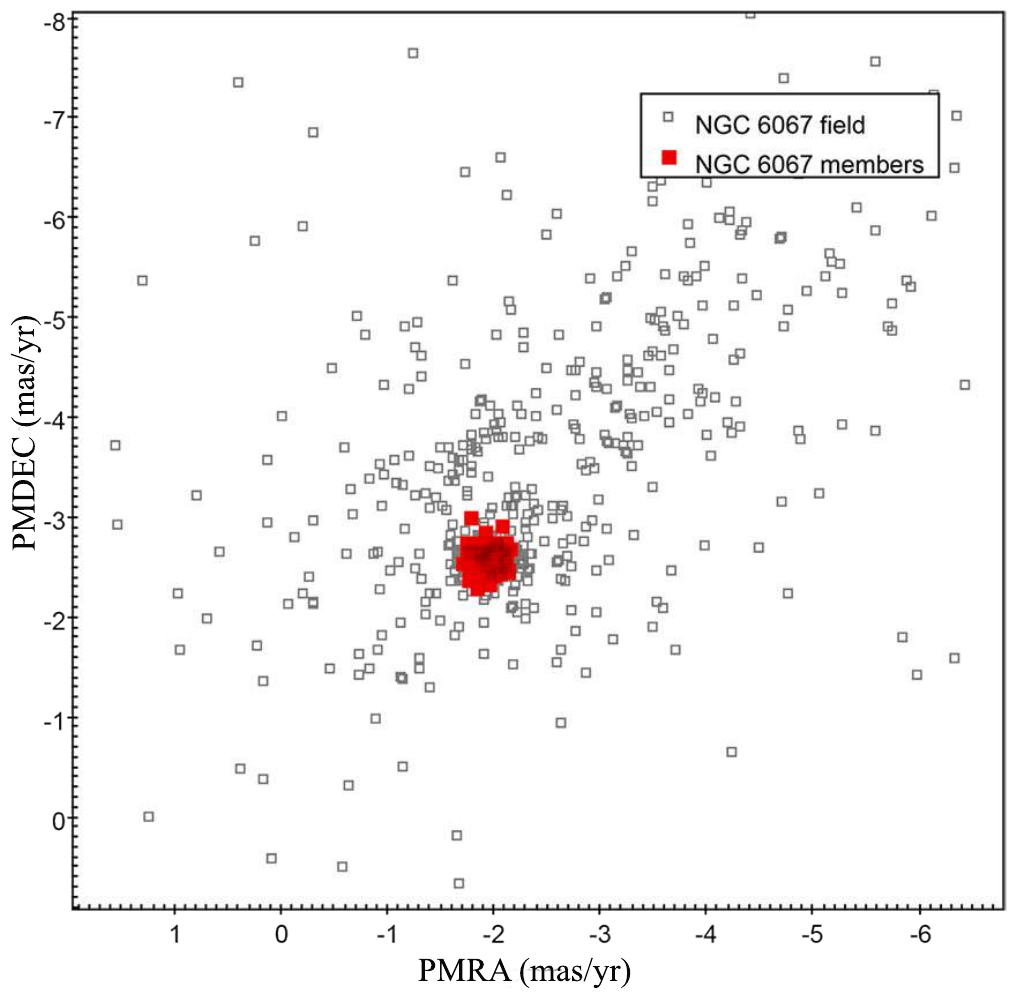}
   \caption{PMs diagram for NGC~6067.}
             \label{fig:97}
    \end{figure}
    
     \begin{figure} [htp]
   \centering
   \includegraphics[width=0.9\linewidth, height=7cm]{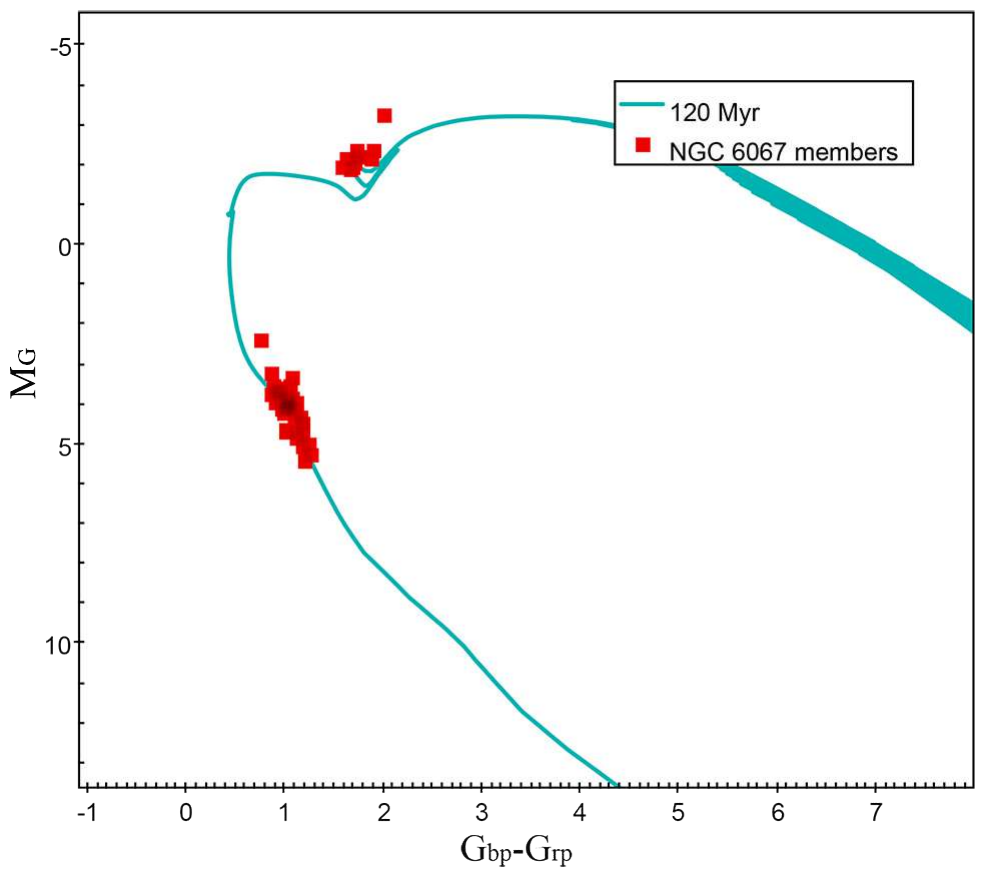}
   \caption{CMD for NGC~6067.}
             \label{fig:98}
    \end{figure}
    
      \begin{figure} [htp]
   \centering
 \includegraphics[width=0.8\linewidth]{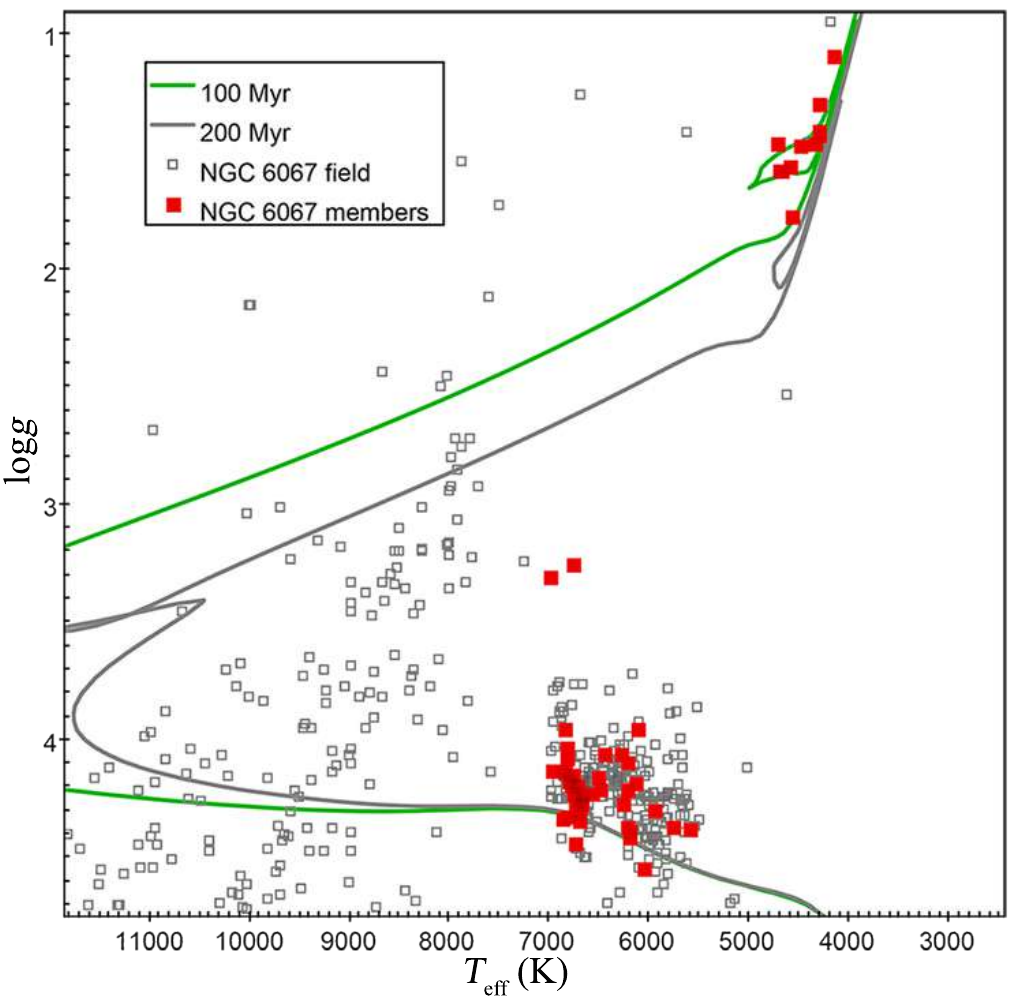} 
\caption{Kiel diagram for NGC~6067.}
             \label{fig:99}
    \end{figure}

  \begin{figure} [htp]
   \centering
 \includegraphics[width=0.8\linewidth]{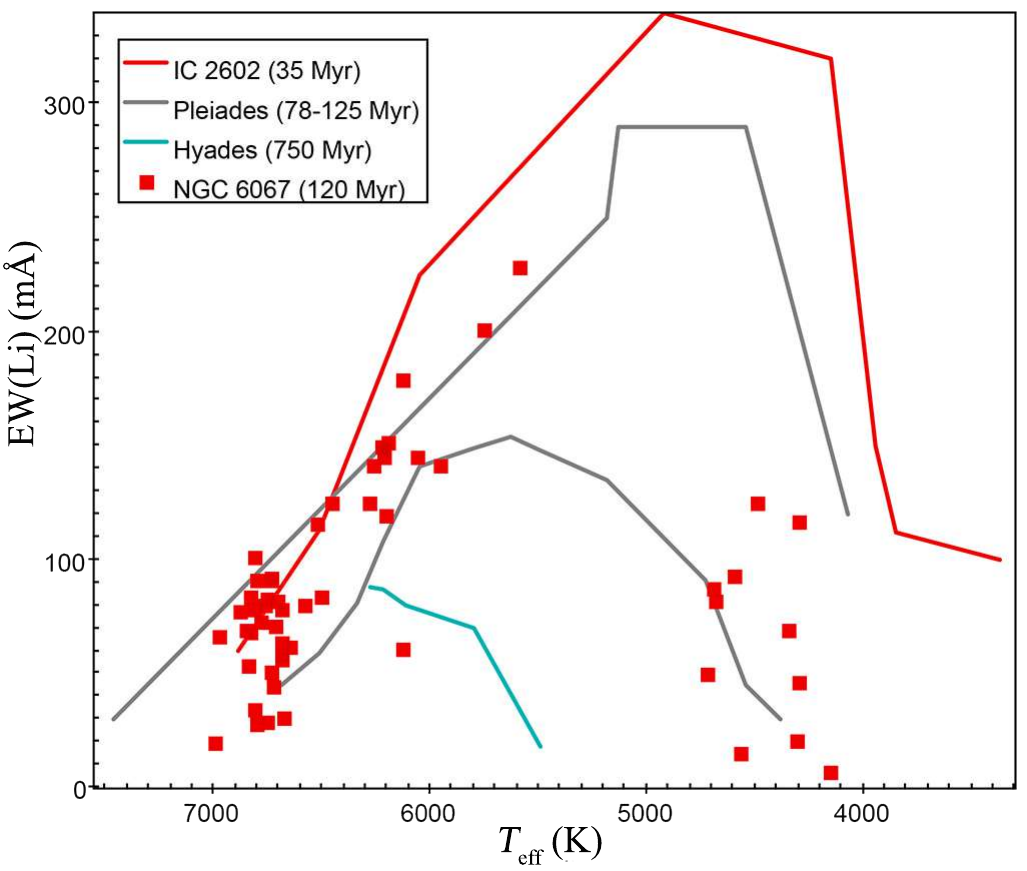} 
\caption{$EW$(Li)-versus-$T_{\rm eff}$ diagram for NGC~6067.}
             \label{fig:100}
    \end{figure}
    
    \clearpage

\subsection{NGC~6649}

               \begin{figure} [htp]
   \centering
   \includegraphics[width=0.9\linewidth]{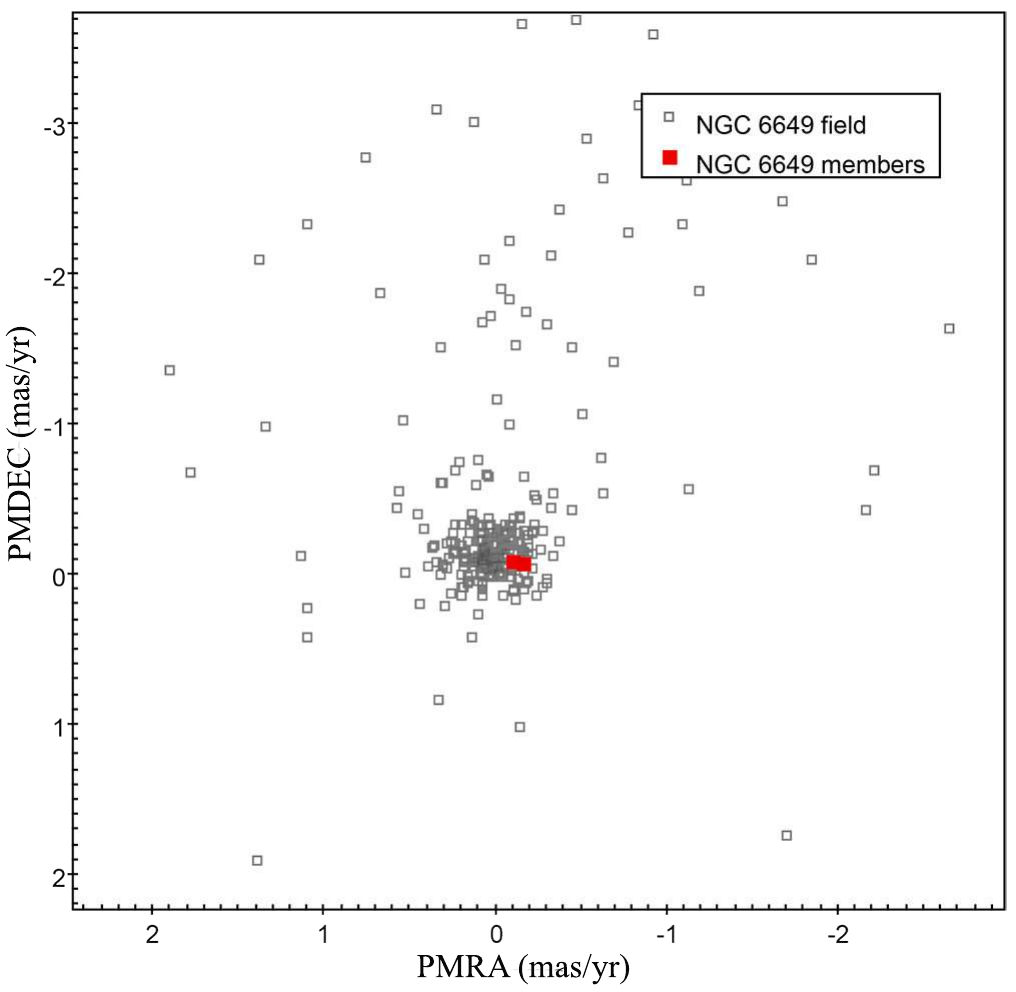}
   \caption{PMs diagram for NGC~6649.}
             \label{fig:101}
    \end{figure}
    
     \begin{figure} [htp]
   \centering
   \includegraphics[width=0.9\linewidth, height=7cm]{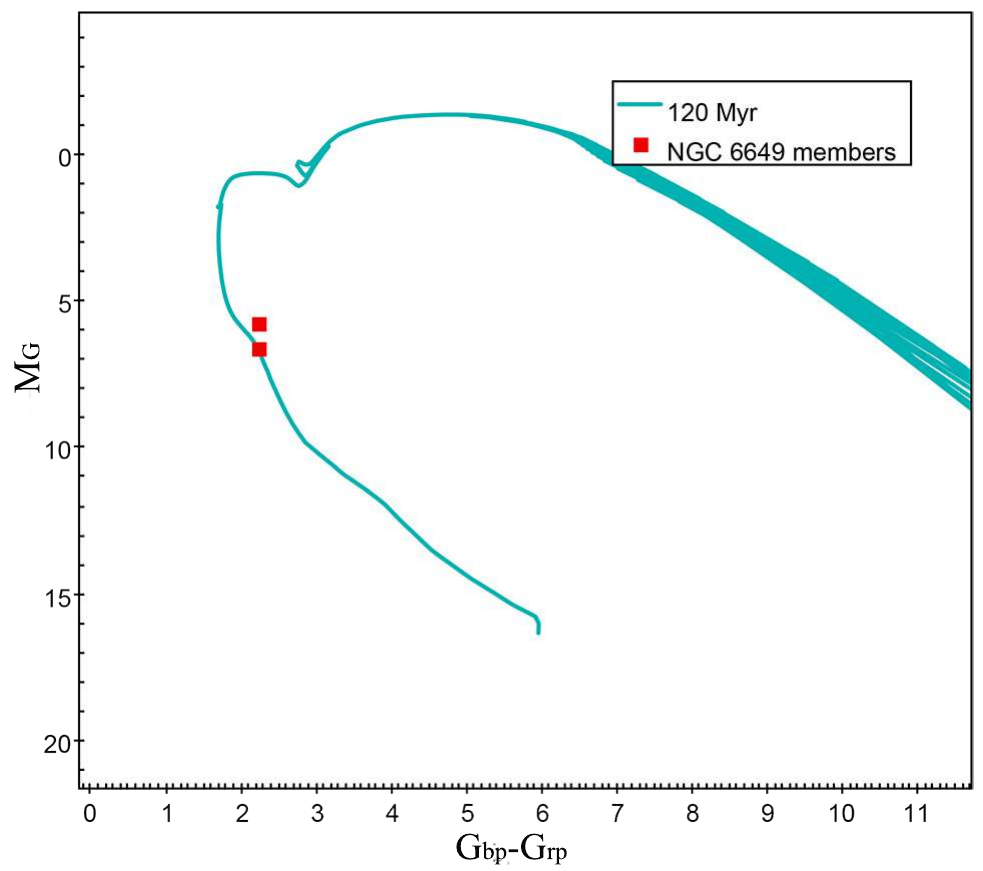}
   \caption{CMD for NGC~6649.}
             \label{fig:102}
    \end{figure}
    
      \begin{figure} [htp]
   \centering
 \includegraphics[width=0.9\linewidth]{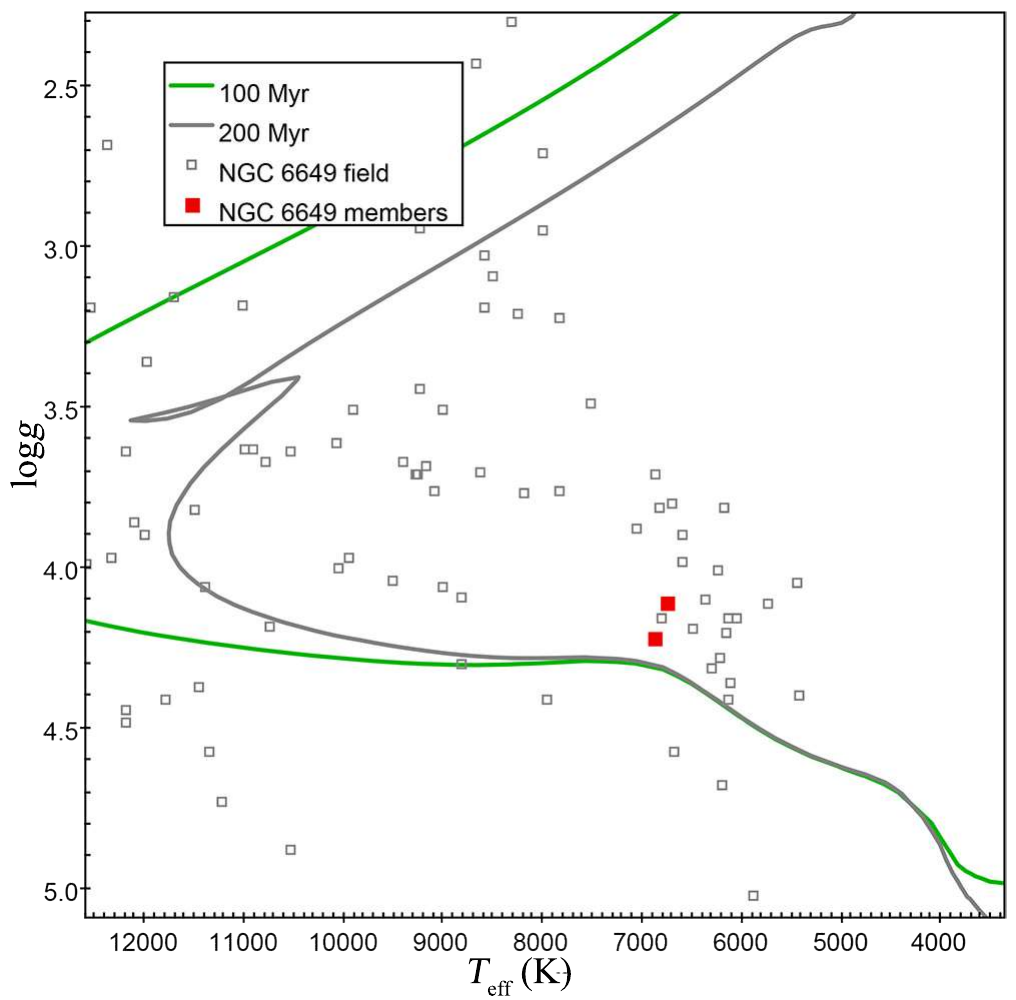} 
\caption{Kiel diagram for NGC~6649.}
             \label{fig:103}
    \end{figure}

  \begin{figure} [htp]
   \centering
 \includegraphics[width=0.9\linewidth]{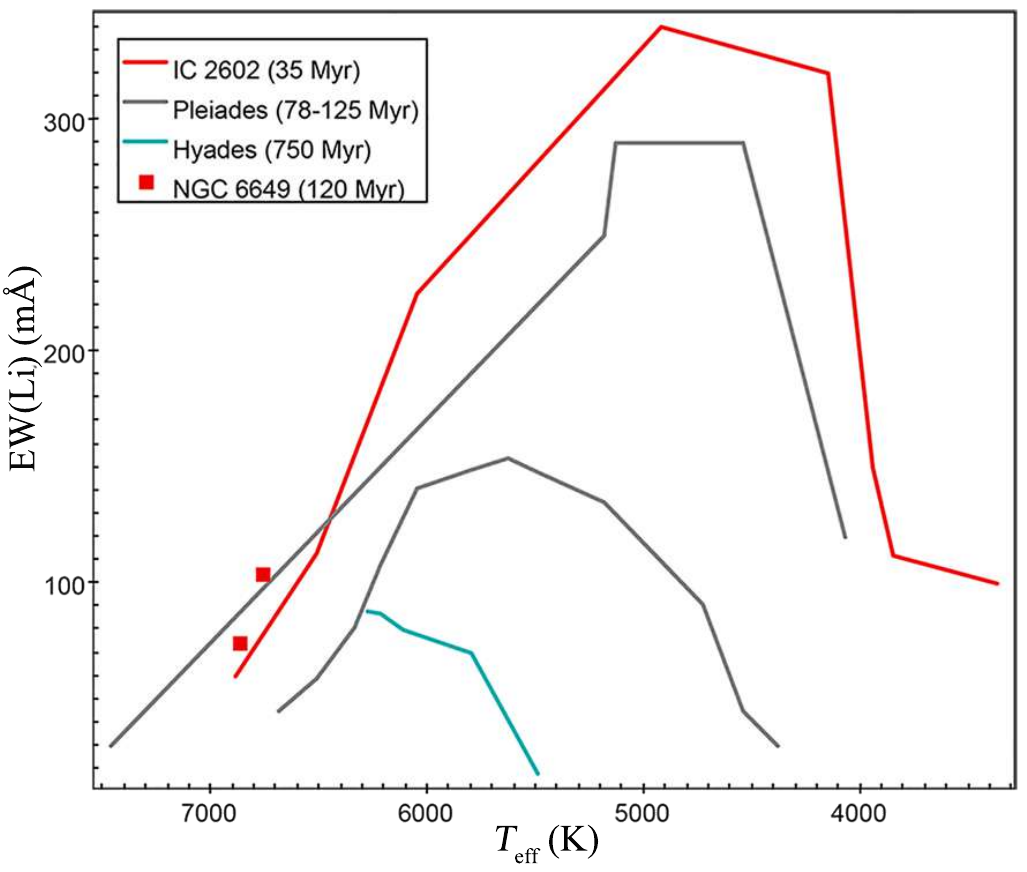} 
\caption{$EW$(Li)-versus-$T_{\rm eff}$ diagram for NGC~6649.}
             \label{fig:104}
    \end{figure}
    
    \clearpage

\subsection{NGC~2516}

 \begin{figure} [htp]
   \centering
\includegraphics[width=0.9\linewidth, height=5cm]{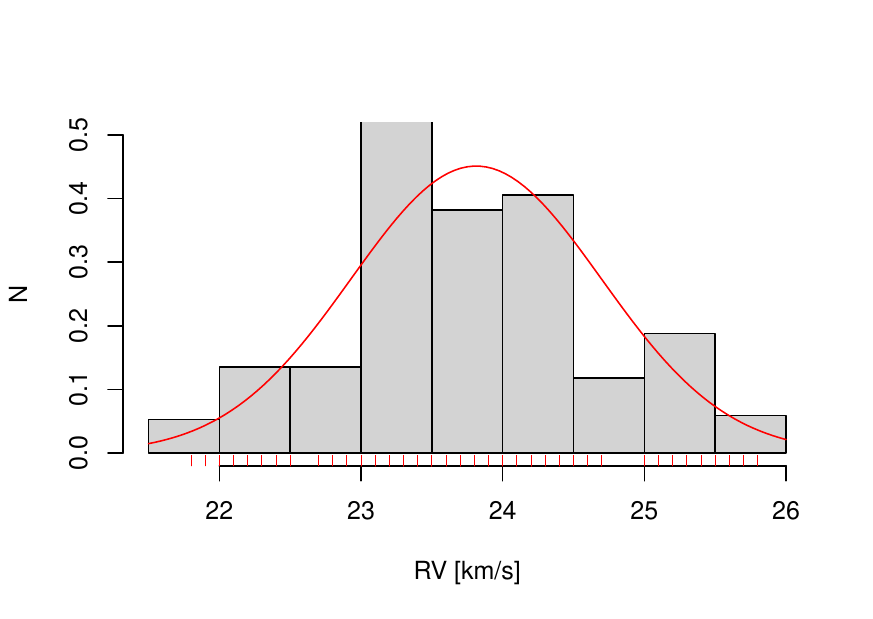}
\caption{$RV$ distribution for NGC~2516.}
             \label{fig:105}
    \end{figure}
    
           \begin{figure} [htp]
   \centering
\includegraphics[width=0.9\linewidth, height=5cm]{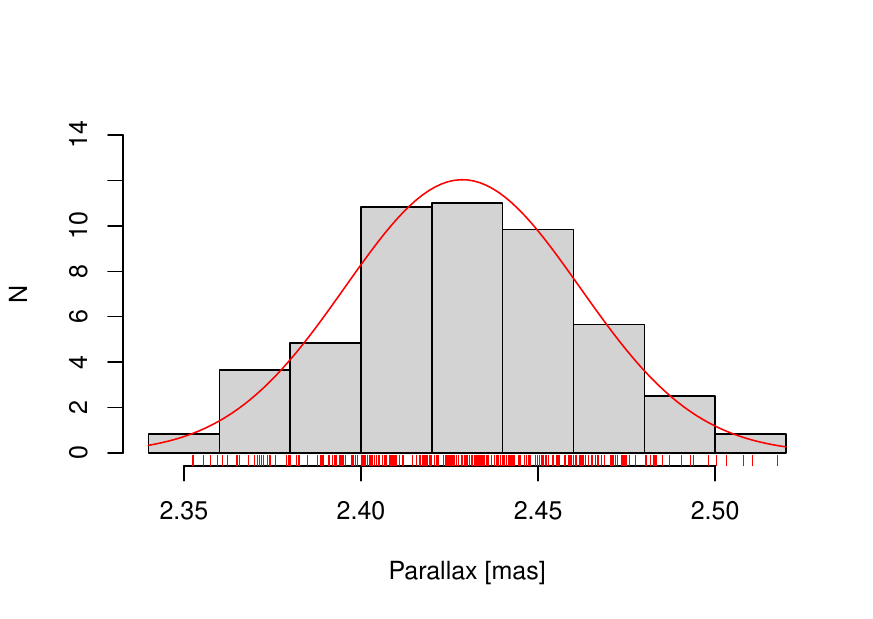}
\caption{Parallax distribution for NGC~2516.}
             \label{fig:106}
    \end{figure}

               \begin{figure} [htp]
   \centering
   \includegraphics[width=0.9\linewidth]{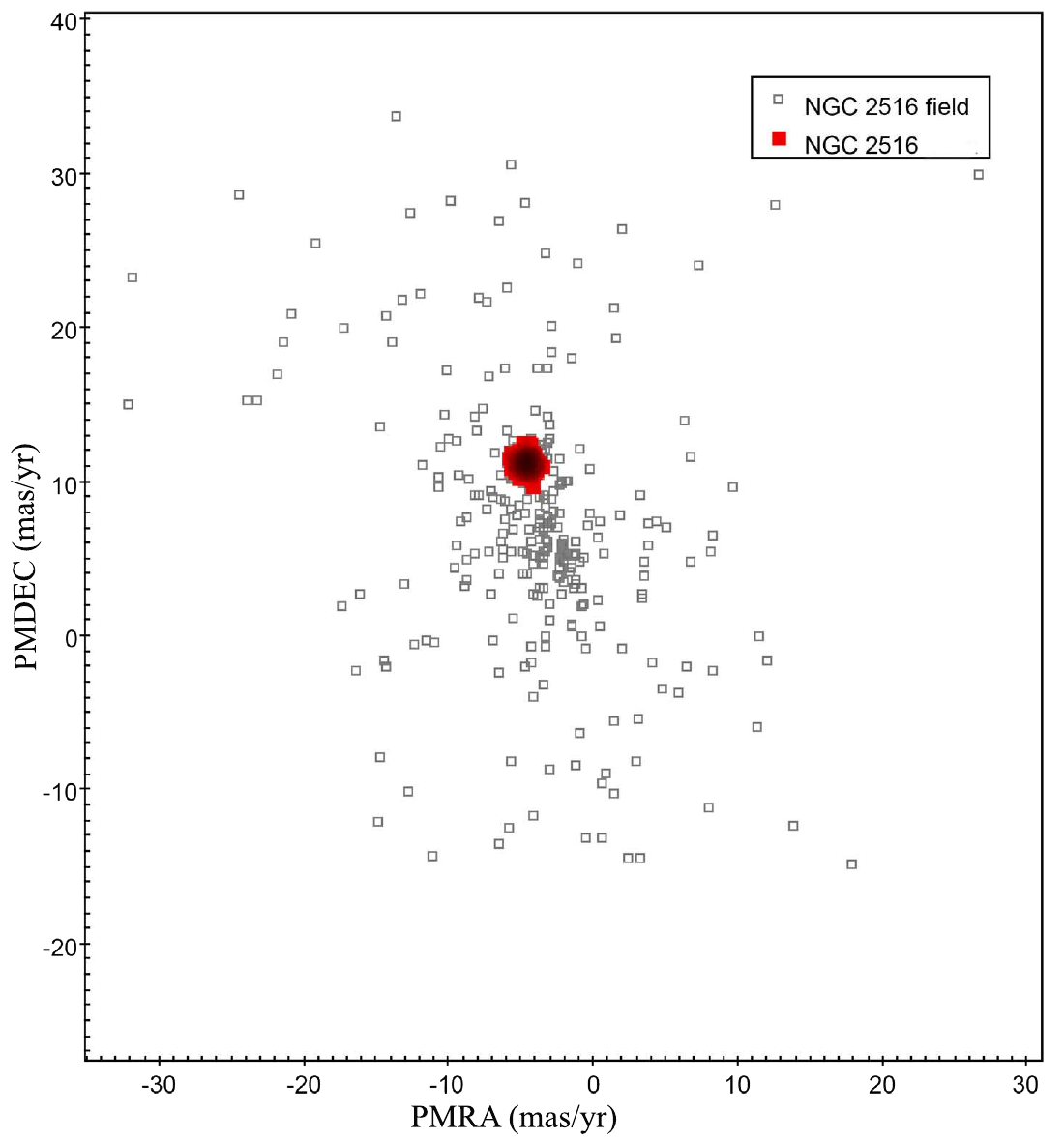}
   \caption{PMs diagram for NGC~2516.}
             \label{fig:106}
    \end{figure}
    
     \begin{figure} [htp]
   \centering
   \includegraphics[width=0.8\linewidth, height=7cm]{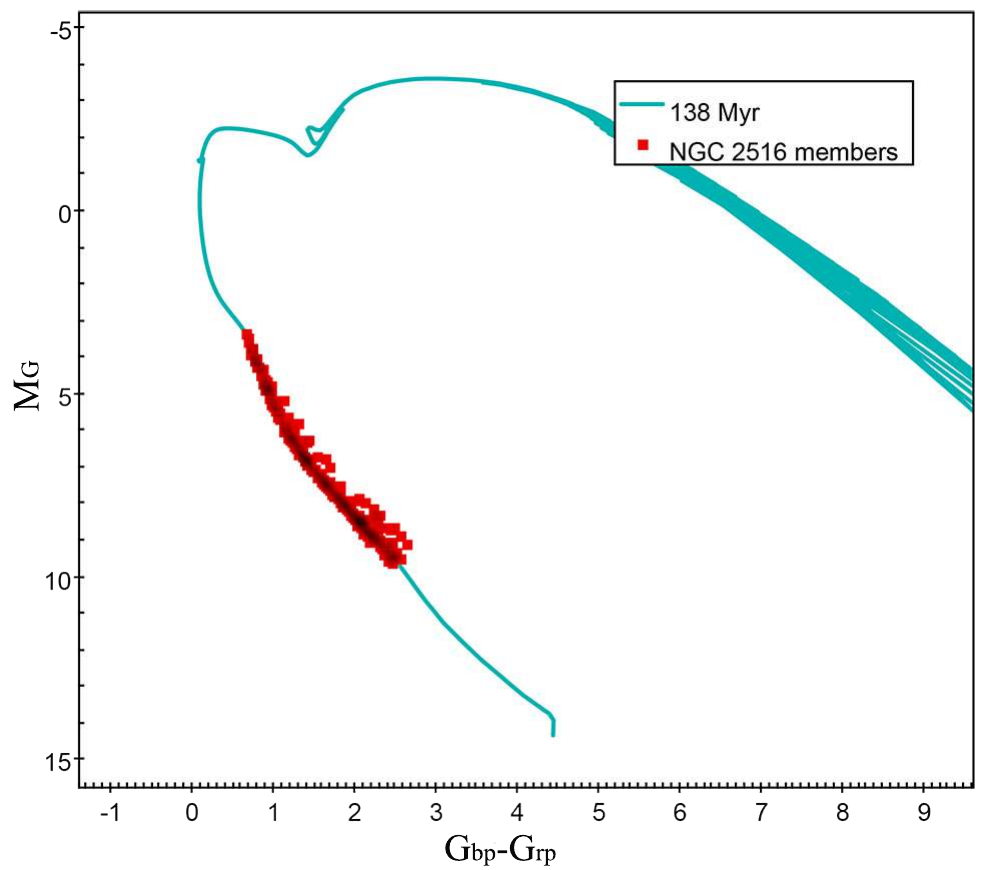}
   \caption{CMD for NGC~2516.}
             \label{fig:107}
    \end{figure}
    
      \begin{figure} [htp]
   \centering
 \includegraphics[width=0.8\linewidth]{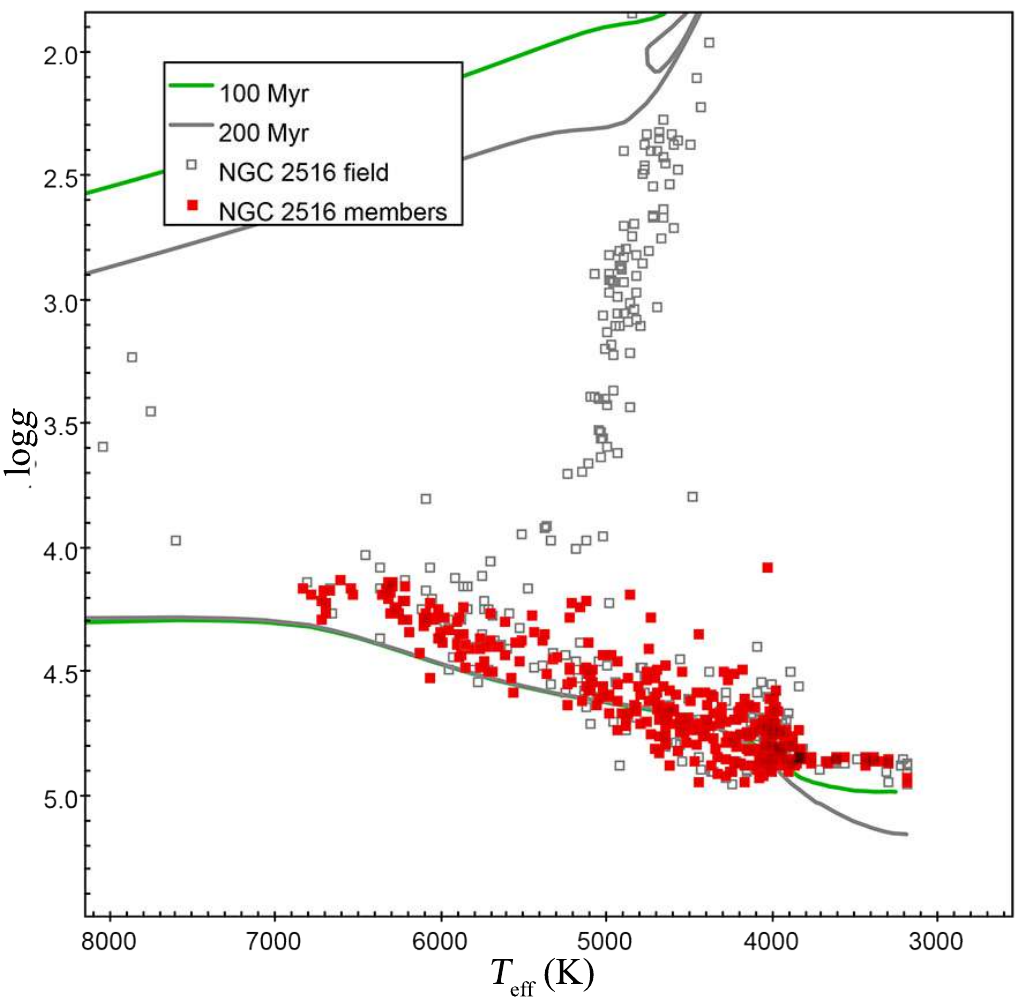} 
\caption{Kiel diagram for NGC~2516.}
             \label{fig:108}
    \end{figure}

  \begin{figure} [htp]
   \centering
 \includegraphics[width=0.8\linewidth]{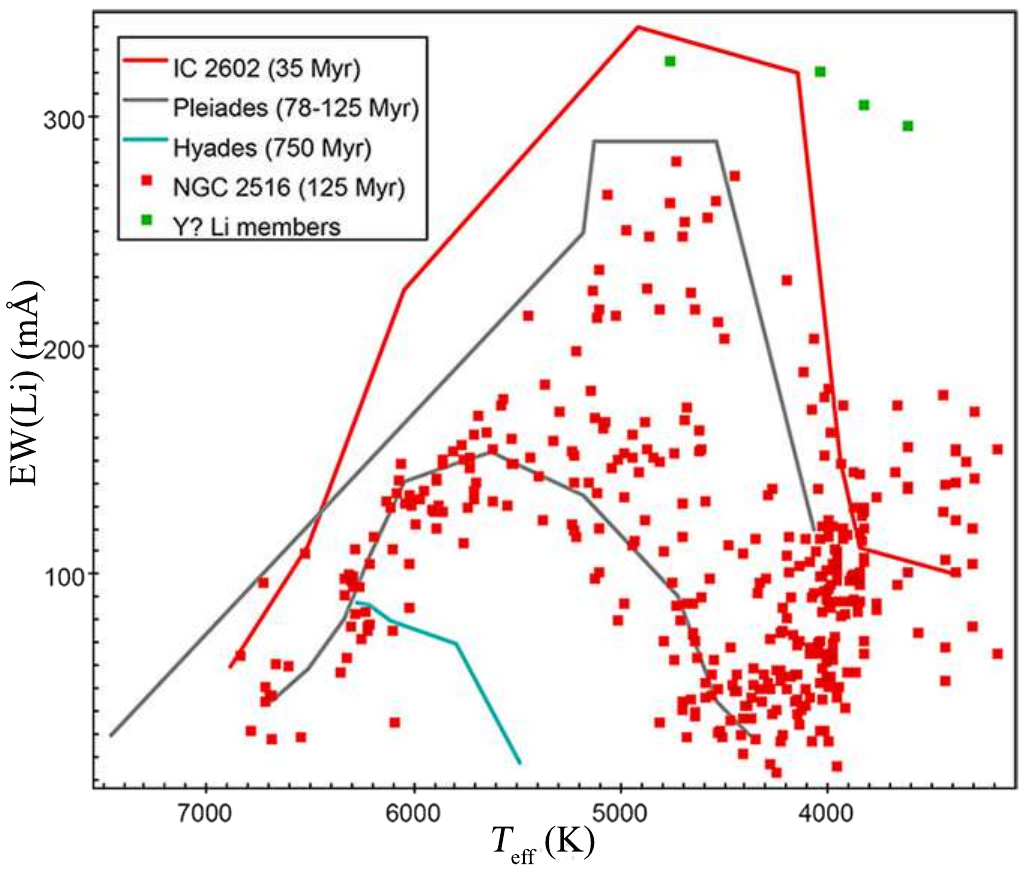} 
\caption{$EW$(Li)-versus-$T_{\rm eff}$ diagram for NGC~2516.}
             \label{fig:109}
    \end{figure}
    
    \clearpage

\subsection{NGC~6709}

 \begin{figure} [htp]
   \centering
\includegraphics[width=0.9\linewidth, height=5cm]{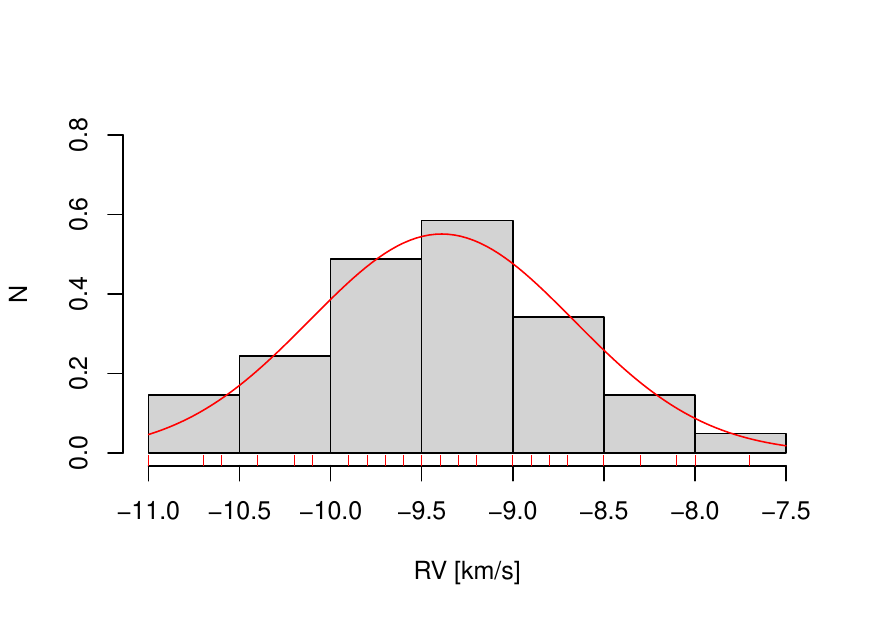}
\caption{$RV$ distribution for NGC~6709.}
             \label{fig:110}
    \end{figure}
    
           \begin{figure} [htp]
   \centering
\includegraphics[width=0.9\linewidth, height=5cm]{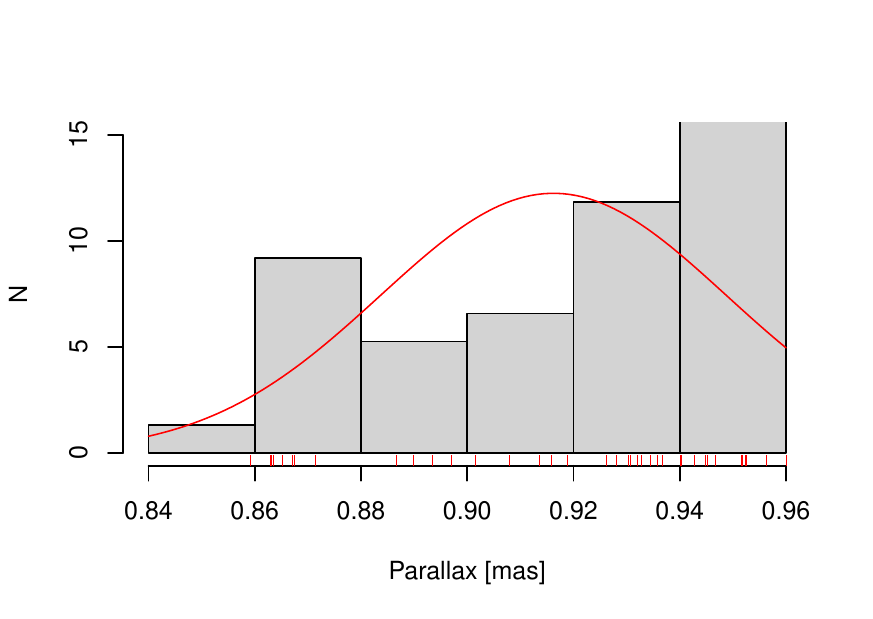}
\caption{Parallax distribution for NGC~6709.}
             \label{fig:111}
    \end{figure}

               \begin{figure} [htp]
   \centering
   \includegraphics[width=0.9\linewidth]{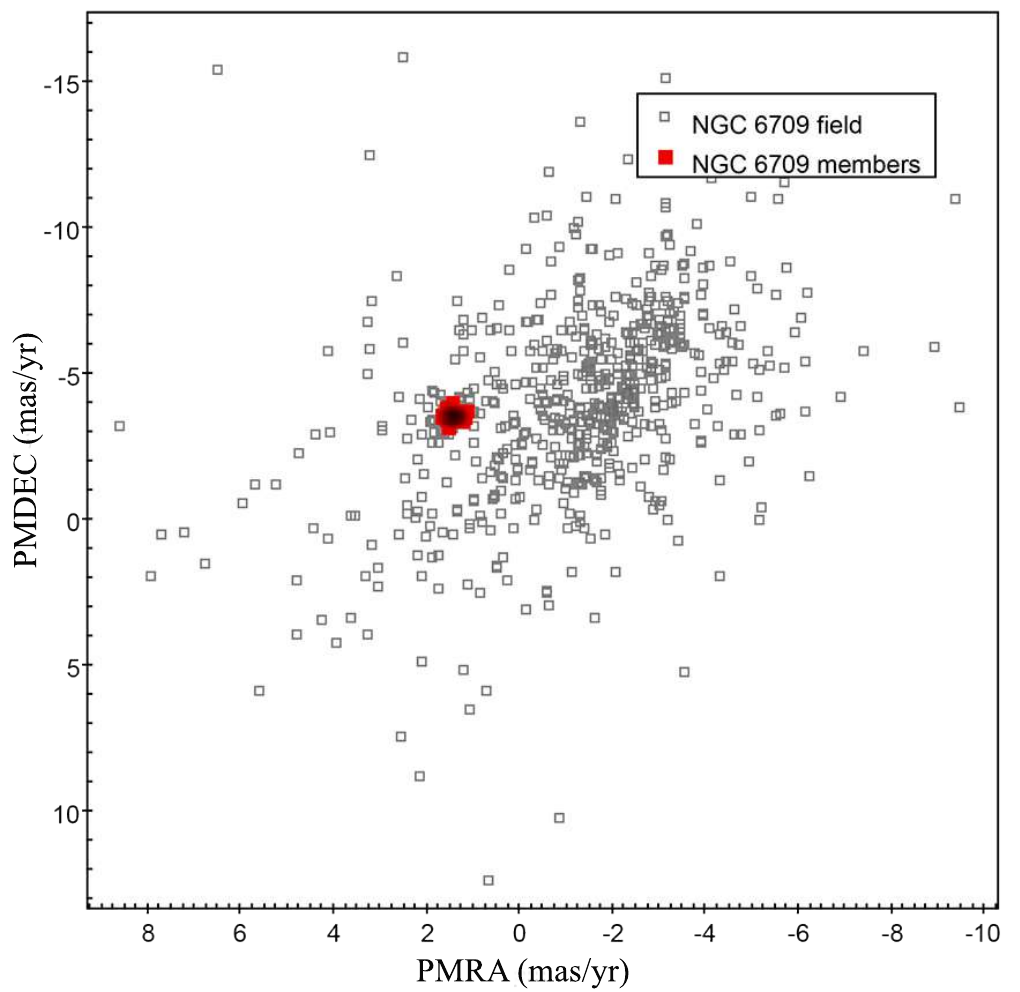}
   \caption{PMs diagram for NGC~6709.}
             \label{fig:112}
    \end{figure}
    
     \begin{figure} [htp]
   \centering
   \includegraphics[width=0.8\linewidth, height=7cm]{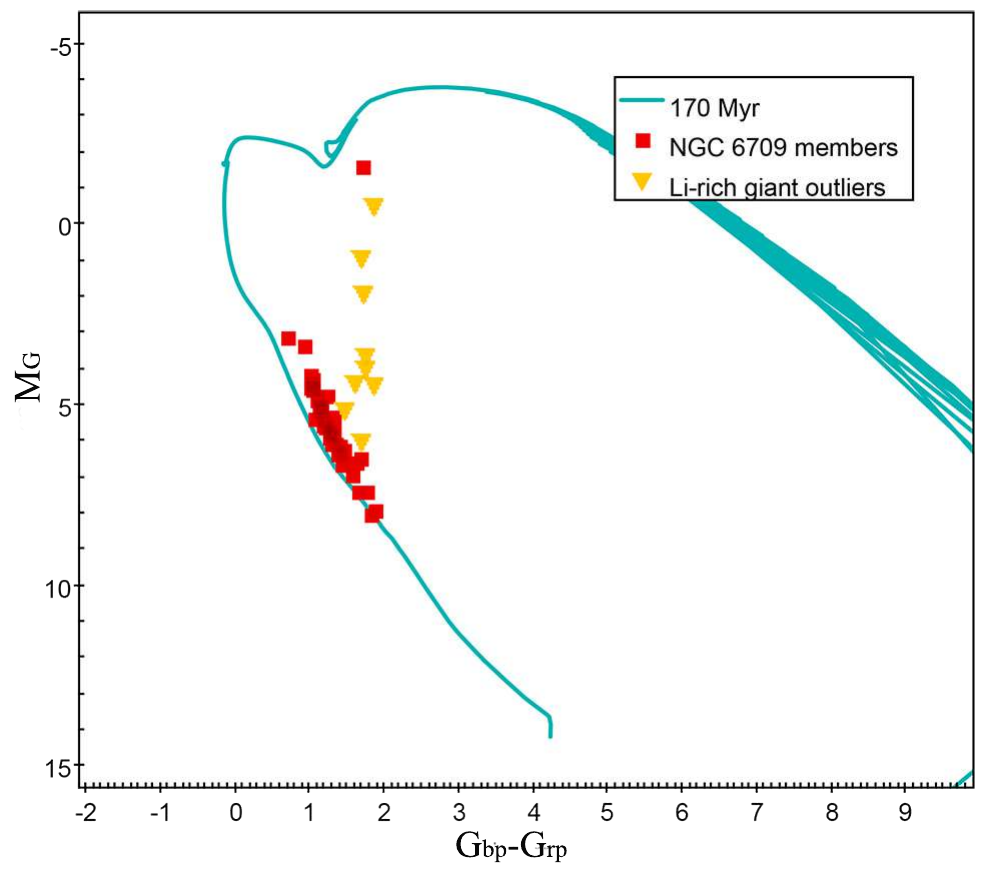}
   \caption{CMD for NGC~6709.}
             \label{fig:113}
    \end{figure}
    
      \begin{figure} [htp]
   \centering
 \includegraphics[width=0.8\linewidth]{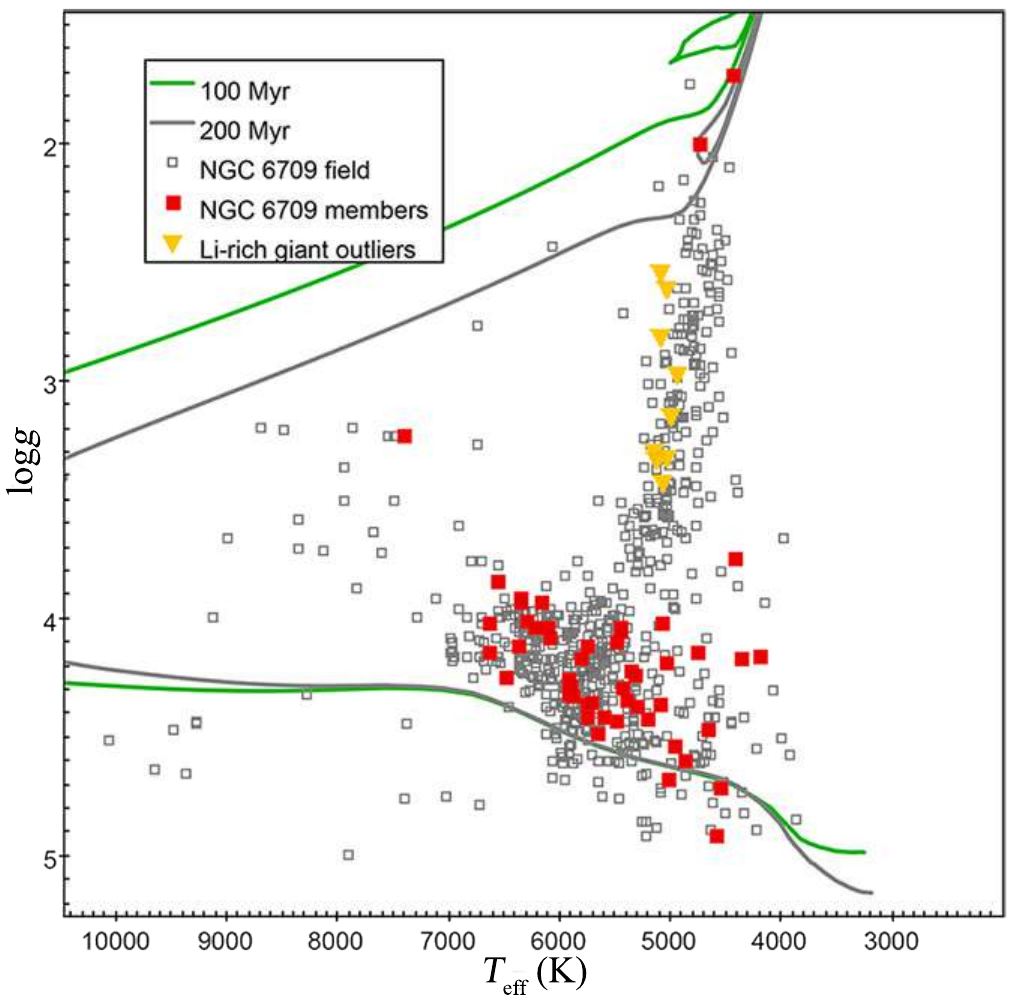} 
\caption{Kiel diagram for NGC~6709.}
             \label{fig:114}
    \end{figure}

  \begin{figure} [htp]
   \centering
 \includegraphics[width=0.8\linewidth]{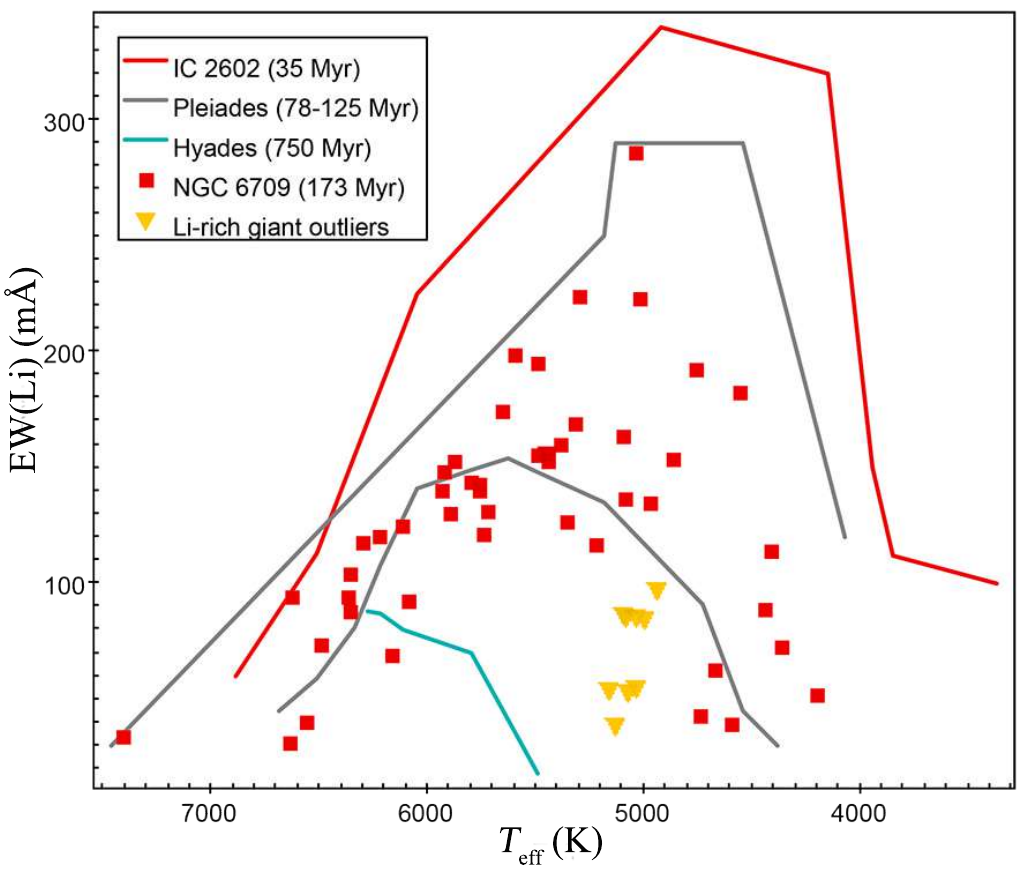} 
\caption{$EW$(Li)-versus-$T_{\rm eff}$ diagram for NGC~6709.}
             \label{fig:115}
    \end{figure}
    
    \clearpage

\subsection{NGC~6259}

 \begin{figure} [htp]
   \centering
\includegraphics[width=0.9\linewidth, height=5cm]{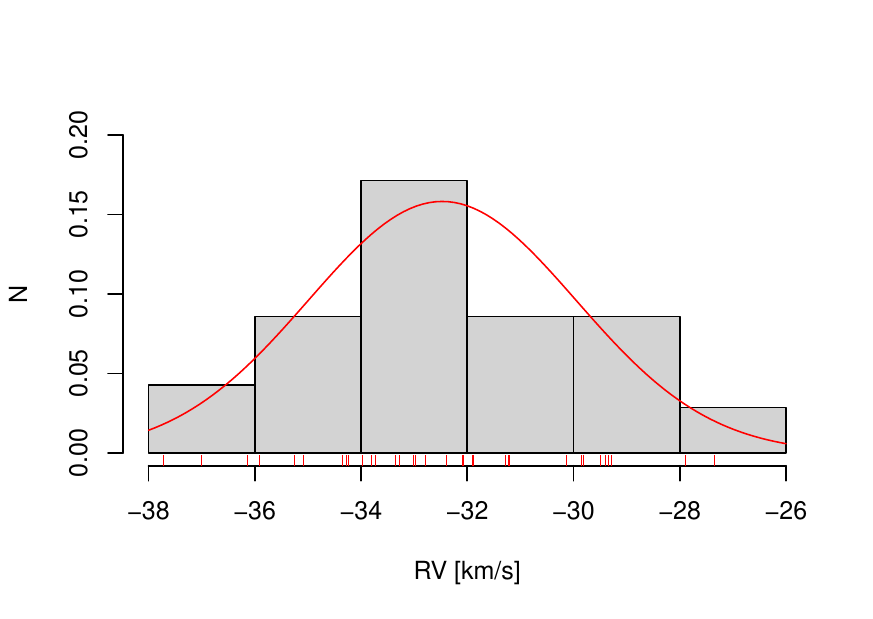}
\caption{$RV$ distribution for NGC~6259.}
             \label{fig:116}
    \end{figure}
    
           \begin{figure} [htp]
   \centering
\includegraphics[width=0.9\linewidth, height=5cm]{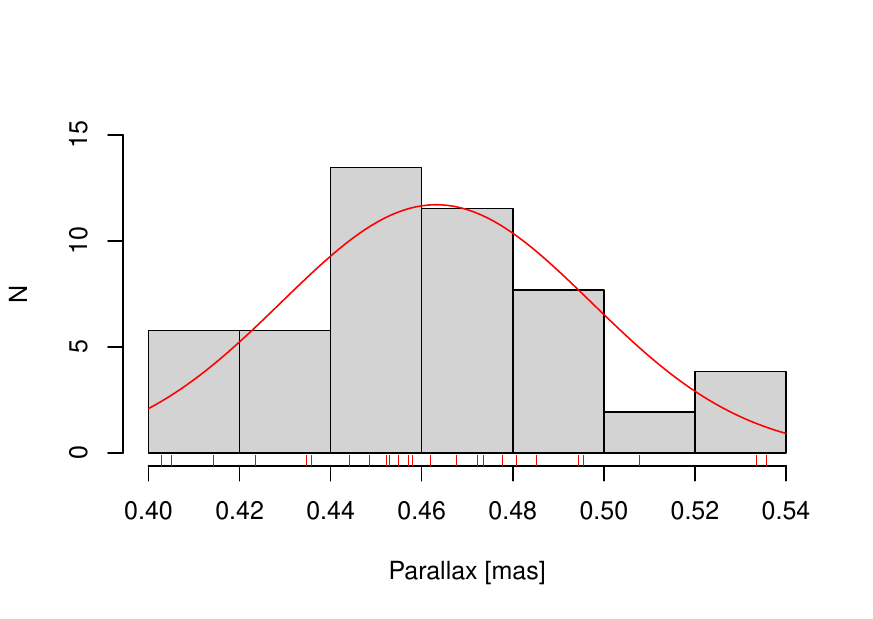}
\caption{Parallax distribution for NGC~6259.}
             \label{fig:117}
    \end{figure}

               \begin{figure} [htp]
   \centering
   \includegraphics[width=0.9\linewidth]{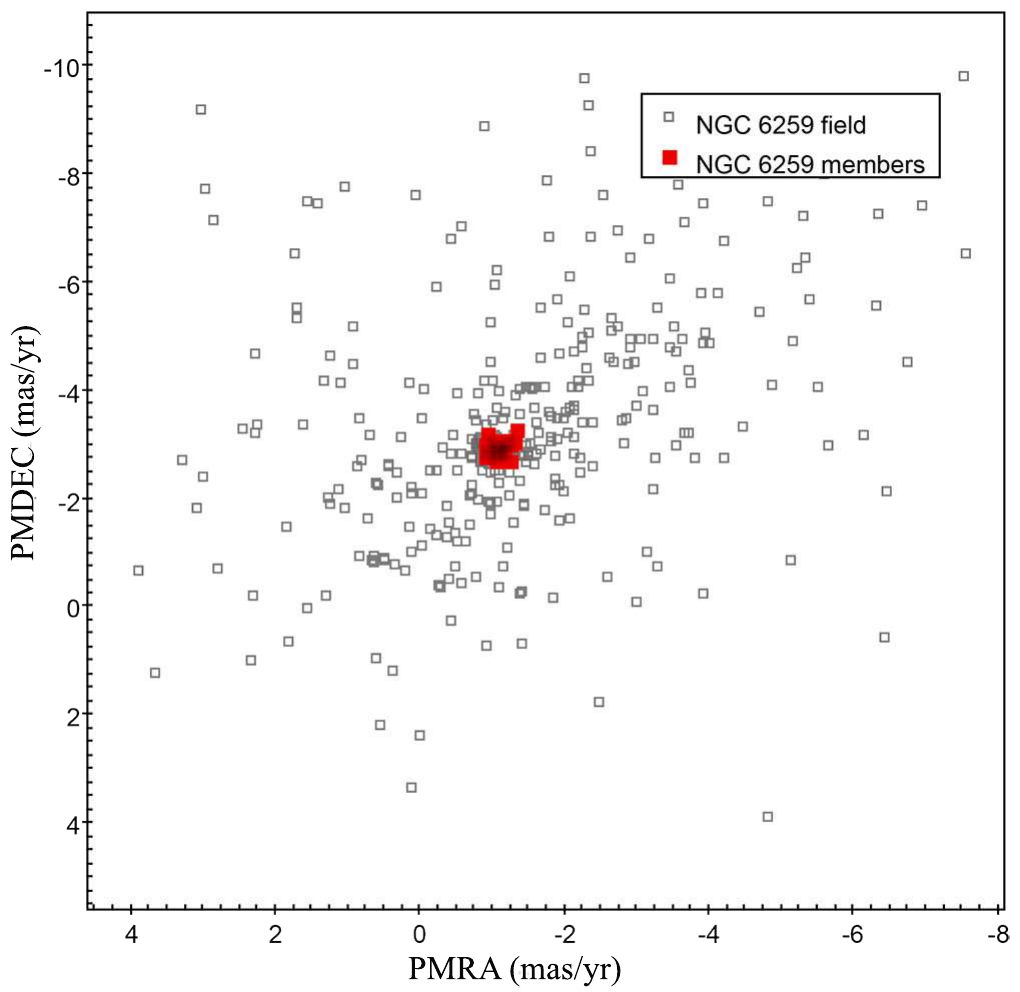}
   \caption{PMs diagram for NGC~6259.}
             \label{fig:118}
    \end{figure}
    
     \begin{figure} [htp]
   \centering
   \includegraphics[width=0.8\linewidth, height=7cm]{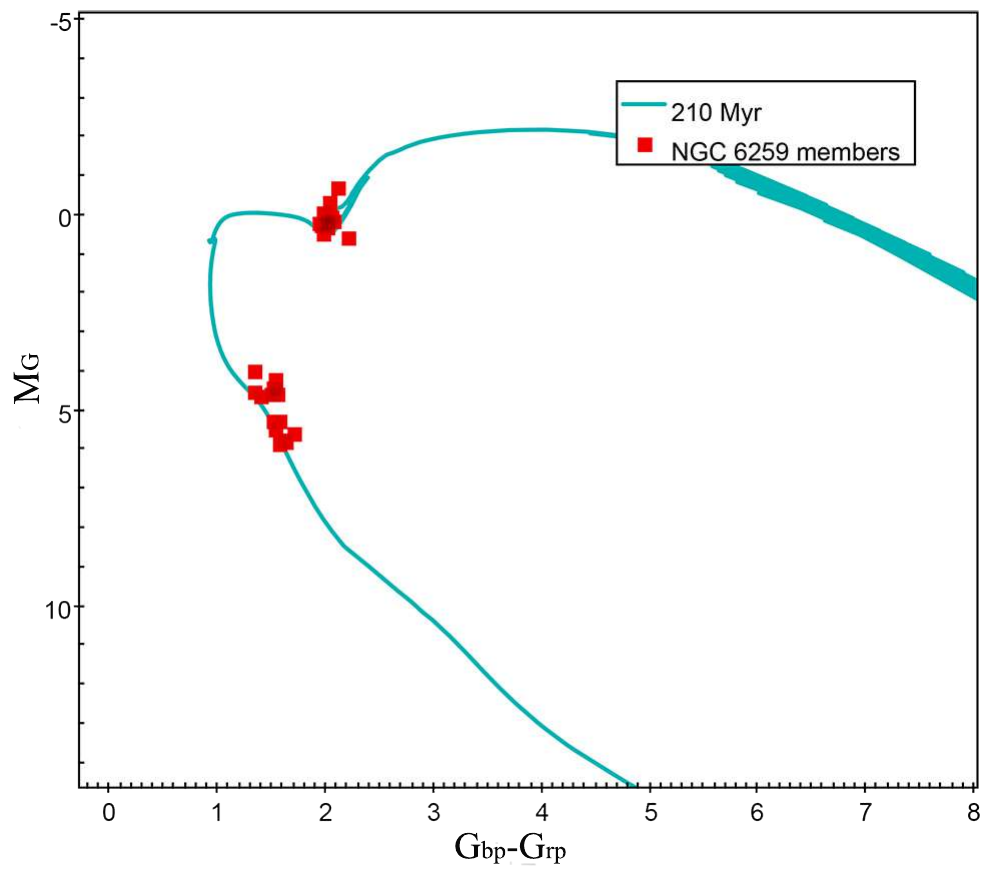}
   \caption{CMD for NGC~6259.}
             \label{fig:119}
    \end{figure}
    
      \begin{figure} [htp]
   \centering
 \includegraphics[width=0.8\linewidth]{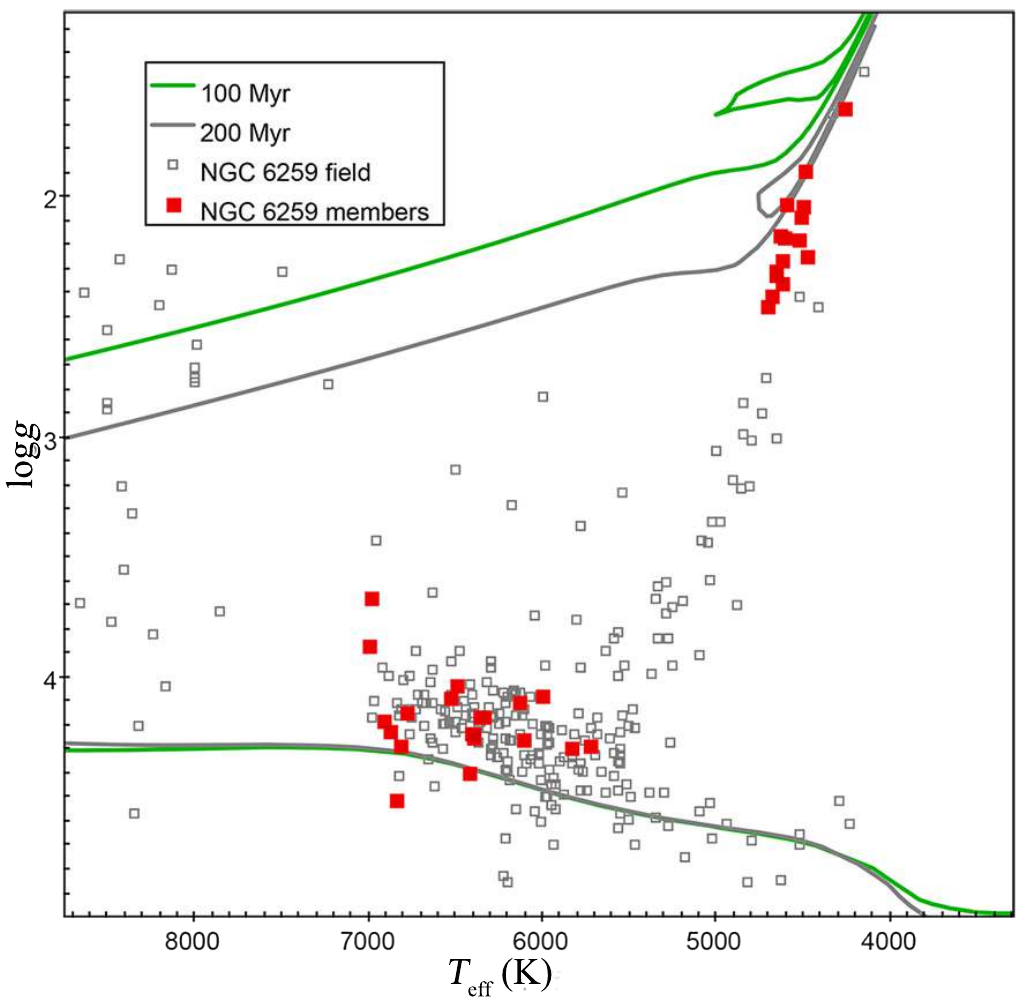} 
\caption{Kiel diagram for NGC~6259.}
             \label{fig:120}
    \end{figure}

  \begin{figure} [htp]
   \centering
 \includegraphics[width=0.8\linewidth]{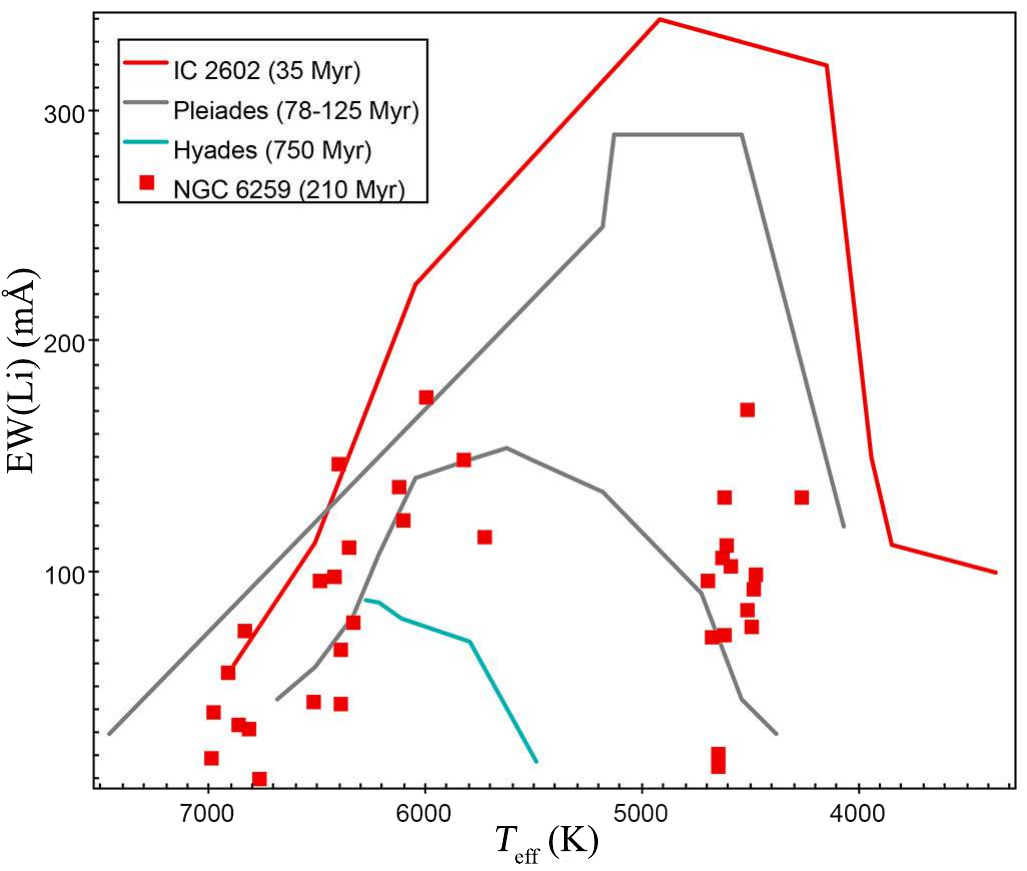} 
\caption{$EW$(Li)-versus-$T_{\rm eff}$ diagram for NGC~6259.}
             \label{fig:121}
    \end{figure}
    
    \clearpage

\subsection{NGC~6705}

 \begin{figure} [htp]
   \centering
\includegraphics[width=0.9\linewidth, height=5cm]{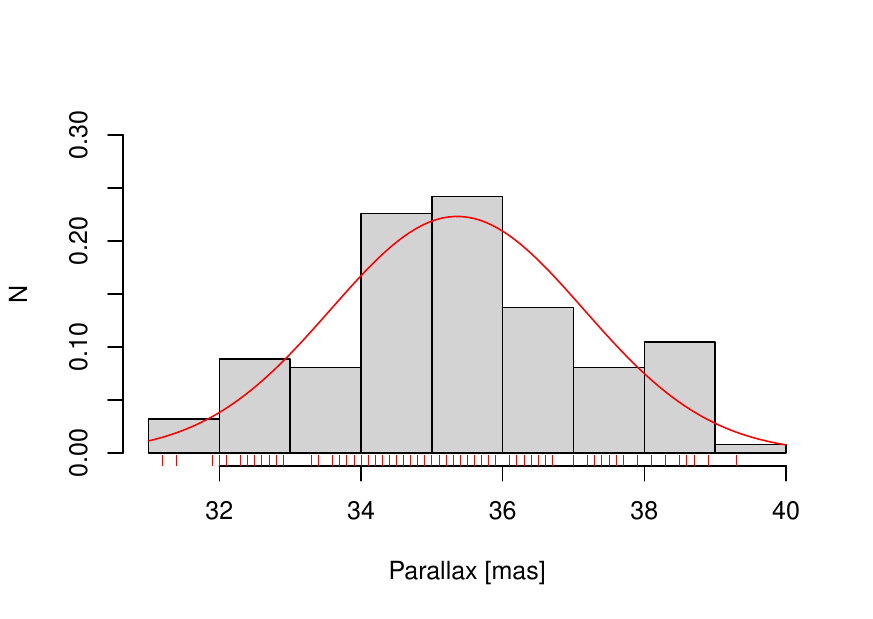}
\caption{$RV$ distribution for NGC~6705.}
             \label{fig:122}
    \end{figure}
    
           \begin{figure} [htp]
   \centering
\includegraphics[width=0.9\linewidth, height=5cm]{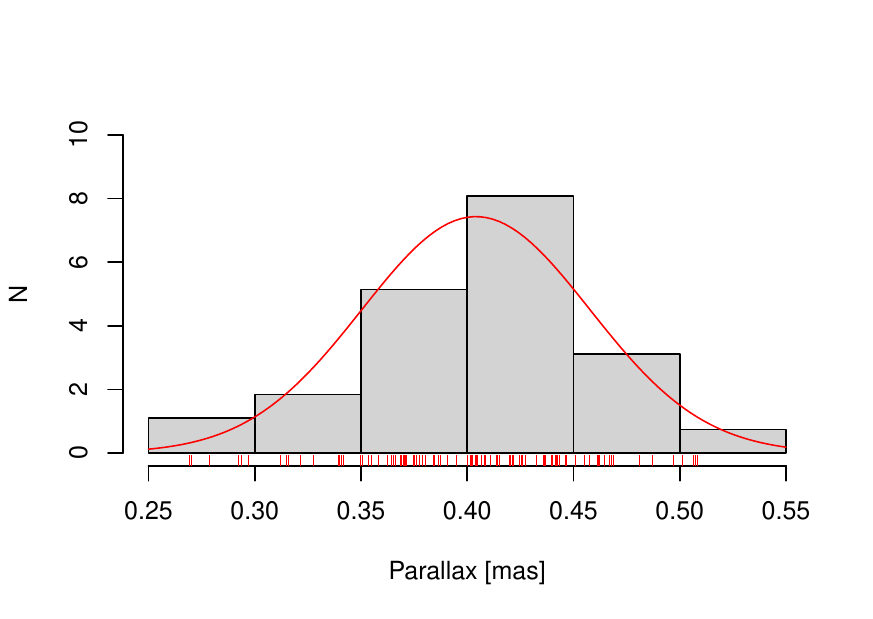}
\caption{Parallax distribution for NGC~6705.}
             \label{fig:123}
    \end{figure}

               \begin{figure} [htp]
   \centering
   \includegraphics[width=0.9\linewidth]{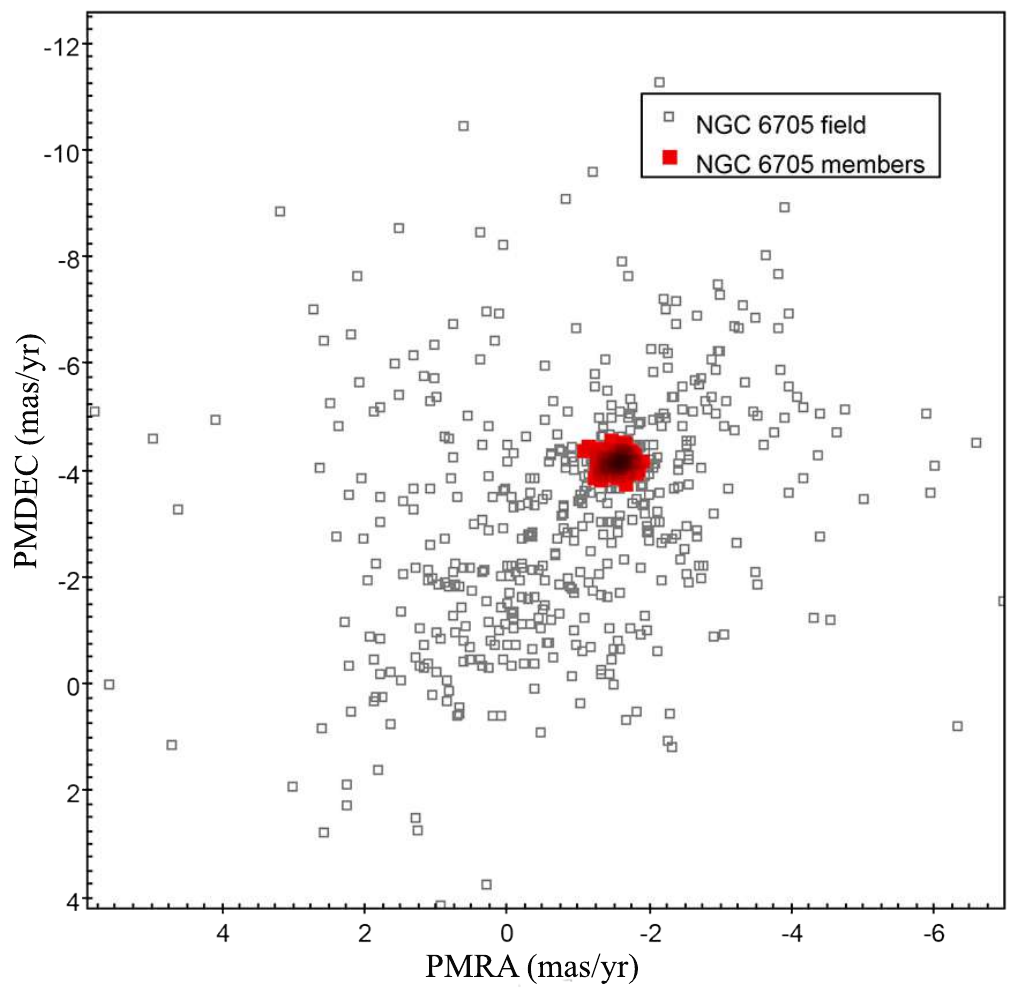}
   \caption{PMs diagram for NGC~6705.}
             \label{fig:124}
    \end{figure}
    
     \begin{figure} [htp]
   \centering
   \includegraphics[width=0.9\linewidth, height=7cm]{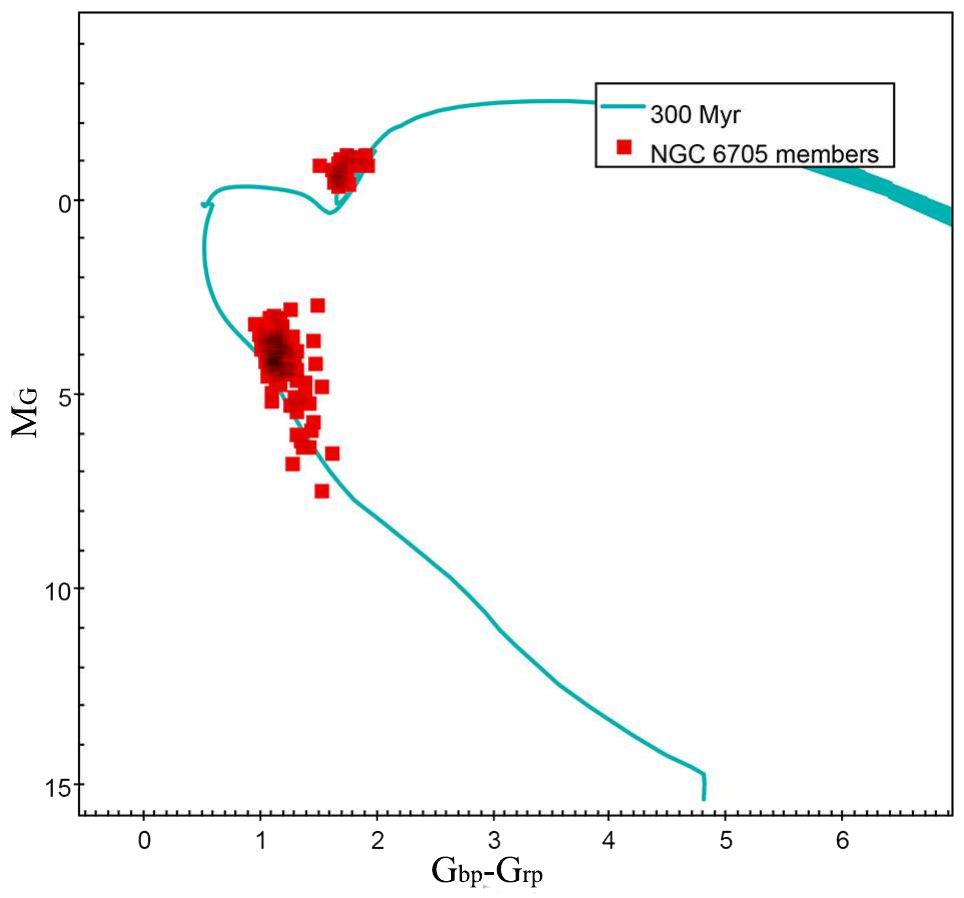}
   \caption{CMD for NGC~6705.}
             \label{fig:125}
    \end{figure}
    
      \begin{figure} [htp]
   \centering
 \includegraphics[width=0.9\linewidth]{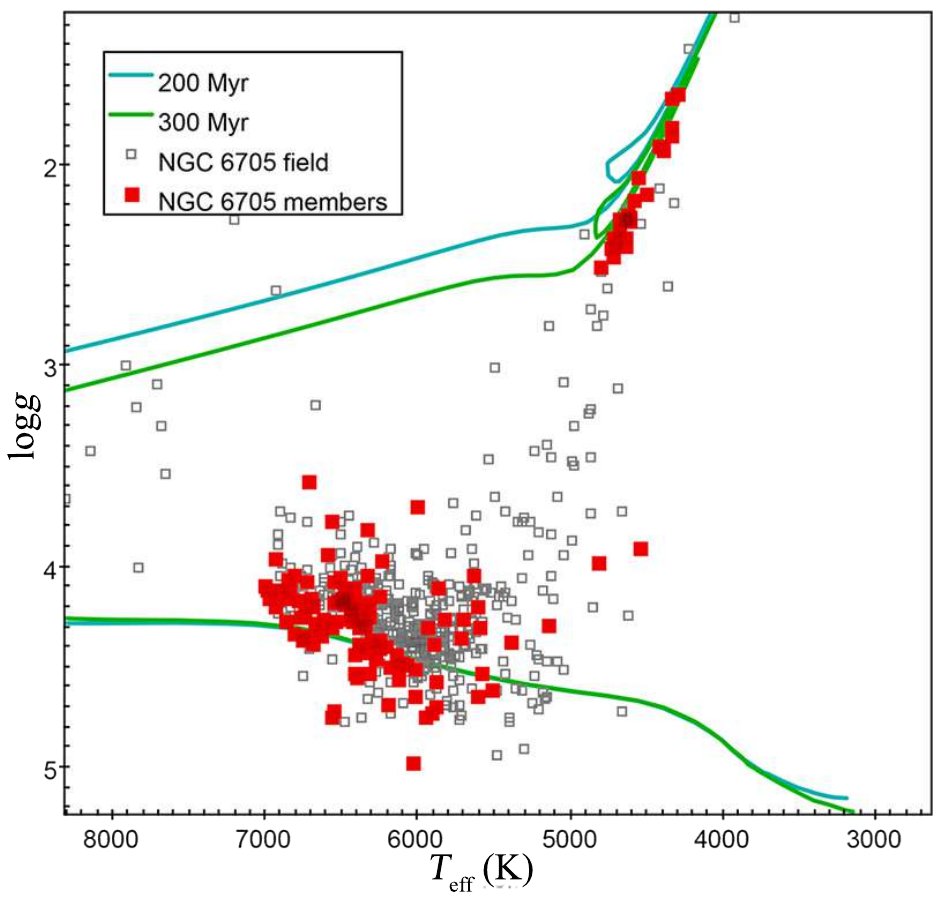} 
\caption{Kiel diagram for NGC~6705.}
             \label{fig:126}
    \end{figure}

  \begin{figure} [htp]
   \centering
 \includegraphics[width=0.9\linewidth]{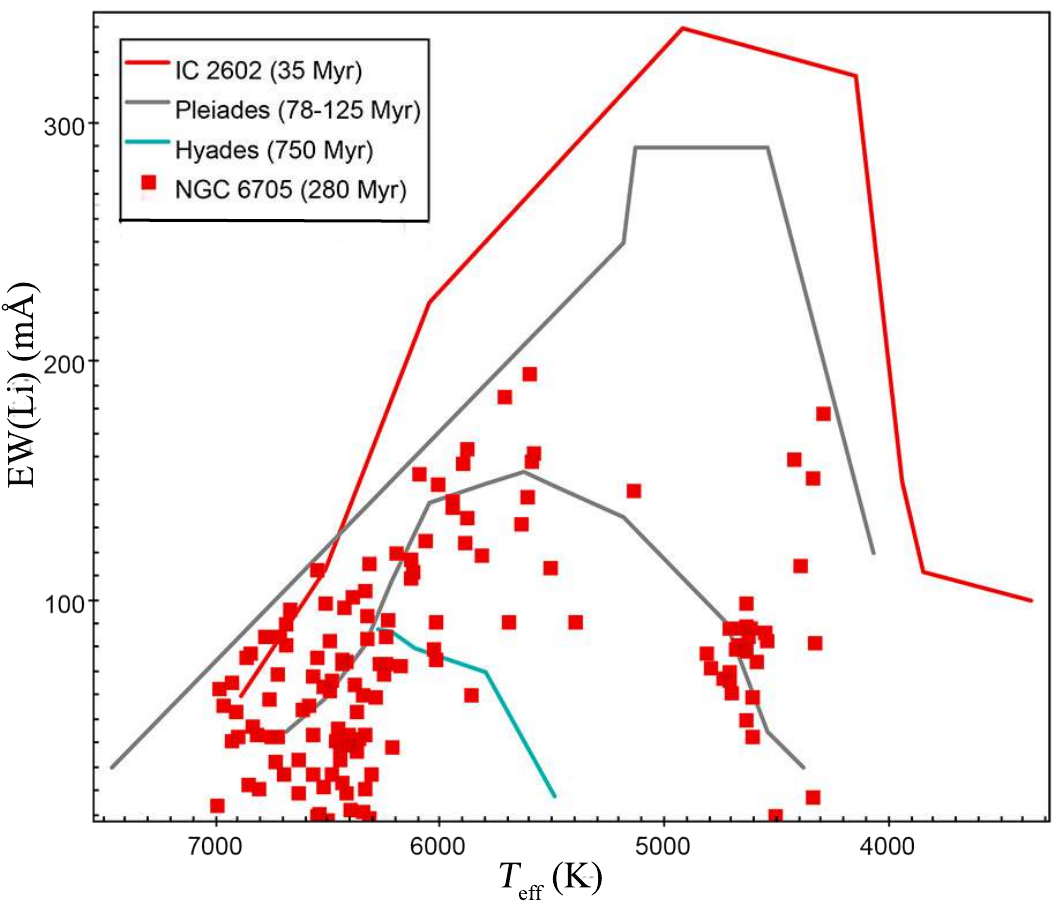} 
\caption{$EW$(Li)-versus-$T_{\rm eff}$ diagram for NGC~6705.}
             \label{fig:127}
    \end{figure}
    
    \clearpage

\subsection{Berkeley~30}

 \begin{figure} [htp]
   \centering
\includegraphics[width=0.9\linewidth, height=5cm]{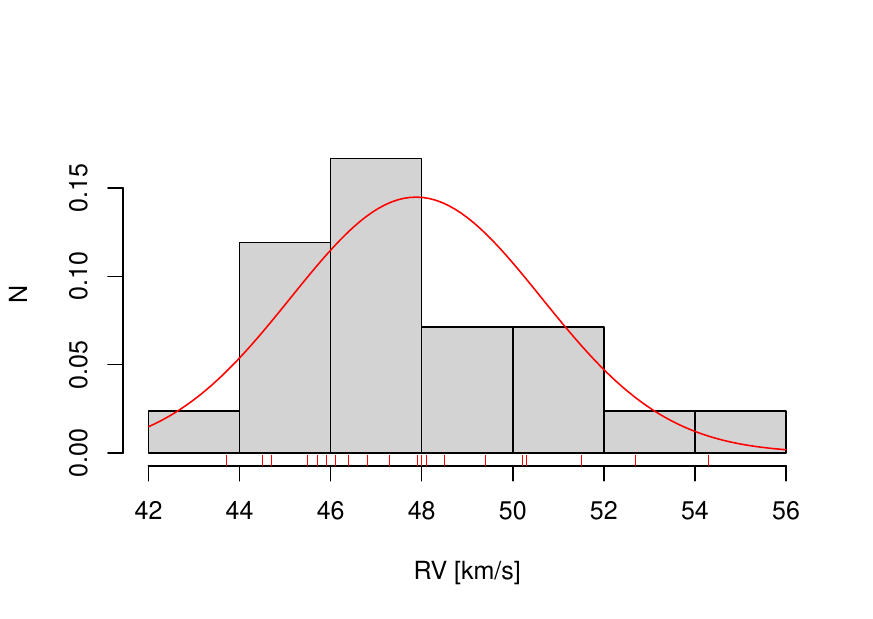}
\caption{$RV$ distribution for Berkeley~30.}
             \label{fig:128}
    \end{figure}
    
           \begin{figure} [htp]
   \centering
\includegraphics[width=0.9\linewidth, height=5cm]{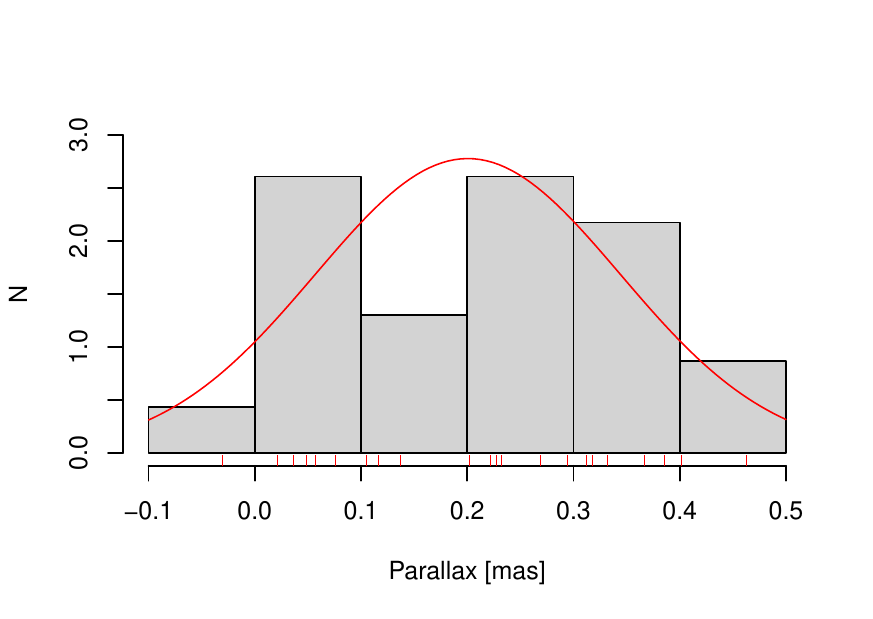}
\caption{Parallax distribution for Berkeley~30.}
             \label{fig:129}
    \end{figure}

               \begin{figure} [htp]
   \centering
   \includegraphics[width=0.9\linewidth]{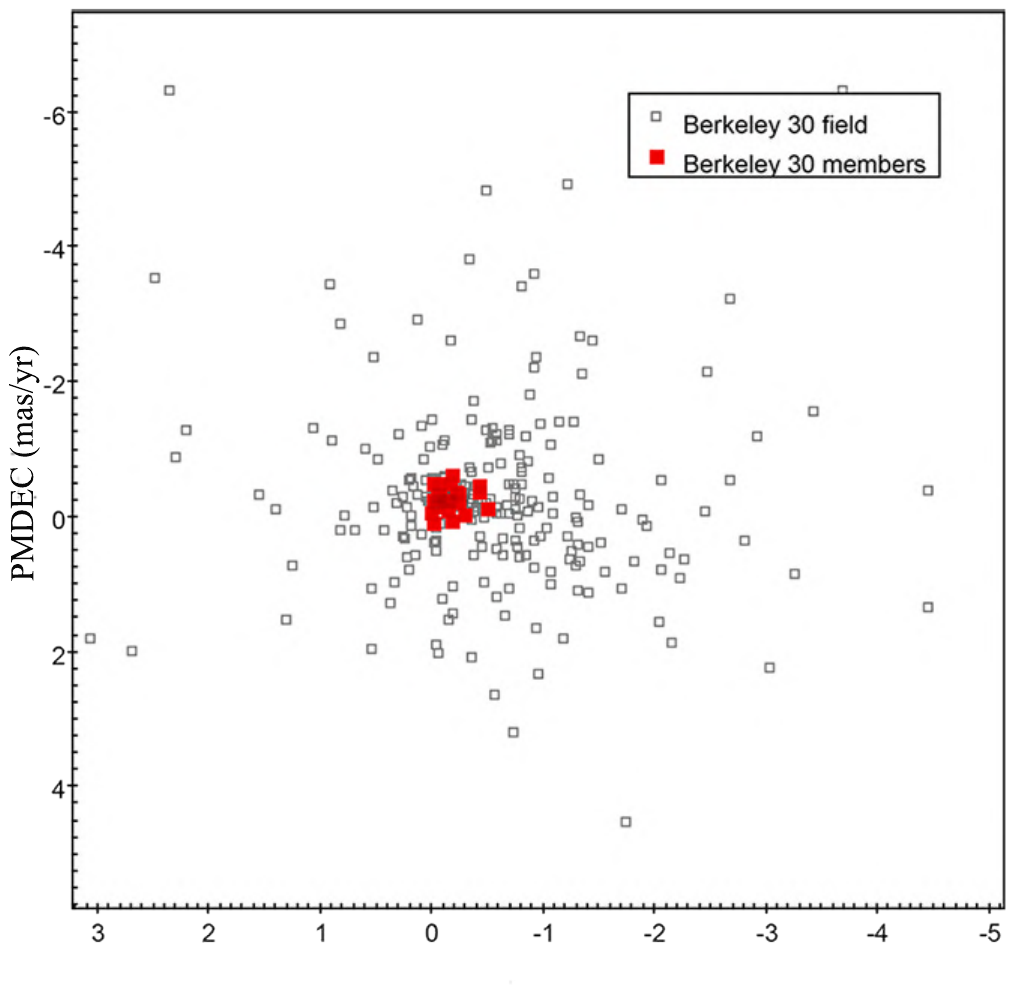}
   \caption{PMs diagram for Berkeley~30.}
             \label{fig:130}
    \end{figure}
    
     \begin{figure} [htp]
   \centering
   \includegraphics[width=0.9\linewidth, height=7cm]{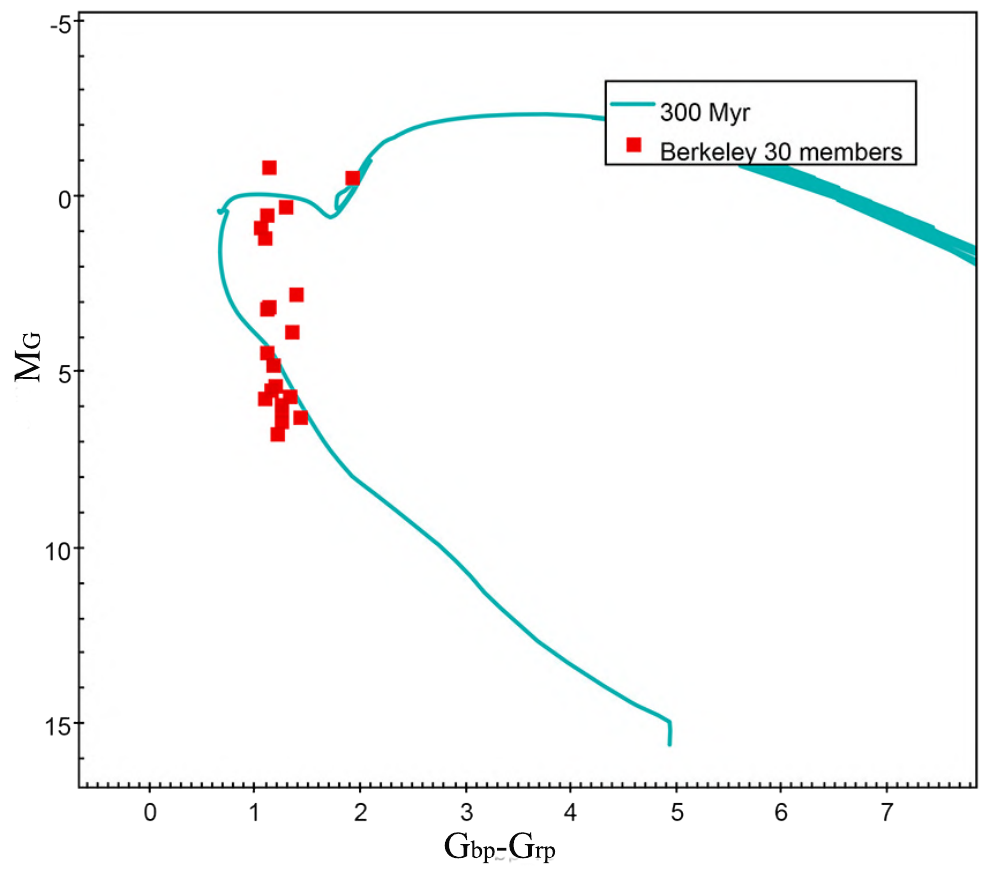}
   \caption{CMD for Berkeley~30.}
             \label{fig:131}
    \end{figure}
    
      \begin{figure} [htp]
   \centering
 \includegraphics[width=0.9\linewidth]{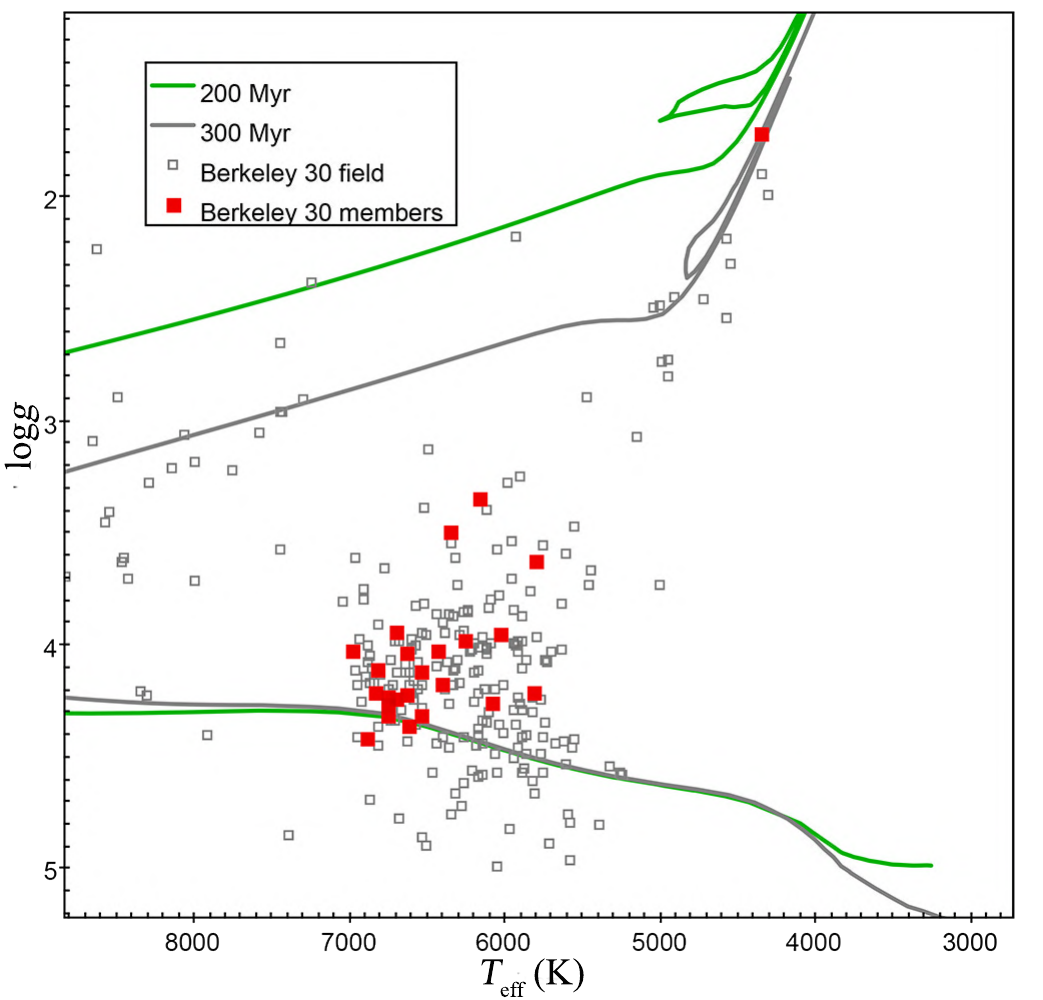} 
\caption{Kiel diagram for Berkeley~30.}
             \label{fig:132}
    \end{figure}

  \begin{figure} [htp]
   \centering
 \includegraphics[width=0.9\linewidth]{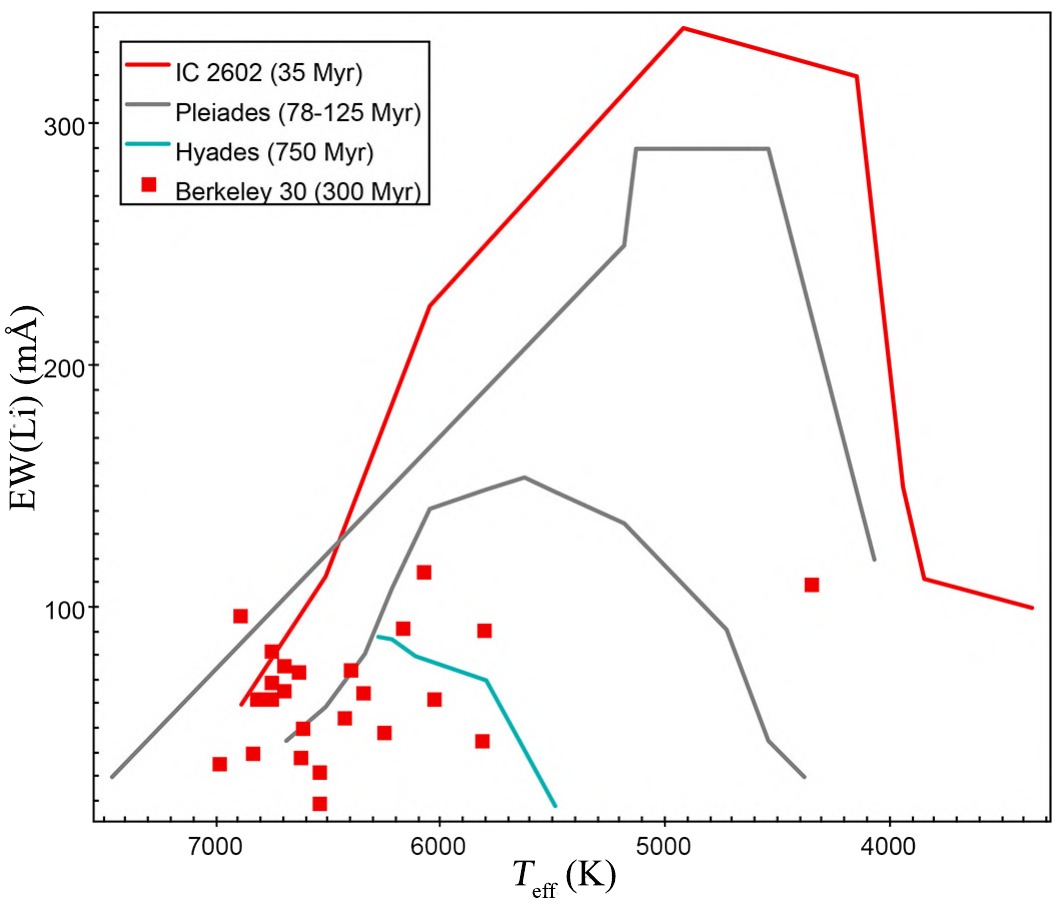} 
\caption{$EW$(Li)-versus-$T_{\rm eff}$ diagram for Berkeley~30.}
             \label{fig:133}
    \end{figure}
    
    \clearpage

\subsection{NGC~6281}

 \begin{figure} [htp]
   \centering
\includegraphics[width=0.9\linewidth, height=5cm]{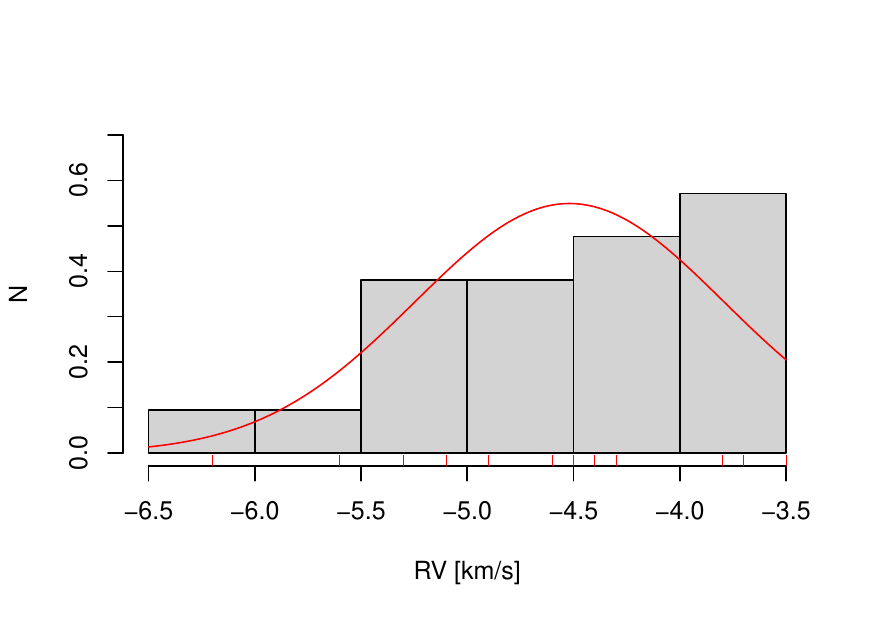}
\caption{$RV$ distribution for NGC~6281.}
             \label{fig:134}
    \end{figure}
    
           \begin{figure} [htp]
   \centering
\includegraphics[width=0.9\linewidth, height=5cm]{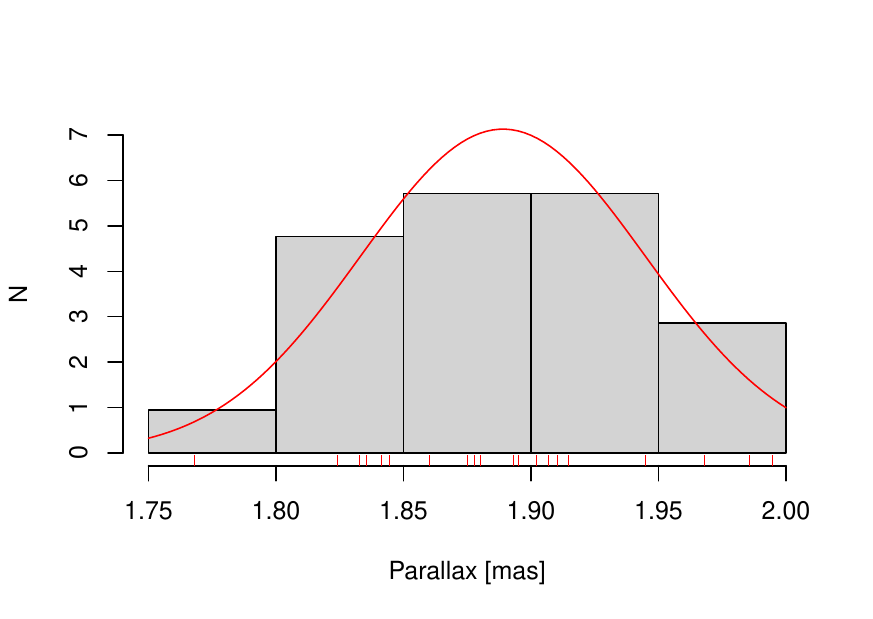}
\caption{Parallax distribution for NGC~6281.}
             \label{fig:135}
    \end{figure}

               \begin{figure} [htp]
   \centering
   \includegraphics[width=0.9\linewidth]{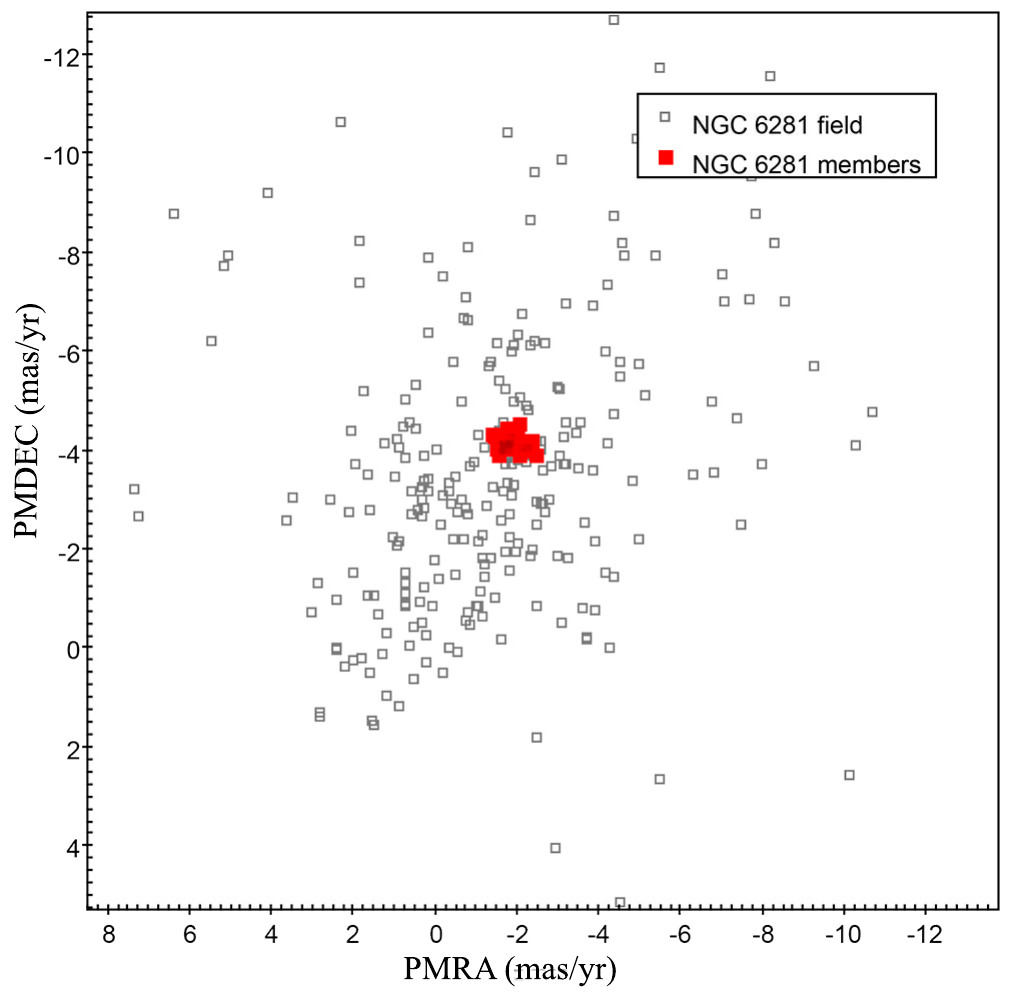}
   \caption{PMs diagram for NGC~6281.}
             \label{fig:136}
    \end{figure}
    
     \begin{figure} [htp]
   \centering
   \includegraphics[width=0.8\linewidth, height=7cm]{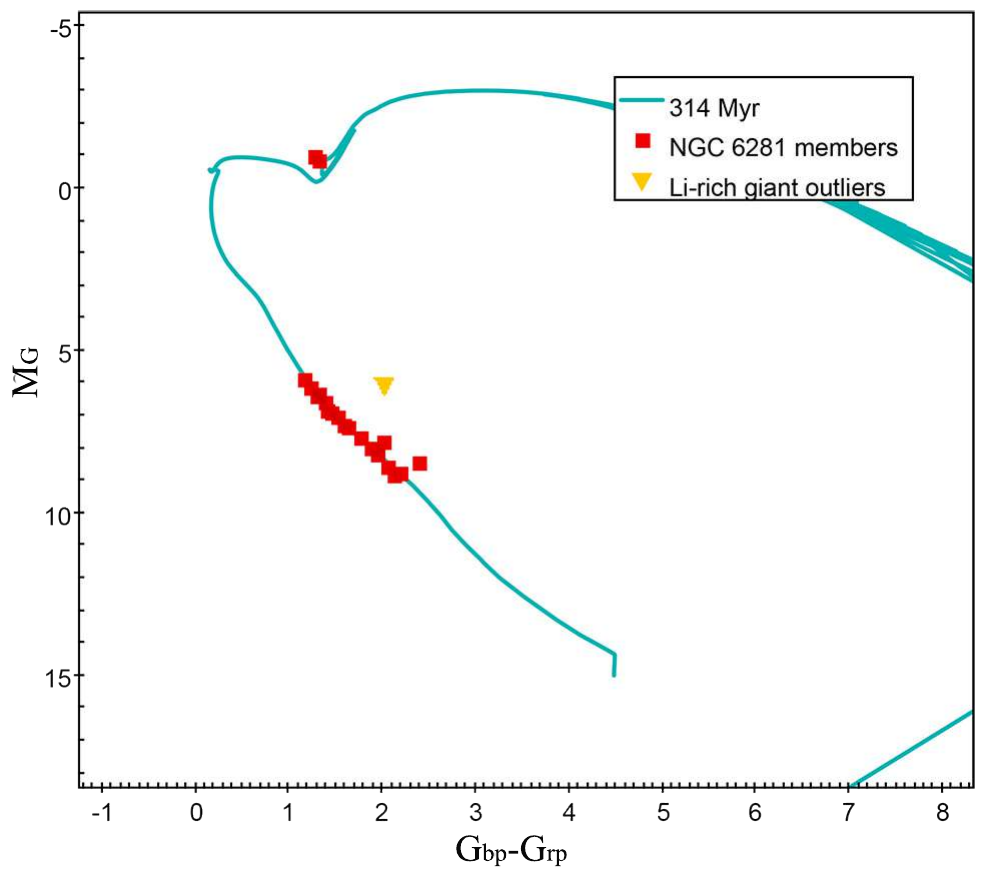}
   \caption{CMD for NGC~6281.}
             \label{fig:137}
    \end{figure}
    
      \begin{figure} [htp]
   \centering
 \includegraphics[width=0.8\linewidth]{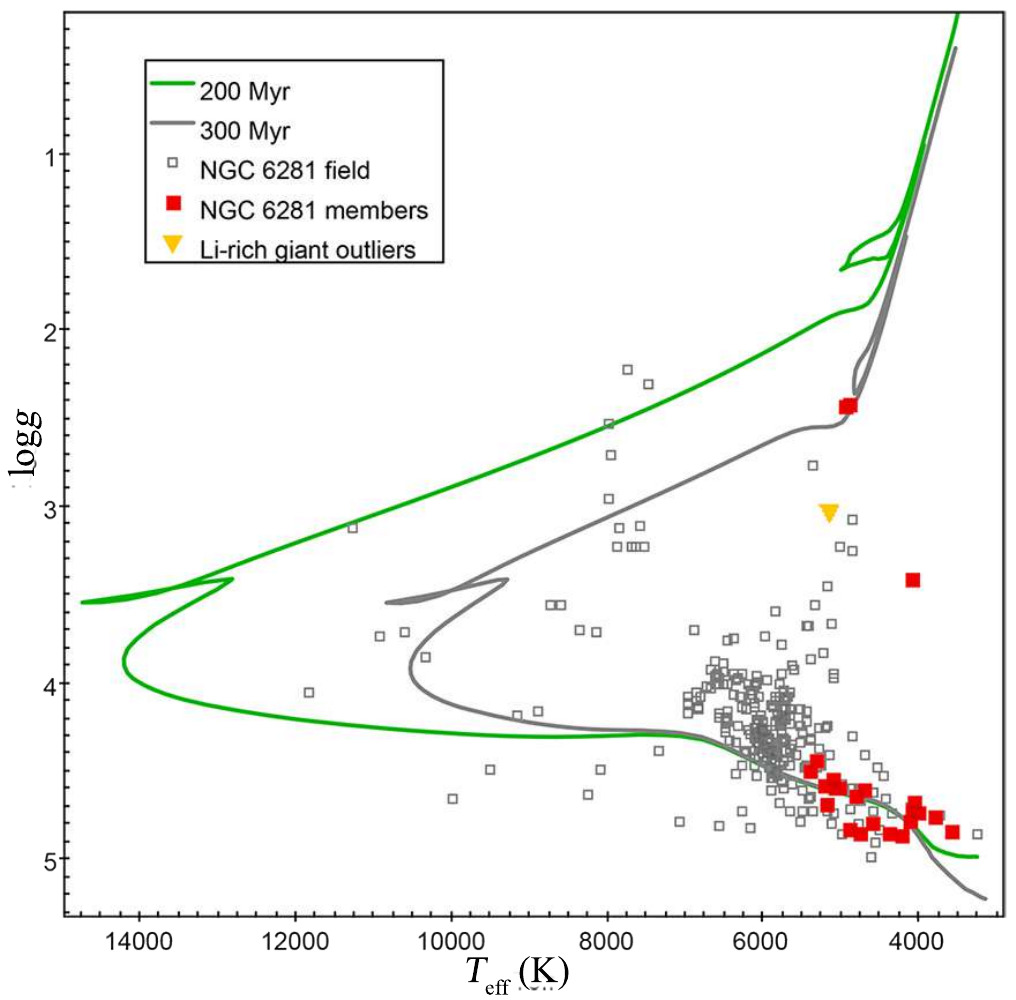} 
\caption{Kiel diagram for NGC~6281.}
             \label{fig:138}
    \end{figure}

  \begin{figure} [htp]
   \centering
 \includegraphics[width=0.8\linewidth]{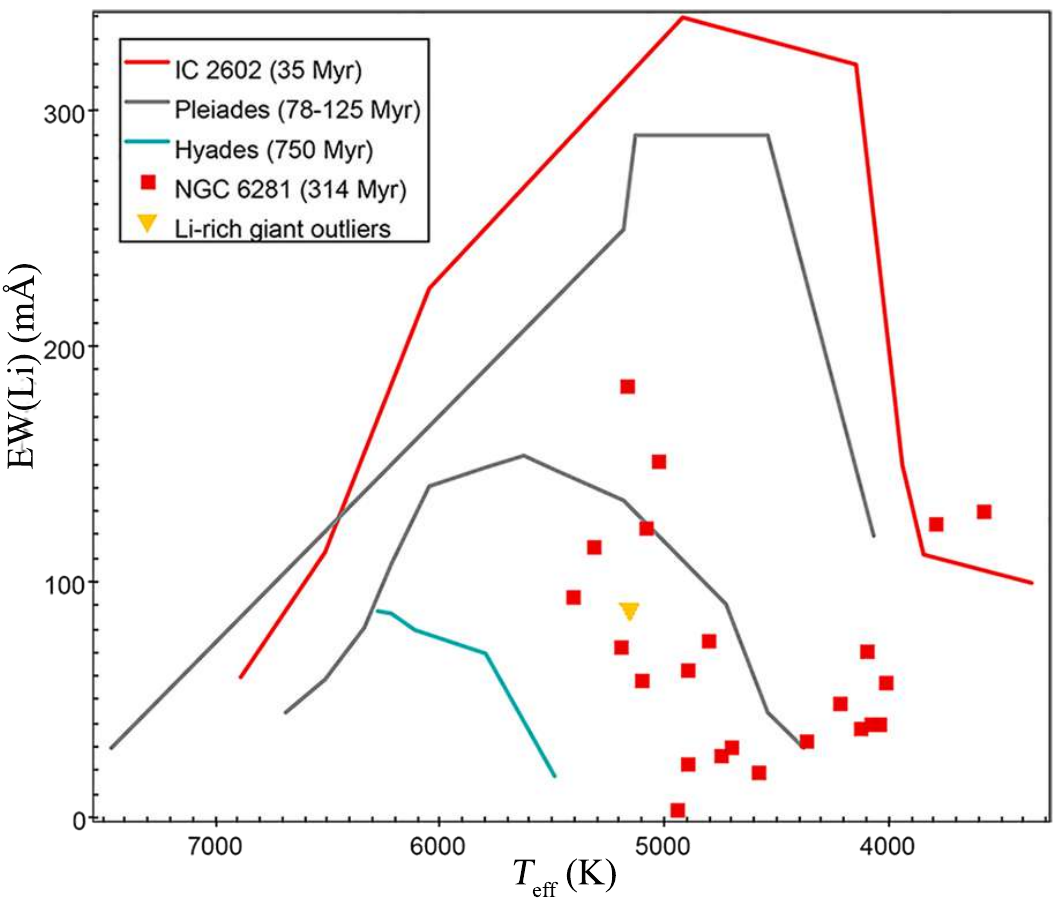} 
\caption{$EW$(Li)-versus-$T_{\rm eff}$ diagram for NGC~6281.}
             \label{fig:139}
    \end{figure}
    
    \clearpage

\subsection{NGC~3532}

 \begin{figure} [htp]
   \centering
\includegraphics[width=0.9\linewidth, height=5cm]{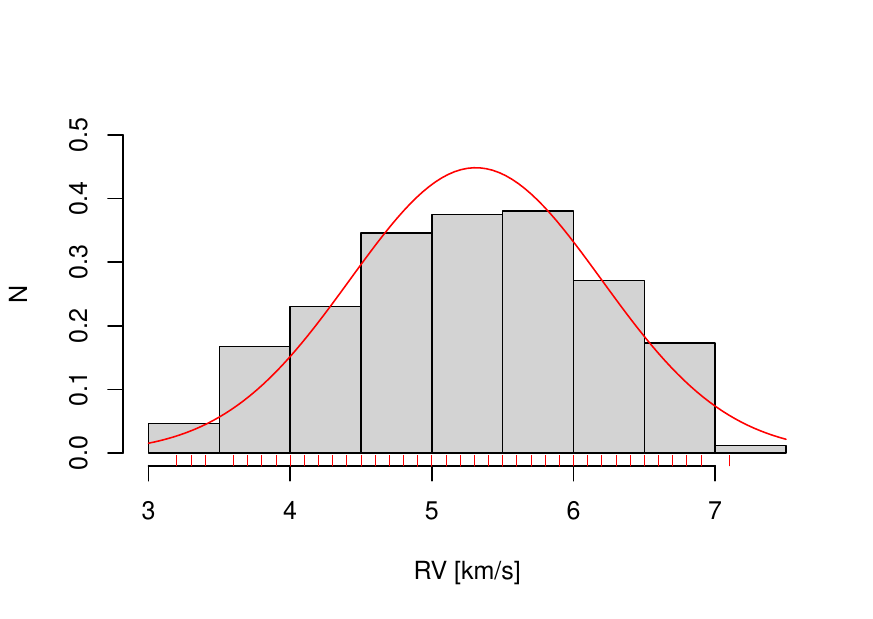}
\caption{$RV$ distribution for NGC~3532.}
             \label{fig:140}
    \end{figure}
    
           \begin{figure} [htp]
   \centering
\includegraphics[width=0.9\linewidth, height=5cm]{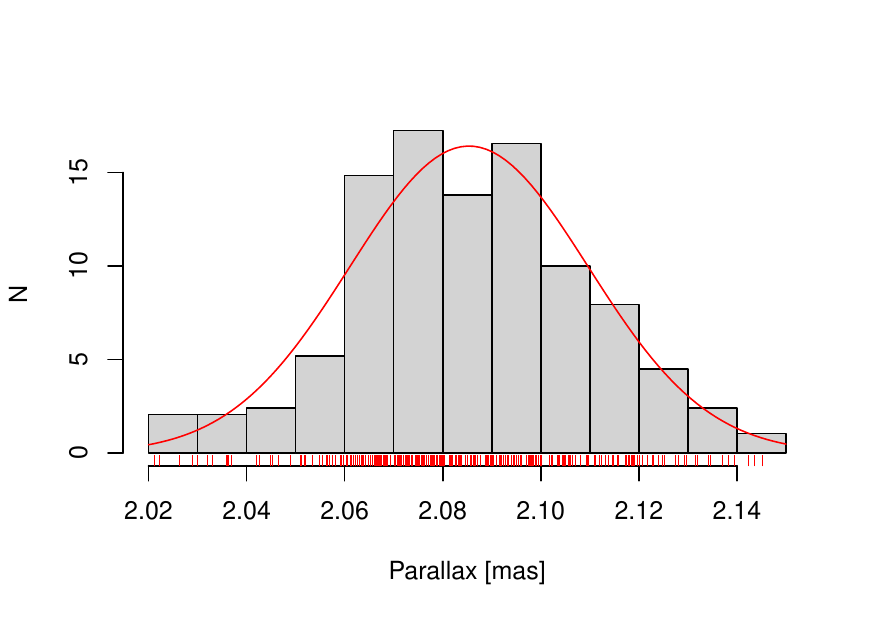}
\caption{Parallax distribution for NGC~3532.}
             \label{fig:141}
    \end{figure}

               \begin{figure} [htp]
   \centering
   \includegraphics[width=0.9\linewidth]{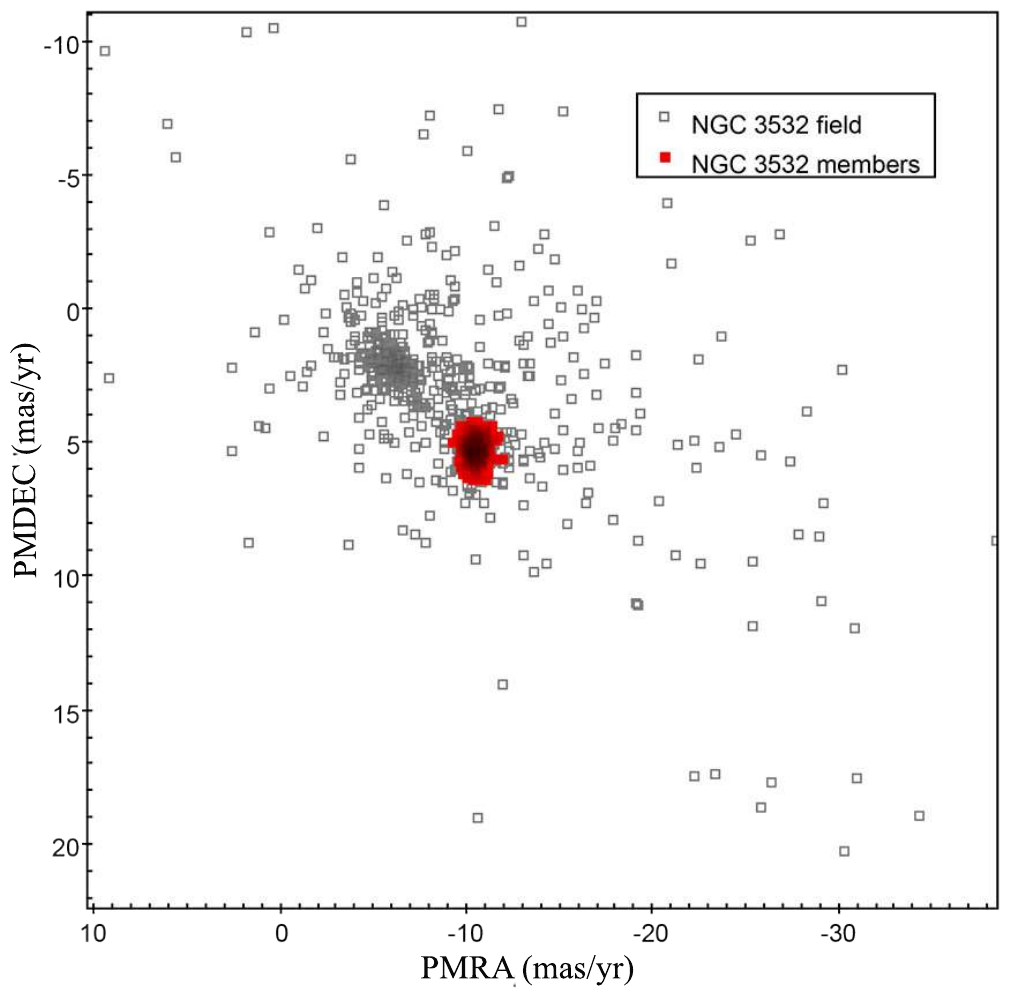}
   \caption{PMs diagram for NGC~3532.}
             \label{fig:142}
    \end{figure}
    
     \begin{figure} [htp]
   \centering
   \includegraphics[width=0.8\linewidth, height=7cm]{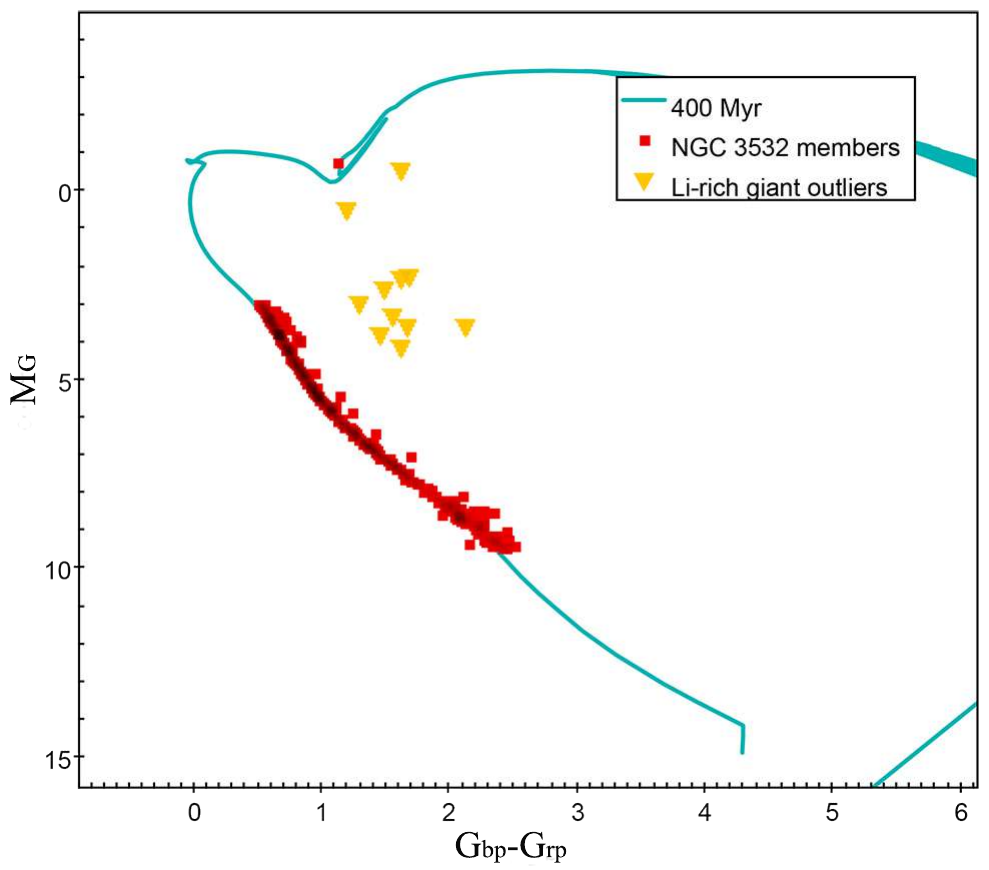}
   \caption{CMD for NGC~3532.}
             \label{fig:143}
    \end{figure}
    
      \begin{figure} [htp]
   \centering
 \includegraphics[width=0.8\linewidth]{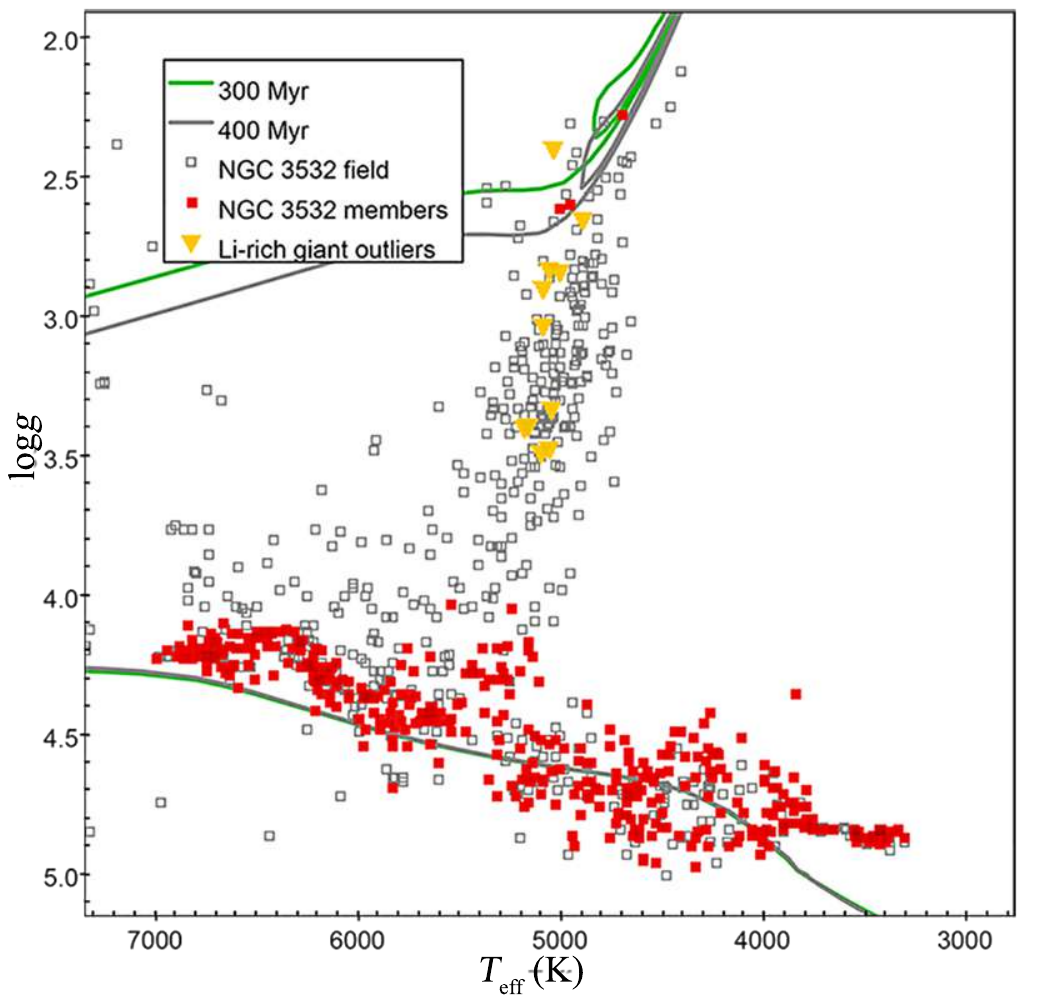} 
\caption{Kiel diagram for NGC~3532.}
             \label{fig:144}
    \end{figure}

  \begin{figure} [htp]
   \centering
 \includegraphics[width=0.8\linewidth]{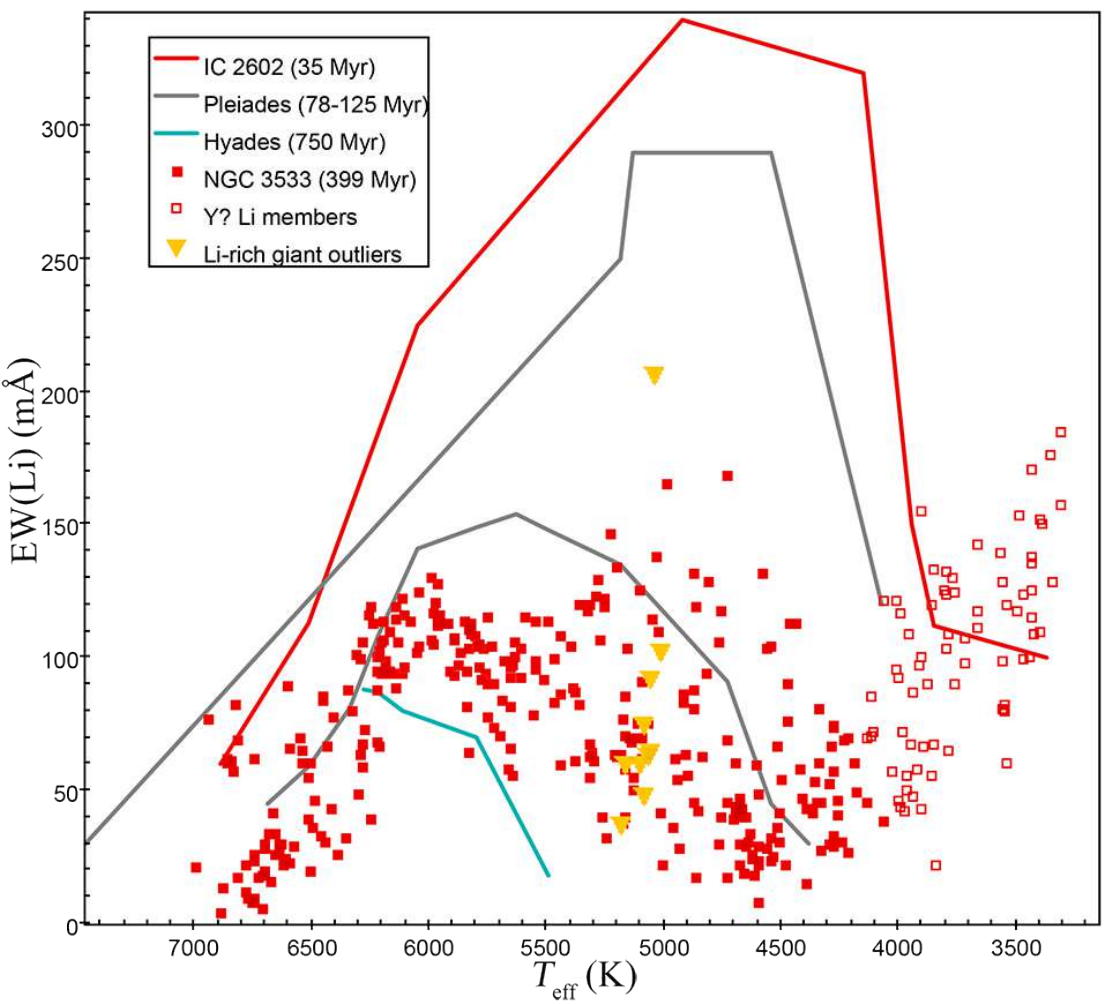} 
\caption{$EW$(Li)-versus-$T_{\rm eff}$ diagram for NGC~3532.}
             \label{fig:145}
    \end{figure}
    
    \clearpage

\subsection{NGC~4815}

 \begin{figure} [htp]
   \centering
\includegraphics[width=0.9\linewidth, height=5cm]{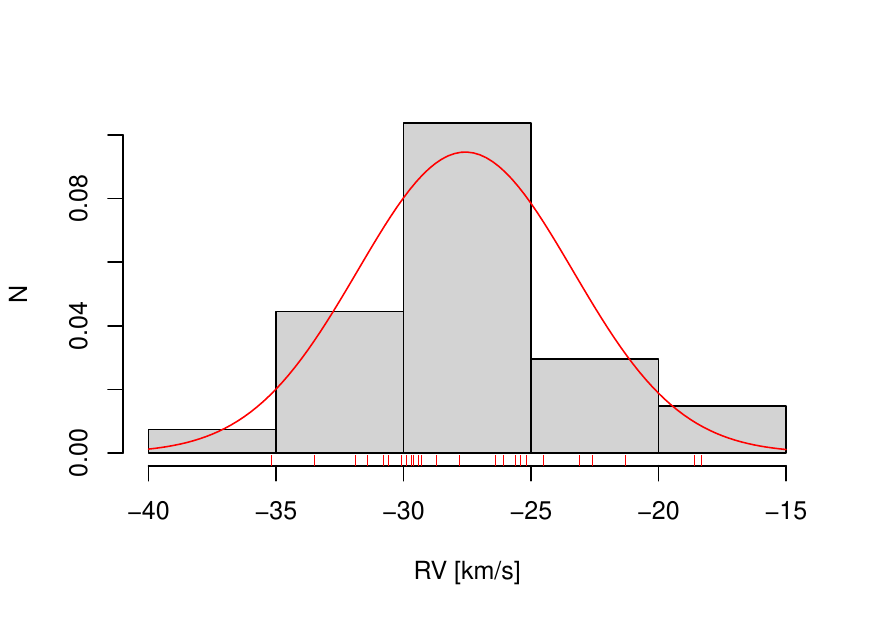}
\caption{$RV$ distribution for NGC~4815.}
             \label{fig:146}
    \end{figure}
    
           \begin{figure} [htp]
   \centering
\includegraphics[width=0.9\linewidth, height=5cm]{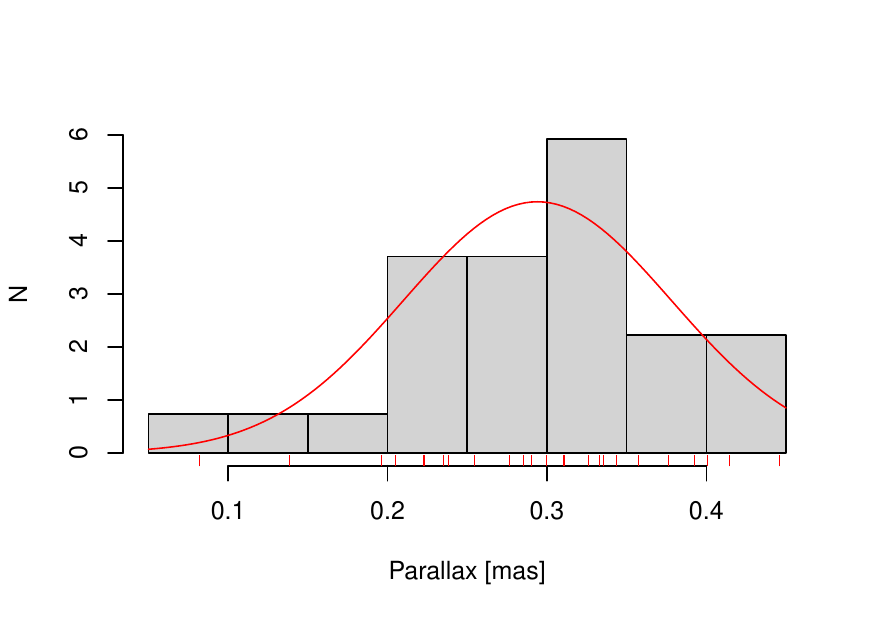}
\caption{Parallax distribution for NGC~4815.}
             \label{fig:147}
    \end{figure}

               \begin{figure} [htp]
   \centering
   \includegraphics[width=0.9\linewidth]{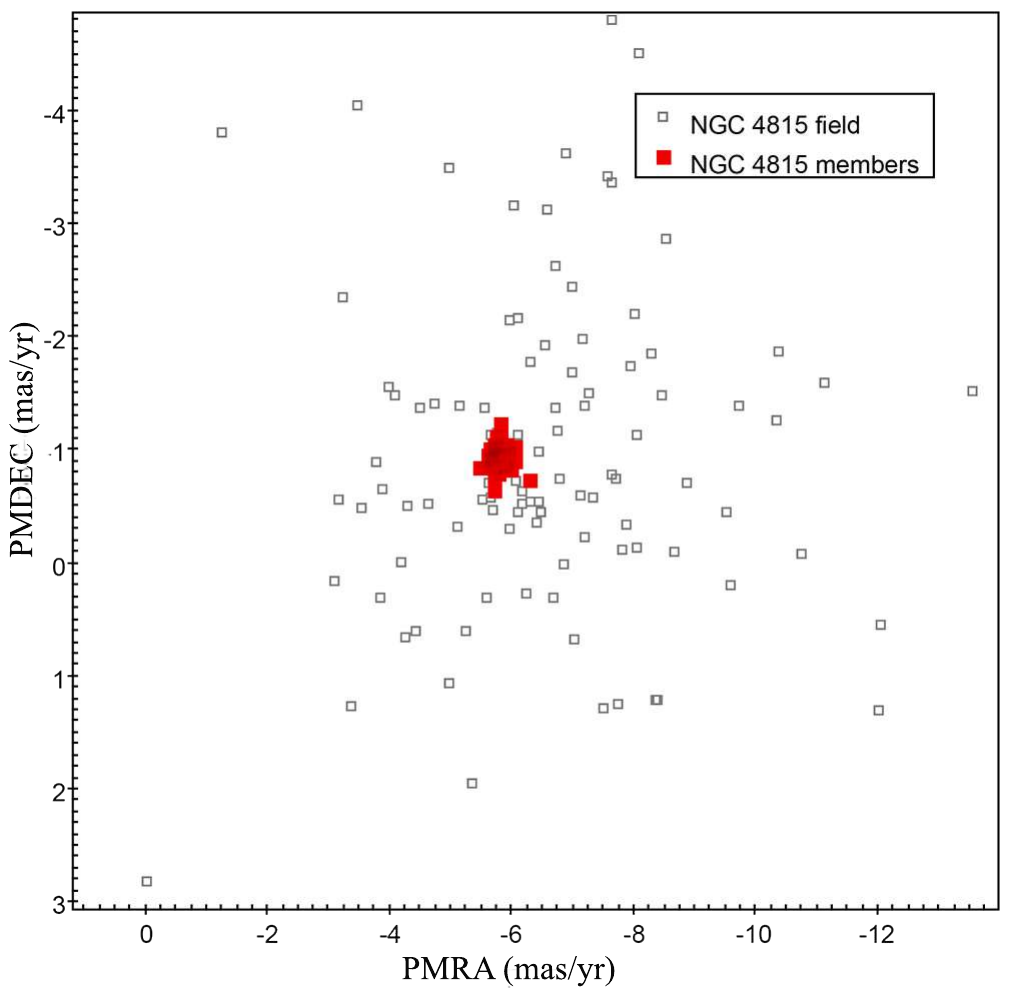}
   \caption{PMs diagram for NGC~4815.}
             \label{fig:148}
    \end{figure}
    
     \begin{figure} [htp]
   \centering
   \includegraphics[width=0.8\linewidth, height=7cm]{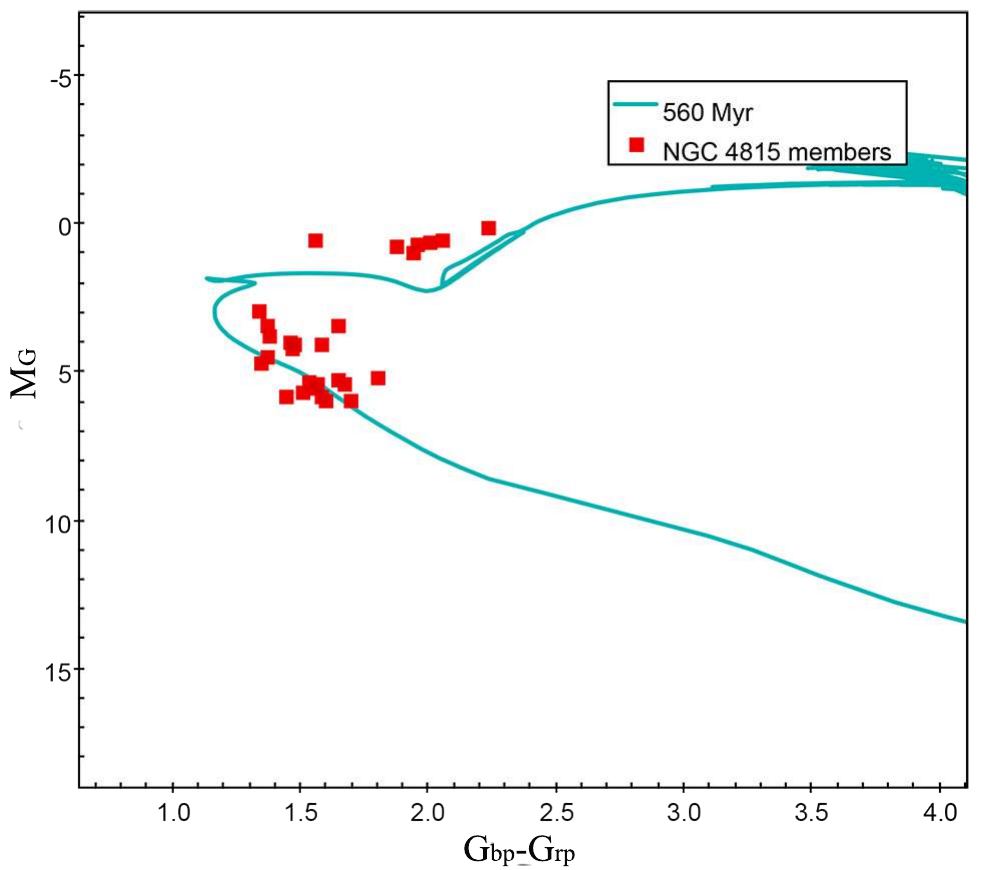}
   \caption{CMD for NGC~4815.}
             \label{fig:149}
    \end{figure}
    
      \begin{figure} [htp]
   \centering
 \includegraphics[width=0.8\linewidth]{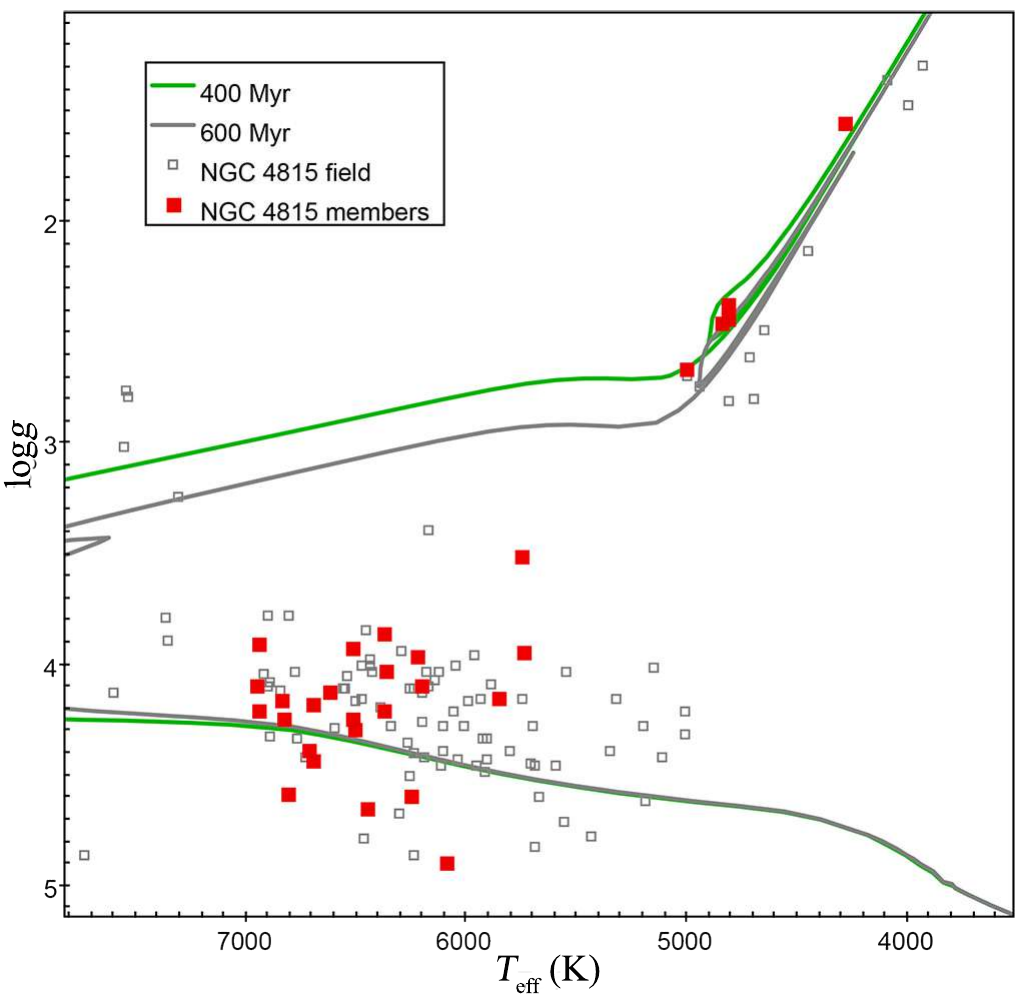} 
\caption{Kiel diagram for NGC~4815.}
             \label{fig:150}
    \end{figure}

  \begin{figure} [htp]
   \centering
 \includegraphics[width=0.8\linewidth]{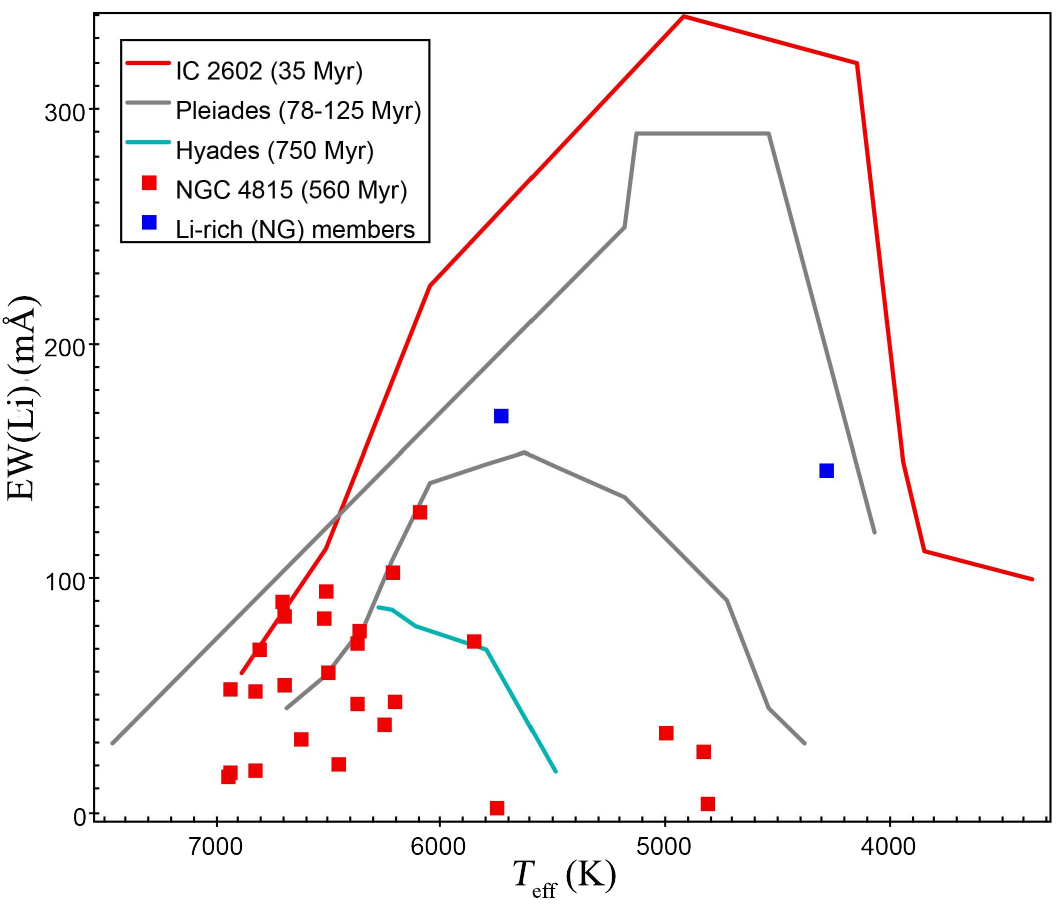} 
\caption{$EW$(Li)-versus-$T_{\rm eff}$ diagram for NGC~4815.}
             \label{fig:151}
    \end{figure}
    
    \clearpage

\subsection{NGC~6633}

 \begin{figure} [htp]
   \centering
\includegraphics[width=0.9\linewidth, height=5cm]{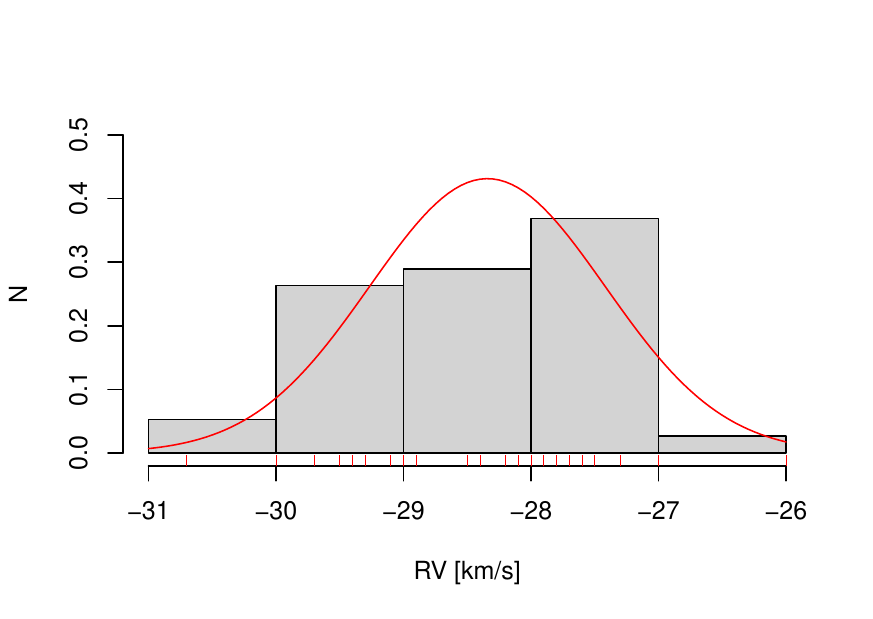}
\caption{$RV$ distribution for NGC~6633.}
             \label{fig:152}
    \end{figure}
    
           \begin{figure} [htp]
   \centering
\includegraphics[width=0.9\linewidth, height=5cm]{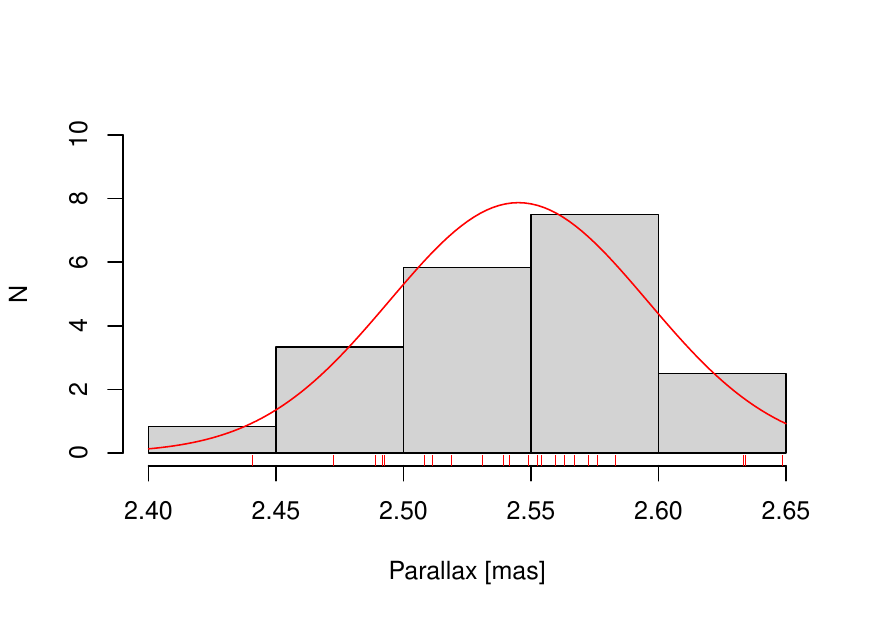}
\caption{Parallax distribution for NGC~6633.}
             \label{fig:153}
    \end{figure}

               \begin{figure} [htp]
   \centering
   \includegraphics[width=0.9\linewidth]{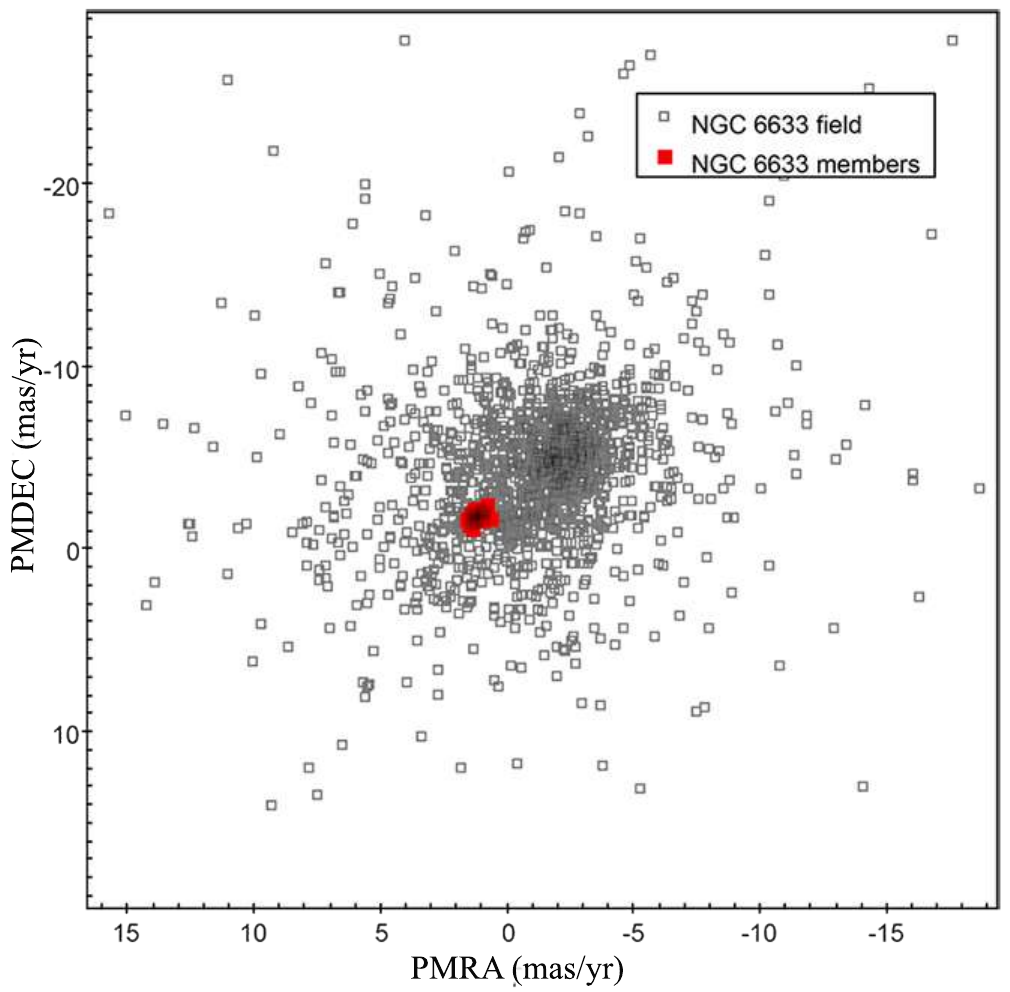}
   \caption{PMs diagram for NGC~6633.}
             \label{fig:154}
    \end{figure}
    
     \begin{figure} [htp]
   \centering
   \includegraphics[width=0.8\linewidth, height=7cm]{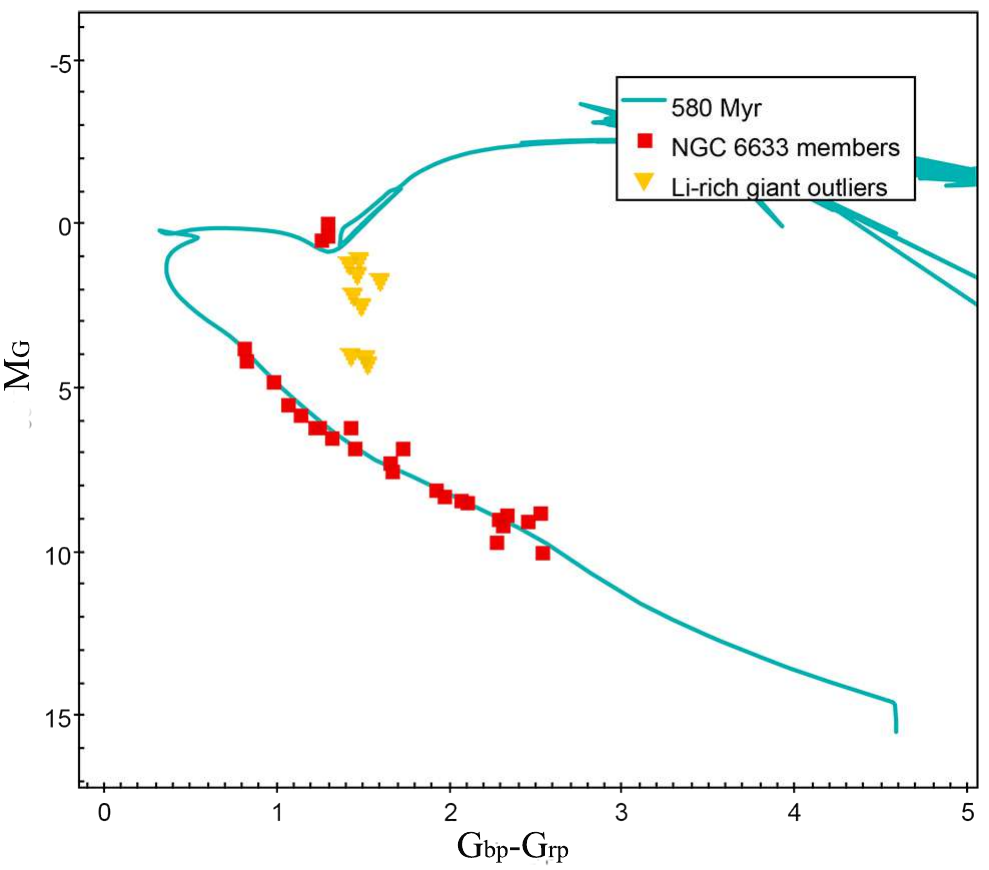}
   \caption{CMD for NGC~6633.}
             \label{fig:155}
    \end{figure}
    
      \begin{figure} [htp]
   \centering
 \includegraphics[width=0.8\linewidth]{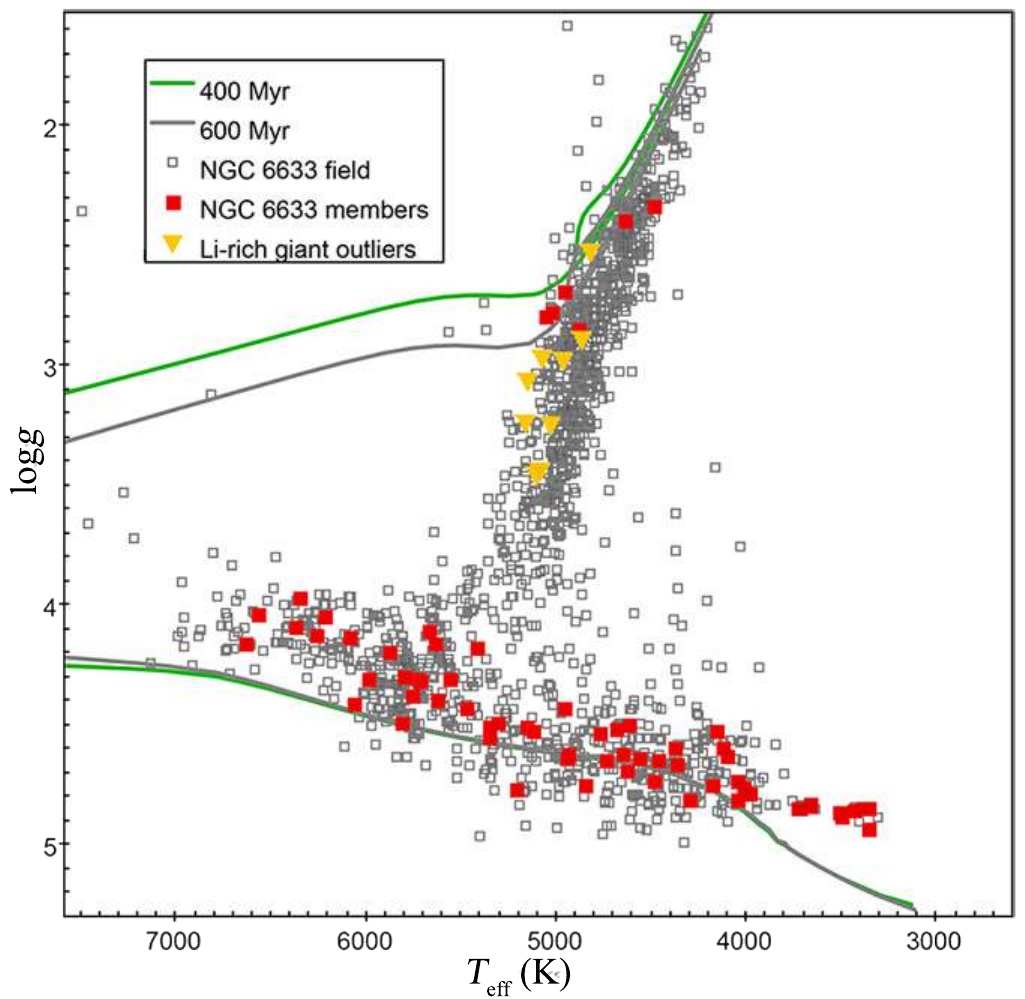} 
\caption{Kiel diagram for NGC~6633.}
             \label{fig:156}
    \end{figure}

  \begin{figure} [htp]
   \centering
 \includegraphics[width=0.8\linewidth]{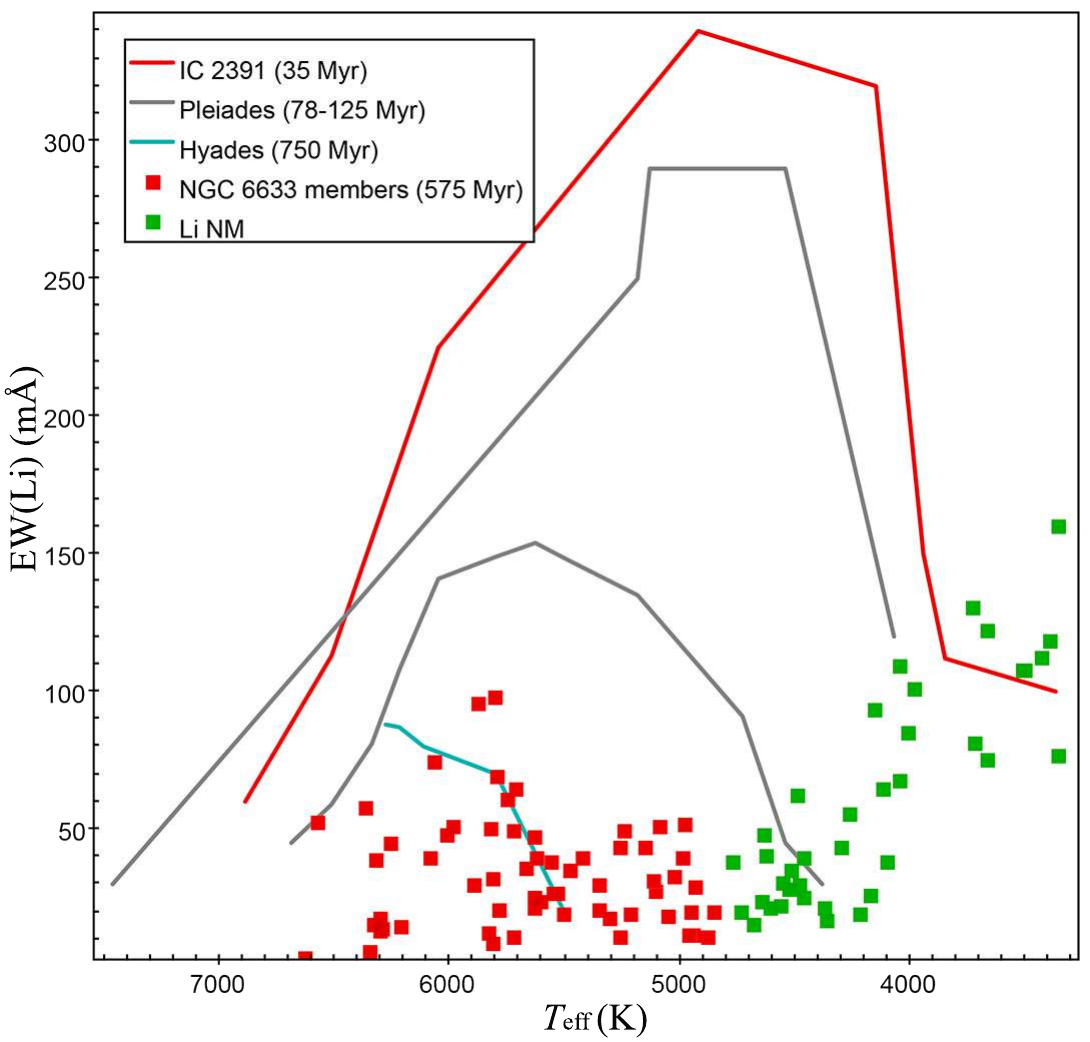} 
\caption{$EW$(Li)-versus-$T_{\rm eff}$ diagram for NGC~6633.}
             \label{fig:157}
    \end{figure}
    
    \clearpage

\subsection{NGC~2477}

 \begin{figure} [htp]
   \centering
\includegraphics[width=0.9\linewidth, height=5cm]{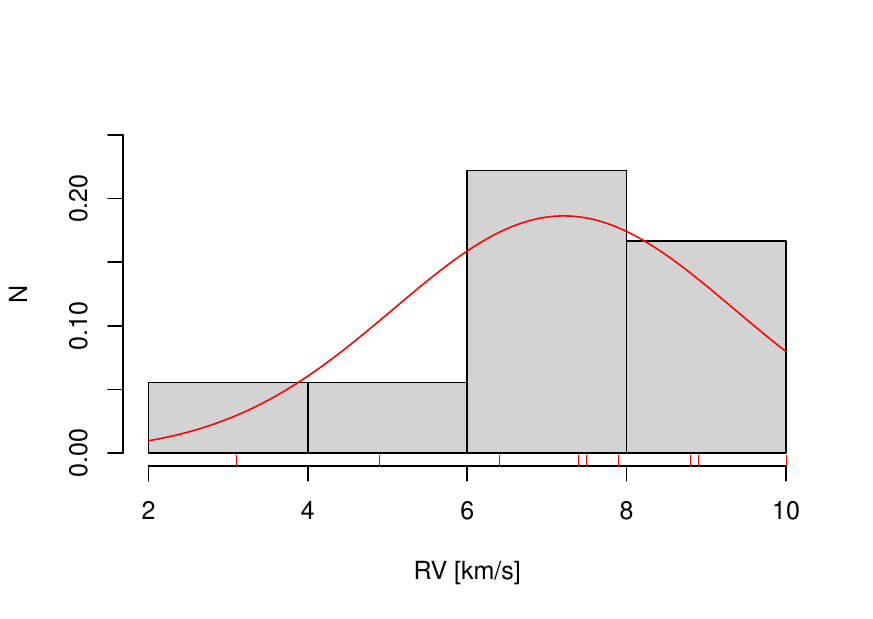}
\caption{$RV$ distribution for NGC~2477.}
             \label{fig:158}
    \end{figure}
    
           \begin{figure} [htp]
   \centering
\includegraphics[width=0.9\linewidth, height=5cm]{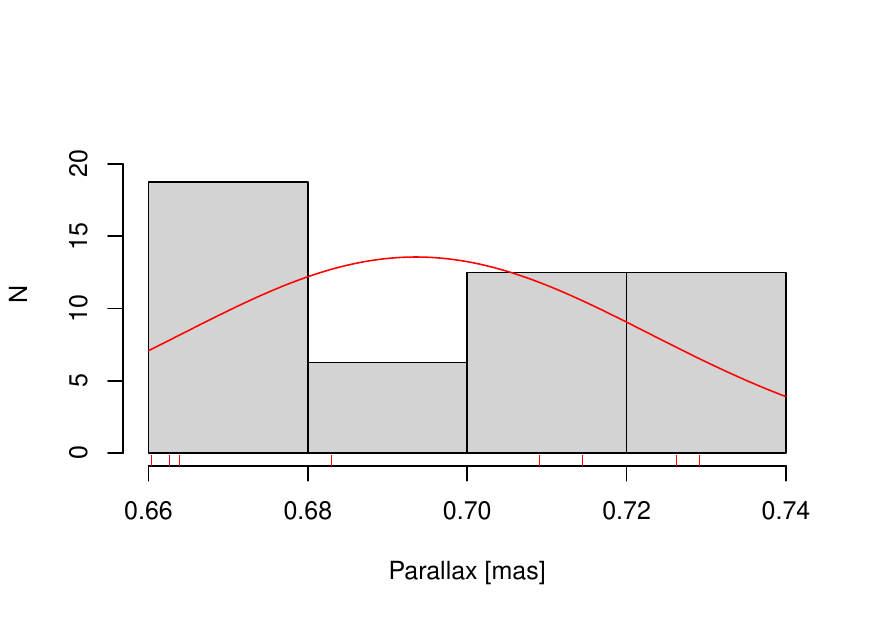}
\caption{Parallax distribution for NGC~2477.}
             \label{fig:159}
    \end{figure}

               \begin{figure} [htp]
   \centering
   \includegraphics[width=0.9\linewidth]{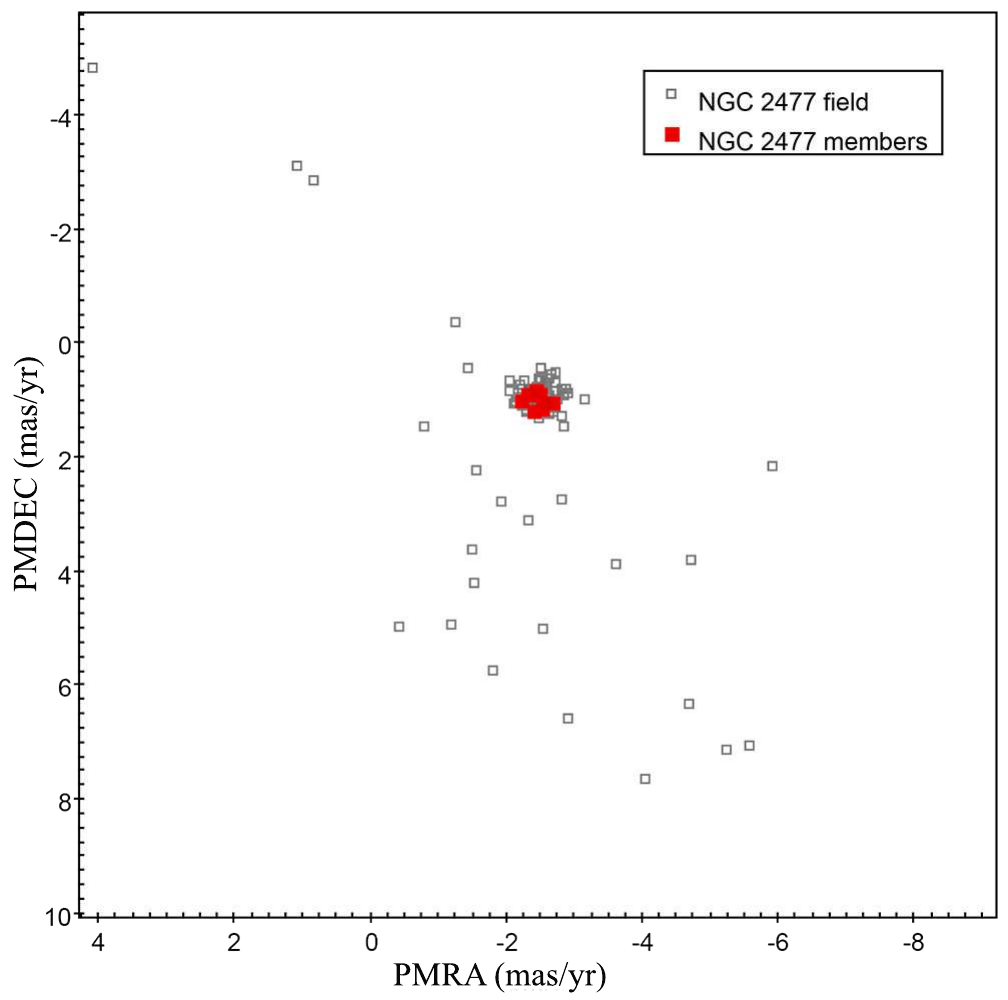}
   \caption{PMs diagram for NGC~2477.}
             \label{fig:160}
    \end{figure}
    
     \begin{figure} [htp]
   \centering
   \includegraphics[width=0.8\linewidth, height=7cm]{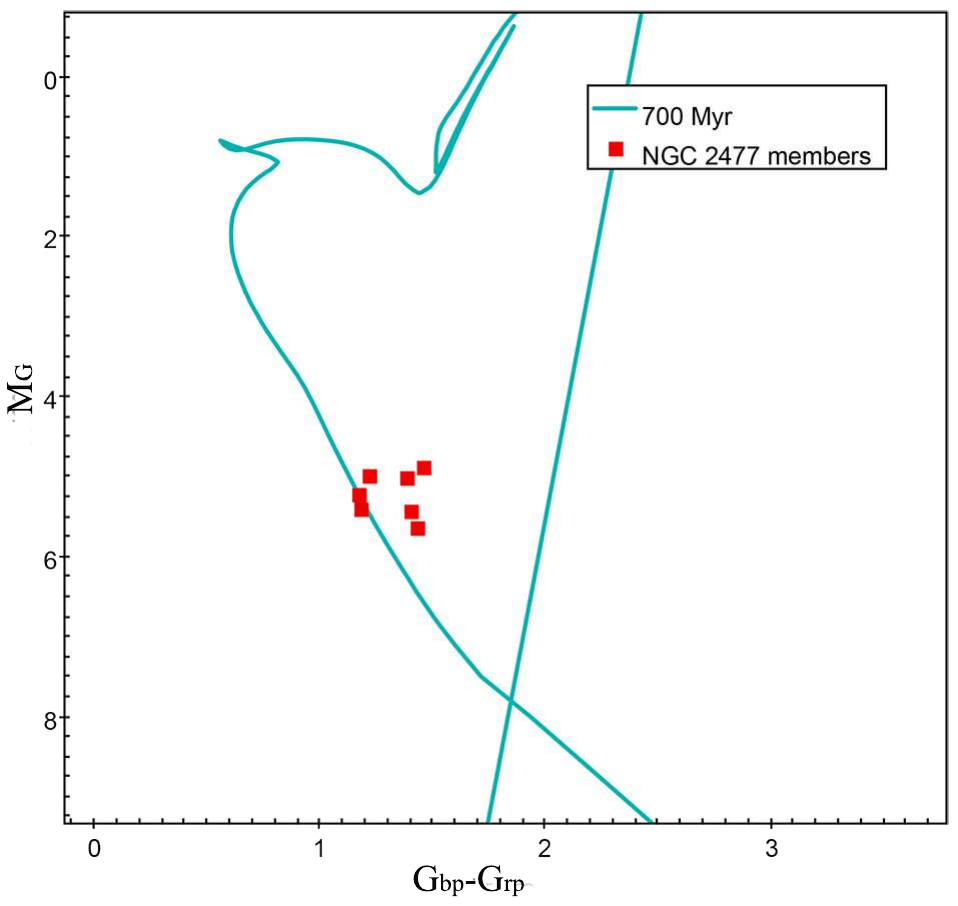}
   \caption{CMD for NGC~2477.}
             \label{fig:161}
    \end{figure}
    
      \begin{figure} [htp]
   \centering
 \includegraphics[width=0.8\linewidth]{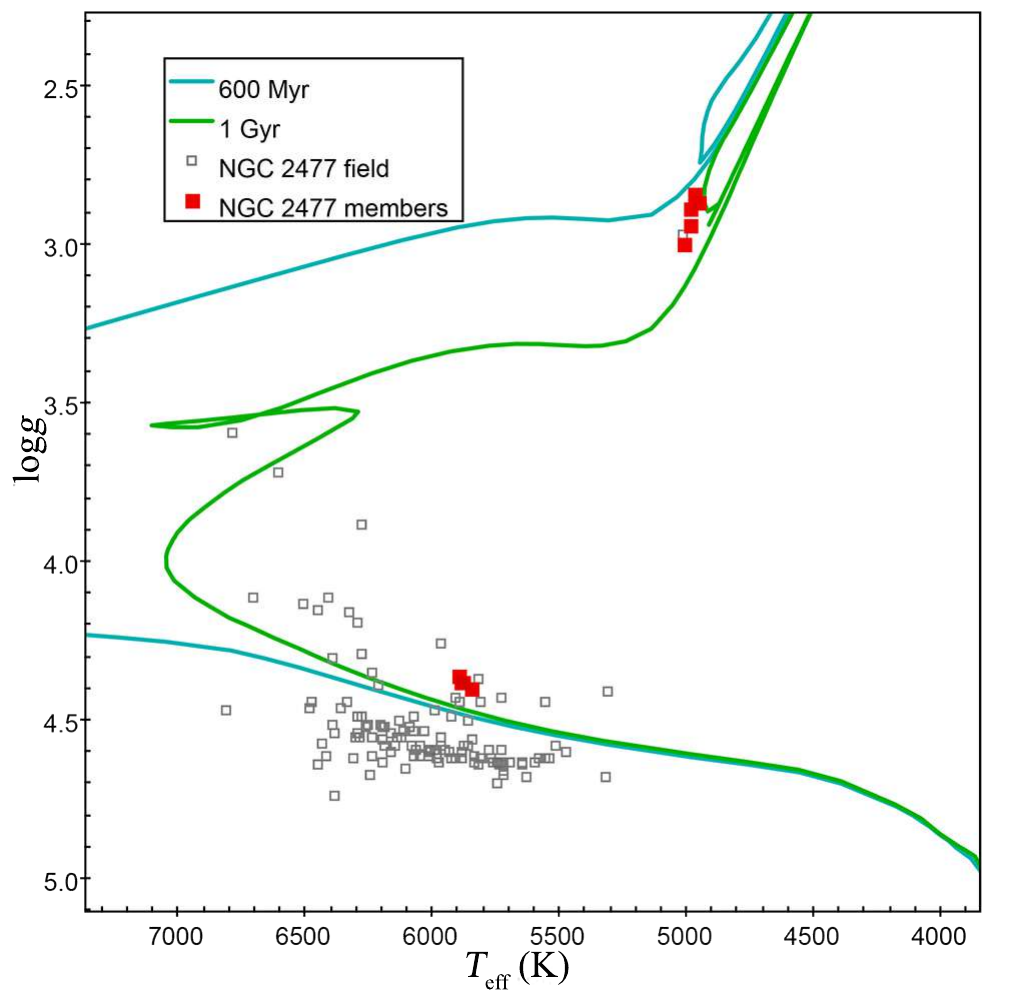} 
\caption{Kiel diagram for NGC~2477.}
             \label{fig:162}
    \end{figure}

  \begin{figure} [htp]
   \centering
 \includegraphics[width=0.8\linewidth]{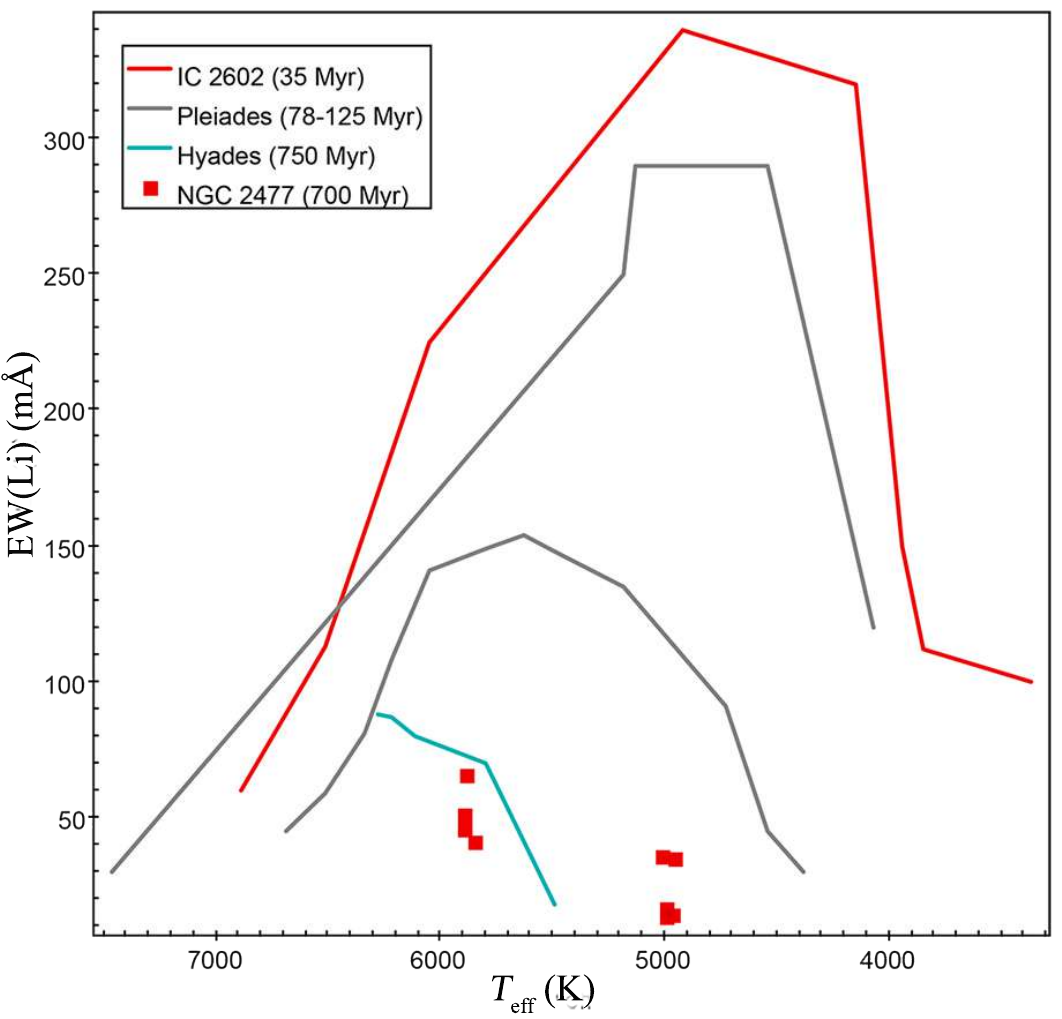} 
\caption{$EW$(Li)-versus-$T_{\rm eff}$ diagram for NGC~2477.}
             \label{fig:163}
    \end{figure}
    
    \clearpage

\subsection{Trumpler~23}

 \begin{figure} [htp]
   \centering
\includegraphics[width=0.9\linewidth, height=5cm]{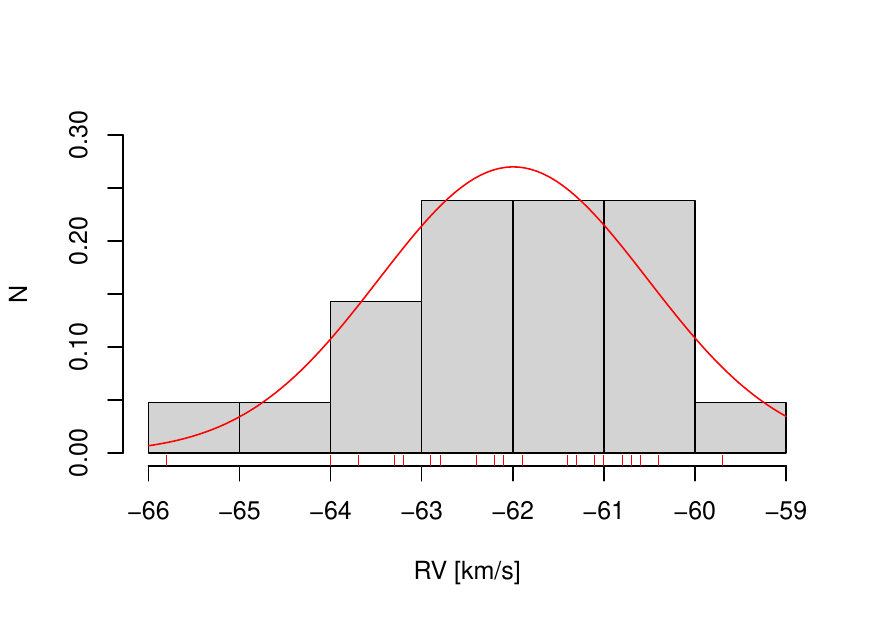}
\caption{$RV$ distribution for Trumpler~23.}
             \label{fig:164}
    \end{figure}
    
           \begin{figure} [htp]
   \centering
\includegraphics[width=0.9\linewidth, height=5cm]{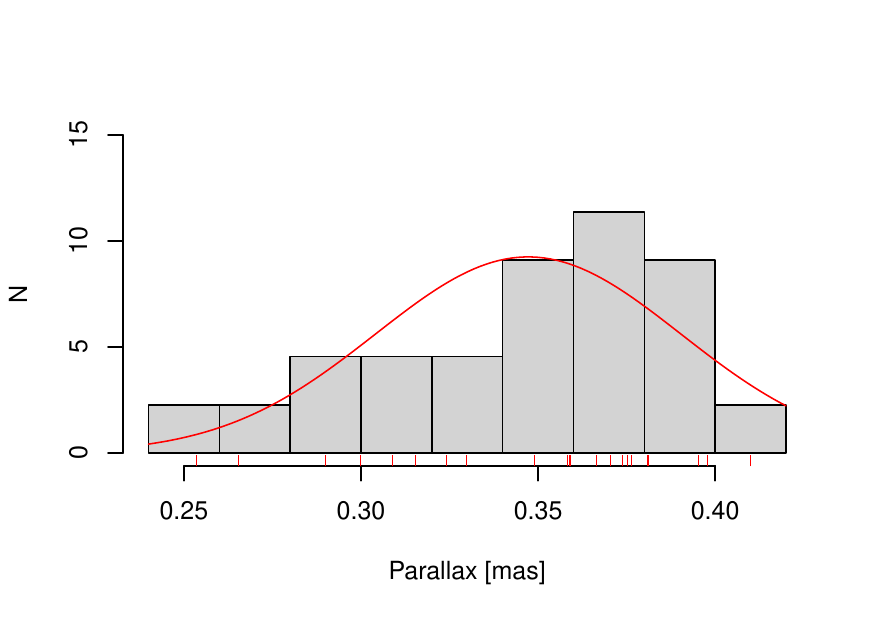}
\caption{Parallax distribution for Trumpler~23.}
             \label{fig:165}
    \end{figure}

               \begin{figure} [htp]
   \centering
   \includegraphics[width=0.9\linewidth]{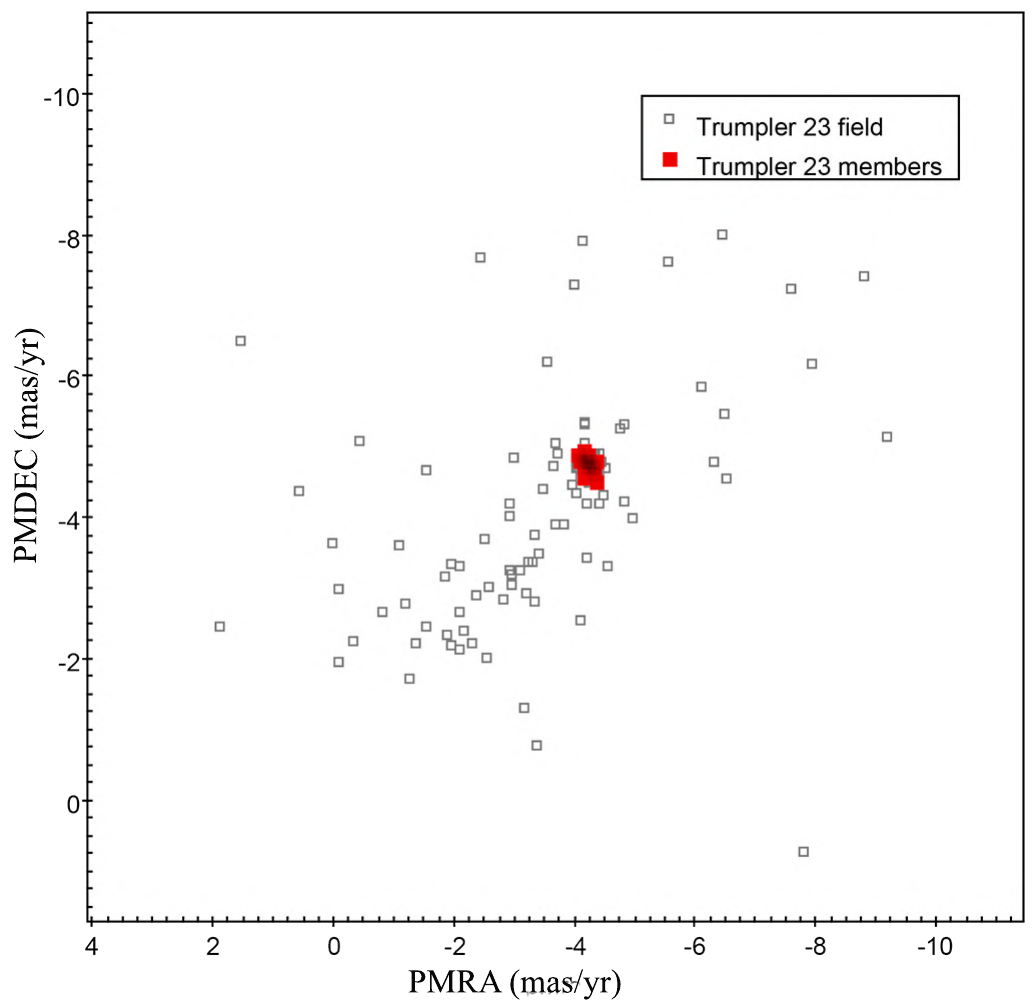}
   \caption{PMs diagram for Trumpler~23.}
             \label{fig:166}
    \end{figure}
    
     \begin{figure} [htp]
   \centering
   \includegraphics[width=0.8\linewidth, height=7cm]{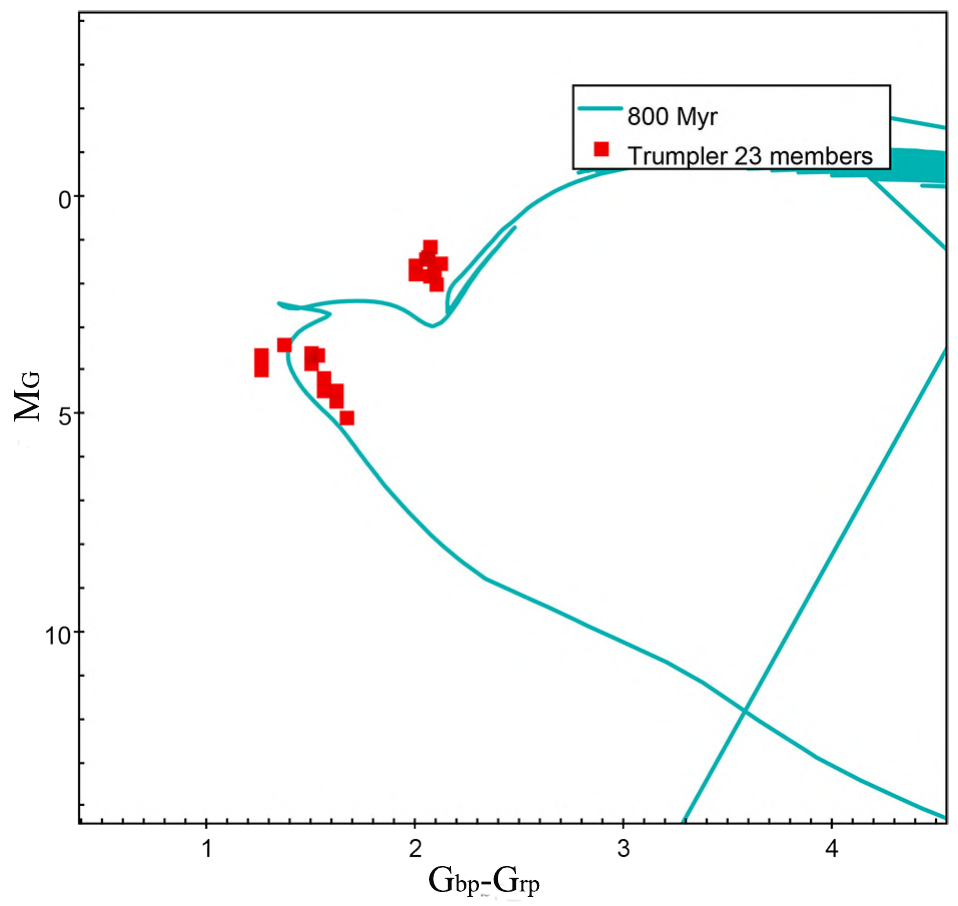}
   \caption{CMD for Trumpler~23.}
             \label{fig:167}
    \end{figure}
    
      \begin{figure} [htp]
   \centering
 \includegraphics[width=0.8\linewidth]{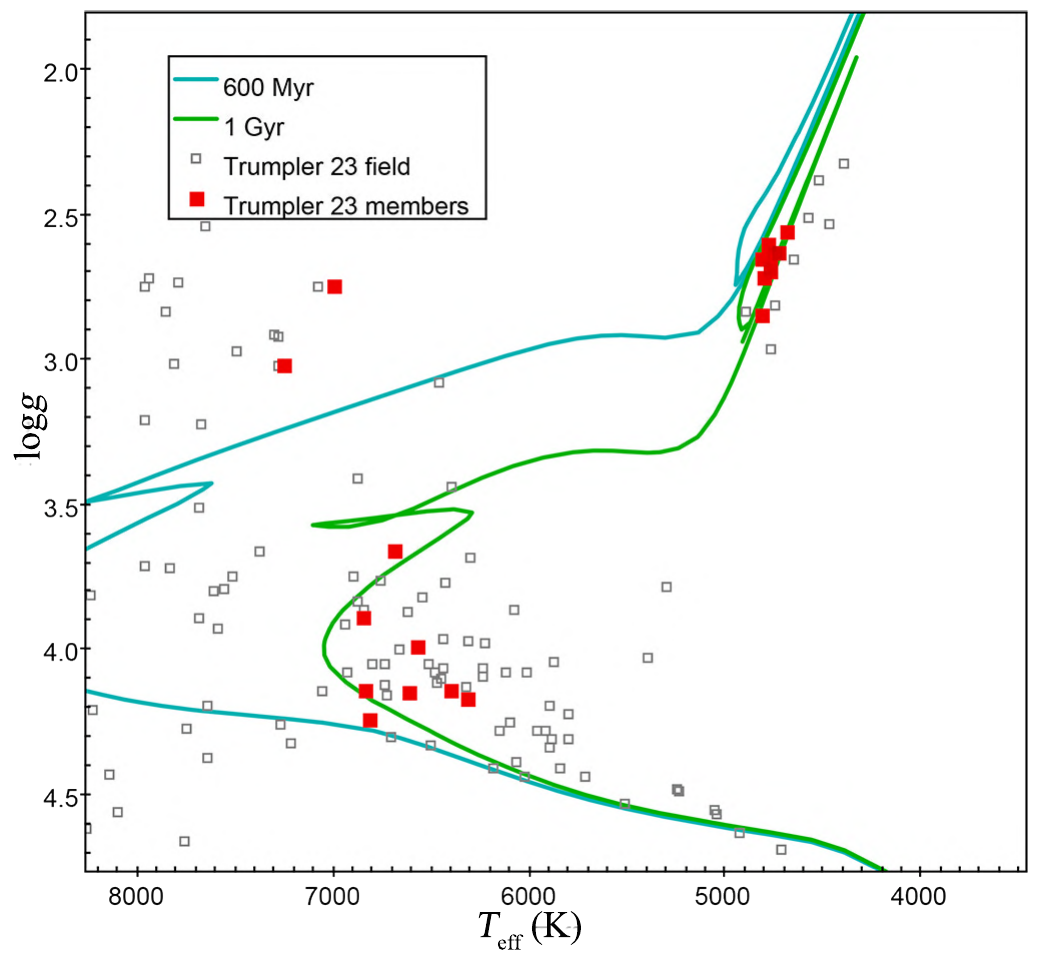} 
\caption{Kiel diagram for Trumpler~23.}
             \label{fig:168}
    \end{figure}

  \begin{figure} [htp]
   \centering
 \includegraphics[width=0.8\linewidth]{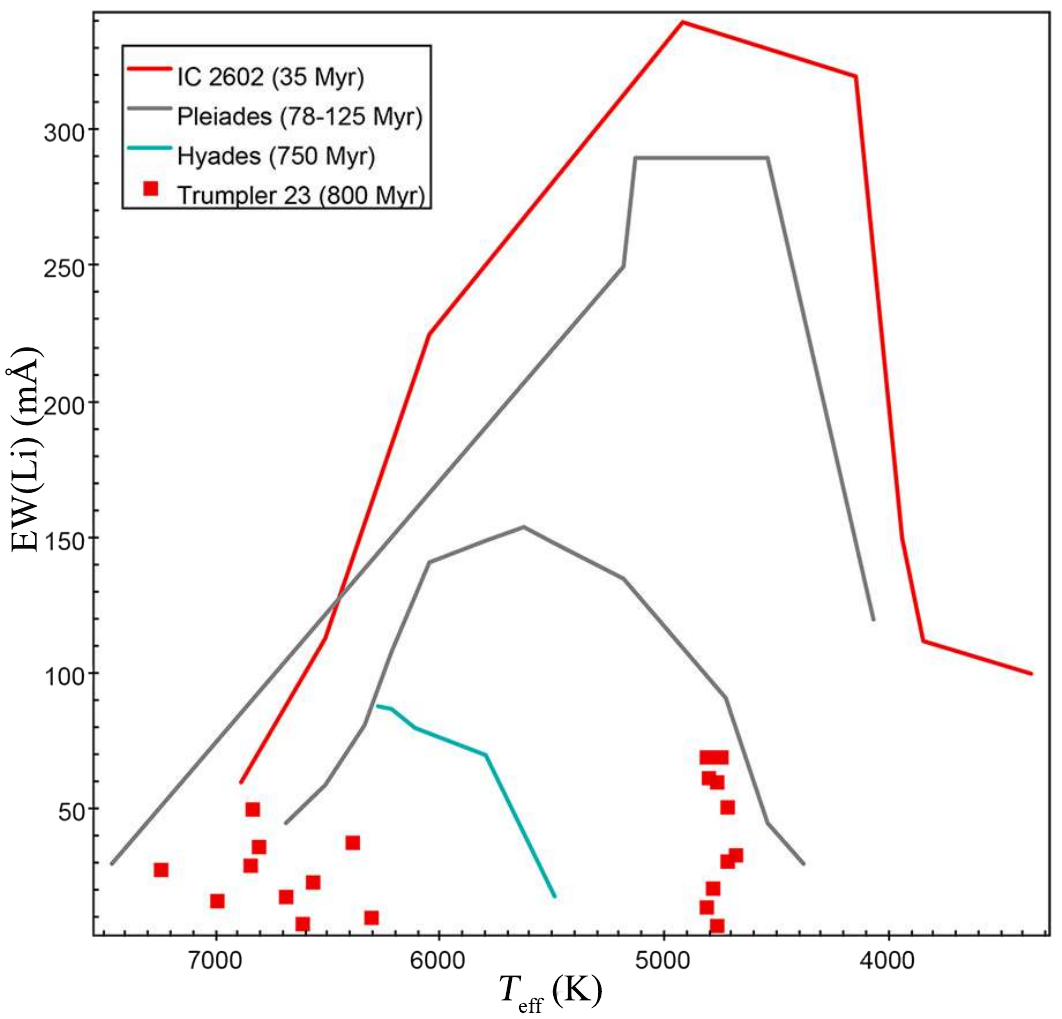} 
\caption{$EW$(Li)-versus-$T_{\rm eff}$ diagram for Trumpler~23.}
             \label{fig:169}
    \end{figure}
    
    \clearpage

\subsection{Berkeley~81}

 \begin{figure} [htp]
   \centering
\includegraphics[width=0.9\linewidth, height=5cm]{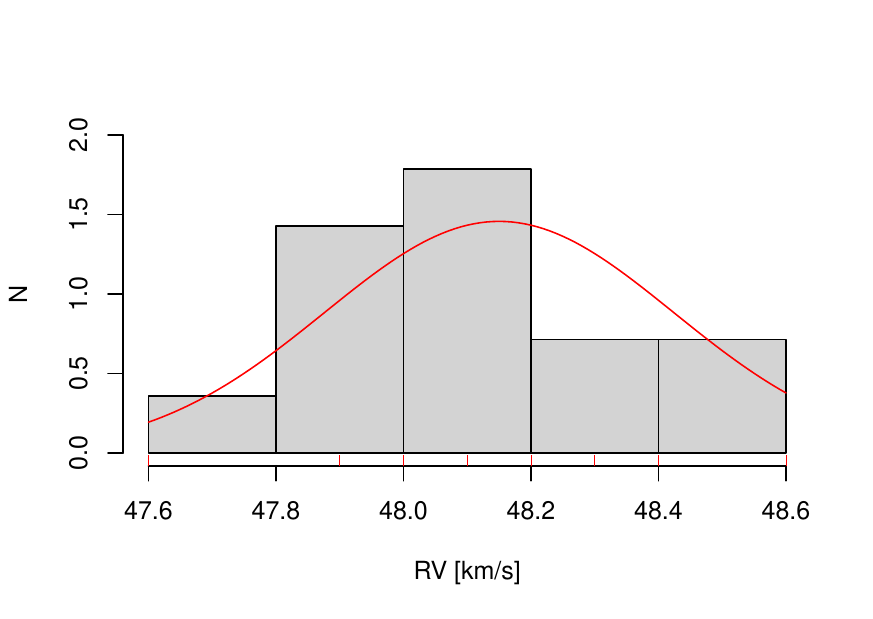}
\caption{$RV$ distribution for Berkeley~81.}
             \label{fig:170}
    \end{figure}
    
           \begin{figure} [htp]
   \centering
\includegraphics[width=0.9\linewidth, height=5cm]{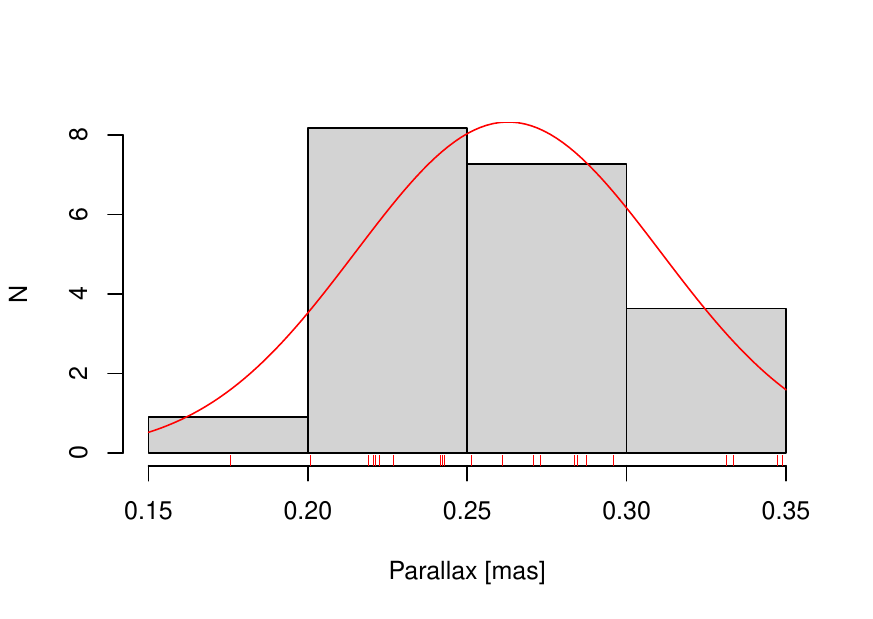}
\caption{Parallax distribution for Berkeley~81.}
             \label{fig:171}
    \end{figure}

               \begin{figure} [htp]
   \centering
   \includegraphics[width=0.9\linewidth]{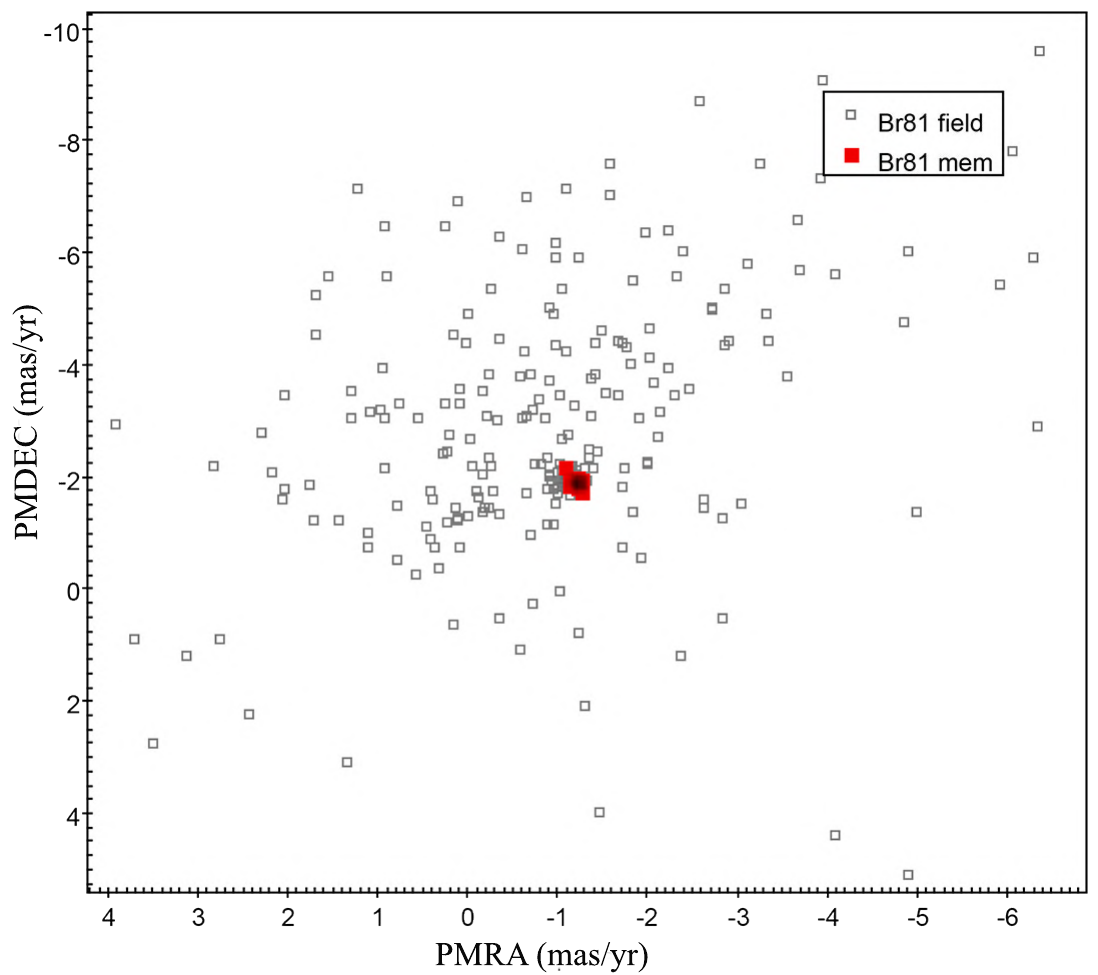}
   \caption{PMs diagram for Berkeley~81}
             \label{fig:172}
    \end{figure}
    
     \begin{figure} [htp]
   \centering
   \includegraphics[width=0.8\linewidth, height=7cm]{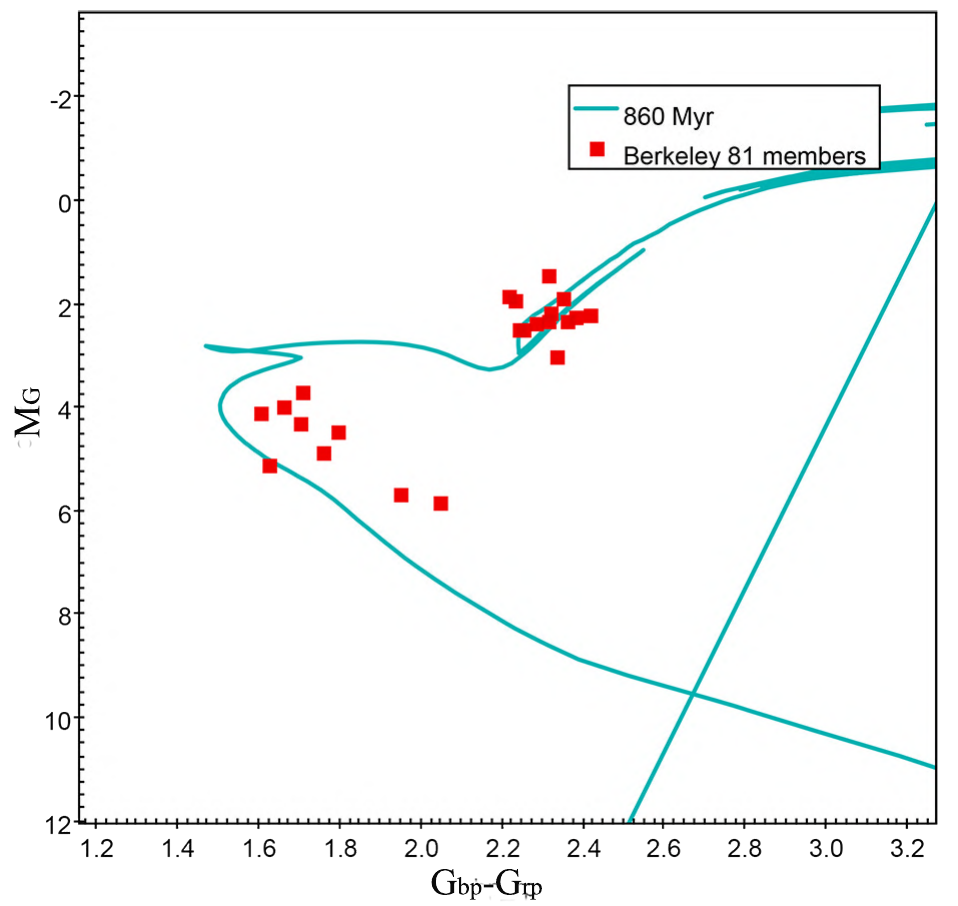}
   \caption{CMD for Berkeley~81.}
             \label{fig:173}
    \end{figure}
    
      \begin{figure} [htp]
   \centering
 \includegraphics[width=0.8\linewidth]{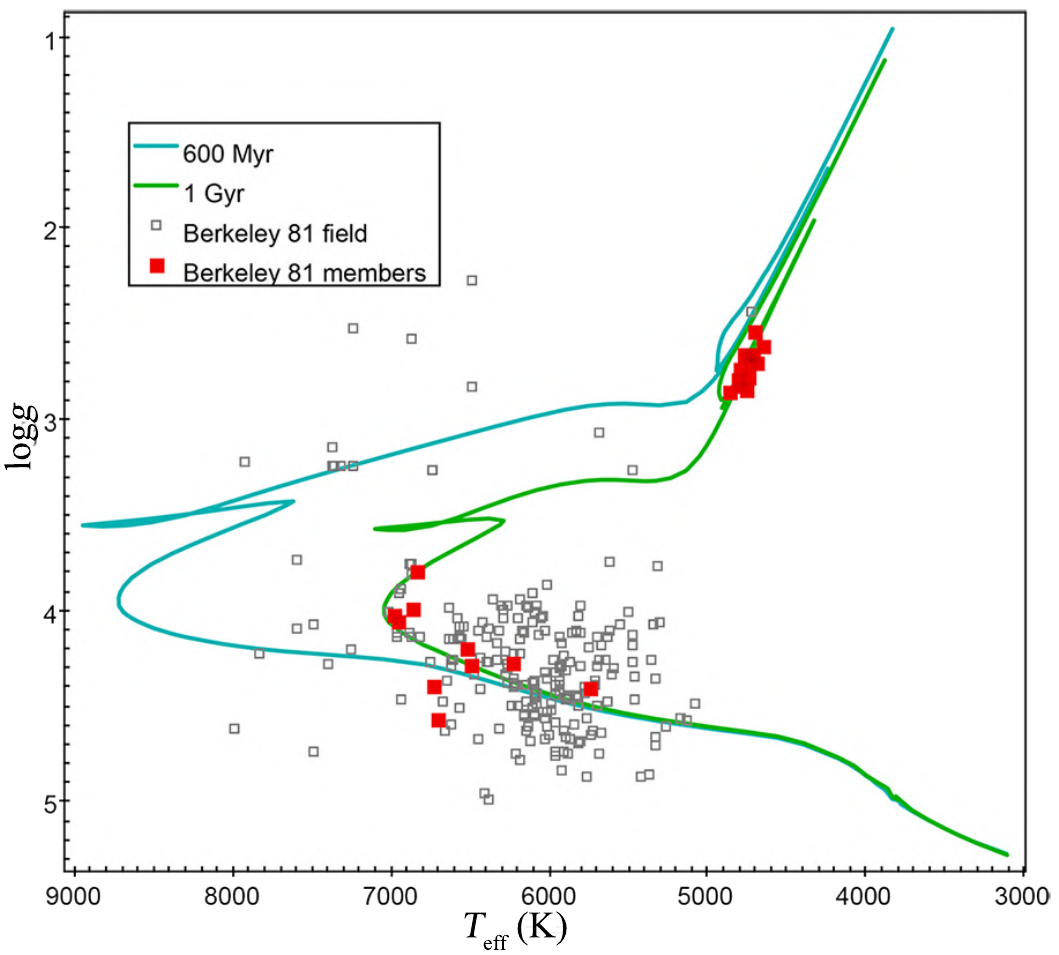} 
\caption{Kiel diagram for Berkeley~81.}
             \label{fig:174}
    \end{figure}

  \begin{figure} [htp]
   \centering
 \includegraphics[width=0.8\linewidth]{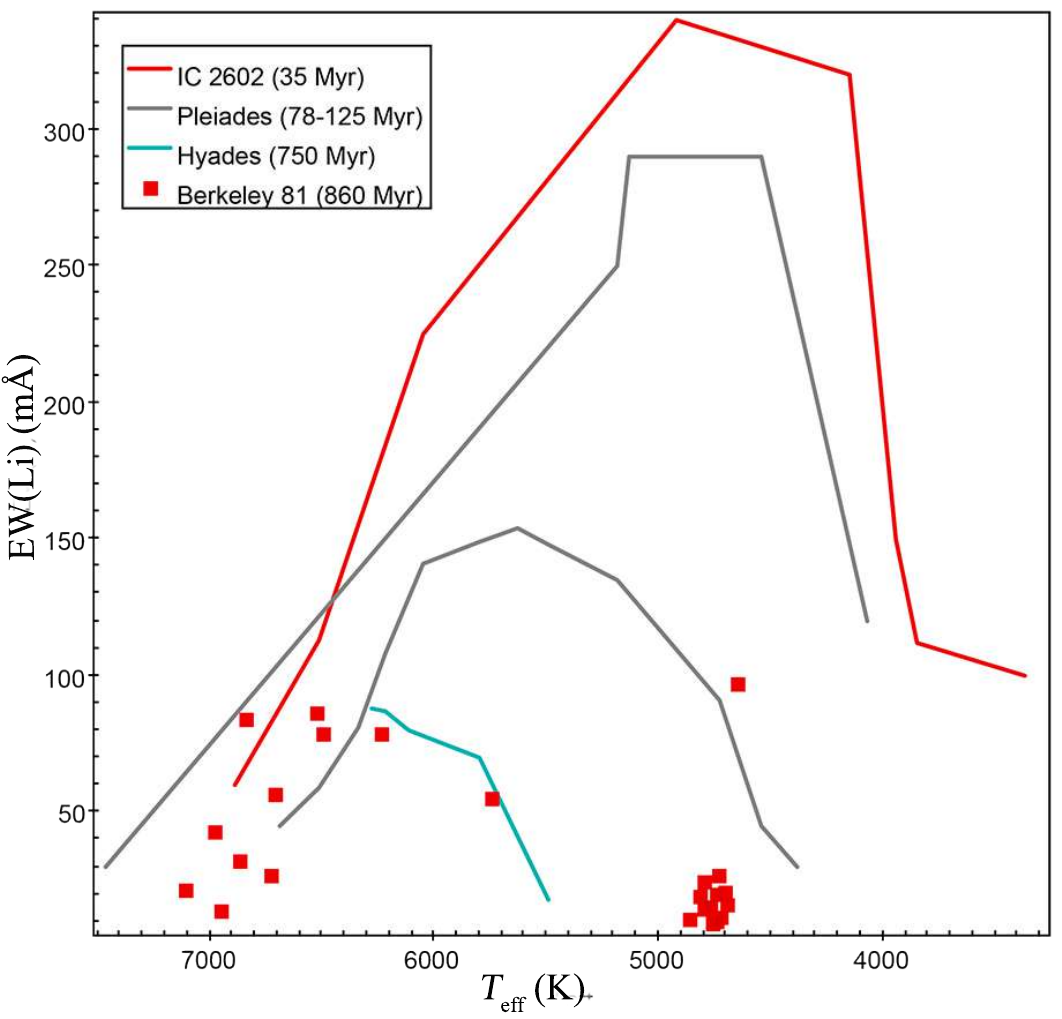} 
\caption{$EW$(Li)-versus-$T_{\rm eff}$ diagram for Berkeley~81.}
             \label{fig:175}
    \end{figure}
    
    \clearpage

\subsection{NGC~2355}

 \begin{figure} [htp]
   \centering
\includegraphics[width=0.9\linewidth, height=5cm]{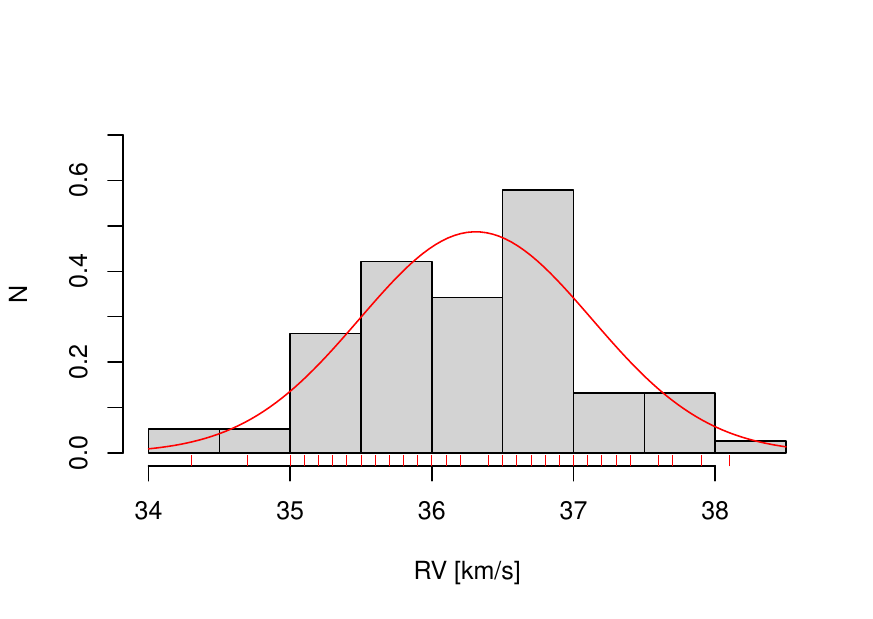}
\caption{$RV$ distribution for NGC~2355.}
             \label{fig:176}
    \end{figure}
    
           \begin{figure} [htp]
   \centering
\includegraphics[width=0.9\linewidth, height=5cm]{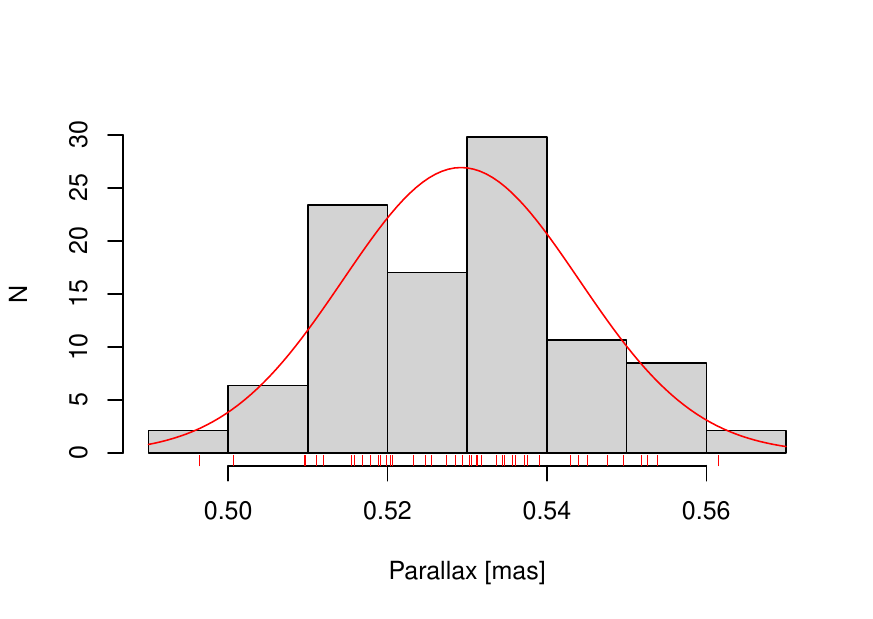}
\caption{Parallax distribution for NGC~2355.}
             \label{fig:177}
    \end{figure}

               \begin{figure} [htp]
   \centering
   \includegraphics[width=0.9\linewidth]{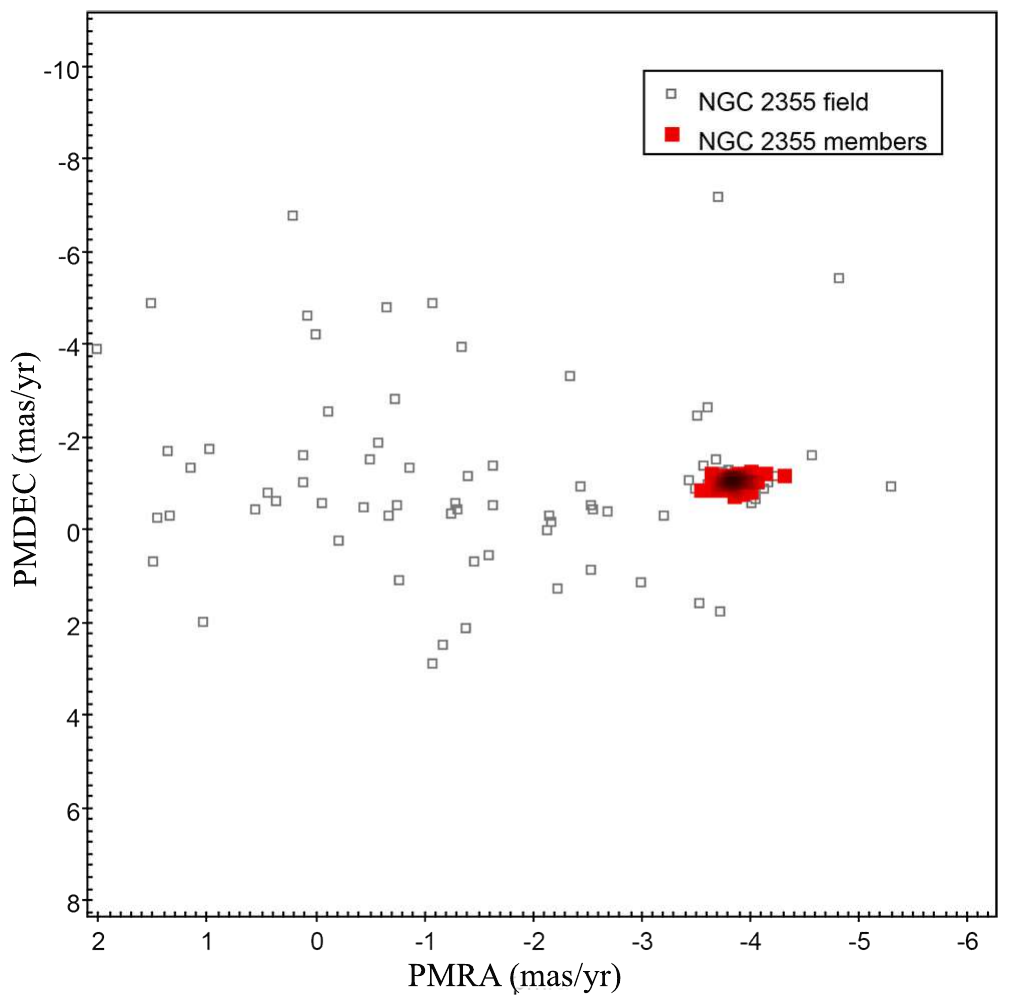}
   \caption{PMs diagram for NGC~2355.}
             \label{fig:178}
    \end{figure}
    
     \begin{figure} [htp]
   \centering
   \includegraphics[width=0.8\linewidth, height=7cm]{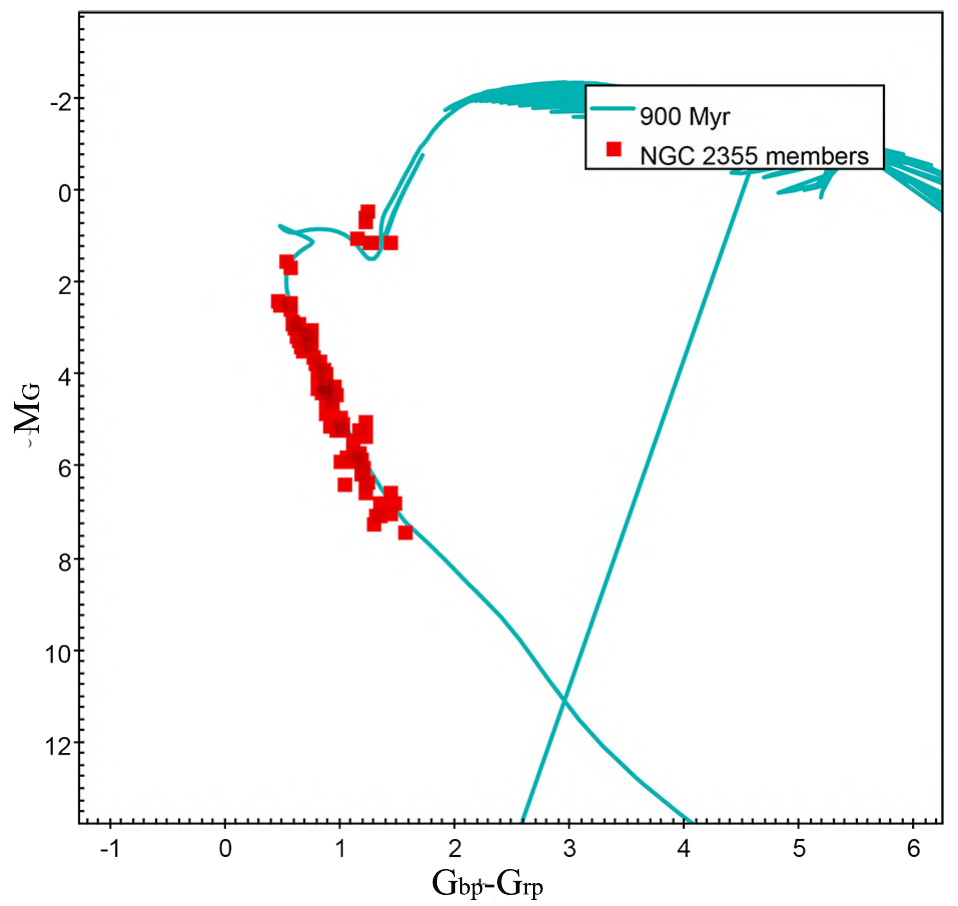}
   \caption{CMD for NGC~2355.}
             \label{fig:179}
    \end{figure}
    
      \begin{figure} [htp]
   \centering
 \includegraphics[width=0.8\linewidth]{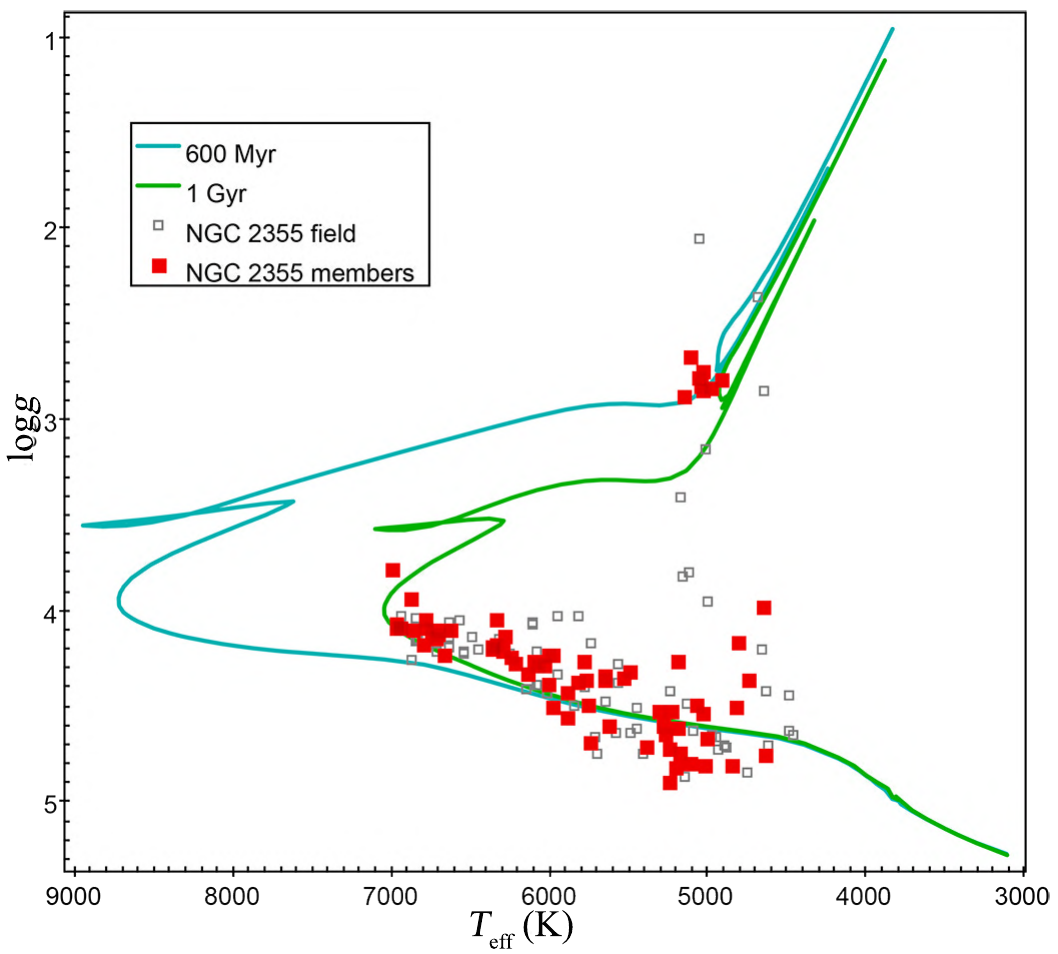} 
\caption{Kiel diagram for NGC~2355.}
             \label{fig:180}
    \end{figure}

  \begin{figure} [htp]
   \centering
 \includegraphics[width=0.8\linewidth]{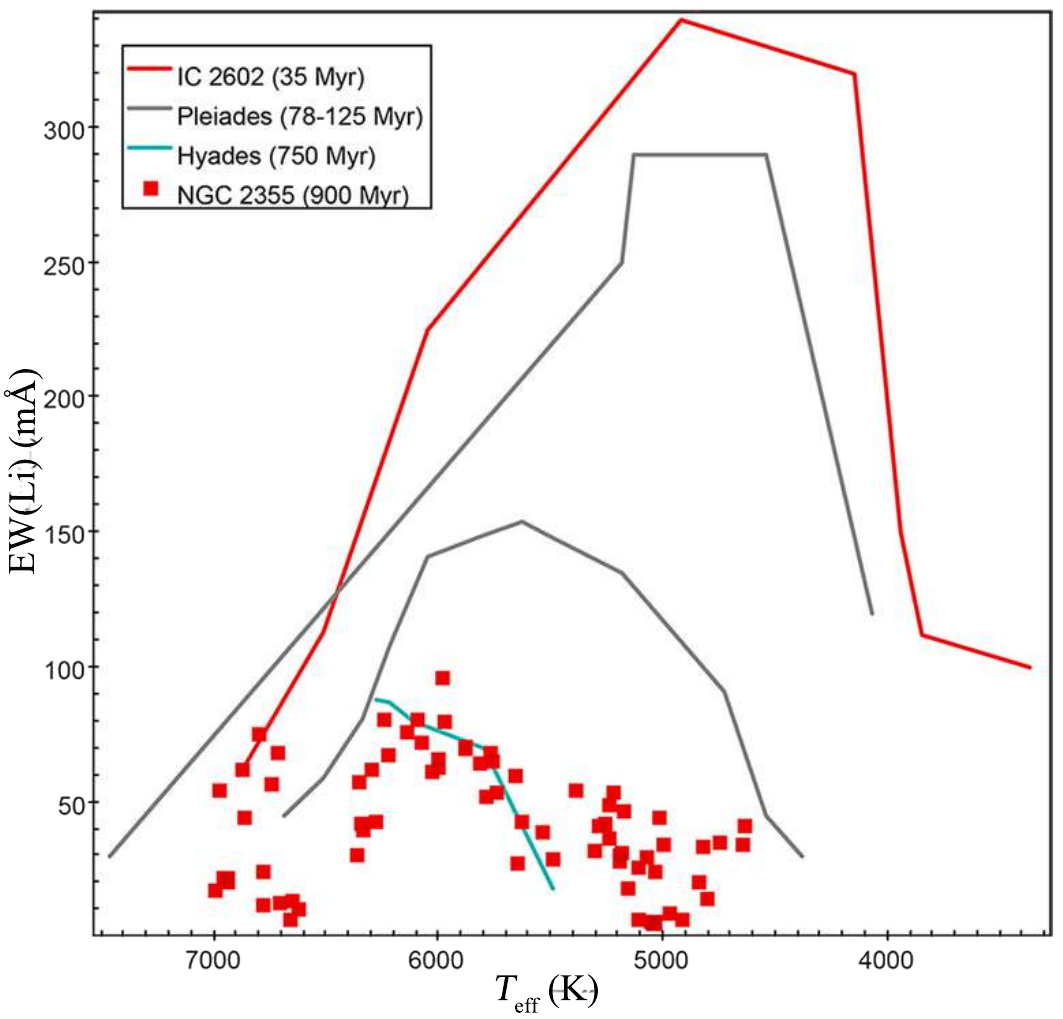} 
\caption{$EW$(Li)-versus-$T_{\rm eff}$ diagram for NGC~2355.}
             \label{fig:181}
    \end{figure}
    
    \clearpage

\subsection{NGC~6802}

 \begin{figure} [htp]
   \centering
\includegraphics[width=0.9\linewidth, height=5cm]{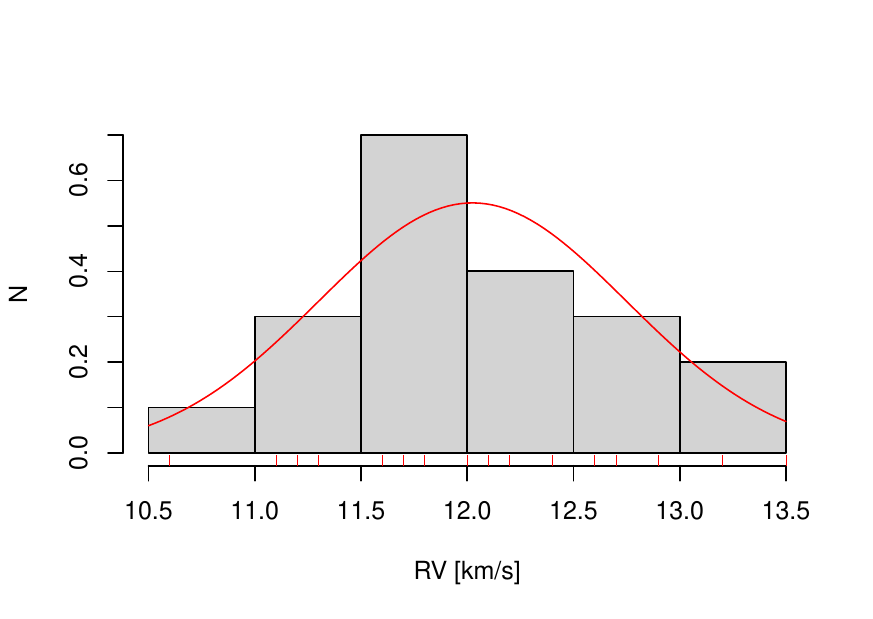}
\caption{$RV$ distribution for NGC~6802.}
             \label{fig:182}
    \end{figure}
    
           \begin{figure} [htp]
   \centering
\includegraphics[width=0.9\linewidth, height=5cm]{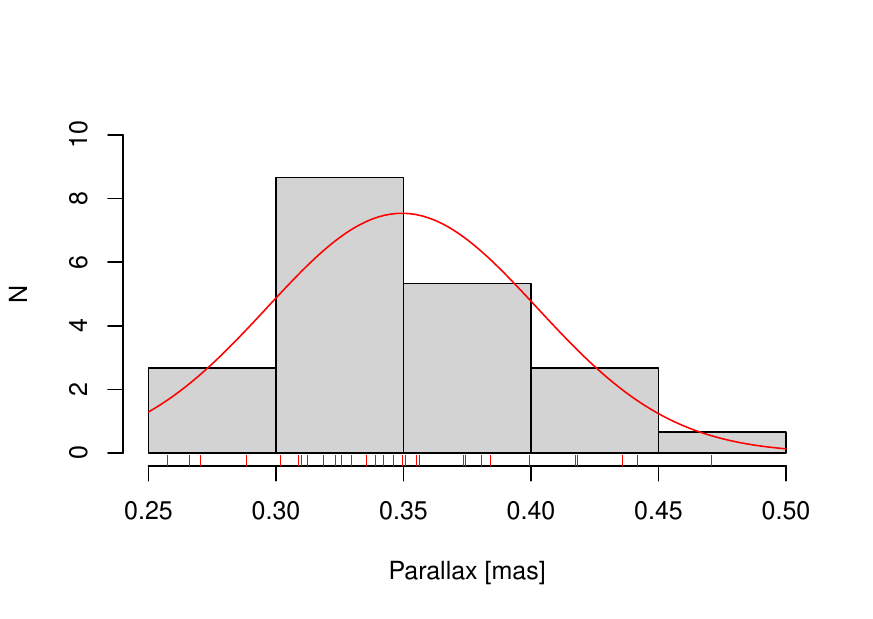}
\caption{Parallax distribution for NGC~6802.}
             \label{fig:183}
    \end{figure}

               \begin{figure} [htp]
   \centering
   \includegraphics[width=0.9\linewidth]{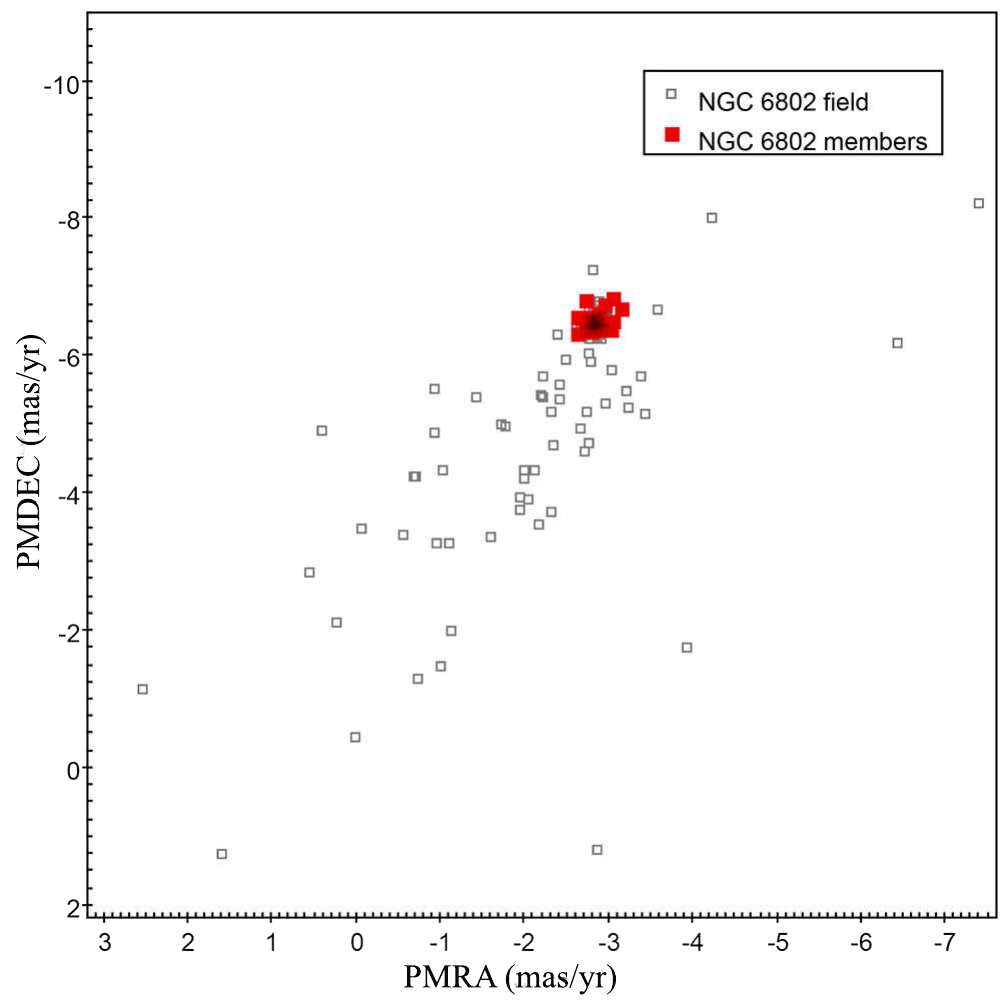}
   \caption{PMs diagram for NGC~6802.}
             \label{fig:184}
    \end{figure}
    
     \begin{figure} [htp]
   \centering
   \includegraphics[width=0.8\linewidth, height=7cm]{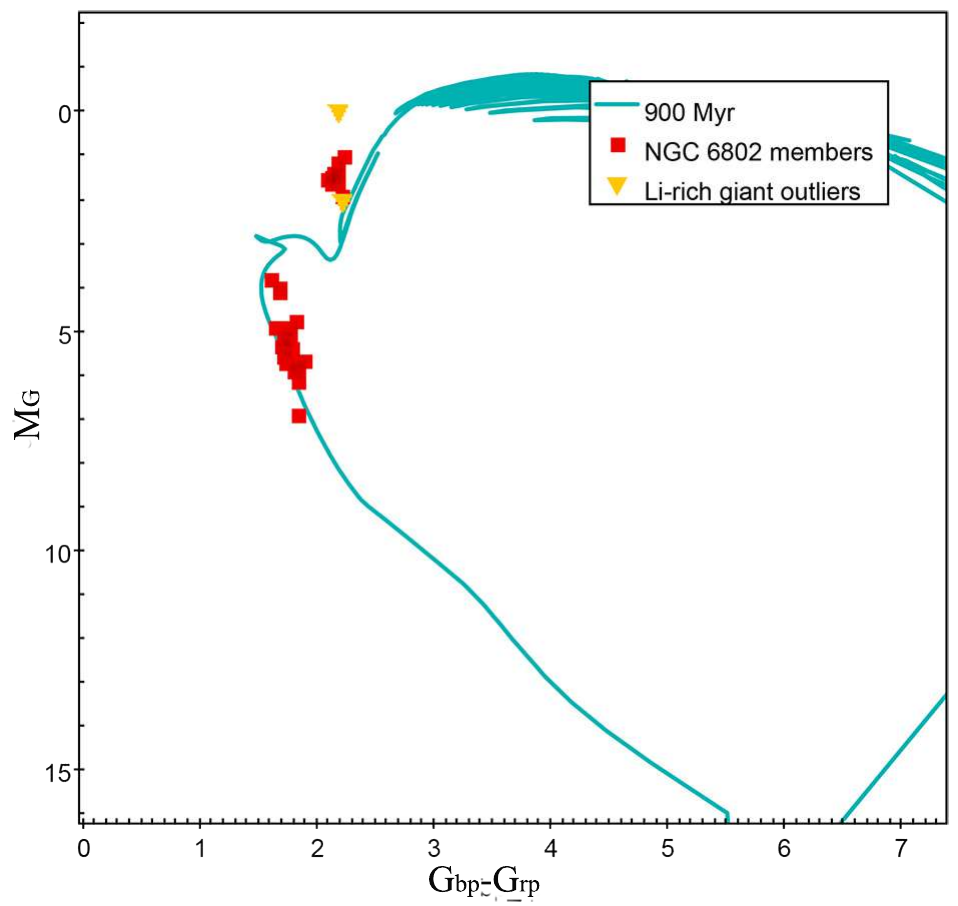}
   \caption{CMD for NGC~6802.}
             \label{fig:185}
    \end{figure}
    
      \begin{figure} [htp]
   \centering
 \includegraphics[width=0.8\linewidth]{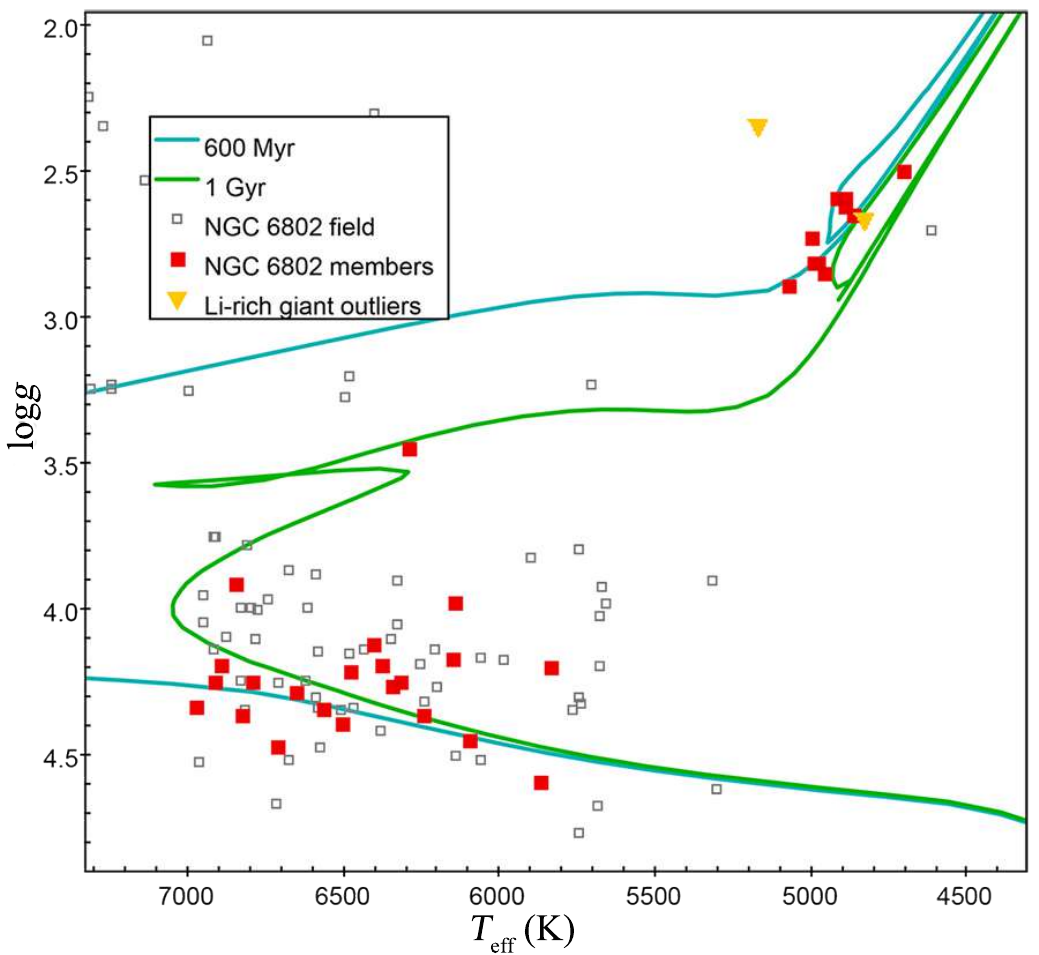} 
\caption{Kiel diagram for NGC~6802.}
             \label{fig:186}
    \end{figure}

  \begin{figure} [htp]
   \centering
 \includegraphics[width=0.8\linewidth]{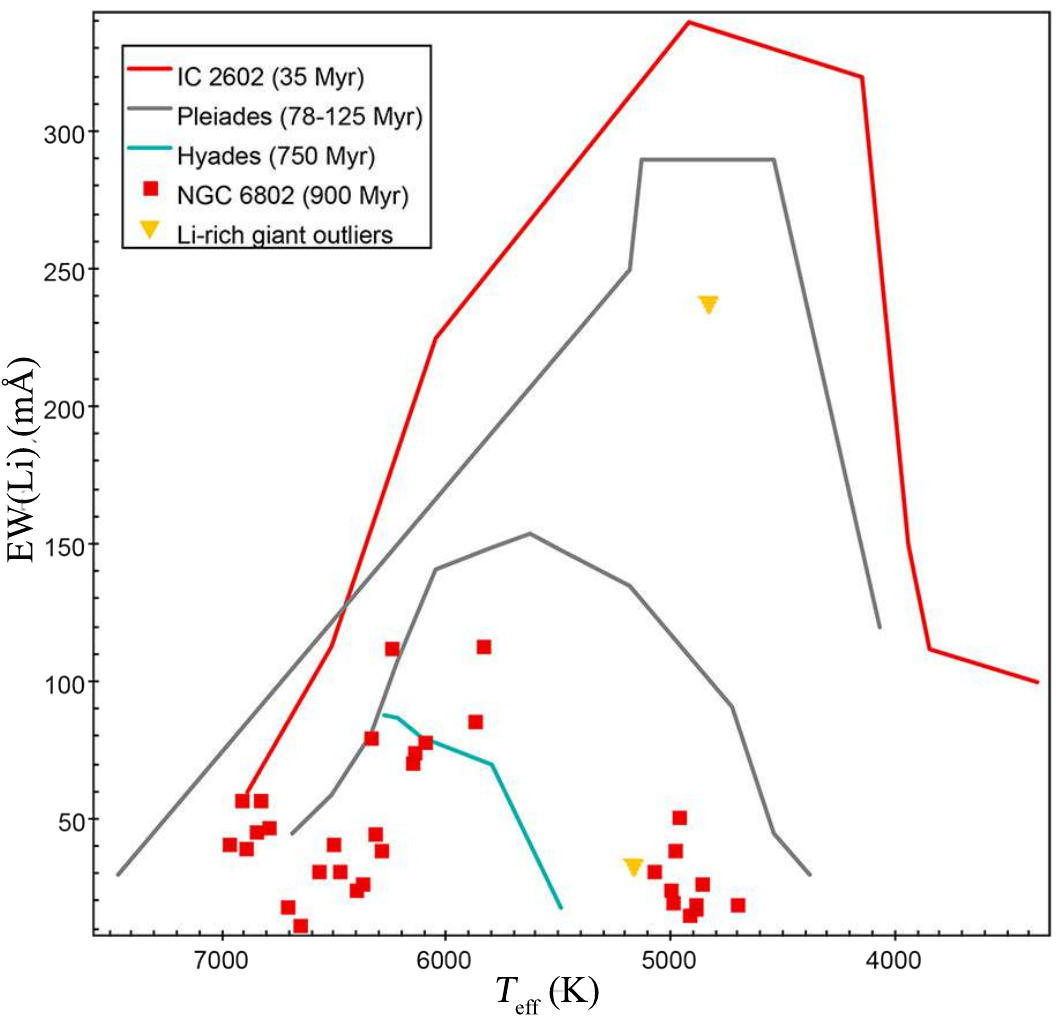} 
\caption{$EW$(Li)-versus-$T_{\rm eff}$ diagram for NGC~6802.}
             \label{fig:187}
    \end{figure}
    
    \clearpage

\subsection{NGC~6005}

\begin{figure} [htp]
   \centering
\includegraphics[width=0.9\linewidth, height=5cm]{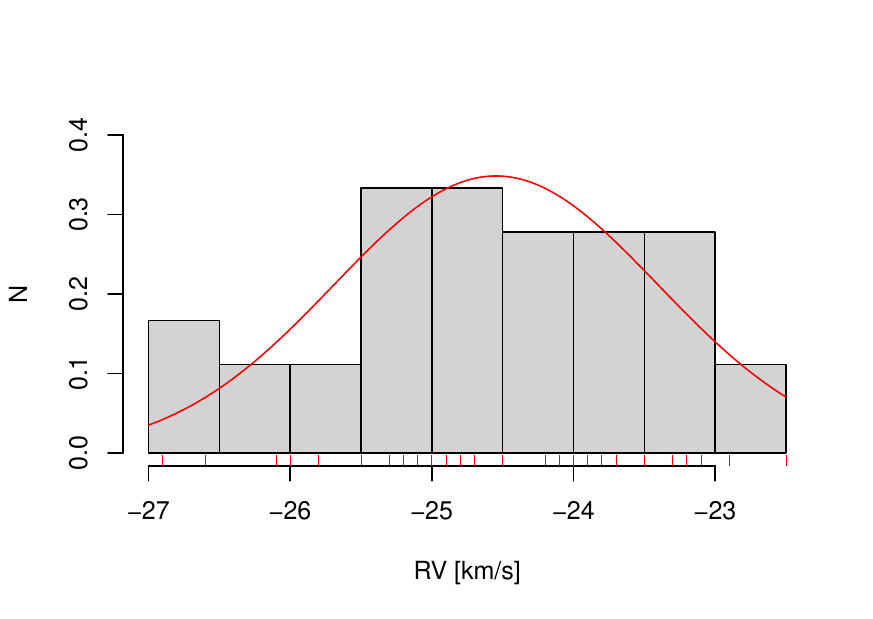}
\caption{$RV$ distribution for NGC~6005.}
             \label{fig:188}
    \end{figure}
    
           \begin{figure} [htp]
   \centering
\includegraphics[width=0.9\linewidth, height=5cm]{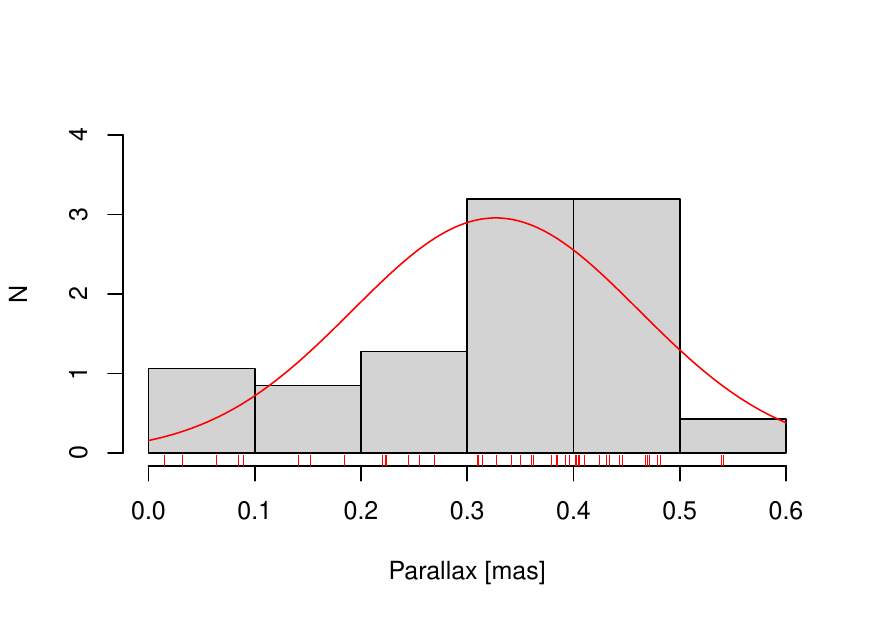}
\caption{Parallax distribution for NGC~6005.}
             \label{fig:189}
    \end{figure}

               \begin{figure} [htp]
   \centering
   \includegraphics[width=0.9\linewidth]{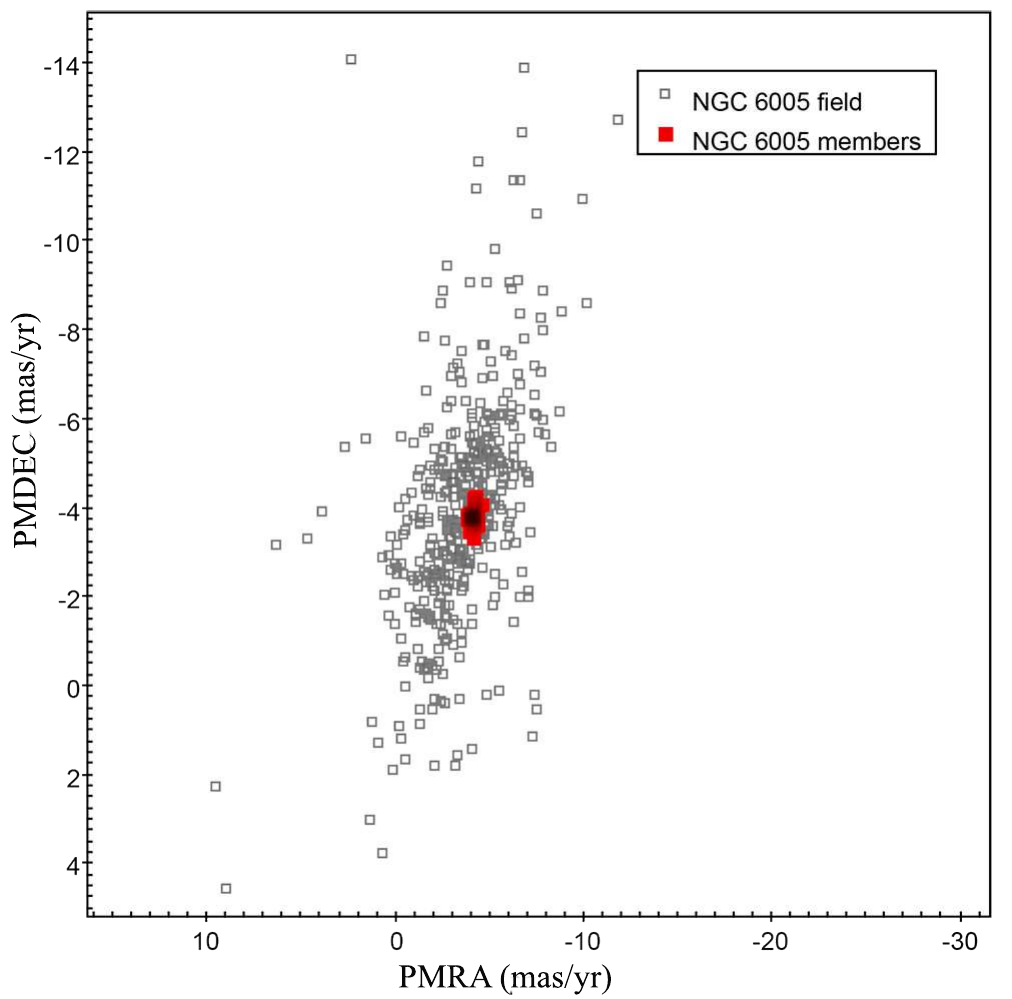}
   \caption{PMs diagram for NGC~6005.}
             \label{fig:190}
    \end{figure}
    
     \begin{figure} [htp]
   \centering
   \includegraphics[width=0.8\linewidth, height=7cm]{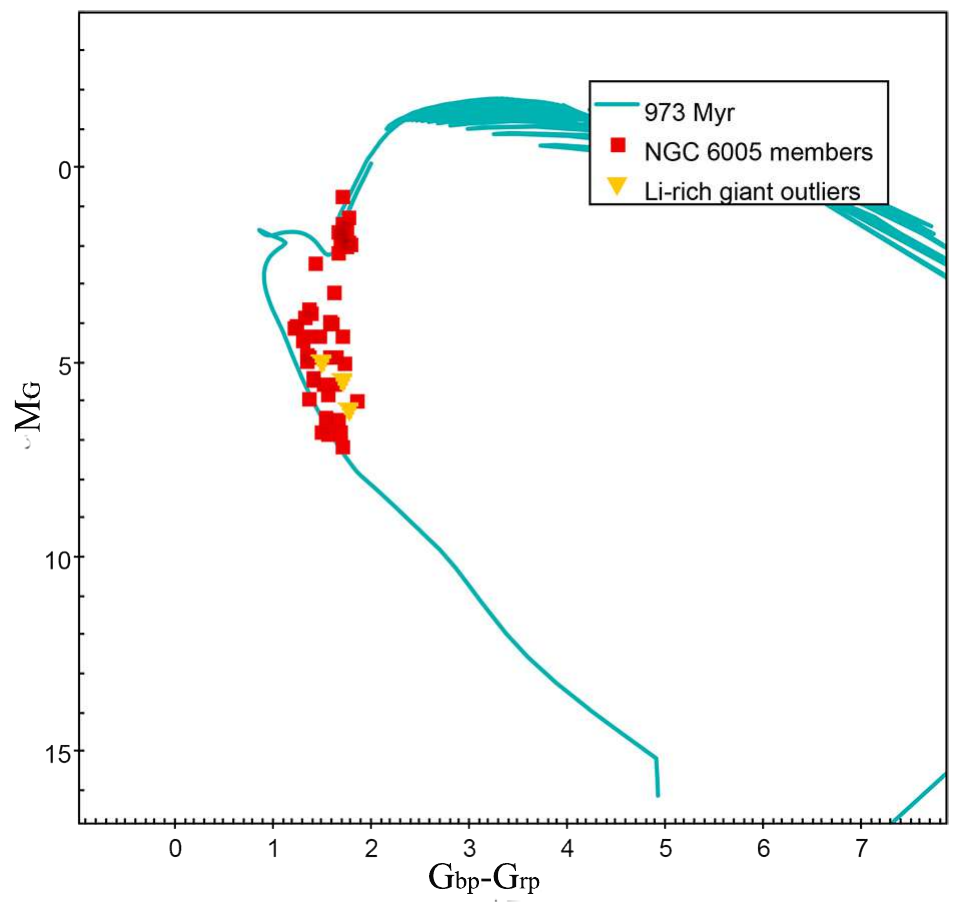}
   \caption{CMD for NGC~6005.}
             \label{fig:191}
    \end{figure}
    
      \begin{figure} [htp]
   \centering
 \includegraphics[width=0.8\linewidth]{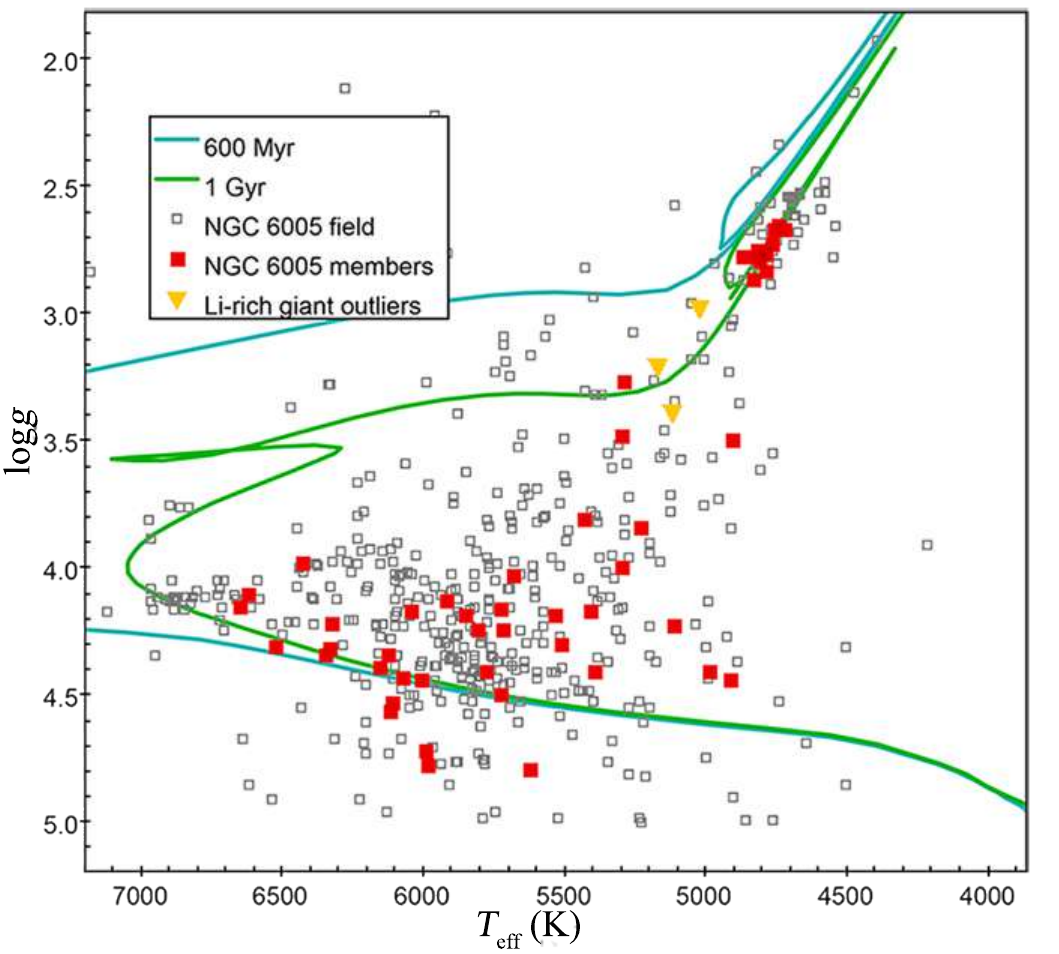} 
\caption{Kiel diagram for NGC~6005.}
             \label{fig:192}
    \end{figure}

  \begin{figure} [htp]
   \centering
 \includegraphics[width=0.8\linewidth]{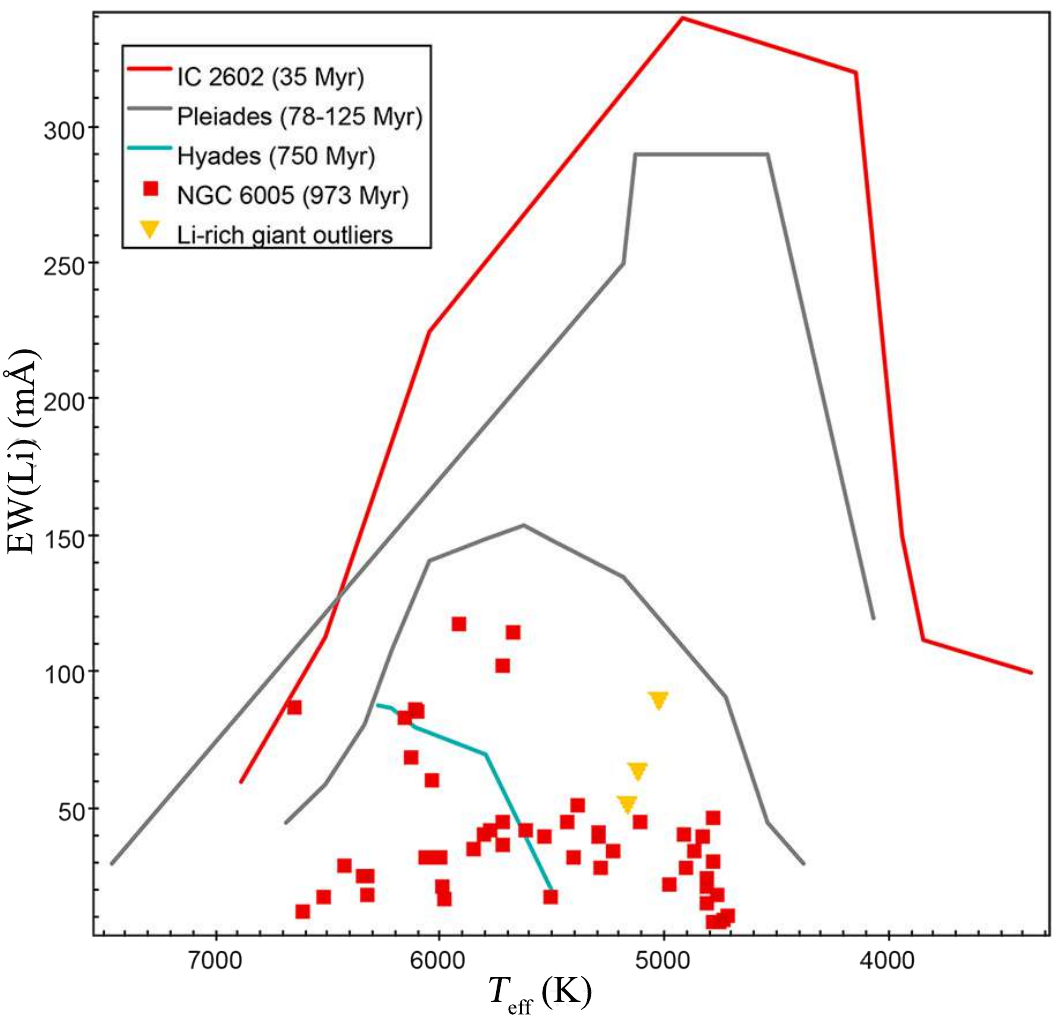} 
\caption{$EW$(Li)-versus-$T_{\rm eff}$ diagram for NGC~6005.}
             \label{fig:193}
    \end{figure}
    
    \clearpage

\subsection{Pismis~18}

\begin{figure} [htp]
   \centering
\includegraphics[width=0.9\linewidth, height=5cm]{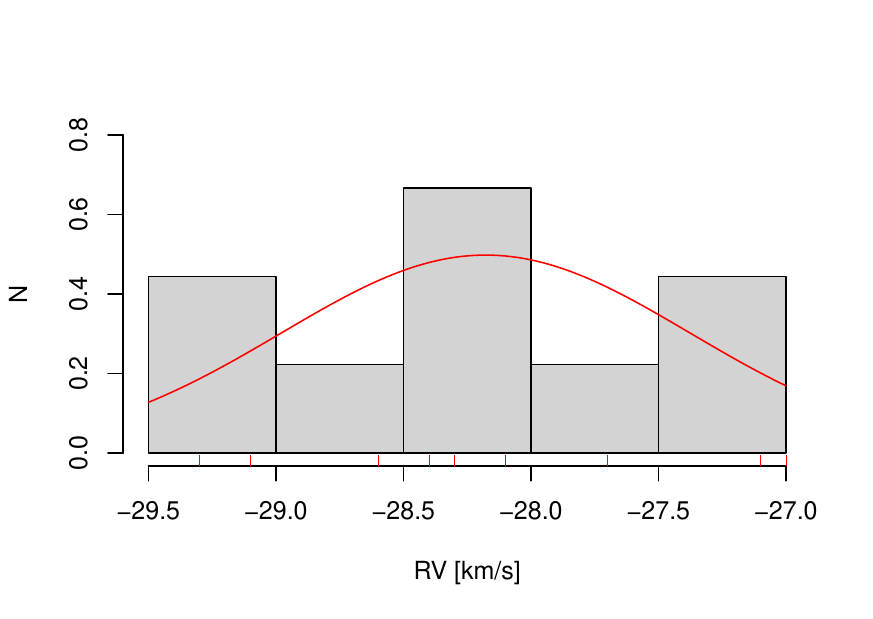}
\caption{$RV$ distribution for Pismis~18.}
             \label{fig:194}
    \end{figure}
    
           \begin{figure} [htp]
   \centering
\includegraphics[width=0.9\linewidth, height=5cm]{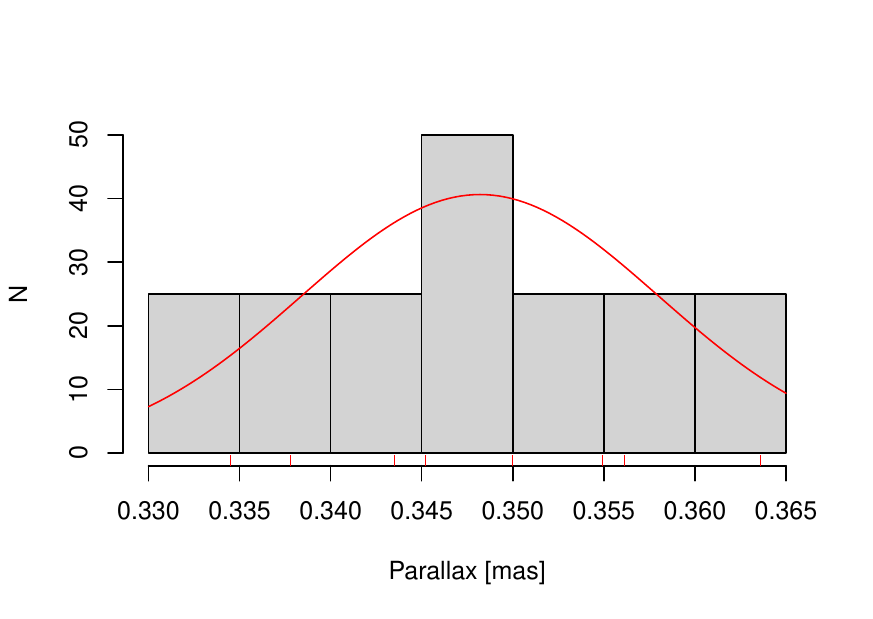}
\caption{Parallax distribution for Pismis~18.}
             \label{fig:195}
    \end{figure}

               \begin{figure} [htp]
   \centering
   \includegraphics[width=0.9\linewidth]{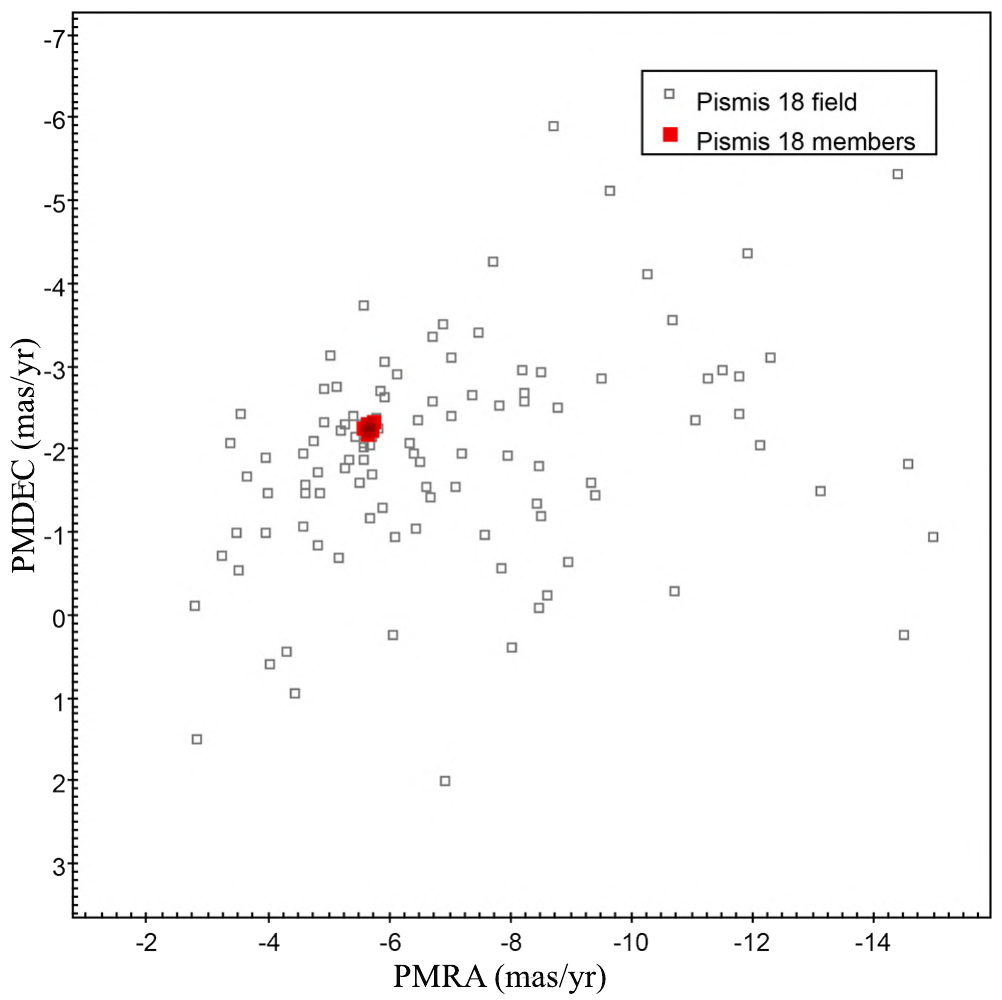}
   \caption{PMs diagram for Pismis~18.}
             \label{fig:196}
    \end{figure}
    
     \begin{figure} [htp]
   \centering
   \includegraphics[width=0.8\linewidth, height=7cm]{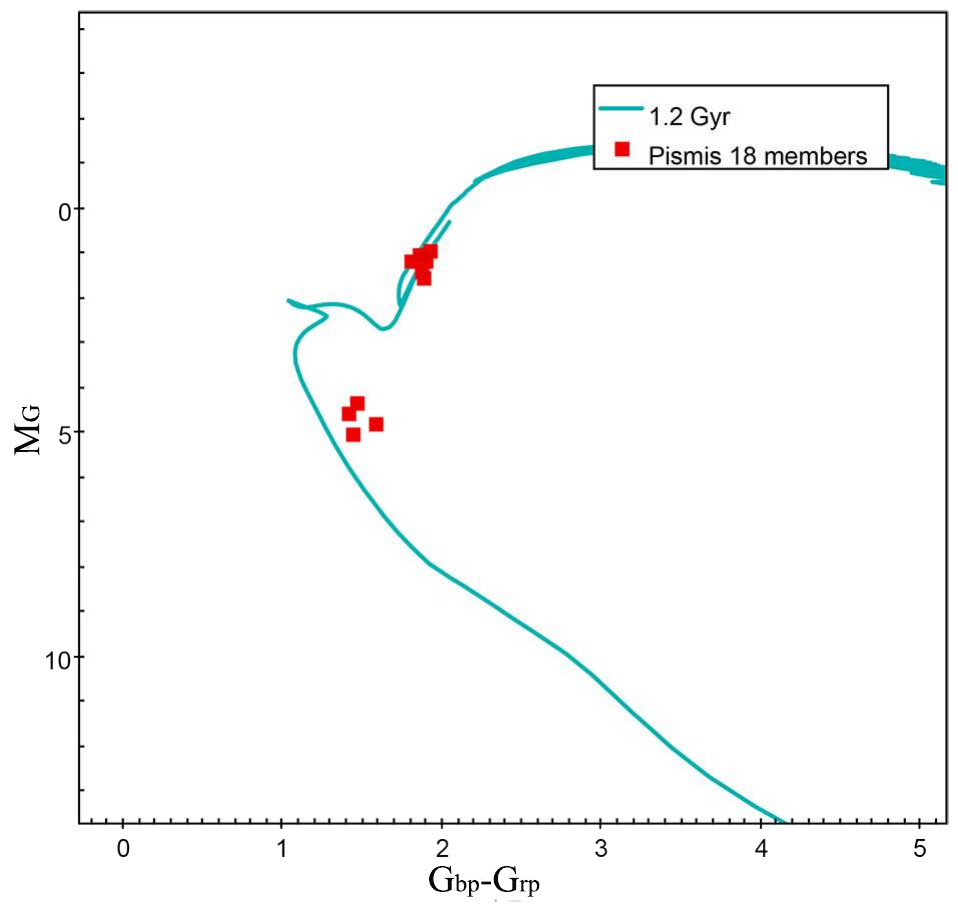}
   \caption{CMD for Pismis~18.}
             \label{fig:197}
    \end{figure}
    
      \begin{figure} [htp]
   \centering
 \includegraphics[width=0.8\linewidth]{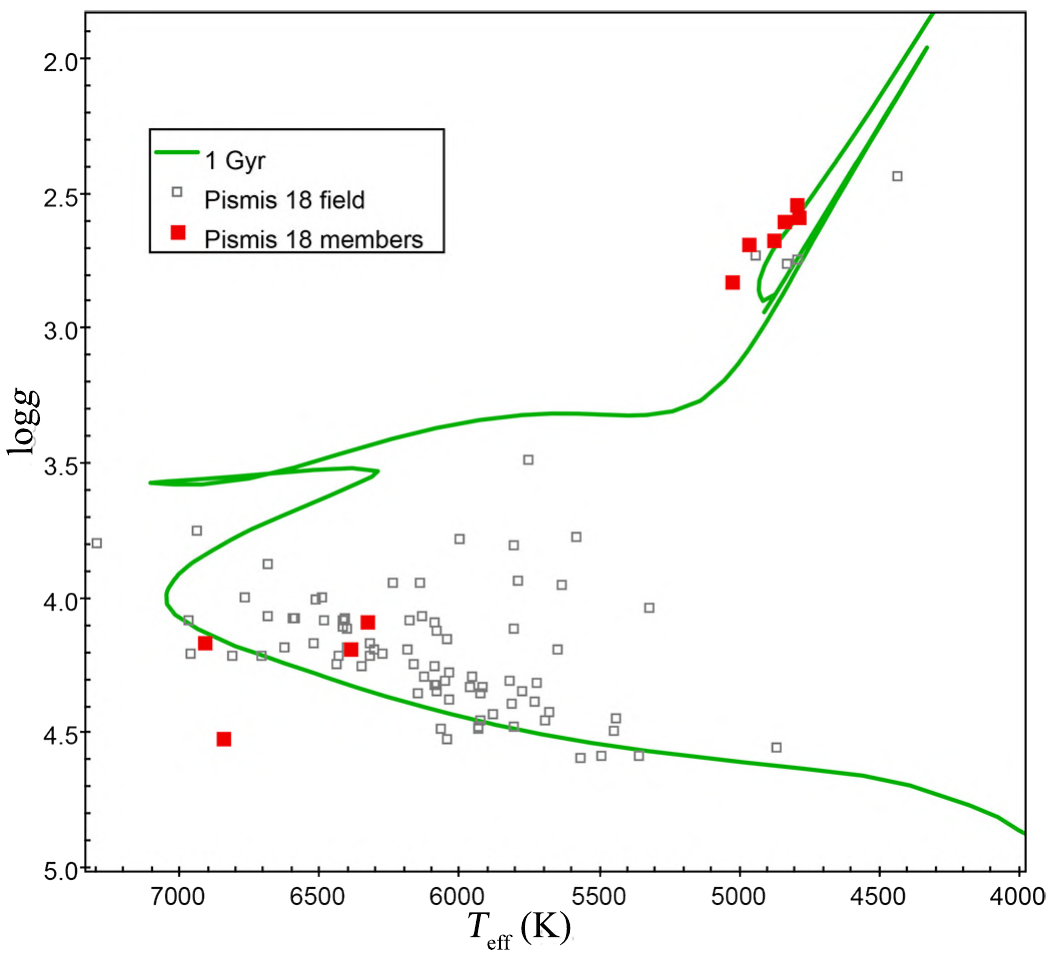} 
\caption{Kiel diagram for Pismis~18.}
             \label{fig:198}
    \end{figure}

  \begin{figure} [htp]
   \centering
 \includegraphics[width=0.8\linewidth]{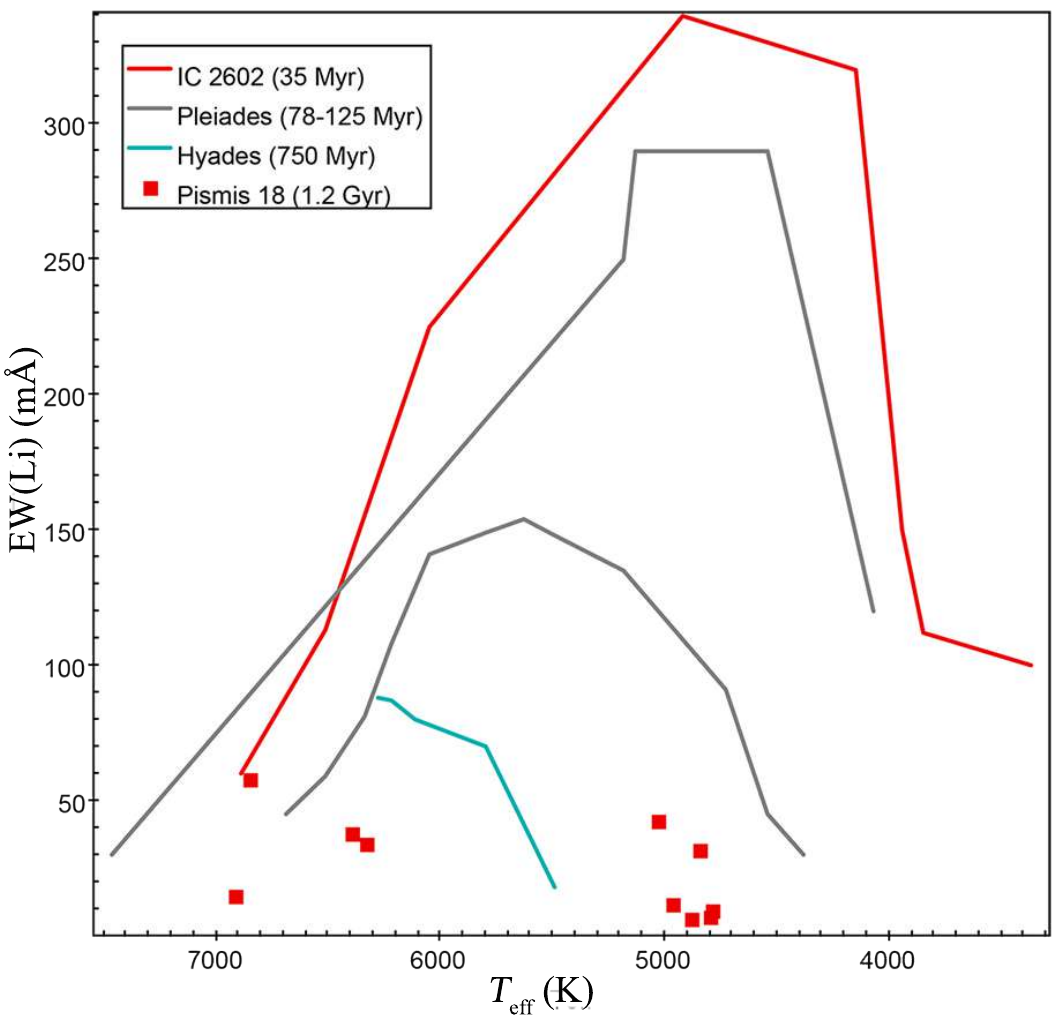} 
\caption{$EW$(Li)-versus-$T_{\rm eff}$ diagram for Pismis~18.}
             \label{fig:199}
    \end{figure}
    
    \clearpage

\subsection{Melotte~71}

\begin{figure} [htp]
   \centering
\includegraphics[width=0.9\linewidth, height=5cm]{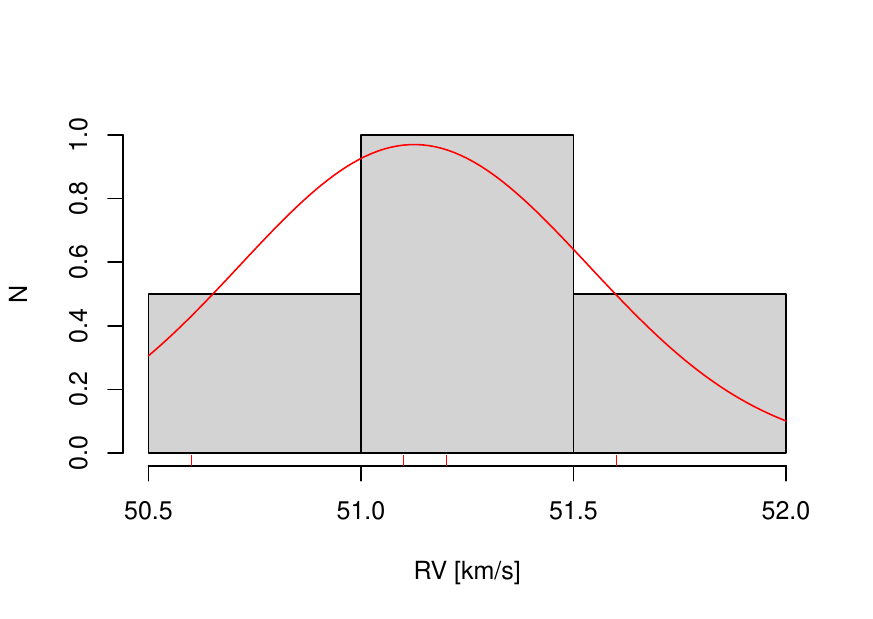}
\caption{$RV$ distribution for Melotte~71.}
             \label{fig:200}
    \end{figure}
    
           \begin{figure} [htp]
   \centering
\includegraphics[width=0.9\linewidth, height=5cm]{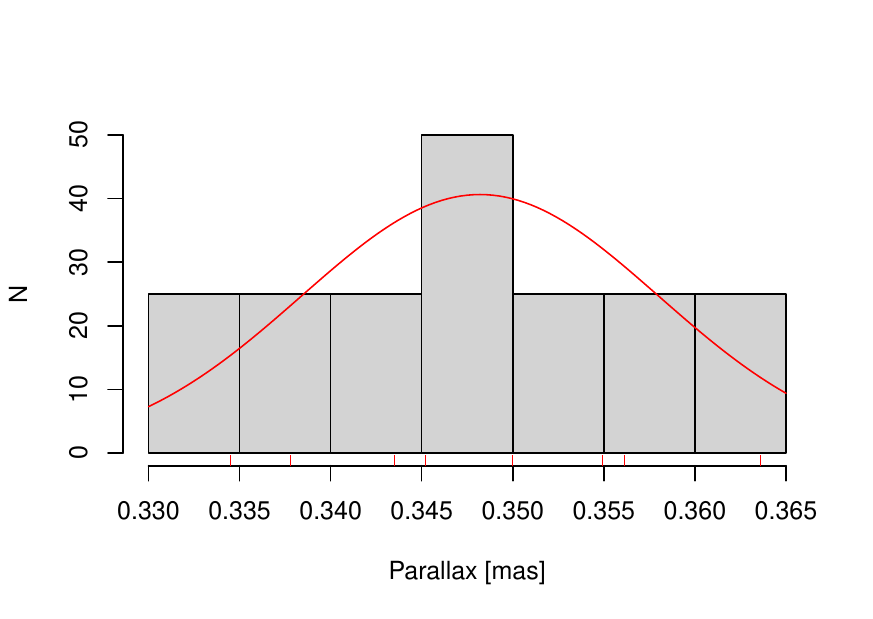}
\caption{Parallax distribution for Melotte~71.}
             \label{fig:201}
    \end{figure}

               \begin{figure} [htp]
   \centering
   \includegraphics[width=0.9\linewidth]{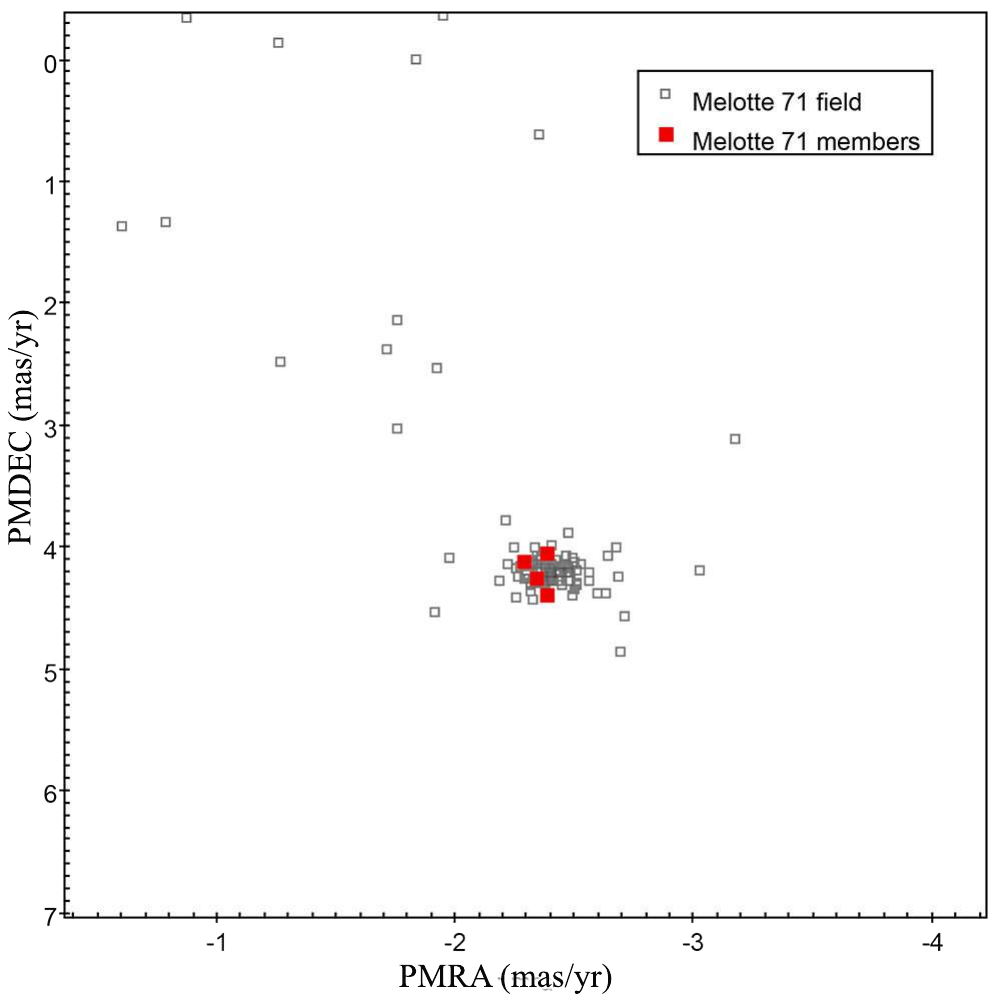}
   \caption{PMs diagram for Melotte~71.}
             \label{fig:202}
    \end{figure}
    
     \begin{figure} [htp]
   \centering
   \includegraphics[width=0.8\linewidth, height=7cm]{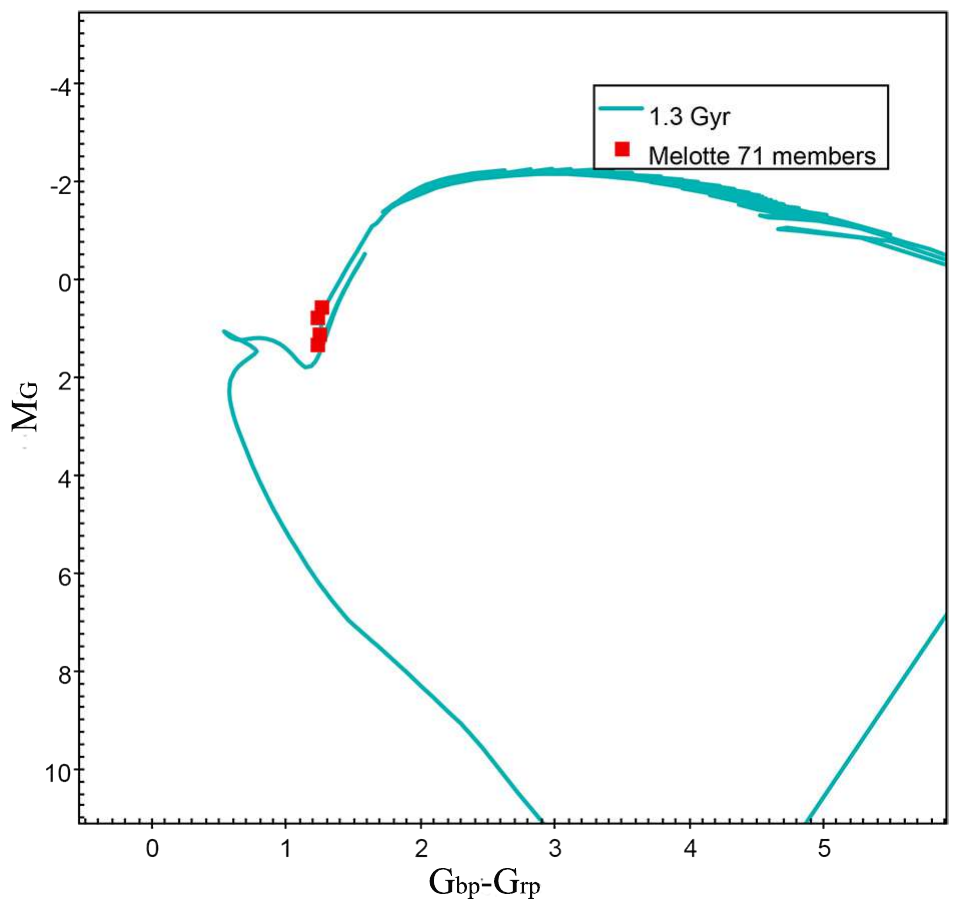}
   \caption{CMD for Melotte~71.}
             \label{fig:203}
    \end{figure}
    
      \begin{figure} [htp]
   \centering
 \includegraphics[width=0.8\linewidth]{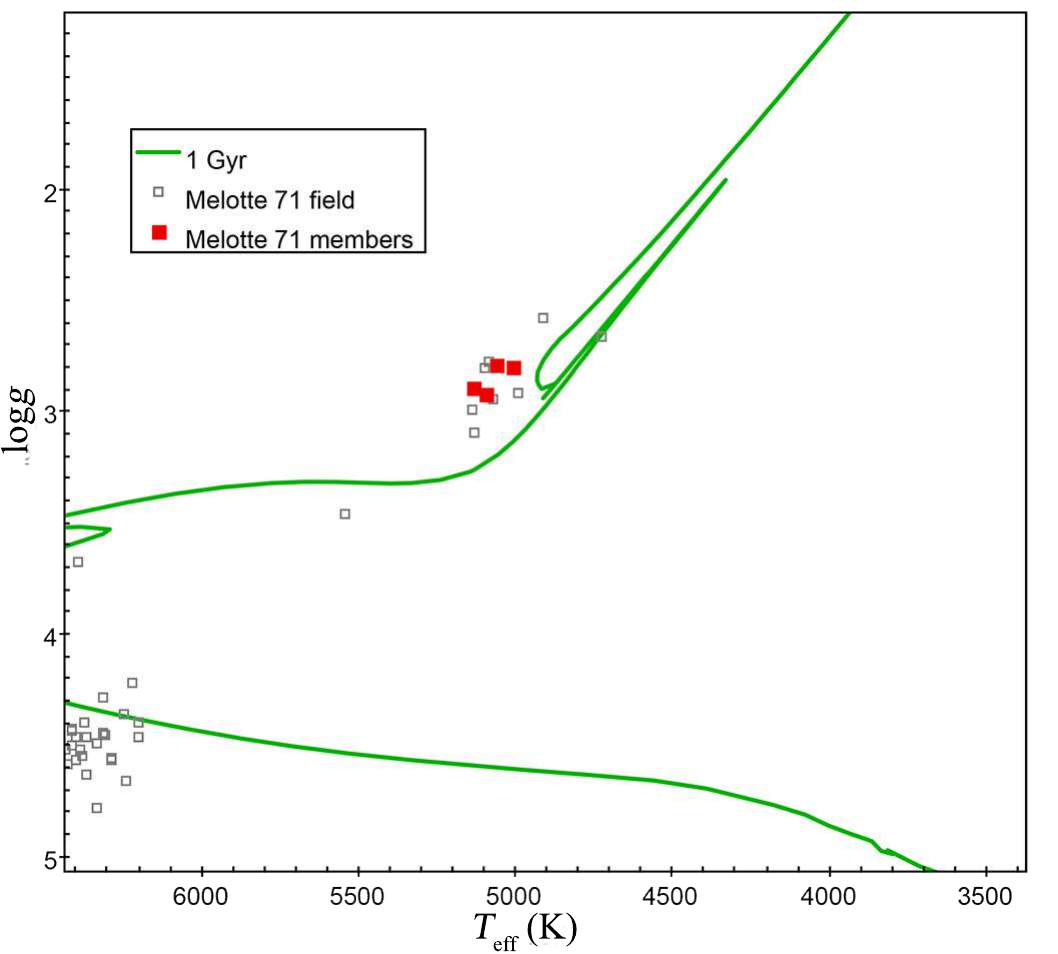} 
\caption{Kiel diagram for Melotte~71.}
             \label{fig:204}
    \end{figure}

  \begin{figure} [htp]
   \centering
 \includegraphics[width=0.8\linewidth]{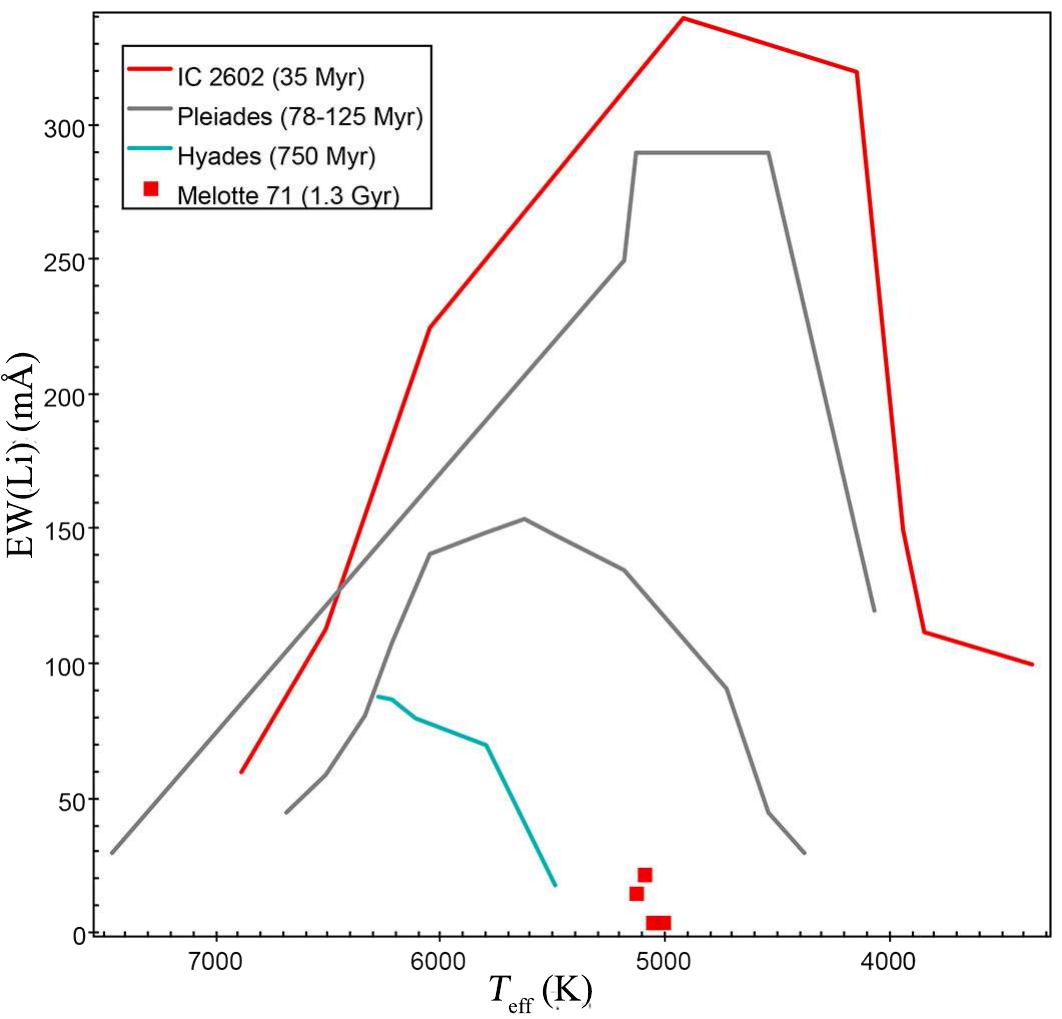} 
\caption{$EW$(Li)-versus-$T_{\rm eff}$ diagram for Melotte~71.}
             \label{fig:205}
    \end{figure}
    
    \clearpage

\subsection{Pismis~15}

\begin{figure} [htp]
   \centering
\includegraphics[width=0.9\linewidth, height=5cm]{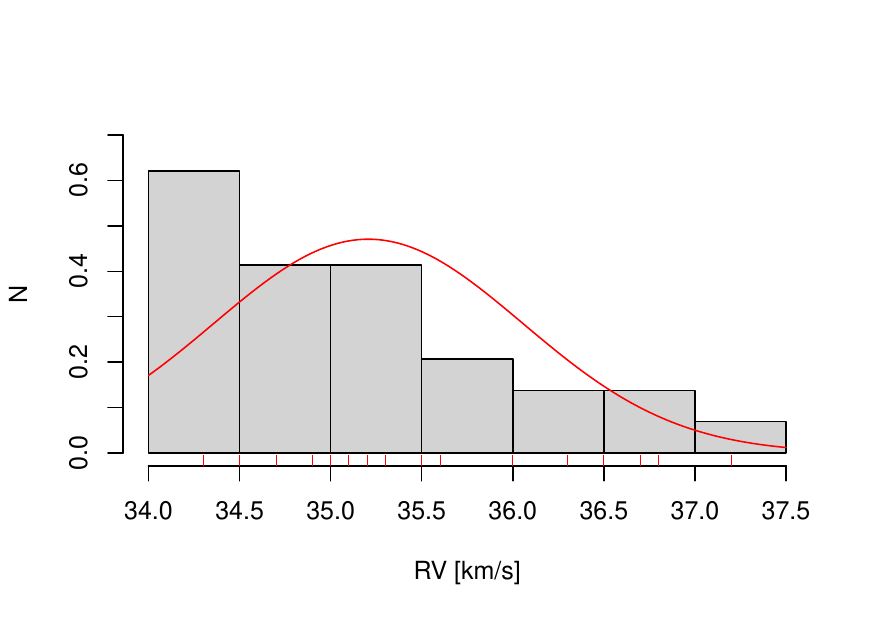}
\caption{$RV$ distribution for Pismis~15.}
             \label{fig:206}
    \end{figure}
    
           \begin{figure} [htp]
   \centering
\includegraphics[width=0.9\linewidth, height=5cm]{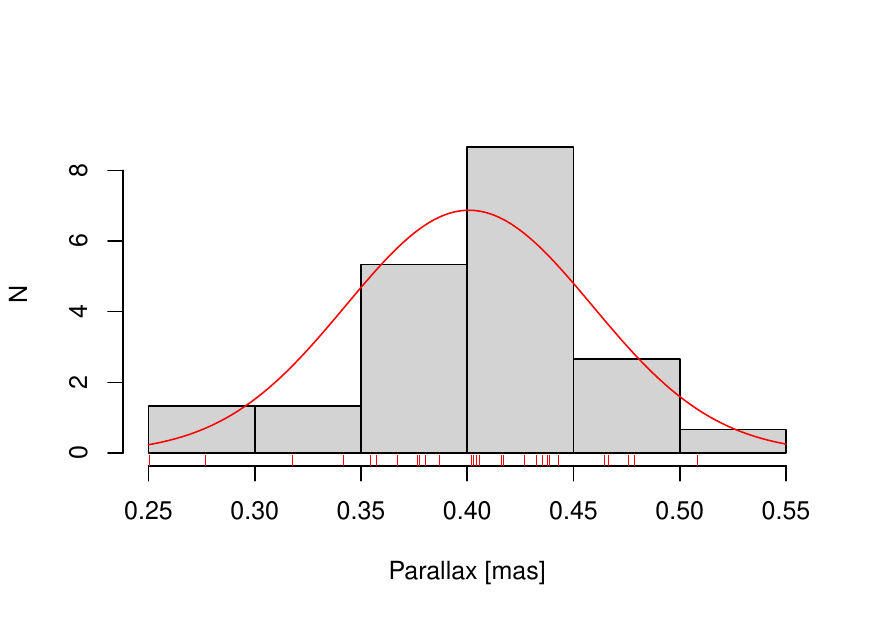}
\caption{Parallax distribution for Pismis~15.}
             \label{fig:207}
    \end{figure}

               \begin{figure} [htp]
   \centering
   \includegraphics[width=0.9\linewidth]{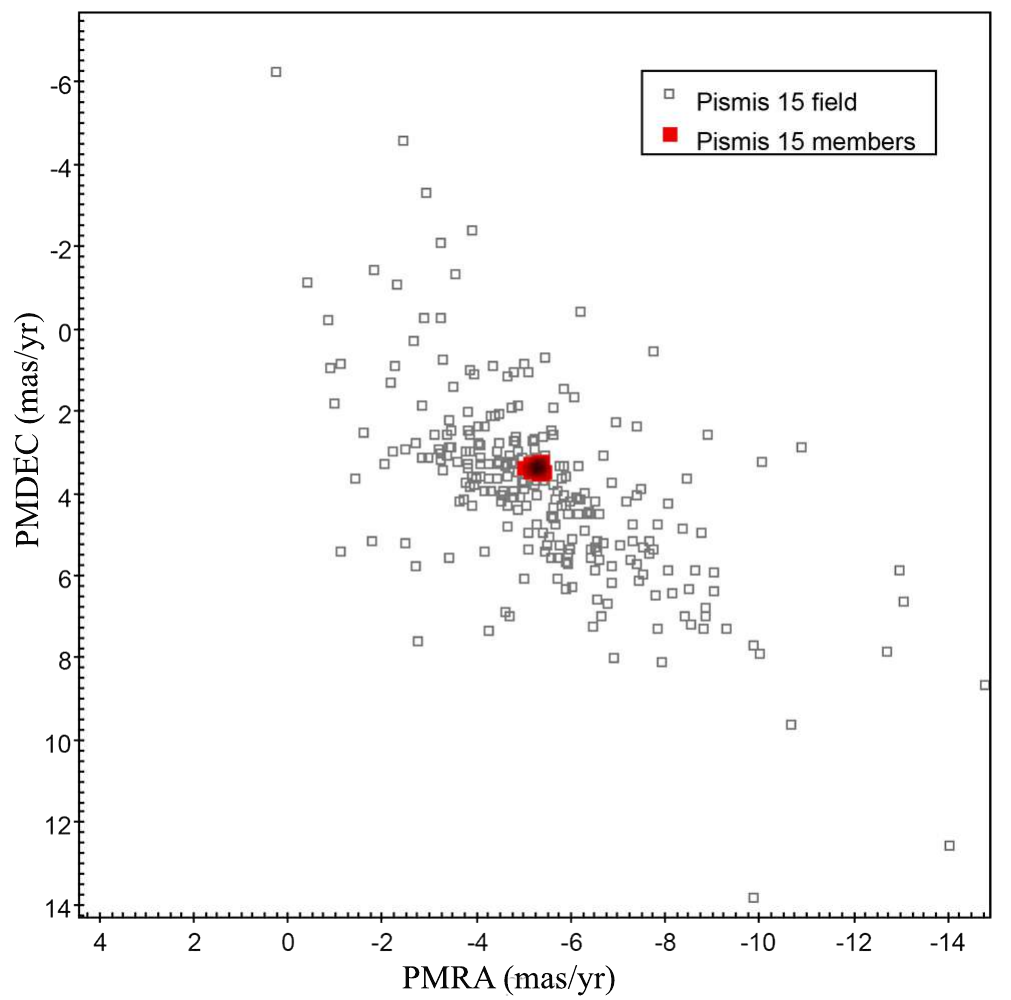}
   \caption{PMs diagram for Pismis~15.}
             \label{fig:208}
    \end{figure}
    
     \begin{figure} [htp]
   \centering
   \includegraphics[width=0.8\linewidth, height=7cm]{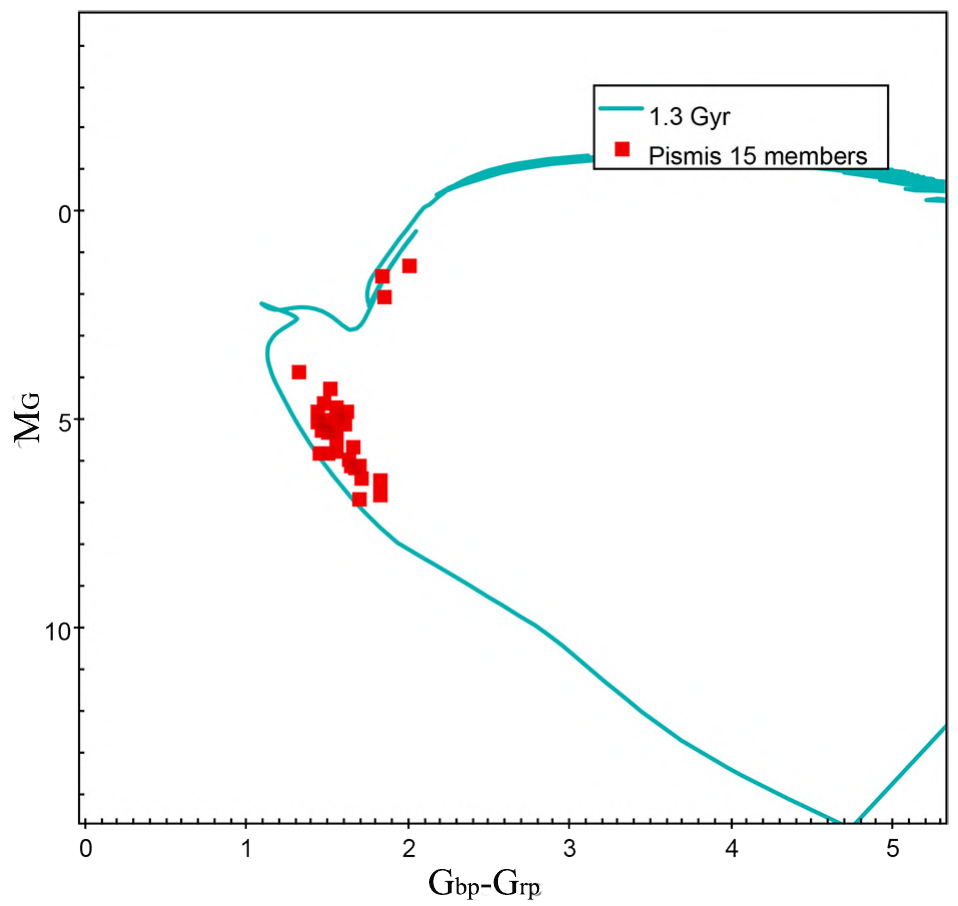}
   \caption{CMD for Pismis~15.}
             \label{fig:209}
    \end{figure}
    
      \begin{figure} [htp]
   \centering
 \includegraphics[width=0.8\linewidth]{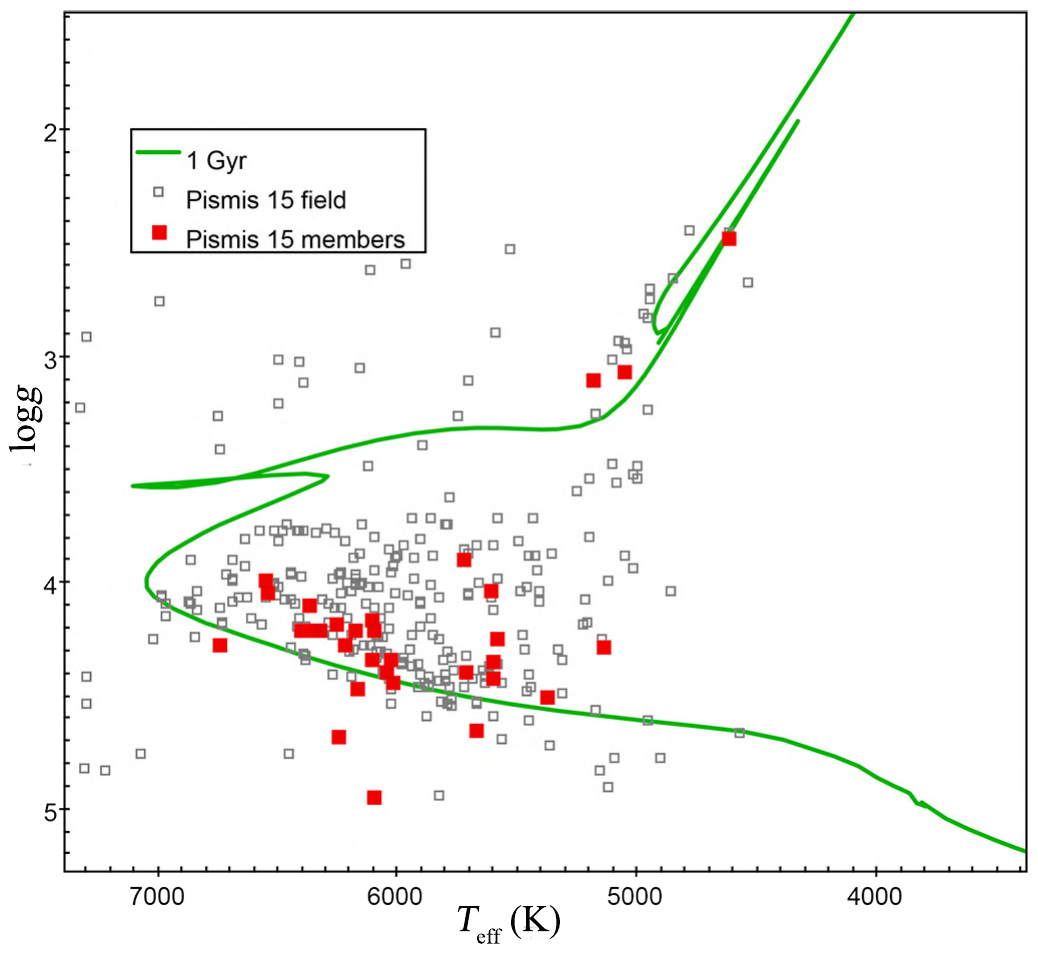} 
\caption{Kiel diagram for Pismis~15.}
             \label{fig:210}
    \end{figure}

  \begin{figure} [htp]
   \centering
 \includegraphics[width=0.8\linewidth]{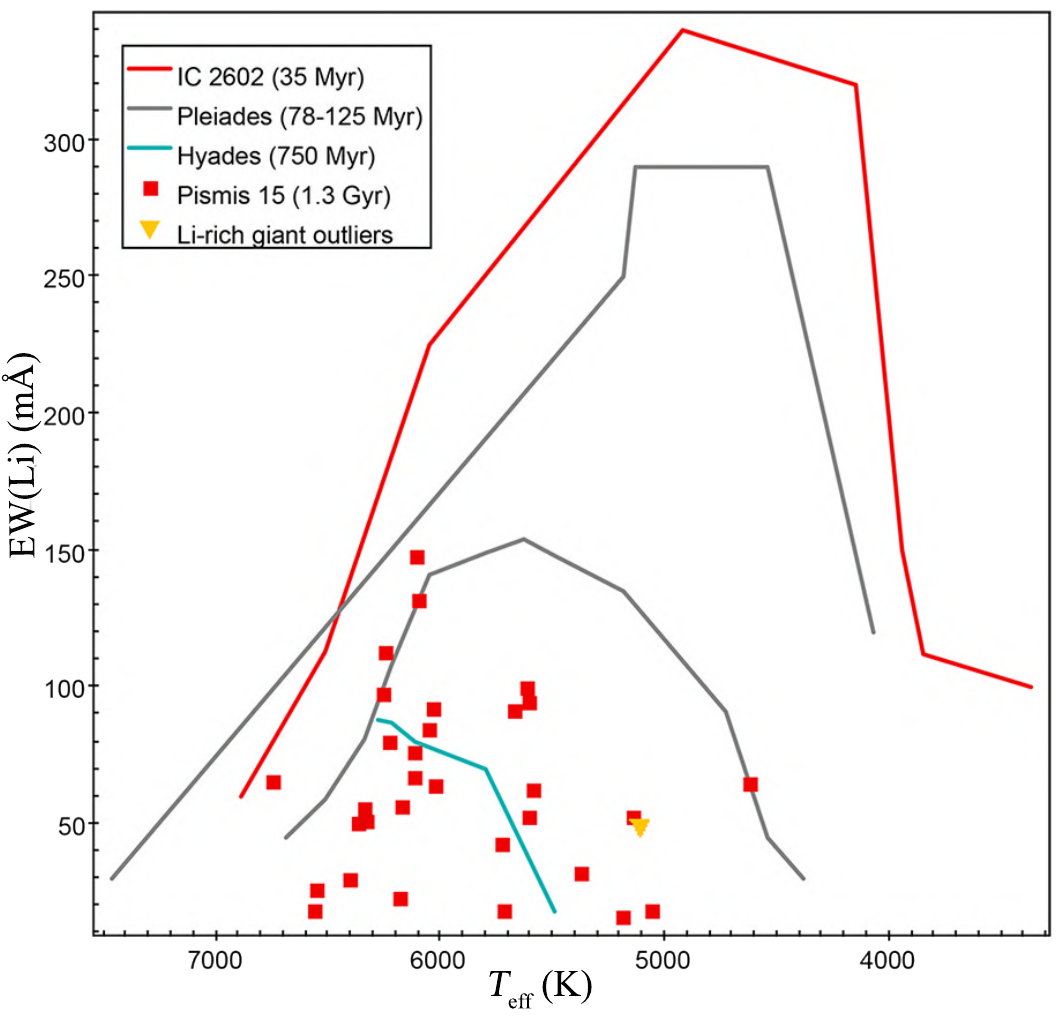} 
\caption{$EW$(Li)-versus-$T_{\rm eff}$ diagram for Pismis~15.}
             \label{fig:211}
    \end{figure}
    
    \clearpage

\subsection{Trumpler~20}

\begin{figure} [htp]
   \centering
\includegraphics[width=0.9\linewidth, height=5cm]{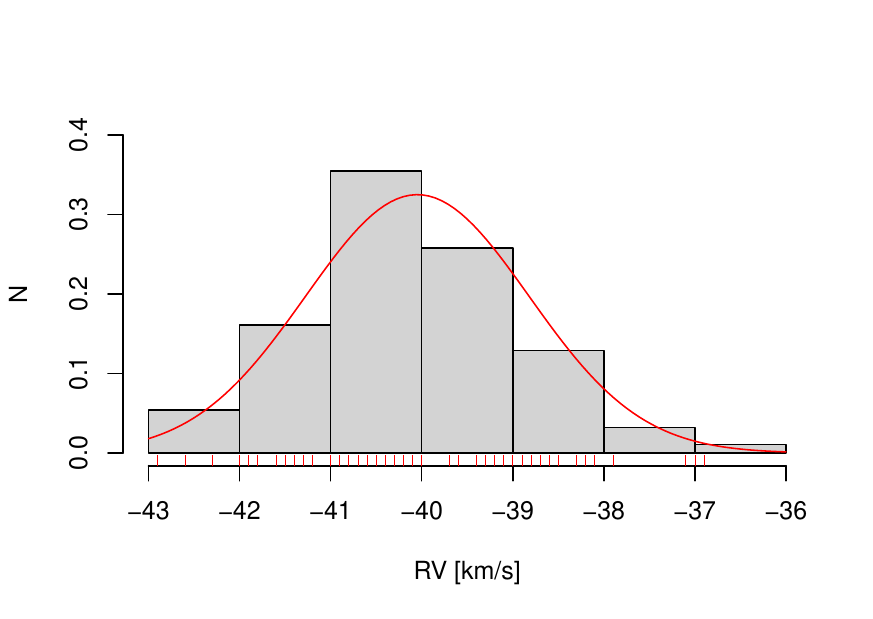}
\caption{$RV$ distribution for Trumpler~20.}
             \label{fig:212}
    \end{figure}
    
           \begin{figure} [htp]
   \centering
\includegraphics[width=0.9\linewidth, height=5cm]{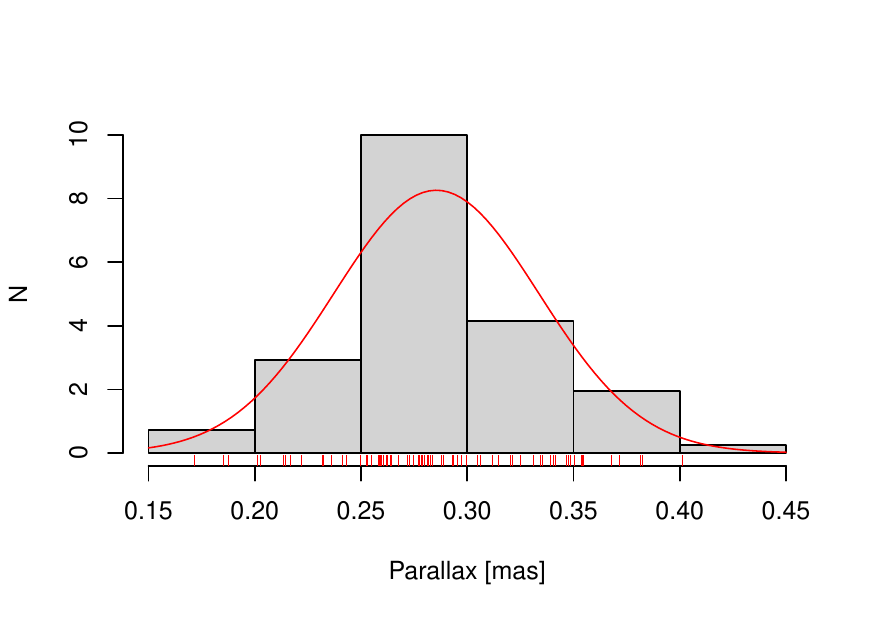}
\caption{Parallax distribution for Trumpler~20.}
             \label{fig:213}
    \end{figure}

               \begin{figure} [htp]
   \centering
   \includegraphics[width=0.9\linewidth]{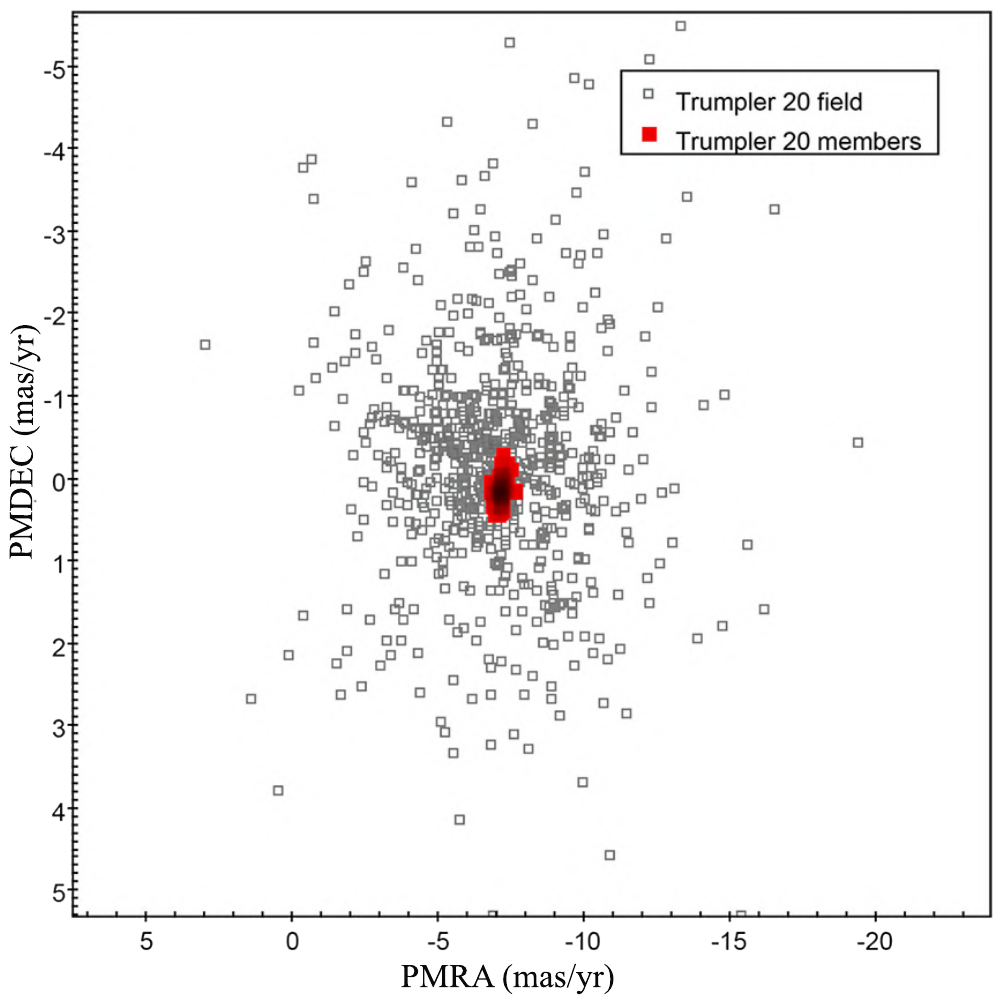}
   \caption{PMs diagram for Trumpler~20.}
             \label{fig:214}
    \end{figure}
    
     \begin{figure} [htp]
   \centering
   \includegraphics[width=0.8\linewidth, height=7cm]{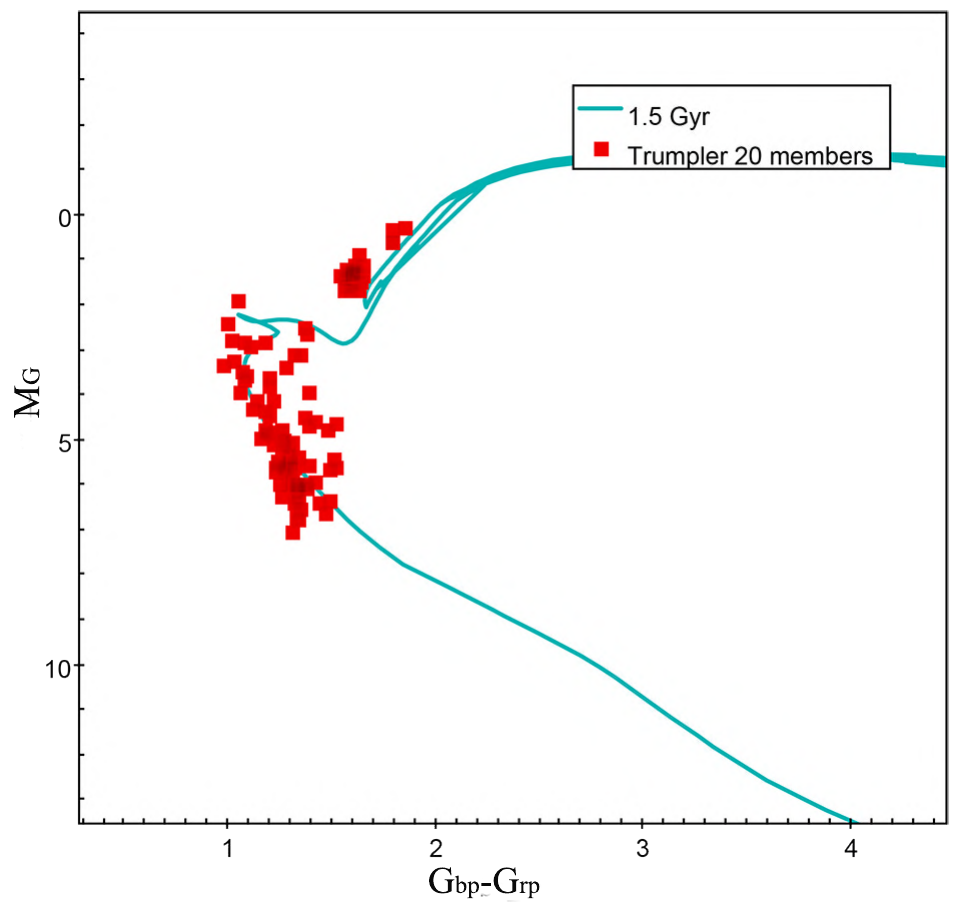}
   \caption{CMD for Trumpler~20.}
             \label{fig:215}
    \end{figure}
    
      \begin{figure} [htp]
   \centering
 \includegraphics[width=0.8\linewidth, height=7cm]{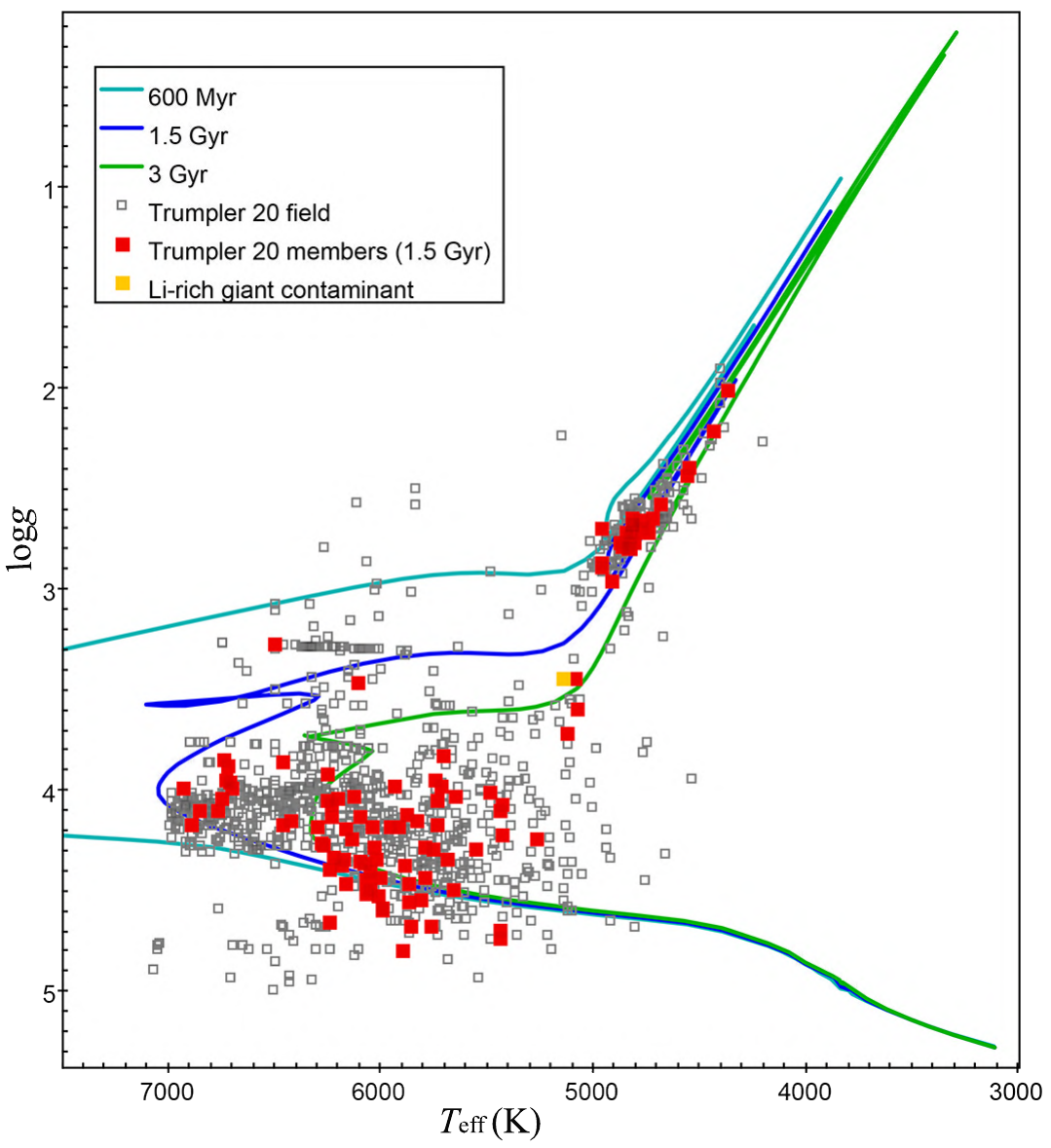} 
\caption{Kiel diagram for Trumpler~20.}
             \label{fig:216}
    \end{figure}

  \begin{figure} [htp]
   \centering
 \includegraphics[width=0.8\linewidth]{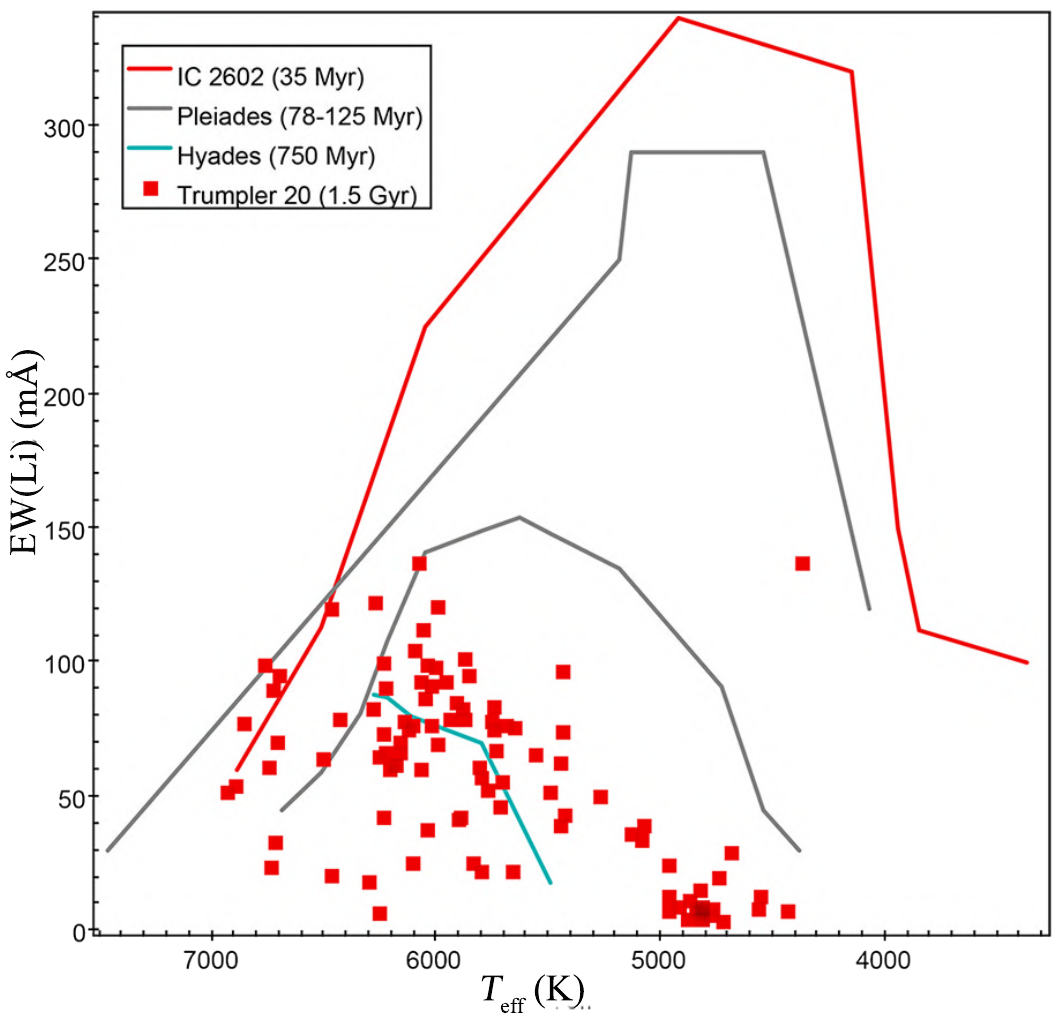} 
\caption{$EW$(Li)-versus-$T_{\rm eff}$ diagram for Trumpler~20.}
             \label{fig:217}
    \end{figure}
    
    \clearpage

\subsection{Berkeley~44}

\begin{figure} [htp]
   \centering
\includegraphics[width=0.9\linewidth, height=5cm]{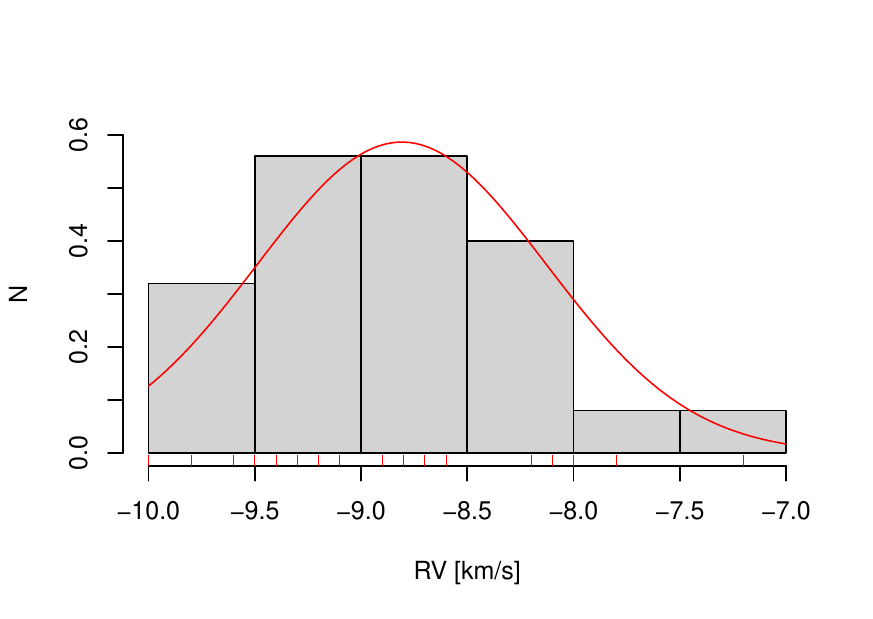}
\caption{$RV$ distribution for Berkeley~44.}
             \label{fig:218}
    \end{figure}
    
           \begin{figure} [htp]
   \centering
\includegraphics[width=0.9\linewidth, height=5cm]{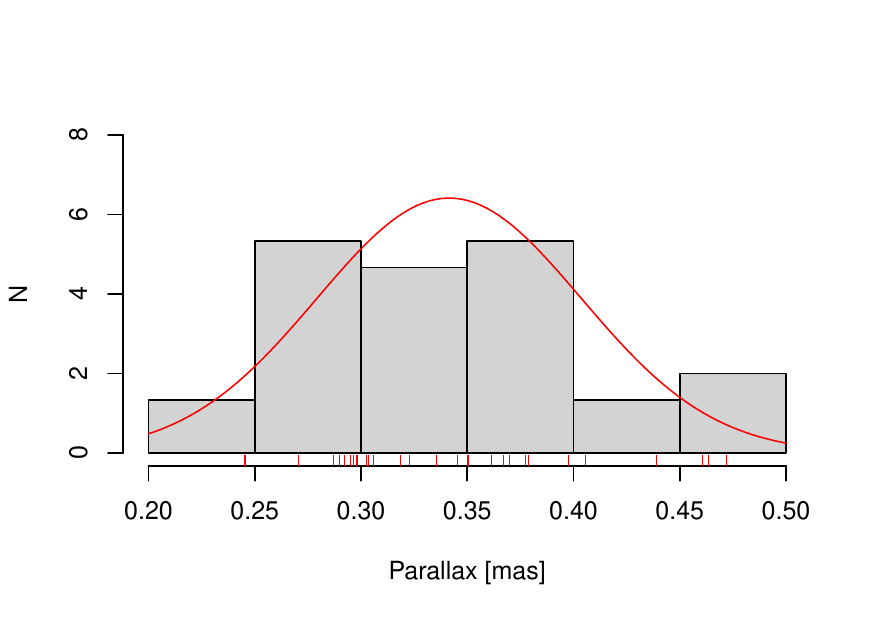}
\caption{Parallax distribution for Berkeley~44.}
             \label{fig:219}
    \end{figure}

               \begin{figure} [htp]
   \centering
   \includegraphics[width=0.9\linewidth]{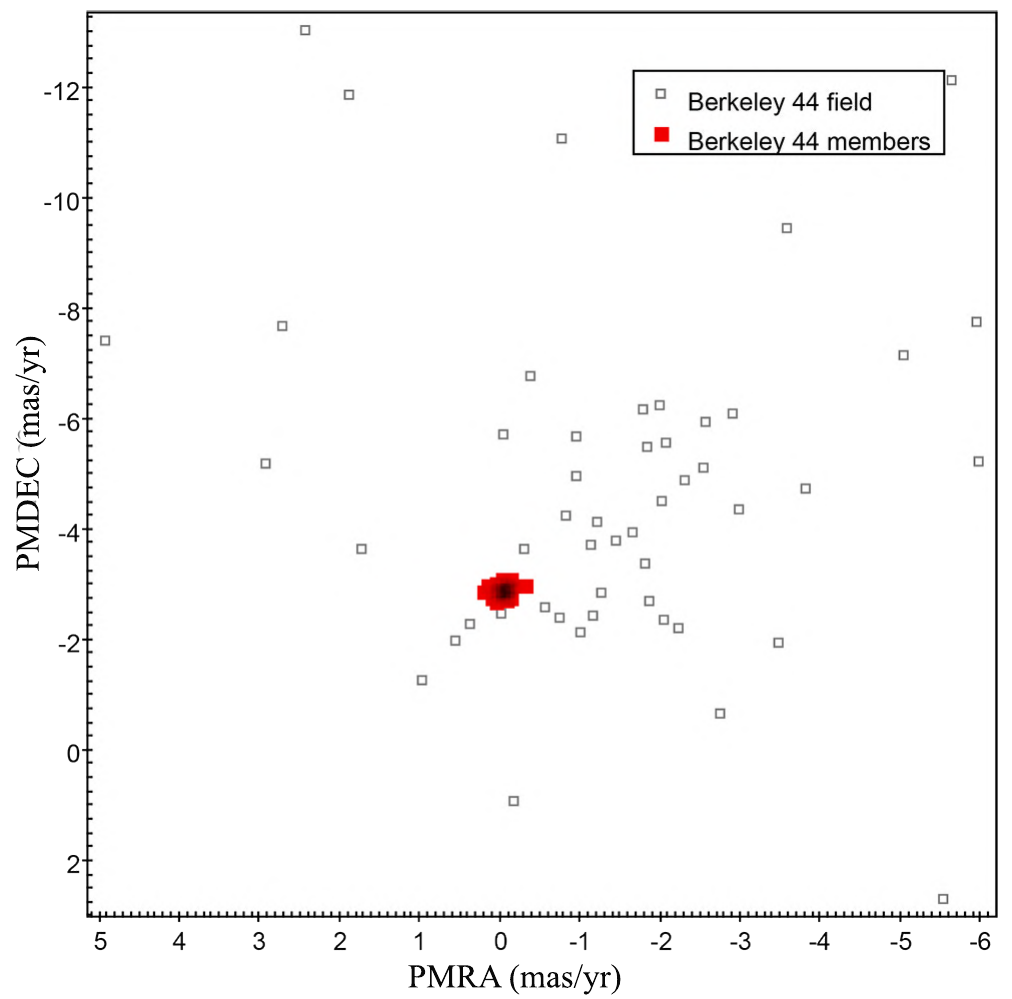}
   \caption{PMs diagram for Berkeley~44.}
             \label{fig:220}
    \end{figure}
    
     \begin{figure} [htp]
   \centering
   \includegraphics[width=0.8\linewidth, height=7cm]{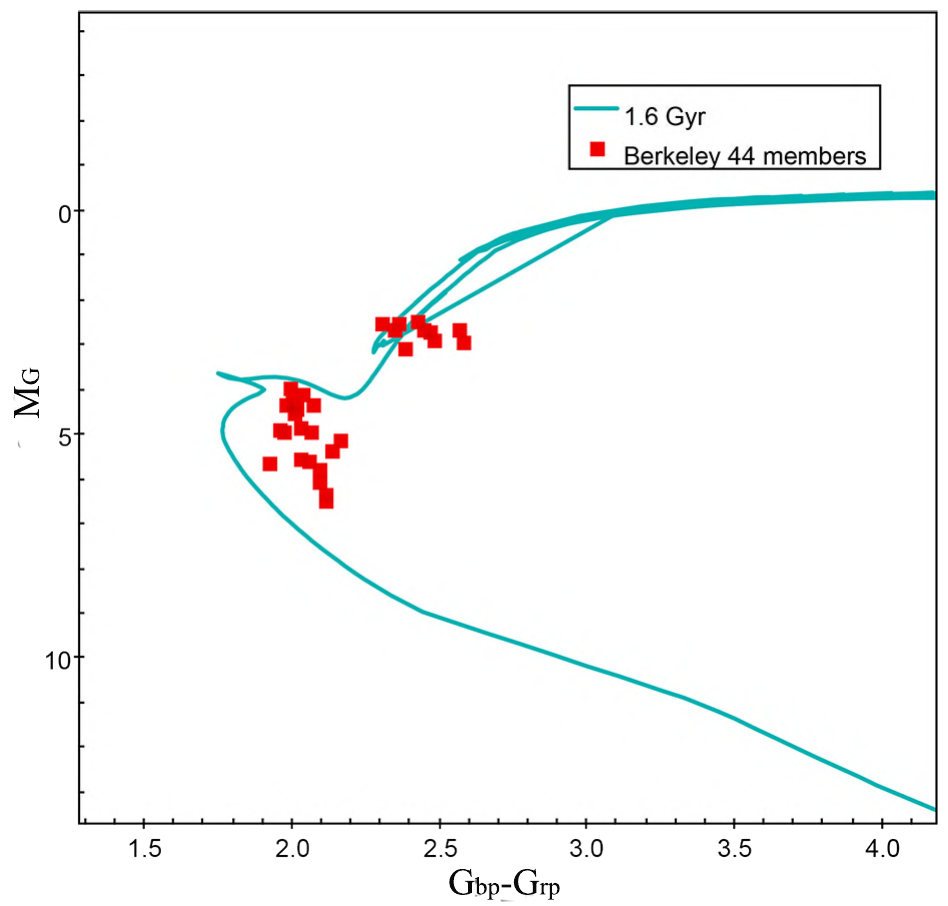}
   \caption{CMD for Berkeley~44.}
             \label{fig:221}
    \end{figure}
    
      \begin{figure} [htp]
   \centering
 \includegraphics[width=0.8\linewidth]{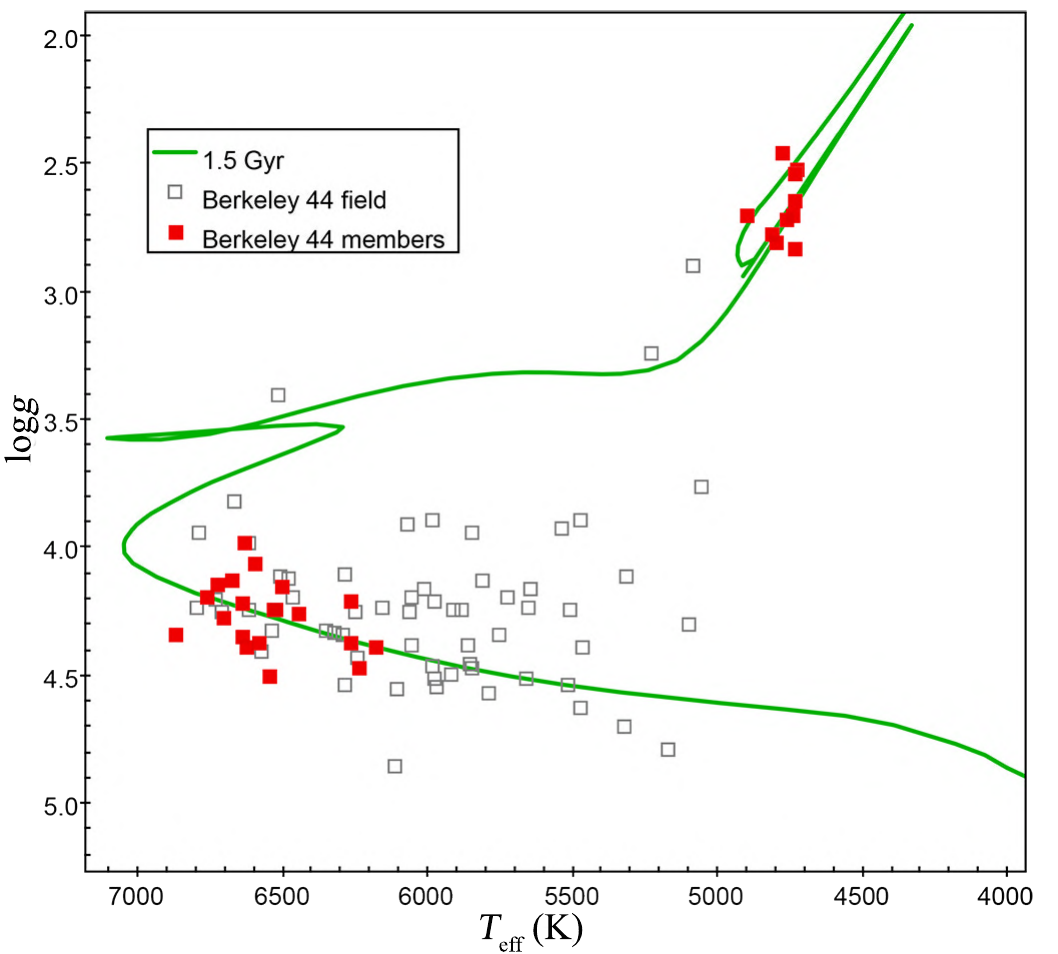} 
\caption{Kiel diagram for Berkeley~44.}
             \label{fig:222}
    \end{figure}

  \begin{figure} [htp]
   \centering
 \includegraphics[width=0.8\linewidth]{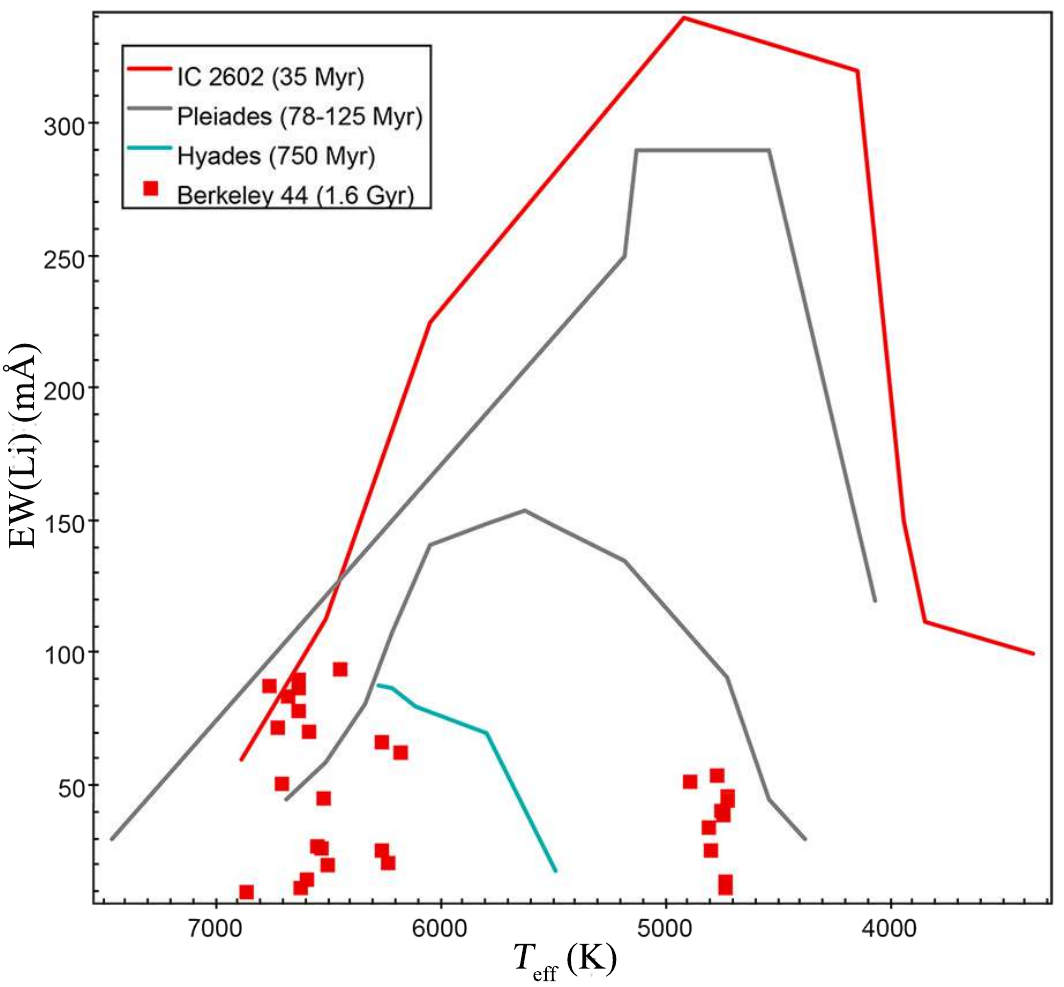} 
\caption{$EW$(Li)-versus-$T_{\rm eff}$ diagram for Berkeley~44.}
             \label{fig:223}
    \end{figure}
    
    \clearpage

\subsection{NGC~2243}

\begin{figure} [htp]
   \centering
\includegraphics[width=0.9\linewidth, height=5cm]{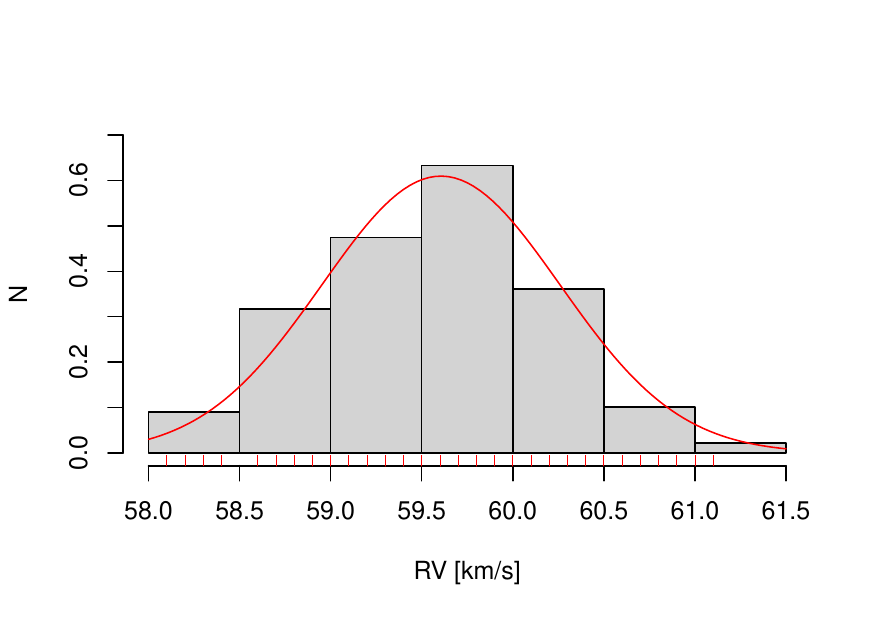}
\caption{$RV$ distribution for NGC~2243.}
             \label{fig:224}
    \end{figure}
    
           \begin{figure} [htp]
   \centering
\includegraphics[width=0.9\linewidth, height=5cm]{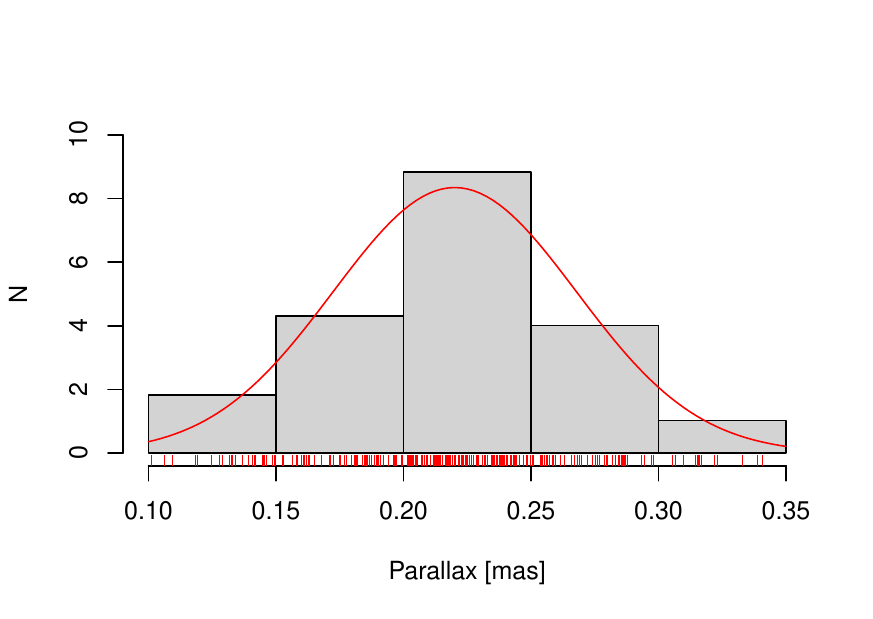}
\caption{Parallax distribution for NGC~2243.}
             \label{fig:225}
    \end{figure}

               \begin{figure} [htp]
   \centering
   \includegraphics[width=0.9\linewidth]{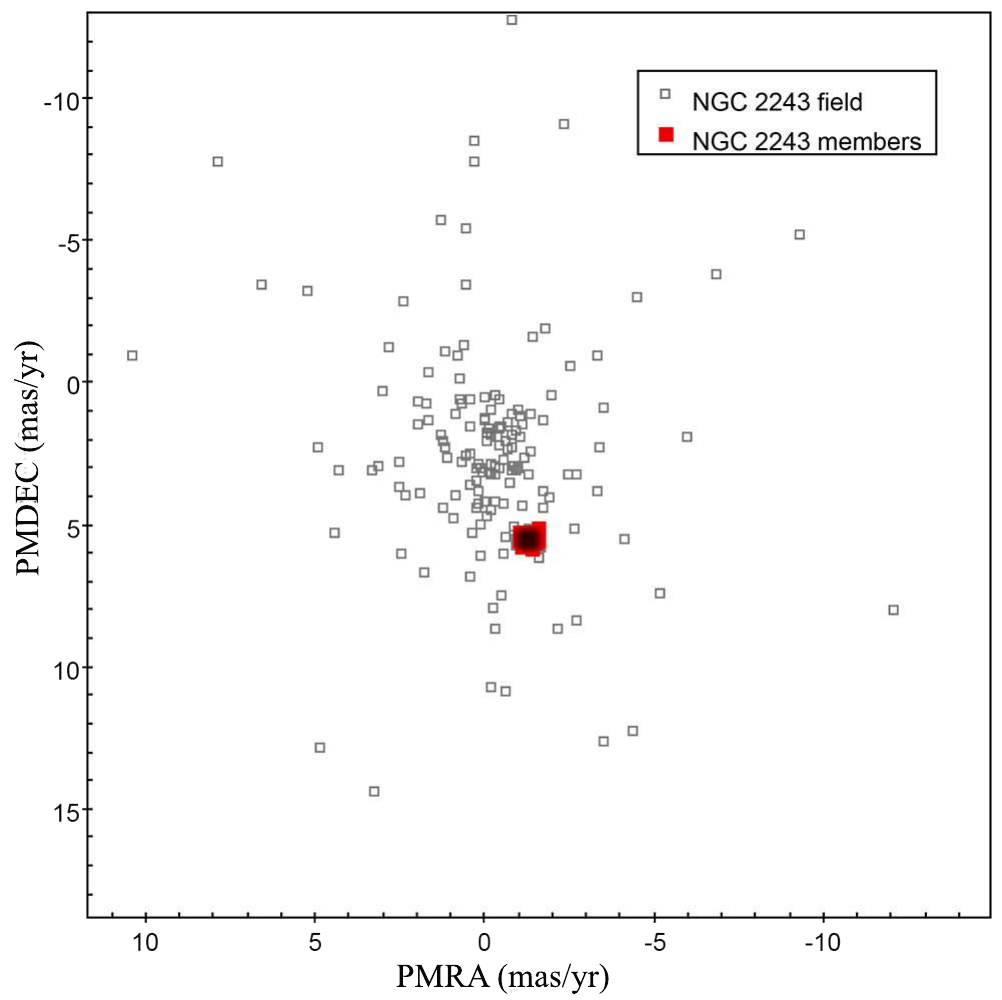}
   \caption{PMs diagram for NGC~2243.}
             \label{fig:226}
    \end{figure}
    
     \begin{figure} [htp]
   \centering
   \includegraphics[width=0.8\linewidth, height=7cm]{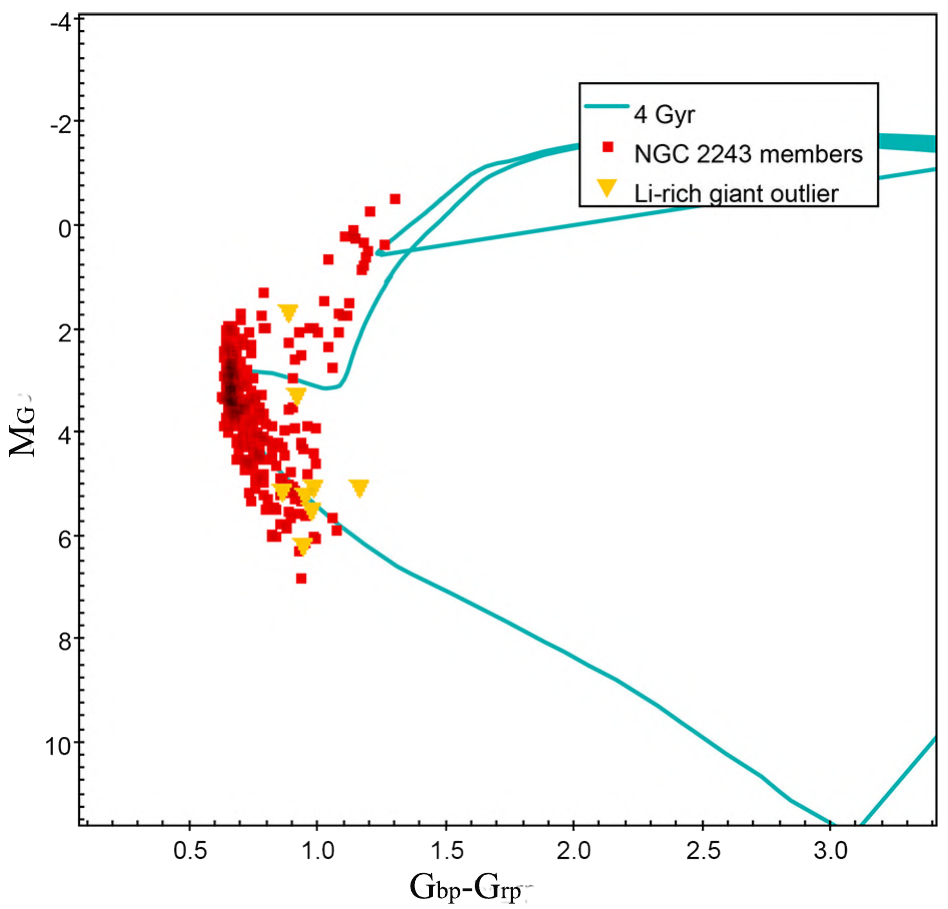}
   \caption{CMD for NGC~2243.}
             \label{fig:227}
    \end{figure}
    
      \begin{figure} [htp]
   \centering
 \includegraphics[width=0.8\linewidth]{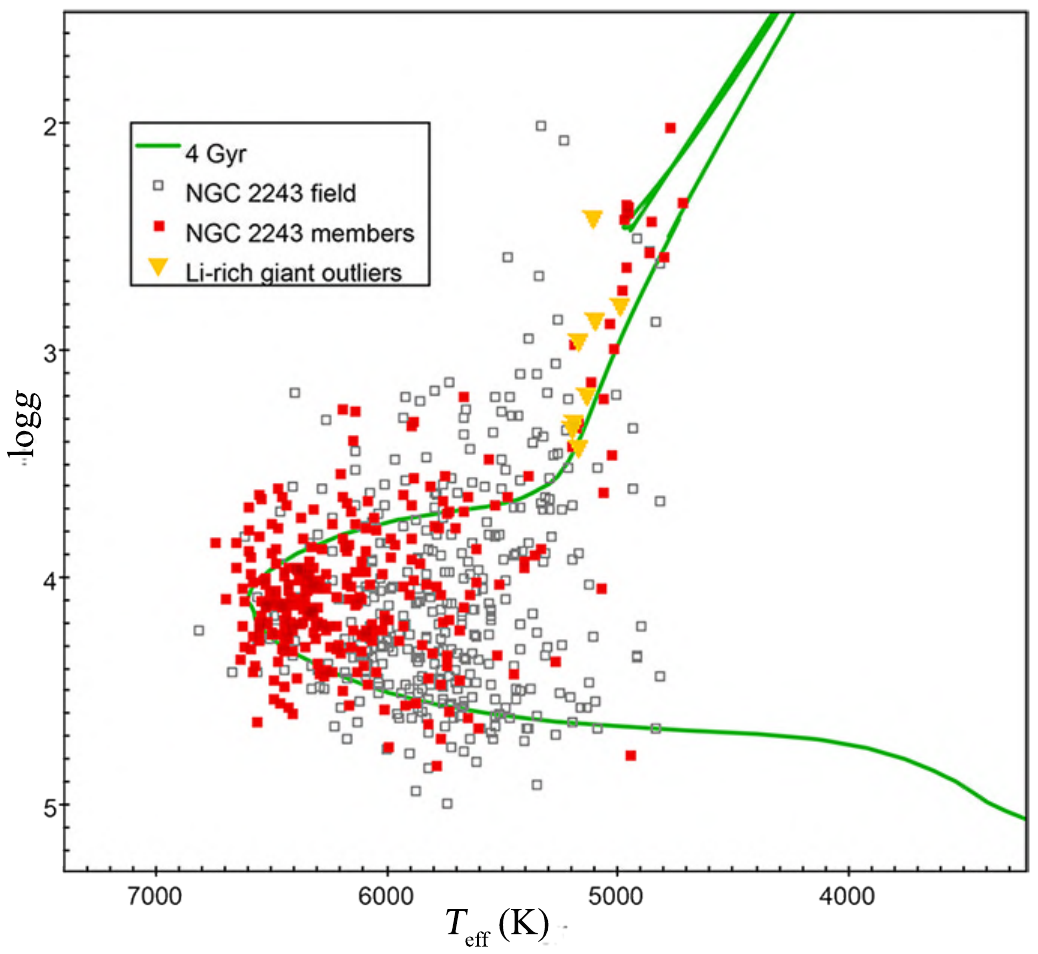} 
\caption{Kiel diagram for NGC~2243.}
             \label{fig:228}
    \end{figure}

  \begin{figure} [htp]
   \centering
 \includegraphics[width=0.8\linewidth]{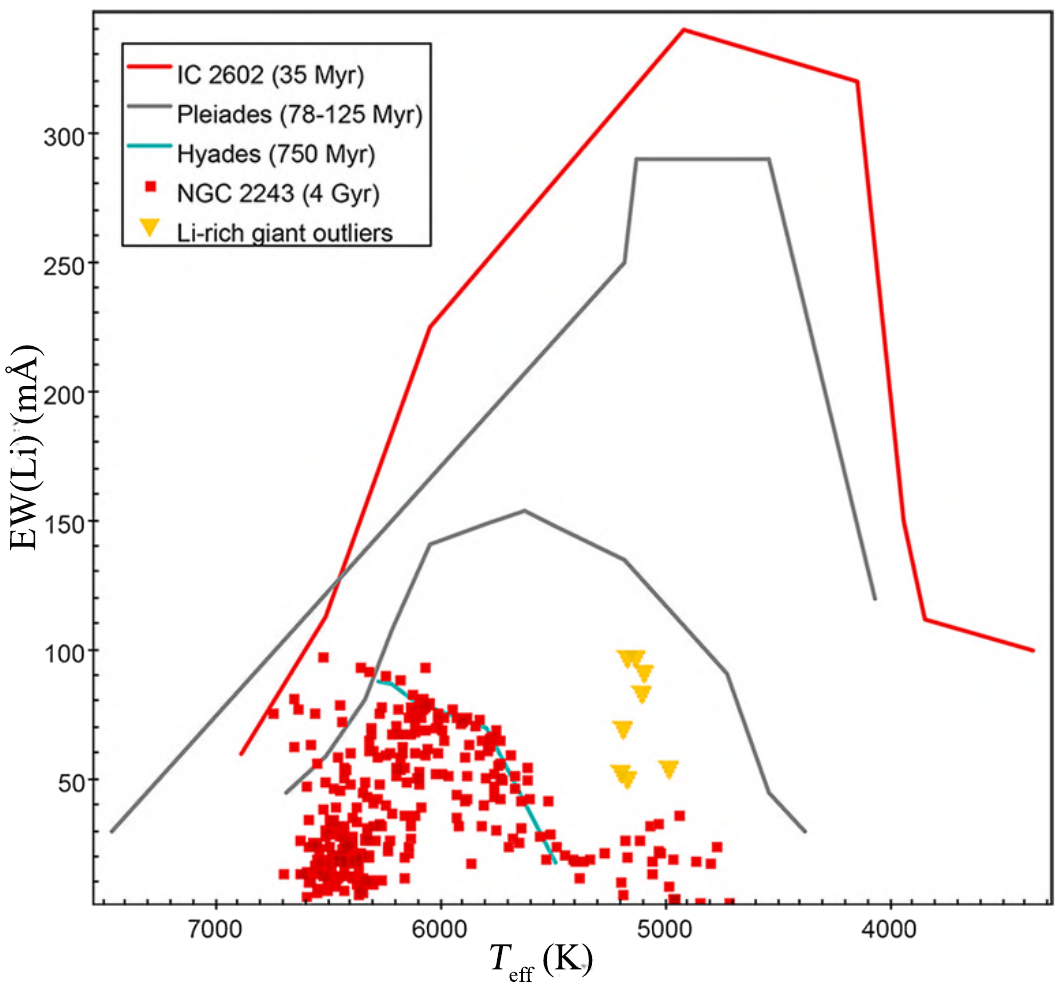} 
\caption{$EW$(Li)-versus-$T_{\rm eff}$ diagram for NGC~2243.}
             \label{fig:229}
    \end{figure}
    
    \clearpage

\subsection{M67}

\begin{figure} [htp]
   \centering
\includegraphics[width=0.9\linewidth, height=5cm]{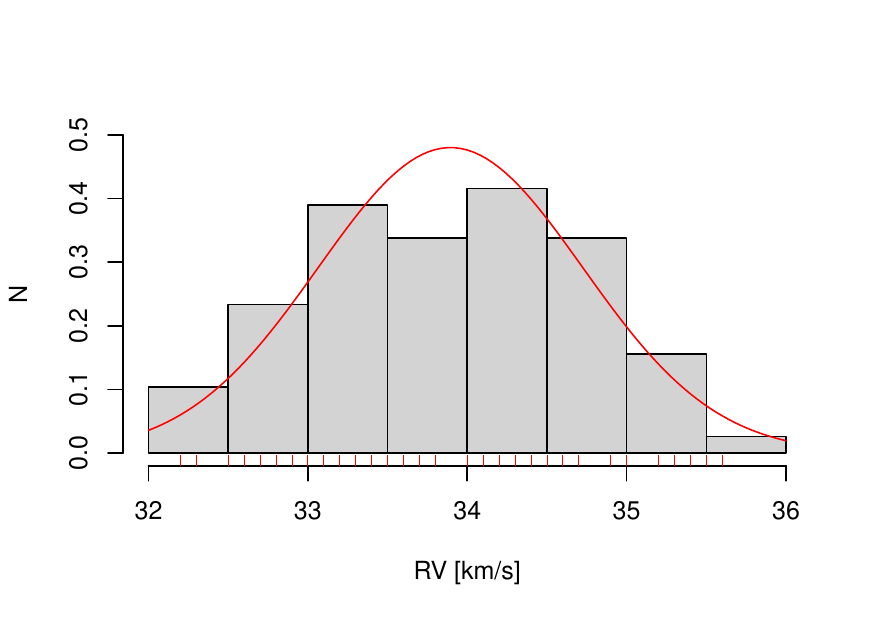}
\caption{$RV$ distribution for M67.}
             \label{fig:230}
    \end{figure}
    
           \begin{figure} [htp]
   \centering
\includegraphics[width=0.9\linewidth, height=5cm]{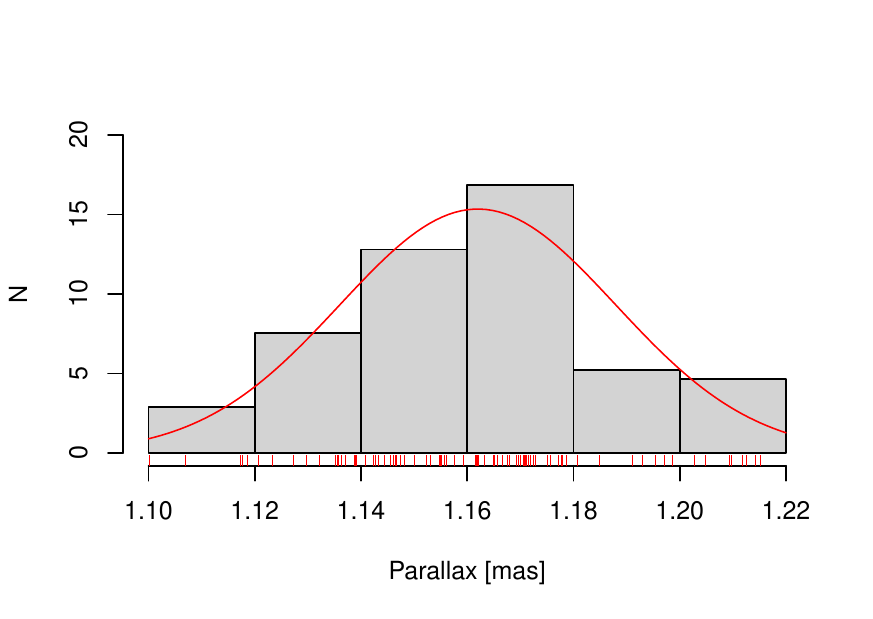}
\caption{Parallax distribution for M67.}
             \label{fig:231}
    \end{figure}

               \begin{figure} [htp]
   \centering
   \includegraphics[width=0.9\linewidth]{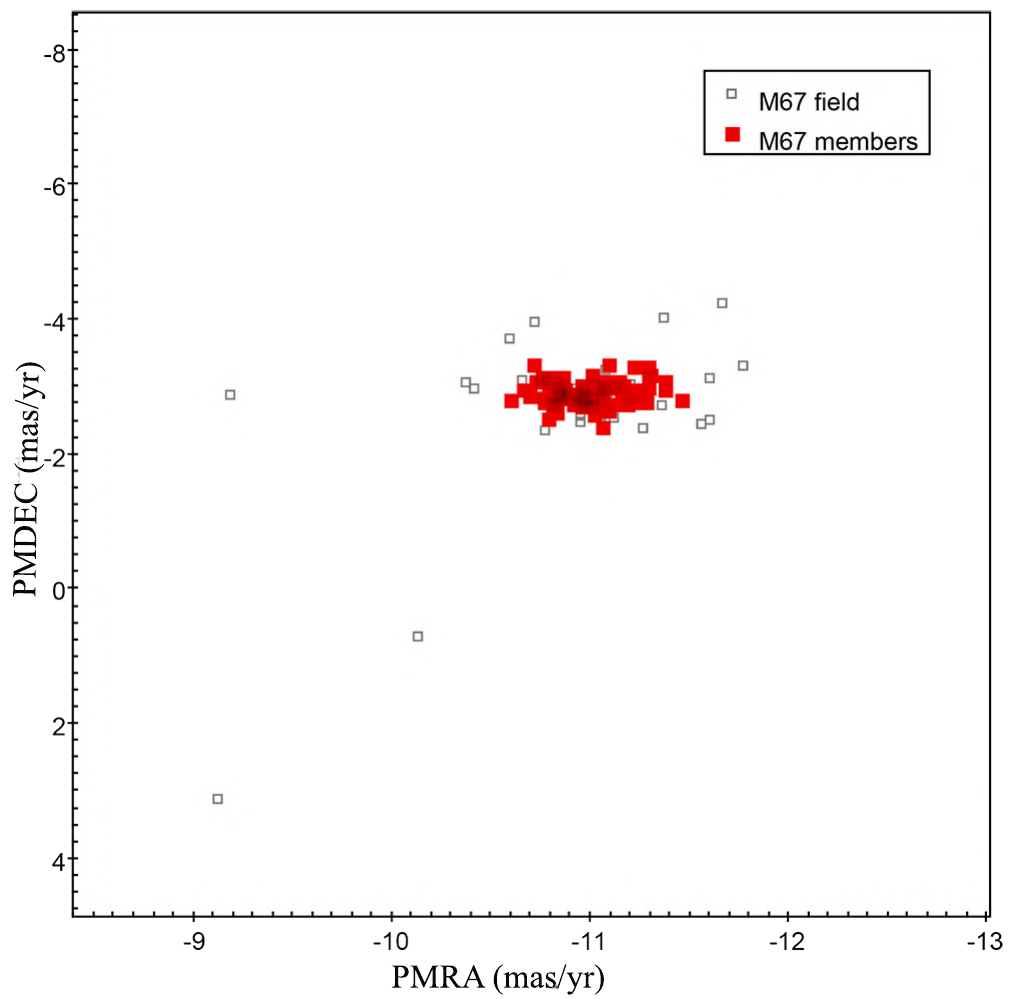}
   \caption{PMs diagram for M67.}
             \label{fig:232}
    \end{figure}
    
     \begin{figure} [htp]
   \centering
   \includegraphics[width=0.8\linewidth, height=7cm]{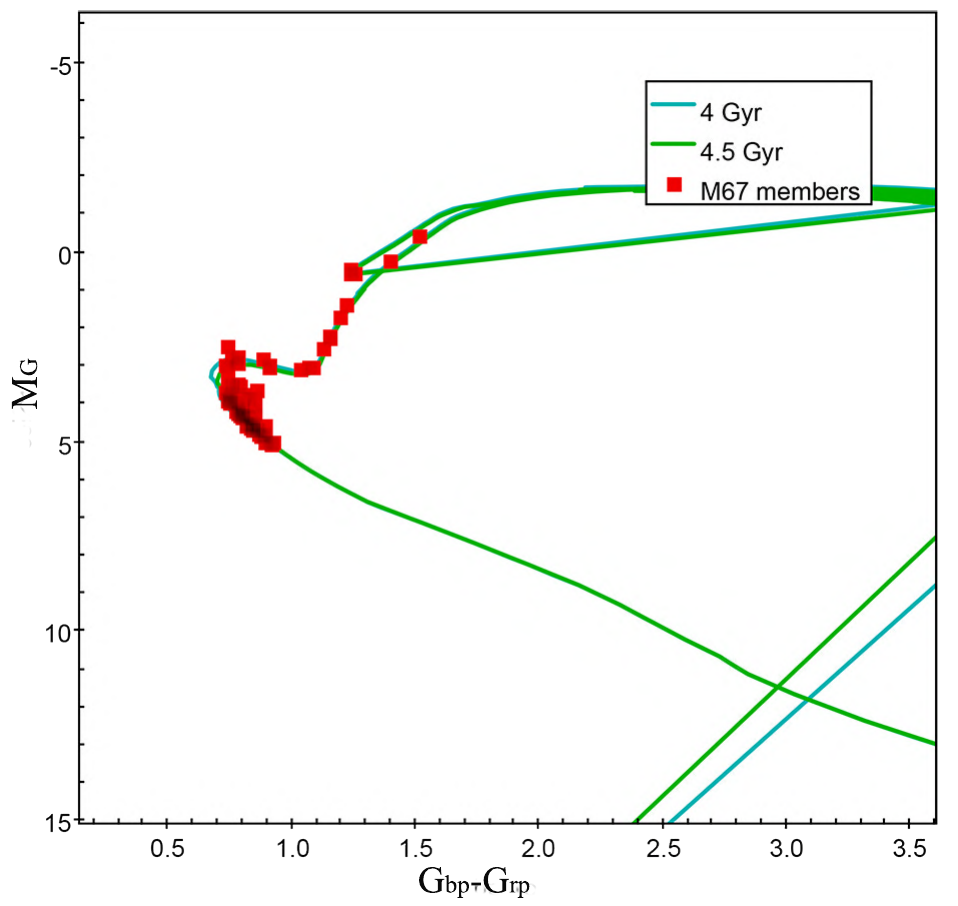}
   \caption{CMD for M67.}
             \label{fig:233}
    \end{figure}
    
      \begin{figure} [htp]
   \centering
 \includegraphics[width=0.8\linewidth]{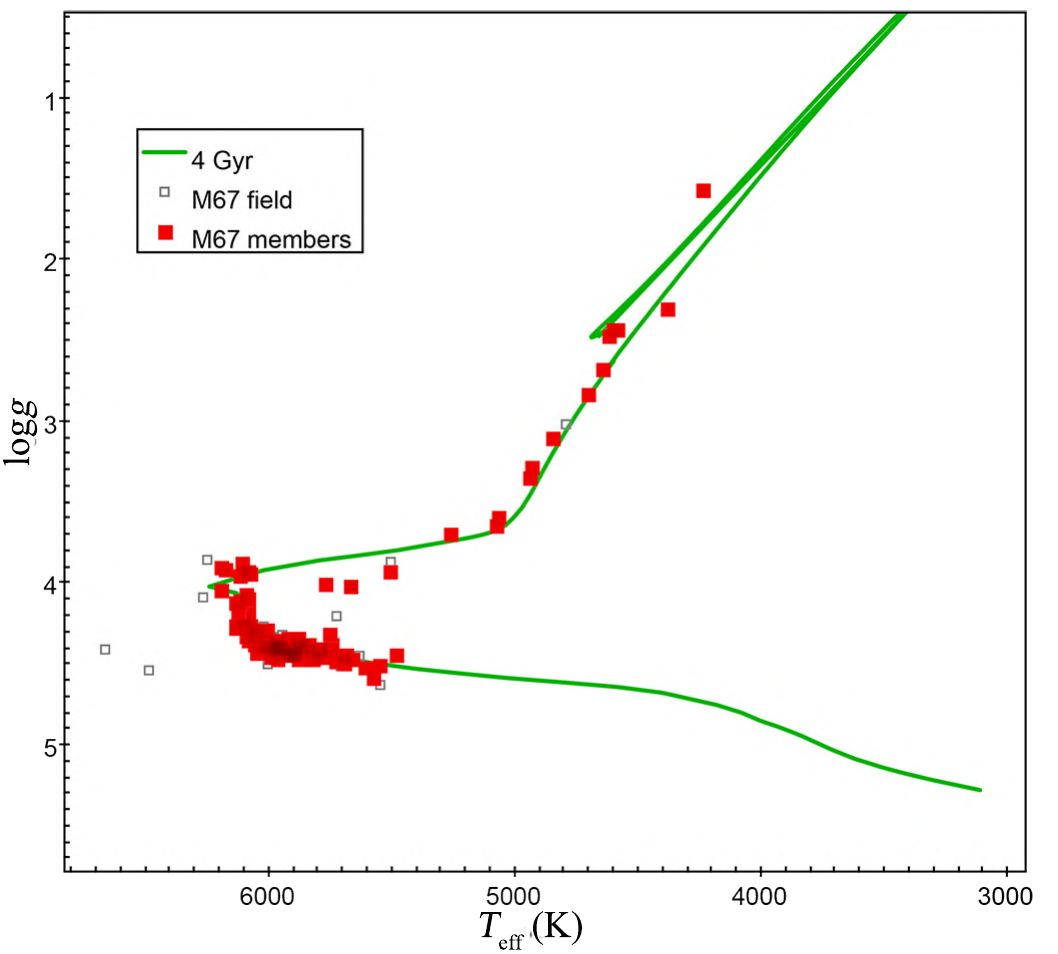} 
\caption{Kiel diagram for M67.}
             \label{fig:234}
    \end{figure}

  \begin{figure} [htp]
   \centering
 \includegraphics[width=0.8\linewidth]{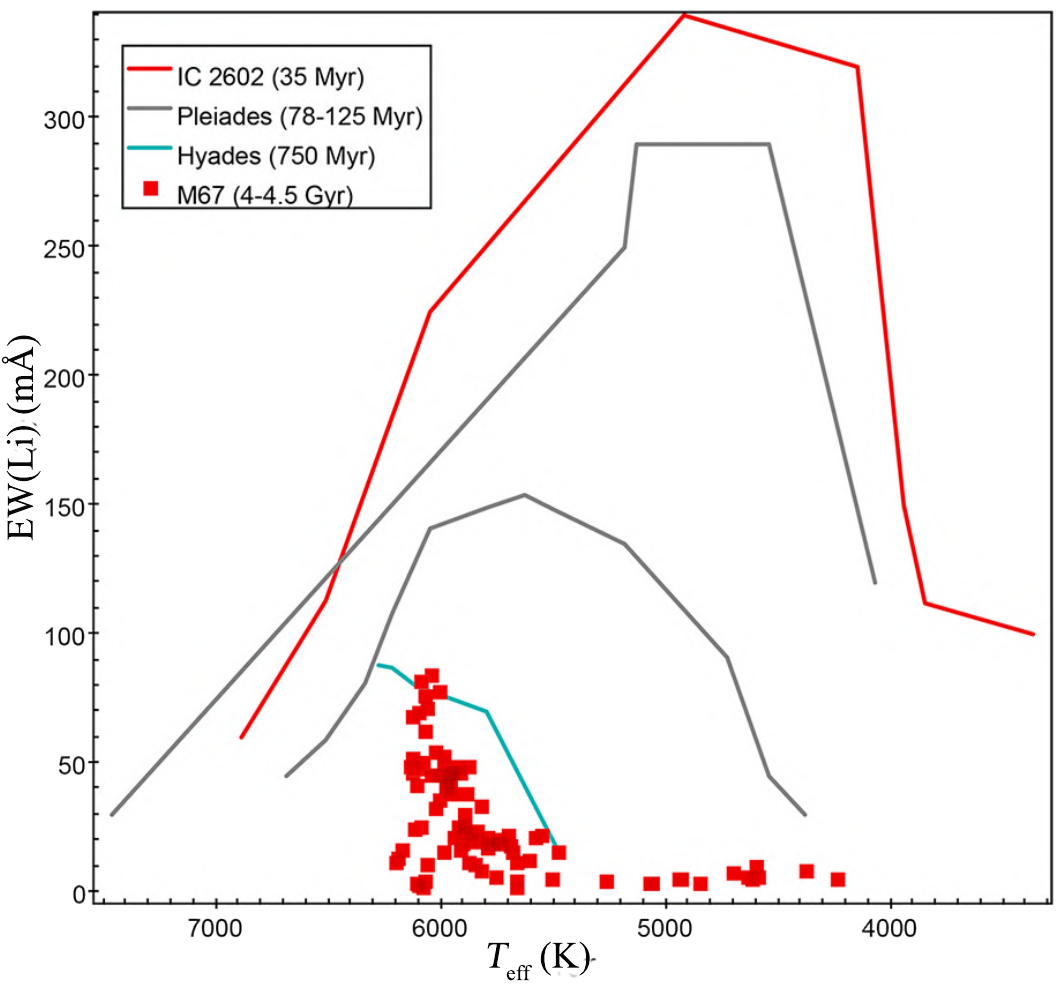} 
\caption{$EW$(Li)-versus-$T_{\rm eff}$ diagram for M67.}
             \label{fig:235}
    \end{figure}
    
    \clearpage


%
%

\section{Cluster membership and final selections: Long tables}
\label{ap1:AppendixD}
  
  
  This appendix describes the long tables containing all the parameters from GES and \textit{Gaia} which we have used throughout this work, as well as the tables summarizing the membership analysis of Sect.~\ref{analysis} and the final cluster selections for each of the $42$ clusters in the sample. The full tables for all sample clusters are available in electronic form at the CDS via anonymous ftp to \url{http://cdsarc.u-strasbg.fr/} (130.79.128.5) or via \url{http://cdsweb.u-strasbg.fr/cgi-bin/qcat?J/A+A/}. The columns for each table are described as follows:\\
  
   \begin{table*} [htp]
 \begin{center}
 \caption[Tables of Sect.~\ref{analysis}: GES parameters]{\textbf{GES parameters} (see the exemplified Tables~\ref{table1} and~\ref{table2} for cluster NGC~2516).}
 \label{GES_parameters}  
 \begin{tabular}{c c} 
\hline
\hline
\noalign{\smallskip}
 Column name & Description \\ [1 ex] 
\hline
\noalign{\smallskip}
CNAME	&	GES object name from coordinates	\\
RA	&	Right ascension in J2000 (hrs)	\\
DEC	&	Declination in J2000 (deg) \\
VRAD	&	Radial velocity (km s$^{-1}$)	\\
E\textunderscore VRAD	&	Error in radial velocity (km s$^{-1}$) \\
VSINI	&	$v$sin$i$ rotational velocity (km s$^{-1}$)	\\
E\textunderscore VSINI	& Error in $v$sin$i$ (km s$^{-1}$)	\\
TEFF	&	Effective temperature $T_{\rm eff}$ (K)	\\
E\textunderscore TEFF	&	Error in $T_{\rm eff}$ (K)	\\
TEFF\textunderscore IRFM	& Infrared photometric temperature (K)\tablefootmark{a}	\\
E\textunderscore TEFF\textunderscore IRFM & Error in infrared photometric temperature (K)	\\
LOGG	& log$\textit{g}$ surface gravity (dex)\tablefootmark{b}	\\
E\textunderscore LOGG	& Error in log$\textit{g}$ (dex) \\
GAMMA	& $\gamma$ index\tablefootmark{c} (dex) \\
E\textunderscore GAMMA	& Error in $\gamma$ index (dex) \\
FEH	& [Fe/H] metallicity (dex) \\
E\textunderscore FEH	& Error in [Fe/H] metallicity (dex) \\
EWC\textunderscore LI & Blends-corrected values of $EW$(Li) (m$\r{A}$) \\
E\textunderscore EWC\textunderscore LI & Error in $EW$(Li) (m$\r{A}$) \\
LIM\textunderscore EWC\textunderscore LI & Flag for blends-corrected $EW$(Li) error\tablefootmark{d} \\
EW\textunderscore LI\textunderscore UNVEIL & The primarily-used veiling-corrected values of $EW$(Li) (m$\r{A}$) \\
E\textunderscore EW\textunderscore LI\textunderscore UNVEIL & Error in $EW$(Li) (m$\r{A}$) \\
LIM\textunderscore EW\textunderscore LI\textunderscore UNVEIL & Flag for veiling-corrected $EW$(Li) errors\tablefootmark{d} \\
 LI1 &  \textit{A}(Li) neutral Li abundances in log$\epsilon$(X) (dex) \\
 E\textunderscore LI1  & Error in \textit{A}(Li) (dex) \\
 EW\textunderscore HA\textunderscore ACC  &  H$\alpha$ EW: accretion ($\r{A}$) \\
 E\textunderscore EW\textunderscore HA\textunderscore ACC  & Error in H$\alpha$ EW: accretion ($\r{A}$)  \\
 HA10 & H$\alpha$10\%, width of the H$\alpha$ emission line at 10$\%$ peak intensity (km s$^{-1}$) \\
  E\textunderscore HA10 & Error in H$\alpha$10\% (km s$^{-1}$) \\
  EW\textunderscore HA\textunderscore CHR & H$\alpha$ chromospheric activity ($\r{A}$)  \\
  E\textunderscore EW\textunderscore HA\textunderscore CHR & Error in H$\alpha$ chromospheric activity ($\r{A}$)  \\
 \noalign{\smallskip}
 \hline
 \end{tabular}  
  \tablefoot{
\tablefoottext{a}{Not shown in Table~\ref{table1}, primarily measured for young clusters.}
\tablefoottext{b}{Used in this work for intermediate-age and old clusters.}
\tablefoottext{c}{The empirical gravity indicator defined by \cite{damiani} and used in the case of the young clusters.}
\tablefoottext{d}{0=no flag necessary; 1=$EW$(Li) corrected by blends contribution using models; 2=$EW$(Li) measured separately --- Li line resolved, UVES only; and 3=Upper limit --- no error for $EW$(Li) is given.}
}
 \end{center}
\end{table*}
   
      \begin{table*} [h]
 \begin{center}
 \caption[Tables of Sect.~\ref{analysis}: \textit{Gaia} parameters and $P_{rot}$ measurements]{\textbf{\textit{Gaia} parameters and $P_{rot}$ measurements} (see the exemplified Tables~\ref{table3} and~\ref{table4} for cluster NGC~2516).}
 \label{Gaia_parameters}  
 \begin{tabular}{c c} 
\hline
\hline
\noalign{\smallskip}
 Column name & Description \\ [1 ex] 
\hline
\noalign{\smallskip}
CNAME	&	GES object name from coordinates	\\
\textit{parallax}	&	$\pi$ parallax (mas)	\\
\textit{parallax\textunderscore error}	&	Error in parallax (mas) \\
\textit{pmra}	& $pmra$, PM in RA (mas yr$^{-1}$)	\\
\textit{pmra\textunderscore error}	&	Error in $pmra$ (mas yr$^{-1}$) \\
\textit{pmdec}	&	$pmdec$, PM in DEC (mas yr$^{-1}$)	\\
\textit{pmdec\textunderscore error}	& Error in $pmdec$ (mas yr$^{-1}$)	\\
\textit{ruwe} &	RUWE renormalized unit weight error \\
\textit{phot\textunderscore g\textunderscore mean\textunderscore mag}	&	 \textit{G} photometric band (mag) \\
\textit{phot\textunderscore bp\textunderscore mean\textunderscore mag}	&	  $G_{BP}$ photometric band (mag) \\
\textit{phot\textunderscore rp\textunderscore mean\textunderscore mag}	&	  $G_{RP}$ photometric band (mag) \\
\textit{phot\textunderscore g\textunderscore mean\textunderscore mag\textunderscore error} & Error in \textit{G} photometric band (mag) \\
\textit{phot\textunderscore bp\textunderscore mean\textunderscore mag\textunderscore error} &	  Error in $G_{BP}$ photometric band (mag) \\
\textit{phot\textunderscore rp\textunderscore mean\textunderscore mag\textunderscore error}	&	  Error in $G_{RP}$ photometric band (mag) \\
\textit{bp\textunderscore rp} & Obtained \textit{Gaia} $G_{bp-rp}$ colour index (mag)	\\
\textit{Mg}	& Obtained absolute magnitude $M_{G}$ (mag) \\
\textit{Prot} & $P_{rot}$ measurements, when available (d) \\
 \noalign{\smallskip}
 \hline
 \end{tabular}  
\newline
 \centering
  \end{center}
\end{table*}

      \begin{table*} [h]
 \begin{center}
 \caption[Tables of Sect.~\ref{analysis}: Membership analysis and candidate selections]{\textbf{Membership analysis and candidate selections} (see the exemplified Table~\ref{table5} for cluster NGC~2516).}
 \label{selection_analysis}  
 \begin{tabular}{c c} 
\hline
\hline
\noalign{\smallskip}
 Column name & Description \\ [1 ex] 
\hline
\noalign{\smallskip}
CNAME	&	GES object name from coordinates	\\
\textit{RV\textunderscore mem}	&	Membership: $RV$  \\
\textit{PM\textunderscore mem}	&	Membership: Proper motions \\
\textit{Parallax\textunderscore mem}	&	Membership: Parallax \\
\textit{CMD\textunderscore mem}	&	Membership: CMD \\
\textit{logg\textunderscore mem}	&	Membership: $log\textit{g}$ (intermediate-age and old clusters) \\
\textit{gamma\textunderscore mem}	&	Membership: $\gamma$ index (young clusters) \\
\textit{Met\textunderscore mem}	&	Membership: [Fe/H] metallicity \\
\textit{Li\textunderscore mem}	&	Membership: $EW$(Li) \\
\textit{Cantat\textunderscore Gaudin\textunderscore 2018} & Members from \cite{cantatgaudin_gaia}$^a$ \\
\textit{Randich\textunderscore 2018} & Members from \cite{randich_gaia}\tablefootmark{a} \\
\textit{Jackson\textunderscore 2021\textunderscore MEM3D} & Members from \cite{jackson2021}\tablefootmark{a}, \tablefootmark{b} \\
\textit{Jackson\textunderscore 2021\textunderscore MEMQG} & Members from \cite{jackson2021}\tablefootmark{a}, \tablefootmark{c}  \\
\textit{Final}	& Final candidate members\tablefootmark{d} \\
\textit{Particular\textunderscore cases}	& Final column listing particular cases\tablefootmark{e}  \\
 \noalign{\smallskip}
 \hline
 \end{tabular}  
   \tablefoot{
\tablefoottext{a}{Additional columns, whenever possible, listing candidates according to relevant \textit{Gaia} studies from the literature.}
\tablefoottext{b}{\cite{jackson2021} --- MEM3D refers to the membership using the full data set.}
\tablefoottext{c}{\cite{jackson2021} --- MEMQC refers to the probability computed using data set filtered to remove targets with suspect \textit{Gaia} data.}
\tablefoottext{d}{`Y'=members; `n'=Non-members.}
\tablefoottext{e}{Particular cases include the Li-rich giant outliers (listed as `Li-rich G'), strong accretors (`Strong accretor'), and the SB1 and SB2 binary stars (`SB1' and `SB2') listed by GES iDR6 and \cite{merle, merle2}.}
}
 \end{center}
\end{table*} 


\begin{landscape}

\centering

\begin{longtable}{ccccccccc}
\caption{\label{NGC2516} NGC~2516~GES~parameters (I)\tablefootmark{a}.}\\
    \hline
    \hline
    \noalign{\smallskip}
CNAME	& RA & DEC &  $RV$  & $v$sin$i$  & $T_{\rm eff}$  & $\log g$ &  $\gamma$ & [Fe/H]  \\
	& (hrs) & (deg) & (km s$^{-1}$) & (km s$^{-1}$) &  (K) & (dex) & (dex)  \\
\noalign{\smallskip}
\hline
\noalign{\smallskip}
\endfirsthead
\caption{continued.}\\
    \hline
    \hline
    \noalign{\smallskip}
CNAME	& RA & DEC &  $RV$  & $v$sin$i$  & $T_{\rm eff}$  & $\log g$ &  $\gamma$ & [Fe/H]  \\
	& (hrs) & (deg) & (km s$^{-1}$) & (km s$^{-1}$) &  (K) & (dex) & (dex)  \\
\noalign{\smallskip}
\hline
\noalign{\smallskip}
\endhead
\noalign{\smallskip}
\hline
\endfoot
07515457-6047568	&	117.98	&	-60.80	&	6.8	$\pm$	0.4	&		11.1	$\pm$	1.2	&	4369	$\pm$	76	&	4.90	$\pm$	0.18	&	0.920	$\pm$	0.009	&	0.00	$\pm$	0.09  \\
07515966-6047220	&	118.00	&	-60.79	&	25.8	$\pm$	0.3	&	$<$	7.0			&	4233	$\pm$	78	&	4.69	$\pm$	0.18	&	0.907	$\pm$	0.005	&	0.06	$\pm$	0.10  \\
07520129-6043233	&	118.01	&	-60.72	&	14.5	$\pm$	0.8	&		\dots			&	\dots			&	\dots			&	0.830	$\pm$	0.023	&	\dots	\\
07520389-6050116	&	118.02	&	-60.84	&	25.5	$\pm$	0.3	&	$<$	7.0			&	4256	$\pm$	76	&	4.90	$\pm$	0.18	&	0.892	$\pm$	0.005	&	0.05	$\pm$	0.09 \\
07521002-6044245	&	118.04	&	-60.74	&	25.5	$\pm$	0.3	&	$<$	7.0			&	4153	$\pm$	76	&	4.74	$\pm$	0.18	&	0.893	$\pm$	0.005	&	0.00	$\pm$	0.09 \\
07521382-6047151	&	118.06	&	-60.79	&	24.2	$\pm$	0.3	&	$<$	7.0			&	3993	$\pm$	77	&	4.75	$\pm$	0.18	&	0.861	$\pm$	0.007	&	-0.02	$\pm$	0.10  \\
07521911-6039241	&	118.08	&	-60.66	&	-7.4	$\pm$	0.5	&		14.6	$\pm$	2.5	&	3189	$\pm$	96	&	4.95	$\pm$	0.23	&	0.824	$\pm$	0.015	&	-0.38	$\pm$	0.09  \\
07522464-6043006	&	118.10	&	-60.72	&	-4.1	$\pm$	0.6	&		15.5	$\pm$	2.6	&	3845	$\pm$	77	&	4.56	$\pm$	0.18	&	0.876	$\pm$	0.014	&	-0.14	$\pm$	0.09  \\
07523060-6049134	&	118.13	&	-60.82	&	23.4	$\pm$	0.4	&	$<$	7.0			&	3444	$\pm$	108	&	4.88	$\pm$	0.24	&	0.812	$\pm$	0.012	&	-0.29	$\pm$	0.09  \\
07523629-6046405	&	118.15	&	-60.78	&	31.2	$\pm$	0.5	&	$<$	7.0			&	3883	$\pm$	107	&	4.85	$\pm$	0.24	&	0.815	$\pm$	0.013	&	-0.08	$\pm$	0.08  \\
07524604-6039537	&	118.19	&	-60.66	&	23.7	$\pm$	0.3	&	$<$	7.0			&	4141	$\pm$	76	&	4.76	$\pm$	0.18	&	0.888	$\pm$	0.005	&	0.00	$\pm$	0.09  \\
07525994-6053288	&	118.25	&	-60.89	&	23.4	$\pm$	0.3	&		10.0	$\pm$	1.1	&	4053	$\pm$	76	&	4.92	$\pm$	0.18	&	0.844	$\pm$	0.008	&	-0.03	$\pm$	0.10  \\
07530001-6042137	&	118.25	&	-60.70	&	5.7	$\pm$	0.4	&		9.0	$\pm$	1.2	&	3921	$\pm$	77	&	4.69	$\pm$	0.18	&	0.856	$\pm$	0.009	&	-0.01	$\pm$	0.10  \\
07530057-6048094	&	118.25	&	-60.80	&	24.1	$\pm$	0.3	&		9.0	$\pm$	1.0	&	3913	$\pm$	75	&	4.80	$\pm$	0.18	&	0.838	$\pm$	0.007	&	-0.08	$\pm$	0.09  \\
07530259-6050259	&	118.26	&	-60.84	&	28.4	$\pm$	1.0	&		84.3	$\pm$	1.3	&	\dots			&	\dots			&	0.962	$\pm$	0.004	&	\dots		\\
07531031-6046400	&	118.29	&	-60.78	&	23.3	$\pm$	0.4	&		13.3	$\pm$	1.2	&	3623	$\pm$	110	&	4.87	$\pm$	0.23	&	0.830	$\pm$	0.012	&	-0.27	$\pm$	0.09  \\
07531177-6041390	&	118.30	&	-60.69	&	23.4	$\pm$	0.4	&		8.4	$\pm$	1.3	&	3841	$\pm$	108	&	4.84	$\pm$	0.23	&	0.836	$\pm$	0.010	&	-0.11	$\pm$	0.09  \\
07531326-6043422	&	118.31	&	-60.73	&	3.6	$\pm$	0.3	&	$<$	7.0			&	3972	$\pm$	77	&	4.79	$\pm$	0.18	&	0.858	$\pm$	0.007	&	0.01	$\pm$	0.09  \\
07532107-6058131	&	118.34	&	-60.97	&	22.4	$\pm$	0.5	&		10.5	$\pm$	1.1	&	3879	$\pm$	108	&	4.85	$\pm$	0.23	&	0.838	$\pm$	0.014	&	-0.08	$\pm$	0.09  \\
07532163-6102129	&	118.34	&	-61.04	&	23.5	$\pm$	0.4	&		9.3	$\pm$	1.6	&	3620	$\pm$	108	&	4.84	$\pm$	0.23	&	0.826	$\pm$	0.011	&	-0.22	$\pm$	0.09 \\
\dots	&	\dots	&	\dots	&	\dots	&	\dots	&	\dots	&	\dots	&	\dots	&	\dots\\
\dots	&	\dots	&	\dots	&	\dots	&	\dots	&	\dots	&	\dots	&	\dots	&	\dots\\
\dots	&	\dots	&	\dots	&	\dots	&	\dots	&	\dots	&	\dots	&	\dots	&	\dots\\

\label{table1}
\end{longtable} 
\tablefoot{
\tablefoottext{a}{This is a truncated table showing the first 20 rows.}
}
\end{landscape}

\begin{landscape}

\centering
\setlength\LTleft{-1.2in}  
\setlength\LTright{-1in} 

\begin{longtable}{ccccccccc}
\caption[]{\label{NGC2516} NGC~2516~GES~parameters (II)\tablefootmark{a}.}\\
\hline
\hline
\noalign{\smallskip}
 CNAME & $EW$(Li)  & $EW$(Li)  &  $EW$(Li) &  $EW$(Li)  & \textit{A}(Li) & H$\alpha$ & H$\alpha$10\% & H$\alpha$ \\
	 & (blends-corrected) (m$\r{A}$) &  error flag  & (veiling-corrected) (m$\r{A}$)	 &  error flag $^c$  & (dex) & (accretion) ($\r{A}$) & (km s$^{-1}$) & (activity) ($\r{A}$) \\
\noalign{\smallskip}
\hline
\noalign{\smallskip}
\endfirsthead
\caption[]{continued.}\\
\hline
\hline
\noalign{\smallskip}
 CNAME & $EW$(Li)  & $EW$(Li)  &  $EW$(Li) &  $EW$(Li)  & \textit{A}(Li) & H$\alpha$ & H$\alpha$10\% & H$\alpha$ \\
	 & (blends-corrected) (m$\r{A}$) &  error flag  & (veiling-corrected) (m$\r{A}$)	 &  error flag $^c$  & (dex) & (accretion) ($\r{A}$) & (km s$^{-1}$) & (activity) ($\r{A}$) \\
\noalign{\smallskip}
\hline
\noalign{\smallskip}
\endhead
\noalign{\smallskip}
\hline
\endfoot
07515457-6047568	&	$<$	22			&	3	&	$<$	22			&	3	&	$<$	0.2			&	\dots			&	\dots			&	\dots	 \\	
07515966-6047220	&		\dots			&	\dots	&		75	$\pm$	5	&	0	&		0.6	$\pm$	0.1	&	0.82	$\pm$	0.07	&	94.81	$\pm$	1.79	&	1.78	$\pm$	0.14 \\
07520129-6043233	&		\dots			&	\dots	&		108	$\pm$	24	&	0	&		\dots			&	\dots			&	\dots			&	\dots \\		
07520389-6050116	&	$<$	13			&	3	&		13	$\pm$	6	&	0	&	$<$	-0.4			&	\dots			&	\dots			&	0.78	$\pm$	0.09 \\
07521002-6044245	&		\dots			&	\dots	&		39	$\pm$	6	&	0	&		-0.1	$\pm$	0.3	&	0.16	$\pm$	0.07	&	106.61	$\pm$	6.58	&	0.92	$\pm$	0.09 \\
07521382-6047151	&		\dots			&	\dots	&		86	$\pm$	7	&	0	&		0.4	$\pm$	0.2	&	0.19	$\pm$	0.11	&	119.19	$\pm$	9.95	&	1.08	$\pm$	0.14  \\
07521911-6039241	&		\dots			&	\dots	&		127	$\pm$	15	&	0	&	$<$	-1.0			&	\dots			&	\dots			&	\dots  \\		
07522464-6043006	&		\dots			&	\dots	&		84	$\pm$	15	&	0	&		0.0	$\pm$	0.3	&	0.42	$\pm$	0.32	&	\dots			&	1.71	$\pm$	0.35  \\
07523060-6049134	&		\dots			&	\dots	&		128	$\pm$	13	&	0	&		-0.7	$\pm$	0.5	&	0.76	$\pm$	0.14	&	\dots			&	1.61	$\pm$	0.16  \\
07523629-6046405	&		\dots			&	\dots	&		117	$\pm$	14	&	0	&		0.4	$\pm$	0.3	&	\dots			&	\dots			&	0.18	$\pm$	0.16  \\
07524604-6039537	&		\dots			&	\dots	&		35	$\pm$	6	&	0	&		-0.3	$\pm$	0.4	&	0.25	$\pm$	0.07	&	96.33	$\pm$	2.59	&	1.22	$\pm$	0.12  \\
07525994-6053288	&		\dots			&	\dots	&		70	$\pm$	8	&	0	&		0.4	$\pm$	0.2	&	0.26	$\pm$	0.09	&	114.25	$\pm$	33.15	&	1.12	$\pm$	0.16  \\
07530001-6042137	&		\dots			&	\dots	&		80	$\pm$	9	&	0	&		0.2	$\pm$	0.2	&	\dots			&	\dots			&	0.09	$\pm$	0.07  \\
07530057-6048094	&		\dots			&	\dots	&		57	$\pm$	8	&	0	&		-0.2	$\pm$	0.3	&	\dots			&	\dots			&	0.49	$\pm$	0.08  \\
07530259-6050259	&		\dots			&	\dots	&	$<$	9			&	3	&		\dots			&	0.45	$\pm$	0.11	&	241.81	$\pm$	37.44	&	1.51	$\pm$	0.13  \\
07531031-6046400	&		\dots			&	\dots	&		98	$\pm$	13	&	0	&		-0.3	$\pm$	0.5	&	1.47	$\pm$	0.19	&	97.84	$\pm$	3.38	&	2.59	$\pm$	0.21  \\
07531177-6041390	&		\dots			&	\dots	&		126	$\pm$	10	&	0	&		0.5	$\pm$	0.2	&	0.22	$\pm$	0.11	&	92.52	$\pm$	2.90	&	1.13	$\pm$	0.16  \\
07531326-6043422	&		\dots			&	\dots	&		74	$\pm$	7	&	0	&		0.2	$\pm$	0.2	&	\dots			&	\dots			&	0.08	$\pm$	0.07 \\
07532107-6058131	&		\dots			&	\dots	&		129	$\pm$	14	&	0	&		0.6	$\pm$	0.2	&	1.99	$\pm$	0.27	&	132.89	$\pm$	10.14	&	3.46	$\pm$	0.35 \\
07532163-6102129	&		\dots			&	\dots	&		101	$\pm$	12	&	0	&		-0.5	$\pm$	0.6	&	2.14	$\pm$	0.19	&	119.94	$\pm$	6.70	&	3.28	$\pm$	0.36 \\
\dots	&	\dots	&	\dots	&	\dots	&	\dots	&	\dots	&	\dots	&	\dots	&	\dots\\
\dots	&	\dots	&	\dots	&	\dots	&	\dots	&	\dots	&	\dots	&	\dots	&	\dots\\
\dots	&	\dots	&	\dots	&	\dots	&	\dots	&	\dots	&	\dots	&	\dots	&	\dots\\
\label{table2}
\end{longtable} 
\centering
\tablefoot{
\tablefoottext{a}{This is a truncated table showing the first 20 rows.}
}
\end{landscape}

\begin{landscape}
\setlength\LTleft{-1in}  
\setlength\LTright{0.8in} 
\centering

\begin{longtable}{cccccccccc}
\caption[NGC~2516~\textit{Gaia}~parameters]{\label{NGC2516} NGC~2516~\textit{Gaia}~parameters\tablefootmark{a}.}\\
\hline
\hline
\noalign{\smallskip}
CNAME	& $\pi$ & $pmra$ &  $pmdec$ & RUWE & $G$  & $G_{BP}$ &  $G_{RP}$ & $G_{BP-RP}$ &  $M_{G}$ \\
	& (mas) &  (mas~yr$^{-1}$) &  (mas~yr$^{-1}$) & error &  (mag) & (mag) & (mag) & (mag) & (mag)  \\
\noalign{\smallskip}
\hline
\noalign{\smallskip}
\endfirsthead
\caption[]{continued.}\\
\hline
\hline
\noalign{\smallskip}
CNAME	& $\pi$ & $pmra$ &  $pmdec$ & RUWE & $G$  & $G_{BP}$ &  $G_{RP}$ & $G_{BP-RP}$ &  $M_{G}$ \\
	& (mas) &  (mas~yr$^{-1}$) &  (mas~yr$^{-1}$) & error &  (mag) & (mag) & (mag) & (mag) & (mag) \\
\noalign{\smallskip}
\hline
\noalign{\smallskip}
\endhead
\noalign{\smallskip}
\hline
\endfoot
07515457-6047568	&	2.560	$\pm$	0.027	&	-5.73	$\pm$	0.04	&	10.74	$\pm$	0.03	&	1.0	&	15.752	$\pm$	0.003	&	16.573	$\pm$	0.004	&	14.868	$\pm$	0.004	&	1.71	&	7.79 \\
07515966-6047220	&	2.435	$\pm$	0.032	&	-3.93	$\pm$	0.04	&	11.10	$\pm$	0.04	&	1.0	&	15.965	$\pm$	0.003	&	16.848	$\pm$	0.006	&	15.037	$\pm$	0.004	&	1.81	&	7.90 \\
07520129-6043233	&	2.111	$\pm$	0.061	&	-24.60	$\pm$	0.09	&	28.61	$\pm$	0.08	&	1.0	&	17.420	$\pm$	0.003	&	18.653	$\pm$	0.017	&	16.347	$\pm$	0.006	&	2.31	&	9.04 \\
07520389-6050116	&	2.462	$\pm$	0.028	&	-4.19	$\pm$	0.04	&	11.49	$\pm$	0.03	&	1.0	&	15.928	$\pm$	0.003	&	16.807	$\pm$	0.005	&	15.012	$\pm$	0.004	&	1.79	&	7.88 \\
07521002-6044245	&	2.403	$\pm$	0.032	&	-4.83	$\pm$	0.04	&	11.49	$\pm$	0.04	&	1.1	&	16.087	$\pm$	0.003	&	16.998	$\pm$	0.006	&	15.149	$\pm$	0.004	&	1.85	&	7.99 \\
07521382-6047151	&	2.457	$\pm$	0.042	&	-4.62	$\pm$	0.06	&	10.61	$\pm$	0.05	&	1.0	&	16.604	$\pm$	0.003	&	17.642	$\pm$	0.008	&	15.599	$\pm$	0.005	&	2.04	&	8.56 \\
07521911-6039241	&	3.613	$\pm$	0.073	&	-16.21	$\pm$	0.11	&	2.75	$\pm$	0.09	&	1.0	&	17.731	$\pm$	0.003	&	19.182	$\pm$	0.020	&	16.564	$\pm$	0.006	&	2.62	&	10.52 \\
07522464-6043006	&	3.367	$\pm$	0.357	&	-1.59	$\pm$	0.50	&	0.76	$\pm$	0.45	&	6.0	&	17.814	$\pm$	0.003	&	18.607	$\pm$	0.016	&	16.330	$\pm$	0.006	&	2.28	&	10.45 \\
07523060-6049134	&	2.401	$\pm$	0.068	&	-4.60	$\pm$	0.10	&	11.72	$\pm$	0.08	&	1.0	&	17.514	$\pm$	0.003	&	18.877	$\pm$	0.024	&	16.376	$\pm$	0.006	&	2.50	&	9.42 \\
07523629-6046405	&	1.731	$\pm$	0.063	&	-3.34	$\pm$	0.09	&	-0.66	$\pm$	0.10	&	0.9	&	17.441	$\pm$	0.003	&	18.664	$\pm$	0.022	&	16.370	$\pm$	0.006	&	2.29	&	8.63 \\
07524604-6039537	&	2.443	$\pm$	0.030	&	-3.98	$\pm$	0.04	&	10.94	$\pm$	0.04	&	1.0	&	16.127	$\pm$	0.003	&	17.047	$\pm$	0.007	&	15.183	$\pm$	0.005	&	1.86	&	8.07 \\
07525994-6053288	&	2.500	$\pm$	0.045	&	-4.32	$\pm$	0.06	&	11.44	$\pm$	0.06	&	1.0	&	16.696	$\pm$	0.003	&	17.794	$\pm$	0.009	&	15.666	$\pm$	0.005	&	2.13	&	8.69 \\
07530001-6042137	&	2.030	$\pm$	0.043	&	-14.78	$\pm$	0.06	&	13.56	$\pm$	0.06	&	1.1	&	16.685	$\pm$	0.003	&	17.722	$\pm$	0.009	&	15.687	$\pm$	0.004	&	2.03	&	8.22 \\
07530057-6048094	&	2.441	$\pm$	0.048	&	-4.25	$\pm$	0.07	&	10.88	$\pm$	0.07	&	1.1	&	16.779	$\pm$	0.003	&	17.899	$\pm$	0.008	&	15.711	$\pm$	0.005	&	2.19	&	8.72 \\
07530259-6050259	&	2.435	$\pm$	0.028	&	-4.12	$\pm$	0.04	&	10.96	$\pm$	0.04	&	1.0	&	15.852	$\pm$	0.003	&	16.709	$\pm$	0.006	&	14.935	$\pm$	0.006	&	1.77	&	7.78 \\
07531031-6046400	&	2.531	$\pm$	0.060	&	-3.41	$\pm$	0.09	&	12.14	$\pm$	0.09	&	0.9	&	17.420	$\pm$	0.003	&	18.765	$\pm$	0.022	&	16.297	$\pm$	0.006	&	2.47	&	9.44 \\
07531177-6041390	&	2.328	$\pm$	0.049	&	-4.75	$\pm$	0.07	&	12.04	$\pm$	0.07	&	1.0	&	17.006	$\pm$	0.003	&	18.124	$\pm$	0.012	&	15.956	$\pm$	0.006	&	2.17	&	8.84 \\
07531326-6043422	&	2.142	$\pm$	0.037	&	-2.38	$\pm$	0.05	&	7.98	$\pm$	0.06	&	1.0	&	16.508	$\pm$	0.003	&	17.506	$\pm$	0.008	&	15.521	$\pm$	0.004	&	1.99	&	8.16 \\
07532107-6058131	&	2.412	$\pm$	0.048	&	-4.36	$\pm$	0.07	&	11.44	$\pm$	0.06	&	1.0	&	16.858	$\pm$	0.003	&	18.016	$\pm$	0.014	&	15.795	$\pm$	0.005	&	2.22	&	8.77 \\
07532163-6102129	&	2.392	$\pm$	0.053	&	-4.02	$\pm$	0.07	&	11.32	$\pm$	0.08	&	1.0	&	17.175	$\pm$	0.003	&	18.430	$\pm$	0.017	&	16.080	$\pm$	0.006	&	2.35	&	9.07 \\
\dots	&	\dots	&	\dots	&	\dots	&	\dots	&	\dots	&	\dots	&	\dots	&	\dots &	\dots \\
\dots	&	\dots	&	\dots	&	\dots	&	\dots	&	\dots	&	\dots	&	\dots	&	\dots &	\dots \\
\dots	&	\dots	&	\dots	&	\dots	&	\dots	&	\dots	&	\dots	&	\dots	&	\dots &	\dots \\
\label{table3}
\end{longtable} 
\centering
\tablefoot{
\tablefoottext{a}{This is a truncated table showing the first 20 rows.}
}
\end{landscape}

\begin{landscape}
\centering

\begin{longtable}{ccccc}
\caption[NGC~2516~$P_{rot}$~measurements]{\label{NGC2516} NGC~2516~$P_{rot}$~measurements\tablefootmark{a}.}\\
\hline
\hline
\noalign{\smallskip}
CNAME	& $P_{rot}$ (d) & $P_{rot}$ (d) &  $P_{rot}$ (d) & $P_{rot}$ (d) \\
	& \cite{irwin} & \cite{wright2011} & \cite{jackson} & \cite{fritzewski2020} \\
\noalign{\smallskip}
\hline
\noalign{\smallskip}
\endfirsthead
\caption[]{continued.}\\
\hline
\hline
\noalign{\smallskip}
CNAME	& $P_{rot}$ & $P_{rot}$ &  $P_{rot}$ & $P_{rot}$ \\
	& \cite{irwin} & \cite{wright2011} & \cite{jackson} & \cite{fritzewski2020} \\
\noalign{\smallskip}
\hline
\noalign{\smallskip}
\endhead
\noalign{\smallskip}
\hline
\endfoot
07515457-6047568	&	\dots	&	\dots	&	\dots	&	\dots \\
07515966-6047220	&	\dots	&	\dots	&	\dots	&	\dots \\
07520129-6043233	&	\dots	&	\dots	&	\dots	&	\dots \\
07520389-6050116	&	\dots	&	\dots	&	\dots	&	\dots \\
07521002-6044245	&	\dots	&	\dots	&	\dots	&	\dots \\
07521382-6047151	&	\dots	&	\dots	&	\dots	&	\dots \\
07521911-6039241	&	\dots	&	\dots	&	\dots	&	\dots \\
07522464-6043006	&	\dots	&	\dots	&	\dots	&	\dots \\
07523060-6049134	&	\dots	&	\dots	&	\dots	&	\dots \\
07523629-6046405	&	\dots	&	\dots	&	\dots	&	\dots \\
07524604-6039537	&	\dots	&	\dots	&	\dots	&	\dots \\
07525994-6053288	&	5.66	&	\dots	&	5.66	&	5.66 \\
07530001-6042137	&	\dots	&	\dots	&	\dots	&	\dots \\
07530057-6048094	&	11.61	&	\dots	&	11.61	&	11.61 \\
07530259-6050259	&	\dots	&	\dots	&	4.58	&	\dots \\
07531031-6046400	&	\dots	&	\dots	&	2.52	&	\dots \\
07531177-6041390	&	4.58	&	\dots	&	4.58	&	4.58 \\
07531326-6043422	&	\dots	&	\dots	&	\dots	&	\dots \\
07532107-6058131	&	2.52	&	\dots	&	2.52	&	2.52 \\
07532163-6102129	&	2.34	&	\dots	&	2.34	&	2.34 \\
\dots	& \dots 	&	\dots	&	\dots	&	\dots \\
\dots	& \dots 	&	\dots	&	\dots	&	\dots \\
\dots	& \dots 	&	\dots	&	\dots	&	\dots \\
\label{table4}
\end{longtable} 
\centering
\tablefoot{
\tablefoottext{a}{This is a truncated table showing the first 20 rows.}
}
\end{landscape}

\begin{landscape}
\setlength\LTleft{-1in}  
\setlength\LTright{1in} 
\centering

\begin{longtable}{cccccccccccccc}
\caption[NGC~2516~membership analysis and selection]{\label{NGC2516} NGC~2516~membership analysis and selection\tablefootmark{a}.}\\
\hline
\hline
\noalign{\smallskip}
CNAME	&  \multicolumn{7}{c}{Membership} & Cantat-Gaudin & Randich & \multicolumn{2}{c}{Jackson et al (2021)} & Final & Particular \\
& $\gamma$ & $RV$ & PMs & $\pi$ & CMD & [Fe/H] & $EW$(Li) & et al (2018) & et al (2018) &  MEM3D & MEMQG  & members & cases \\
\noalign{\smallskip}
\hline
\noalign{\smallskip}
\endfirsthead
\caption[]{continued.}\\
\hline
\hline
\noalign{\smallskip}
CNAME	&  \multicolumn{7}{c}{Membership} & Cantat-Gaudin & Randich & \multicolumn{2}{c}{Jackson et al (2021)} & Final & Particular \\
& $RV$ & PMs & $\pi$ & CMD & log$g$ & [Fe/H] & $EW$(Li) & et al (2018) & et al (2018) &  MEM3D & MEMQG  & members & cases \\
\noalign{\smallskip}
\hline
\noalign{\smallskip}
\endhead
\noalign{\smallskip}
\hline
\endfoot
07515457-6047568	&	N	&	\dots	&	\dots	&	\dots	&	\dots	&	\dots	&	\dots	&	\dots	&	N	&	0.5901	&	0.5542	&	n	&	\dots	\\
07515966-6047220	&	Y	&	Y	&	Y	&	Y	&	Y	&	Y	&	Y	&	Y	&	Y	&	0.9991	&	0.9991	&	Y	&	\dots	\\
07520129-6043233	&	N	&	\dots	&	\dots	&	\dots	&	\dots	&	\dots	&	\dots	&	\dots	&	Y	&	0	&	0	&	\dots	&	\dots	\\
07520389-6050116	&	Y	&	Y	&	Y	&	Y	&	Y	&	Y	&	Y	&	Y	&	Y	&	0.9997	&	0.9997	&	Y	&	\dots	\\
07521002-6044245	&	Y	&	Y	&	Y	&	Y	&	Y	&	Y	&	Y	&	Y	&	Y	&	0.9998	&	0.9998	&	Y	&	\dots	\\
07521382-6047151	&	Y	&	Y	&	Y	&	Y	&	Y	&	Y	&	Y	&	N	&	Y	&	0.9999	&	0.9999	&	Y	&	\dots	\\
07521911-6039241	&	N	&	\dots	&	\dots	&	\dots	&	\dots	&	\dots	&	\dots	&	\dots	&	N	&	0	&	0	&	n	&	\dots	\\
07522464-6043006	&	N	&	\dots	&	\dots	&	\dots	&	\dots	&	\dots	&	\dots	&	\dots	&	N	&	0	&	-1	&	n	&	\dots	\\
07523060-6049134	&	Y	&	Y	&	Y	&	Y	&	Y	&	Y	&	Y	&	\dots	&	Y	&	0.9999	&	0.9999	&	Y	&	\dots	\\
07523629-6046405	&	N	&	\dots	&	\dots	&	\dots	&	\dots	&	\dots	&	\dots	&	\dots	&	Y	&	0	&	0	&	n	&	\dots	\\
07524604-6039537	&	Y	&	Y	&	Y	&	Y	&	Y	&	Y	&	Y	&	Y	&	Y	&	0.9999	&	0.9999	&	Y	&	\dots	\\
07525994-6053288	&	Y	&	Y	&	Y	&	Y	&	Y	&	Y	&	Y	&	\dots	&	Y	&	1	&	1	&	Y	&	\dots	\\
07530001-6042137	&	N	&	\dots	&	\dots	&	\dots	&	\dots	&	\dots	&	\dots	&	\dots	&	N	&	0	&	0	&	n	&	\dots	\\
07530057-6048094	&	Y	&	Y	&	Y	&	Y	&	Y	&	Y	&	Y	&	\dots	&	Y	&	0.9999	&	1	&	Y	&	\dots	\\
07530259-6050259	&	N	&	\dots	&	\dots	&	\dots	&	\dots	&	\dots	&	\dots	&	Y	&	Y	&	0.9967	&	0.9971	&	\dots	&	\dots	\\
07531031-6046400	&	Y	&	N	&	\dots	&	\dots	&	\dots	&	\dots	&	\dots	&	\dots	&	Y	&	0.99	&	0.9854	&	n	&	\dots	\\
07531177-6041390	&	Y	&	Y	&	Y	&	Y	&	Y	&	Y	&	Y	&	\dots	&	Y	&	0.9998	&	0.9999	&	Y	&	\dots	\\
07531326-6043422	&	N	&	\dots	&	\dots	&	\dots	&	\dots	&	\dots	&	\dots	&	\dots	&	N	&	0	&	0	&	n	&	\dots	\\
07532107-6058131	&	Y	&	Y	&	Y	&	Y	&	Y	&	Y	&	Y	&	\dots	&	Y	&	0.9999	&	0.9999	&	Y	&	\dots	\\
07532163-6102129	&	Y	&	Y	&	Y	&	Y	&	Y	&	Y	&	Y	&	Y	&	Y	&	0.9999	&	0.9999	&	Y	&	\dots	\\
\dots	&	\dots	&	\dots	&	\dots	&	\dots	&	\dots	&	\dots	&	\dots	&	\dots	&	\dots	&	\dots	&	\dots	&	\dots	&	\dots	\\
\dots	&	\dots	&	\dots	&	\dots	&	\dots	&	\dots	&	\dots	&	\dots	&	\dots	&	\dots	&	\dots	&	\dots	&	\dots	&	\dots	\\
\dots	&	\dots	&	\dots	&	\dots	&	\dots	&	\dots	&	\dots	&	\dots	&	\dots	&	\dots	&	\dots	&	\dots	&	\dots	&	\dots	\\

\label{table5}
\end{longtable} 
\centering
\tablefoot{
\tablefoottext{a}{This is a truncated table showing the first 20 rows.}
}
\end{landscape}

\label{ap1:journals}


\end{appendix}

\end{document}